\documentclass[a4paper, 11pt]{memoir}
\usepackage{color}
\usepackage[english]{babel} 
\usepackage{amsmath,amsfonts,amsthm,amssymb,gensymb} 
\usepackage{graphicx}
\usepackage{subfig}
\usepackage{hyperref}
\usepackage{lipsum}
\usepackage{geometry}
\usepackage{standalone}
\usepackage{appendix}
\usepackage[figurewithin=none]{caption}
\allowdisplaybreaks

\numberwithin{equation}{section} 
\numberwithin{figure}{section}
\numberwithin{table}{section} 
\makeatletter
\renewcommand\@memmain@floats{%
  \counterwithin{figure}{section}
  \counterwithin{table}{section}}
\makeatother

\newcommand{\horrule}[1]{\rule{\linewidth}{#1}}

\begin{document}


\frontmatter

\begin{center}
\textsc{\Large Universit\'e Paris Diderot}\\
{\Large{} \vspace{0.2cm}}

\par\end{center}{\Large \par}

\begin{center}
\begin{figure}[ht]
\centering
{\includegraphics[width=.08\columnwidth]{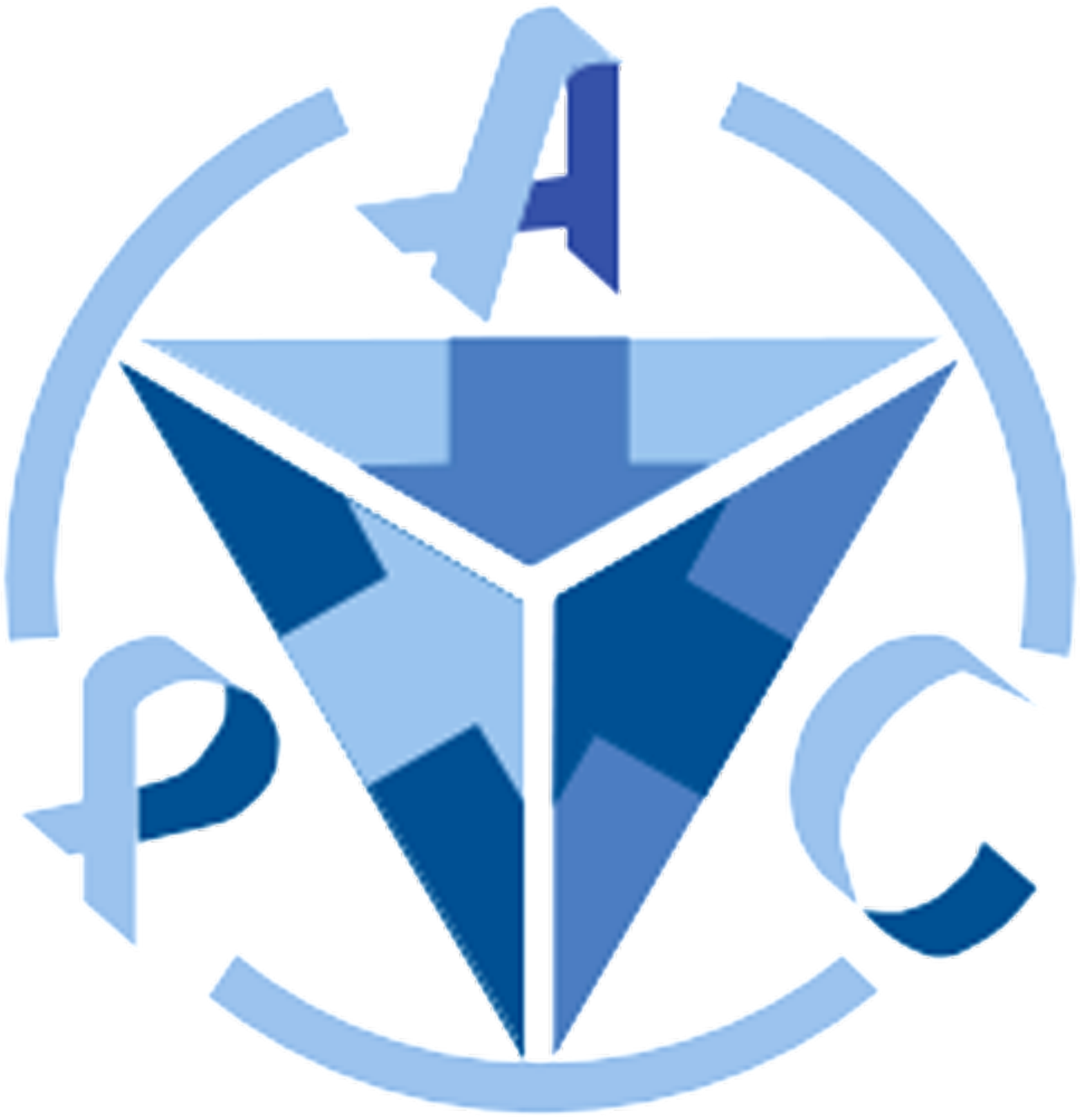}} \quad
{\includegraphics[width=.10\columnwidth]{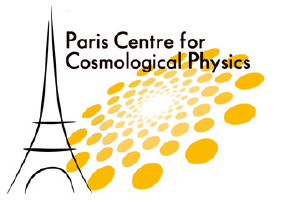}} \quad 
{\includegraphics[width=.15\columnwidth]{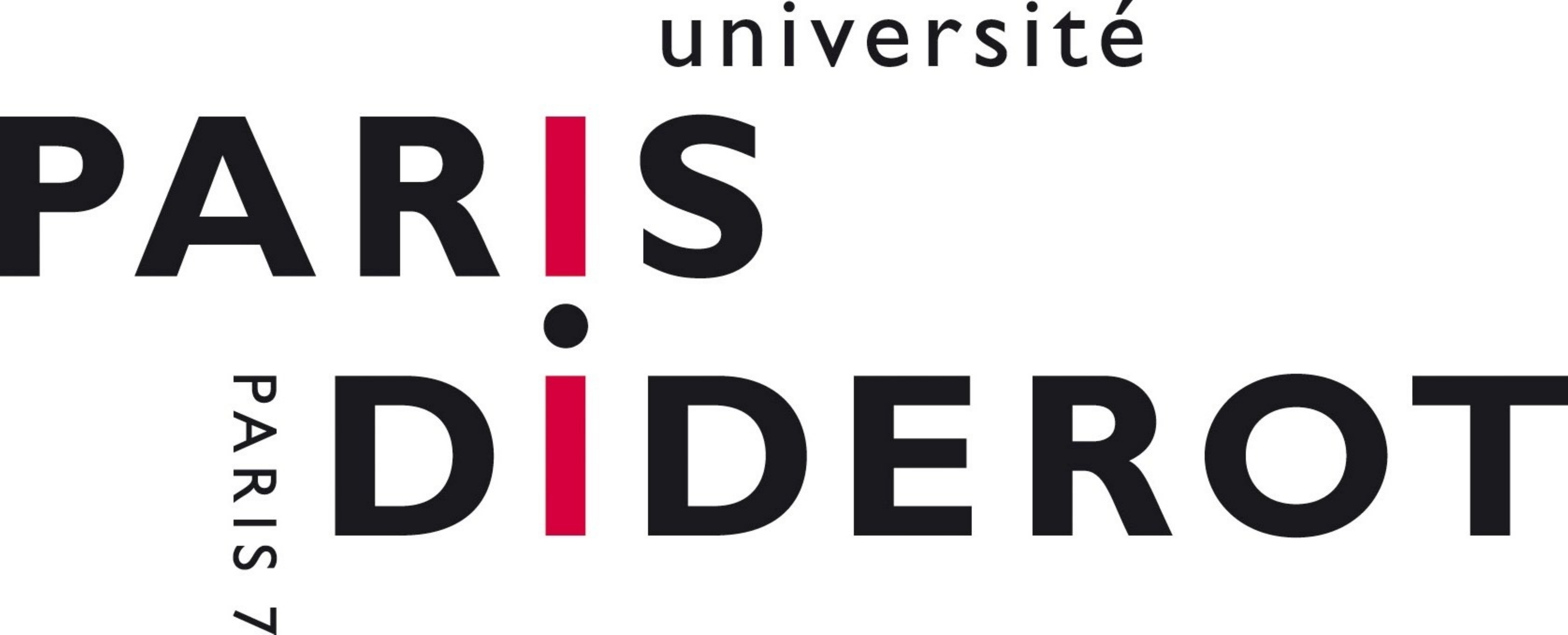}}
\end{figure}
\par\end{center}

\begin{center}
{\large \vspace{1cm}
}\textbf{\Large Classification of inflationary models \\ \vspace{0.2cm}
and constraints on fundamental physics.}{\large \vspace{0.2cm}
}
\par\end{center}{\large \par}

\begin{center}
{\Large 28 September 2016}\\
{\large \vspace{0.2cm}
 }\textsc{\LARGE Ph.D. Thesis}{\LARGE{} \vfill{}
}
\par\end{center}{\LARGE \par}

\begin{center}
{\Large Candidate}\\
{\large{} \vspace{0.07cm}
 }\textbf{\Large Mauro Pieroni}{\large }\\
{\large{} \vspace{-0cm}
 }\texttt{mauropieroni@gmail.com}\texttt{\large{} }
\par\end{center}{\large \par}

{\large \vspace{0.8cm}
}{\large \par}

\makebox[1\columnwidth]{%
{\large }%
\begin{minipage}[c]{0.45\linewidth}%
 \centering {\large Supervisors:}\\
 \vspace{0.5cm}
 \textbf{\large Prof. P. Bin\'etruy} \\
  \vspace{0.1cm}
 {\large Laboratoire APC}\\
  \vspace{0.3cm}
 \textbf{\large Dr. C. Rosset} \\
  \vspace{0.1cm}
 {\large Laboratoire APC}

\end{minipage}{\large{} \hfill{}}%
\begin{minipage}[c]{0.45\linewidth}%
 \centering {\large Referees:}\\
 \vspace{0.1cm}
 \textbf{\large Dr. C. Caprini} \\
 {\large IPhT/CEA}\\
 \vspace{0.3cm}
 \textbf{\large Dr. F. Vernizzi}\\
 {\large IPhT/CEA}\\
 \vspace{0.5cm}
 \centering {\large Other members of the committee:}\\
  \vspace{0.3cm}
 \textbf{\large Dr. F. Bouchet} \\
  \vspace{0.1cm}
 {\large Institut d'astrophysique de Paris}\\
 \vspace{0.3cm}
 \textbf{\large Prof. A. Davis}\\
  \vspace{0.1cm}
 {\large DAMPT, Cambridge University}
\end{minipage}%
}
\vspace{1.5cm}
\begin{center}
\textsc{\Large Accademic year 2015/2016}
\par
\end{center}

{\large 
\chapter*{Abstract.}
This work is focused on the study of early time cosmology and in particular on the study of inflation. We start our treatment with an introductory Chapter where we draw the main lines of standard cosmology. In this Chapter we present the standard Big Bang theory, we discuss the physics of CMB and we explain how its observations can be used to set constraints on cosmological models. \\

\noindent
The second Chapter of this work is dedicated to a general discussion of inflation. We start by presenting the reasons that led to its introduction and we explain why an early phase of exponential expansion would solve the so-called ``shortcomings'' of the standard Big Bang theory. We discuss the observables and the experimental constraints and we present a set of models that implement the simplest realization of inflation also known as slow-roll inflation. We conclude the chapter by presenting some possible generalizations of the minimal setup and by discussing the prospect of observing primordial gravitational wave (GW) produced during inflation. \\

\noindent
In Chapter~\ref{chapter:beta} we present the $\beta$-function formalism for inflation (introduced in~\cite{Binetruy:2014zya}). We start by presenting the reasons to define a classification of inflationary models and show that cosmological evolution of a scalar field in its potential can be described in terms of a renormalization group equation. We explain how this naturally leads to definition of a set of universality classes for inflationary models and we carry out the formulation of inflation in terms of the formalism. We present a series of examples are we conclude by discussing our results. \\

\noindent
Theoretical motivations that support the formulation of inflation in terms of the $\beta$-function formalism are presented in Chapter~\ref{chapter:holographic_universe}, where we discuss the possibility of applying holography to cosmology. The extension of the $\beta$-function formalism for inflation to models with non-standard kinetic terms and with non-minimal couplings are discussed in Chapter~\ref{chapter:generalized_models}. At the end of this Chapter we reproduce \emph{in extenso} the analysis of~\cite{Pieroni:2015cma}.\\

\noindent
In Chapter~\ref{chapter:pseudoscalar} we discuss the framing of inflation in the context of early time cosmology. The Chapter is focused on the study of models where the inflaton (which is considered to be a pseudo-scalar) is non-minimally coupled to some Abelian gauge fields that can be present during inflation. The analysis of the problem is carried out by using a characterization of inflationary models in terms of their asymptotic behavior (accordingly with the discussion of Chapter~\ref{chapter:beta} and of~\cite{Domcke:2016bkh}). A wide set of theoretical aspects and of observational consequences is discussed.

 {\large \par}}

{\large \newpage{}\tableofcontents{}}{\large \par}


\mainmatter
\setcounter{page}{1}  
\setcounter{secnumdepth}{3}  
\pagenumbering{arabic}
{\large 
\chapter*{Notation and Conventions.}
The Planck mass $M_P$ is defined as:
\begin{equation*}
	M_P \equiv \sqrt{\hbar c / G_N} \simeq 1.2 \times 10^{19} \,\text{GeV/c}^2 \ .
\end{equation*}
In the context of cosmology it is customary to use the reduced Planck mass $m_P$ which is defined as:
\begin{equation*}
 	m_P \equiv M_P/\sqrt{8 \pi} \simeq 2.4 \times 10^{18} \,\text{GeV/c}^2 \ .
 \end{equation*}
As usual in the context of theoretical physics we work in natural units that corresponds to set:
\begin{equation*}
 	c = \hbar = 1 \ .
 \end{equation*} 
The Planck length $l_P$ is defined as: 
\begin{equation*}
	l_P \equiv M_P^{-1} = \sqrt{G_N} \simeq 1.6 \times 10^{-35} \text{m} \ . 
\end{equation*}
In this work we use the length scale $\kappa^{-1}$, defined as:
\begin{equation*}
 	\kappa \equiv m_p^{-1} = \sqrt{8 \pi} l_P \simeq 5 l_P \simeq 8.0 \times 10^{-35} \text{m} \ .
 \end{equation*}  
Given a function $f(\vec{x},t)$ in a $d+1$-dimensional spacetime, we define $\tilde{f}(\vec{k},t)$, spatial Fourier transform of $f(\vec{x},t)$, as:
\begin{equation*}
  f(\vec{x},t) = \int \frac{\mathrm{d}^d \vec{k}}{(2 \pi)^{d/2}} e^{i \vec{k} \cdot \vec{x}}\tilde{f}(\vec{k},t) \ .
\end{equation*}

 {\large \par}}
{\large 
\chapter{Standard cosmology, CMB and Planck.}

\label{chapter:introduction}
\horrule{0.1pt} \\[0.5cm]

\begin{abstract} 
\noindent
  In this Chapter we present the Standard Model of cosmology also known as $\Lambda$CDM (for $\Lambda$ Cold Dark Matter) model. This is the simplest model based on General Relativity (GR) that gives a reasonable explanation to the basic properties of the observable Universe. The accidental discovery of the Cosmic Microwave Background (CMB) by Arno Penzias and Robert Wilson~\cite{Penzias:1965wn} provided an extremely powerful observable quantity to observe the Universe. After the pioneering COBE mission~\cite{Mather:1991pc,Smoot:1992td} has revealed the presence of small fluctuations around the black body spectrum of the CMB several high precision experiments have been realized in order to study these fluctuations. An extremely accurate measurement of the CMB spectrum and of its fluctuations has been recently provided by the Planck mission, helping to understand the physics of the early Universe.
\end{abstract}

\horrule{0.1pt} \\[0.5cm] 

\noindent
The understanding of the characteristics and of the evolution of the Universe has always been a main subject of study for mankind. Over the centuries several theories have been formulated, but it is only in 1915, when Einstein formulated the theory of General Relativity (GR), that Modern cosmology was actually born. The idea of applying the equations of GR to describe the Universe has led to several notable discoveries such as the revelation of extra-galactic objects. Observing the velocity of the structures outside of our galaxy in 1929, Hubble~\cite{Hubble:1929ig} has noticed a difference in the relative velocities of these objects. In particular, Hubble has shown that farther objects from our galaxy are receding with greater velocities with respect to closer objects. Such a striking observation has led to the conclusion that the Universe is expanding. This major result has thus led to the formulation of the Big Bang theory that predicts a hot and dense early Universe that expands and cools down. \\

\noindent The Standard Model of cosmology is a model that aims at describing the origin and evolution of the Universe. The formulation of this model is based both on GR and on the Standard Model of particle physics\cite{Glashow:1961tr,Weinberg:1967tq,salam,Higgs:1964pj,1964PhRvL..13..321E} that describes non-gravitational interactions. As we will explain in this Chapter, ideas coming from particle physics are useful in the definition of a background solution and in the study of perturbations around this background. One of the main successes of the Standard Model of cosmology that actually differentiates it from previous cosmological models is the possibility of giving quantitative predictions. In particular the $\Lambda$CDM model of cosmology gives reasonable explanation for:
\begin{itemize}
  \item The Hubble diagram that shows the expansion of the Universe.
  \item The abundance of light elements explained by the Big Bang Nucleosynthesis.
  \item The black body spectrum and the isotropy of the CMB.
\end{itemize} 
Once inflation\footnote{As inflation is a main topic of this work we postpone its treatment to Chapter~\ref{chapter:inflation} where it will be properly introduced and explained in detail.} is included, the $\Lambda$CDM is also capable of explaining the observations of the small perturbations in the CMB.\\

\noindent Since its accidental discovery by Arno Penzias and Robert Wilson~\cite{Penzias:1965wn} in 1965 the Cosmic Microwave Background (CMB) has been extensively studied. This thermal radiation provides an extremely powerful observable quantity to get information on the history of the observable Universe. Its homogeneity and isotropy are well explained by the Big Bang cosmology and precious information on early time cosmology can be extracted by studying its order $10^{-4} \div 10^{-5}$ inhomogeneities. The first detection of these inhomogeneities was realized by the COBE mission~\cite{Mather:1991pc,Smoot:1992td} and earned George Fitzgerald Smoot the Nobel Prize in Physics in 2006. This major discovery actually opened the era of modern observational cosmology that led to the realization of several CMB experiments. In this context it is worth mentioning the Wilkinson Microwave Anisotropy Probe (WMAP) as the second space-based CMB mission that gave an accurate measure of the high angular-scale CMB fluctuations~\cite{Spergel:2003cb,Spergel:2006hy,Komatsu:2008hk,Komatsu:2010fb,Hinshaw:2012aka}. Finally in May 2009 a third space-based mission called Planck was launched. Planck measurements give an extremely accurate mapping of the small angular-scale CMB fluctuations~\cite{Ade:2013sjv,Ade:2013zuv,Planck:2013jfk,Ade:2013ydc,Ade:2015xua,Ade:2015lrj}. These measurements have actually helped to get a better understanding of the structure and evolution of the Universe and of the mechanism driving the inflationary epoch. \\

\noindent In this Chapter we proceed as follows. We start by presenting the Cosmological principle and by introducing the Friedmann-Lema\^itre-Robertson-Walker (FLRW) metric. Once this quantity is introduced, we discuss the matter content of the observable Universe and we derive the Einstein Equations that govern its evolution. In Sec.~\ref{sec_introduction:energy_content_and_history} we discuss the energy content of Universe, the scaling behaviors for the species of energy and the history of the Universe. In Sec.~\ref{sec_introduction:CMB} we present the CMB, and we discuss its main properties. Finally, in Sec.~\ref{sec_introduction:experiments} the discussion is focused on CMB observations. In particular in this Section we give some details on the Planck mission and we explain how CMB measurements can be used to set constraints on the cosmological parameters.

\section{Cosmological principle and FLRW Universe.}
\label{sec_introduction:FLRW_metric}
The standard model of cosmology is based on the so called \emph{cosmological principle}, that can be formulated as:
\begin{quote}
``\emph{When observed on sufficiently large scale, the properties of the Universe are the same for all observers}.''
\end{quote}
A different formulation of the cosmological principle states that the Universe is homogeneous and isotropic on large scales. At the time when it was introduced, the cosmological principle was intended as an assumption to base the study of the Universe. Nowadays, direct observations can be used to test the homogeneity and isotropy of the Universe. In particular, we find that the Universe appears to be homogeneous and isotropic on scales\footnote{It can be interesting to compare this length scale with other typical length units that are commonly used in physics. $100\,$Mpc are approximatively equal to $3.26 \times 10^{8}\,$light-years that actually correspond to $3.09 \times10^{24}\,$m or equivalently to $1.9 \times 10^{59}\,l_P \simeq 3.8 \times 10^{58}\,\kappa$.} of order $100\,$Mpc. As it is supported by observational evidences, the cosmological principle should not be intended as a principle in the strict sense but more as an observational fact. The most general ansatz that solves Einstein Equations for a homogeneous and isotropic spacetime is the well known FLRW metric:
\begin{equation}
\label{eq_intro:FLRW}
\textrm{d}s^2 = g_{\mu\nu}\textrm{d}x^\mu \textrm{d}x^\nu = -\textrm{d}t^2 + a^2(t) \gamma_{ij} \textrm{d}x^i \textrm{d}x^j \ ,
\end{equation}
where $\gamma_{ij}$ is the spatial part of the metric, which can be expressed as:
 \begin{equation}
\label{eq_intro:FLRW_spatial_1}
\gamma_{ij} \textrm{d}x^i \textrm{d}x^j = \frac{\textrm{d}r^2}{1 - k r^2} + r^2 \left(\textrm{d}\theta^2 + \sin^2 \theta \textrm{d}\varphi^2\right) \ ,
\end{equation}
where the constant $k$ may take the three values $-1, 0 $ and $ +1$\footnote{In our convention $a$ is dimensionless and $r$ is dimensionful. As a consequence, to make $k r^2$ dimensionless, we can use $\kappa \equiv m_P^{-1}$.}. As we will see in the following, the constant $ k$ fixes the scalar curvature of the 3-dimensional surfaces at constant $t$. It is possible to show that the three values $k = -1, 0, +1$ correspond to an open, flat or closed space respectively. \\

\noindent 
In GR\footnote{The formal definitions of the typical quantities that appear in GR are given in Appendix~\ref{appendix_GR:General_equations}.} the evolution of a system is fixed by Einstein Equations :
\begin{equation}
  \label{eq_intro:general_EE}
  G_{\mu \nu} \equiv R_{\mu \nu} -\frac{1}{2}g_{\mu \nu} R = - \Lambda g_{\mu \nu} + 8 \pi G_N \ T_{\mu \nu} \ ,
\end{equation}
where $\Lambda$ is a cosmological constant term, $T_{\mu \nu}$ is the stress-energy tensor and $G_{\mu \nu}$ is the Einstein tensor. Given the Christoffel symbols, the Einstein tensor can be expressed in terms of the Ricci tensor and of the Ricci scalar. In particular it is possible to show that for the FRLW metric, the only non-zero components of the Christoffel symbols are:
\begin{equation}
\Gamma^{0}_{\ i j } = a \dot{a} \gamma_{ij} \ , \qquad  \Gamma^{i}_{\ 0 j} = \delta^{i}_{j} \frac{\dot{a}}{a} \ , \qquad \Gamma^i_{\ j k } = \Gamma^i_{\ j k }(\gamma) \ ,
\end{equation}
where $\Gamma^i_{\ j k }(\gamma)$ is used to denote the standard Christoffel symbols computed for the 3-dimensional metric $\gamma_{ij}$. The only non-zero components of the the Ricci tensor (defined accordingly with Eq.~\eqref{appendix_GR:Ricci_tensor}) are:
\begin{equation}
\label{eq_intro:Ricci_tensor}
R_{00} = -3 \frac{\ddot{a}}{a} \ , \qquad \qquad R_{ij} = a^2 \gamma_{ij} \left[ \frac{2 k}{a^2} + \left( \frac{\dot{a}}{a} \right)^2+ \frac{\ddot{a}}{a} \right] \ ,
\end{equation}
and thus the Ricci scalar reads:
\begin{equation}
\label{eq_intro:Ricci_scalar}
R = 6 \left[ \frac{   k}{a^2} + \left( \frac{\dot{a}}{a} \right)^2+ \frac{\ddot{a}}{a} \right] \ ,
\end{equation}
Notice that setting $a(t)$ to be a constant, the scalar curvature simply reads $R = 12 k/a^2$. As already anticipated, the constant $ k$ is thus directly related with the curvature of the 3-dimensional surfaces at constant $t$. \\

\noindent Finally we can use Eq.~\eqref{eq_intro:Ricci_tensor} and Eq.\eqref{eq_intro:Ricci_scalar} to compute the Einstein tensor. In particular it is possible to show that its only non-zero components are:
\begin{equation}
\label{eq_intro:Einstein_tensor}
G_{00} = 3\left[ \left( \frac{\dot{a}}{a} \right)^2 + \frac{ k}{a^2} \right] \ , \qquad G_{ij} = -\gamma_{ij} \left(  k + 2a \ddot{a} + \dot{a^2 }\right) \ .
\end{equation}
Notice that both $G_{00}$ and $G_{ij}$ have the dimension of the inverse of a length squared.  \\

\noindent To be able to write the Einstein equations we still need to specify the right hand side of Eq.~\eqref{eq_intro:general_EE}, \textit{i.e.} the energy content of the Universe. Following the assumption of homogeneity and isotropy, the Universe energy content can be expressed using the stress-energy tensor for a perfect fluid at rest and in thermodynamic equilibrium:
\begin{equation}
\label{eq_intro:energy_content_tensor}
T_{\mu \nu} = p g_{\mu \nu} + (p + \rho) U_{\mu} U_{\nu} \ , 
\end{equation}
where $U^{\mu}=(1,0,0,0)$ is the four velocity of the fluid and $p$ and $\rho$ are respectively its pressure and energy density. Notice that homogeneity and isotropy imply that these quantities may only depend on time. Moreover Eq.~\eqref{eq_intro:energy_content_tensor} directly implies:
\begin{equation}
\label{eq_intro:energy_content_rho_p}
T_{00} = \rho \ , \qquad \qquad T_{ij} = p \ a^2 \ \gamma_{ij} \ .
\end{equation}
In order to give an appropriate description of the different species of energy it is useful to introduce the equation of state parameter $w$ as:
\begin{equation}
w \equiv \frac{p}{\rho} \ .
\end{equation}
As we explain in the following Section different forms of energy correspond to different values of the equation of state parameter and as consequence they induce different evolutions for the scale factor $a(t)$. \\

\noindent Finally we can substitute Eq.~\eqref{eq_intro:Einstein_tensor} and Eq.~\eqref{eq_intro:energy_content_rho_p} into Eq.~\eqref{eq_intro:general_EE} to get the system of differential equations:
\begin{eqnarray}
\label{eq_intro:EE_1}
3\left[ \left( \frac{\dot{a}}{a} \right)^2 + \frac{ k}{a^2} \right]  & = &  \Lambda + 8 \pi G_N \rho \ , \\ 
\label{eq_intro:EE_2}
-\gamma_{ij} \left(  k + 2 a \ddot{a} + \dot{a^2 }\right)  & = & - a^2 \Lambda \gamma_{ij} + a^2 \gamma_{ij} 8 \pi G_N p \ .
\end{eqnarray}
 It is now useful to introduce the Hubble parameter $H \equiv \dot{a} / a$. Once the system is expressed in terms of this quantity, Eq.~\eqref{eq_intro:EE_1} and Eq.~\eqref{eq_intro:EE_2} are usually referred to as Friedmann equation. In terms of $H$, the first of these two equations reads:
 \begin{equation}
\label{eq_intro:friedmann}
3 H^2   =   \Lambda + 8 \pi G_N \rho  -3  \frac{ k}{a^2} \ .
\end{equation}
On the contrary, after some algebraic manipulations, the second equation can be expressed as:
\begin{equation}
\label{eq_intro:friedmann_der}
2 \dot{H }   =    - 8 \pi G_N ( p + \rho)  + 2  \frac{ k}{a^2} \ .
\end{equation}
Once the matter content of the observable Universe is specified, \textit{i.e.} when we fix the value of the the equation of state parameter $w$ for the different matter species that populate the Universe, these two equations can be used to determine the evolution of the scale factor $a(t)$. Fixing the evolution of this parameter actually corresponds to determine the history of the Universe. \\

\noindent
It is interesting to point out that to completely specify the system, we can also impose the conservation of the stress-energy tensor\footnote{Eq.~\eqref{eq_intro:covariant_stress_energy} expresses a \emph{local} conservation of the energy-momentum of matter. However, in general this is not leading to a \emph{global} conservation law~\cite{wald:1984}.}:
\begin{equation}
\label{eq_intro:covariant_stress_energy}
\nabla^{\mu} T_{\mu \nu} = 0 \ .
\end{equation}
Using the definition of covariant derivative, this equation can be expressed as:
\begin{equation}
\nabla^{\mu} T_{\mu \nu} = \partial^{\mu} T_{\mu \nu}  - g^{\mu \sigma} \Gamma^{\rho}_{\mu \sigma} T_{\rho \nu} - g^{\mu \sigma} \Gamma^{\rho}_{\nu \sigma} T_{\rho \mu} = 0 \ .
\end{equation}
Finally we can substitute the expressions of $T_{\mu\nu}$ and of the Christoffel symbols to get:
\begin{equation}
\label{eq_intro:conservation}
\dot{\rho} = - 3 H (p + \rho) \ .
\end{equation}
This equation has a rather simple interpretation: similarly to the case of an expanding gas where temperature decreases during an expansion, the spatially expanding spacetime causes a decrease of the energy density. This equation could also have been obtained by taking a derivative of Eq.~\eqref{eq_intro:friedmann} with respect to time and using Eq.~\eqref{eq_intro:friedmann_der}.

\section{The energy content and the history of the Universe.}
\label{sec_introduction:energy_content_and_history}
As explained in the previous Section, Eq.~\eqref{eq_intro:friedmann} and Eq.~\eqref{eq_intro:friedmann_der} can be used to determine the evolution of the scale factor $a(t)$. For this purpose we should thus specify the matter content of the observable Universe. Inspired by particle physics, we can think of at least two energy species that can give a contribution to these equations:
\begin{itemize}
  	\item \textbf{Radiation}: \textit{i.e.} ultra-relativistic matter such as photons. As these particles are massless, their four-momentum $p^{\mu}$ can be expressed as $p^{\mu} = (|p|,\vec{p})$. The stress-energy tensor of a homogeneous gas of ultra-relativistic particles can be expressed as:
  	\begin{equation}
		T_{\mu \nu} = \textrm{diag}\left( \rho, \frac{p a^2(t)}{3}, \frac{p a^2(t)}{3}, \frac{p a^2(t)}{3} \right)\ ,
	\end{equation}
	\textit{i.e.} the equation of state parameter is $w = 1/3$. 
  	\item \textbf{Cold matter}: \textit{i.e.} non-relativistic matter such as baryons. These are massive particles whose four-momentum $p^{\mu}$ can be expressed as $p^{\mu} = (m^2,0)$. The stress-energy tensor of a homogeneous gas of non-relativistic particles can be expressed as:
  	\begin{equation}
		T_{\mu \nu} = \textrm{diag}( \rho, 0,0,0 )\ ,
	\end{equation}
	\textit{i.e.} the equation of state parameter is $w = 0$. 
\end{itemize}
It is interesting to notice that it is possible to redefine the stress-energy tensor in order to include the cosmological constant contributions. In particular this is parametrized as a form of energy with equation of state parameter\footnote{Actually this is not the only form of energy that gives this particular value for $w$. As we will see in details in Chapter~\ref{chapter:inflation} and more generally in the rest of this work, it is possible to consider some forms of energy that approach $w \simeq -1$ dynamically.} $w=-1$. Similarly, an ``energy density'' $\rho_k$ associated with the spatial curvature can be defined as\footnote{While formally we can proceed with definition, its important to stress that it is misleading to interpret the curvature as a form of energy. In particular, the spatial curvature is an intrinsic property of spacetime and thus it should not be considered as a form of energy that fills the Universe.}:
\begin{equation}
	\rho_k \equiv - \frac{3 k }{8 \pi G_N a^2} \ .
\end{equation}
Notice that this equation implies that the ``energy density'' associated with curvature scales with with $a^{-2}$. To conclude this discussion we should also stress that it is possible to consider forms of energy with different values for $w$. For example cosmic strings~\cite{Kibble:1976sj,Hindmarsh:1994re} have an equation of state parameter $w=-1/3$. However, for the purpose of this Chapter, we ignore these possibilities and proceed with our discussion.\\

\noindent
Defining $\rho_{tot}$ and $p_{tot}$ as the total energy density and pressure, we can express Eq.~\eqref{eq_intro:friedmann} and Eq.~\eqref{eq_intro:friedmann_der} as:
\begin{equation}
	\label{eq_intro:mod_friedmann_tot}
	3 H^2 = 8 \pi G_N \rho_{tot} \ , \qquad \qquad -2 \dot{H} = 8 \pi G_N (p_{tot} + \rho_{tot}) \ .
\end{equation}
Let us proceed with our analysis by assuming that the contribution due to one of these species (with equation of state parameter equal to $w$) dominates over the others. In this limit we approximate $p_{tot}$ and $\rho_{tot}$ with $p_w$ and $\rho_w$. Eqs.~\eqref{eq_intro:mod_friedmann_tot} can thus be expressed as:
\begin{equation}
	\label{eq_intro:mod_friedmann}
	3 H^2 = 8 \pi G_N \rho_{w} \ , \qquad \qquad -2 \dot{H} = 8 \pi G_N (1 + w) \rho_{w} \ ,
\end{equation}
and we can thus solve this system to get an explicit expression for $a(t)$. With some computations it is possible to show that for $w \neq -1$ we get:
\begin{equation}
	a(t) \simeq a_0 \left(\frac{t}{t_0}\right)^{\frac{2}{3(1 + w)}} \ , \qquad \rho \simeq \frac{1}{G_N t_0^2} \left[ \frac{a(t)}{a_0}\right]^{-3(1+w)}  \ ,
\end{equation}
while for $w = -1 $ we get:
\begin{equation}
  \label{eq_intro:lambda_dominated_universe}
	a(t) \simeq a_0 e^{H_0 t } \ , \qquad \rho \simeq \frac {H_0^2}{G_N}  \ .
\end{equation}
These equations imply that, given the equation of state parameter, the scaling solution for the energy density for the different species can be easily obtained. In particular we can show that radiation scales with $a^{-4}$, cold matter scales with $a^{-3}$ and that the cosmological constant, by definition, remains constant. It is also crucial to notice that a Universe dominated by the cosmological constant matches with the de Sitter (dS) spacetime (discussed in Appendix~\ref{appendix_GR:dS_spacetime}).\\

\noindent It is interesting to notice that for all of these components we have $H>0$ \textit{i.e.} an increasing scale factor. Moreover, to get a better understanding of the properties of each species, it is useful to introduce the \emph{deceleration parameter} $q$ as:
\begin{equation}
	\label{eq_intro:deceleration_param}
	q \equiv - \frac{\ddot{a}a}{\dot{a}^2} = - \frac{\ddot{a}}{a} H^{-2} = - 1 - \frac{\dot{H}}{H^2} \ .
\end{equation}
This parameter is proportional to $\ddot{a}$, implying that for an accelerated expansion \textit{i.e.} $\ddot{a}>0$, we get $q<0$. Using the asymptotic expressions for $a(t)$ for the different energy species, it is easy to prove that $q$ can be also expressed as:
\begin{equation}
  \label{eq_intro:deceleration_param_2}
	q = \frac{2}{3} \left( \frac{1}{3} + w \right) \ ,
\end{equation}
so that for both matter or radiation-dominated Universe the expansion is decelerating. It is also interesting to notice that components with $w < - 1/3$ give an accelerated expansion. 

\subsection{The history of the observable Universe.}
\label{sec_introduction:universe_history}
\noindent As explained in the previous paragraph, the history of the Universe can be studied by using Eqs.~\eqref{eq_intro:mod_friedmann_tot}. As we already know the scaling behavior of the different components, in order to get the correct solution of these equations we only need to specify the initial conditions. To set the initial conditions we can use the experimental measurements taken at present time. For historical reasons the value of the Hubble parameter is usually expressed in units of $100\,$km$\,$s${}^{-1}\,$Mpc${}^{-1}$. Although there's still some tension between different measurements of its exact value\footnote{In particular it is worth mentioning the recent $3.3$ sigma tension between the value of $H_0$ measured by Planck~\cite{Ade:2015xua} \textit{i.e.} $H_0 = (67.8 \pm 0.9 ) \times 100 \,$km$\,$s${}^{-1}\,$Mpc${}^{-1}$  and the value measured by Riess et. al~\cite{Riess:2016jrr} observing the Cepheids \textit{i.e.} $H_0 = (73.00\pm1.75) 100 \,$km$\,$s${}^{-1}\,$Mpc${}^{-1}$.}, the present measurements give:
\begin{equation}
	\label{eq_intro:Hubble_param_value}
	h_0 \equiv \frac{H_0}{100 \text{km}\,\text{s}^{-1}\,\text{Mpc}^{-1}} = 0.7 \pm 0.1 \ .
\end{equation}
The energy densities of the different species are usually normalized in terms of the \emph{critical density} defined as:
\begin{equation}
\label{eq_intro:critical_density}
	\rho_c \equiv \frac{3 H^2}{8 \pi G_N} \ .
\end{equation}
The normalized energy densities are thus defined as:
\begin{equation}
\label{eq_intro:energy_densities}
	\Omega_M \equiv \frac{\rho_M}{\rho_c} \ , \qquad \Omega_R \equiv \frac{\rho_R}{\rho_c} \ , \qquad \Omega_\Lambda \equiv \frac{\rho_\Lambda}{\rho_c} \ , \qquad \Omega_k \equiv - \frac{3 k }{8 \pi G_N a^2 \rho_c} \equiv \frac{\rho_k}{\rho_c} \ ,
\end{equation}
where the subscripts $M, \ R, \ \Lambda$ and $k$ are used to denote respectively matter, radiation, cosmological constant and curvature. Let us define $t_0$ the value of $t$ today. A measurement of the normalized energy densities at present time $t = t_0$, gives the approximate values:
\begin{equation}
\label{eq_intro:energy_densities_val}
	\Omega_{M}( t_0) \simeq 0.3 \ , \qquad \Omega_R( t_0) \simeq 10^{-4}\ , \qquad \Omega_\Lambda ( t_0) \simeq 0.7 \ , \qquad \Omega_k ( t_0)\lesssim 10^{-3} \ .
\end{equation}
It is crucial to stress that the measured value of $\Omega_M ( t_0)$ is not consistent with the observed density of ordinary matter, \textit{i.e.} the contribution of baryons to the normalized energy density at present time is $\Omega_b( t_0) \simeq 0.04$. The solution of this problem proposed by the $\Lambda$CDM is the introduction of a new form of matter \textit{i.e.} Cold Dark Matter (CDM), that has the same equation of state parameter of baryons \textit{i.e.} $w = 0$ but is not interacting with ordinary matter and electromagnetic radiation. \\

\noindent A backward evolution of Eqs.~\eqref{eq_intro:mod_friedmann_tot} can finally be performed by using the initial conditions of Eq.~\eqref{eq_intro:energy_densities_val}. For this purpose we call $t_0$ the value of $t$ today and we set the initial condition $a_0 = a(t_0) = 1$. Using the scaling behaviors of the different species $\rho_R \propto a ^{-4}$, $\rho_M \propto a^{-3}$, $\rho_k \propto a^{-2}$, $\rho_c \propto const$ we can thus study the history of the Universe. As a backwards evolution corresponds to a shrinking scale factor and today $\Omega_\Lambda$ starts to dominate over $\Omega_M$, we expect the Universe to pass first through a phase of matter domination and then through a phase that is dominated by radiation. \\

\noindent
To conclude this Section, we give an alternative description of the history of the observable Universe. In fact, instead of using cosmic time, another natural parametrization can be given in terms of the mean temperature of the photons. As discussed in the previous paragraphs, the energy density of photons scales as $a^{-4}$ and by definition this is an energy divided by a volume. Volume scales with $a^{-3}$, implying that the energy must scale as $a^{-1}$. We can thus introduce a thermodynamic temperature $T$ for the gas of photons that is proportional to its mean energy, implying that it scales as $a^{-1}$. In this picture in very early times the Universe was extremely hot and dense and, expanding it has cooled down. As a consequence during the evolution different matter species have progressively decoupled. For example, at $T_{\nu} \simeq 1\,$MeV neutrinos have decoupled from the rest of matter. A crucial event in the Universe history occurs at $T_{BBN} \simeq  0.1\,$MeV when the mean energy of photons has become insufficient to break a neutron-proton bound state, leading to the production of light elements. At $T_{EQ} \simeq 2.6\,$eV the energy densities of matter and radiation have become equal. This moment is usually referred to as the time of matter-radiation equality. After this moment we progressively switch from the radiation-dominated to the matter-dominated epoch. At $T_{rec} \simeq 0.23\,$eV, photons have become unable to break the bound state between electrons and nuclei. This event is usually called ``recombination''. After this moment free electrons have progressively disappeared from the Universe and at $T_{dec} \simeq 0.23\,$eV photons have completely decoupled from the rest of ordinary matter. This moment is usually referred to as ``decoupling''. The evolution of the distribution of photons after decoupling has thus only been affected by gravity and corresponds to the CMB observed today. The temperature of the CMB photons at present time is $T_0 \simeq 2.3 \times 10^{-4}\,$eV which corresponds to the usual $T_0 \simeq 2.7 \,$K.

\section{Cosmic Microwave Background.}
\label{sec_introduction:CMB}
In this Section, we give a review of the main properties of the CMB. Before starting with this discussion, it is important to stress that up to this point we have only considered the background picture of the Universe. In particular we have considered the Universe to be smooth and homogeneous (and at equilibrium). In this picture the Universe is expanding and thus it cools down leading to the progressive decoupling of some particles. In the following we are interested in describing the evolution of the perturbations over this background. In particular, we are interested in describing the processes affecting photons before (and after) decoupling and in discussing how these processes may leave observable signatures in the CMB. To give an accurate description of the evolution of the Universe, we thus need to keep into account the effects of the dynamics at microscopical level. In particular this description is required in order to describe some stages (such as recombination) where the Universe is expected to be out of equilibrium. As usual in the framework of statistical mechanics, the evolution of the Universe should thus be expressed in terms of a Boltzmann Transport Equation (BTE). More details on the definition of BTEs are given in Sec.~\ref{sec_introduction:Boltzmann_eq} and the methods to solve these equations are discussed in Sec.~\ref{sec_introduction:Boltzmann_code}. 

\subsection{The basic picture.}
\label{sec_introduction:CMB_general}
Depending on the energy, different interactions between photons and matter take place. At high energy we have for example: creation of particle-antiparticle pairs by photons in presence of heavy neutral particles; annihilation of particle-antiparticle pairs that generates photons; bremsstrahlung \textit{i.e.} production of a photon due to the interaction between two charged particles; Compton scattering \textit{i.e.} inelastic scattering of photons on electrons; production of photons due to radiative (or double) Compton scattering. Note that most of these processes (with the sole exception of Compton scattering) variate the number of photons that are present in the early Universe. As the temperature drops, all of these processes subsequently stop to occur. For example, the production of electron-positron pairs may only take place if the energy of the photon is larger than $1\,$MeV. On the other hand, bremsstrahlung and radiative Compton scattering may continue to take place until the temperature drops under $T \lesssim 0.5\,$keV. After this moment, the main interaction between photons and electrons is the scattering of photons on electrons \textit{i.e.} Compton scattering. If this scattering occurs at sufficiently low energy\footnote{In particular, if in the reference frame where the electron is at rest, the energy of the photon is smaller than the rest mass of the electron ($m_e \simeq 0.5\,$MeV).} the energy lost by the photon is negligible and the process is well approximated by Thomson scattering that is the elastic scattering of a photon on an electron. More details on Thomson scattering are given in Sec.~\ref{sec_introduction:polarization} where we discuss CMB polarization. \\

\noindent
As explained in the previous Section, after recombination free electrons progressively disappear from the Universe. In this phase of the evolution, interactions between photons and matter become progressively less frequent. In particular, at decoupling \textit{i.e.} at $T \simeq 0.2\,$eV or equivalently at $a \simeq 10^{-3}$ or $t \simeq 4 \cdot 10^5\,$yrs, most of the free electrons have disappeared from the Universe and Thomson scattering stops. This moment is usually referred to as the ``time of last scattering'' or ``last scattering surface''. As after this moment the main effect that affects the distribution of CMB photons is their interaction with gravity, CMB provides a natural method to probe the physics of the Universe at early times. \\

\noindent
At this point it is important to discuss the spectral distribution of CMB photons. As explained in the previous paragraphs until the temperature is above $T \sim 0.5\,$keV the interactions between photons and matter are highly efficient. In particular, the efficiency of these interactions ensures thermal equilibrium between photons and electrons. In this regime the photons are thus well described by a blackbody spectrum, so that the occupation number $n_{\nu}$ of photons at a given frequency $\nu$ is:
\begin{equation}
  n_{\nu} = \frac{1}{\exp \left( \frac{h\nu}{k T} \right)  - 1 } \ .
\end{equation}
It is important to stress that until $T \gtrsim 0.5\,$keV any perturbation that may distort the spectrum is expected to be quickly smoothed by the interactions. However, this is not expected to be true for $T \lesssim 0.5\,$keV. In particular, if some perturbations are introduced, they may not be efficiently smoothed and they may leave an observable signature (\textit{i.e.} distortions) in the spectrum\footnote{More on this topic is said in Chapter~\ref{chapter:pseudoscalar}, where we discuss the case of $\mu$-distortions.}. The minimal version of the $\Lambda$CDM does not predict a significant amount of these perturbations and thus we are not expecting a significant deviation from a pure black body spectrum.\\

\noindent
As explained in the previous paragraphs, at $T \simeq 0.2\,$eV the photons decouple from the rest of matter and after this moment they are only affected by gravity. As a consequence, an observation of the spectrum of CMB photons at present time gives important information on the accuracy of the predictions of the $\Lambda$CDM\footnote{In particular this observation may be used to set constraints on the presence of perturbations that modify the spectrum between $T \sim  0.2\,$eV and $ T \sim 0.5\,$keV.}. An accurate measurement of the CMB spectrum (shown in Fig.~\ref{fig_introduction:firas}) was given by the FIRAS instrument onboard the COBE satellite~\cite{Mather:1991pc}. In particular, no significant deviation from a pure black body spectrum was observed. Such a result is a robust evidence that supports the $\Lambda$CDM and it earned John C. Mather the Nobel Prize in physics in 2006.
\vspace{-0.3cm}
\begin{figure}[h!]
\centering
{\includegraphics[width=0.7 \columnwidth]{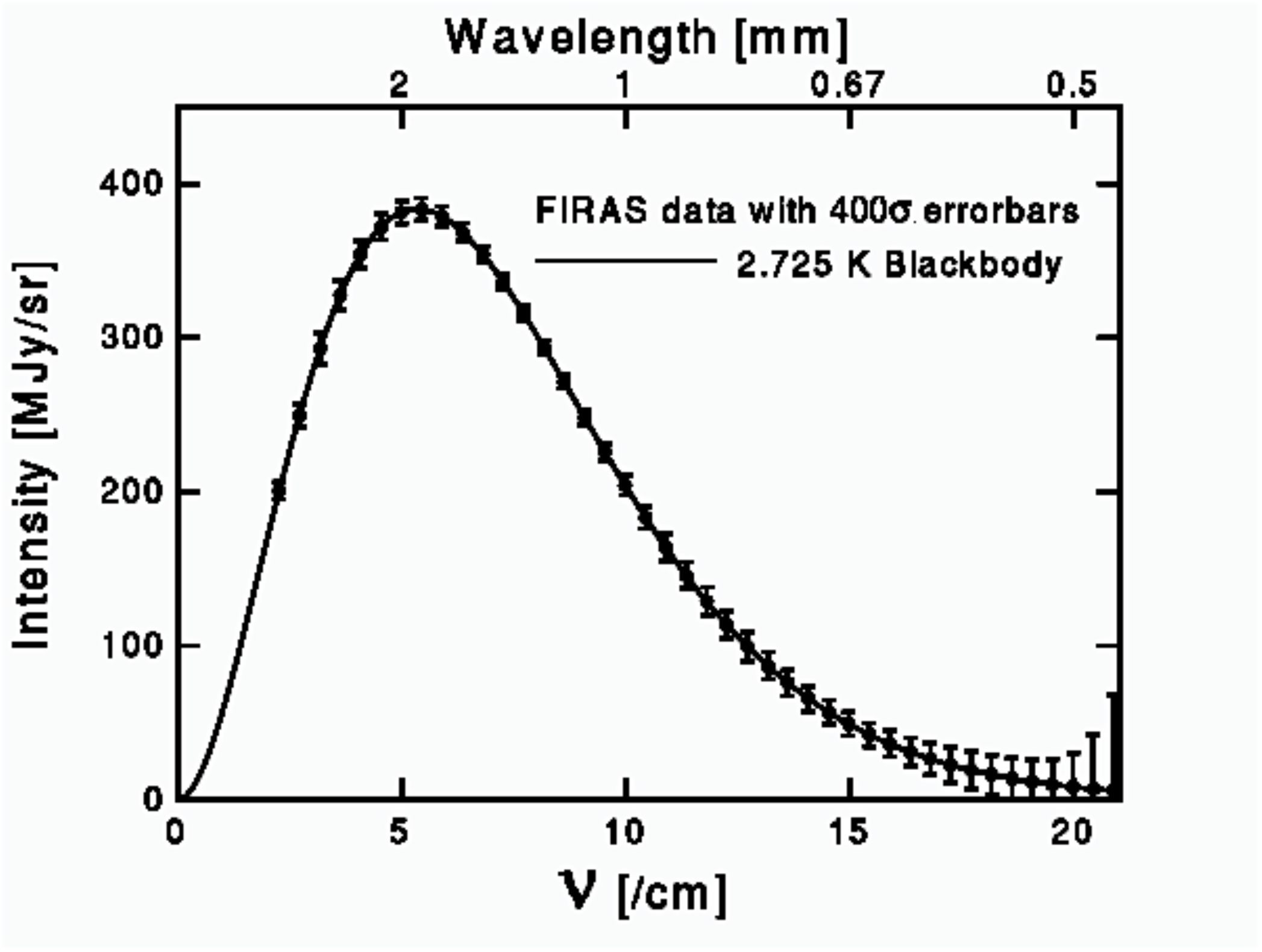}}\\
\caption{ \label{fig_introduction:firas}  CMB blackbody spectrum as observed by the instrument FIRAS onboard the COBE satellite~\cite{Mather:1991pc}.}
\end{figure}

\noindent
Around this homogeneous and isotropic background there are fluctuations of order $10^{-5}$ that were measured by COBE~\cite{Smoot:1992td}. The origin and the distribution of these fluctuations are of great theoretical interest as they may give accurate information on the Universe at early times. In particular, the study of these fluctuations may reveal important details on high energies physics at scales that are not accessible at colliders. For this reason, since the days of COBE, several CMB experiments have aimed at giving an accurate measurement of these fluctuations. More details on some of these experiments and in particular on Planck are given in Sec.~\ref{sec_introduction:experiments}. Moreover, as inflation provides an elegant mechanism to explain the presence of these fluctuations, more details on their generation are given in Chapter~\ref{chapter:inflation} where inflation is discussed. 

\subsection{The angular power spectrum.}
\label{sec_introduction:Spectrum}
An interesting quantity that is usually measured by CMB experiments such as Planck is the difference in temperature for photons received by two antennas pointing in two different directions of the sky. This quantity is usually referred to as the temperature two-point correlation function and is defined as:
\begin{equation}
	\label{eq_intro:two_point}
	C(\vec{n}_1,\vec{n}_2) \equiv \langle \Delta T(\vec{n}_1)  \Delta T(\vec{n}_2 ) \rangle \ ,
\end{equation}
where $\vec{n}_1,\vec{n}_2$ denote the directions of the two antennas and the brackets $\langle \ \cdot \ \rangle$ are used to denote the mean over all the possible statistical realization of the Universe. As the Universe is isotropic, this quantity is expected to depend only on the angle $\theta$ between the directions $\vec{n}_1$ and $\vec{n}_2$ (and thus $\theta$ is defined by $\vec{n}_1 \cdot \vec{n}_2 = \cos \theta$). As a consequence, a useful description of this quantity can be obtained by decomposing the fluctuations in the basis of the spherical harmonics $Y^m_l(\vec{n})$. We start by computing the coefficients of the expansion:
\begin{equation}
	a^T_{lm} \equiv \int \Delta T(\vec{n}) Y^m_l(\vec{n}) \textrm{d} \vec{n} \ ,
\end{equation}
so that the fluctuations can be expressed as:
\begin{equation}
	\Delta T(\vec{n}) = \sum_{l = 1}^{\infty} \sum_{m = - l}^{l}  a^T_{lm} Y^m_l(\vec{n}) \ .
\end{equation}
Substituting this expansion for the fluctuations into Eq.~\eqref{eq_intro:two_point}, we can express the two-point correlation function as:
\begin{equation}
	C(\vec{n}_1,\vec{n}_2) =  \left \langle \sum_{l_{1} = 1}^{\infty} \sum_{m_{1} = - l_{1}}^{l_{1}}  a^T_{l_{1} m_{1}} Y^{m_{1}}_{l_{1}}(\vec{n}_1) \sum_{l_2 = 1}^{\infty} \sum_{m_{2} = - l_2}^{l_2}  a^{T*}_{l_2 m_{2}} Y^{m_{2}*}_{l_2}(\vec{n}_2) \right \rangle \ ,
\end{equation}
As the spherical harmonics form an orthogonal and normalized set\footnote{Meaning that $\int  Y^{m^\prime}_{l^\prime}(\vec{n}) Y^{m*}_l(\vec{n}) \, \textrm{d}\vec{n} = \delta_{l l^\prime} \delta_{m m^\prime}$.}, we can multiply the two sides of these equations by $Y^{m*}_l(\vec{n}_1)$, $Y^{m^\prime}_{l^\prime}(\vec{n}_2)$ and integrate over $\vec{n}_1$ and $\vec{n}_2$ to get:
\begin{equation}
\label{eq_intro:angular_spectrum_1}
	\left \langle a^T_{l m} a^{*T}_{l^\prime m^\prime} \right \rangle = \int  \int  C(\vec{n}_1,\vec{n}_2) Y^{m^\prime}_{l^\prime}(\vec{n}_1) Y^{m*}_l(\vec{n}_2)\textrm{d} \vec{n}_1 \textrm{d} \vec{n}_2 \ .
\end{equation}
As $C(\vec{n}_1,\vec{n}_2)$ only depends on $\theta$, we can proceed by using its decomposition in terms of the Legendre polynomials $P_l(\cos(\theta))$. In particular we have:
\begin{equation}
  \label{eq_intro:Legendre_decomposition}
  C(\vec{n}_1,\vec{n}_2) = C(\theta) = \sum_{l = 0}^{\infty} \frac{(2l + 1) }{4 \pi} C_l  P_l(\cos(\theta))  \ ,
\end{equation}
where, for reasons that will be clear in the following, we have defined $(2l + 1) C_l/(4 \pi)$ the coefficients of the decomposition. The Legendre polynomials $P_l(\cos(\theta))$ can then be expressed in terms of the spherical harmonics using the spherical harmonics addition theorem:
\begin{equation}
  \label{eq_intro:harmonics_theorem}
   P_l(\cos(\theta)) =  \frac{4 \pi}{2 l + 1} \sum_{m = -l}^l Y^{m*}_l(\vec{n}_1) Y^{m}_l(\vec{n}_2) \ .
\end{equation} 
Substituting Eq.~\eqref{eq_intro:harmonics_theorem} into Eq.~\eqref{eq_intro:Legendre_decomposition}, we then get the decomposition of $C(\vec{n}_1,\vec{n}_2)$ on the basis of the spherical harmonics:
\begin{equation}
 	C(\vec{n}_1,\vec{n}_2) = \sum_{l = 0}^{\infty} C_l \sum_{m = -l}^l Y^{m*}_l(\vec{n}_1) Y^{m}_l(\vec{n}_2) \ .
\end{equation} 
Finally we can thus substitute into Eq.~\eqref{eq_intro:angular_spectrum_1} to get:
\begin{equation}
	\label{eq_intro:angular_ps}
	\left \langle a^T_{l m} a^{*T}_{l^\prime m^\prime} \right \rangle = C_l \delta_{l l^\prime} \delta_{m m^\prime}  \ .
\end{equation}
The coefficients $C_l$ are usually referred to as angular power spectrum and the parameter $l$ is usually referred to as multipole. The multipoles are actually associating a component of the angular power spectrum to a given angular scale\footnote{For each value of $l$ there are $2l+1$ values of $m$ \emph{i.e.} we divide the azimuthal angle into $2l$ parts.} $\theta = 180\degree / l$ for the fluctuations. \\

\noindent
As we only have one observable Universe, we should define an estimator\footnote{In statistics, an estimator is a function that given a sample, defines the estimate of a certain parameter using the data of the sample.} for $C_l$. As $C_l$ is basically a variance (is a sum of $\Delta T$ in different directions), a proper estimator\footnote{Actually this is the Maximum Likelihood Estimator (MLE) for the variance. Few more details on MLE are given in Sec.~\ref{sec_introduction:Planck_constraints}.} for $C_l$ is:
\begin{equation}
  \tilde{C}_l = \frac{1}{2l + 1} \sum_{m = -l }^{l} \left| a^T_{l m} \right|^2 \ ,
\end{equation}
where $\tilde{C}_l$ denotes the estimator for $C_l$.\\

\begin{figure}[ht!]
\centering
{\includegraphics[width=0.88 \columnwidth]{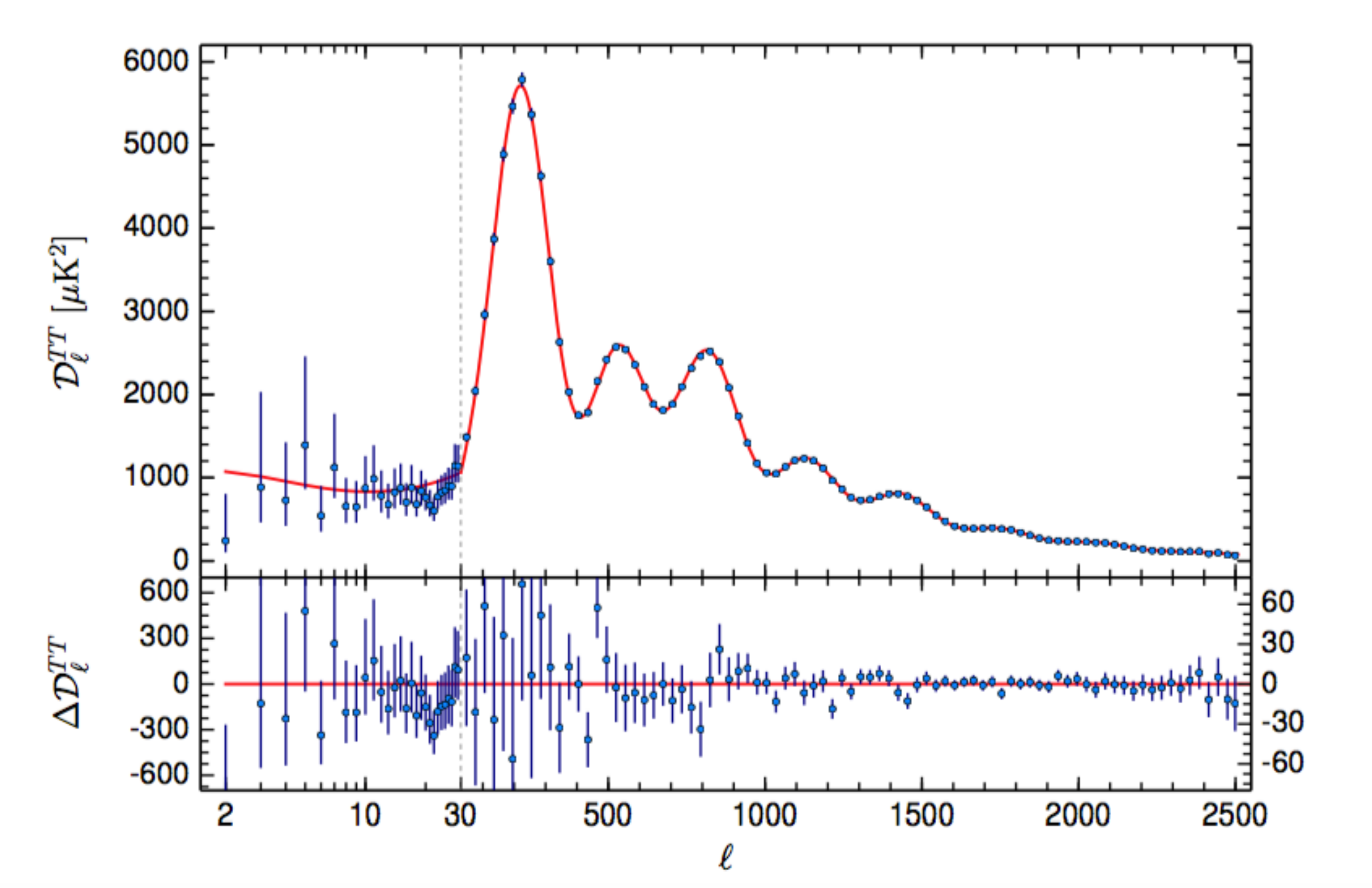}}\\
\caption{ \label{fig_introduction:planck_TT}  Comparison between observed angular power spectrum (blue dots) measured from Planck~\cite{Ade:2015xua}, and the $\Lambda$CDM best fit (red line). The quantity $\mathcal{D}^{TT}_l$ shown in this plot is defined as $\mathcal{D}^{TT}_l \equiv l(l+1)C_l/(2 \pi)$ and is expressed in units of $\mu$K${}^2$.}
\end{figure}

\noindent
CMB measurements give an extremely accurate measurement of angular power spectrum. In particular, in Fig.~\ref{fig_introduction:planck_TT} we show the comparison between the data measured by Planck~\cite{Ade:2015xua} and the best fit produced with the $\Lambda$CDM model. More details on the procedure used to obtain this plot are given through this Section and in Sec.~\ref{sec_introduction:experiments}.

\subsubsection{Acoustic peaks.}
A phenomenon that induces a major effect that characterizes the angular power spectrum shown in Fig.~\ref{fig_introduction:planck_TT} starts to take place right after the matter-radiation equality. At this point of the evolution of the Universe large scale structures begin to form. The basic mechanism that takes place during this process may be depicted as follows: Gravity tends to attract matter towards regions of higher density, the growth of the density causes an increase in the temperature that consequently induces an increase of the radiation pressure. As gravity tends to pull matter towards the higher density region and pressure tends to push it away from it, the interplay of these two effects induces a series of acoustic oscillations. As these oscillations are only affecting regions of space that are compatible with causality, in order to give a quantitative characterization of these oscillations we need to quantify the size of these regions. \\

\noindent
In order to give a more quantitative description of acoustic oscillations, we start by considering the FLRW metric of Eq.\eqref{eq_intro:FLRW} and recalling that massless particles (and in particular photons) travel along null geodesics ($\textrm{d} s^2 = 0$). Assuming the trajectory to be radial, the speed of light at a given time $t$ is simply given by $\textrm{d} r / \textrm{d} t= 1/a(t)$. As a consequence we can introduce two useful length scales:
\begin{itemize}
  \item The comoving distance $d_{p}(t)$ at a given time $t$, defined as the distance that a light ray emitted at time $t=t_i$, has traveled at the instant $t$. Assuming the trajectory to be radial and $r_0 = 0$ this can be expressed as:
  \begin{equation}
    \label{sec_introduction:comoving_distance}
    d_{p}(t) \equiv \int_{t_i}^t \frac{d\hat{t}}{a(\hat{t})} \ .
  \end{equation}
  Notice that this quantity corresponds to the radius of the region that at time $t$ is causally connected with the point $r_0$.
  \item As the inverse of Hubble parameter $H^{-1}(t)$ defines a natural timescale for cosmology, we define the comoving Hubble radius $R_H$ as:
  \begin{equation}
    \label{sec_introduction:comoving_Hubble}
    R_H \equiv (a H)^{-1} \ , 
  \end{equation}
  as the size of a region that at time $t$, can have casual intercourse during a time interval $H^{-1}(t)$.
\end{itemize}
Using these quantities we can finally describe acoustic oscillations.\\

\noindent
The causality of the oscillations\footnote{In order to perform a more accurate estimate of the typical length scale of the oscillations, we should account for the speed of sound $c_s$ of the fluid (defined as $c_s^2 = \delta p/\delta \rho$). As the fluid is almost completely dominated by photos, we have $c_s \simeq 1/\sqrt{3}$ and thus causality is ensured if the wavelength of the oscillations is smaller than $c_s R_H$.} is thus ensured if the wavelength $\lambda$ is smaller than $R_H$. As the typical length of the largest oscillation is $\lambda_1 \simeq R_H$, other oscillations may take place at higher frequency (shorter wavelengths) giving $\lambda_n \simeq R_H/n $, where $n$ is a natural number. To compute the angular scales $\theta_n$ associated with these oscillations, we should divide this quantity by the comoving distance $d(t_{CMB})$ between the observer (at $r_0$) and the surface at which presently observable CMB photons were emitted (\textit{i.e.} the last scattering surface):
\begin{equation}
    d(t_{CMB}) \equiv d_{p}(t_0 - t_{CMB}) = \int^{t_0}_{t_{CMB}} \left( \frac{\hat{t}}{t_0} \right)^{-\frac{2}{3}} \textrm{d}\hat{t} \simeq 3 t_ 0 \ ,
  \end{equation}
where we have used that for $t>t_{CMB}$ the Universe is dominated by matter\footnote{For a more accurate estimate we should also consider the domination of $\Lambda$ at late times. It is possible to show that neglecting its contribution to the evolution of the scale factor, we get a slightly larger value for $l$. However, it is fair to point out that neglecting the effect of $c_s < 1$, we get a slightly smaller value of $l$. As the effects of these the approximations will affect the estimate in opposite ways, we can proceed by neglecting both.} \textit{i.e.} $a(t) \simeq \left( t/t_0 \right)^{2/3}$. The angle $\theta_n$ is thus given by the ratio:
\begin{equation}
  \theta_{\lambda_n} = \frac{1}{n} \frac{R_H(t_{CMB})}{d(t_{CMB})} =  \frac{(aH)^{-1}(t_{CMB})}{3 n t_0}= \frac{ \dot{a}^{-1}(t_{CMB})}{3 n t_ 0} = \frac{a^{1/2}(t_{CMB})}{2 n } \ .
 \end{equation}
Finally we can use $a(t_{CMB}) \simeq 10^{-3} $ and compute the angular scale $\theta_{\lambda_1}$ under which we observe the first peak in the angular power spectrum:
\begin{equation}
  \theta_{\lambda_1} \simeq 0.015 \simeq 1 \degree \ .
\end{equation}
To convert this angle into a value of $l$, we should then use $\theta \simeq 180 \degree \ / l$ so that the angle $\theta_{\lambda_1}$ is finally converted into $l \simeq 180$. As we can see from Fig.~\ref{fig_introduction:planck_TT}, the first peak in the angular power spectrum actually occurs roughly at this value. \\

\noindent
Before concluding this Section, we discuss two more effects that play a role in the generation (and affect the shape) of the acoustic peaks. These effects are due to:
\begin{itemize}
   \item The presence of a dark matter component.\\
    To clarify this point, a more detailed description of the mechanism of oscillations is required. During compressions gravity pulls both baryonic and dark matter towards the regions of higher density (which is basically a potential well). On the other hand, the pressure due to the presence of photons only affects baryons and thus dark matter is not experiencing oscillations, but it is directly falling towards the higher density region increasing the depth of the gravitational well. As a consequence, dark matter increases the amplitude of the oscillations affecting the height of the acoustic peaks. \\
   \item Different scales oscillate at different times.\\
   As explained in this Section, in order to respect causality, the largest size that can start to oscillate at a given time $t$ is of order $R_H(t)\propto t^{1/3}$. Smaller scales (with a typical size roughly equal to $\lambda_n \simeq R_H/n$ with $n>1$) start to oscillate earlier than larger scales ($n=1$). As small scales are starting to oscillate slightly before decoupling, and the distribution of photons freezes at decoupling the corresponding photons (associated with scales roughly equal to $\lambda_n$) may still experience Thomson scattering. This scattering smooths the anisotropies and leads to an exponential suppression of the peaks. This effect, usually referred to as ``Silk damping'', was firstly described by Silk in~\cite{Silk:1967kq}.
 \end{itemize} 
The shape of the angular power spectrum of Fig.~\ref{fig_introduction:planck_TT} is affected by several physical processes. In the definition of a theoretical model which fits the observed angular power spectrum, we should thus model all of these processes. 

\subsection{Theoretical model and predictions.}
\label{sec_introduction:model_and_prediction}
In the previous Section we have described the generation of the acoustic peaks and we have discussed the dependence of this physical process on some parameters (such as $a(t_{CMB}), \Omega_{\textrm{b}}, \Omega_{\textrm{c}}$). More in general, in order to get quantitative predictions \textit{i.e.} to produce theoretical curves that may hopefully fit the data, we should proceed with two steps:
\begin{itemize}
  \item Define a theoretical model. 
  \item Compute the theoretical predictions. 
 \end{itemize} 
The first of these steps corresponds to choosing a certain number of parameters and equations to describe the evolution of the Universe. On the other hand, the second step consists in solving the equations for a given set of parameters. In this Section we present more details on these two steps.

\subsubsection{Beyond Equilibrium and BTE.}
\label{sec_introduction:Boltzmann_eq}
As we have discussed at the beginning of this Section, in order to describe the evolution of the perturbations over the homogeneous and isotropic background it is necessary to define a BTE. As customary in the context of statistical mechanics where we aim at describing many-body systems, we are not interested in considering the motion of every single component. On the contrary, we are interested in defining a set of probability distribution functions $f_{i}(t,\vec{x},\vec{p})$ (where the $i$ denotes different particle species) whose integrals over a certain region $\mathcal{V}$ in the phase space define the number of particles of the species $i$ contained in $\mathcal{V}$. As a consequence, the evolution of the system is encoded in the evolution of the probability distribution functions.\\

\noindent
As usual in the context of statistical mechanics, the total variation of the $f_i$ with respect to the time is defined in terms of a set of BTE. In general a BTE contains a ``free'' part (that is set by Liouville's Theorem) and a ``collision'' term which keeps into account for the interactions between the different species. In the following we discuss the contributions that appear in the BTE for CMB photons. Similar equations should be derived for electrons, neutrinos and in general for all the different particle species. For a detailed review on the definition of BTEs in the context of cosmology see for example~\cite{Hu:1994uz,Hu:1995em}. \\

\noindent
The free part of the BTE is set by the condition that particles move along geodetics. In the context of general relativity this term keeps into account for the non-trivial structure of the spacetime. In particular, in the case of cosmology the free part contains both the gravitational redshifts due to the expansion and the higher order effects due to the metric fluctuations. On the other hand, the collision term should carry the information on the interactions. As already explained through this Chapter, the main interaction experienced by CMB photons before decoupling is Compton Scattering. This process is both driving thermalization and smoothing inhomogeneities before decoupling. \\

\noindent 
The cross Section associated with a Compton scattering $\gamma(\vec{p}_i) + e^-(\vec{q}_i) \rightarrow \gamma(\vec{p}_f) + e^-(\vec{q}_f)$ depends on the momenta $\vec{p}_i$ and $\vec{q}_i$ of the incoming photon and electron, and on the $\vec{p}_f$ and $\vec{q}_f$ of the scattered particles. As a consequence, in order to express the collision term, we should compute the so-called ``Collision Integral''. This quantity depends on the distribution functions of photons and electrons and gives the scattering into and out of a state at a given momentum $\vec{p}$. 

\subsubsection{Cosmological parameters.}
\label{sec_introduction:cosmological_parameters}
In this Section we present a set of cosmological parameters that specifies the theoretical model. In particular, we both define these parameters from a theoretical point of view, and we give a physical interpretation of their effect on the angular power spectrum. \\

\noindent
The minimal set of parameters that can be used to give an acceptable fit of the current observations is six. In particular these parameters are\footnote{We describe the parameters that are used by the Planck collaboration. }:
\begin{itemize}
  \item The parameter $\theta_{MC}$, is related to the \emph{position} of the acoustic peaks. Given the comoving size of the sound horizon at last scattering $r_s(t_{CMB})$ and the angular distance\footnote{For a given object with size (diameter) $D$, that is seen from Earth under an angle $\theta_D$, the angular distance $d_\theta$ is defined as:
\begin{equation}
  \label{sec_introduction:angular_distance}
   d_{\theta} \equiv \frac{D}{\theta_D} \ .
 \end{equation}} $d_\theta(t_{CMB})$ at which we observe the fluctuations, the observed angular size $\theta(t_{CMB})$ is defined as $ \theta(t_{CMB}) \equiv \left. r_s / d_\theta(t_{CMB}) \right|_{CMB}$. The parameter $\theta_{MC}$ is defined as the sampled\footnote{Details on the sampling procedure are given in Sec.~\ref{sec_introduction:Bayesian_inference}.} value of $\theta(t_{CMB})$. The estimate of this parameter is quite robust and basically depends on normalized density parameters.
  \item The parameters $\Omega_{\textrm{b}}h^2$ and $\Omega_{\textrm{c}}h^2$, are defined in terms of the normalized baryon and cold dark matter densities ($\Omega_{\textrm{b}}$,$\Omega_{\textrm{c}}$) and of dimensionless Hubble parameter $h$. These parameters are basically affecting the relative height of the acoustic peaks. Because of a degeneracy between $\Omega_{\textrm{b}}, \Omega_{\textrm{c}}$ and $h$, it is reasonable to put constraints on these combinations. 
  \item After recombination, light elements start to populate the universe. At later times these light elements start to condensate leading to the emission of photons that can reionize free hydrogen atoms. This process is usually called \emph{reionization} and occurs at $1.6\times 10^{-3}\,$eV $ \lesssim T \lesssim 4.8\times 10^{-3}\,$eV. The parameter $\tau$ is defined as the optical depth at reionization and it induces a $e^{-\tau}$ suppression on acoustic peaks that correspond to modes with wavelength smaller than the Hubble radius at reionization.  
  \item The parameter $A_s$ is directly related with the amplitude of scalar fluctuations at $k_* = 0.05 \,$Mpc${}^{-1}$. 
  \item Finally, the parameter $n_s$ (typically called \emph{scalar spectral index}) is used to quantify the variation in the amplitude of the scalar fluctuations according with the variation in the scale $k$ at which we observe the CMB.
\end{itemize}
The constraints that we use in the following Chapters are typically obtained by enlarging this set of parameters. In particular, we typically include in the model a parameter $r$ (called \emph{tensor-to-scalar ratio}) that is used to parametrize the presence of a gravitational waves background. In some cases we also include a further parameter $\alpha_s$ which is used to quantify the scale dependence (running) of the scalar spectral index. Notice that if $\alpha_s$ is introduced in the model, $n_s$ depends on the scale $k$ and its value should thus be defined at a certain scale.

\subsubsection{Boltzmann Codes.}
\label{sec_introduction:Boltzmann_code}
As we have discussed in Sec.~\ref{sec_introduction:model_and_prediction}, the definition of a theoretical model both consists in the definition of a set of equations to describe the evolution of the Universe (BTE) and in the choice of a certain set of parameters (cosmological parameters). Once the theoretical model is specified, we can then proceed with the computation of the corresponding observable quantities. Direct observations can be used to set constraints on the parameters of the model\footnote{This procedure is actually based on the application of Bayesian inference. More details on this procedure are presented in Sec.~\ref{sec_introduction:Bayesian_inference}.} by comparing theoretical predictions with direct measurements. In particular, in order to set these constraints we need to compute theoretical predictions for several different choices of the parameters of the model. As a consequence, it becomes crucial to have a method to compute predictions as efficiently as possible. A solution to this problem is offered by cosmological Boltzmann codes.\\ 

\noindent
Boltzmann codes are computer codes to find numerical solutions for BTEs. The definition of these codes stands on a rather simple procedure originally defined by Bertschinger and Ma in~\cite{Ma:1995ey}. We start by considering the full BTEs that should be defined according to the explanation of Sec.~\ref{sec_introduction:Boltzmann_eq}. The distribution functions that are appearing in the BTEs are then expanded in a series of Legendre polynomials $P_l(\cos(\theta))$ according to their angular dependence. In particular, we find that the expansion up to order $l$ depends on terms of order $l +1$. We then truncate the BTEs at some maximum multipole $l_{max}$ and we numerically solve the system of coupled differential equations. \\

\noindent
The code released by Bertschinger and Ma in 1995, called COSMICS, was used to compute the angular power spectra up $l \simeq 2500$. A major improvement in this context came in 1996 when Seljak and Zaldarriaga released the CMBFAST code~\cite{Seljak:1996is}. CMBFAST is based on COSMICS but it contains some new functions and it highly improves the efficiency in the computations. In particular, with CMBFAST the time to compute the angular power spectra dropped from several days to few minutes. After this moment several further developments of the codes were proposed. Nowadays the two codes that are used the most are:
\begin{itemize}
  \item CAMB, developed by Antony Lewis and Anthony Challinor. CAMB is a reorganized and updated version of CMBFAST. The source code is in Fortran 90 but it can be called by a Python wrapper in order to make it simpler to be used.
  \item CLASS, developed by Julien Lesgourge~\cite{2011arXiv1104.2932L}. In order to make it faster, the code is completely written in C. However, the modules are organized in order to reproduce an object oriented programming and in particular the C++/Python classes. For these reasons, the code has the high performances of C and the readability and user friendliness of C++/Python.
 \end{itemize} 
The Planck collaboration uses both these Boltzmann codes.

\subsection{CMB polarization.}
\label{sec_introduction:polarization}
Another interesting quantity characterizing CMB photons is their polarization. As we explain in the following, this feature can be used to infer important information on the physical processes that take place in the very early Universe and in particular on the process that induces the presence of fluctuations in the CMB \textit{i.e.} on inflation. We start this Section by defining the formalism to describe polarized radiation in terms of the Stokes parameters. Following the proposal of Zaldarriaga and Seljak~\cite{Zaldarriaga:1996xe}, we show that convenient description can be given in terms of the so-called ``E'' and ``B'' modes that as we discuss in Sec.~\ref{sec_introduction:EB_modes}, correspond to a projection of the $Q$ and $U$ parameters (defined in Sec.~\ref{sec_introduction:stokes_parameters}) on a sphere. The mechanisms that polarize the CMB are explained and in particular we explain why the study of this feature is relevant for the scope of this work.

\subsubsection{Stokes parameters.} 
\label{sec_introduction:stokes_parameters}
Let us consider an electromagnetic wave with frequency $\omega$ and with wave-vector $\vec{k}$. For simplicity and without loss of generality, we consider $\vec{k}$ to be in the $z$ direction, so that in the complex notation the electric field $\vec{E}$ can be expressed as:
\begin{equation}
	\vec{E} =\textrm{Re}\left[  E_x(t) \ \hat{x}  + E_y(t) \  \hat{y} \right] \ , 
\end{equation}
where $\hat{x}$ and $\hat{y}$ are unit vectors along the $x$ and $y$ directions respectively and $\textrm{Re}(z)\equiv (z + z^*)/2 $ is the real part of a complex number $z$. The polarization of the electromagnetic wave is specified by the correlation between the $x$ and $y$ components of the field. If there is no correlation between the two components the wave is said to be \emph{unpolarized}.Otherwise if they oscillate in phase the polarization is \emph{linear} and, if they oscillate with a phase shift equal to $\pm \pi$ the polarization is said to be \emph{circular}. To completely specify the state of the wave, we can introduce the four Stokes parameters\footnote{For a detailed review of the topic see for example~\cite{ob:bornwolf}.} that are schematically represented in Fig.~\ref{fig_introduction:stokes}.\\

\begin{figure}[ht!]
\centering
{\includegraphics[width=0.6 \columnwidth]{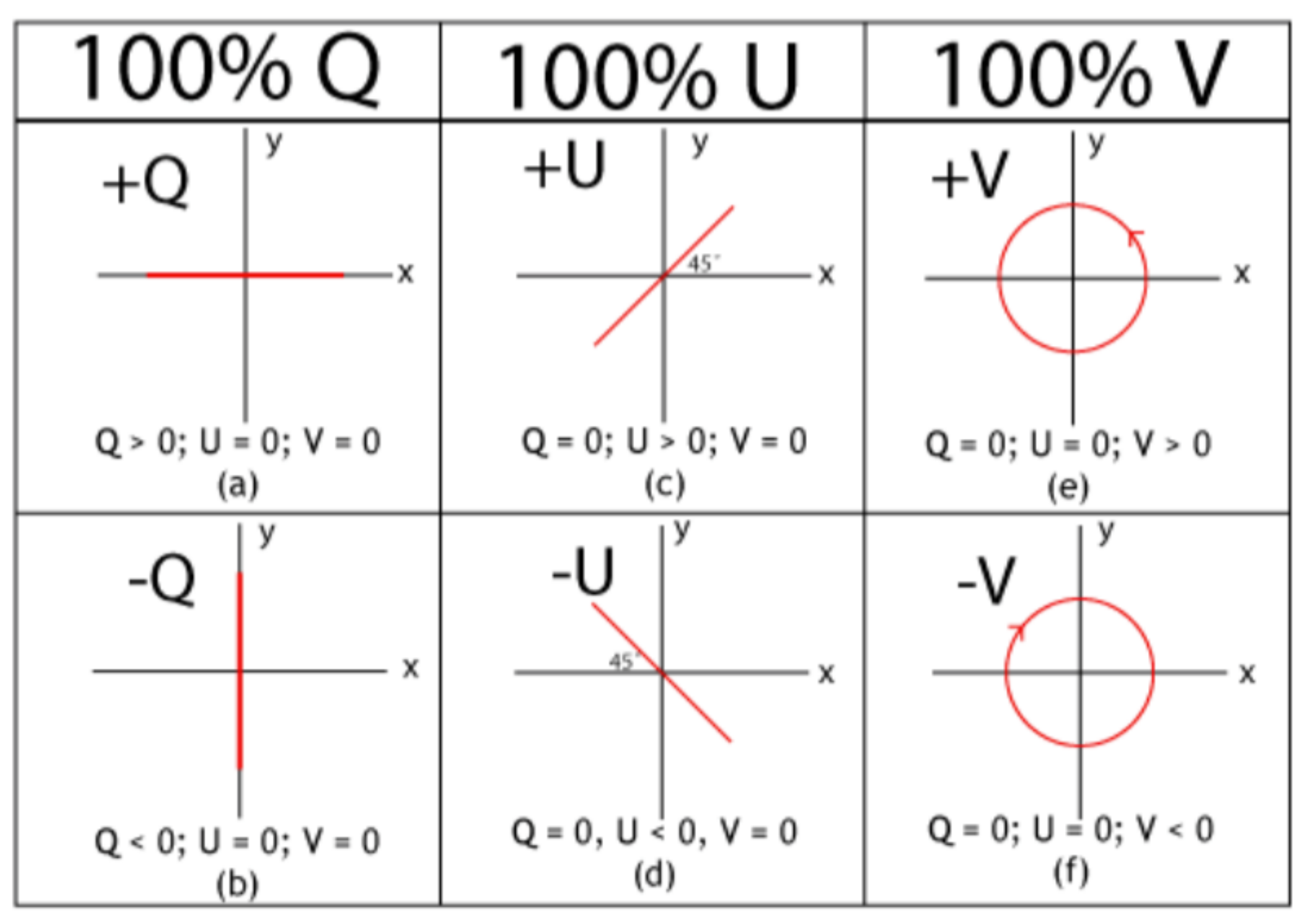}}\\
\caption{   \label{fig_introduction:stokes} A schematic representation of the values of the Stoke parameters for the different states of polarization of an electromagnetic wave.}
\end{figure}

\begin{itemize}
	\item The first parameter is the intensity of the wave and is defined as:
	\begin{equation}
		\label{eq_intro:stokes_intensity}	
		I \equiv \langle \left| E_x \right| ^2 \rangle + \langle \left| E_y  \right|^2 \rangle \ ,
	\end{equation}
	where the brackets $\langle \ \cdot \ \rangle $ denote an average over many oscillation periods. 
	\item Linear polarization may induce (if for example one of the two components is zero) a difference between $\left| E_x \right|^2$ and $\left| E_y \right|^2$. We can thus introduce the second Stokes parameter:
	\begin{equation}
		\label{eq_intro:stokes_difference}	
		Q \equiv \langle \left| E_x \right| ^2 \rangle - \langle \left| E_y  \right|^2 \rangle \ ,
	\end{equation}
	in order to measure this difference. 
	\item As the wave can be linearly polarized at $45\degree$ with respect to the $\hat{x}$ and $\hat{y}$ directions, we introduce the third Stokes parameter:
	\begin{equation}
		\label{eq_intro:stokes_45}	
		U \equiv \langle E_x  E_y^* \rangle + \langle E_x^* E_y \rangle \ ,
	\end{equation}
	as the difference of the intensities of the two components along the directions $\hat{x}^\prime$, $\hat{y}^\prime$ that are rotated by $45\degree$ with respect to the directions $\hat{x}$, $\hat{y}$.
	\item Finally, in order to identify circular polarization, we introduce a fourth Stokes parameter: 
	\begin{equation}
		\label{eq_intro:stokes_circular}	
		V \equiv i \left( \langle E_x  E_y^* \rangle - \langle E_x^* E_y \rangle \right) \ .
	\end{equation}
	As for circular polarization the two components have a phase shift of $\pi/2$, for a clockwise circular polarizations this quantity is negative and for an anticlockwise circular polarizations it is positive.
\end{itemize}
Introducing a phase shift of $\pi/2$ in $E_x$, a circular polarization is turned into a linear polarization at $45\degree$. As a consequence, the parameter $V$ can be defined as the value of $U$ after the introduction of a phase shift of $\pi/2$ in $E_x$.

\subsubsection{Thomson Scattering.} 
\label{sec_introduction:thomson_scattering}
As the four Stokes parameters can be used to completely specify the polarization of the wave, CMB polarization can be finally discussed. As already explained in this Chapter, the main interaction between photons and matter before decoupling is the Thomson scattering. Defining $\vec{k}_i$ and $\vec{\varepsilon}_{\textrm{i}}$ to be the wave-vector and the polarization vector of the incident light, it is possible to show\footnote{For a detailed treatment see for example~\cite{jackson_classical_1999}.} that the differential cross Section can be expressed as:
\begin{equation}
	\frac{\textrm{d} \sigma}{\textrm{d} \Omega} =r_e^2 \left| \vec{\varepsilon}_{\textrm{i}} \cdot \vec{\varepsilon}_{\textrm{f}} \right|^2 \ =  \frac{3 \sigma_T}{8 \pi }  \left| \vec{\varepsilon}_{\textrm{i}} \cdot \vec{\varepsilon}_{\textrm{f}} \right|^2 \ ,
\end{equation}
where $\vec{\varepsilon}_{\textrm{f}}$ is the polarization vector of the scattered photon and where we have introduced the \emph{classical electron radius} $r_{e} = 2.82 \times 10^{-15}\,$m and $\sigma_T = 6.65 \times 10^{-29}\,$m${}^2$ is the \emph{total Thomson scattering cross-section}. This formula implies that the intensity of the scattered radiation peaks in the direction normal to the incident polarization. Given $\vec{k}_f$, wave-vector of the scattered light, incident unpolarized light is scattered into light that is linearly polarized along the direction $\vec{k}_i \times \vec{k}_f$. Notice that if the incident light arrives from all the directions with same intensity, the scattered light is unpolarized. On the contrary, if there is a difference in the intensity in different directions, the scattered light has a linear polarization at $45\degree$ with respect to the $x$ and $y$ axes. It is also crucial to stress that a circular polarization cannot be generated through Thomson scattering.\\

\begin{figure}[ht!]
\centering
{\includegraphics[width=0.6 \columnwidth]{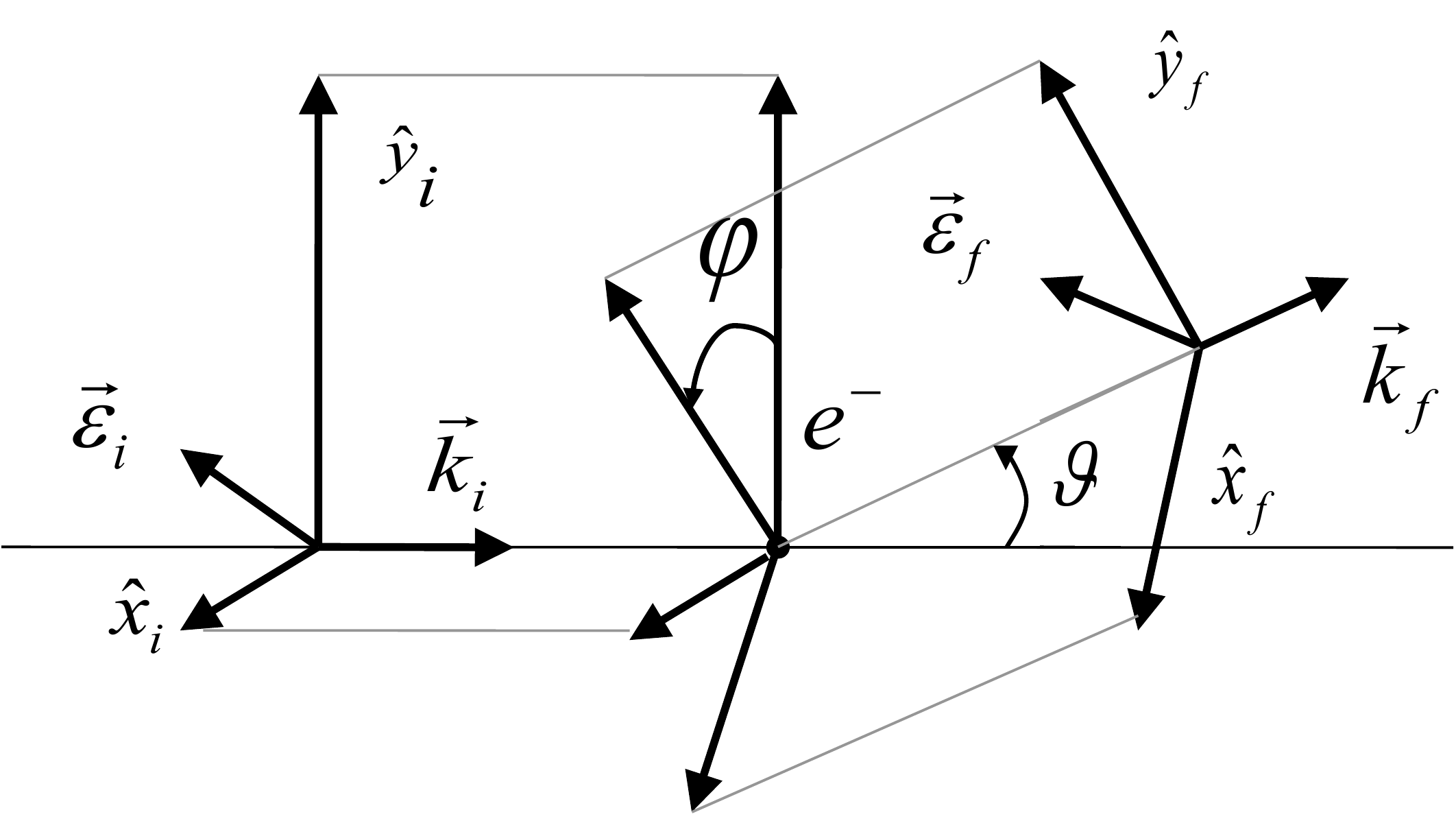}}\\
\caption{   \label{fig_introduction:thomson_scattering} Schematic representation of the Thomson scattering in the rest frame of the electron.}
\end{figure}

\noindent
Stokes parameters are extremely useful to characterize the process of Thomson scattering. Using the notation of Fig.~\ref{fig_introduction:thomson_scattering}, we can express the Stokes parameters in the final reference frame $\hat{x}_{\textrm{f}}$, $\hat{y}_{\textrm{f}}$. For this purpose, we define $I_{x_\textrm{i}} \equiv E_{x_\textrm{i}}^2$, $I_{y_\textrm{i}} \equiv E_{y_\textrm{i}}^2$ as the intensities along the directions $\hat{x}_{\textrm{i}}$, $\hat{y}_{\textrm{i}}$ in the initial reference frame. Assuming that $I_{x_\textrm{i}} = I_{y_\textrm{i}} = I/2$, we can express $I_{x_\textrm{f}} , I_{y_\textrm{f}}$ intensities along $\hat{x}_{\textrm{f}}$, $\hat{y}_{\textrm{f}}$ as:
\begin{equation}
	\begin{aligned}
	I_{x_\textrm{f}} & = \frac{3 \sigma_T }{16 \pi} I \left( \left| \hat{x}_{\textrm{i}} \cdot \hat{x}_{\textrm{f}} \right|^2 + \left| \hat{y}_{\textrm{i}} \cdot \hat{x}_{\textrm{f}} \right|^2 \right) = \frac{3 \sigma_T }{16 \pi} I \left( \cos^2(\varphi) + \cos^2(\theta) \sin^2(\varphi)  \right) \\ 
	I_{y_\textrm{f}} & = \frac{3 \sigma_T }{16 \pi} I \left( \left| \hat{x}_{\textrm{i}} \cdot \hat{y}_{\textrm{f}} \right|^2 + \left| \hat{y}_{\textrm{i}} \cdot \hat{y}_{\textrm{f}} \right|^2 \right) = \frac{3 \sigma_T }{16 \pi} I \left( \sin^2(\varphi) + \cos^2(\theta) \cos^2(\varphi)  \right) \ .
	\end{aligned}
\end{equation}
Notice that in general $I$ is a function of $\theta$ and $\varphi$. Using this expression for the two intensities, we can compute the first two Stokes parameters in the reference frame defined by $\hat{x}_{\textrm{f}}$, $\hat{y}_{\textrm{f}}$:
\begin{equation}
	I_\textrm{f} = \frac{3 \sigma_T }{16 \pi} I \left[ 1 + \cos^2(\theta) \right] \ , \qquad  Q_\textrm{f} = \frac{3 \sigma_T }{16 \pi} I \sin^2(\theta) \cos(2 \varphi) \ .
\end{equation}
As the third Stokes parameter $U$ is defined as the value of the second parameter in a reference frame that is rotated by $45\degree$ with respect to the frame where we measure $Q$, it is easy to get:
\begin{equation}
	U_\textrm{f} = - \frac{3 \sigma_T }{16 \pi} I \sin^2(\theta) \sin(2 \varphi) \ .
\end{equation}
As Thomson scattering does not generate circular polarizations, the fourth Stokes parameter $V_\textrm{f}$ is identically zero. Finally, by integrating over the solid angle, we get the observed value for $I_\textrm{f,TOT}$, $Q_\textrm{f,TOT}$, $U_\textrm{f,TOT}$. In particular, expressing $I_\textrm{i}$, $Q_\textrm{i}$, $U_\textrm{i}$ in terms of the three spherical harmonics $Y^0_0$, $Y^0_2$ and $Y^2_2$, and using the orthogonality of the spherical harmonics we get:
\begin{equation}
\label{eq_intro:spherical_harmonics_decomposition}
\begin{gathered}
	I_\textrm{f,TOT}  = \frac{3 \sigma_T }{16 \pi} \left[ \frac{8}{3}\sqrt{\pi} \, a_{00} + \frac{4}{3}\sqrt{\frac{\pi}{5}}\, a_{20}  \right] \ ,  \\
	Q_\textrm{f,TOT} =  \frac{3 \sigma_T }{4 \pi} \sqrt{\frac{2 \pi}{15}} \, \textrm{Re}(a_{22}) \ , \qquad   U_\textrm{f,TOT} = - \frac{3 \sigma_T }{4 \pi} \sqrt{\frac{2 \pi}{15}} \, \textrm{Im}(a_{22}) \ ,
\end{gathered}
\end{equation}
where the coefficients $a_{lm}$ are the coefficient of the decomposition of $I_\textrm{i}$, $Q_\textrm{i}$, $U_\textrm{i}$ in spherical harmonics. As a consequence, we can conclude that only an incoming quadrupole moment (\textit{i.e.} $a_{22} \neq 0$) may generate a linear polarization for the scattered light.

\subsubsection{E and B modes.}
\label{sec_introduction:EB_modes}
In the case of the CMB a quadrupole moment is present in the intensity of the incident light. This corresponds to the local quadrupole that is seen by an electron in its reference frame. A schematic representation of the corresponding mechanism is shown in Fig.~\ref{fig_introduction:local_quadrupole}. This figure shows a distribution of electrons receding (left) from a region of higher pressure (hot spot) and falling (right) towards a region of lower pressure (cold spot). Let us discuss the second of this two processes. As the electrons fall towards the cold spot, the radial velocity of the electrons is progressively increasing along the radial direction. Effectively, the electrons thus recede from one another along the radial direction, and they approach along the angular direction. This mechanism is thus inducing a local quadrupole in the electron reference frame. The situation is clearly reversed if we consider electrons receding from a hot spot.\\

\begin{figure}[h!]
\centering
{\includegraphics[width=0.6 \columnwidth]{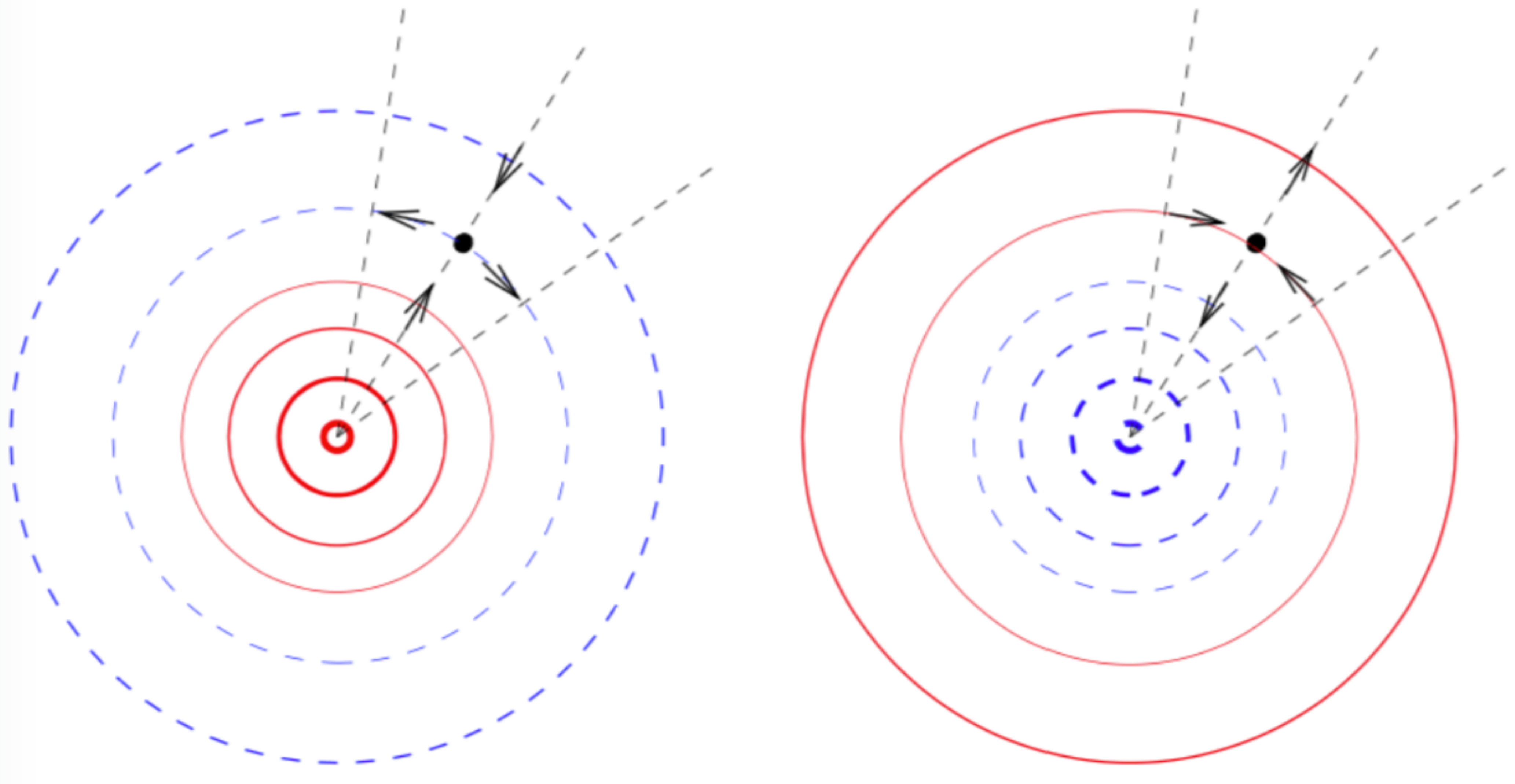}}\\
\caption{   \label{fig_introduction:local_quadrupole} Mechanisms that induces the electron to see a local quadrupole in its reference frame. On the left show a photons departing from a hot spot, where the pressure is higher, and on the right we show an electron falling towards a cold spots, where the radiation pressure is lower.}
\end{figure}

\noindent
Following the proposal of Zaldarriaga and Seljak~\cite{Zaldarriaga:1996xe}, we introduce the two scalar fields $E(\hat{n})$ and $B(\hat{n})$ where as usual $\hat{n}$ denotes a unit vector in the direction $\theta,\varphi$. Expressing $Q_{rad}$ and $U_{rad}$ as the second and third Stokes parameters expressed in polar coordinates, $E$ and $B$ are defined as:
\begin{equation}
\begin{gathered}
	E(\hat{n}) \equiv - \int \textrm{d}\hat{n}^\prime w(\hat{n},\hat{n}^\prime) Q_{rad}(\hat{n}^\prime) \ , \\
	B(\hat{n}) \equiv - \int \textrm{d}\hat{n}^\prime w(\hat{n},\hat{n}^\prime) U_{rad}(\hat{n}^\prime) \ , 
\end{gathered}
\end{equation}
where we have introduced a function $w(\hat{n},\hat{n}^\prime)$ (usually called \emph{weight} function) that does not depend on the radial coordinate. Following the proposal of Zaldarriaga~\cite{Zaldarriaga:2001st} this function is usually chosen to be $w = 1 /(\bar{\theta})^2$ where $\bar{\theta}$ is the angle between $\hat{n},\hat{n}^\prime$. Notice that while $E$ is a scalar, $B$ is a pseudoscalar. Moreover, scalar fluctuations (\textit{i.e.} the two cases shown in Fig.~\ref{fig_introduction:local_quadrupole}) can only induce the presence of a primordial $E$ polarization. On the contrary, tensor fluctuations may generate both $E$ and $B$ polarizations. For this reason a detection of primordial $B$ modes would correspond to an evidence for the presence of primordial tensor fluctuations. \\

\noindent
As pointed out by Zaldarriaga and Seljak in~\cite{Zaldarriaga:1998ar}, gravitational lensing mixes $E$ and $B$ modes: in particular, given a primordial signal with $E \neq 0$ and $B = 0$, the effect of lensing induces a non-zero $B$ pattern. Clearly, the lensing-induced $B$ modes are not a signal of primordial tensor modes and thus it is important to quantify this component in order to measure the primordial $B$ modes.  \\

\noindent
To conclude this Section, we define the correlations between $E$, $B$ and $T$. Decomposing $E$ and $B$ fields in terms of the spherical harmonics we get:
\begin{equation}
	E(\hat{n}) = \sum_{l = 1}^{\infty} \sum_{m = - l}^{l}  a^E_{lm} Y^m_l(\vec{n}) \ , \qquad B(\hat{n}) = \sum_{l = 1}^{\infty} \sum_{m = - l}^{l}  a^B_{lm} Y^m_l(\vec{n}) \ ,
\end{equation}
so that the power spectra can be defined as:
\begin{equation}
	\label{eq_intro:EE_BB}
	\left \langle a^E_{l m} a^{E*}_{l m} \right \rangle = C_l^{EE} \ , \qquad \left \langle a^B_{l m} a^{B*}_{l m} \right \rangle = C_l^{BB} \ .
\end{equation}
Similarly we can also define three more correlation functions \textit{i.e.} $C_l^{TE}$, $C_l^{TB}$, $C_l^{EB}$. However, as $B$ is a pseudoscalar, it is possible to show (see~\cite{Zaldarriaga:1996xe}) that the $C_l^{TB}$ and $C_l^{EB}$ spectra are zero. On the contrary, as both $E$ and $T$ are scalars, we expect a non-zero correlation between $T$ and $E$. The $TE$ and $EE$ spectra measured by Planck are shown in Fig.\ref{fig_introduction:TE_EE_spectra}.

\begin{figure}[h]
\centering
\subfloat[][\emph{TE angular spectrum.}]
{\includegraphics[width=.60\columnwidth]{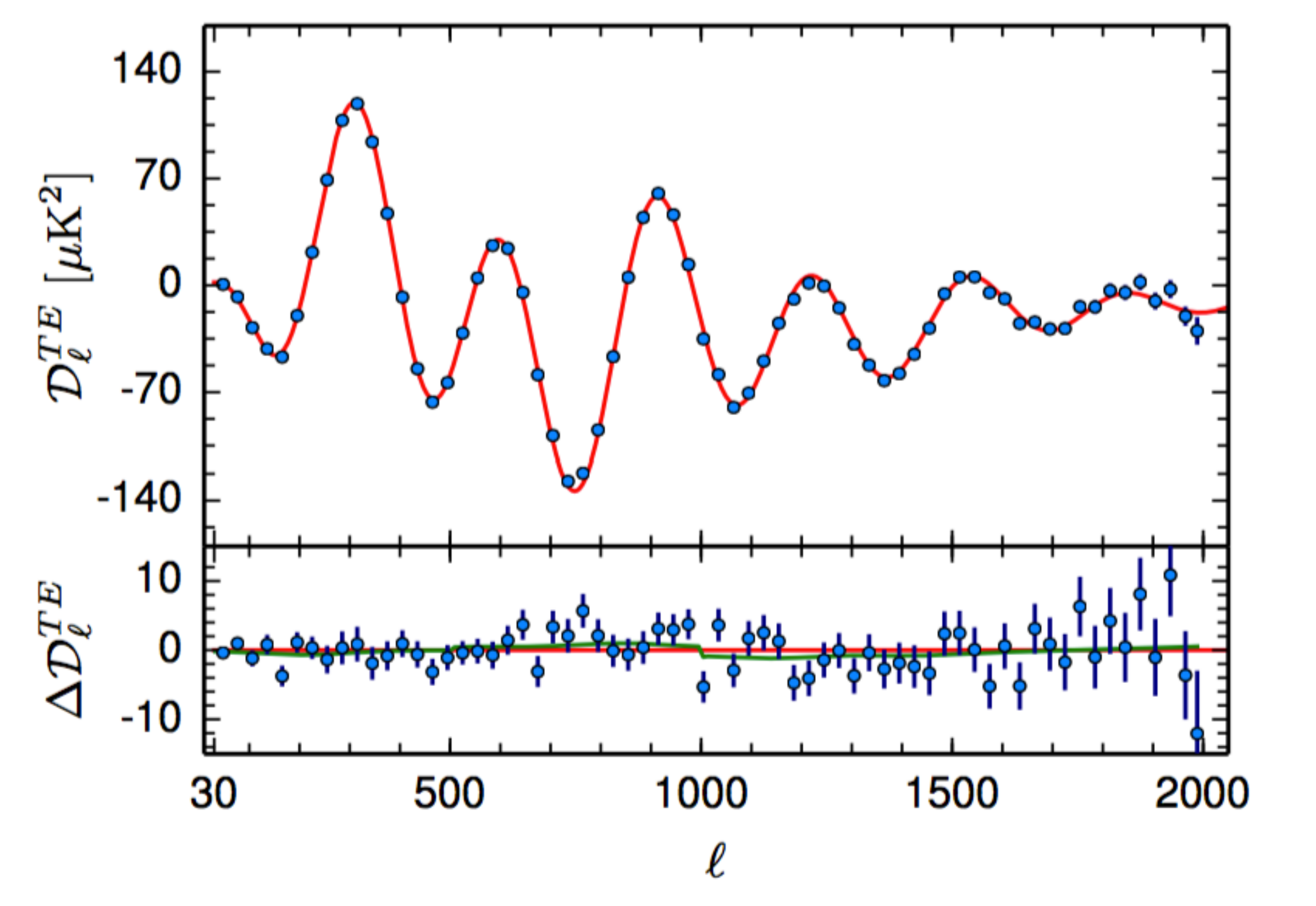}} \\
\subfloat[][\emph{EE angular spectrum.}]
{\includegraphics[width=.60\columnwidth]{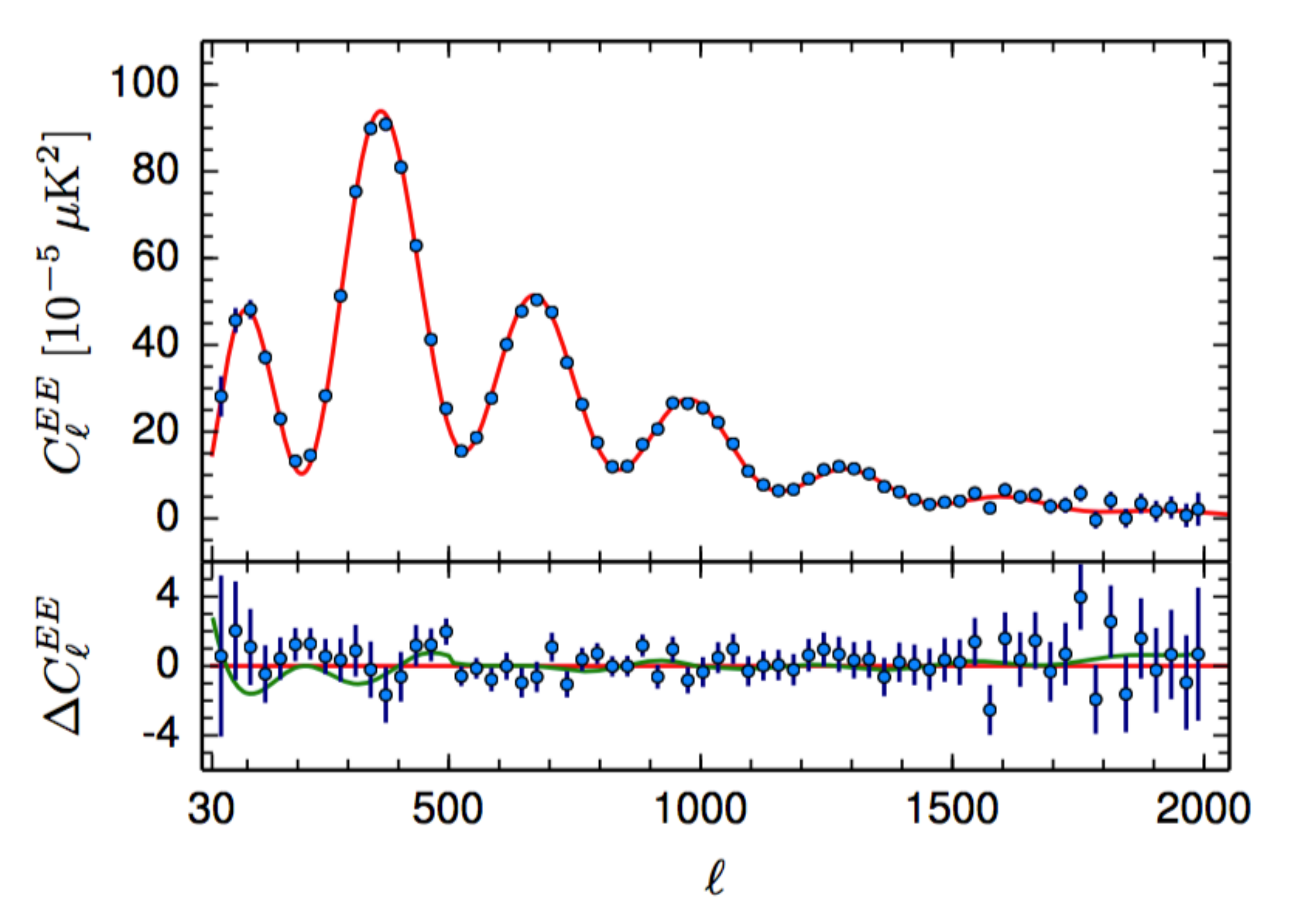}} \\
\caption{TE and EE angular spectra observed by Planck~\cite{Ade:2015xua}. The red curves correspond to the best $\Lambda$CDM model fit of Fig.~\ref{fig_introduction:planck_TT}. The green curve (for details see~\cite{Ade:2015xua}) also keeps into account for additional sources (leakage from temperature to polarization) of systematic error.} 
\label{fig_introduction:TE_EE_spectra}
\end{figure}

\section{CMB observations and Planck constraints.}
\label{sec_introduction:experiments}
As the CMB offers a picture of the early Universe, its observations can be used to get information on the history of the Universe. In particular, as we discuss in the next Chapter, observing CMB fluctuations is a very efficient method to probe high energy physics and the mechanism behind inflation. Since the first measurements performed by the COBE mission~\cite{Smoot:1992td} several CMB experiments have been realized. To present a quick overview, we can start by performing a classification into \emph{ground-based}, \emph{balloons} and \emph{space-based} experiments. Among the ground-based CMB experiments we can mention: \begin{itemize}
	\item BICEP 2/Keck Array, that in 2014 has claimed a detections of $B$ modes~\cite{Ade:2014xna}. A joint analysis with the Planck collaboration proved this signal to be mostly due to galactic dust~\cite{Ade:2015tva}.
	\item POLARBEAR, that at the end of 2013~\cite{Ade:2013hjl,Ade:2014afa} obtained an evidence of the presence of $B$ modes induced by Gravitational Lensing in the CMB. 
\end{itemize} 
Among the balloons for example we can mention:
\begin{itemize}
 	\item BOOMERanG, that in 2000~\cite{deBernardis:2000sbo} measured the position of the first acoustic peak to be at $l = (197 \pm 6)$ at $1\sigma$ level. 
 	\item MAXIMA, that in 2000 (few months later with respect to BOOMERanG) obtained results~\cite{Hanany:2000qf} that are consistent with the ones obtained by BOOMERanG~\cite{deBernardis:2000sbo}, but with higher precision on small angular scales. In particular, it measured the position of the first three acoustic peaks.
 \end{itemize}  
Among the space-based experiments after COBE we have:
\begin{itemize}
	\item WMAP~\cite{Spergel:2003cb,Spergel:2006hy,Komatsu:2008hk,Komatsu:2010fb,Hinshaw:2012aka}, that operated between 2001 and 2010 produced an accurate full-sky temperature map. Its results consist in tights constraints on the parameters of the $\Lambda$CDM model.
  \item Planck~\cite{Ade:2013sjv,Ade:2013zuv,Ade:2015xua,Planck:2013jfk,Ade:2015lrj,Ade:2013ydc}, that operated between 2009 and 2013. More details on the instrumental apparatus and the results of the Planck mission are discussed below and in the rest of this Chapter.  
\end{itemize}
Clearly each type of experiment presents its own merits and flaws. Ground-based are cheaper and can have bigger dimensions with respect to the other experiments. While space-based experiments are more expensive they present three advantages with respect to the ground-based ones:
\begin{itemize}
	\item Whole sky coverage: which is required in order to minimize the cosmic variance. As a consequence they can measure the angular spectrum at low $l$ with greater precision. 
	\item No atmospheric contamination: being outside of the atmosphere they remove its contamination on the CMB photons. For this reason, they are able to measure a cleaner signal.
	\item Wider range of frequencies: measurements of the sky at several frequencies are required in order to remove the foregrounds\footnote{Some more information on this point is given in Sec.~\ref{sec_introduction:planck}, where we give some of the ideas that guide the component separation.}. Once again, being outside of the atmosphere is important to avoid contaminations. 
\end{itemize}
Balloons experiments can be considered as a reasonable compromise between the merits and the flaws of the other two types. For example they are cheaper than space-based missions and as they are flying (for example MAXIMA flew at around $40\,$km) they can remove a part of the atmosphere contaminations. However, we should point out that balloons present several drawbacks such as: the shortness of the period of activity (10 days for BOOMERanG), less precision with respect to space-based missions and less control with respect to ground-based experiments. 

\subsection{Planck.} 
\label{sec_introduction:planck}
The Planck Satellite is a space-based experiment that observes the radiation in the infrared radiation. Planck was realized by the European Space Agency (ESA) and it operated between 2009 and 2013. One of the main goals of the experiment was the production of high precision measurement of temperature and polarization anisotropies from large to small scales with a single instrument. In particular Planck measures the temperature anisotropies from $l = 2$ to $l \simeq 2500$ and the polarization anisotropies from $l \simeq 30$ to $l \simeq 2000$. In order to determine the spectrum at different angular scales the measurements are taken at different frequencies. In particular, measurements at different frequencies are required in order to identify the different components of the signal observed in the sky (through the process of component separation on which we give some details in the following). \\

\noindent
A crucial element in reducing systematics and in producing a high precision measurement is the scanning strategy. We may identify two main points that guided in the definition of this strategy: 
\begin{itemize}
	\item \textbf{The choice of the orbit and of the spin axis}. The orbit and the spin axis were chosen in order to avoid contaminations due to the Earth and to the Sun emissions. For this reason the spin axis is chosen to be on the Ecliptic plane and the satellite observes the sky in a direction that is at $85\degree$ with respect to this axis.
	\item \textbf{The redundancy of the measurements}. In order to reduce noise, the satellite should map the whole sky and each direction should be observed several times. For this purpose Planck observed each portion of the sky for roughly an hour (in which it made around 60 cycles) before changing the rotation axis. 
\end{itemize}

\noindent
The satellite was composed of two different instruments: the Low Frequency Instrument (LFI) and the High Frequency Instrument (HFI), designed to observe the sky at different frequencies. The LFI was composed by 22 antenna that were kept at a temperature of $20 \,$K and that observed the sky at $30 \,$GHz, $44 \,$GHz and $70 \,$GHz. On the other hand, the HFI was composed by 52 bolometers (in practice only 48 of these bolometers where used) that were kept at $100 \,$mK and that observed the sky from $100\,$GHz to $857\,$GHz. In the following, we give a schematic overview of the HFI data processing scheme which, starting from raw data, leads to the production of the well known maps of the sky (shown in Fig.~\ref{fig_introduction:planck_TT_CMB_map}), and to the determination of cosmological parameters.

\begin{figure}[h]\vspace{-0.4cm}
\centering
\subfloat[\emph{Full-sky image.}]{\includegraphics[width=.48\columnwidth, trim = 0cm -2cm 0cm 0cm]{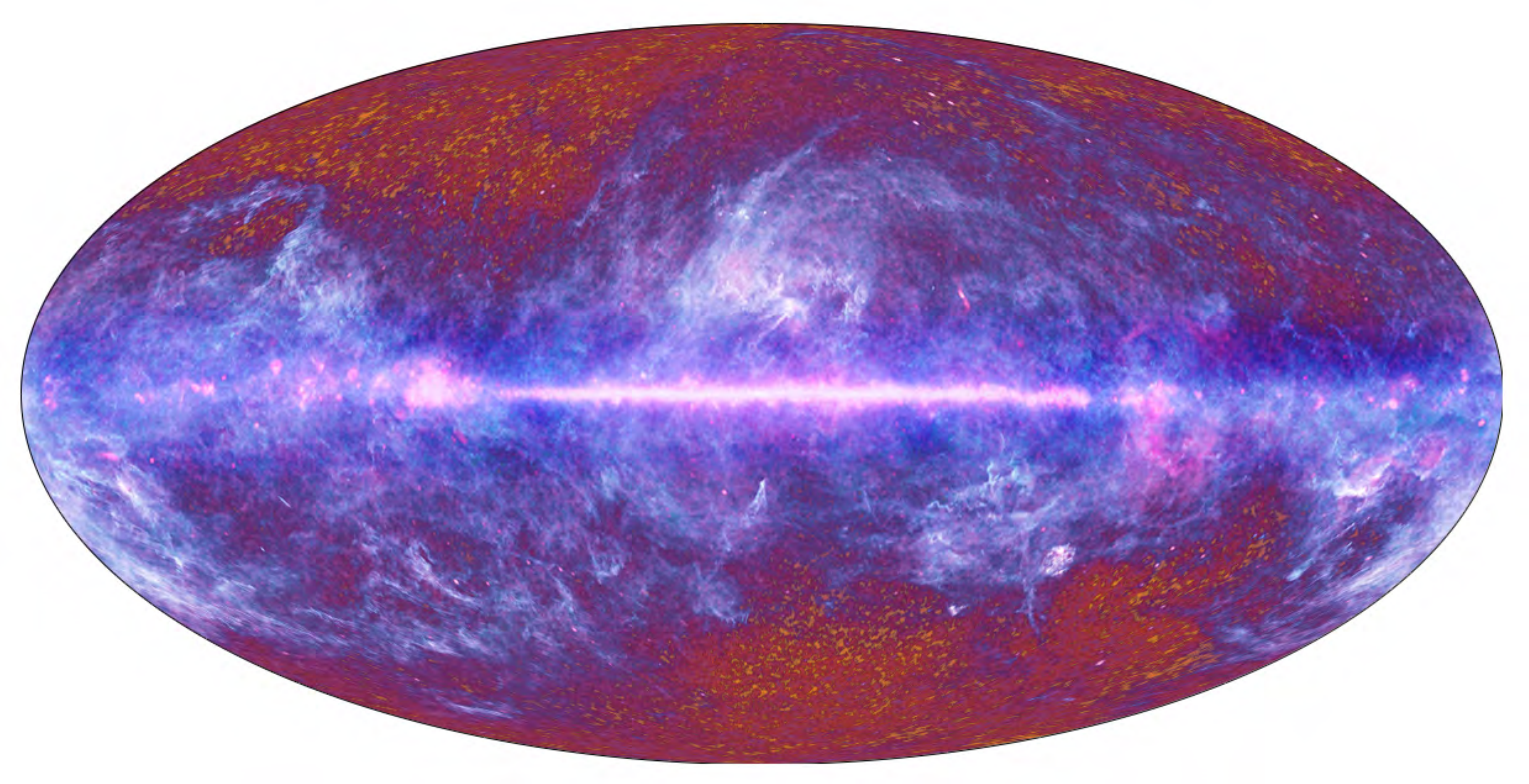}  } 
\subfloat[\emph{CMB anisotropies.}]{\includegraphics[width=.48\columnwidth]{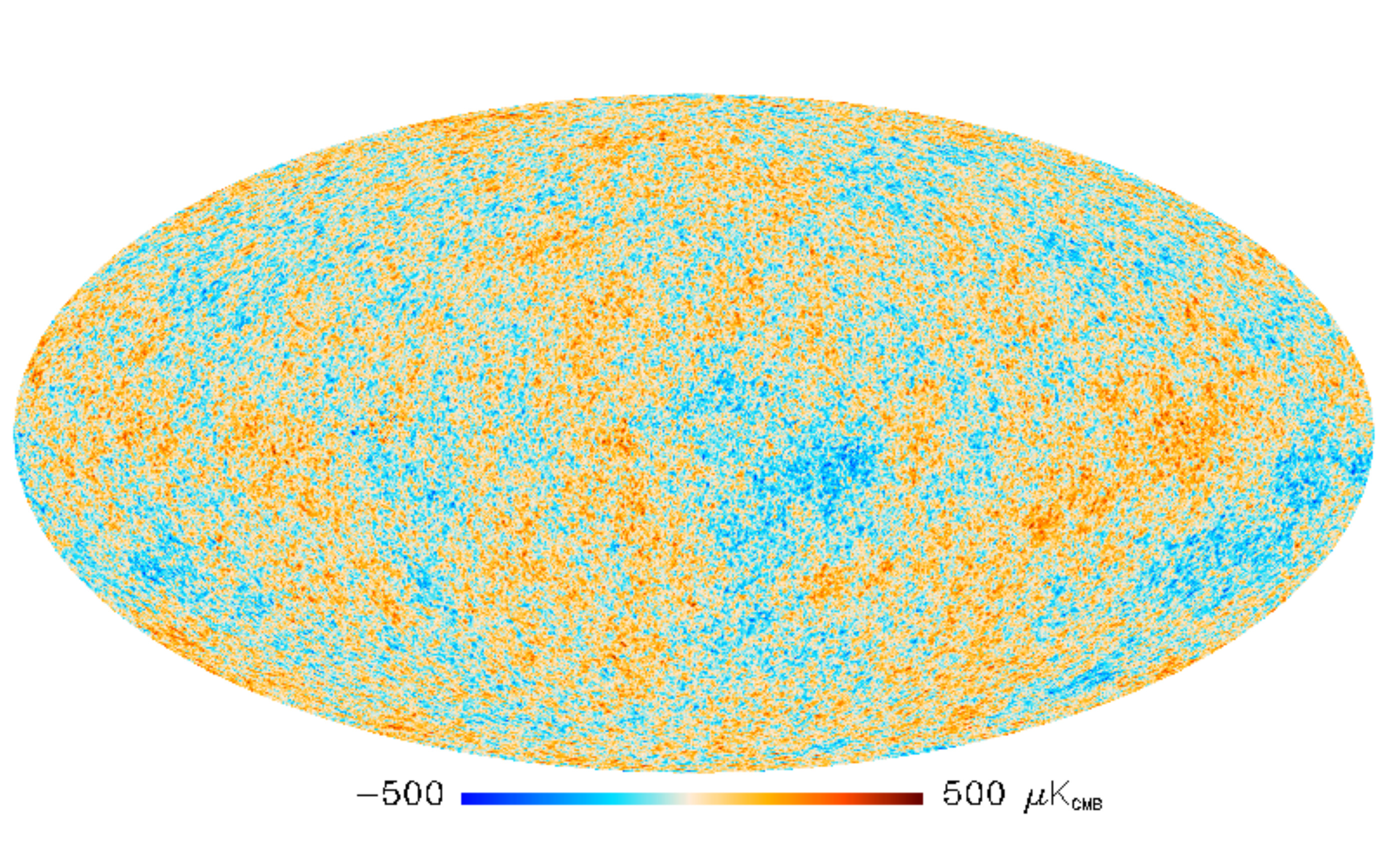} }
\caption{   \label{fig_introduction:planck_TT_CMB_map} The figure shows the maps of the sky produced by Planck. On the left we have a map that shows the complete signal (CMB is red and foregrounds are blue/white) observed by Planck. On the right we have an image of the CMB seen by Planck. More details on the methods to produce these maps are given in the text.} 
\end{figure}

\noindent
The HFI data process can be divided into three steps usually called:
\begin{itemize}
  \item Level 1 (L1): The raw data observed by the instruments are saved into a database. The database has the information on the time (and thus on the pointing direction) of the measurements. 
  \item Level 2 (L2): The time-ordered information (TOI) are processed. This process includes several sub-steps in which the signal is progressively cleaned (for example glitches and the 4K cooler line are removed). Maps of the sky at the different frequencies are created. 
  \item Level 3 (L3): the data are finally used to generate different products, such as all-frequency maps of separate astrophysical components (something more on this process is explained below) and a likelihood code (more on this point is said in Sec.~\ref{sec_introduction:Bayesian_inference} and in Sec.~\ref{sec_introduction:Planck_constraints}) to compare the data with the theoretical model.
\end{itemize}
We conclude this Section by giving few more details on the process of component separation. The signal observed by Planck is a superposition of several physical components (between 10 and 20 for temperature and basically 3 for polarization~\cite{Adam:2015wua}) due to different astrophysical processes. The study and extraction of the different components is required in order to determine the real CMB signal. Among the components that constitute the observed signal~\cite{Adam:2015wua} we have: 
\begin{itemize}
  \item The CMB. A nearly perfect blackbody spectrum described by a single parameter $T_{CMB}$.
  \item Synchrotron radiation. Emitted by free relativistic electrons accelerated by magnetic fields.
  \item Thermal dust. It corresponds to thermal emissions from the interstellar dust. It dominates the foreground for frequencies $\gtrsim 70\,$GHz.
  \item Free-free emission (Thermal bremsstrahlung). Due to the electron-ion scattering in the interstellar plasma.
  \item Thermal Sunyaev-Zeldovich\footnote{The Sunyaev-Zeldovich effect~\cite{1972CoASP...4..173S} is the result of the scattering of CMB photons on clusters of high energy electrons.}. Alters the spectrum of the CMB photons, and imparts a small (negligible~\cite{Adam:2015wua}) linear polarization to the photons.
\end{itemize}
The first three of these components are important for temperature and polarization, the latter two are important for only temperature~\cite{Adam:2015wua}. In addition to the astrophysical components even some instrumental effects (Relative calibration and Bandpass errors) are present~\cite{Adam:2015wua}. After component separation the all-frequency maps are obtained by performing a (weighted) linear superposition of the maps at different frequencies.

\subsection{Bayesian Inference.}
\label{sec_introduction:Bayesian_inference}
The methods to use direct observations of the Universe to set constraints on the Cosmological parameters are actually based on the application of \emph{Bayesian inference}, which corresponds to the application of the \emph{Bayes' theorem} to data analysis\footnote{For a formal introduction to statistics see for example the well known book of Kolmogorov~\cite{kolmogorov1950foundations}. An approach that is more oriented to data analysis for physicists can be found in more modern books~\cite{Bohm:12793423,Bevan_Stat}.}. In this Section we give a brief review of some elements of the theory of probability that are necessary in order to formulate the Bayes theorem. After this review we proceed by explaining how Monte Carlo Markov Chain (MCMC) methods are involved in order to estimate the probability distributions for the cosmological parameters.

\subsubsection{Elements of probability.}
\label{sec_introduction:elements_of_probability}
Let us start by considering a set $S$ and a measure $\mu : S \rightarrow \mathbb{R}$. The space $S$ is usually referred to as \emph{sample space} and an \emph{event} is associated with a subset $X \subseteq S$. The probability $P(X)$ of the event $X$ is defined as $P(X)\equiv \mu(X)/\mu(S) $. In particular, the total probability $P(S)$ satisfies $P(S) = 1$. It should be clear that, given two events $A$ and $B$, the probability for both these events to occur together is given by $P(A \cap B)$. At this point we can define the \emph{conditional probability} $P(A|B)$, probability of the event $A$ given $B$ (assuming $P(B)\neq 0$ ), as:
\begin{equation}
  \label{eq_intro:conditional} 
  P(A|B) \equiv \frac{ P(A \cap B)}{ P(B) } \ ,
\end{equation}
and similarly we have $P(B|A) = P(B \cap A) / P(A)$. Let us assume that $S$ can be expressed as the disjoint union of a certain number $n$ of subsets $B_i$ of $S$ \textit{i.e.} 
\begin{equation}
  \sum_{i = 1}^n P(B_i) = 1 \ , \qquad \qquad  P(B_i \cap B_j) = 0 \ , \forall i \neq j \ .
 \end{equation} 
We can thus use Eq.~\eqref{eq_intro:conditional} to express $P(B_i | A)$ and, substituting $P(B_i \cap A)$, we directly get the Bayes theorem:
\begin{equation}
  \label{eq_intro:Bayes}
  P(B_i | A) = \frac{ P(A|B_i) \cdot P(B_i)}{P(A)} \ .
\end{equation}
While the derivation of this equation is rather trivial, its interpretation is way more interesting. In particular this equation leads to the definition of the Bayesian approach to statistics as an alternative to standard frequentist approach.\\

\noindent
To clarify the meaning of the different quantities appearing in Eq.~\eqref{eq_intro:Bayes}, it is useful to consider an example. Let us consider three boxes $b_1$, $b_2$ and $b_3$ containing two balls each. The first ball contains two red balls, the second box contains a red ball and a blue ball, the third box contains two blue balls. Let us choose a box randomly and extract one ball. We define $A$ the event of extracting a red ball and $B_i$ the event of picking the box $b_i$. It is thus trivial to compute the probability $P(A) = 0 + \frac{1}{2} \cdot \frac{1}{3} + \frac{1}{3}$. Assuming that we have extracted a red ball, we are now interested in computing the probability $P(B_i | A)$ that the ball was extracted from the box $b_i$. This can clearly be computed using Eq.~\eqref{eq_intro:Bayes} giving $P(B_1|A) = 2/3$, $P(B_2|A) = 1/3$ and $P(B_3|A) = 0$. Guided by this example, we can interpret the quantities appearing in this equation:
\begin{itemize}
  \item The $P(B_i | A)$ are usually called \emph{Posterior probabilities}.
  \item The $P(B_i)$ are the so-called \emph{Prior probabilities}.
  \item The $P(A | B_i)$ is the \emph{Likelihood}.
  \item The $P(A)$ is usually called \emph{Model Evidence}. 
\end{itemize}
Notice that the model evidence is the same for all the different cases and thus, modulo the introduction of a normalization constant, it is safe to express the posterior probabilities as $P(B_i|A) \propto P(A|B_i) \cdot P(B_i)$. It is also important to stress that in general both $A$ and $B_i$ can be vectors \textit{i.e.} we may have a set of observations $\vec{A}$ and we may have a set of parameters $\vec{b}_i$ that characterize each box.

\subsubsection{Monte Carlo Markhov Chains.}
\label{sec_introduction:MCMC}
With a real experiment, we are interested in constraining the values of a given set of parameters (usually denoted with $\theta$) of a given theoretical model that describes the corresponding physical process. In this case the prior probabilities for the $P(\theta)$ are usually taken to be constant. All the information from the measured data, is thus encapsulated in the definition of the likelihood. As in general it is not possible to get an analytical expression for the posterior probabilities, these probabilities are usually estimated using MCMC methods. Let us explain in detail how this process works. Monte Carlo methods are methods that use the definition of random samples in order to solve numerical problems. In particular it is well known that these methods provide an extremely powerful tool to perform numerical integrations. However, in order to reproduce the shape of a certain probability distribution, we should be able to generate a set of random points that follow this distribution. It should be clear that in general this cannot be realized. However, it is possible to elude this problem by recurring to the definition of Markov Chains. \\

\noindent
A Markhov Chain is a series of random variables $X_1, X_2, \dots$ with the property that the value $x_{i+1}$ of $X_{i+1}$ only depends on $x_{i}$ value of $X_{i}$. With MCMC we thus refer to Monte Carlo methods to define a Markhov Chains. For a review on MCMC see for example~\cite{gamerman2006markov}. As the MCMC that are relevant for the scope of this work are based on the Metropolis-Hastings algorithms, we only focus on this particular choice. In Metropolis-Hastings algorithms a new point $x_{n+1}$ is randomly generated with a \emph{proposal density} distribution $q(x_n,X_{n+1})$ and it is then accepted with a certain probability $\alpha(x_n,X_{n+1})$. The idea is to make $\alpha(x_n,X_{n+1})$ depend on the posterior that we are actually able to compute at a given parameter point. While we are unable to directly generate data that follow the posterior probability distribution, we can generate random data and reject the ones that are unlikely with respect to the posterior. As a result we can thus generate samples of data that are actually following the posterior probability distribution. These samples can finally be used to set constraints on the parameters of the model.

\subsection{Planck constraints.}
\label{sec_introduction:Planck_constraints}
In this Section we give a schematic explanation of the procedure used by the Planck collaboration to set constraints on the cosmological parameters. After the level L1 and L2 of the data processing explained in Sec.~\ref{sec_introduction:planck}, the data can be used at the level L3 to define a code that computes the likelihood associated with a given theoretical model. The cosmological parameters thus correspond to the $B_i$ of Sec.~\ref{sec_introduction:elements_of_probability} (or equivalently to the $\theta$ of Sec.~\ref{sec_introduction:MCMC}) and the data correspond to the $A$ of Sec.~\ref{sec_introduction:elements_of_probability}. The MCMC used by the Planck collaboration is the CosmoMC package. This method has been defined by Lewis and Bridle in~\cite{Lewis:2002ah} and it uses a Metropolis-Hastings algorithm to generate the Markhov Chains.\\

\noindent
The minimal set of cosmological parameters used by the the Planck collaboration has been explained in Sec.~\ref{sec_introduction:cosmological_parameters}. In particular a flat prior is imposed on these parameters. The MCMC are then generated using the Likelihood to compute the acceptance/rejection probability. All the other parameters of the model are considered as derived parameters that are determined using their Maximum likelihood Estimators (MLE)\footnote{MLE estimate the values of the parameters by selecting the values that maximize the likelihood function. For more details on the definition and on the properties of MLE see~\cite{Bohm:12793423,Bevan_Stat}.}. The values of the minimal set of cosmological parameters used by Planck is shown in Table~\ref{table_intro:parameters}.

\begin{table}[hb!]\vspace{0.4cm}\hspace{-0.6cm}
\begin{tabular}{|c|c|c|c|}
\hline
\textbf{Parameter} & \textbf{Planck TT+low P} & \textbf{Planck TT,TE,EE} & \textbf{Planck TT,TE,EE+low P}\\
 & \textbf{$68 \%$ CL} & \textbf{+low P $68 \%$ CL} & \textbf{+lensing $68 \%$ CL}  \\
\hline\hline
$100 \, \theta_{MC}$ & $1.04085 \pm 0.00047$ &  $1.04077 \pm 0.00032$ & $1.04087 \pm 0.00032$ \\
\hline
$\Omega_{\textrm{b}} h^2$ & $0.02222 \pm 0.00023$ & $ 0.02225 \pm 0.00016$ & $0.02226 \pm 0.00016$\\
\hline
$\Omega_{\textrm{c}} h^2$ & $0.1197 \pm 0.0022$ & $ 0.1198 \pm 0.0015$ & $0.1193 \pm 0.0014$\\
\hline
$\tau$ & $0.078 \pm 0.019$ &  $0.079 \pm 0.017$ & $0.063 \pm 0.014$ \\
\hline
$\ln\left( 10^{10} A_s \right)$ & $3.089 \pm 0.036$ &  $3.094 \pm 0.034$ & $3.059 \pm 0.025$ \\
\hline
$n_s$ & $0.9655 \pm 0.0062$ &  $0.9645 \pm 0.0049$ & $0.9653 \pm 0.0048$\\
\hline
\end{tabular} \caption{Base $\Lambda$CDM model parameters $68\%$ confidence limits (CL) from Planck CMB power spectra (TT,TE and EE), in combination with lensing reconstruction~\cite{Ade:2015xua}.}
\label{table_intro:parameters}
\end{table}

 {\large \par}}
{\large 
\chapter{Inflation and Inflationary models.}
\label{chapter:inflation}

\horrule{0.1pt} \\[0.5cm]

\begin{abstract} 
\noindent
	In this chapter we present inflation. This is a supposed early phase of exponential expansion of our Universe. Inflation has firstly been introduced by Alan Guth~\cite{Guth:1980zm}, Andrei Linde~\cite{Linde:1981mu}, Andreas Albrecht and Paul Steinhardt~\cite{Albrecht:1982wi} in order to solve some problems of the early time behavior of standard cosmology. As inflation elegantly solves these problems, and also provides a mechanism to explain the $10^{-5}$ CMB fluctuations, it is nowadays commonly accepted as a natural extension of $\Lambda$CDM. After the first proposals of Guth and Linde, several models have been proposed in order to give a proper description of inflation. 
\end{abstract}

\horrule{0.1pt} \\[0.5cm] 

\noindent As we have discussed in Chapter~\ref{chapter:introduction}, although it is defined in terms of a small amount of parameters, the $\Lambda$CDM model offers a proper description of our Universe. However, in its simplest realization, which we have introduced in Chapter~\ref{chapter:introduction}, this model is plagued by a certain amount of problems that are related with the early time Universe. Among the main issues of the standard cosmological model it is worth mentioning the flatness, the horizon and the monopole problems that we discuss in detail in this Chapter.\\

\noindent The introduction of an early phase of exponential expansion, which is usually referred to as Cosmic Inflation or simply Inflation, has actually been proposed~\cite{Guth:1980zm, Linde:1981mu,Albrecht:1982wi} in order to solve these problems. A remarkable result that has subsequently been obtained in the independent works of several physicists~\cite{Mukhanov:1981xt,Guth:1982ec,Starobinsky:1982ee,Hawking:1982cz}, concerns the evolution of the quantum fluctuations of the inflaton field and of the metric. As we explain in this Chapter, and through an explicit computation presented in Appendix~\ref{appendix_perturbations:Cosmological_perturbations}, it is possible to show that the initial fluctuations of quantum vacuum, that are generated during inflation, are stretched on macroscopic scales by the exponential expansion. In particular, this mechanism naturally provides an explanation to the presence of the $10^{-5}$ anisotropies in the CMB.\\

\noindent The first full model of inflation, was proposed by Guth~\cite{Guth:1980zm}, even if it is also worth mentioning the model proposed by Alexei Starobinsky~\cite{Starobinsky:1980te}. However, the original Guth's model also known as \emph{old inflation} had some problems with the definition of a mechanism that ensures a smooth ending of the inflationary phase. This problem has been solved by the independent proposals of Linde~\cite{Linde:1981mu} and Albrecht and Steinhardt~\cite{Albrecht:1982wi} that are usually called \emph{new inflation} or \emph{slow-roll inflation} models. A further step has been done with the introduction of \emph{Chaotic inflation}, proposed by Linde in~\cite{Linde:1983gd}. After these pioneering works, several other inflationary models have been proposed. A fairly complete review of these models has been proposed in the fairly recent work of Martin, Ringeval and Vennin~\cite{Martin:2013tda}.\\

\noindent In this Chapter we proceed as follows. We start by presenting the main problems of the $\Lambda$CDM model and in Sec.~\ref{sec_inflation:solution_LCMD} we explain how they may be solved by an early phase of exponential expansion. In Sec.~\ref{sec_inflation:simplest_Inflation} we discuss the simplest realization of inflation in terms of a single slow-rolling scalar field. In Sec.~\ref{sec_inflation:Inflationary_models} we give a brief review of some inflationary models that will be relevant for the scope of this work.

\section{The shortcomings of the $\Lambda$CDM model.}
\label{sec_inflation:problems_LCDM}
As we have already argued in the introduction of this Chapter, the $\Lambda$CDM model presents some problems related with its early time behavior. Good reviews on the problems of standard cosmology can be found in~\cite{Dodelson-Cosmology-2003,Mukhanov:991646,weinberg2008cosmology} and also in the accurate review of Linde~\cite{Linde:2005ht}. In this Section we discuss three of these problems \textit{i.e.} the horizon, the flatness and the monopole problems. We start by stating these problems and then we explain how they can all be solved by inflation.

\subsection{The horizon problem.} 
	\label{sec_inflation:horizon_problem}
	As explained in Chapter~\ref{chapter:introduction}, the standard model of cosmology is based on the assumption that the Universe is homogeneous and isotropic on large scales. Indeed this is only an assumption, but using the CMB observations this can be proved to be a factual evidence. As we show in this Section, a problem arises if we compare the size of the observable Universe at present time with the size of the causally connected regions at earlier times. To give a precise statement of this problem, we can start by computing the comoving distance $d_{p}(t)$ (according to the definition of Eq.~\eqref{sec_introduction:comoving_distance}) traveled by a photon emitted at a given instant $t_i$. In the following we always assume trajectories to be radial and $r_0 = 0$. For this purpose we assume the Universe to be only filled with a single matter energy species with equation of state parameter equal to $w$ (that moreover is supposed to be constant in time):
	\begin{equation}
		\label{sec_inflation:comoving_distance}
		d_{p}(t) = \int_{t_i}^t  \ \left(\frac{\hat{t}}{\hat{t}_0}\right)^{\ -\frac{2}{3(1 + w)}} \textrm{d} \hat{t} \simeq \left(\frac{t}{t_0}\right)^{\frac{1 + 3w} {3(1 + w)}} \simeq  a(t)^{\frac{1 + 3w}{2}}\ t_0 \ ,
	\end{equation}
	where we have set $a(t_0) \equiv a_0 = 1$ and with this expression we can both consider $w = 0, 1/3$ for matter or radiation dominated Universe. As this quantity grows with time (for $w > -1/3$), it is possible to find several areas of the sky, that are in causal contact today, that were not in causal contact in the past. Clearly this result may be in conflict with the homogeneity and isotropy that is observed at present time.\\

	\noindent
	To be more quantitative, let us consider CMB photons, \textit{i.e.} photons that were emitted at $t_{CMB}$ satisfying $a(t_{CMB})/a_0 \sim 10^{-3}$. As a first step we compute the comoving distance at $t_{CMB}$:
	\begin{equation}
		d_{p}(t_{CMB}) = a(t_{CMB})^{\frac{1 + 3w}{2}} \ t_0  \simeq 10^{-3} \ t_0,
	\end{equation}
	where we have used $t_i \ll t_0 $ and in the last step we have reasonably assumed the early Universe to be dominated by radiation. This quantity should be compared with the comoving distance between $r = 0$ and the surface (\textit{i.e.} the last scattering surface) at which CMB photons that are presently observable were emitted:
	\begin{equation}
		d_{p}(t_0 - t_{CMB}) = t_0  \ \left[ a(t_{0})^{\frac{1 + 3w}{1}} - a(t_{CMB})^{\frac{1 + 3w}{2}} \right] \simeq t_ 0 \ a(t_{0})^{\frac{1 + 3w}{2}} = t_0 \ ,
	\end{equation}
	The ratio between the area of the last scattering surface and the area of a causally connected surface at $t_{CMB}$ is thus proportional to $10^6$. As a consequence CMB observation proves that at $t=t_{CMB}$ the Universe was homogeneous on $10^6$ regions that were not causally connected. As there is no symmetry that enforces homogeneity over these regions, there is no reason to justify the global isotropy and homogeneity if these regions have never been in causal contact in the past.

\subsection{The flatness problem.} 
	\label{sec_inflation:flatness_problem}
	The flatness problem is based on the naturalness principle and thus can be rephrased as a ``fine-tuning'' problem. In theoretical physics, a theory is said to respect naturalness if all the dimensionless free parameters of the theory take values ``of order one''. A fine-tuning problem is thus faced when one or more parameters of the theory take ridiculously big or ridiculously small values without a symmetry enforcing it. In the case of standard cosmology, a fine tuning problem arises if we consider the present value of $\Omega_k$ defined in Eq.\eqref{eq_intro:energy_densities}. An accurate measurement of this value is given by Planck~\cite{Ade:2015xua}:
	\begin{equation}
	\label{eq_inflation:omega_k_now}
 		\Omega_k(t_0) = 0.000 \pm 0.005 \ (95\%, \text{Planck TT+lowP+lensing+BAO}) ,
 	\end{equation} 
 	where $t_0$ denotes present time. To compute the value of $\Omega_k(t) \equiv \rho_k (t)/ \rho_c(t)$ at a given time $t$, we can thus use the scaling behaviors for the different energy species using the values of the normalized densities at present time (given in Eq.~\eqref{eq_intro:energy_densities_val}) as the initial conditions for the backward evolution:
 	\begin{equation}
 		\label{eq_inflation:omega_k_evolution}
 		\Omega_k(t) = \frac{\Omega_k ( t_0) a^{-2}(t)} {\left[  \Omega_R(t_0) a^{-4}(t) +  \Omega_M(t_0) a^{-3}(t) + \Omega_k( t_0) a^{-2}(t) + \Omega_\Lambda \right]} \ .
 	\end{equation}
	For example we can evaluate this quantity at the epoch $t_{\text{GUT}}$ of Grand Unification Theory (GUT) :
	 \begin{equation}
	\label{eq_inflation:omega_k_evolution_GUT}
	\Omega_k(t_{\text{GUT}}) = \left[ 140 (a_{\text{GUT}})^2 + 1 +60 (a_{\text{GUT}})^{-1} + 2 \cdot 10^{-2} (a_{\text{GUT}})^{-2}\right]^{-1} \ ,
	\end{equation} 
	where $a_{\text{GUT}} = a(t_{\text{GUT}})$. We can then proceed by using the definition of temperature given in Chapter~\ref{chapter:introduction} to express the scale factor as $a(t)\simeq T_0/T$ where $T_0$ is the temperature of the Universe at present time \textit{i.e.} $T_0 \simeq 2.3 \cdot 10^{-4}\,$eV. At the epoch of GUT, the temperature $T$ is expected to be of order $10^{16}\,$GeV so that the ratio $T_0/T \simeq 10^{-29}$. We can thus substitute into Eq.~\eqref{eq_inflation:omega_k_evolution_GUT} to get:
	 \begin{equation}
	\label{eq_inflation:omega_k_value}
	\Omega_k(t_{\text{GUT}}) \simeq 10^{-56} .
	\end{equation} 
	As no symmetry prefers a flat Universe with respect to an open or a closed one, there is no reason to impose such a small number for $\Omega_k(t_{\text{GUT}})$. This choice clearly corresponds to an extreme fine-tuning for this parameter of the $\Lambda$CDM model.

\subsection{The monopole problem.} 
	\label{sec_inflation:monopoles_problem}
	The mechanism of Spontaneous Symmetry Breaking (SSB) is one of the main concepts in modern theoretical physics. This phenomenon occurs when a classical symmetry is broken at a quantum level and in particular it is possible to show that this is realized when the potential of the theory has degenerate minima. When this condition is satisfied, the symmetry group $G$ is broken to a subgroup $H$, that is usually called ``little group''. Moreover, it is possible to define the so-called vacuum manifold $\mathcal{M} = G/H$ as the manifold containing all the physically different vacua of the broken theory. A typical example of SSB is the breaking of the Electroweak symmetry in the standard model of particle physics (SM):
	\begin{equation}
	G_{SM} = SU(3)_C \otimes SU(2)_L  \otimes U(1)_Y \rightarrow H_{SM} = SU(3)_C  \otimes U(1)_{EM}.
	\end{equation}
	When a SSB occurs, depending on the topological properties of $\mathcal{M}$, it may lead to the formation of topological defects~\cite{Kibble:1976sj,Hindmarsh:1994re}. In particular we may have three kind of defects: monopoles (point-like defects), cosmic strings (one-dimensional defects) and domain walls (two-dimensional defects).\\

	\noindent
	As explained in Chapter~\ref{chapter:introduction}, at very early times our Universe is expected to be very compact and hot. At this stage the interactions of the SM are expected to unify into a GUT, described by certain gauge group $G_{\text{GUT}}$. As the Universe expands and cools down, the theory is expected to pass through certain number of phase transitions \textit{i.e.} SSB, so that the original gauge groups break into successive subgroups:
	\begin{equation}
	G_{\text{GUT}} \rightarrow H_{1} \rightarrow H_{2} \rightarrow \dots \rightarrow G_{SM}  \rightarrow H_{SM} \ . 
	\end{equation}
	During the several stages of this process, a high density of topological defects may be generated. An explicit computation of their energy density evaluated at present time would give $\rho_{Monop}/\rho_c \sim 10^{15}$. This implies that these monopoles are expected to dominate the evolution of our Universe. As this behavior is unobserved, theoretical expectations are in contradiction with direct observations, leading to the so-called monopole problem.

\section{Inflation.}
\label{sec_inflation:definition_inflation}
As we explain in this Section, a simple and elegant solution to the three problems stated in the previous Section is provided by the introduction of an early phase of exponential expansion of our Universe. This phase is usually referred to as cosmic inflation or simply inflation. We start this Section with a brief review of the main lines of the physics of inflation and then (in Sec.~\ref{sec_inflation:solution_LCMD}) we explain how it solves the three problems of Sec.~\ref{sec_inflation:problems_LCDM}. \\

\noindent
As already argued across this Chapter, inflation is an early phase of exponential expansion of the Universe. In practice, this implies that the scale factor $a(t)$ appearing in the FLRW metric (Eq.~\eqref{eq_intro:FLRW}) is exponentially growing with time. As explained in Sec.~\ref{sec_introduction:energy_content_and_history} (in particular see Eq.~\eqref{eq_intro:lambda_dominated_universe}), this condition can be realized if the Universe is filled by a form of energy with equation of state parameter $w = -1$. However, as already explained in Sec.~\ref{sec_introduction:energy_content_and_history}, this particular configuration matches\footnote{If $a(t) \propto \exp (C t)$, the FLRW metric matches with the dS Metric of Eq.~\eqref{appendix_GR:dS_metric}.} with the de Sitter (dS) spacetime described in~\ref{appendix_GR:dS_spacetime}. dS spacetime is a static solution of the Einstein Equations~\eqref{eq_intro:general_EE} that actually describes an eternally inflating Universe. As a consequence, this particular configuration cannot be included into the evolution of the Universe because it lacks a graceful exit from the early phase of exponential expansion. \\

\noindent
In order to be consistent with the direct observations of the Universe, we have to define a mechanism that enforces a graceful exit from inflation. A solution to this problem, is the definition of an energy species with a time dependent equation of state parameter $w(t)$. In particular, we start with $w(t) \simeq -1$ that actually leads to a nearly exponential growth of the scale factor, and $w(t)$ is required to grow. As explained in Sec.~\ref{sec_introduction:energy_content_and_history} (in particular see Eq.~\eqref{eq_intro:deceleration_param_2}), the accelerated expansion stops as soon as $w(t)$ becomes larger than $ -1/3$. \\

\noindent
In order to realize this configuration we thus need a form of energy that mimics the cosmological constant for a certain (finite) period of time. Guth~\cite{Guth:1980zm} realized that this scenario can be implemented by a (scalar) field with non-zero potential energy that is usually called ``inflaton''. The main difference between the old inflation scenario of Guth~\cite{Guth:1980zm} and the new inflation scenarios of Linde~\cite{Linde:1981mu} Albrecht and Steinhardt~\cite{Albrecht:1982wi} lays in the mechanism to implement this configuration. In the first case, the field was assumed to be trapped in a false (metastable) vacuum of the potential, a quantum tunneling process is thus required in order to generate bubbles of true vacuum which rapidly expand putting an end to the inflationary phase. On the other hand, with the definition of new inflation models, it was realized that under particular conditions\footnote{In particular, the field must be ``slow-rolling'' down its potential. A more quantitative statement of this condition is given in Sec.~\ref{sec_inflation:simplest_Inflation}.} inflation could be implemented by a (scalar) field that rolls down its potential. \\

\noindent
It is crucial to stress that an exponential expansion of the scale factor leads to an exponential decrease of the temperature ($T \propto a^{-1}$). Moreover, an exponentially growing scale factor is also leading to an exponential decrease of the energy densities\footnote{Actually the contribution $\rho_\Lambda$ due to the cosmological constant, is not decreasing. However, this contribution is expected to be much smaller than the one associated with the inflaton and thus, for the scope of this discussion, we can safely ignore it.} ($\rho_k \propto a^{-2}$, $\rho_M\propto a^{-3}$, $\rho_R \propto a^{-4}$). As a consequence, at the end of inflation the Universe is extremely cold and it is only filled by the inflaton (and $\Lambda$ which anyway is irrelevant for this discussion). For this reason, right after the end of inflation, it is necessary to have phase transition where the inflaton decays into the other species (matter and radiation) repopulating the Universe. As discussed by Kofman, Linde and Starobinsky~\cite{Kofman:1994rk,Kofman:1997yn}, during this process, that is usually called \emph{reheating}, it is possible to distinguish three different stages. In a first step, that is usually called \emph{pre-heating}, the inflaton quickly (explosively) decays into (massive) bosons. At this stage, the production of fermions (which is affected by the Pauli principle) is significantly smaller. It should be clear that, because of its explosive nature, this process is expected to be strongly out of equilibrium. In a second step, the huge amount of bosons that was produced during pre-heating decays into other particles. Methods to describe this phase were proposed in~\cite{Dolgov:1982th,Abbott:1982hn}, however, as pointed out in~\cite{Kofman:1994rk} they should not be applied to the decay of the inflaton itself but rather to the decay of bosons produced during pre-heating. Finally, during the third stage, the particles that were produced thermalize. The temperature of the Universe at this stage is called reheating temperature $T_{rh}$ and it is basically determined by the efficiency of the decay (parametrized by the decay constant) of the inflaton. Notice that in general the second and third stage may occur simultaneously.

\subsection{A solution to the problems of $\Lambda$CDM.}
\label{sec_inflation:solution_LCMD}
In the rest of this Section we show that inflation solves the three problems of Sec.~\ref{sec_inflation:solution_LCMD}. In order to get a quantitative description of this process, we start by making two assumptions:
	\begin{itemize}
	 	\item We assume the reheating temperature to be $T_{rh} \lesssim 10^{15}\,$GeV. As we discuss in Sec.~\ref{sec_inflation:solve_monopoles}, this is actually required in order to solve the monopole problem.
	 	\item We assume the reheating to be highly efficient. To be more precise, given $a_E$, value of the scale factor at the end of inflation, and $a_{rh} \equiv a(T_{rh})$ value of the scale factor at $T\simeq T_{rh}$, we have $a_E \simeq a(T_{rh})$.
	 \end{itemize} 
Using the first of these assumptions (in particular we fix $T_{rh} \simeq 10^{15}\,$GeV) we compute $a_{rh} \propto T_0/T_{rh} \simeq 10^{-28} $, where $T_0$ is the temperature of the Universe at present time. Notice that an exponential growth of the scale factor is realized if the Hubble parameter (denoted with $H_I$) is nearly constant. As a consequence, we conclude that:
\begin{equation}
	 	\frac{a_E}{a_0} \simeq  \frac{T_0}{T_E} \simeq 10^{-28} ,
\end{equation} 
where $T_E$ denotes the value of $T$ at the end of inflation. Finally we can conclude that the scale factor during inflation ($t < t_E$) can be expressed as:
\begin{equation}
	\label{eq_inflation:scale_factor}
 	a(t) \simeq a_E \exp\left[ H_I (t - t_E) \right]
\end{equation} 
In the following, we simplify the problem by assuming the Universe to be dominated by radiation between $t_E$ and $t_0$. With this assumption we can directly get $H_I \simeq 1/(2 t_E) \simeq 10^{56}/(2 t_0)$.

\subsubsection{A solution to the horizon problem.}
	\label{sec_inflation:solve_horizon}
	In this Section we show that inflation offers a solution to the horizon problem. As explained in Sec.~\ref{sec_introduction:Spectrum}, the comoving distance at a given time defines the size of a causally connected patch at that time. As a consequence, in order to solve the horizon problem, we need the comoving distance at the end of inflation to be larger than the size of the observable Universe at present time. As stated at the beginning of this Section, we assume the Universe to be dominated by radiation between $t_E$ and $t_0$. Under this assumption the comoving distance $d_{p}(t)$ traveled by a photon emitted at a given time $t<t_E$ is:
	\begin{equation}
	\label{eq_inflation:comoving_distance_inflation}
	\begin{aligned}
		d_{p}(t) & =  \int_{t}^{t_{E}}  \ \frac{ \textrm{d}\hat{t} }{a(\hat{t})} + \int_{t_E}^{t_0}  \ \frac{ \textrm{d}\hat{t} }{a(\hat{t})} =  \\
		& = \frac{1}{a_E H_I} \left\{ \exp\left[ H_I (t_E - t)\right] \right\}+ t_0 \left[ 1 - \left( \frac{t_E}{t_0} \right)^{\frac{1}{2}}  \right] \simeq \\
		& \simeq 10^{-28} \left\{ \exp\left[ H_I (t_E - t)\right] \right\}+ t_0 \ ,
	\end{aligned}
	\end{equation}
	where in the last line we have used $t_E/t_0 \simeq 10^{-56}$, $a_E/a_0 \simeq 10^{-28}$ and $H_I \simeq 10^{56}/(2t_0)$. For a sufficiently long period of inflation, the first term on the right hand side becomes larger than $t_0$, the comoving distance at the end of inflation thus becomes larger than the size of the observable Universe at present time and the horizon problem is solved. To quantify the minimal duration of inflation in order to solve horizon problem, it is useful to introduce the number $N(t)$ of e-foldings from the end of inflation, defined as:
	\begin{equation}
		\label{eq_inflation:e_folds_problems}
		N(t) \equiv - \ln\left( \frac{a(t)}{a(t_E)}\right) = H_I \left( t_E - t \right) \ ,
	\end{equation}
	where the minus is introduced in order to have $N > 0$ during inflation. As $10^{28} \simeq e^{64}$, we can directly conclude that for $N(t) \gtrsim 64$ the first term on the right hand side of Eq.~\eqref{eq_inflation:comoving_distance_inflation} becomes larger than $t_0$. Notice that this result matches with the well know requirement of around $60$ e-foldings.

\subsubsection{A solution to the flatness problem.}
	\label{sec_inflation:solve_flatness}
  	Let us reconsider the argument of Sec.~\ref{sec_inflation:flatness_problem}. Starting from the present constraints on the value of $\Omega_k$, and assuming $T_{rh} \simeq 10^{15}\,$GeV, we can use Eq.~\eqref{eq_inflation:omega_k_evolution} to get $\Omega_k(t_E) \simeq 10^{-54}$. The evolution should now continue through the inflationary phase where $a(t)$ is given by Eq.~\eqref{eq_inflation:scale_factor}. We can thus use the definition of $N(t)$ given in Eq.~\eqref{eq_inflation:e_folds_problems} to express $a(t)$ as:
  	\begin{equation}
  		a(t) \simeq a_E \exp\left[- N(t) \right] \ .
  	\end{equation}
  	Let us assume that at a given time $t < t_E$ we have $\Omega_k(t) \simeq 0.1 $ and $\Omega_I(t) \simeq 0.9 $, where $\Omega_I(t) \equiv \rho_I/\rho_c$ is the normalized energy density associated with inflation. As inflation typically lasts for at $64$ e-foldings the scale factor increases by a factor $10^{28}$. It should then be clear that at the end of inflation we have:
  	\begin{equation}
  		\Omega_k(t_E) \simeq \frac{\Omega_k(t)/a^2(t_E)}{\Omega_k(t)/a^2(t_E) + \Omega_I(t)} \simeq 10^{-57} .
  	\end{equation}
	As the value of $\Omega_k$ is exponentially decreasing during inflation, this provides a dynamical mechanism to explain the small value of $\Omega_k(t_E)$. A nice physical interpretation of this effect can be provided by considering the case of an inflating balloon. Until the size of the balloon is small, its curvature can be appreciated even locally. As the balloon inflates, it flattens and locally it is not possible to appreciate its curvature anymore.

\subsubsection{A solution to the monopole problem.}
	\label{sec_inflation:solve_monopoles}
	The solution to the monopole problem is actually similar to the solution to the flatness problem discussed in the previous Section. As during inflation the scale factor increases by a factor $10^{26}$, the energy density of monopoles that are generated before inflation drops by a factor $10^{78}$. This clearly implies that even if a large amount of monopoles is generated before inflation, they cannot affect the evolution of the observable Universe. In this picture, topological defects are thus only generated before inflation. In particular, monopoles are generated during phase transitions at GUT scales ($T\gtrsim 10^{15}\,$GeV). As a consequence, if the reheating temperature $T_{rh} $ happens to be smaller than $10^{15}\,$GeV we generate particles at high energy, avoiding the generation of monopoles.

\section{The simplest realization of inflation.}
\label{sec_inflation:simplest_Inflation}
As explained in Sec.~\ref{sec_inflation:solution_LCMD}, an early phase of exponential expansion solves the problems of the $\Lambda$CDM model stated in Sec.~\ref{sec_inflation:problems_LCDM}. As discussed in Sec.~\ref{sec_introduction:energy_content_and_history}, an exponentially increasing scale factor can be obtained if the Universe is dominated by the cosmological constant \textit{i.e.} by a matter species with $w = -1 $. Actually, if the only contribution to the stress-energy tensor of the Universe is given by this term, we obtain a dS spacetime\footnote{More details on dS spacetime are given in Appendix~\ref{appendix_GR:dS_spacetime}}. As this is a static solution of Einstein Equations, this configuration clearly cannot correspond to a phase of the evolution of our Universe. In order to realize inflation we thus need a graceful exit from the phase of exponential expansion and the simple way to implement this feature is to consider a matter species with varying equation of state parameter $w_I(t)$. In particular we want $w_I\simeq -1$ at very early times, and $w \gtrsim 0$ approaching the end of inflation. 

\subsection{The background dynamics.}
The simplest way to realize this configuration is by considering a scalar field $\phi$, with a canonical kinetic term and a minimal coupling with gravity, that as usual is described by an Einstein-Hilbert term:
\begin{equation}
  \label{eq_inflation:action}
    \mathcal{S}= \int\mathrm{d}t\mathrm{d}^3x \sqrt{|g|}\left( \frac{R}{2 \kappa^2} -  X - V(\phi) \right),
  \end{equation}
where $X \equiv g^{\mu\nu} \partial_\mu \phi \partial_\nu \phi/2$ and $ \kappa^{2} \equiv 8 \pi G_N$. As usual we consider a FLRW metric given in Eq.~\eqref{eq_intro:FLRW}, and for simplicity we assume the curvature to be zero. Under these assumptions the metric simply reads: $g_{\mu\nu} = \text{diag}(-1,a^2(t),a^2(t),a^2(t))$. Using the definition of stress-energy tensor (see Eq.~\eqref{appendix_GR:Stress_energy_tensor}) it is possible to show that the energy density $\rho_{\phi}$ and pressure $p_{\phi}$ associated with the scalar field $\phi$ are:
\begin{equation}
\label{eq_inflation:p_and_rho_gener}
\rho_{\phi} = \frac{\dot{\phi}^2}{2} + \frac{a^{-2}}{2} \left( \vec{\nabla} \phi \right)^2 +  V(\phi) \ ,\qquad \qquad p_{\phi} =  \frac{\dot{\phi}^2}{2} - \frac{a^{-2}}{2} \left( \vec{\nabla} \phi \right)^2 - V(\phi)\ ,
\end{equation}
where $\vec{\nabla} \equiv \partial / (\partial \vec{x}) $ is the ordinary flat space gradient operator. The Friedmann Equations describing the system are again given by Eq.~\eqref{eq_intro:mod_friedmann}, and the equation of motion for the scalar field $\phi$ can be expressed as: 
\begin{equation}
\label{eq_inflation:scalar_eq_motion}
\ddot{\phi} + 3 H \dot{\phi} - \frac{\nabla^2 \phi}{a^2} + \frac{\partial V}{\partial \phi} = 0 \ .  
\end{equation}
To proceed with our treatment we assume the scalar field $\phi$ to be homogeneous. Under this assumption we have $\vec{\nabla} \phi=0$, and thus $\rho_{\phi}$ and $p_{\phi}$ can be expressed as:
\begin{equation}
\label{eq_inflation:p_and_rho}
\rho_{\phi} = \frac{\dot{\phi}^2}{2} + V(\phi) \ ,\qquad \qquad p_{\phi} =  \frac{\dot{\phi}^2}{2} - V(\phi) \ ,
\end{equation}
and thus Eqs.~\eqref{eq_intro:mod_friedmann} for this system reduce to:
\begin{equation}
\label{eq_inflation:friedmann_scalar}
3 \kappa^{-2} H^2 = \frac{\dot{\phi}^2}{2} + V(\phi) \ , \qquad \qquad - 2 \kappa^{-2} \dot{H} = \dot{\phi}^2 \ .  
\end{equation}
The equation of state for this scalar field thus reads:
\begin{equation}
\label{eq_inflation:eq_of_state}
-\frac{2}{3}\frac{\dot{H}}{H^2} = \frac{ p_{\phi} + \rho_{\phi} }{\rho_{\phi} } =  1 + w_{\phi} =  \frac{\dot{\phi}^2}{\frac{\dot{\phi}^2}{2} + V(\phi)}  \ .
\end{equation}
It is clear that for $\dot{\phi}^2/V \ll 1$ we get $w \simeq -1$, that realizes an exponentially growing scale factor.  Moreover, for a homogeneous scalar field, the equation of motion of Eq.~\eqref{eq_inflation:scalar_eq_motion} reduces to:
\begin{equation}  
\label{eq_inflation:eq_motion}
\ddot{\phi} + 3 H \dot{\phi} + \frac{\partial V}{\partial \phi} = 0 \ .  
\end{equation}
Actually these three equations are not independent and the system is completely specified by Eq.~\eqref{eq_inflation:friedmann_scalar}. A solution for this system of differential equations sets the so-called background solution, that actually corresponds to the evolution of the scale parameter $a(t)$ and of the homogeneous scalar field $\phi(t)$.  In order to produce an appropriate description of inflation it is useful to introduce the number $N(t)$ of e-foldings from the end of inflation (that occurs at $t = t_f$) as:
\begin{equation}
\label{eq_inflation:number_of_efoldings}
N(t) \equiv  - \int_{a_f}^{a(t)}  \textrm{d} \ln \hat{a}  =  - \int_{t_f}^{t} H(\hat{t}) \textrm{d}\hat{t} = - \ln\left( \frac{a(t)}{a(t_f)}\right).
\end{equation}
This quantity measures the number of Hubble times that passed from the end of inflation to a given instant $t$. The minus sign is introduced in order to have $N(t)>0$ for $t<t_f$ \textit{i.e.} this quantity is positive during inflation and it is monotonically increasing as we go deeper ( \textit{i.e.} back in time) into the inflationary phase. 

\subsection{Scalar and tensor fluctuations.}
\label{sec_inflation:scalar_and_tensor}
It is now interesting to consider small inhomogeneous perturbations around the homogeneous background solution. In this Section we only give a brief review of the procedure and we discuss the most interesting results. A detailed analysis of this problem is given in Appendix~\ref{appendix_perturbations:Cosmological_perturbations}. We start by considering the action of Eq.~\eqref{eq_inflation:action}, and instead of directly fixing $\phi$ and $g_{\mu\nu}$ to be homogeneous, we consider the expansion:
\begin{equation}
\label{eq_inflation:perturbations}
  g_{\mu\nu}(t,\vec{x}) = {}^{(0)}g_{\mu\nu}(t) + \delta g_{\mu\nu}(t,\vec{x}) \ , \qquad \qquad \phi(t,\vec{x}) =  {}^{(0)}\phi(t) + \delta \phi (t,\vec{x}) \ , 
\end{equation}
where the background evolution described in the previous paragraph is now described in terms of quantities with a superscript ${}^{(0)}$. After a decomposition of the metric perturbations into \emph{scalar}, \emph{vector} and \emph{tensor} perturbations, we proceed with a gauge fixing procedure in order to express the problem in terms of two physically relevant quantities:
\begin{itemize}
  \item The comoving curvature perturbation, $\zeta(t,\vec{x})$. This quantity is defined as a combination of the scalar field perturbation and of the scalar part of the metric perturbation. It is possible to show that at $\delta \phi (t,\vec{x}) = 0 $ this quantity is proportional to the perturbation of the scalar curvature. 
  \item The traceless traverse spatial tensor $\gamma_{ij}(t,\vec{x})$. Following the discussion of Appendix~\ref{appendix_GR:GW}, it is natural to interpret this quantity as a propagating GW.
\end{itemize}
As $\gamma_{ij}$ contains two independent degrees of freedom, that corresponds to the two polarizations of the GW, it is useful to express it as $\gamma_{ij} = h_\alpha(t,\vec{x}) \ e^{\alpha}_{ij}$ where $e^{\alpha}_{ij}$ are the two polarization vectors and $\alpha = +,\times$. As usual, the observable quantities $\Delta^2_s (k,\tau) $ and $\Delta^2_t(k,\tau) $ are defined in terms of correlators: 
\begin{equation}
\begin{aligned}
  \label{eq_inflation:power_spectra}
  \langle  \zeta(\tau,x_1) \zeta(\tau,x_2)  \rangle &  \equiv \int \frac{\textrm{d}^3 \vec{k} }{4\pi}  \frac{\Delta^2_s (k,\tau)}{ k^3} e^{-i \vec{k} (\vec{x}_1 - \vec{x}_2)} \ , \\
  \langle  h_\alpha(\tau,x_1) h_\alpha(\tau,x_2) \rangle & \equiv \  \int \frac{\textrm{d}^3 \vec{k} }{4\pi}  \frac{\Delta^2_t (k,\tau)}{ k^3} e^{-i \vec{k} (\vec{x}_1 - \vec{x}_2)} \ ,
\end{aligned}
\end{equation}
where $\vec{q}$ is the comoving wave-vector and $\Delta^2_s (k,\tau) $ and $\Delta^2_t(k,\tau) $ are the (dimensionless) \emph{scalar} and \emph{tensor power spectra} respectively. As shown in Fig.~\ref{fig_inflation:scheme_fluctuations}, quantum fluctuations are generated during inflation, where the Universe is a nearly dS spacetime, and they grow until they become super-horizon. As discussed in Appendix~\ref{appendix_perturbations:Cosmological_perturbations}, when perturbations reach this regime, they freeze out and their evolution becomes classical. During radiation and matter domination $a(t)$ scales as $t^{1/2}$ and $t^{2/3}$ and thus fluctuations may re-enter the horizon. This actually happens when $a(t)/k $ becomes equal to $H^{-1}$. The scalar and tensor power spectra defined in Eq.~\eqref{eq_inflation:power_spectra} should thus be evaluated when they re-enter the horizon \textit{i.e.} at horizon crossing ($ k c_s \tau = 1$ for scalar perturbations and $ k \tau = 1$ for tensor perturbations).\\

\begin{figure}
\centering
\includegraphics[width=.8\textwidth]{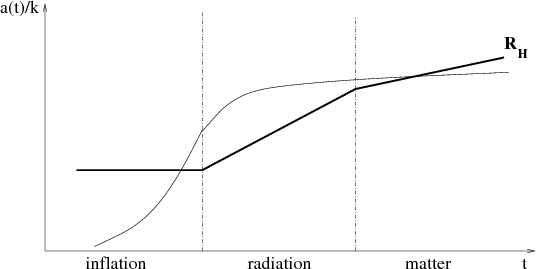}
\caption{Schematic representation of the evolution of the scale of a comoving physical perturbation (thin line), with respect to the Hubble radius $R_H \simeq H^{-1}$ (thick line). The evolution is shown during inflation ($R_H \simeq const$), radiation ($R_H \simeq 2 t$) and matter ($R_H \simeq 3 t /2$) domination. \label{fig_inflation:scheme_fluctuations}}
\end{figure} 

\noindent Before giving the explicit expressions for the spectra it is useful to introduce the Hubble slow-roll parameters:
\begin{equation}
\label{eq_inflation:slow_roll_H}
\epsilon_{H} \equiv - \frac{\textrm{d} \ln (H/H_f)}{\textrm{d} \ln a} =  - \frac{\dot{H}}{H^2} \ , \qquad \qquad  \eta_{H} \equiv -  \frac{\textrm{d} \ln( \dot{\phi} / \dot{\phi}_f )}{\textrm{d} \ln a} = - \frac{\ddot{\phi}}{\dot{\phi} H } \ ,
\end{equation}
where $H_f$ and $\dot{\phi}_f$ are the values of $H$ and $\phi$ at the end of inflation. More details on these parameters and on their properties are given in the following Section. For the moment we can just use them as they are extremely useful to parametrize the spectrum. The explicit expressions for $\Delta^2_s (k)$ and $\Delta^2_t (k)$ at horizon crossing are given by Eq.~\eqref{appendix_perturbations:scalar_power_spectrum_final} and Eq.~\eqref{appendix_perturbations:tensor_power_spectrum_final} respectively:
\begin{equation}
\begin{aligned}
\label{eq_inflation:power_spectra_final}
\left.   \Delta^2_s (k,\tau) \right|_{\tau = (k c_s)^{-1}} & =   \frac{ 1 }{8 \pi^2 } \frac{H^2 \kappa^2}{  c_s \ \epsilon_H }\ ,\\
\left.   \Delta^2_t (k,\tau) \right|_{\tau = k^{-1}} & = 2  \left( \frac{ \kappa H}{ \pi } \right)^2  \ ,
\end{aligned}
\end{equation}
where we have defined the speed of sound $c_s^2$ as:
\begin{equation}
\label{eq_inflation:speed_of_sound}
c_s^2 \equiv {}^{(0)}\left( \left . \frac{\delta p }{\delta \rho}\right|_{\delta \phi = 0} \right) ={}^{(0)}\left( \frac{p_{,X}}{\rho_{,X}} \right) = {}^{(0)}\left(  \frac{p + \rho}{2 X \rho_{,X}} \right) \ ,
\end{equation}
where, consistently with the notation of Appendix~\ref{appendix_perturbations:Cosmological_perturbations}, we have defined ${p}_{,X}\equiv \partial p/\partial X$ and ${\rho}_{,X}\equiv \partial \rho/\partial X$. Notice that for the case discussed in this Section we have $c_s =1$. Using these expressions for the spectra, it is actually possible to define a set of constraints on the different models for inflation. In this work we are interested in discussing the constraints on the spectra that come from CMB observations.\\

\noindent
In order to characterize the scalar power spectrum of Eq.~\eqref{eq_inflation:power_spectra_final}, it is useful to introduce the \emph{scalar spectral index} $n_s(k)$ as (in the following we set $c_s =1$):
\begin{equation}
\label{eq_inflation:scalar_spectral_index}
n_s(k) \equiv 1 + \frac{\textrm{d} \ln \Delta^2_s (k)}{\textrm{d} \ln k} = 1 -  \left(  2  \eta_H - 4 \epsilon_H \right) \left( 1 - \epsilon_H \right)^{-1} \ .
\end{equation}
The scalar spectral index quantifies the scale ($k$) dependence of the scalar power spectrum. A scale-invariant power spectrum corresponds to $n_s = 1$ and thus a measurement of this quantity gives important information on inflation. Similarly we define the \emph{tensor spectral index} $n_t(k)$ as:
\begin{equation}
\label{eq_inflation:tensor_spectral_index}
n_t(k) \equiv \frac{\textrm{d} \ln \Delta^2_t (k)}{\textrm{d} \ln k} =- \frac{2  \epsilon_H}{  1 - \epsilon_H } \ .
\end{equation}
Notice that this quantity is only depending on $ \epsilon_H $. Actually we can define another quantity, called \emph{tensor-to-scalar ratio}:
\begin{equation}
\label{eq_inflation:tensor_to_scalar_ratio}
r \equiv \frac{\Delta^2_t }{\Delta^2_s } = 16 \epsilon_H \ ,
\end{equation}
that measures the amplitude of tensor perturbations with respect to the amplitude of scalar perturbations, which is also depending on $ \epsilon_H $ only. In the simplest realization of inflation discussed so far, the tensor-to-scalar ratio $r$ and the tensor spectral index $n_t$ are thus related by the so-called \emph{consistency relation}:
\begin{equation}
	r = - 8 n_t \ .
\end{equation}
A direct measurement of primordial tensor fluctuations would thus offer a way to falsify single-field slow-roll inflation. While a detection of primordial tensor fluctuations is still missing, it is possible to set constraints on the upper value for $r$. The constraint on this quantity set by Planck~\cite{Ade:2015lrj} is reported in Sec.~\ref{sec_inflation:CMB_constraints}. \\

\noindent
Finally, to quantify the scale ($k$) dependence of the scalar spectral index we define the \emph{running of the scalar spectral index}, usually denoted with $\alpha_s$, as:
\begin{equation}
\label{eq_inflation:running}
\alpha_s \equiv \frac{\textrm{d} \ln n_s (k)}{\textrm{d} \ln k} .
\end{equation} 
As we discuss in Sec.~\ref{sec_inflation:slow_roll}, in the case of slow-roll inflation this quantity is expected to be small (second-order in the slow-roll parameters).

\subsection{CMB constraints.}
\label{sec_inflation:CMB_constraints}
In order to present the constraints set by CMB observations, it is useful to report an alternative parametrization of the power spectra~\cite{Ade:2015lrj}:
\begin{equation}
\label{eq_inflation:spectra_parametrization}
\begin{aligned}
 \Delta^2_s (k) &  = A_s \left( \frac{k}{k_*}\right)^{n_s|_{k_*} -1 + \frac{1}{2}  \left. \frac{\textrm{d} n_s}{\textrm{d}\ln k}\right|_{k_*} \ln(k/k_*) + \dots }  \ , \\ 
  \Delta^2_s (k) & = A_t \left( \frac{k}{k_*}\right)^{n_t|_{k_*} + \frac{1}{2} \left.  \frac{\textrm{d} n_t}{\textrm{d}\ln k}\right|_{k_*} \ln(k/k_*) + \dots }  \ ,
  \end{aligned}
\end{equation}
where $k_*$ is usually called the pivot scale. Using this parametrization, we are actually expanding the power spectrum in powers of $(k/k_*)$ around the pivot scale.

\noindent
The first constraint that we can set on the power spectra is the so called \emph{COBE Normalization}. This constraint sets the value of the scalar power spectrum at the pivot scale $k_* = 0.05\,$Mpc${}^{-1}$. In particular, using the parametrization of Eq.~\eqref{eq_inflation:spectra_parametrization}, the COBE Normalization reads~\cite{Ade:2015lrj}:
\begin{equation}
\label{eq_inflation:COBE_normalization}
\left. \Delta^2_s (k) \right|_{k = k_*} \equiv A_s = (2.21 \pm 0.07) \cdot 10^{-9} \ .
\end{equation}
Comparing this constraint with the expression for the power spectrum given in Eq.~\eqref{eq_inflation:power_spectra_final} and with Eq.~\eqref{eq_inflation:friedmann_scalar}, it is clear that this constraint basically sets the scale of inflation. \\

\noindent
We can proceed by discussing the constraint set on the scalar spectral index defined in Eq.~\eqref{eq_inflation:scalar_spectral_index}. Comparing this definition with the parametrization of Eq.~\eqref{eq_inflation:spectra_parametrization}, it is clear that $n_s$ measures the scale-dependence of the scalar power spectrum. A main result of the Planck mission~\cite{Ade:2015lrj} is the observation that at the pivot scale $k_* = 0.05 \,$Mpc${}^{-1}$ the power spectrum is nearly, but not exactly, scale-invariant:
\begin{equation}
\label{eq_inflation:scalar_spectral_index_planck}
n_s(k)|_{k = k_*} = 0.9677 \pm 0.0060  \ , \qquad (68 \% \text{ CL, Planck TT+lowP+lensing}) \ ,
\end{equation}
The Planck mission~\cite{Ade:2015lrj} is also setting constraints on the value of the running $\alpha_s$ (defined in Eq.~\eqref{eq_inflation:running}). In particular the value of $\alpha_s$ at $k_* = 0.05 \,$Mpc${}^{-1}$ is found to be compatible with zero~\cite{Ade:2015lrj}.\\

\noindent
Using the measurements of the CMB polarization Planck is also setting constraints on the generation of primordial tensor modes~\cite{Ade:2015xua,Ade:2015lrj}. Usually these constraints are expressed as an upper bound on the value of the tensor-to-scalar ratio $r$. The $95\%$ CL for $r$ given by the Planck mission~\cite{Ade:2015lrj} is usually defined at the pivot scale $k_* = 0.002 \,$Mpc${}^{-1}$:
\begin{equation}
\label{eq_inflation:tensor_to_scalar_planck}
r_{0.002} \equiv r(k)|_{k=k_*} < 0.11 \ ,  \qquad (95 \% \text{ CL, Planck TT+lowP+lensing}) \ .
\end{equation}
The marginalized $68\%$ and $95\%$ confidence level regions for $n_s$ and $r_{0.002}$ set by the Planck mission~\cite{Ade:2015lrj} are shown in Fig.~\ref{fig_inflation:planck_ns_r}.  

\begin{figure}[h]
\centering
\includegraphics[width=.95\textwidth]{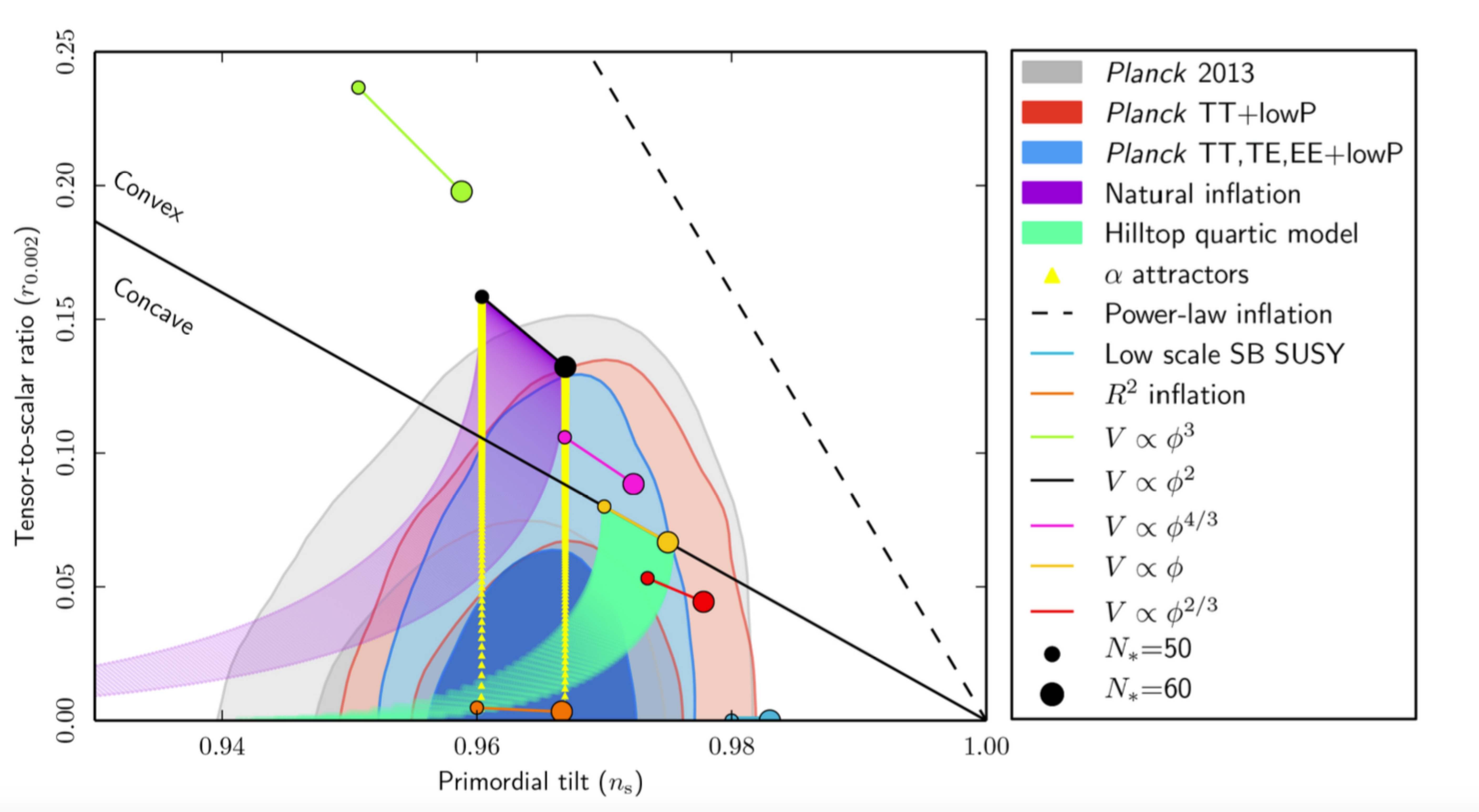}
\caption{ Planck~\cite{Ade:2015lrj} plot of the predictions for different models on the $(n_s,r)$ plane, compared with the marginalized joint $68 \%$ and $95 \%$ CL regions for $n_s$ (evaluated at the pivot scale $k_* = 0.05 \,$Mpc${}^{-1}$) and $r_{0.002}$ (\textit{i.e.} $r(k)$ evaluated at the pivot scale $k_* = 0.002 \,$Mpc${}^{-1}$). \label{fig_inflation:planck_ns_r}}
\end{figure}

\noindent
In the plot of Fig.~\ref{fig_inflation:planck_ns_r} we also have the predictions for the values of $n_s$ and $r$ given by some inflationary models. Notice that in this plot, the predictions for the different models are expressed in terms of $N_* = N(k_*)$,the number of e-foldings at which the scale $k_*$ leaves the horizon. While the definition of some of these models shown in Fig.~\ref{fig_inflation:planck_ns_r} is postponed to Sec.~\ref{sec_inflation:Inflationary_models}, we conclude this section by explaining the procedure to compute $N(k_*)$. 

\begin{figure}[h]
\centering
\includegraphics[width=0.9\textwidth]{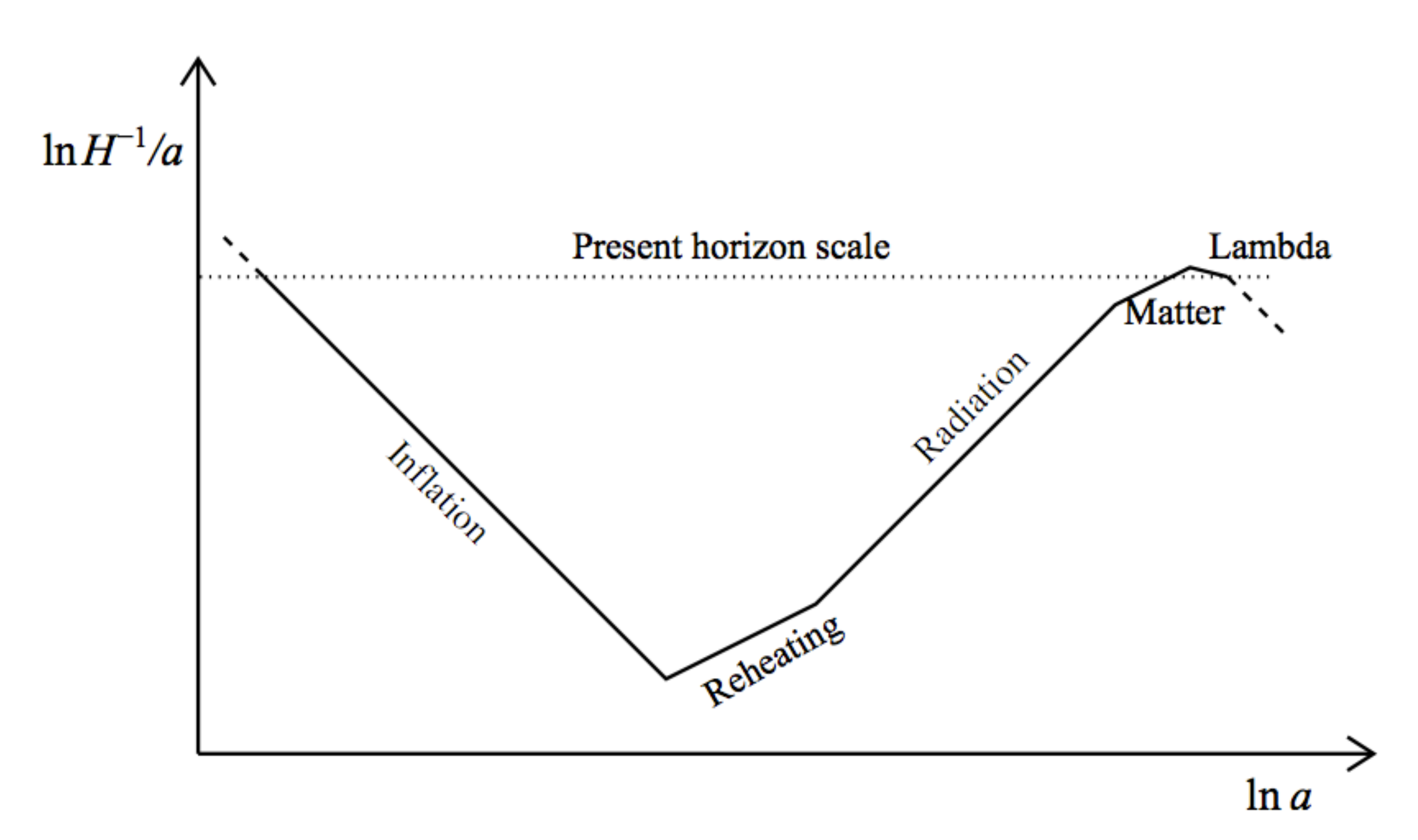}
\caption{Evolution of the Hubble radius $R_H = (aH)^{-1}$ during the different epochs that are relevant for the calculation of $N_*$~\cite{Liddle:2003as}. \label{fig_inflation:scale_inflation}}
\end{figure}

\noindent
As explained at the beginning of Sec.~\ref{sec_inflation:scalar_and_tensor} (and as shown in Fig.~\ref{fig_inflation:scheme_fluctuations}), scalar and tensor fluctuations are generated at small scales during inflation, they grow until they leave the horizon, they freeze out on super-horizon scales and they finally re-enter the horizon during radiation and matter domination. We start by giving a rough estimate of the value of $N_* = N(k_*)$ at which the scale $k_* = 0.05 \,$Mpc${}^{-1}$ leaves the horizon, we then proceed by giving the accurate formula for $N_*$. A picture of the different epochs that should be kept into account for the calculation of $N_*$ is shown in Fig.~\ref{fig_inflation:scale_inflation}. \\

\noindent
To determine the value of $N$ at which a given scale $k$ leaves the horizon we simply use $k = aH$. For this purpose we thus need an explicit expression of $a$. Assuming $a_E \simeq a_{rh}$, $T_{rh} \simeq 10^{15}\,$GeV and $H_I \simeq 10^{15}\,$GeV this is simply given by Eq.~\eqref{eq_inflation:scale_factor}: 
\begin{equation}
	1.3 \times 10^{-58} \, \kappa^{-1} \simeq 0.05 \,\text{Mpc}^{-1}  \simeq  k_* = aH \simeq a_E H_I \exp\left[ H_I (t - t_E)\right] \ .
\end{equation}
Using the definition of $N$ given in Eq.~\eqref{eq_inflation:e_folds_problems} we get: 
\begin{equation}
	H_I (t_E - t) \equiv N_* =- \ln\left( \frac{k_*}{a_E H_I} \right) =- \ln\left( 1.4 \times 10^{-27}\right) \simeq 61.8 \ ,
\end{equation}
where we have used $a_e/a_0 \simeq 2.3 \times 10^{-28}$ and $H_I \simeq 0.4 \times 10^{-3} \kappa^{-1}$. To get this value for $N$ we have made several assumptions both on inflation and on the physics of reheating. For example we have fixed the value of the Hubble parameter (the energy scale) during inflation ($H_I \simeq 10^{15}\,$GeV), we have assumed reheating to be instantaneous ($a_E \simeq a_{rh}$) and we have fixed the temperature during reheating ($T_{rh} \simeq 10^{15}\,$GeV). In general, there is no reason to make these assumptions and thus we need a more general formula to compute $N_*$.\\

\noindent
In order to compute the correct expression for $N_*$, we need to define a certain number of parameters that specify the physics of inflation and reheating. An accurate formula to compute $N_*$ was computed by Liddle and Leach in~\cite{Liddle:2003as}. To derive this formula we start by expressing $k_* = a(k_*)H(k_*)$ as:
\begin{equation}
	\label{sec_inflation:liddle_N}
	\frac{k_*}{a_0 H_0} = \frac{a(k)H(k)}{a_0 H_0} = \frac{a(k)}{a_{E}} \, \frac{a_E}{a_{rh}} \, \frac{a_{rh}}{a_{eq}} \, \frac{H(k)}{H_{eq}} \, \frac{a_{eq}H_{eq}}{a_0 H_0} \ ,
\end{equation}
where $a_0$, $a_{eq}$,$a_{rh}$, $a_{E}$ denote the values of the scale factor at present time, at matter radiation equality, at the end of reheating and at the end of inflation respectively. The notation for $H$ is analogous. In order to compute $N_*$ we can use the definition of $N$ (given in Eq.~\eqref{eq_inflation:e_folds_problems}) to express $a(k)/a_{E}$ as $e^{-N_*}$. As a consequence, we can substitute into Eq.~\eqref{sec_inflation:liddle_N} to get:
\begin{equation}
	\label{sec_inflation:liddle_N_2}
	N_* = -\ln \left( \frac{k_*}{a_0 H_0}\right) + \ln \left( \frac{a_E}{a_{rh}} \right)+ \ln \left(  \frac{a_{rh}}{a_{eq}} \right) +\ln \left( \frac{H(k)}{H_{eq}}\right) +\ln \left( \frac{a_{eq}H_{eq}}{a_0 H_0} \right) \ .
\end{equation}
The value of $N_*$ can thus be computed by specifying all the different contributions that appear in Eq.~\eqref{sec_inflation:liddle_N_2}. A convenient method to express these contribution (in particular of the terms depending on reheating) was proposed by Martin and Ringeval in~\cite{Martin:2010kz}. In this work, the evolution of the scale factor during reheating is expressed in terms of an (effective) equation of state parameter $w_{rh}$ for reheating. Moreover, in order to specify the duration of reheating (it terminates when radiation dominates the evolution), we need to specify both the total energy density $\rho_{rh}$ and the radiation energy density $\rho_{r,rh}$ during reheating. The latter, can actually be expressed~\cite{Martin:2010kz} in terms of $g_{rh}$, effective number of massless degrees of freedom~\cite{Dodelson-Cosmology-2003,Mukhanov:991646,weinberg2008cosmology} defined as:
\begin{equation}
	\rho_r(T) = \sum_i \rho_i(T) = \frac{\pi^2}{30} g(T) T^4 \ .
\end{equation}
Finally, we should specify the ratio between the energy density at which the scale $k_*$ leaves the horizon (that assuming slow-roll can be expressed in terms of \textit{i.e.} $V_*$, see Sec.~\ref{sec_inflation:slow_roll}) and the energy density $\rho_{end}$ at the end of inflation. \\

\noindent
Using these definitions, the number of e-foldings $N_*$ at which the scale $k_*$ leaves the horizon can finally be expressed as~\cite{Liddle:2003as,Planck:2013jfk,Ade:2015lrj}:
\begin{equation}
\label{eq_inflation:e_fold_k}
N_* \simeq 67 -\ln \left( \frac{k_*}{a_0 H_0}\right) + \frac{1}{4}\ln \left(  \frac{V_*^2 \kappa^4 }{\rho_{end}}\right)+   \frac{1 - w_{rh} }{12(1 + w_{rh} ) } \ln \left( \frac{\rho_{rh} } {\rho_{end} } \right) -\frac{1}{12} \ln(g_{rh}) \ .
\end{equation}
It is interesting to notice that for $0.002\,$Mpc${}^{-1} \lesssim k_* \lesssim 0.05) \,$Mpc${}^{-1}$, $V_*^2 \kappa^4 / \rho_{end} \simeq 1$ (\textit{i.e.} scale-invariant inflation at $V_* \simeq \kappa^{-1}$), $w_{rh}\simeq [0,1/3]$ (\textit{i.e.} reheating effectively dominated by matter or radiation), $\left( 10^3 \, \text{GeV}\right)^4  \lesssim \rho_{rh} \lesssim \rho_{end} $ (\textit{i.e.} reasonable range for the energy density during reheating) and $g_{rh}\simeq 10^3$ (large number of bosonic degrees of freedom during reheating) we recover the usual $50 \lesssim N\lesssim 60$. As we discuss in the following, for a wide class of models it is possible to express $n_s$, $r$ and $\alpha_s$ in terms of the small quantity $1/N$. It is thus useful to notice that keeping constant all the parameters except for $k_*$, the difference $\Delta N$ between $N(k_* = 0.002 \,$Mpc${}^{-1})$ and $N(k_* = 0.05 \,$Mpc${}^{-1})$ is $\Delta N \simeq 3$. As $\Delta N $ is much smaller than $50 \lesssim N\lesssim 60$, at the lowest order it can be safely neglected.

\subsection{The slow-rolling regime.}
\label{sec_inflation:slow_roll}
A convenient description of inflation is obtained by assuming the inflaton to be ``slow-rolling'' in its potential. A proper definition of the requirement is usually given in terms of the slow-roll parameters of Eq.~\eqref{eq_inflation:slow_roll_H}. In particular, the field is said to be in the slow-rolling regime if the condition $\epsilon_{H} , \eta_{H} \ll 1$ is satisfied. The parameters $\epsilon_H$ and $\eta_H$ defined in Eq.~\eqref{eq_inflation:slow_roll_H} are usually referred to as Hubble slow-roll parameters\footnote{It is interesting to point out that it is actually possible to give a slightly different parametrization for the slow-roll parameters:
\begin{equation}
\label{eq_inflation:slow_roll_hierarchy}
\epsilon_{0} \equiv \frac{H }{H_f} \ , \qquad \qquad \epsilon_{i+1} \equiv - \frac{\textrm{d} \ln| \epsilon_{i} | }{\textrm{d} \ln a} \ ,
\end{equation} 
where $H_f$ is the value of $H$ at the end of inflation. With this definition it is actually possible to define a whole hierarchy of slow-roll parameters. As we will see in Chapter~\ref{chapter:beta}, this definition for the slow-roll parameters is particularly convenient when we discuss inflation in terms of the $\beta$-function formalism.} and they are usually denoted as $\epsilon_{H}$ or $\eta_{H}$. As we discuss in the following, in the slow-rolling regime these parameters can actually be related to the shape of the inflationary potential. Notice that the Hubble slow-roll parameters should not be confused with the potential slow-roll parameters $\epsilon_{V}$ and $\eta_{V}$, that we define in Eq.~\eqref{eq_inflation:slow_roll_V}. \\

\noindent
We can proceed by computing the lowest order approximations of Eqs~\eqref{eq_inflation:friedmann_scalar} and of the equation of motion for the inflaton~\eqref{eq_inflation:eq_motion}:
\begin{equation}
\label{sec_intro:slow_roll_background}
3 \kappa^{-2} H^2 \simeq V(\phi) \ , \qquad \qquad - 2 \kappa^{-2}  \dot{H} = \dot{\phi}^2 \ , \qquad \qquad 3 H \dot{\phi} + V_{,\phi} \simeq 0 \ ,
\end{equation}
where the subscript $ {}_{,\phi}$ is used to denote differentiation with respect to $\phi$. Using these approximations we can express the slow-roll parameters as:
\begin{equation}
\begin{aligned}
\label{eq_inflation:slow_roll_V}
&\epsilon_{H} = - \frac{\dot{H}}{H^2} \simeq \frac{1}{2 \kappa^2} \left( \frac{ V_{,\phi}}{V} \right)^2 \equiv \epsilon_V \ , \\
& \eta_{H} \equiv - \frac{\textrm{d} \ln(\dot{\phi}/\dot{\phi}_f )}{\textrm{d} \ln a} \simeq  \frac{1}{\kappa^2} \frac{V_{, \phi \phi}}{V} + \frac{\textrm{d} \ln( H/H_f)}{\textrm{d} \ln a} \equiv \eta_V - \epsilon_H \simeq \eta_V - \epsilon_V,
\end{aligned}
\end{equation}
where we have defined $\epsilon_V$ and $\eta_V$, potential slow-roll parameters. 
Notice that in the slow-roll approximation $\epsilon_{H}$ and $\epsilon_{V}$ are almost equivalent. However, in general this is not true, and in particular this will be relevant for the discussion of Chapter~\ref{chapter:pseudoscalar}. We can thus get the approximated expressions for $n_s, \ r$ and $\alpha_s$ in terms of these quantities:
\begin{eqnarray}
\label{eq_inflation:slow_roll_r}
r &\simeq& 16 \epsilon_V \ , \\ 
\label{eq_inflation:slow_roll_ns}
n_s &\simeq& 1 + \left( 2 \eta_V -6 \epsilon_V \right)\left( 1 -\epsilon_V \right)^{-1}\ \simeq 1 + 2 \eta_V -6 \epsilon_V  \ , \\ 
\label{eq_inflation:slow_roll_alpha}
\alpha_s &\simeq & - 24  \epsilon_V^2 + 16 \epsilon_V \eta_V -2 \xi_V \ ,
\end{eqnarray}
where we have defined the second-order slow roll parameter $\xi_V$ as:
\begin{equation}
\label{eq_inflation:slow_roll_xi}
\xi_V \equiv \frac{1}{\kappa^4} \frac{V_{,\phi} V_{,\phi \phi \phi}}{V^2} \ .
\end{equation}
As $n_s, \ r$ and $\alpha_s$ are usually expressed in terms of the number of e-foldings $N$, it is useful to compute its approximated expression in the slow-roll regime:
\begin{equation}
N(t) =   - \int_{t_f}^{t} H(\hat{t}) \textrm{d}\hat{t} \ \simeq  \int_{\phi_f}^{\phi} \kappa^2 \frac{V(\hat{\phi})}{V_{,\phi}(\hat{\phi})} \textrm{d}\hat{\phi}
\end{equation}
Before concluding this Section it is also useful to give the approximate expression of the scalar power spectrum of Eq.~\eqref{eq_inflation:power_spectra_final} in the slow-roll regime:
\begin{equation}
\label{eq_inflation:slow_roll_spectrum}
\Delta^2_s (\phi) \simeq  \frac{ 1 }{24 \pi^2 } \frac{V(\phi) \kappa^4}{ \ \epsilon_V }   \ ,
\end{equation}
it should thus be clear that the COBE normalization sets a constraint on the ratio $V(\phi) \kappa^4/ \epsilon_V $. As $V(\phi) \kappa^4$ is a pure number that expresses the scale of the potential for the inflaton with respect to the Planck scale, this constraint can actually be used to set the scale of inflation.

\section{Inflationary models.}
\label{sec_inflation:Inflationary_models}
Since the first models have been defined, a huge amount of models has been proposed. Giving a complete review of all these models is beyond the scope of this work. A comprehensive discussion of the major part of the existing inflationary models can be found in~\cite{Martin:2013tda}. Slow-roll models of inflation can be roughly divided into three classes: \emph{small-field models}, \emph{large-field models}, \emph{hybrid models}. This classification is set by the field excursion $\Delta \phi$ during inflation. In small-field models the field excursion during inflation is much smaller than the reduced Planck mass (\textit{i.e.} $\kappa \Delta \phi \ll 1$). On the other hand, in large-field models we have $\kappa \Delta \phi \gtrsim 1$ implying that the value of $\phi$ during inflation is larger than the Planck mass $\kappa^{-1}$ (more on the consequences of this point is said in Sec.~\ref{sec_inflation:large_eta_prob}). Finally, hybrid inflation models may arise from multi-field models (more details on multi-field models are given in Sec.~\ref{sec_inflation:generalized_models}) where all the fields, except for one, are heavy and frozen. In this sense, hybrid models can effectively result in single-field models where the minimum of the potential is different from zero. \\

\noindent
Before starting our discussion of inflationary model building, we discuss an interesting (and rather model-independent) constraint on the tensor-to-scalar ratio which is directly related with the field excursion. This constraint is typically referred to as ``Lyth bound'', since it was derived by Lyth in~\cite{Lyth:1996im}. We start by substituting $- 2 \kappa^{-2}  \dot{H} = \dot{\phi}^2$ (Eq.~\eqref{sec_intro:slow_roll_background}) into the definition of $\epsilon_H$ (Eq.\eqref{eq_inflation:slow_roll_H}) to get:
\begin{equation}
	\label{eq_inflation:Lyth_deriv}
	\epsilon_H = - \frac{\dot{H}}{H^2} = \left( \frac{\kappa \dot{\phi}}{2 H}\right)^2 = \frac{\kappa^2 }{4 } \left(\frac{\textrm{d} \phi}{\textrm{d} N} \right)^2 , 
\end{equation}
where we have used $\dot{\phi} = \textrm{d} \phi/\textrm{d} t = - H \textrm{d} \phi/\textrm{d} N$ (from Eq.~\eqref{eq_inflation:number_of_efoldings}). We can proceed by integrating Eq.~\eqref{eq_inflation:Lyth_deriv} to get:
\begin{equation}
	\kappa \Delta \phi = 2 \int_{0}^{N_*} \sqrt{\epsilon_H} \textrm{d} N \simeq 2 \sqrt{\epsilon_H} N_* = \sqrt{\frac{r}{8}} N_* \ , 
\end{equation}
where we have assumed $\epsilon_H$ to be nearly constant during inflation and we have substituted $r = 16 \epsilon_H $ (Eq.~\eqref{eq_inflation:tensor_to_scalar_ratio}). As a consequence we can directly relate the value of $r$, with the field excursion $\Delta \phi$ during inflation. In particular, we find that small-field models typically give a small (unobservable) value for $r$.

\subsection{Large-field models and $\eta$-problem.}
\label{sec_inflation:large_eta_prob}
As explained in the introduction of this Chapter, the first concrete model of inflation, known as \emph{Chaotic inflation} (whose predictions are discussed in~\ref{sec_inflation:chaotic}), was introduced by Linde in~\cite{Linde:1983gd}. In this model the inflationary potential is $V(\phi) = m^2 \phi^2 /2 $ and thus in the slow-roll approximation we have:
\begin{equation}
\epsilon_V = \frac{1}{2 \kappa^2} \left( \frac{ V_{,\phi}}{V} \right)^2 \simeq \frac{2}{(\kappa \phi)^2} \ ,
\end{equation} 
so that slow-rolling is reached for $\phi > \sqrt{2}/\kappa$ \text{i.e.} a super-Planckian regime.\\

\noindent
As explained in the introduction of this section, this feature is typical of a whole set of models (large-field models) where both the field $\phi $ and the field excursion $ \Delta \phi$ are larger than the reduced Planck mass $m_P = \kappa^{-1}$. At this point it is important to stress that a super-Planckian field does not imply that Quantum Gravity (QG) is required! Actually QG is required if the \emph{energy density}, that for slow-roll inflation models is basically given the inflaton potential $V$, is super-Planckian \text{i.e.} $\rho \kappa^4 \gtrsim 1$. In general, this condition is not necessarily satisfied by a super-Planckian field and thus a QG treatment is not required. For example, in the case of Chaotic inflation we have:
\begin{equation}
	\rho_{\phi}  \kappa^4 \simeq V \kappa^4 = \frac{1}{2} ( \kappa m)^2 (\phi \kappa)^2 \ .
\end{equation}
As a consequence, even if the field is super-Planckian, QG is not required until:
\begin{equation}
	\kappa \phi \ll \frac{\sqrt{2}}{ \kappa m} \ .
\end{equation}
While this condition is sufficient to ensure that QG is not required, a super-Planckian field still gives rise to some problems in the definition of the theory. Let us explain this point in detail. \\

\noindent
Since a natural embedding of inflation in the standard model of particle physics is still lacking\footnote{Actually there are several proposals that invoke Physics beyond the Standard Model (BSM). More on this point is said in Sec.~\ref{sec_inflation:generalized_models}, where we discuss some generalizations of the simplest realization of inflation.}, the definition of inflationary models is usually carried out in the context of effective field theories\footnote{The definition of EFT is deeply connected with the concept of renormalization group (RG) introduced by Kenneth G. Wilson~\cite{Wilson:1970ag,Wilson:1971bg,Wilson:1973jj,Wilson:1974mb}. In this context, the theory that describe a physical system is expected to change according to energy scale at which we observe the system.} (EFTs). In order to define a EFT we should start by specifying a cut-off scale, usually denoted with $\Lambda$, that sets the maximal energy at which the EFT is valid. Once the cut-off is set, all the degrees of freedom with energy larger than $\Lambda$ are integrated out. The resulting theory is thus expected to give an accurate description of the physical processes that take place at energy smaller than $\Lambda$, and it is expected to break down when the energy is of order $\Lambda$. In the context of EFT, the effects of high-energy physics are usually described in terms of higher-dimensional (non-renormalizable) operators which are suppressed by powers of $\Lambda$. Clearly, the theory is affected by these higher-dimensional operators, but low energy physics (that take place at energy scales much smaller than $\Lambda$) is expected to be insensitive to the introduction of these operators. However, when we consider large-field inflationary models, the field is expected to be super-Planckian, and thus problems may arise.\\

\noindent
As explained in Sec.~\ref{sec_inflation:slow_roll}, the condition to ensure slow-rolling are related with the flatness of the inflationary potential. In particular, in the simplest realization of inflation discussed so far, in order to define a viable model for inflation we should specify a potential (for example $V(\phi) = m^2 \phi^2 /2 $) that satisfies:
\begin{equation}
	\frac{1}{\kappa} \frac{V_{,\phi}}{V} \ll 1 \ , \qquad \qquad \frac{1}{\kappa^2} \frac{V_{,\phi \phi}}{V} \ll 1 \ .
\end{equation}
Since for energy densities larger than $\kappa^{-4}$ QG is needed, a natural cut-off scale for inflation is given by $m_P \simeq \kappa^{-1}$. As a consequence, if the theory is not protected by some symmetry, higher-dimensional operators of general form:
\begin{equation}
	\mathcal{O} = \mathcal{O}_\Delta \kappa^{\Delta - 4} \ ,
\end{equation}
where $\Delta$ is the mass dimension of the operator $\mathcal{O}_\Delta$, are allowed. Among these operators we have for example:
\begin{equation}
 C_2 \kappa^{-4} (\kappa \phi)^2 \ , \qquad C_3 \kappa^{-4} (\kappa \phi)^3 \ , \dots \ , C_n \kappa^{-4} (\kappa \phi)^n \ ,
 \end{equation}
where $C_i$ are constants, that in order to respect the principle of naturalness should be of order one. Notice that for example the first of these operators induces a order-one correction in the second slow-roll parameter $\eta_V$, defined in Eq.~\eqref{eq_inflation:slow_roll_V} leading to the so-called ``$\eta$-problem''.

\subsection{Possible solutions to the $\eta$-problem.}
\label{sec_inflation:eta_solutions}
In order to solve the $\eta$-problem, we need to specify a mechanism that ensures that radiative corrections (described by higher-dimensional operators) are under control. In particular, we need to make sure that the effects of these corrections are not spoiling the conditions to achieve inflation. As a consequence, we can attempt different strategies in order to solve the $\eta$-problem:
\begin{itemize}
	\item Small-field models.\\
	As explained in Sec.~\ref{sec_inflation:large_eta_prob}, in EFTs higher-dimensional operators are suppressed by the cut-off scale. Choosing a field that satisfies $\kappa \phi \ll 1 $, radiative corrections are not allowed to be arbitrarily large. This thus ensures the stability of the inflationary potential. 
	\item Embedding in a UV-complete theory.\\
	Higher-dimensional operators that are not forbidden by symmetries are rather unavoidable in the context of EFTs. Discussing the embedding of the model into some high energy theory we may invoke the presence of symmetries to protect the potential from (large) radiative corrections. 
	\item Go beyond slow-roll inflation.\\
	While slow-roll inflation provides the simplest realization of the condition to obtain inflation, in general different realizations can be proposed. For example in Sec.~\ref{sec_inflation:generalized_models}, we briefly present models with non-trivial kinetic terms which are discussed in more detail in Chapter~\ref{chapter:generalized_models}.
\end{itemize}
In addition to these solutions, we should also mention that inflationary model building can be discussed with a completely phenomenological approach. In this prospect, instead of focusing on the theoretical implications, we can concentrate our study on the constraints imposed by direct observations. However, this kind of study is not aimed at defining an exhaustive model for inflation but rather at ruling out existing models.

\subsection{Some models and their predictions.}
\label{sec_inflation:some_models}
In this section we present some inflationary models that are relevant for this work. For each model we start by presenting the potential and by summarizing the physical reasons that lead to its introduction. We then proceed by giving an explicit expression for the associated observable quantities.

\subsubsection{Chaotic potentials.}
\label{sec_inflation:chaotic}
As already explained through this Chapter, Chaotic inflation is a large-field model that was introduced by Linde in~\cite{Linde:1983gd}. In the original model the inflaton is rolling down a potential:
\begin{equation}
\label{eq_inflation:Chaotic_Linde}
V(\phi)  = \frac{1}{2} m^2 \phi^2 \ .
\end{equation}
However, this model can be generalized in order to define a broader class of models:
\begin{equation}
\label{eq_inflation:Chaotic_inflation}
V(\phi)  = \Lambda^4 \left( \kappa \phi  \right)^p \ ,
\end{equation}
where $\Lambda$ is a mass scales (which for $p=2$ is identified with the mass of the inflaton). The value of $\Lambda$ is fixed by the COBE normalization of Eq.~\eqref{eq_inflation:COBE_normalization}. As a consequence these models are completely specified by a single dimensionless parameter $p \in \mathbb{R}^+$. To give an explicit expression of the predictions for $n_s$, $r$ and $\alpha_s$, we start by computing the number of e-foldings:
\begin{equation}
\label{eq_inflation:Chaotic_N}
N(\phi) \simeq \frac{\kappa^2 \phi^2 }{2 p} .
\end{equation}
It is possible to show that the predictions for the values of $n_s$, $r$ and $\alpha_s$ for these models are given by:
\begin{equation}
n_s = 1 - \frac{p+2}{2N} \ , \qquad \qquad r = \frac{4 p }{ N } \ , \qquad \qquad \alpha_s = -\frac{p+2}{2N^2}
\end{equation}
An embedding of this class of models in the context of Supergravity has recently been discussed by Kallosh and Linde in~\cite{Kallosh:2010ug}. In particular it has been shown that generalizations of these models, for example introducing a non-minimal coupling between the inflaton and gravity, may lead to the existence of a certain number of cosmological attractors~\cite{Kallosh:2013tua,Galante:2014ifa}. These generalized models and the corresponding attractors are discussed in Chapter~\ref{chapter:generalized_models}.

\subsubsection{Plateau-like potentials.}
\label{sec_inflation:plateau}
As explained in Sec.~\ref{sec_inflation:eta_solutions}, one of the possible solutions to the $\eta$-problem is the discussion of the embedding of the model in a UV-complete theory. As string theory is one of the most promising candidates to define a UV-complete theory, it is natural to consider the possibility of defining inflationary models in this context~\cite{Stewart:1994ts,Dvali:1998pa,Burgess:2001vr,Cicoli:2008gp,Silverstein:2008sg}. In particular several high energy models give rise to an effective potential for the inflaton of form: 
\begin{equation}
\label{eq_inflation:general_plateau}
V(\phi) = \Lambda^4 \left[ 1 - \exp\left( - \gamma \kappa \phi \right) \right]^2  \ .
\end{equation}
As inflation takes place in the region where the exponential is small, we proceed by approximating the potential as:
\begin{equation}
\label{eq_inflation:general_plateau_app}
V(\phi) = \Lambda^4 \left[ 1 - 2 \exp\left( - \gamma \kappa \phi \right) \right]  \ ,
\end{equation}
and thus the number of e-foldings can be expressed as:
\begin{equation}
\label{eq_inflation:plateau_N}
N \simeq \frac{\exp\left(\gamma \kappa \phi \right) }{2\gamma^2} .
\end{equation}
It is straightforward to show that the predictions for $n_s$, $r$ and $\alpha_s$ for models with potential of Eq.~\eqref{eq_inflation:general_plateau} are thus given by:
\begin{equation}
\label{eq_inflation:plateau_predictions}
n_s = 1 - \frac{2}{N} \ , \qquad r =  \frac{8}{\gamma^2 N^2}\ , \qquad \alpha_s = -\frac{2}{N^2}\ .  
\end{equation}
It is worth mentioning two models: the Starobinsky model~\cite{Starobinsky:1982ee} and the Higgs inflation model~\cite{Bezrukov:2007ep,Bezrukov:2009db}. In the strict sense\footnote{In particular, the action for the Starobinsky model contains an higher order term for gravity proportional to $R^2$, and the action for the Higgs inflation contains a non-minimal coupling between the inflaton and gravity proportional to $\xi R \phi^2$. More on modified gravity and on non-minimal couplings is said in Sec.~\ref{sec_inflation:generalized_models}.}, these models are not described by the action of Eq.~\eqref{eq_inflation:action}. However, after some manipulations, their defining actions can be reduced to form of Eq.~\eqref{eq_inflation:action}. In particular, as the potentials for these models have the form shown in Eq.~\eqref{eq_inflation:general_plateau}, it is appropriate to include these models in this class.

\subsubsection{Hilltop potentials.}
\label{sec_inflation:hilltop}
Hilltop models~\cite{Boubekeur:2005zm} are small-field models where inflation takes place in a neighborhood of an unstable maximum of the potential. During inflation the scalar field departs from this unstable configuration and slow-rolls towards the true minimum of the potential. In particle physics these models may appear in correspondence with a SSB. The potential for these models can be expressed as: 
\begin{equation}
\label{eq_inflation:general_hilltop}
V(\phi) = \Lambda^4 \left[ 1 - \left( \frac{\phi}{v}\right)^p \right]^2  \ ,
\end{equation}
where $\Lambda$ and $v$ are constant with the dimension of a mass and $p>0$\footnote{The case with $p = 2$ is different from the other models of this class. The origin of this difference will be clarified in Chapter~\ref{chapter:beta}.}. The COBE normalization fixes the value of $\Lambda$. Inflation can actually take place for $0 < \phi <  v$. In this Section we only consider models where inflation takes place at $\phi/v \ll 1$\footnote{ The limit $\phi/v \simeq 1$ is also interesting but we postpone its discussion to Chapter~\ref{chapter:beta} where we treat this case in terms of the $\beta$-function formalism for inflation}. In this limit, the potential can be approximated by:
\begin{equation}
\label{eq_inflation:general_hilltop_app}
V(\phi) \simeq \Lambda^4 \left[ 1 - 2 \left( \frac{\phi}{v}\right)^p \right]  \ .
\end{equation}
More in general higher order terms in $ \phi/v$ may appear in the potential, without contributing to inflation. We start by considering $p \neq 2$ which is special and has to be treated separately. The number of e-foldings can be expressed as:
\begin{equation}
\label{eq_inflation:Hilltop_N}
N \simeq \frac{\kappa^2 v^2}{2 p (p-2)} \left( \frac{\phi}{v} \right)^{2-p} \ .
\end{equation}
It is possible to show that, neglecting higher order in $1/N$, the approximated expressions for $n_s$, $r$ and $\alpha_s$ are:
\begin{equation}
n_s \simeq 1 - \frac{2(p-1)}{(p -2 )N} \ , \qquad r \simeq \frac{32 p^2 }{\kappa^2 v^2} \left[ \frac{2p(p-2)}{\kappa^2 v^2} N \right]^{\frac{2p-2}{2-p}} \ , \qquad  \alpha_s \simeq -\frac{2(p-1)}{(p-2)N^2} \ .
\end{equation}
Notice that for these models it is possible to define an extremely small value for $r$ while keeping $n_s$ fixed. In particular this is realized by taking $\kappa v \ll1$ \textit{i.e.} assuming the scale $v$ and consequently the inflaton to be much smaller than the Planck mass. For this reason these models are usually referred to as small field models. \\

\noindent Finally let us consider the case with $p =2$. In this case the number of e-foldings can be expressed as:
 \begin{equation}
\label{eq_inflation:Hilltop_N2}
N \simeq - \frac{v^2 \kappa^2}{4} \ln \left(\frac{\phi}{\phi_f}\right) \ ,
\end{equation}
where $\phi_f$ is the value of $\phi$ at the end of inflation. It is possible to show that the lowest order expressions for $n_s$, $r$ and $\alpha_s$ are:
\begin{equation}
\label{eq_inflation:Hilltop_prediction2}
\begin{aligned}
n_s \simeq 1 - \frac{8 }{\kappa^2 v^2} \ ,&  \qquad r \simeq \frac{256 }{\kappa^2 v^2} \left( \frac{\phi_f}{v} \right)^2 \exp\left( - \frac{8 N }{\kappa^2 v^2} \right) \ , \\
  \alpha_s \simeq &- \frac{1024 }{\kappa^4 v^4} \left( \frac{\phi_f}{v} \right)^2 \exp\left( - \frac{8 N }{\kappa^2 v^2} \right) \ .
\end{aligned}
\end{equation}
Notice that in this case $n_s$ is not depending on $N$.

\subsubsection{Natural inflation.}
\label{sec_inflation:natural}
Natural inflation was originally introduced by Freese and others in~\cite{Freese:1990rb,Adams:1992bn}. In this model the inflationary potential is protected from radiative corrections by imposing shift symmetry $\phi \rightarrow \phi + \text{constant}$ for the inflaton. This can be obtained by considering the inflaton to be a Nambu-Goldstone Boson (NGB) which arises from the breaking of a global symmetry. However, the potential of a NGB is exactly flat, and thus they are not suitable to describe inflation (if the potential is exactly flat there is no graceful exit from inflation). Nevertheless, the situation changes if we consider pseudo Nambu-Goldstone Boson (pNGB) where the continuous shift symmetry is broken to a discrete subset. In this case the potential has some periodicity (which without loss of generality can be assumed to be of period $\pi$) that leads to the definition of potentials of form:
\begin{equation}
\label{eq_inflation:natural_inflation}
V(\phi) = \Lambda^4 \left[ 1 + \cos\left( \frac{\phi}{v}\right) \right] \ ,
\end{equation}
where $\Lambda$ and $v$ are constants with the dimension of a mass. As usual $\Lambda$ is fixed by COBE normalization. The number of e-foldings can be expressed as:
\begin{equation}
\label{eq_inflation:natural_N}
N \simeq - 2 \kappa^2 v^2 \ln\left[ \sin \left( \frac{\phi}{2v} \right) \right].
\end{equation}
For this model we also give the explicit expressions for the first and second slow roll parameters:
\begin{equation}
\label{eq_inflation:slow_roll}
\epsilon_V \simeq \frac{1}{2 \kappa^2 v^2 } \left[ \exp\left( \frac{N}{\kappa^2 v^2}\right) - 1 \right]^{-1} \ , \qquad \eta_V \simeq - \frac{1}{2 \kappa^2 v^2 } \frac{\exp\left( \frac{N}{\kappa^2 v^2}\right) - 2 }{ \exp\left( \frac{N}{\kappa^2 v^2}\right) - 1  } \ , 
\end{equation}
and the lowest order expression for $\xi_V$ simply reads $\xi_V \simeq -  \eta_V \ \sqrt{2 \epsilon_V}/(\kappa v )$.
The expression for $n_s$, $r$ thus read:
\begin{equation}
\label{eq_inflation:natural_inflation_predictions}
n_s \simeq 1 - \frac{1}{ \kappa^2 v^2 } \frac{\exp\left( \frac{N}{\kappa^2 v^2}\right) + 1 }{ \exp\left( \frac{N}{\kappa^2 v^2}\right) - 1  } \ , \qquad r \simeq \frac{8}{ \kappa^2 v^2 } \left[ \exp\left( \frac{N}{\kappa^2 v^2}\right) - 1 \right]^{-1} \ ,  
\end{equation}
while the explicit expression for $\alpha_s$ can be computed using Eq.~\eqref{eq_inflation:slow_roll_alpha}. It is worth stressing that these expressions admit two different limits:
\begin{itemize}
  \item $N/(\kappa v)^2 \gg 1 $. In this limit we get:
	\begin{equation}
		\label{eq_inflation:natural_inflation_small}
		n_s = 1 - \frac{1}{ \kappa^2 v^2 }\ , \qquad \qquad r = \frac{8}{ \kappa^2 v^2 } \exp\left( - \frac{ N}{\kappa^2 v^2}\right)  \ .
	\end{equation}
	Notice that via a redefinition of the parameter $v$ these expressions can be modified in order to match with Eq.~\eqref{eq_inflation:Hilltop_prediction2}. In this limit the predictions are thus matching with the ones of an Hilltop model with $p = 2$.
  \item  $N/(\kappa v)^2 \ll 1 $. In this limit we get:
  	\begin{equation}
		\label{eq_inflation:natural_inflation_large}
		n_s \simeq 1 - \frac{2}{ N} \ ,  \qquad r \simeq \frac{8}{ N }  \ .
	\end{equation}
	That match with the predictions for a chaotic model with $p = 2$.
\end{itemize}

\subsubsection{Power law inflation.}
\label{sec_inflation:power_law}
The last model that we present in this Section are the so-called ``power law'' inflation model introduced by Lucchin and Matarrese in~\cite{Lucchin:1984yf}. The inflationary potential is:
\begin{equation}
\label{eq_inflation:power_law}
V(\phi) = \Lambda^4 \exp\left( - \lambda \kappa \phi \right) \ ,
\end{equation}
where $\Lambda$ is a constant set by the COBE normalization and $\lambda$ is a dimensionless constant. This particular model is interesting because with the potential of Eq.~\eqref{eq_inflation:power_law} an exact solution for Eqs.~\eqref{eq_inflation:friedmann_scalar} and Eq.~\eqref{eq_inflation:eq_motion} exists even without imposing slow-rolling. Notice that this model is extremely different from the ones discussed so far because it is not predicting the scale factor $a(t)$ to grow exponentially with time but they are predicting $a(t) \propto t^{2/\lambda^2}$. As a consequence, in this case we do not have an exponential expansion of the scale factor, but rather an accelerated expansion which can still offer~\cite{Abbott:1984fp} a viable alternative to standard exponential inflation. In this model the slow roll parameters are not depending on $N$, but they are only depending on $\lambda$. The predictions for $n_s$, $r$ and $\alpha_s$ are:
\begin{equation}
n_s = 1 - \lambda^2 \ , \qquad \qquad r = 8 \lambda^2 \ , \qquad \qquad \alpha_s = 0 \end{equation}
It is important to point out that as power law inflation is an exact solution of Eqs.~\eqref{eq_inflation:friedmann_scalar} and Eq.~\eqref{eq_inflation:eq_motion}, there is no natural end to the expansion $a(t) \propto t^{2/\lambda^2}$. As a consequence this model is not complete as it lacks a mechanism to exit from the inflationary phase.

\section{Generalized models.}
\label{sec_inflation:generalized_models}
As explained in Sec.~\ref{sec_inflation:eta_solutions}, in order to define a concrete and theoretically well-motivated model of inflation several extensions of the simplest realization presented in Sec.~\ref{sec_inflation:simplest_Inflation}, were proposed. As anticipated in Sec.~\ref{sec_inflation:simplest_Inflation}, the simplest realization of inflation discussed so far, basically relies on four assumptions: 1. Single-field models, 2. Gravity described by a standard Einstein-Hilbert term, 3. Canonical kinetic terms, 4. Minimal coupling between the inflaton and gravity. In the following we explain why and how we can reconsider these assumptions and then we explain some possible observational consequences of the relaxation of these assumptions.
\begin{enumerate}
  	\item 
  	The first generalization that we discuss is the possibility of dropping the assumption that a single-field is present in the Universe during inflation. In order to obtain the standard single-field models, we usually assume that all the other fields are frozen and that their energy densities and their interactions with the inflaton are negligible. In general, there is no reason to make these assumptions and generalized models of inflation should thus admit the presence of several fields during inflation.\\

	\noindent
	A well known example of multi-field inflation is the classical hybrid inflation model~\cite{Linde:1991km,Linde:1993cn,Copeland:1994vg}. In this two-field ($\phi$ and $\sigma$) theory the potential is given by:
	\begin{equation}
		V(\phi, \sigma) = \frac{\left( M^2 - \lambda \sigma^2\right)^2}{4 \lambda} + \frac{m^2 \phi^2}{2} + \frac{g^2}{2} \phi^2 \sigma^2 \ ,
	\end{equation}
	where $\lambda$ and $M$ are respectively an effective coupling constant and a mass term for $\sigma$ and $g$ is the coupling constant that parametrizes the strength of the interactions between $\phi$ and $\sigma$. In this model until $\phi>\phi_c = M/g$ the only minimum of the potential is at $\sigma = 0$ and thus the field $\sigma$ is stabilized at this value. When $\phi$ becomes smaller than $\phi_c$, the point $\sigma = 0 $ is turned into an unstable maximum so that the field $\sigma$ becomes active and rolls towards is true minimum. Once $\phi$ becomes smaller than $\phi_c$ and $\sigma$ becomes active inflation stops almost instantaneously.\\

	\noindent
	Hybrid inflation models are extremely interesting because of their interpretation in the context of particle physics. These models typically have a rather natural embedding in the context of SUSY. In particular they may arise from F-term~\cite{Copeland:1994vg,Dvali:1994ms,Linde:1997sj} or D-terms~\cite{Binetruy:1996xj,Halyo:1996pp} in the context of supergravity\footnote{For a review of SUSY and supergravity see for example~\cite{Binetruy:2006ad}.}.
  	
  	\item 
	Theories of Modified Gravity are based on the assumption that General Relativity (GR) is not the correct theory for gravity. For example, we can consider the possibility that GR is modified at high energies. The simplest example of these high energy modifications of gravity is the well known case of $f(R)$ theories where gravity is described by~\cite{Bergmann:1968ve,1969JETP...30..372R,1970ZhETF..59..288B,Capozziello:2011et}: 
	\begin{equation}
	  \label{eq_inflation:action_fr}
	    \mathcal{S}_G=  \int\mathrm{d}t\mathrm{d}^3x \sqrt{|g|} \frac{f(R)}{2 \kappa^2} \ ,
	  \end{equation}
	where $f(R)$ is a generic function of the Ricci scalar $R$ (for the definition see Eq.~\eqref{appendix_GR:Ricci_scalar}). $f(R)$ theories are also known for their application to 	
	provide alternatives to the cosmological constant in order to explain the dark energy (see for example~\cite{Capozziello:2002rd,Capozziello:2003tk}). \\

	\noindent
	It is important to point out that $f(R)$ theories are just one of the possible extensions of GR. Another interesting possibility is the case of scalar-tensor theories formulated in the works of Bergmann~\cite{Bergmann:1968ve} Nordtvedt~\cite{Nordtvedt:1970uv} Wagoner~\cite{Wagoner:1970vr} generalizing the original Brans-Dicke\footnote{The Brans-Dicke theory was originally proposed as an alternative (a competitor) of GR where gravity is not only described by the metric but also by an additional degree of freedom (a scalar field) which effectively changes the gravitational coupling (the Planck mass).} theory~\cite{Brans:1961sx}. More on this models is said when we discuss models where the inflaton has a non-minimal coupling with gravity. \\

	\noindent
	A well known case of $f(R)$ theory is the Starobinsky model~\cite{Starobinsky:1980te}, proposed by Starobinsky in 1980. In this model we have:
	\begin{equation}
		f(R) = R + \alpha R^2 \ , \qquad \qquad (\alpha > 0) \ .
	\end{equation}
	This model was originally proposed in order to re discuss the conditions that lead to the presence of a primordial singularity. It is well known that, because of the presence of the $\alpha R^2$ term, these models may lead to the accelerated expansion of the Universe which is suitable to describe inflation.

  	\item 
  	Models with non-standard kinetic terms may naturally arise in the context of string theory and supergravity. In the context of supergravity these terms naturally arise from the definition of the kinetic term in terms of the  K\"ahler potential (see ~\cite{Binetruy:2006ad}). In the context of string theory non-standard kinetic terms may arise when we consider the degrees of freedom of a D-brane which are effectively described~\cite{Leigh:1989jq} by a Dirac-Born-Infeld (DBI) action~\cite{Dirac:1962iy,Born:1934gh}. As discussed by Armendariz-Picon, Damour and Mukhanov in~\cite{ArmendarizPicon:1999rj}, in models with non-standard kinetic terms inflation may be realized at a finite value of $X \equiv g^{\mu\nu} \partial_\mu \phi \partial_\nu \phi/2$. Two well known examples of inflation models with non-standard kinetic terms are the case of Dirac-Born-Infeld (DBI) inflation proposed by Eva Silverstein and David Tong in~\cite{Silverstein:2003hf} and the case of Tachyonic inflation proposed by Gibbons in~\cite{Gibbons:2002md}. More details on DBI inflation are given in Chapter~\ref{chapter:generalized_models} (see Sec.~\ref{sec_inflation:Non_standard_kinetic}).\\

	\noindent
	In general, the action for a homogeneous classical scalar field $\phi$ with non-standard kinetic term can be expressed as:
	\begin{equation}
		\label{eq_inflation:action_non_standard_kinetic}
		\mathcal{S}=\int\mathrm{d}^4x\sqrt{|g|}\left(\frac{1}{2\kappa^2}R + p(\phi,X)\right),
	\end{equation}
	where as usual we have defined $X \equiv g^{\mu\nu} \partial_\mu \phi \partial_\nu \phi /2 $. A first difference between the models discussed so far and models with non-standard kinetic terms is the expression for the speed of sound $c_s^2$ (defined in Eq.~\eqref{eq_inflation:speed_of_sound}). While $c_s^2$ is equal to one for standard kinetic terms, this condition is no longer true for generalized kinetic terms.

  	\item
  	Finally, we discuss the possibility of considering models where the inflaton has a non-minimal coupling with gravity. A first suggestion on the possibility of exploring these models in the context of inflationary model building was given by Salopek, Bond and Bardeen in~\cite{Salopek:1988qh}. Further interest on the topic came with the works of Futamase~\cite{Futamase:1989hb,Futamase:1987ua}, who discussed the viability of these models, and in particular with the works of Fakir and Unruh~\cite{Fakir:1990eg}, who noticed the possibility of considering large positive non-minimal couplings.\\ 

 	\noindent
	Interesting developments on this topic came with the works of Gasperini and Veneziano~\cite{Gasperini:1992em,Gasperini:1993hu} who embedded inflation into scalar-tensor theories (inspired by string theory) for early time cosmology. However, these models are predicting a blue spectrum~\cite{Gasperini:2002bn} \textit{i.e.} $n_s > 1$, and are thus ruled out by present CMB observations\footnote{While the interest in these models to describe inflation was damped by this evidence, these models were extensively studied as possible models for quintessence~\cite{Uzan:1999ch,Chiba:1999wt,Amendola:1999qq,Perrotta:1999am,Bertolami:1999dp,Boisseau:2000pr,EspositoFarese:2000ij}.}. Scalar-tensor theories recently achieved a renewed interest in the context of early time cosmology with the rediscovery~\cite{Deffayet:2011gz,Kobayashi:2011nu} of a more general formulation of these theories firstly proposed by Horndeski\footnote{Using the works of Lovelock on the generalization of the Einstein tensor~\cite{Lovelock:1971yv}.} in~\cite{Horndeski:1974wa}. In the context of scalar-tensor we mention the case of generalized Higgs inflation~\cite{Germani:2010gm,Germani:2010ux,Germani:2011ua} where the role of the inflaton is played by standard models Higgs field that has a non-minimal \emph{kinetic} coupling with gravity. \\

	\noindent
	Another interesting case of theories where the inflaton has a \emph{constant} non-minimal coupling with gravity was proposed by Bezrukov and Shaposhnikov~\cite{Bezrukov:2007ep,Bezrukov:2009db} who discussed the possibility of using the standard model Higgs field as the inflaton. In general, non-minimal couplings naturally arise from radiative corrections in the context of QFT in curved spacetime\footnote{For a review of the topic see for example~\cite{birrell1984quantum}.}. Moreover, using a conformal transformation and a field redefinition, in these models we typically obtain extremely flat potential (of the same for of Eq.~\eqref{eq_inflation:general_plateau}) for the inflaton. As a consequence, these models offer a natural mechanism to define flat potentials which thus are suitable to predict values for $n_s$ and $r$ in agreement with the constraints set by CMB observations (shown in Sec.~\ref{sec_inflation:CMB_constraints}). Several proposals to embedded similar models in the context of supergravity can be found in the literature\footnote{See for example the work of Einhorn and Jones~\cite{Einhorn:2009bh} and the works of Kallosh and Linde~\cite{Kallosh:2010ug,Kallosh:2010xz}.}. More on models with non-minimal coupling between the inflaton and gravity is said in Chapter~\ref{chapter:generalized_models}.\\

  	\noindent
	The model of Bezrukov and Shaposhnikov~\cite{Bezrukov:2007ep,Bezrukov:2009db}, usually known as ``Higgs inflation'', corresponds to a chaotic model~\cite{Linde:1983gd} with potential $V(\phi) = \lambda \phi^4$ where a non-minimal coupling $\xi \phi^2 R/2$ between the inflaton and gravity is introduced. Via a conformal transformation and a field redefinition it is possible to show that in the limit of large $\xi$ (in particular for $\xi \gtrsim 10^4$) the inflationary potential can be expressed as in Eq.~\eqref{eq_inflation:general_plateau} with $\gamma = \sqrt{2/3}$. The corresponding predictions for $n_s$ and $r$ are thus well into the sweet spot of the Planck CMB constraints.
  \end{enumerate} 
The generalizations presented in this Section may lead to several interesting consequences on the observables quantities related with inflation (for example $c_s^2 \neq 1 $). Before concluding this Section, we present one of the main differences between some of these generalized models and the simplest realization of inflation discussed in Sec.~\ref{sec_inflation:simplest_Inflation}.

\noindent
As discussed in Chapter~\ref{chapter:beta} and as explicitly shown in Appendix~\ref{appendix_perturbations:Cosmological_perturbations}, scalar fluctuations over the background solution are described in terms of the comoving curvature perturbation $\zeta$. While so far we have only focused on the calculation, on the predictions and on the constraints on the two-point function $\langle \zeta \zeta \rangle$, we can now focus on the generation of the so-called ``non-Gaussianities'' \textit{i.e.} deviations from a pure Gaussian spectrum. In order to compute the theoretical predictions for non-Gaussianities, we should push the perturbative expansion of the action up to some higher order. In particular, if we are only interested in computing the three-point function, it is sufficient to expand the action up to the second order in the perturbations. \\

\noindent
Constraints on the generation of non-Gaussianities during inflation are usually set on the so-called ``bispectrum'' defined as:
\begin{equation}
	\label{eq_inflation:def_bispectrum}
	\langle \zeta(\tau,\vec{k}_1) \zeta(\tau,\vec{k}_2)\zeta (\tau,\vec{k}_3) \rangle = \left( 2\pi \right)^3 B(k_1,k_2,k_3)\delta^3 (\vec{k}_1 + \vec{k}_2 + \vec{k}_3) \ .
\end{equation}
The value of this quantity in single-field models was firstly computed by Maldacena in~\cite{Maldacena:2002vr}. It is possible to show (more on this topic is said in Sec.~\ref{sec_inflation:Non_standard_kinetic}) that the amount of non-Gaussianities produced in the simplest realization of inflation (of Sec.~\ref{sec_inflation:simplest_Inflation}) is expected to be highly suppressed~\cite{Maldacena:2002vr,Acquaviva:2002ud}. However, this result may change if we consider some generalized models (for example multi-fields or models with non-standard kinetic terms). Some details on this possibility are given in Chapter~\ref{chapter:generalized_models} (see Sec.~\ref{sec_inflation:Non_standard_kinetic}) and in Chapter~\ref{chapter:pseudoscalar} (see Sec.~\ref{sec_pseudoscalar:non_gaussianities}). For the scope of this Chapter it is sufficient to mention that Non-Gaussianities are strongly constrained at CMB scales (resulting in constraints on generalized models of inflation) for example by Planck measurements~\cite{Ade:2015lrj,Ade:2015ava}.

\section{Primordial GW and direct GW detectors.}
\label{sec_inflation:direct_GW_detection}
Gravitational waves (GWs) are one of the main predictions of General Relativity and their recent first detection by the LIGO/VIRGO collaboration~\cite{Abbott:2016blz} is a major result for modern physics. A peculiar characteristic of GW is the weakness of their interactions which in practice make them travel freely through the Universe. As a consequence GWs can give us precise information on the very early times of our Universe and in particular on Inflation. As discussed in this Chapter a primordial GW background is expected to be generated during inflation. As a consequence it is interesting to discuss the possibility of observing this primordial GW background at direct GW detectors. \\

\noindent
The intensity of GW backgrounds at a given instant $\tau$ (expressed in terms of the conformal time) is typically characterized by the dimensionless quantity:
\begin{equation}
	\label{eq_inflation:GW_spectrum}
	\Omega_{GW}(f,\tau) \equiv \frac{1}{\rho_{c}}\frac{\textrm{d} \rho_{GW}(f,\tau)} {\textrm{d} \ln f} \ ,
\end{equation}
where $\rho_{GW}$ is the energy density of the GW background and $\rho_{c}$ is the critical energy density (defined in Eq.~\eqref{eq_intro:critical_density}). As we discuss in the following, we are interested in computing this quantity at present time and usually GW backgrounds are characterized by $h_0^2 \, \Omega_{GW}(f,\tau_0)$ where $h_0$ is the dimensionless Hubble parameter at present time (see Eq.~\eqref{eq_intro:Hubble_param_value}). A this point it is useful to introduce the characteristic amplitude $h_c^2(f,\tau)$ of the GW, defined as:
\begin{equation}
	\label{eq_inflation:characteristic_amplitude_GW}
	\langle h_{ij}(\tau) h^{ij}(\tau) \rangle = 2 \int_{f = 0}^{f = \infty} \textrm{d}\ln f \, h_c^2(f,\tau) \ ,
\end{equation}
where the brackets $\langle \cdot \rangle$ are thus used to denote an average over the Fourier amplitudes. Notice that $f$ is the \emph{physical} frequency that is related to $k$ (comoving wave-vector) by $2 \pi f = k/a(\tau)$. A comparison between Eq.~\eqref{eq_inflation:characteristic_amplitude_GW} and Eq.~\eqref{eq_inflation:power_spectra} leads to:
	\begin{equation}
		\Delta_t^2(k,\tau) = 2 \frac{\textrm{d} \ln f}{\textrm{d} \ln k} \, \tilde{h}_c^2(k,\tau) \ ,
	\end{equation}
where $\tilde{h}_c^2(k,\tau) \equiv h_c^2(f(k,\tau),\tau)$. The energy density of the GW background $\rho_{GW}$ can then be expressed~\cite{Maggiore:1900zz} as an integral over the frequency of characteristic amplitude:
\begin{equation}
	\label{eq_inflation:GW_energy_density}
	\rho_{GW}(\tau) = \frac{\kappa^{-2}}{2} \int_{f = 0}^{f = \infty} \textrm{d}\ln f \, (2 \pi f)^2 h_c^2(f,\tau) \ .
\end{equation}
Finally, we can substitute into Eq.~\eqref{eq_inflation:GW_spectrum} to get :
\begin{equation}
	\label{eq_inflation:GW_spectrum_tau}
	\Omega_{GW}(f,\tau) = \frac{\kappa^{-2}}{ 2 \rho_{c}} (2 \pi f)^2 h_c^2(f,\tau) = \frac{\kappa^{-2}}{ 4 \rho_{c}} \frac{\textrm{d} \ln k}{\textrm{d} \ln f} \, (2 \pi f)^2 \tilde{\Delta}^2_t(f,\tau) \ ,
\end{equation}
where $\tilde{\Delta}^2_t(f) \equiv \Delta^2_t(k(f))$. In order to express this quantity at present time $\tau_0$, we should express $f$ at present time \textit{i.e.} $2 \pi f = k/a(\tau_0) = k$ and we should also express $\tilde{\Delta}^2_t(f,\tau_0)$. For GW that are produced during inflation we can use the spectrum at horizon crossing (\textit{i.e.} when they re-enter the horizon) is set by Eq.~\eqref{eq_inflation:power_spectra_final}. As a consequence, we need to define the so-called ``transfer function'' $T_f(f,\tau)$ that expresses the evolution of the GW after they have re-entered the horizon. In terms of this quantity, the GW spectrum at present time reads:
\begin{equation}
	\label{eq_inflation:GW_spectrum_now}
	\Omega_{GW}(f,\tau_0) = \frac{\kappa^{-2}}{ 4 \rho_{c}} (2 \pi f)^2 T_f(f,\tau_0) \tilde{\Delta}^2_t(f,\tau = k) \ .
\end{equation}
It is possible to show~\cite{Boyle:2005se,Maggiore:1900zz,weinberg2008cosmology} that for modes that re-enter the horizon during radiation domination, $T_f(f,\tau_0)$ scales like $(2 \pi f)^{-2}$ while for modes that re-enter during matter domination $T_f(f,\tau_0) \propto (2 \pi f)^{-4}$.  A schematic representation of this spectrum is shown in Fig.~\ref{fig_inflation:detectors_schematic}. \\

\begin{figure}[h]
	\centering
	\includegraphics[width=.9 \textwidth]{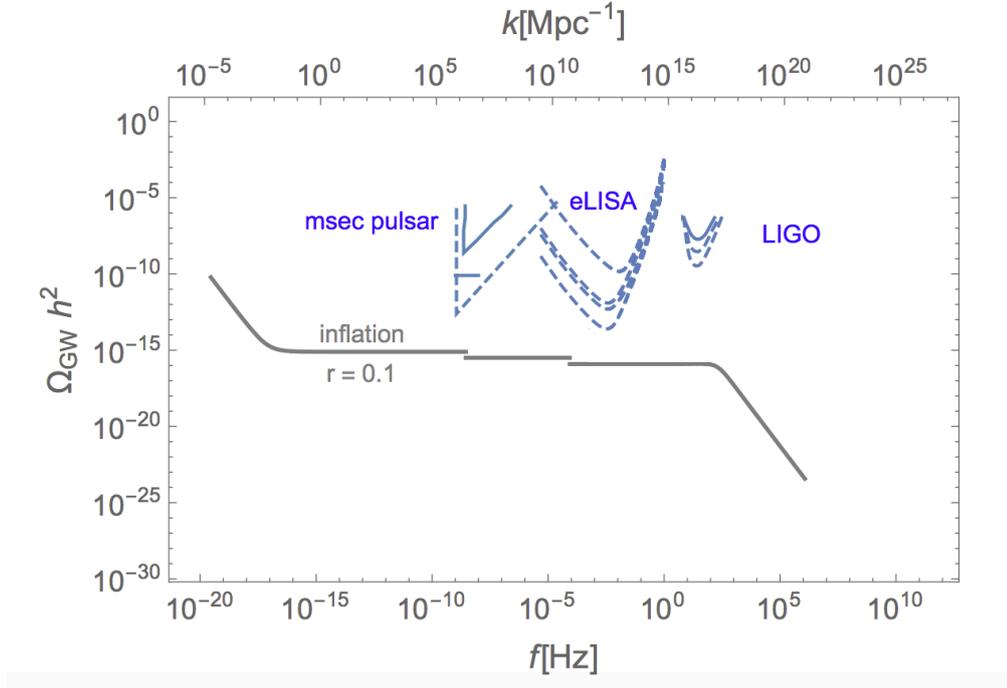}
	\caption{Schematic view of the spectrum of primordial GWs compared with the sensitivity curves of present and future direct GW detectors. Current bounds are denoted by solid lines, expected sensitivities of upcoming experiments by dashed lines. More details on this curves are given in the text and in Chapter~\ref{chapter:pseudoscalar} (in particular Sec.~\ref{sec_pseudoscalar:models}). \label{fig_inflation:detectors_schematic} }
\end{figure}

\noindent
In Fig.~\ref{fig_inflation:detectors_schematic} we compare the GW spectrum (we have fixed $r \simeq 0.1$) with the sensitivity curves of present (solid lines) and future (dashed lines) direct GW detectors. The first set of curves on the left represents the millisecond pulsar timing arrays covering frequencies around $10^{-10}$~Hz. In particular we show the constraint depicted in Ref.~\cite{Smith:2005mm}, the update from EPTA~\cite{vanHaasteren:2011ni} and the expected sensitivity of SKA~\cite{Kramer:2004rwa}. The two other sets are respectively the space-based GW interferometers in the milli-Hz range (eLISA~\cite{Caprini:2015zlo}) and the ground-based detectors which are sensitive at a larger frequencies \textit{i.e.} up to few 10 Hz (LIGO/VIRGO~\cite{TheLIGOScientific:2016wyq}). More on the sensitivity curves for eLISA and advanced LIGO is said in Chapter~\ref{chapter:pseudoscalar} (in particular see Sec.~\ref{sec_pseudoscalar:numerical_results}). It this plot we are not showing the expected sensitivity curves for Big Bang Observatory (BBO)/Deci-Hertz Interferometer Gravitational wave Observatory (DECIGO)\cite{Yagi:2009zz,Yagi:2011wg} and for Einstein Telescope (ET)~\cite{Hild:2008ng,Punturo:2010zz}. The reason for this choice is that differently from the other detectors shown in Fig.~\ref{fig_inflation:detectors_schematic}, these missions are still not approved and their realization is still under evaluation.\\

\noindent
As it is possible to see from Fig.~\ref{fig_pseudoscalar:Omega_schematic}, the signal produced by standard slow-roll inflation models is expected to be below the range of current and upcoming direct GW detectors. However, this picture can change dramatically if we consider generalized models of inflation~\cite{Cook:2011hg}. In particular, some models where an observable GW signal is produced are presented in Chapter~\ref{chapter:pseudoscalar}.

 {\large \par}}
{\large \chapter{$\beta$-function formalism.}

\label{chapter:beta}
\horrule{0.1pt} \\[0.5cm]

\begin{abstract} 
\noindent
	In this Chapter we present the $\beta$-function formalism for inflation. In this framework inflation is described by means of a renormalization group (RG) equation. In this context the slow-rolling regime is interpreted as the slow evolution in the neighborhood of a fixed point for the RG flow. In analogy with statistical mechanics, it is natural to use this formalism to define a set of universality classes for inflationary models. These universality classes should be intended as a collection of models sharing a single scale invariant limit. This formalism thus gives a partial explanation to the degeneracy in the predictions of different inflationary models.
  \end{abstract}

\horrule{0.1pt} \\[0.5cm] 

\noindent 
As explained in Chapter~\ref{chapter:inflation}, inflation is nowadays considered as one of the cornerstones of early time cosmology. Since the definition of the first models~\cite{Guth:1980zm,Linde:1981mu,Albrecht:1982wi} a huge amount of inflationary models has been defined. While some of these models are presented in Chapter~\ref{chapter:inflation}, a comprehensive review can be found in~\cite{Martin:2013tda}. CMB experiments~\cite{Planck:2013jfk,Ade:2015lrj} have fixed several constraints on the physics of inflation and of course this has helped to restrict the possibility to a smaller set of preferred models. However, despite the enormous progresses, a convincing and theoretically well-motivated model for inflation is still lacking.\\

\noindent
As already stated several times during this work, an enormous set of inflationary models already exists. Moreover, in some cases observable quantities are not sufficient to distinguish between different models. A recent and topical example of this degeneracy between different high energy models is the well known case of the $R^2$ model of Starobinsky~\cite{Starobinsky:1980te} and the non-minimal Higgs inflation~\cite{Bezrukov:2007ep}. For these reasons during the last years, several methods have been proposed in order to produce a systematic classification of inflationary models. In this context it is worth mentioning the work of Mukhanov~\cite{Mukhanov:2013tua}, who proposed a classification of inflationary models based on the parameterization of the equation of state, and the one of Roest~\cite{Roest:2013fha} and Garcia-Bellido~\cite{Garcia-Bellido:2014gna}, who proposed a parameterization of the slow-roll parameters using an expansion in terms of the small quantity $1/N$ where $N$ is the number of e-foldings.\\

\noindent
The $\beta$-function formalism for inflation defined in~\cite{Binetruy:2014zya} relies on the simplest property of inflation: the approximate scale invariance, typical of a nearly dS spacetime. As we explain in detail through this Chapter using the Hamilton-Jacobi formalism of Salopek and Bond~\cite{Salopek:1990jq}, the solutions for the evolution of the inflaton in its potential can be parameterized in terms of a superpotential. Once this quantity is defined, we can find a formal resemblance between the equations describing the evolution of this system and a RG group equation in the context of quantum field theory (QFT). Following this analogy, the different inflation scenarios are interpreted as different evolutions for a system that is slowing approaching or leaving a critical (fixed) point. As usual this process may be described in terms of the $\beta$-function that parameterizes the RG flow. In this framework it is thus natural to define universality classes of inflationary models. \\

\noindent
The structure of this Chapter is as follows. In the first section we discuss some issues related with the standard method to define inflationary models by specifying potentials and we give a brief review of the classifications proposed by Mukhanov in~\cite{Mukhanov:2013tua} and Roest~\cite{Roest:2013fha,Garcia-Bellido:2014gna}. In Sec.~\ref{sec_beta:beta_func} we define the $\beta$-function formalism for inflation of~\cite{Binetruy:2014zya} and we compute the expressions for the observable quantities in this framework. In Sec.~\ref{sec_beta:universality_classes} we present the universality classes for inflationary models introduced in~\cite{Binetruy:2014zya} and we show the predictions for the observable quantities for each class. Finally in Sec.~\ref{sec_beta:comoposite_classes} we discuss some more elaborated classes.

\section{Reasons to classify inflationary models.}
\label{sec_beta:reasons_to_classify}
As already stated in the introduction of this Chapter, a common issue with the standard approach to model building, \textit{i.e.} defining the inflaton potential, is the degeneracy in the predictions for the observable quantities. A clear example of this degeneracy is the well known case of the $R^2$ model of Starobinsky~\cite{Starobinsky:1980te} and the non-minimal Higgs inflation~\cite{Bezrukov:2007ep}. The similarity between these two models is manifest once the problem is described in the Einstein frame, \textit{i.e.} the reference frame where gravity is described by a standard Einstein-Hilbert term\footnote{More details on this procedure are given in Chapter~\ref{chapter:generalized_models}}. In both these cases, in the Einstein frame it is possible to redefine the inflaton field (\textit{i.e.} the degree of freedom associated with higher derivatives in the case $R^2$ inflation and the Higgs boson in the case of non-minimal Higgs inflation) so that the system is now described in terms of a new field $\varphi$. This new field $\varphi$ has by construction a canonically normalized kinetic term and is minimally coupled with gravity. In the large field limit $\kappa \varphi \gg 1$ (\emph{i.e.} the regime where inflation takes place) its potentials read:
\begin{equation}
	V(\varphi) \simeq \Lambda^4 \left[ 1 - \exp\left( - \sqrt{\frac{2}{3}} \kappa \varphi \right) \right]^2 \ .
\end{equation}
As the two models that are extremely different from a theoretical point of view, have a similar potential in the region that is relevant for inflation, they end up predicting the same values for $n_s$, $r$ and $\alpha$. In particular these can be computed using Eq.~\eqref{eq_inflation:plateau_predictions}:
\begin{equation}
	n_s = 1 - \frac{2}{N} \ , \qquad r =  \frac{12}{ N^2}\ , \qquad \alpha_s = -\frac{2}{N^2}\ .
\end{equation}
It is fair to point out that with a more detailed analysis (see~\cite{Bezrukov:2011gp}) that keeps into account the physics of reheating\footnote{In particular, following the treatment of~\cite{Bezrukov:2008ut,Asaka:2005an} it is possible to show that in the case of non-minimal Higgs inflation the reheating temperature is:
\begin{equation}
	T^{reh} \simeq 6 \times 10^{13} \text{GeV} \ ,
\end{equation}
with an order one uncertainty factor. On the other hand, in the case of Starobinsky model the it is possible to show~\cite{Starobinsky:1980te,Vilenkin:1985md,Gorbunov:2010bn} that the reheating temperature is significantly lower:
\begin{equation}
	T^{reh} \simeq 3.1 \times 10^9 \text{GeV} \ .
\end{equation}} it is possible to find some slight differences in the predictions for these two models. In particular as the two models predict different reheating temperature, the values of $n_s$, $r$ and $\alpha_s$ should be evaluated at slightly different values for of $N$ (see Eq.~\eqref{eq_inflation:e_fold_k}). \\

\noindent
The reason for this degeneracy is due to the fact that direct observations are only exploring a small part of the inflationary potential. As a consequence, different high energy models may therefore end up predicting the same values for the observable quantities. As a wide set of viable inflationary models has already been defined~\cite{Martin:2013tda}, during the last years several physicist (guided by this observation) have proposed methods to define a classification of inflationary models. In general this classification can be specified by quantities that are different from the scalar potential. In this context it is worth mentioning the proposal of Mukhanov in~\cite{Mukhanov:2013tua} who classified inflationary landscapes in terms of the equation of state for the scalar field. This proposal (that stands on solid phenomenological grounds) chooses an equation of state of the type:
\begin{equation} 
	\frac{(p+\rho)}{\rho} \simeq \frac{\beta}{(N+1)^\alpha} \ ,
\end{equation}
where $p$ and $\rho$ are the pressure and energy density associated with the inflation field, $N$ is the number of e-foldings and $\alpha$ and $\beta$ are order one dimensionless parameters. This parameterization only assumes the equation of state to be smooth, to approach zero during the inflation and to be of order one at the end of inflation. \\

\noindent
Another interesting method is the one introduced by Roest in~\cite{Roest:2013fha}. In this work the slow-roll parameters are parameterized as:
\begin{equation}
	\epsilon \simeq \frac{\alpha}{N^p} \ , \qquad \qquad \eta \simeq \frac{\beta}{N} \ ,
\end{equation}
where again $N$ is the number of e-foldings and the parameters $\alpha$, $\beta$, $p$ are dimensionless parameters that are usually assumed to be of order one. This parameterization has subsequently been developed in~\cite{Garcia-Bellido:2014gna}, where different terms in the $1/N$ expansion have been introduced. Assuming $N$ to be large during the inflation, this parameterization only requires the slow-roll parameters to approach zero as $N$ goes to the infinity, and to grow towards the end of inflation. \\

\noindent As argued in the above paragraphs, Mukhanov and Roest parameterizations stand on solid phenomenological grounds as they are both based on a small set of reasonable assumptions. The $\beta$-function formalism for inflation of~\cite{Binetruy:2014zya} is actually based on a similar logic. The idea that stands behind this formalism is the possibility of characterizing inflationary models by parameterizing the departure from the nearly scale invariant regime, that is a defining property of inflation. It is interesting to point out that, as explained in detail in Chapter~\ref{chapter:holographic_universe}, the possibility of describing inflation by means of a $\beta$-function naturally appears in the context of AdS/CFT~\cite{McFadden:2009fg,McFadden:2010na,Kiritsis:2013gia}. For this reason our proposal is not only standing on phenomenological arguments, but it is supported by interesting theoretical reasons.

\section{The $\beta$-function formalism.}
\label{sec_beta:beta_func}
As discussed in Chapter~\ref{chapter:inflation}, inflation can be realized in terms of a classical field $\phi(t)$ in its potential $V(\phi)$. In this Chapter we restrict the discussion to the case of single field models where the inflaton in minimally coupled with gravity, that as usual is described by a standard Einstein-Hilbert term. This system is defined by the action of Eq.~\eqref{eq_inflation:action}:
\begin{equation}
  \label{eq_beta:action}
    \mathcal{S}= \int\mathrm{d}t\mathrm{d}^3x \sqrt{|g|}\left( \frac{R}{2 \kappa^2} -  X - V(\phi) \right),
  \end{equation}
Assuming a FLRW Universe with metric given by Eq.~\eqref{eq_intro:FLRW} with zero curvature, the metric simply reads $g_{\mu\nu} = \text{diag}(-1,a^2(t),a^2(t),a^2(t))$. We can follow the treatment of Chapter~\ref{chapter:inflation} and assume the scalar field $\phi$ to be homogeneous so that Friedmann Equations read:
\begin{equation}
\label{eq_beta:friedmann_scalar}
3 \kappa^{-2} H^2 = \frac{\dot{\phi}^2}{2} + V(\phi) \ , \qquad \qquad - 2 \kappa^{-2} \dot{H} = \dot{\phi}^2 \ ,  
\end{equation}
where $p_{\phi}$ and $\rho_{\phi}$ are the pressure and energy density associated with the scalar field:
\begin{equation}
\label{eq_beta:p_and_rho}
\rho_{\phi} = \frac{\dot{\phi}^2}{2} + V(\phi) \ ,\qquad \qquad p_{\phi} =  \frac{\dot{\phi}^2}{2} - V(\phi) \ .
\end{equation}
The equation of motion for the scalar field is then given by Eq.~\eqref{eq_inflation:eq_motion} \textit{i.e.}
\begin{equation}
\label{eq_beta:eq_motion}
\ddot{\phi} + 3 H \dot{\phi} + \frac{\partial V}{\partial \phi} = 0 \ .  
\end{equation}
It is interesting to notice that this system of differential equations is redundant and the system can be completely specified by Friedmann equations~\eqref{eq_beta:friedmann_scalar} or alternatively by the first of these two equations plus the equation of motion for the inflaton~\eqref{eq_beta:eq_motion}. Once these equations are solved, the inflationary trajectory is uniquely fixed. Instead of describing the system in terms of the usual formalism, we proceed by using the Hamilton-Jacobi approach defined by Salopek and Bond~\cite{Salopek:1990jq} (see also~\cite{Lidsey:1995np}). In this framework, under the reasonable assumption of a (piece-wise) monotonic field $\phi(t)$, we invert $\phi(t)$ to get $t(\phi)$ and we parameterize the evolution of the system by using the field as a clock.\\

\noindent As a first step, we express the Hubble parameter as a function of $\phi$, and we define the \emph{superpotential} $W(\phi)$ as:
\begin{equation}
	\label{eq_beta:superpotential}
	H (\phi) = \frac{\dot{a}}{a} (\phi ) \equiv -\frac{1}{2} W(\phi) \ .
\end{equation}
The reason for this definition and for the choice of the calling $W(\phi)$ superpotential will be clear in the following. Notice that using the definition of $W(\phi)$ we can express Eq.~\eqref{eq_beta:friedmann_scalar} as:
\begin{equation}
	\label{eq_beta:friedmann_superpotential}
	\frac{3}{4}  W^2 (\phi)= \kappa^{2} \rho \ , \qquad \qquad  \dot{\phi} \  W_{,\phi} = \kappa^{2} \left( p + \rho \right)  = \kappa^{2} \dot{\phi}^2 \ ,
\end{equation}
where as usual the subscript ${}_{,\phi}$ is used to denote differentiation with respect to $\phi$. Notice that the second of these equations implies that $\dot{\phi}$ can be expressed as:
\begin{equation}
	\label{eq_beta:phisuperpot}
	\dot{\phi} = \frac{ W_{,\phi} }{\kappa^2} \ .
\end{equation}
Finally we use the definition of $\rho$ given in Eq.~\eqref{eq_beta:p_and_rho} to express the potential in terms of the superpotential and of its first derivative:
\begin{equation}
	\label{eq_beta:potsuperpot}
	2\kappa^2 V = \frac{3}{2} W^2 - \frac{W_{,\phi}^2}{\kappa^2} \ .
 \end{equation}
Notice that this equation has a formal equation with the parameterization of the potential in terms of the superpotential in the context of supersymmetry\footnote{For a review of the topic see for example~\cite{Binetruy:2006ad}}. This is exactly the reason that motivates our definition of $W(\phi)$. \\

\noindent Before proceeding with the definition of the $\beta$-function formalism for inflation, it is interesting to discuss the new system of equations that we have derived. We started with a system of a first order differential equation plus a second order differential equation. The solution of the original system is thus completely specified by three constants of integrations \textit{i.e.} one for the scale factor and two for the scalar field. The new system is composed by three first order differential equations, \textit{i.e.} Eq.~\eqref{eq_beta:superpotential}, Eq.~\eqref{eq_beta:phisuperpot} and Eq.~\eqref{eq_beta:potsuperpot} and thus we have to specify three constants of integrations \textit{i.e.} one for the scale factor, one for $\phi$ and one for the superpotential\footnote{As explained in the following, the constant of integration for the superpotential is related with the scale of inflation and thus it is fixed by the COBE normalization.}. Notice that in general, given a potential, there is an infinite number of superpotentials that solve Eq.~\eqref{eq_beta:potsuperpot}. However, it is possible to show~\cite{Gursoy:2007cb,Gursoy:2007er,Gursoy:2008za} that only a discrete number of these solutions, and in particular this number is typically equal to one give regular solutions of Eq.~\eqref{eq_beta:potsuperpot} and Eq.~\eqref{eq_beta:phisuperpot}. All others have curvature singularities and for the scope of this work we can ignore them.\\

\noindent As discussed in Chapter~\ref{chapter:inflation}, inflation is realized in correspondence of a zero of the equation of state for this scalar field. Using Eq.~\eqref{eq_beta:friedmann_superpotential} we can express it in terms of the superpotential as:
\begin{equation}
	\label{eq_beta:eq_of_state}
	\frac{ p_{\phi} + \rho_{\phi} }{\rho_{\phi} } = \frac{4}{3 \kappa^2 } \left( \frac{W_{,\phi}}{W^2} \right)^2  \ .
\end{equation}
This equation implies that inflation is realized by approaching of a zero of $W_{,\phi}/W$. By taking a derivative of Eq.~\eqref{eq_beta:potsuperpot} with respect to $\phi$, it is possible to show that, if $W(\phi)$ and $W_{,\phi\phi}(\phi)$ are finite, it corresponds to a stationary point of the potential. It is interesting to notice that:
\begin{equation}
\label{eq_beta:RG}
	\kappa \frac{\textrm{d} \phi}{ \textrm{d}\ln a} = \kappa \frac{ \dot{\phi}}{ H} = -\frac{2}{\kappa} \frac{W_{,\phi}}{W} .
\end{equation}
This equation implies that $W_{,\phi}/W$ has exactly the form of a RG equation: 
\begin{equation}
\label{eq_beta:RG_QFT}
\beta (g) \equiv \frac{\textrm{d} g}{ \textrm{d}\ln \mu}  \ ,
\end{equation}
where the role of the renormalized coupling $g$ is played by the field $\phi$, and the role of the renormalization scale $\mu$ is played by the scale factor $a$. In the QFT context, given the $\beta$-function, this equation describes the evolution of the renormalized coupling in terms of the renormalization scale $\mu$. Moreover, in the context of statistical mechanics, the specification of the $\beta$-function (and correspondingly of the RG flow) leads to the definition of Wilsonian~\cite{Wilson:1970ag,Wilson:1971bg,Wilson:1973jj,Wilson:1974mb} picture of fixed points, scaling regions and critical exponents that explains the observed universality that may be originated in correspondence of phase transitions. Given the formal resemblance between Eq.~\eqref{eq_beta:RG} and Eq.~\eqref{eq_beta:RG_QFT}, it seems natural to define:
\begin{equation}
	\label{eq_beta:beta}
	\beta(\phi) \equiv \kappa \frac{\textrm{d} \phi}{ \textrm{d}\ln a} = -\frac{2}{\kappa} \frac{W_{,\phi}}{W}  = \pm \sqrt{3 \frac{p + \rho}{ \rho } } \ ,	
\end{equation}
where we have used Eq.~\eqref{eq_beta:eq_of_state} to express $W_{,\phi}/W$ in terms of $p$ and $\rho$. This equation clearly implies that inflation is realized when we approach a zero of the $\beta$-function. Guided by the analogy with the statistical mechanics, it seems thus natural to interpret the cosmological evolution during inflation in terms of a standard RG equation. In particular, we should stress that once the $\beta$-function is fixed, we can solve Eq.~\eqref{eq_beta:beta} and compute the superpotential. Finally, once the constants of integration are given, the solution of the system is completely specified. In this framework we are naturally lead to a classification of inflationary models in terms of universality classes, that are actually defined by the characterization of $\beta$ in terms of the critical exponents. In particular, given a value $\phi_*$ for the field $\phi$ where $\beta(\phi_*)= 0$, this characterization is realized by specifying the asymptotic expression for $\beta$ in the vicinity of $\phi_*$.\\

\noindent
Notice that given the definition of $N$ number of e-foldings:
\begin{equation}
	\label{eq_beta:N}
	N = - \ln (a/a_f) \ ,
\end{equation}
where $a_f$ is the value of $a$ at the end of inflation, it is possible to express $N$ as a function of $\phi$. It should thus be clear that specifying the ratio $(p + \rho) /\rho$ ( or equivalently $\beta$) in terms of $N$ we can actually solve~\eqref{eq_beta:beta}. In particular, this can be done by following the proposal of Mukhanov~\cite{Mukhanov:2013tua}\footnote{It is however crucial to stress that even if rather general, this parameterization is still special. In particular, in the following sections of this Chapter we show that this parameterization does not contain all the universality classes defined in~\cite{Binetruy:2014zya}.}:
\begin{equation}
	\label{eq_beta:Mukh1}
	\beta(\phi) = \pm \sqrt{3\beta_{\alpha}}\frac{1}{(N+1)^{\alpha/2}} \ , 
\end{equation}
where $\alpha$ and $\beta_{\alpha}$ are dimensionless constants of order one. This proposal can thus be framed in the $\beta$-function formalism for inflation that furthermore allows to define more general parameterizations.

\subsection{Useful formulae.}
\label{sec_beta:formulas}
As explained in the first part of this section, once the $\beta$-function is fixed the system is completely specified. In this section we compute all the useful quantities to describe inflation in terms of the newly defined $\beta$-function formalism. We start by giving the expression for the number of e-foldings:
\begin{equation}
\label{eq_beta:Nbeta}	
	N \equiv - \kappa\int_{\ln a_{\textrm{f}} }^{\ln a } \textrm{d} \ln \hat{a} = -\kappa\int_{\phi_{\textrm{f}}}^{\phi}\frac{\mathrm{d}\hat{\phi}}{\beta(\hat{\phi})} \ ,
\end{equation}
where $a_{\textrm{f}}$ and $\phi_{\textrm{f}}$ denote respectively the value of the scale factor and the value of $\phi$ at the end of inflation. It is crucial to stress that the value of $\phi_{\textrm{f}}$ can be fixed by:
\begin{equation}
	\label{eq_beta:condition_end}
 	\left| \beta(\phi_{\textrm{f}}) \right| \simeq 1 \ .
\end{equation} 
It is interesting to point out that imposing this condition corresponds to giving an initial condition for Eq.~\eqref{eq_beta:phisuperpot}. Moreover, as inflation is realized in region where $\left| \beta(\phi) \right| \ll 1$, it is rather natural to fix the end of inflation when the $\beta$-function becomes of order one\footnote{It is possible to show that $\ddot{a}>0$ corresponds to $\beta^2 < 2$.}. \\

\noindent
Before giving the explicit expressions for the scalar and tensor power spectra of Eq.~\eqref{eq_inflation:power_spectra_final} in terms of our formalism, it is useful to discuss the parameterization of the slow-roll parameters. The definition of the Hubble slow-roll parameters in the context of the Hamilton-Jacobi formalism can be found in~\cite{Lidsey:1995np}. These definitions are usually used in the context of the horizon-flow approach of Hoffman and Turner~\cite{Hoffman:2000ue,Kinney:2002qn,Liddle:2003py}. It is important to stress that, even if this approach is similar in practice to the one described in this work, from a theoretical point of view the two approaches are different. To compute the expressions of Hubble slow-roll parameters in terms of the $\beta$-function formalism, it is sufficient to use the definition of the superpotential $H(\phi) = -W(\phi)/2$: 
\begin{eqnarray}
\label{eq_beta:epsilonH}
	\epsilon_H &\equiv& \frac{2}{\kappa^2}\left(\frac{H_{,\phi}}{H}\right)^{2} =\frac{\beta^2}{2} , \\
\label{eq_beta:etaH}
	\eta_H &\equiv& \frac{2}{\kappa^2}\frac{H_{,\phi \phi}}{H}=\frac{\beta^2}{2}-\frac{\beta_{,\phi}}{\kappa} \ , \\
\label{eq_beta:xiH}
	\xi_H^2 &\equiv& \frac{4}{\kappa^4}\left(\frac{H_{,\phi}H_{,\phi\phi\phi}}{H^2}\right)=\frac{\beta^4}{4}-\frac{3}{2\kappa}\beta^2\beta_{,\phi}+\frac{\beta\beta_{,\phi\phi}}{\kappa^2} \ .
\end{eqnarray}
It is also interesting to report the alternative definition of the slow-roll parameters given by Schwarz \emph{et al.} in~\cite{Schwarz:2001vv,Schwarz:2004tz} and also used in~\cite{Vennin:2014xta}:
\begin{equation}
\label{eq_beta:slow_roll_hierarchy}
\epsilon_{0} \equiv \frac{H }{H_f} \ , \qquad \qquad \epsilon_{i+1} \equiv - \frac{\textrm{d} \ln| \epsilon_{i} | }{\textrm{d} \ln a} \ .
\end{equation} 
Notice that in terms of the $\beta$-function formalism this can simply be expressed as:
\begin{equation}
\label{eq_beta:hierarchy_beta}
\epsilon_{0} \equiv \frac{W(\phi) }{W_{\textrm{f}}} \ , \qquad \qquad \epsilon_{i+1} \equiv - \frac{\textrm{d} \phi }{\textrm{d} \ln a} \frac{\textrm{d} \ln| \epsilon_{i} | }{\textrm{d} \phi } = - \frac{\beta(\phi)}{\kappa} \frac{\textrm{d} \ln| \epsilon_{i} | }{\textrm{d} \phi } \ .
\end{equation} 
and the first slow-roll parameters can simply be expressed as:
\begin{equation}
	\epsilon_1 = \frac{\beta^2}{2} \ , \qquad \epsilon_2 = -2\frac{\beta_{,\phi}}{\kappa}, \qquad \epsilon_3 = -\frac{\beta \beta_{,\phi\phi}}{\kappa \beta_{,\phi}} \ .
\end{equation}

\noindent
We proceed by expressing the scalar and tensor power spectra for scalar and tensor perturbations Eq.~\eqref{eq_inflation:power_spectra_final} in terms of our formalism:
\begin{equation}
\begin{aligned}
\label{eq_beta:power_spectra_final}
  \left. \Delta^2_s (k)  \right|_{k = -a W/2} & = \frac{ \kappa^2 }{16 \pi^2 } \frac{W^2 }{  \beta^2 }   , \\
\left. \Delta^2_t (k)  \right|_{k = -a W/2} & =  \frac{ \kappa^2 W^2}{2 \pi^2 }   \ .
\end{aligned}
\end{equation}
Notice that again, for the models considered in this Chapter the value of speed of sound $c_s$ is set equal to one. Finally, assuming to be close to the fixed point \textit{i.e.} $\beta(\phi) \ll 1$, we can compute the lowest order expression for the scalar and tensor spectral indexes given in Eq.~\eqref{eq_inflation:tensor_spectral_index}:
\begin{eqnarray}
	\label{eq_beta:ns}
	\frac{\textrm{d} \Delta^2_s (k)}{\textrm{d} \ln k} &\equiv& n_{s}-1 \simeq -\beta^{2}-\frac{2\beta_{,\phi}}{\kappa} \ , \\
	\label{eq_beta:nt}
	\frac{\textrm{d} \Delta^2_t (k)}{\textrm{d} \ln k} &\equiv & n_{t} \simeq -\beta^{2} \ ,
\end{eqnarray}
notice that to get these equations $k$ should be evaluated at horizon crossing \textit{i.e.} at $k = -a W/2$. This implies that:
\begin{equation}
 	\frac{\textrm{d}\ln k}{\textrm{d}\phi} = \kappa \frac{1- \beta^2/2}{\beta} \ .
\end{equation} 
Similarly we get the running of the scalar and tensor spectral indexes:
\begin{eqnarray}
\label{eq_beta:dns/dlnk}
\frac{\mathrm{d}n_{s}}{\mathrm{d}\ln k} & \simeq &
-\frac{2}{\kappa}\beta^2\beta_{,\phi}-\frac{2}{\kappa^2}\beta\beta_{,\phi\phi} \ , \\
\label{eq_beta:alpha}
\frac{\mathrm{d}n_{t}}{\mathrm{d} \ln k} &\simeq& -\frac{2}{\kappa}\beta^2\beta_{,\phi}
\ ,
\end{eqnarray}
and the tensor-to-scalar ratio:
\begin{equation}
	 \label{eq_beta:r}
	r = 8 \beta^2 \ .
\end{equation}
Before concluding this section it is also useful to give the expressions for the potential and the superpotential in terms of the $\beta$-function:
\begin{equation}
	\label{eq_beta:pot_superpot_beta}
 	W(\phi) = W_{\textrm{f}} \exp\left[\int_{\phi_{\textrm{f}}}^\phi \frac{\kappa \beta(\hat{\phi})}{2} \textrm{d}\hat{\phi}\right]\ , \qquad V(\phi) = \frac{3 W^2(\phi)}{4 \kappa^2} \left[ 1 - \frac{\beta^2(\phi)}{6} \right] \ .
 \end{equation} 
In the next section we define some parameterizations for the $\beta$-function and we use these expressions to compute the predictions for the observable quantities. For simplicity and without loss of generality, when we use this formalism we always assume the field $\phi$ to be positive.

\section{Universality classes.}
\label{sec_beta:universality_classes}

\begin{table}[h]
\begin{center}\begin{tabular}{|c|c|c|}
\hline
\textbf{Class} & \textbf{Name} &$\beta(\phi)$\\
\hline\hline
{\textbf{Ia}(q)} & Monomial & $ \hat{\beta}_{q}(\kappa\phi)^{q}$, $q>1$ \\
\hline
{\textbf{Ia}(1)} & Linear & $ \hat{\beta}_{1}(\kappa\phi)$ \\
\hline
{\textbf{Ib}(p)} & Inverse Monomial &  $-\hat{\beta}_p /(\kappa\phi)^p$, $p>1$ \\
\hline
{\textbf{Ib}(1)} & Chaotic &  $-\hat{\beta}_1 /(\kappa\phi)$ \\
\hline
{\textbf{Ip}(p)} & Fractional &  $-\hat{\beta}_p /(\kappa\phi)^p$, $0<p<1$ \\
\hline
{\textbf{Ib}(0)} & Power Law &  $-\hat{\beta}_0 /(\kappa\phi)^p$, $p=0$ \\
\hline\hline
{\textbf{II}($\gamma$)} & Exponential & $-\hat{\beta}\exp[-\gamma \kappa\phi]$\\
\hline
\end{tabular} \caption{Summary of the universality classes.}
\label{table_beta:classes}
\end{center}
\end{table}

\noindent
In this section we proceed by defining a classification of the different inflationary models in terms of the $\beta$-function formalism. Notice that as $\beta(\phi)$ goes to zero approaching the fixed point, the particular parameterizations of $\beta$ is not required to be identified with the complete $\beta$-function of the system\footnote{Actually, far away from the fixed point, the complete $\beta$-function can also be non-perturbative.}. On the contrary, it must be intended as the leading order in the expansion of $\beta(\phi)$ close to the fixed point. In this sense a parameterization does not only specify a single inflationary model but rather a whole set of theories that share a single scale invariant limit. In the language of statistical mechanics specifying a particular parameterization of the $\beta$-function we are specifying a \emph{Universality class} for inflationary models. Following the treatment of~\cite{Binetruy:2014zya} we proceed by specifying a set of Universality classes and we compute the corresponding predictions for the observable quantities. The universality classes considered in this work are summarized by Table~\ref{table_beta:classes}.\\

\begin{figure}[h!]
\centering
\subfloat[][\emph{Planck 2015~\cite{Ade:2015lrj} marginalized joint $68\%$ and $95\%$ plot for $(n_s,r)$.}]
{\includegraphics[width=.8\columnwidth]{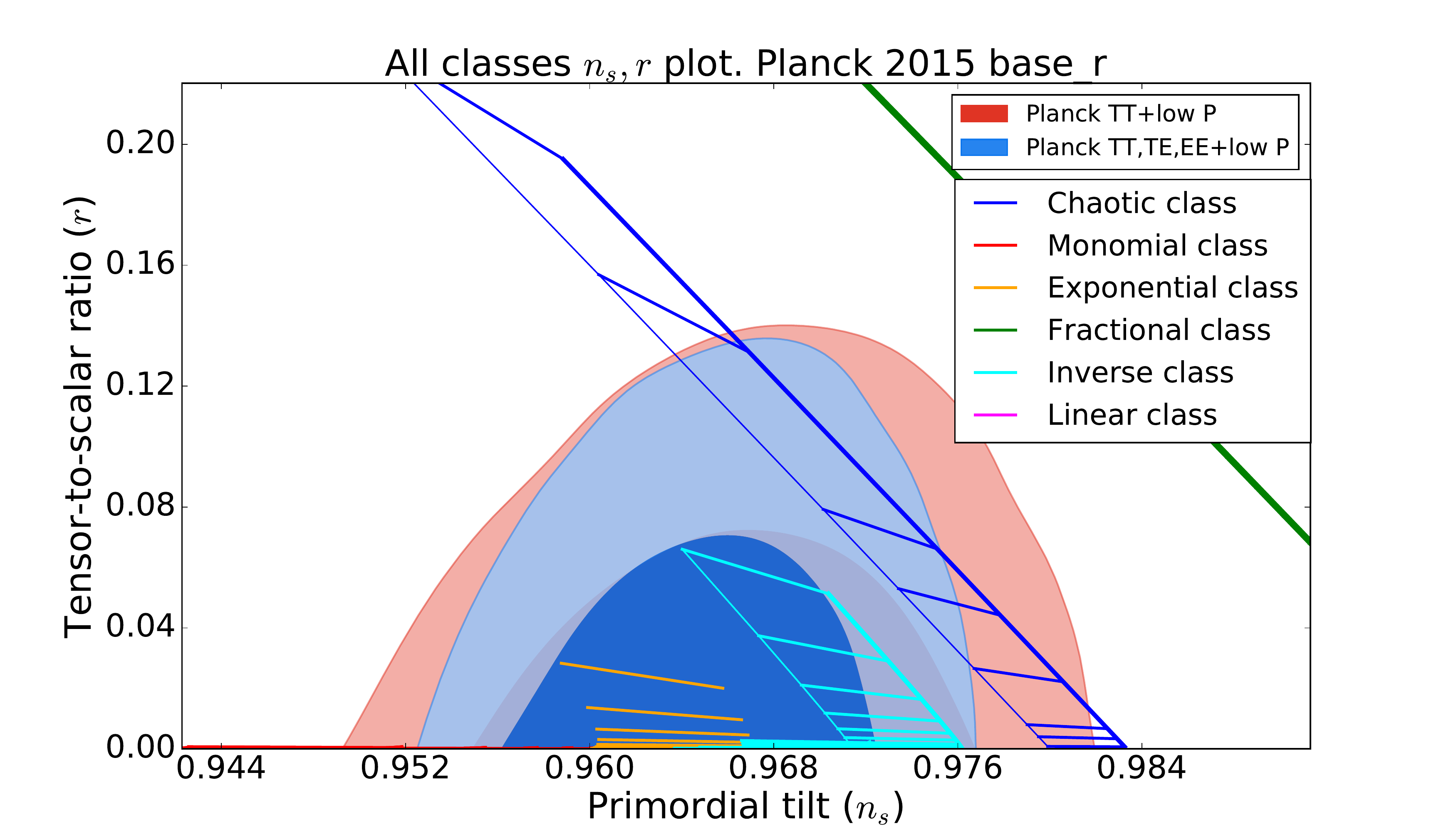}} \quad
\subfloat[][\emph{Planck 2015~\cite{Ade:2015lrj} marginalized joint $68\%$ and $95\%$ plot for $(n_s, r)$. For these plots the running $\alpha_s$ is left as a free parameter.}]
{ \includegraphics[width=.8\columnwidth]{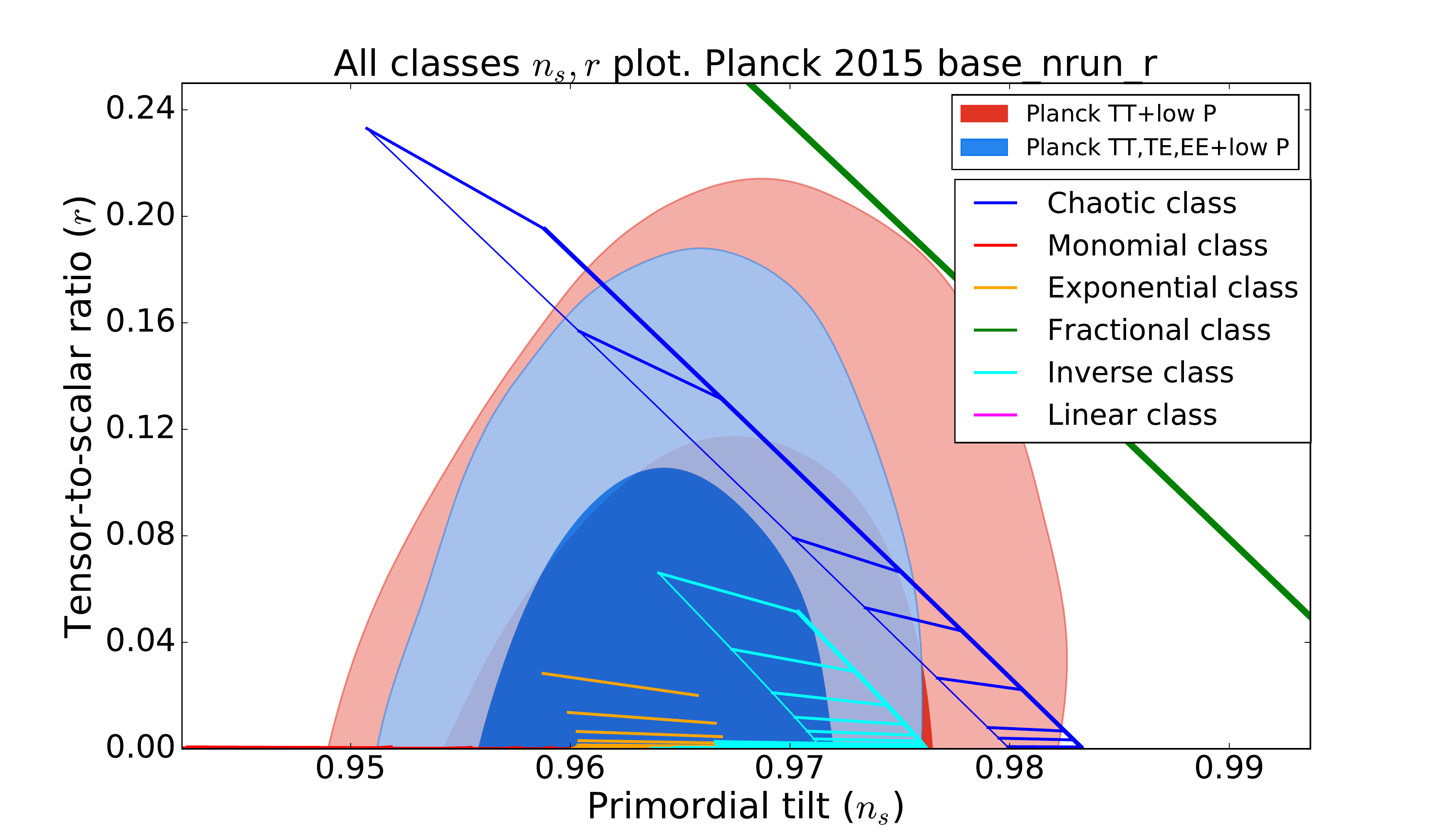}} \\
\caption{Predictions for the different universality classes for models of inflation in the plane $(n_s,r)$. On the background we show the Planck constraints~\cite{Ade:2015lrj} on $n_s$ and $r$.\label{fig_beta:all_Planck}}
\end{figure}

\noindent
Before considering in detail the classes of Table~\ref{table_beta:classes}, we present a set of plot summarizing the results for all the different classes introduced in~\cite{Binetruy:2014zya} and discussed in the present work. We start with the plot of Fig.~\ref{fig_beta:all_Planck}, that shows the predictions for the values of $n_s$ and $r$ for the different universality classes. These predictions are compared to the constraints imposed by using Planck data~\cite{Ade:2015lrj}. To produce these plots we first choose the universality class and then we specify model by making a particular choice for the parameters. Different models are represented using segments that are actually showing the predictions for $n_s$ and $r$ for different values of $N$. For each segment, the value $N =50$ is always on the left end of the segment and the value $N=60$ is always corresponding to the right end. Notice that for greater values for $N$ the segments are always extended to the right.\\

\begin{figure}[h!]
\centering
{\includegraphics[width=.8\columnwidth]{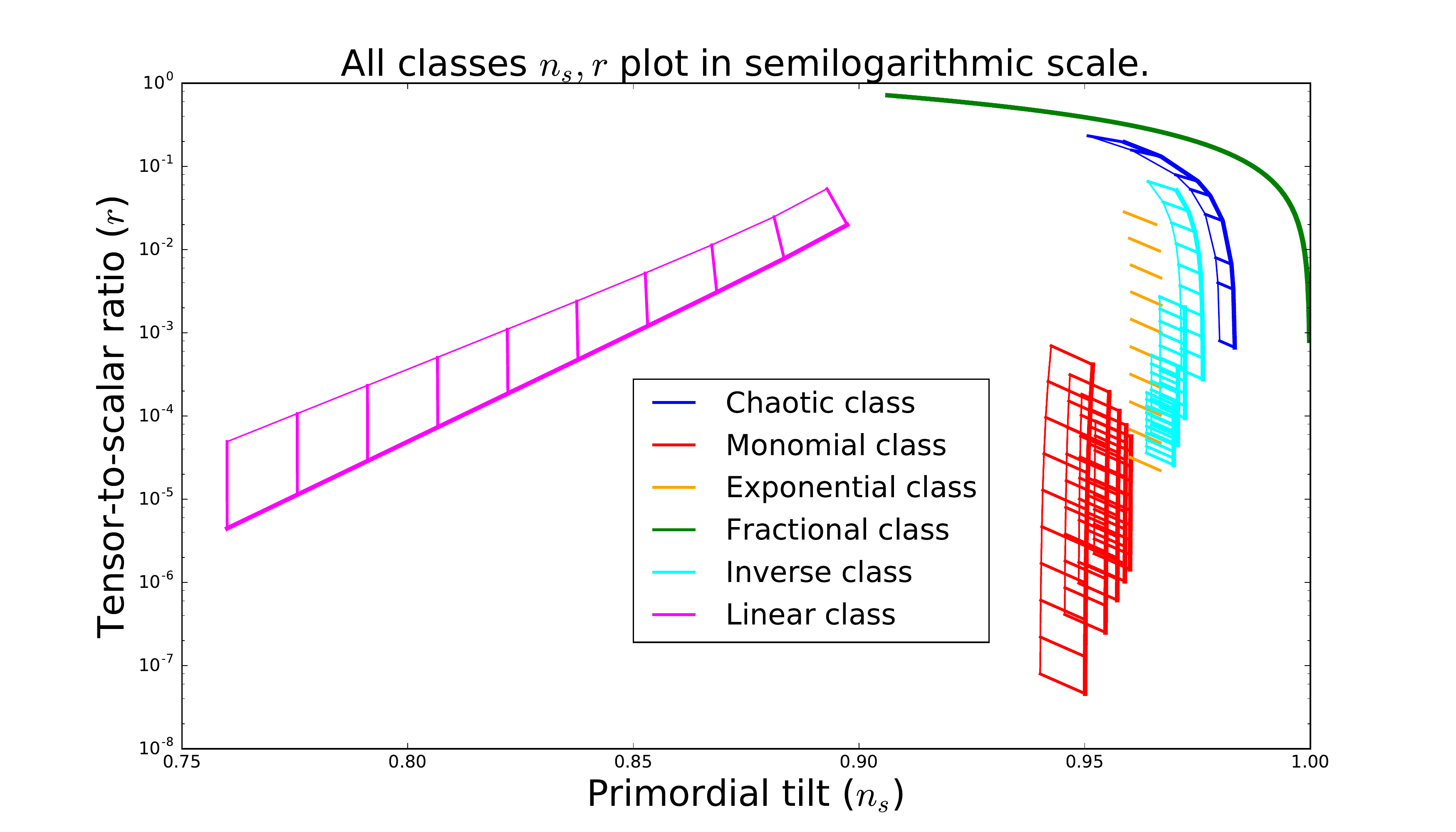}} 
\caption{Semilogarithmic plots of the predictions for $n_s$ and $r$ given by the different universality classes.\label{fig_beta:semilog}}
\end{figure}

\noindent
Another set of plots is shown in Fig.~\ref{fig_beta:semilog} and in Fig.~\ref{fig_beta:log_plots}. In Fig.~\ref{fig_beta:semilog} we present a semilogarithmic plot of the predictions for $n_s$ and $r$ for the different classes. As the values of $r$ are now presented in logarithmic scale, we can have a better picture of the region where $r \ll 1$. In particular, we may notice that as expected small field models\footnote{The Linear class is ignored as completely outside of the region that is preferred by the Planck data.} predict extremely small values of $r$. In the plot of Fig.~\ref{fig_beta:all_classes_log} we show the predictions for $n_s$ and $\alpha_s$ given by the different classes compared with the Planck constraints~\cite{Ade:2015lrj} in the presence of running. Notice that this plot clearly shows that all the classes defined in~\cite{Binetruy:2014zya} are predicting small values for the running. For completeness, in Fig.~\ref{fig_beta:all_running_log} we are also showing the predictions for $\alpha_s$ and $r$. Again the values of $r$ are now presented in logarithmic scale in order to have a better picture of the region where $r \ll 1$.\\

\noindent
In the rest of this section we present in detail all the universality classes of the classes of Table~\ref{table_beta:classes}. We start by presenting small field models, \textit{i.e.} models where the fixed point is reached at a finite value $\phi_0$, that without loss of generality can be set to be equal to zero for $\phi$. We then move to models where the fixed point is reached for an infinite value for the inflaton field. To help the reader to distinguish between small and large field models, the plots showing the $n_s$, $r$ predictions for small field models are presented in purple, while the plots for large field models are in red.

\begin{figure}[ht]
	\centering
	\subfloat[][\emph{Predictions for the different universality classes for models of inflation in the $(n_s,r)$ plane. $r$ is in logarithmic scale.\label{fig_beta:all_classes_log}}]
	{\includegraphics[width=.8\columnwidth]{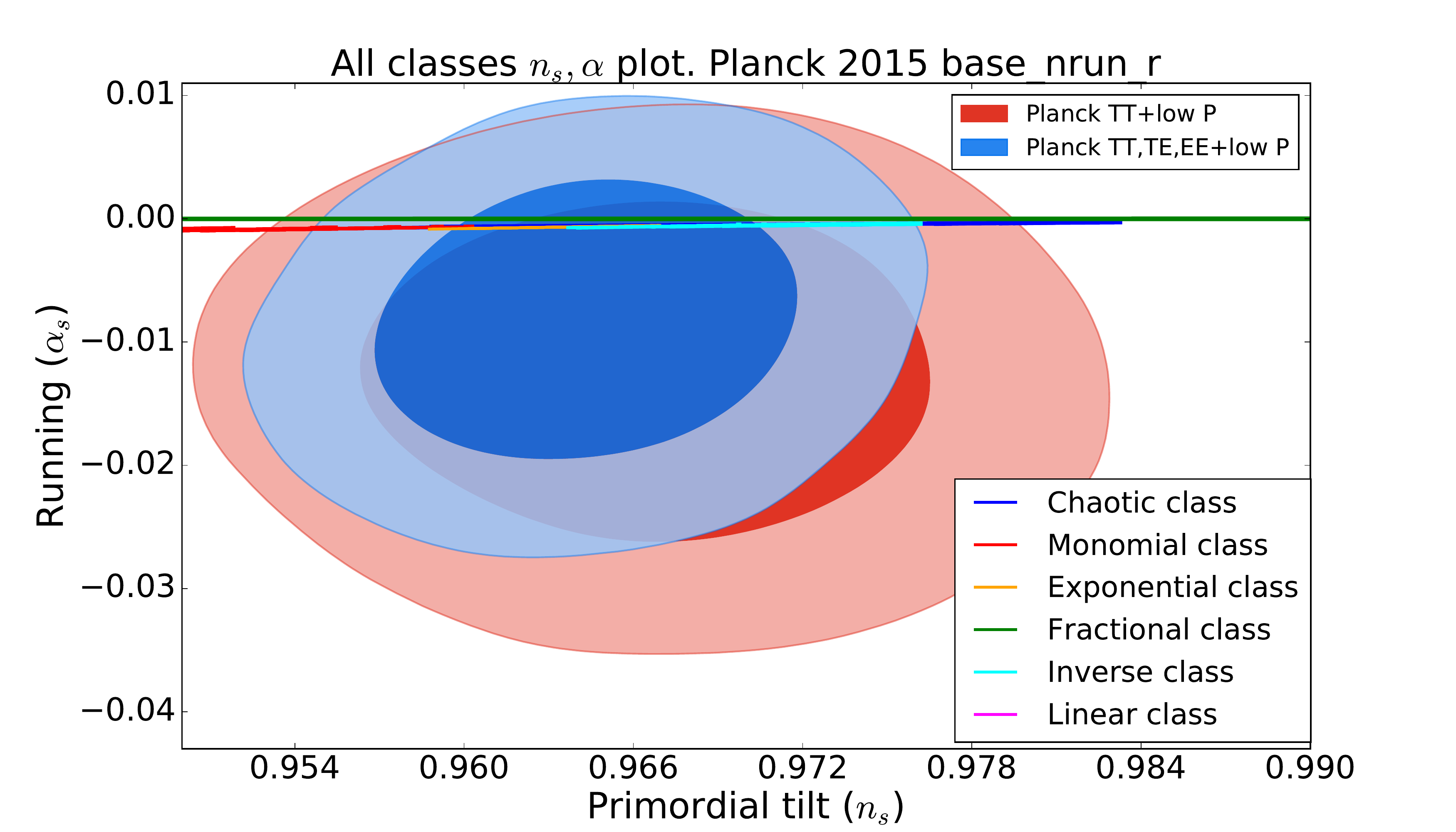}} \quad
	\subfloat[][\emph{Predictions for the different universality classes for models of inflation in the $(\alpha_s,r)$ plane. $r$ is in logarithmic scale.\label{fig_beta:all_running_log}}]
	{\includegraphics[width=.8\columnwidth]{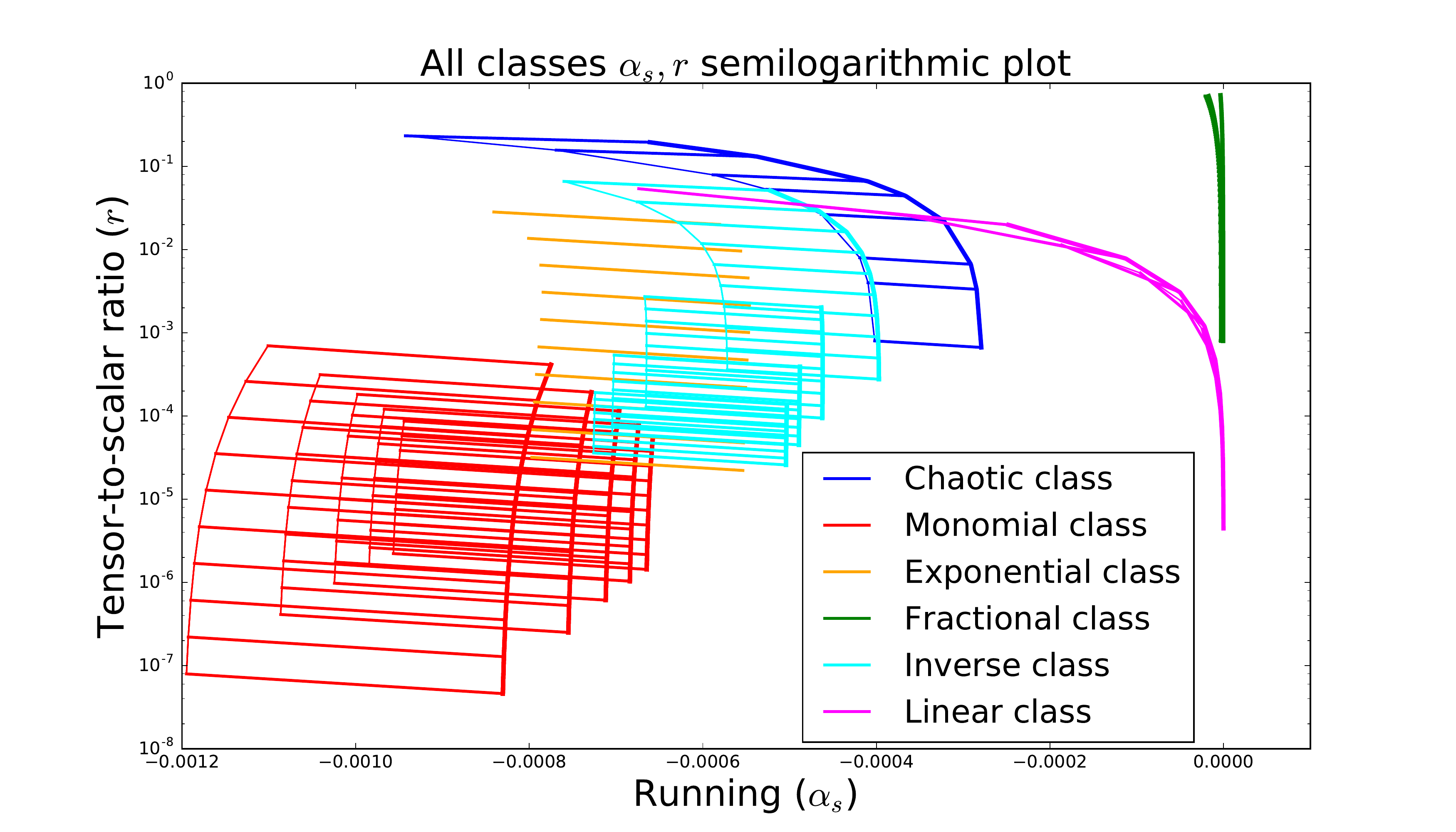}}
	\caption{Logarithmic plots for the predictions of the different universality classes for models of inflation.}
	\label{fig_beta:log_plots}
\end{figure}

\subsection{Small field inflation.}
\label{sec_beta:Small_field}
In this section we present models where the fixed point is reached at a finite value $\phi_0$ for the inflaton field and without loss of generality we fix $\phi_0 = 0$. This corresponds to the models \textbf{Ia(q)} and \textbf{Ia(1)} of Table~\ref{table_beta:classes}. In general, for the universality classes for which the fixed point is defined at a finite value for $\phi$, we can expand the $\beta$-function in a neighborhood of the fixed point as:
\begin{equation}
\label{eq_beta:betaIa}
\beta(\phi) = \beta_q (\kappa \phi )^q \ ,
\end{equation}
where $\beta_q>0$ and $q>0$ are constants of order one. As explained in~\cite{Binetruy:2014zya}, $q < 1$ leads to divergent slow-roll parameters and thus it does not correspond to inflation. We are thus left with $q \geq 1$ that leads to the definition of two different classes called \emph{Linear} and \emph{Monomial} class.

\subsubsection{ Monomial class: Ia(q).}
\label{sec_beta:monomial_class} 

\begin{figure}[ht]
	\begin{center}
	\includegraphics[width=.8\textwidth]{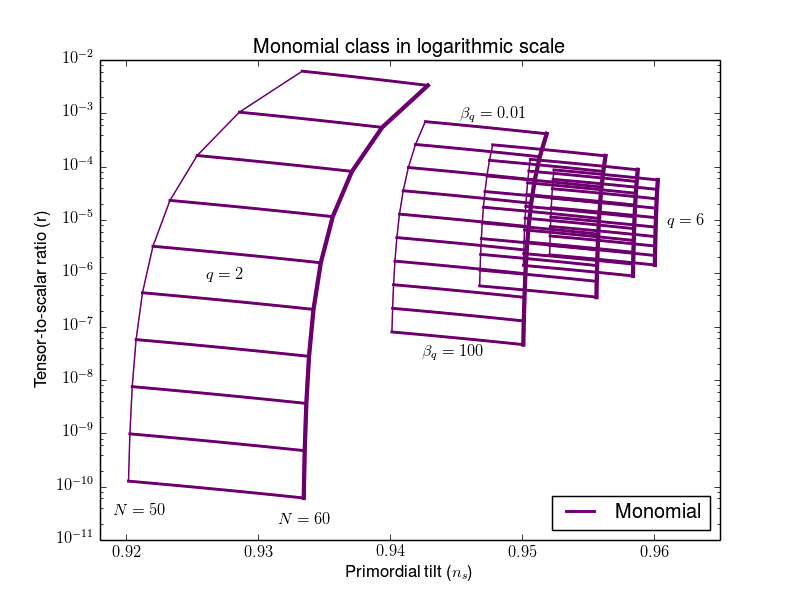}
	\end{center}
	\caption{Prediction for $n_s$ and $r$ for the monomial class. The values of $r$ are in logarithmic scale.}
	\label{fig_beta:monomial_log}
\end{figure}

In this class we have models described by the $\beta$-function of equation Eq.~\eqref{eq_beta:betaIa} with $q>1$\footnote{Notice that the case with $q<1$ does not correspond to inflation. This should be clear by looking at the expression for the potential given in Eq.~\eqref{eq_beta:potential_monomial}. In particular, the second derivative is divergent at the fixed point and thus $\eta_H$ diverges.}. Using Eq.~\eqref{eq_beta:pot_superpot_beta} we can easily compute the potential and the superpotential for the models of this class:
\begin{eqnarray}
	W(\phi) &=& W_{\textrm{f}} \exp \left\{- \frac{\beta_q}{2 (q+1)}  \left[(\kappa\phi)^{q+1} - (\kappa\phi_{\textrm{f}})^{q+1}\right]  \right\} \ , \label{eq_beta:superpotential_monomial} \\
	\label{eq_beta:potential_monomial} 
	V(\phi) &\simeq& \frac{3 W_{\textrm{f}}^2}{4 \kappa^2} \left\{1 -  \frac{\beta_q}{ q+1} \left[(\kappa\phi)^{q+1} - (\kappa\phi_{\textrm{f}})^{q+1}\right] + \mathcal {O}(\phi^{2q}) \right\} \ ,
\end{eqnarray}
where $\phi_{\textrm{f}}$ and $W_{\textrm{f}}$ are respectively the value of the field and of the superpotential at the end of inflation. Notice that Eq.~\eqref{eq_beta:potential_monomial} clearly implies that the value of $W_{\textrm{f}}$ is directly related with the scale on inflation. As a consequence this parameter can be fixed using the COBE normalization. It is also interesting to stress that at the lowest order Eq.~\eqref{eq_beta:potential_monomial} matches with the expressions for the Hilltop potentials of Sec.~\ref{sec_inflation:hilltop} with $p>2$ and $v = \kappa$. We proceed by computing the number of e-foldings:
\begin{equation}
	\label{eq_beta:monomial}
	N = \frac{1 }{ \beta_q (q-1)(\kappa\phi)^{q-1}} - \lambda
	\ ,
\end{equation}
where we defined $\lambda$ as:
\begin{equation}
	\lambda  \equiv\frac{1}{  \beta_q (q-1) \left(\kappa \phi_{\textrm{f}} \right)^{q-1}}\ .
\end{equation}
In particular, using Eq.~\eqref{eq_beta:condition_end}, it is easy to show that $\lambda$ can be expressed as:
\begin{equation}
	\lambda \simeq \frac{\beta_q^{-1/q}}{(q-1)}\ .
\end{equation}
Notice that for $\beta_q $ of order one also $\lambda$ is of order one. Finally, we can express the $\beta$-function in terms of $N$ as:
\begin{equation}
	\label{eq_beta:beta_N_monomial}
	\beta(N) = {1 \over \left[  \beta_q^{\frac{1}{q}} (q-1) (N+\lambda)
\right]^{\frac{q}{q-1}}}
\end{equation}
As $\lambda$ is of order one it can be safely neglected with respect to $N$. The scalar spectral index can thus be expressed as:
\begin{equation}
	\label{eq_beta:scalar_spectral_monomial}
	n_s -1 \simeq - 2\frac{\beta_{,\phi}}{\kappa} = - 2 q \beta_q \left[\kappa(\phi - \phi_{\textrm{f}})\right]^{q-1} \simeq  - \frac{2q}{ q-1} \frac{1 }{ N} \ .
\end{equation}
Similarly we can compute the running of the scalar spectral index
\begin{equation}
	\label{eq_beta:running}
	\alpha_s \simeq - \frac{ 2 \beta \beta_{,\phi\phi} }{\kappa^2} = - 2 q (q-1) \beta_q^2 \left[\kappa(\phi - \phi_{\textrm{f}})\right]^{2(q-1)} =-\frac{2q }{ q-1} \frac{1 }{ N^2} \ ,
\end{equation}
and the tensor-to-scalar ratio:
\begin{equation}
	\label{eq_beta:tensor_monomial}
	r = 8 \beta^2 \simeq \frac{8}{\beta_q^{2/(q-1)} \left[(q-1) N\right]^{2q/(q-1)}} \ .
\end{equation}
The slow-roll parameter can be directly computed from Eq.~\eqref{eq_beta:epsilonH}, Eq.~\eqref{eq_beta:etaH} and Eq.~\eqref{eq_beta:xiH}. The explicit expressions are reported in~\cite{Binetruy:2014zya}. Notice that in the parameterization of Mukhanov~\cite{Mukhanov:2013tua} this class corresponds to the models with $\alpha = 2q/(q-1)$ with $\alpha > 2$. It is interesting to notice that this class actually has three free parameters  $q,\beta_q$ and the number of e-folding. As from CMB experiments we can fix constraints on $n_s, r$ and $\alpha_s$, the three parameters can be independently fixed in order to reproduce the observed values. Numerical predictions for this class are shown in Fig.~\ref{fig_beta:monomial_log}.

\subsubsection{ Linear class: Ia(1).}
\label{sec_beta:linear_class} 

\begin{figure}[ht]
	\begin{center}
	\includegraphics[width=.8\textwidth]{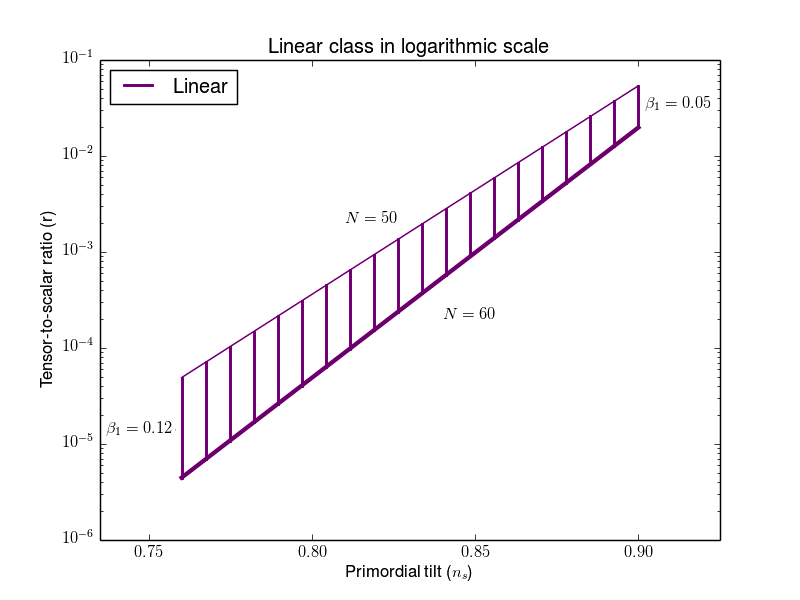}
	\end{center}
	\caption{Prediction for $n_s$ and $r$ for the linear class. The values of $r$ are in logarithmic scale.}
	\label{fig_beta:linear_log}
\end{figure}

\noindent 
Let us consider the $\beta$-function of Eq.~\eqref{eq_beta:betaIa} with $q=1$: 
\begin{equation}
	\label{eq_beta:beta_linear}
	\beta(\phi) = \beta_1 \kappa \phi \ .
\end{equation}
As we explain in the following this case is special and thus it should be considered on his own. The superpotential and potential are given by:
\begin{eqnarray}
	\label{eq_beta:superpotential_linear}
	W(\phi) &=& W_{\textrm{f}} \exp \left\{ - \frac{\beta_1}{ 4} \left[(\kappa\phi)^{2} - (\kappa\phi_{\textrm{f}})^{2}\right] \right\} \ ,  \\
	\label{eq_beta:potential_linear}
	V(\phi) &=& \frac{3 W_{\textrm{f}}^2}{4 \kappa^2} \left\{ 1 -  \frac{\beta_1}{2}  (1 + \beta_1 /3)   \left[(\kappa\phi)^{2} - (\kappa\phi_{\textrm{f}})^{2}\right] + \mathcal{O}\left[ (\kappa\phi)^4 \right]  \right\} .
\end{eqnarray}
It should be clear from Eq.~\eqref{eq_beta:potential_linear}, that at the lowest order this class reproduces the Hilltop potential of Sec.~\ref{sec_inflation:hilltop} with $p=2$ and $v = \kappa$. Using Eq.~\eqref{eq_beta:Nbeta}, it is easy to show that for this class the number of e-folding can be expressed as:
\begin{equation}
	\label{eq_beta:n_linear}
	N = -\frac{1}{\beta_1} \ln \left(\frac{\phi}{\phi_{\textrm{f}}} \right) \ ,
\end{equation}
using Eq.~\eqref{eq_beta:condition_end}, we can thus express the $\beta$-function for this class in terms of $N$ as:
\begin{equation}
	\label{eq_beta:beta_N_linear}
	\beta(N) = e^{-N\beta_1} \ .
\end{equation}
Notice that this particular expression cannot be recovered by using the parameterization of Mukhanov~\cite{Mukhanov:2013tua}. We can proceed by computing the tensor-to-scalar ratio, the scalar spectral index and the running:
\begin{eqnarray}
	\label{eq_beta:tensor_linear}
	r &=& 8 \beta_1^2  \left( \kappa \phi \right)^2 = 8 e^{-2N\beta_1} \ , \\
	\label{eq_beta:scalar_spectral_linear}
	n_s -1 &\simeq&  - 2\beta_1  \ , \qquad \alpha_s = - 2 \beta_1 e^{-2N\beta_1} \ .
\end{eqnarray}
Again the equations for the slow-roll parameters can be computed from Eq.~\eqref{eq_beta:epsilonH}, Eq.~\eqref{eq_beta:etaH} and Eq.~\eqref{eq_beta:xiH}. It is interesting to notice that the second slow-roll parameter $\eta_H$ reads:
\begin{equation}
	\eta_H = - \beta_1   \ , \label{eq_beta:eta_linear}
\end{equation}
and thus to ensure slow-rolling in a neighborhood of $\phi_{\textrm{f}}$ we should also require $\beta_1 \ll 1$. As discussed in Chapter~\ref{chapter:holographic_universe}, this case is also special from the holographic point of view. Notice that this class only has two free parameters, namely $\beta_1$ and $N$. As a consequence, the predictions cannot be arbitrarily adjusted to match with direct observations. Numerical predictions for this class are shown in Fig.~\ref{fig_beta:linear_log}.

\subsection{Large field inflation.} 
\label{sec_beta:large_field}
In this section we discuss models where the fixed point is approached for large values of the inflation field, \textit{i.e.} at $|\phi| \rightarrow \infty$. In this class we consider two different parameterizations for the $\beta$-function:
\begin{itemize}
	\item The first of these parameterizations is:
	\begin{equation}
		\label{eq_beta:beta_Ib}
		\beta(\phi) \simeq - \frac{\hat{\beta_p}}{ \left[\kappa\phi \right]^p} \ ,
	\end{equation}
	where $p\geq 0$ and $\hat{\beta_p}>0$ are constants of order one. This parameterization should be divided into several different classes: 
	\begin{enumerate}
		\item Inverse class \textbf{Ib(p)}, with $p>1$.
		\item Chaotic class \textbf{Ib(1)}, with $p>1$.
		\item Fractional class \textbf{Ib(p)}, with $1<p$.
		\item Power law class \textbf{Ib(0)}, with $p=0$.
	\end{enumerate}
	\item The second parameterizations is:
	\begin{equation}
		\label{eq_beta:beta_exponential}
		\beta(\phi) \simeq \hat{\beta} \exp\left(-\gamma \kappa \phi \right),
	\end{equation}
	where $\hat{\beta} > 0$ and $\gamma > 0$ are again constants of order one.
\end{itemize}
Notice that without loss of generality we have assumed the field $\phi$ to be positive. So that for all of these classes, except for the Power law class \textbf{Ib(0)} that is special, the fixed point is reached at $\phi \rightarrow + \infty$.

\subsubsection{ Inverse monomial class: Ib(p).}
\label{sec_beta:inverse_class} 

\begin{figure}[ht]
	\begin{center}
	\includegraphics[width=.8\textwidth]{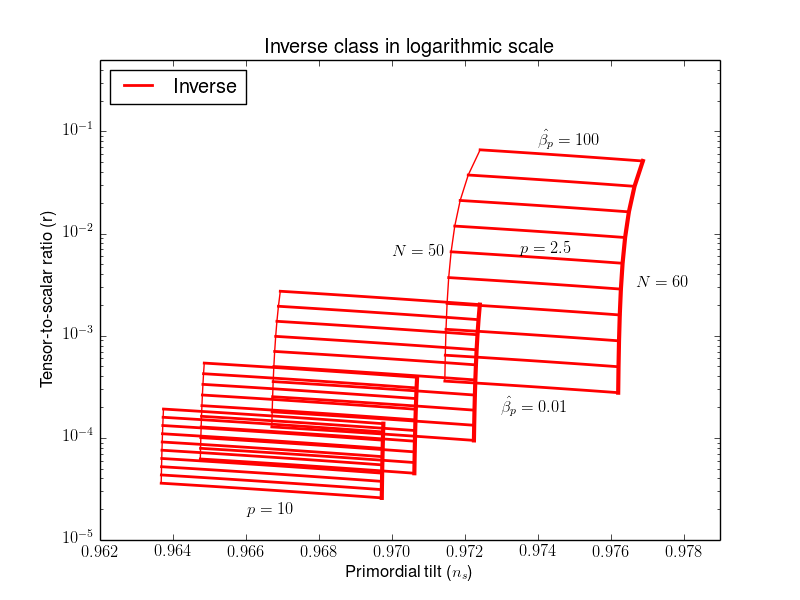}
	\end{center}
	\caption{Prediction for $n_s$ and $r$ for the inverse class. The values of $r$ are in logarithmic scale.}
	\label{fig_beta:inverse_log}
\end{figure}

\noindent In this class the $\beta$-function is parameterized as in Eq.~\eqref{eq_beta:beta_Ib} with $p > 1$. As usual, the superpotential and the potential are given by Eq.~\eqref{eq_beta:pot_superpot_beta}:
\begin{eqnarray}
	W(\phi) &=& W_{\textrm{f}} \exp \left[- {\hat \beta_p \over 2 (p-1)}\frac{1}{ \left[\kappa\phi \right]^{p-1}} \right] \ , \label{eq_beta:superpotential_inverse} \\
	V(\phi) &=& \frac{3 W_{\textrm{f}}^2}{ \kappa^2} \left[1 -  \frac{\hat{\beta}_p}{ (p-1)} \frac{1}{\left[\kappa\phi \right]^{p-1}} + {\mathcal {O}} \left(\phi^{-2(p-1)}\right) \right] \ , \label{eq_beta:potential_inverse}
\end{eqnarray}
We can thus compute the number of e-foldings:
\begin{equation}
	N =  \frac{\left[\kappa\phi \right]^{p+1}}{ (p+1) \hat \beta_p} - \lambda \ , \qquad  \lambda \equiv \frac{\left[\kappa\phi_{\textrm{f}} \right]^{p+1}}{ (p+1) \hat{\beta_p}} \ . \label{eq_beta:N_inverse}
\end{equation}
As usual, we assume the $\beta$-function to be of order one at the end of inflation to get $\lambda \simeq \hat \beta_p^{1/p}/(p+1)$. Again $\lambda$ can be neglected with respect to $N$ and thus the $\beta$-function can be expressed as:
\begin{equation}
	\label{eq_beta:beta_N_inverse}
	\beta(N) = - \frac{\hat{\beta}_p^{\frac{1}{p+1}}}{\left[(p+1) (N+\lambda) \right]^{\frac{p}{p+1}}} \ .
\end{equation}
Finally we compute the scalar spectral index:
\begin{equation}
\label{eq_beta:scalar_spectral_inverse}
	n_s -1 \simeq  - 2 \frac{\beta_{,\phi}}{\kappa} = -2p\frac{\hat{\beta}_p }{ \left[\kappa\phi \right]^{(p+1)}}\simeq - \frac{2p}{p+1} \frac{1}{ N} \ , 
\end{equation}
the running:
\begin{equation}
	\label{eq_beta:running_inverse}
	\alpha_s \simeq -2 \frac{\beta \beta_{,\phi\phi}}{\kappa^2} = -2p(p+1) \frac{\hat{\beta}_p^2 }{ \left[\kappa\phi \right]^{2(p+1)}} \simeq -\frac{2p }{ p+1}\frac{1}{N^2} \ ,
\end{equation}
and the tensor-to-scalar ratio:
\begin{equation}
	\label{eq_beta:tensor_inverse}
	r = 8 \beta^2 \simeq \frac{8 \hat{\beta}_p^{\frac{2}{p+1}}}{\left[ (p+1)  N\right]^{\frac{2p}{p+1}}} \ .
\end{equation}
Note that in the parameterization proposed by Mukhanov~\cite{Mukhanov:2013tua} we can identify $\alpha$ with $2p/(p+1)$. As we have fixed $p>1$, these models correspond to the case $1<\alpha<2$. As usual the slow-roll parameters can be obtained using Eq.~\eqref{eq_beta:epsilonH}, Eq.~\eqref{eq_beta:etaH} and Eq.~\eqref{eq_beta:xiH}. \\

\noindent Similarly to the class \textbf{Ia(q)}, in this class we have three different parameters, namely $q, \beta_q$ and $N$ and thus we can adjust these parameters in order to adjust the values of $n_s$, $r$ and $\alpha_s$. Numerical predictions for this class are shown in Fig.~\ref{fig_beta:inverse_log}.

\subsubsection{ Chaotic class: Ia(1).}
\label{sec_beta:chaotic_class} 

\begin{figure}[ht]
	\begin{center}
	\includegraphics[width=.8\textwidth]{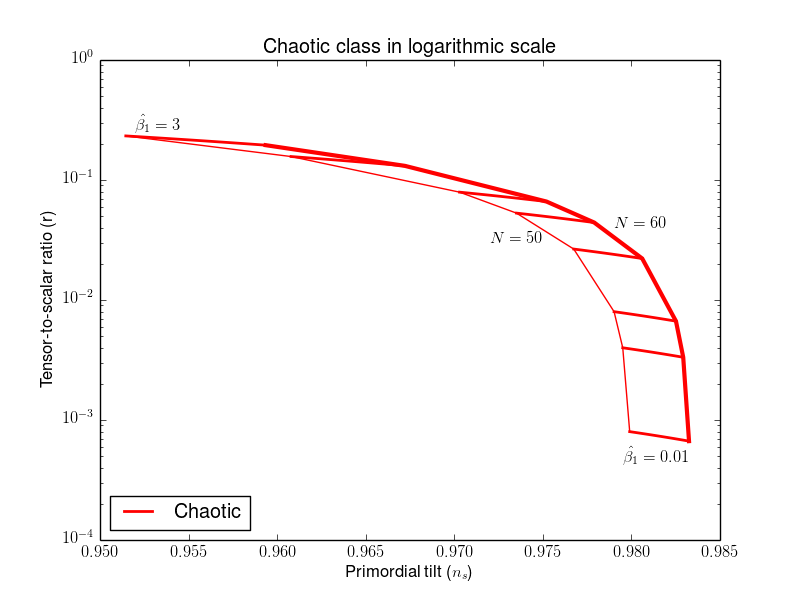}
	\end{center}
	\caption{Prediction for $n_s$ and $r$ for the chaotic class. The values of $r$ are in logarithmic scale.}
	\label{fig_beta:chaotic_log}
\end{figure}

\noindent In this class the $\beta$-function is given by the parameterization of Eq.~\eqref{eq_beta:beta_Ib} with $p = 1$ that, as we show in the following, is special. In this case the superpotential reads:
\begin{equation}
	\label{eq_beta:superpotential_chaotic}
	W(\phi) = W_{\textrm{f}} \left(\kappa\phi \right)^{\frac{\hat{\beta}_1}{2}} \ ,
\end{equation}
where as usual $W_{\textrm{f}}$ is the value of $W(\phi)$ at the end of inflation that is fixed by the COBE normalization. It is then easy to show that the potentials for this class reproduce the chaotic potentials~\cite{Linde:1983gd} of Sec.~\ref{sec_inflation:chaotic}: 
\begin{equation}
	\label{eq_beta:potential_chaotic}
	V(\phi) = \frac{3 W_{\textrm{f}}^2}{4 \kappa^2} \left[ 1 - \frac{\hat{\beta}_1^2}{6 \left(\kappa\phi \right)^2}\right] \left(\kappa\phi \right)^{\hat{\beta}_1} \simeq \frac{3 W_{\textrm{f}}^2}{4 \kappa^2} \left(\kappa\phi \right)^{\hat{\beta}_1} \ .
\end{equation}
It is interesting to notice that in this class the parameter $\hat{\beta}_1$ directly specifies the exponent in the potential. As usual, we proceed by computing the number of e-foldings:
\begin{equation}
N =  \frac{\left(\kappa\phi \right)^2}{2 \hat{\beta}_1} - \lambda \ , \qquad \lambda \equiv \frac{\left(\kappa\phi_{\textrm{f}} \right)^2 }{ 2 \hat{\beta}_1}\ ,
\label{eq_beta:N_chaotic}
\end{equation}
where $\lambda \simeq \beta_1 /2$ is fixed by Eq.~\eqref{eq_beta:condition_end}. Using Eq.~\eqref{eq_beta:N_chaotic} we can thus get the lowest order expression for $\beta(N)$:
\begin{equation}
\label{eq_beta:beta_N_chaotic}
\beta(N) \simeq - \left[ \frac{\hat{\beta}_1}{2(N+ \lambda)}\right]^{\frac{1}{2}} \simeq \left( \frac{\hat{\beta}_1}{2 N}\right)^{\frac{1}{2}} \ ,
\end{equation}
which shows that this model corresponds to $\alpha = 1$ in the parameterization of Mukhanov~\cite{Mukhanov:2013tua}. Finally we compute the scalar spectral index, the running and the tensor-to-scalar ratio:
\begin{eqnarray}
\label{eq_beta:scalar_spectral_chaotic}
n_s -1 &\simeq& - \frac{\hat{\beta}_1}{\left(\kappa\phi  \right)^2} (2+\hat{\beta}_1) \simeq -\frac{1 + \hat{\beta}_1/2}{  N} \ , \\
\label{eq_beta:running_chaotic}
\alpha_s &\simeq& - \frac{\hat{\beta}_1^2}{ \left( \kappa\phi  \right)^4} (2\hat{\beta}_1 + 4) \simeq -(1 + \hat{\beta}_1/2)\frac{1}{ N^2} \ , \\
\label{eq_beta:tensor_chaotic}
r &=& 8 \beta^2 \simeq \frac{4 \hat{\beta}_1}{N} \ .
\end{eqnarray}
As for the other classes the expressions for slow-roll parameters can be found in~\cite{Binetruy:2014zya}. Notice that in this class we only have two free parameters $\hat{\beta}_1$ and $N$. Numerical predictions for this class are shown in Fig.~\ref{fig_beta:chaotic_log}.

\subsubsection{ Fractional class: Ib(p).}
\label{sec_beta:fractional_class} 

\begin{figure}[ht]
	\begin{center}
	\includegraphics[width=.8\textwidth]{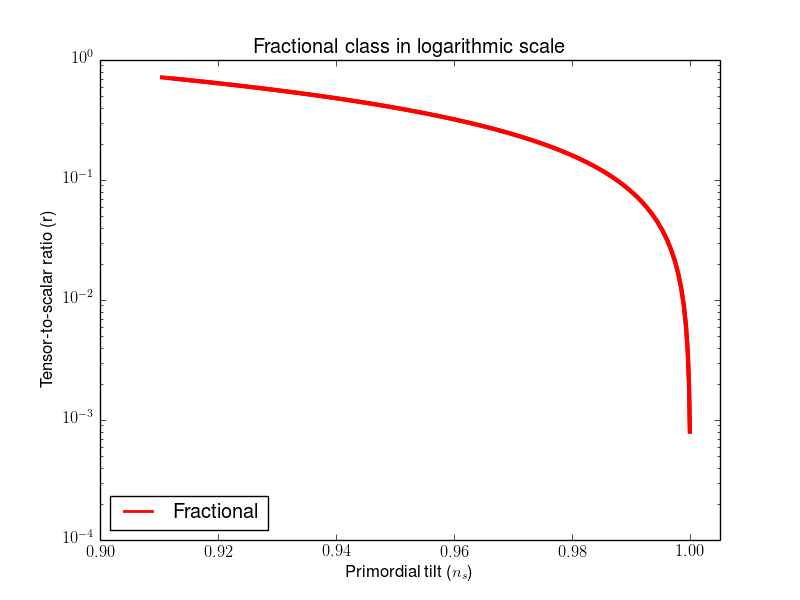}
	\end{center}
	\caption{Prediction for $n_s$ and $r$ for the fractional class. The values of $r$ are in logarithmic scale.}
	\label{fig_beta:fractional_log}
\end{figure}

\noindent
In this class the $\beta$-function is again parameterized as in Eq.~\eqref{eq_beta:beta_Ib}, but the parameter $p$ is in the range $0 <p < 1$. As we show in this section, this class gives different predictions from the Inverse class, and the reason is that close to the fixed point we have: 
\begin{equation}
	\beta_{,\phi} \ll \beta^2 \ .
\end{equation}
As usual we start by computing the superpotential:
\begin{equation}
\label{eq_beta:superpotential_fractional}
	W = W_{\textrm{f}} \exp \left[ \frac{\hat{\beta}_p}{2(1-p)}\left(\kappa\phi \right)^{1-p}\right] \ ,
\end{equation}
where as usual $W_{\textrm{f}}$ can be fixed using the COBE normalization. The potential is now given by:
\begin{equation}
\label{eq_beta:potential_fractional}
	V \simeq \frac{3W_{\textrm{f}}^2}{4\kappa^2} \exp \left[ \frac{\hat{\beta}_p}{1-p} \left(\kappa\phi \right)^{1-p}\right] \ .
\end{equation}
The expressions for the number of e-foldings and for $\beta(N)$ match with the ones obtained for the Inverse class, but the lowest order expressions for $n_s$ and $\alpha_s$ are now given by:
\begin{eqnarray}
\label{eq_beta:scalar_spectral_fractional}
1- n_s = \frac{r}{8} &=& \beta^2 = \frac{\hat{\beta}_p^{2/(p+1)}}{[(p+1)N]^{\frac{2p}{p+1}}} \ , \\
\label{eq_beta:tensor_fractional}
\alpha_s &=& -2\frac{\beta^2 \beta_{,\phi}}{\kappa} =
- 2p \frac{\hat{\beta}_p^{\frac{2}{(p+1)}} }{ [(p+1)N]^{\frac{3p+1}{p+1}}} \ .
\end{eqnarray}
In the parameterization of Mukhanov, this case is again corresponding to $\alpha = 2p/(p+1)$ with $0<\alpha<1$. Notice that Eq.~\eqref{eq_beta:scalar_spectral_fractional} clearly shows that for this class $n_s$ and $r$ are related. Numerical predictions for this class are shown in Fig.~\ref{fig_beta:fractional_log}.

\subsubsection{ Power law class: Ib(0).}
\label{sec_beta:pl_class} 
The limit $p \simeq 0$ of the class \textbf{Ib(p)} is special and should be considered separately. In this case the $\beta$-function is not evolving dynamically but is equal to a certain constant value. As we show in the following, as we approach the fixed point this particular choice is not reproducing a nearly dS spacetime but conversely it leads to the case of power law inflation discussed in Sec.~\ref{sec_inflation:power_law}. The expressions for the superpotential and for the potential are respectively given by:
\begin{eqnarray}
	\label{eq_beta:superpotential_pl}
	W = W_{\textrm{f}} \exp\left(\frac{\hat{\beta_0 }}{2} \kappa \phi \right) \ , \\ 
	\label{eq_beta:potential_pl}
	V = \frac{ W_{\textrm{f}}^2}{8 \kappa^2} \left( 6 -\hat{\beta}_0 \right)  W_{\textrm{f}} \exp\left( \hat{\beta_0 } \kappa \phi \right) \ .
\end{eqnarray}
It is easy to show that in this case the number of e-foldings can be expressed as:
\begin{equation}
	N = \frac{\kappa}{\hat{\beta}_0} \left( \phi - \phi_0\right) \ ,
\end{equation}
integrating Eq.~\eqref{eq_beta:phisuperpot} we can thus show that $a(t) \simeq t^{2/3}$ that actually corresponds to scale factor in the case of power law inflation of Sec.~\ref{sec_inflation:power_law}. It is easy to show that the expressions for $n_s$, $r$ and $\alpha_s$ are simply given by:
\begin{equation}
\label{eq_beta:predictions_pl}
1- n_s = -\frac{\hat{\beta}_0 }{1 - \hat{\beta}_0^2 /2  } \ , \qquad r = 8 \hat{\beta}_0^2 \ , \qquad \alpha_s = 0 \ . 
\end{equation}
This class is controlled by a single parameter, and it is well known that the predictions of this class of models are disfavoured by Planck.

\subsubsection{ Exponential class: (II).}
\label{sec_beta:exponential_class}
\begin{figure}[ht]
	\begin{center}
	\includegraphics[width=.8\textwidth]{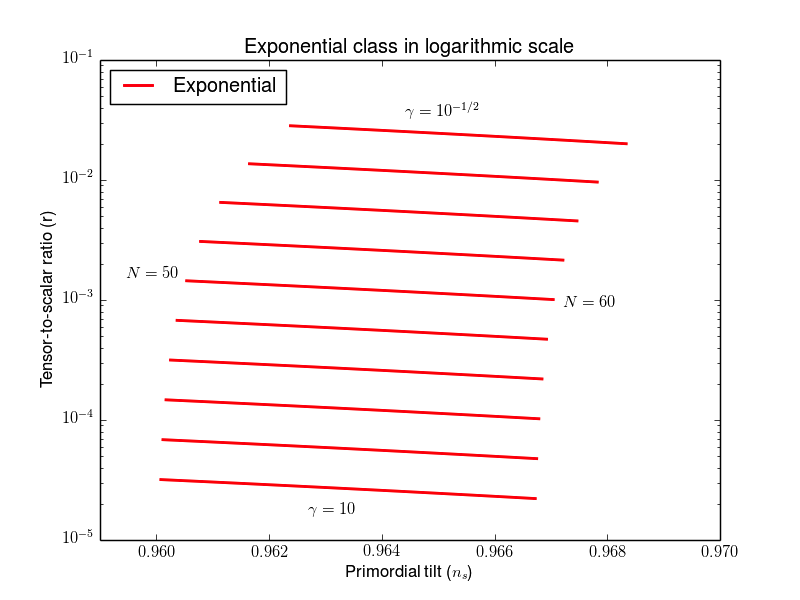}
	\end{center}
	\caption{Prediction for $n_s$ and $r$ for the exponential class. The values of $r$ are in logarithmic scale.}
	\label{fig_beta:exponential_log}
\end{figure}

In this class the parameterization for the $\beta$-function is given by Eq.~\eqref{eq_beta:beta_exponential}. To show the stability of this class under corrections, we can define $Y \equiv \exp\left( -\gamma \kappa \phi \right)$, so that the we can generalize the expression for the $\beta$-function as:
\begin{equation}
	\beta(\phi) = - \sum_{n} \hat{\beta}_n Y^n \ ,
\end{equation}
where the $\hat{\beta}_n \geq 0 $ are constants. As approaching the fixed point the function $Y$ is exponentially suppresed, the leading term is first term with $\hat{\beta}_n \neq 0 $. Actually this parameterization would be redundant and we can thus redefine the parameter $\gamma$ in order the reabsorb $n$ and $\hat{\beta}_n $ into a new parameter $\hat{\beta} $. It is interesting to notice that this class is actually matching with the class of models discussed in~\cite{Kiritsis:2013gia}. The potential and the superpotential for this class are given by:
\begin{eqnarray}
W(\phi) &=& W_{\textrm{f}} \exp \left( - \frac{\hat{\beta}}{2\gamma}  Y  \right)  \ ,
\label{eq_beta:superpotential_exponential} \\
V(\phi) &=& \frac{3 W_{\textrm{f}}^2}{4 \kappa^2} \left[1 - \frac{\hat{\beta} Y}{\gamma } + \mathcal{ O}\left( Y^{2} \right) \right] \ . \label{eq_beta:potential_exponential}
\end{eqnarray}
It is interesting to notice that Eq.~\eqref{eq_beta:potential_exponential} reproduces the potentials of Sec.~\ref{sec_inflation:plateau} and thus both Starobinsky~\cite{Starobinsky:1980te} and non-minimal Higgs inflation model~\cite{Bezrukov:2007ep} belong to this class. Moreover, this corresponds to the case $\alpha = 2$ in Mukhanov's classification. The number of e-foldings can be expressed as:
\begin{equation}
\label{eq_beta:N_exponential}
N =  \frac{1}{ \gamma \hat{\beta}Y} -  \frac{1}{ \gamma \hat{\beta}Y_f}  \simeq -\frac{1}{\gamma \beta(\phi)}  \ ,
\end{equation}
where again we have used Eq.~\eqref{eq_beta:condition_end}. The $\beta$-function can be then expressed in terms of $N$ as:
\begin{equation}
\label{eq_beta:beta_N_exponential}
\beta(N) \simeq -\frac{1}{\gamma N+1} \ .
\end{equation}
Finally, we can compute the scalar spectral index, the tensor-to-scalar ratio and the running:
\begin{eqnarray}
\label{eq_beta:scalar_spectral_exponential}
n_s -1 &\simeq& - 2 \frac{\beta_{,\phi} }{\kappa}\simeq -\frac{2}{N} \ , \\
\label{eq_beta:running_exponential}
\alpha_s &\simeq& - 2 \frac{\beta \beta_{,\phi \phi}}{\kappa^2} \simeq - \frac{2}{N^2}
\ , \\
\label{eq_beta:tensor_exponential}
r & = & 8 \beta^2\simeq \frac{8}{ \gamma^2 N^2} \ .
\end{eqnarray}
Again the expressions for slow-roll parameters are given in~\cite{Binetruy:2014zya}. Notice that Eq.~\eqref{eq_beta:scalar_spectral_exponential}, Eq.~\eqref{eq_beta:running_exponential} and Eq.~\eqref{eq_beta:tensor_exponential} show that for this class the observable quantities are only depending on two free parameters \textit{i.e.} $\gamma$ and $N$. Numerical predictions for this class are shown in Fig.~\ref{fig_beta:exponential_log}. It is also interesting to notice that the models of this class are the most favoured by Planck observations.

\section{Planck constraints on $\beta$.}
\label{sec_beta:beta_constraints}
As the $\beta$-function formalism for inflation provides a useful guide for inflationary model building and a powerful method to classify inflationary models, it seems natural to study the possibility of imposing direct constraints on the typical quantities of this formalism. In this section we discuss the procedure to impose these constraints and, although this work is still in progress, we present some preliminary results that give some hints on the expected outcome. \\

\noindent Our starting point are the lowest order expressions for $n_s$, $r$ and $\alpha_s$ in terms of the $\beta$-function formalism shown in Eq.~\eqref{eq_beta:ns}, Eq.~\eqref{eq_beta:r} and Eq.~\eqref{eq_beta:dns/dlnk}. It should be clear that this system of equations can be inverted in order to express $\beta$ and its first and second derivatives in terms of $n_s$, $r$ and $\alpha_s$ as:
\begin{eqnarray}
	\label{eq_beta:beta_data}
	\beta(\phi) &=& \pm \sqrt{ \frac{ \mathrm{r}}{ 8 }} ,  \\
	\label{eq_beta:beta_1_data}
	\beta_{,\phi}(\phi) & \simeq & \frac{ 1 - n_s}{2} - \frac{ \mathrm{r}}{ 16 }, \\
	\label{eq_beta:beta_2_data}
	\beta_{,\phi \phi}(\phi) &\simeq& \pm \sqrt{\frac{8}{\mathrm{r}}}  \left[ -\frac{\alpha_s}{2} - \frac{\mathrm{r}}{ 8 } \left( \frac{ 1 - n_s}{2} - \frac{ \mathrm{r}}{ 16 } \right) \right] \ .
\end{eqnarray}
At this point, it is worth stressing that the degeneracy in the sign of $\beta$ is only apparent and it can be removed by using Eq.~\eqref{eq_beta:Nbeta}. In fact by differentiating Eq.~\eqref{eq_beta:Nbeta}, we get:
\begin{equation}
\label{eq_beta:Ndiff}	
\mathrm{d }N = - \frac{\mathrm{d}\phi}{\beta(\phi)} \ .
\end{equation}
As $N(\phi)$ increases as we approach the fixed point, we can get a condition on the sign of $\beta$. Let us start by considering the case of big field inflationary models where the fixed point is approached for $\phi \rightarrow \infty$. Approaching the fixed point, we have both $\mathrm{d}N >0$ and $\mathrm{d}\phi>0 $. This directly implies that for this models $\beta(\phi) < 0$. An analogous argument can be produced for the case of small field models, but in this case we conclude that $\beta(\phi) > 0$.\\

\noindent
At this point we can proceed by discussing the method to impose constraints on these quantities using the Planck data. As with these parameters we want to replace $n_s$, $r$ and $\alpha_s$, the correct procedure to impose constraints on $\beta, \beta_{,\phi}$ and $\beta_{,\phi \phi}$ is the one discussed in Sec.~\ref{sec_introduction:Bayesian_inference}. In particular, as a first step we should express the posterior probabilities for $\beta, \beta_{,\phi}$ and $\beta_{,\phi \phi}$ in terms of the likelihood and of the priors. Moreover, following the standard procedure, flat priors on $\beta, \beta_{,\phi}$ and $\beta_{,\phi \phi}$ should be imposed. Once we are able to compute the posterior probabilities, we can generate a new set of Markov Chains that can be used to estimate the values and the confidence levels of the new parameters.\\

\noindent
While this method is well defined and is expected to lead to the sought result, it is also interesting to understand whether it would be possible to obtain these constraints or at least to grasp some information on the expected results without generating new chains. For this purpose we should start by considering the definition of posterior probability $P(B|A) \propto P(A|B) \cdot P(B)$, given in Sec.~\ref{sec_introduction:Bayesian_inference}, for an event $B$ to occur given the event $A$. As the generalization to vector of parameters is trivial, we can now restrict to the case of a single parameter. As explained in Sec.~\ref{sec_introduction:Bayesian_inference} and as reported in the previous paragraph, in order to generate the Markov Chains to reproduce the posterior probability $P(B|A)$ it is customary to impose a flat prior on $B$. At this point it is important to notice that defining a derived parameter $C$ that depends on $B$ (\textit{i.e.} $C(B)$), the posterior probability distribution of $C$ reads $P(C|A) \propto P(A|C) \cdot P(C)$. In particular we should stress that a flat prior on $B$ does not correspond to a flat prior on $C$! \\

\noindent
This can be easily explained with an example: let us consider a random variable $X$ taking values in the interval $[x_0, x_1]$. Let us assume that its normalized probability distribution $f(x)$ is constant \textit{i.e.} $f(x) = 1/(x_1 - x_0)$. If we consider a new variable $Y = X^2$, it should be clear that its probability distribution $f(y)$ is not constant in particular we have $f(y) \propto y^{-1/2}$. Clearly, this is due to the presence of the determinant of the Jacobian matrix of the transformation $| \textrm{d}x/\textrm{d}y|$, that deforms the probability distribution. As a consequence, given a flat prior on a parameter $B$ and the parameter $C(B)$ whose prior is non-flat, we can use the insight given by the example in order to define a procedure to ``flatten'' the prior on C. In particular, this is realized by multiplying the posterior probability distribution of $C$ with the inverse of the determinant of the Jacobian matrix $| \textrm{d}B/\textrm{d}C|^{-1}$. \\

\noindent
Given the Markov Chains released by the Planck collaboration, that are generated with CosmoMC using the standard parameterization in terms of $n_s$, $r$ and $\alpha_s$, this method can actually be implemented by introducing different weights for the points of the chain. In particular, these weights should be fixed by the inverse of the determinant of the Jacobian matrix. While theoretically this method should be well defined and should lead to the correct results, some practical problems, due to statistical fluctuations arise. The introduction of a weight for the points corresponds to the replication of each point in accordance with its weight. As a consequence, if the number of points in the chain is not sufficiently large, single points in the tail of the distribution may be turned into peaks. Actually, as it is possible to notice from Fig.~\ref{fig_beta:beta_2_plots}, this is exactly the case that we meet when we use the chains released by the Planck collaboration. \\

\begin{figure}[h!]
\centering
\subfloat[][\emph{Semilogarithmic plot of the weighted occurrences for $\beta_{,\phi\phi}$.}]
{\includegraphics[width=.8\columnwidth]{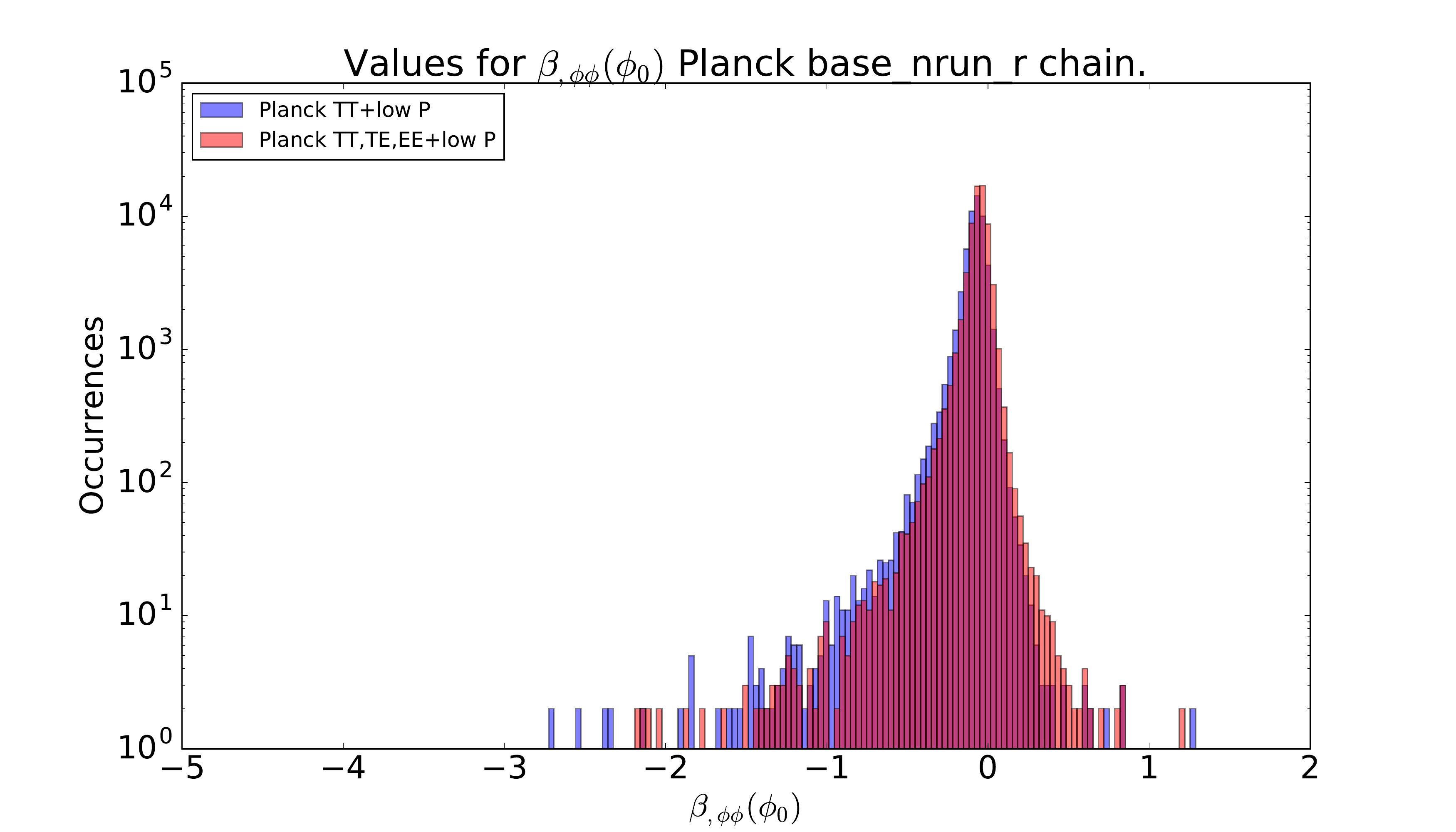}} \quad
\subfloat[][\emph{Linear plot of the marginalized probability distribution function (p.d.f.) for $\beta_{,\phi\phi}$.}]
{\includegraphics[width=.8\columnwidth]{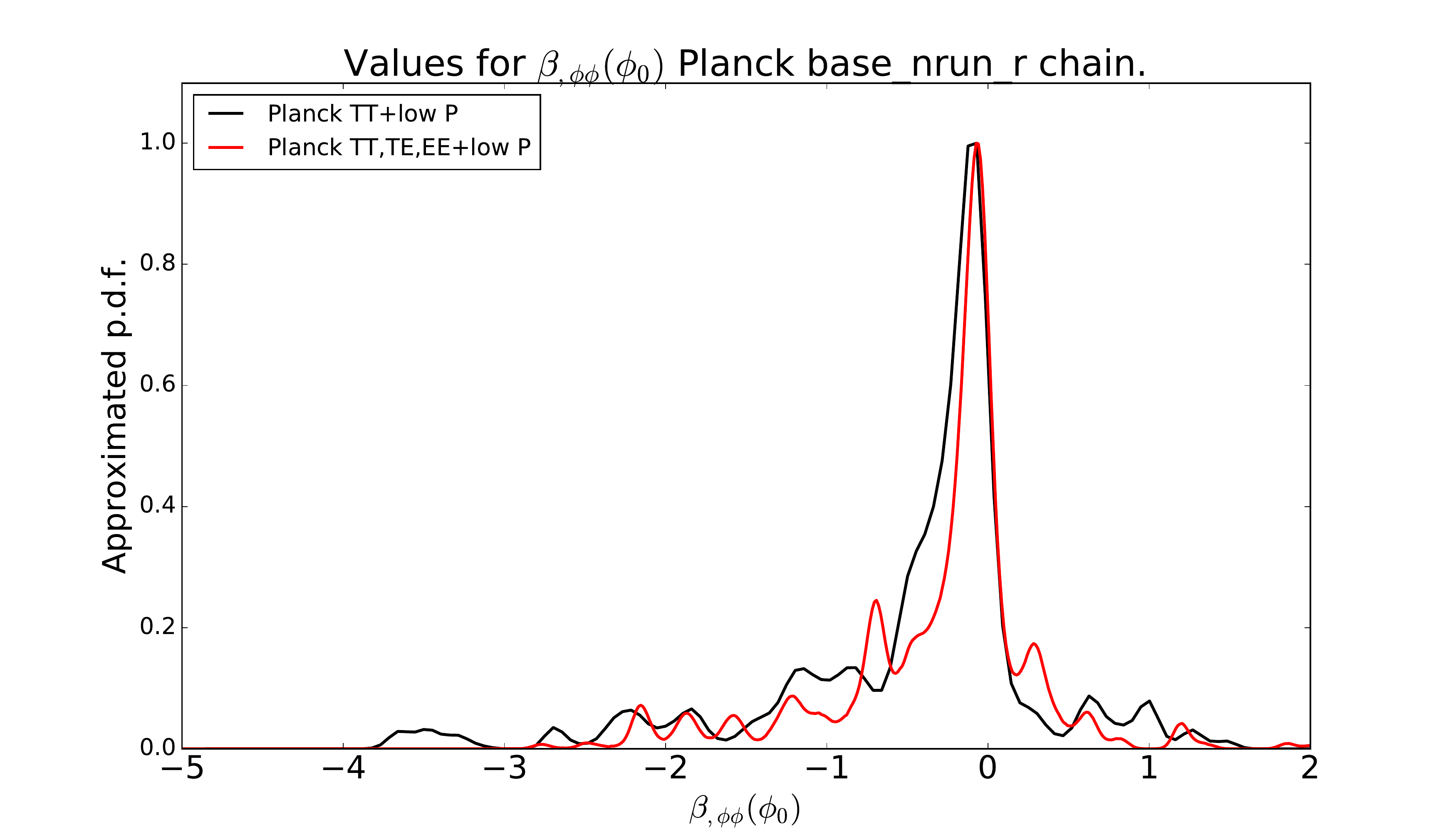}} \\
\caption{Weighted occurrences and marginalized probability distribution for $\beta_{,\phi\phi}$ in assuming large field models. These plots are obtained from the Planck chains that include the running $\alpha_s$ in the list of cosmological parameters with flat prior.}
\label{fig_beta:beta_2_plots}
\end{figure}

\noindent
The plots of Fig.~\ref{fig_beta:beta_2_plots} show the weighted occurrences and the corresponding probability distribution for $\beta_{,\phi\phi}$ in the case of large field models. As expected, the data (and consequently the p.d.f.) are highly peaked around zero, but we can also notice several smaller peaks at larger values for $|\beta_{,\phi\phi}|$. These peaks are not expected and, as explained in the previous paragraph, they are due to statistical fluctuations that are amplified by the introduction of a weight for the points. We can thus conclude that, while plots obtained with this method are not expected to give the exact p.d.f. for $\beta, \beta_{,\phi} $ and $\beta_{,\phi\phi}$, they can still be useful when a rough estimate of the real result is sufficient. In particular, these plots will be useful to check the consistency of the plots obtained when we generate the new chains directly using $\beta, \beta_{,\phi} $ and $\beta_{,\phi\phi}$. As anticipated at the beginning of this section, this work is still in progress and final results on this topic are expected to be published in a future work.

\section{Interpolating models.}
\label{sec_beta:comoposite_classes}
In the previous section we have discussed the possibility of defining a set of universality classes for inflationary models using the $\beta$-function formalism for inflation. This set of universality classes can be thought as a set of fundamental behaviors for the inflationary trajectories. However, when we are dealing with inflationary model building, we may be interested in the definition of more elaborate models. A simple way to implement this procedure is explained in this section. In particular, as we explain in this section, this can be realized by defining models that interpolate between two different classes. \\  

\noindent
As extensively discussed in the previous sections, inflation is realized when $\beta(\phi)$ approaches zero. In the parameterizations considered so far, with the sole exception of class \textbf{Ib(0)}, this is obtained dynamically. However, as for the case of class \textbf{Ib(0)}, inflation can also be realized if the $\beta$-function is nearly constant and the constant is small enough. This can be clear by considering an example. Let us start by considering a generic function $f(\phi)$ that has a zero at a certain value $\phi_0$ of $\phi$. Let us consider a model with $\beta$-function equal to $\beta(\phi) = \epsilon f(\phi)$ were $\epsilon$ is a constant that in a first stage is assumed to be of order one. Clearly, this $\beta$-function has a zero in $\phi_0$, and thus close to this point inflation can be realized. However, we can consider the case $\epsilon \ll 1$. As the system inflates for all the values of $\phi$ that give $\beta(\phi) \ll 1 $, in this case, inflation can be realized at $f(\phi) $ of order one. As a matter of fact, the region with $\beta(\phi) \ll 1 $ is thus stretched by the introduction of a small parameter. Notice that in general the regime with $\epsilon \simeq 1$ and the one with $\epsilon \ll 1$ are not forced to be in the same fundamental class. This actually happens for both the examples considered in this section. \\

\noindent 
In the following we consider two explicit examples to show the realization of the mechanism of interpolation between classes. These are the cases of models inspired to natural inflation introduced in Sec.~\ref{sec_inflation:natural} and to hilltop inflation introduced in Sec.~\ref{sec_inflation:hilltop}. In particular in these models the mechanism of interpolation is realized by introducing a new scale $f$ in the system. The two different regimes are then obtained by considering the two limits $\kappa f \ll 1$ \textit{i.e.} the new scale is small, and $\kappa f \gg 1$ \textit{i.e.} the new scale is large. It is interesting to stress that slow-roll inflation can be obtained in both these limits as well as in the intermediate region.

\subsection{Natural Inflation.}
\label{sec_beta:natural_class}

\begin{figure}
\centering
\subfloat[][\emph{Predictions for $(n_s,r)$ for natural, linear and chaotic class in linear scale.}\label{fig_beta:natural_linear}]
{\includegraphics[width=0.85\columnwidth]{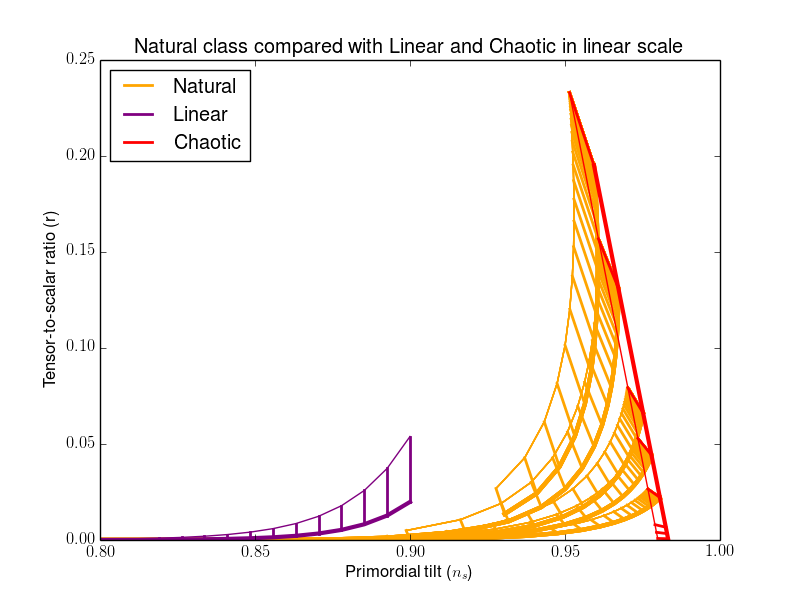}} \quad
\subfloat[][\emph{Predictions for $(n_s,r)$ for natural, linear and chaotic class. $r$ is in logarithmic scale.}\label{fig_beta:Natural_logarithmic}]
{\includegraphics[width=0.85\columnwidth]{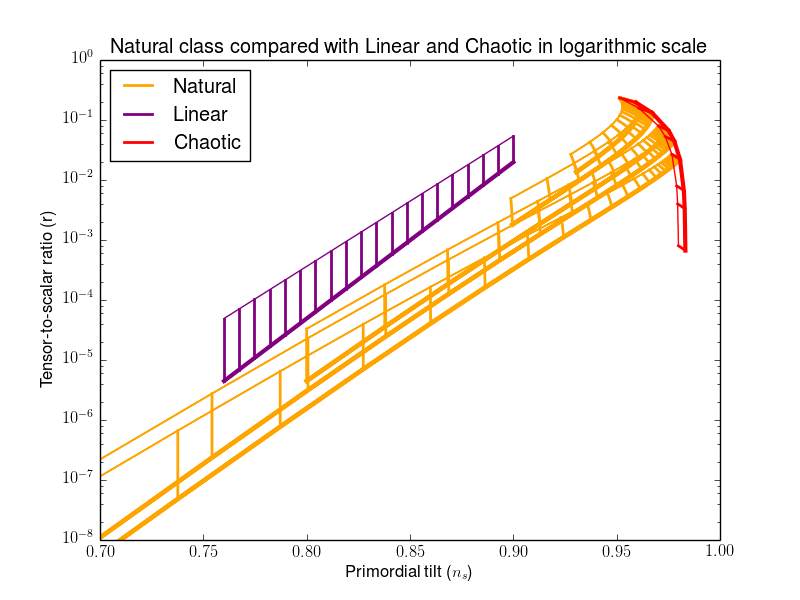}}
\caption{Models of generalized natural inflation (in orange) interpolating between the linear class (in purple) and chaotic class (in red). }
\label{fig_beta:Natural_class}
\end{figure}

In this section we show how natural inflation discussed in Sec.~\ref{sec_inflation:natural} can be considered as a composite model that interpolates between the linear and chaotic classes of Sec.\ref{sec_beta:linear_class} and Sec.\ref{sec_beta:chaotic_class} respectively. For this purpose we start by considering the $\beta$-function:
\begin{equation}
\label{eq_beta:natural_beta}
	\beta(\phi) = \frac{1}{\kappa f} \tan \left(\frac{\phi}{2f} \right) \ ,
\end{equation}
and computing the potential and superpotential which as usual are given by Eq.~\eqref{eq_beta:pot_superpot_beta}:
\begin{equation}
	\label{eq_beta:pot_superpot_natural}
	W(\phi) = W_{\textrm{f}} \ \frac{\cos(\phi/2f)}{\cos(\phi_{\textrm{f}}/2f)} \ , \qquad V(\phi) = \frac{3 W_{\textrm{f}}}{4 \kappa^2 } \ \frac{\cos^2(\phi/2f)}{\cos(\phi_{\textrm{f}}^2/2f)} \ \left[ 1 - \frac{\beta^2(\phi)}{6} \right] \ .
\end{equation}
Using $\cos^2(x) = (1 + \cos(2x))/2 $ and the fact that during inflation $\beta(\phi) \ll 1 $, the potential can be expressed as:
\begin{equation}
	V(\phi) \simeq \frac{3 W_{\textrm{f}}}{8 \kappa^2 } \ \frac{1 + \cos (\phi / f)}{\cos(\phi_{\textrm{f}}^2/2f)}  \ ,
\end{equation}
that actually corresponds to the potential of natural inflation given in Sec.~\ref{sec_inflation:natural}. In the following we present the two limits of small and large scale. Numerical predictions for this class are shown in Fig.~\ref{fig_beta:Natural_class}.

\subsection{Natural inflation: Small scales ($\kappa f \ll 1$).}
\label{sec_beta:natural_small}

In this limit the factor $1/\kappa f$ is large and thus inflation can only be realized approaching a zero of $\tan(\phi/2f)$. We can thus consider the case $\phi/2f \ll 1$. In this limit the lowest order expression for $\beta$-function is simply given by:
\begin{equation}
	\beta(\phi) \simeq \frac{\phi}{2 \kappa f^2} \ ,
\end{equation}
 that clearly corresponds to the Linear class \textbf{Ia(1)} discussed in Sec.~\ref{sec_beta:linear_class}. The predictions in this limit are thus given by Eq.~\eqref{eq_beta:scalar_spectral_linear} and Eq.~\eqref{eq_beta:tensor_linear}.

\subsection{Natural inflation: Large scales ($\kappa f \gg 1$).}
\label{sec_beta:natural_large}
In this limit the factor $1/\kappa f$ is small and thus inflation can be realized with $\tan(\phi/2f) \simeq 1$. To clarify this point let us consider $\phi \simeq \pi f$ and let us express $\phi$ in terms of a new field $\phi^\prime$ as:
\begin{equation}
	\phi = \pi f - \phi^\prime \ .
\end{equation}
In terms of the new field, the limit $\phi/f \simeq \pi$ can thus be expressed as $\phi^\prime / f \ll 1$. However, it is crucial to notice that this limit can be approached satisfying the condition:
\begin{equation}
	\label{eq_beta:natural_large_condition}
	\frac{1}{\kappa f } \ll \frac{\phi^\prime}{f} .
\end{equation}
In terms of the new field $\phi^\prime$ the $\beta$-function reads:
\begin{equation}
	\beta(\phi^\prime) = \frac{1}{\kappa f} \cot\left(\frac{\phi^\prime}{2f}\right) \simeq \frac{2 }{\kappa \phi^\prime} \ .
\end{equation}
Notice that this expression matches with the Chaotic class \textbf{Ib(1)}, discussed in Sec.~\ref{sec_beta:chaotic_class} with $\hat{\beta}_1 = 2$. Moreover, using the condition of Eq.~\eqref{eq_beta:natural_large_condition} it should be clear that the condition to realized inflation, \textit{i.e.} $\beta(\phi^\prime)$ approaching zero, can be satisfied.

\subsection{Hilltop class.}
\label{sec_beta:hilltop_class}

\begin{figure}
\centering
\subfloat[][\emph{Predictions for $(n_s,r)$ for hilltop, linear and chaotic class in linear scale.}\label{fig_beta:hilltop_linear}]
{\includegraphics[width=0.9\columnwidth]{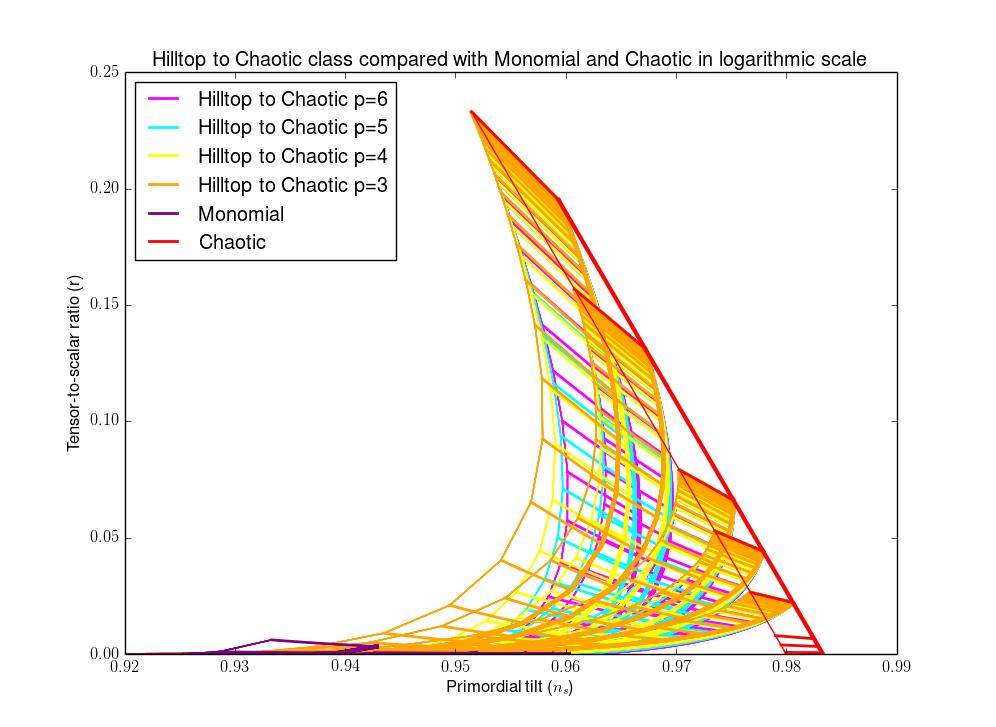}} \quad
\subfloat[][\emph{Predictions for $(n_s,r)$ for hilltop, linear and chaotic class. $r$ is in logarithmic scale.}\label{fig_beta:hilltop_log}]
{\includegraphics[width=0.9\columnwidth]{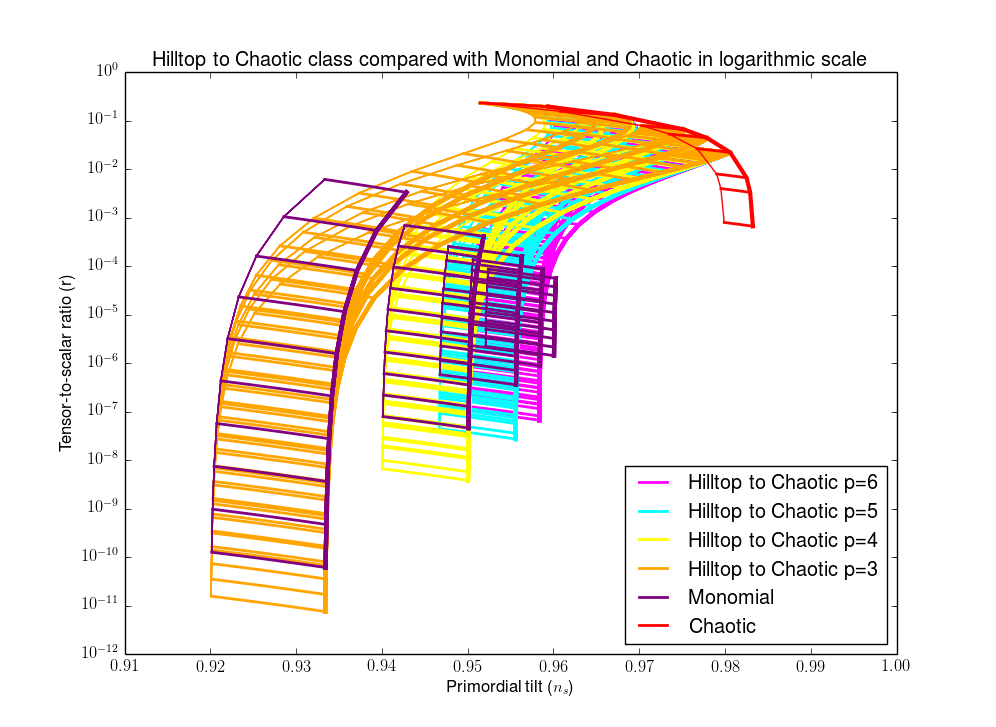}}
\caption{Generalized hilltop models for different values of $p$ interpolating between monomial (in purple) and chaotic class (in red).}
\label{fig_beta:hilltop_class}
\end{figure}

In this section we show how the models of hilltop inflation discussed in Sec.~\ref{sec_inflation:hilltop} can be considered as composite models that interpolate between the monomial and chaotic classes of Sec.\ref{sec_beta:monomial_class} and Sec.\ref{sec_beta:chaotic_class} respectively. For this purpose we start by considering the $\beta$-function:
\begin{equation}
\label{eq_beta:hilltop_beta}
	\beta(\phi) = p \frac{\hat{\beta}_p}{\kappa f} \frac{ \ (\phi/f)^{p-1}}{1 - (\phi/f)^p }  \ ,
\end{equation}
with $p >2 $. Notice that this function has a zero at $\phi/f = 0$ and another zero at $\phi/f \rightarrow \infty$. In the discussion of this section we study the case of inflation taking place in the range $0 < \phi < f$. However, a similar discussion can also be produced for the case $f < \phi $ \textit{i.e.} considering the fixed point to be reached for infinitely large values of $\phi$. \\

\noindent
As usual we start by computing the potential and superpotential using Eq.~\eqref{eq_beta:pot_superpot_beta}:
\begin{equation}
	\label{eq_beta:pot_superpot_hilltop}
	W(\phi) =W_{\textrm{f}} \left[  \frac{ 1 - \left(\frac{\phi}{f} \right)^p  }{ 1 - \left(\frac{\phi_{\textrm{f}}}{f} \right)^p } \right]^{\frac{\hat{\beta}_p}{2}} \ , \qquad  V(\phi) = \frac{3 W_{\textrm{f}}^2 }{4 \kappa^2} \left[  \frac{ 1 - \left(\frac{\phi}{f} \right)^p  }{ 1 - \left(\frac{\phi_{\textrm{f}}}{f} \right)^p } \right]^{\hat{\beta}_p} \left[ 1 - \frac{\beta^2(\phi)}{6}\right] \ .
\end{equation}
As during inflation $\beta(\phi) \ll 1 $ it should be clear that this potential corresponds to the potential of hilltop inflation given in Sec.~\ref{sec_inflation:hilltop}. In the following we present the two limits of small and large scale. Numerical predictions for this class are shown in Fig.~\ref{fig_beta:hilltop_class}.

\subsection{Hilltop inflation: Small scales ($\kappa f \ll 1$).}
\label{sec_beta:hilltop_small}
In this limit the factor $1/\kappa f$ is big and thus inflation can be realized only for $(\phi/f) \ll 1$. We can thus consider the case $\phi/2f \ll 1$. In this limit the lowest order expression for $\beta$-function is simply given by:
\begin{equation}
	\beta(\phi) \simeq p \frac{\hat{\beta}_p}{\kappa f} (\phi/f)^{p-1} \ .
\end{equation}
As we have chosen $p>2$, this parameterization for the $\beta$ function clearly corresponds to the Monomial class \textbf{Ia(q)} discussed in Sec.~\ref{sec_beta:monomial_class}. The predictions in this limit are thus given by Eq.~\eqref{eq_beta:scalar_spectral_linear} and Eq.~\eqref{eq_beta:tensor_linear}. Notice that choosing $p = 2$, we can reproduce the Linear class \textbf{Ia(0)} discussed in Sec.~\ref{sec_beta:linear_class}.

\subsection{Hilltop inflation: Large scales ($\kappa f \gg 1$).}
\label{sec_beta:hilltop_large}
In this limit the factor $1/\kappa f$ is small, and thus inflation can be realized at $(\phi/f) \simeq 1$. Similarly to the large field case for natural inflation we can thus define a new field $\phi^\prime$ as:
\begin{equation}
	\phi = f - \phi^\prime \ ,
\end{equation}
so that the in the limit $\phi/f \simeq$ \textit{i.e.} $(\phi^\prime/f) \ll 1$ the $\beta$-function reads:
\begin{equation}
	\beta(\phi^\prime) = \frac{p \hat{\beta}_p}{\kappa f} \frac{\left(1 - \phi^\prime/f\right)^{p-1}}{1 - \left(1 - \phi^\prime/f\right)^{p}} \simeq \frac{\hat{\beta}_p }{\kappa \phi^\prime} \ .
\end{equation}
To realize inflation we thus need $1 \ll \kappa \phi^\prime \ll \kappa f$. In this limit we recover the Chaotic class \textbf{Ib(1)} discussed in Sec.~\ref{sec_beta:chaotic_class}.

\section{Discussion.} 
\label{sec_beta:discussion}
As explained in Chapter~\ref{chapter:inflation} (in particular at the beginning of Sec.~\ref{sec_inflation:definition_inflation}), inflation is an early phase of (nearly) exponential expansion of the Universe. During this phase the FLRW metric is thus approaching the metric of de Sitter (dS) spacetime (see Eq.~\eqref{appendix_GR:dS_spacetime}). As dS is static and scale invariant (self-similar), the inflating Universe presents an approximate scale invariance. It is crucial to stress, that this can be seen as a defining property of inflation. \\

\noindent
In this Chapter we have developed a new formalism to describe inflation based on the Hamilton-Jacobi approach of Salopek and Bond~\cite{Salopek:1990jq}. With this approach we classify inflationary models by only relying on the approximate scale invariance. In particular, models are characterized according to the way they break the scale invariant regime. The power of the approach thus derives from its generality as it only relies on the simplest property of inflation. \\

\noindent
In analogy with statistical mechanics\footnote{Where nearly scale invariant dynamics is usually characterized in terms of a set of critical exponents.} the parameterization of the $\beta$-function naturally defines universality classes for inflationary models. A major benefit of working with this framework is that we are not forced to specify a single model but, on the contrary, we can consider whole classes of models. In this sense results obtained in this framework are more general than the ones obtained with the standard methods (specifying the potential). \\

\noindent
While in the standard picture the focus is put on the potential, in the $\beta$-function formalism the focus is put on the $\beta$-function. As explained in this Chapter this is the quantity that captures the main features that characterize the dynamics of inflation. While in some cases (beyond the simplest realization of inflation) the description of inflation in terms of its potential may be misleading, the $\beta$-function formalism will still be well-defined. As a consequence the classification of models in terms of this formalism is particularly fitted to describe inflation\footnote{This will be manifest in Chapter~\ref{chapter:generalized_models} where we apply the $\beta$-function formalism to generalized inflationary models. In particular we discuss non-standard kinetic terms and non-minimal coupling between the inflaton and gravity.}. \\

\noindent
For example it is important to notice that under the single (and reasonable) assumption of a piece-wise monotonic field, we have derived an \emph{exact} parameterization for the potential and for the superpotential (Eq.~\eqref{eq_beta:pot_superpot_beta}) in terms of the $\beta$-function. In particular it is worth stressing that the whole discussion of Sec.~\ref{sec_beta:beta_func} was carried out without assuming slow-roll (we have only used $\beta^2 \ll 1$)!\\

\noindent
While in this Chapter the introduction of the $\beta$-formalism is motivated by a formal resemblance (between the equations that describe the inflating universe and RG flows in the context of QFT), in Chapter~\ref{chapter:holographic_universe} we provide some deep theoretical reasons that motivates the introduction of this framework. In particular, the discussion of Chapter~\ref{chapter:holographic_universe} aims at explaining the reasons to consider (and also the methods to implement) the application of holography to cosmology.

 {\large \par}}
{\large 
\chapter{AdS/CFT and Holographic universe.}

\label{chapter:holographic_universe}

\horrule{0.1pt} \\[0.5cm] 

\begin{abstract} 
\noindent
In this Chapter we discuss the application of holography to cosmology proposed by Mc Fadden and Skenderis~\cite{McFadden:2009fg,McFadden:2010na}. This approach is based on the AdS/CFT correspondence formulated by Maldacena in~\cite{Maldacena:1997re}, that conjectures a duality between quantum field theories (QFTs) and theories of gravity. In this context we present some arguments that support the interpretation of the inflationary phase in terms of a Renormalization Group (RG) flow (discussed in Chapter~\ref{chapter:beta}). In particular, we discuss the possibility of describing inflation in terms of the RG flow of the dual QFT. As the AdS/CFT correspondence is a strong/weak duality, it provides powerful tools to study theories in their strongly coupled regime. This can be relevant for the definition of models that go beyond the standard weak gravitational description.
\end{abstract}

\horrule{0.1pt} \\[0.5cm]

\noindent
During the last years several works~\cite{McFadden:2009fg,McFadden:2010na,McFadden:2010jw,McFadden:2010vh,McFadden:2011kk,Bzowski:2012ih} have explored the possibility of applying holography to cosmology and specifically to inflation. This idea is supported by the formal resemblance between domain-wall and cosmological (inflationary) solutions. In particular, by changing the sign of the potential\footnote{Consistently with the discussion of Sec.~\ref{appendix_GR:dS_AdS} (in particular see Eq.~\eqref{appendix_GR:cosmological_constant_sign}), this change in the sign maps a term that plays the role of a positive cosmological constant into a term that plays the role of a negative cosmological constant.}, it is possible to map cosmological solutions that asymptote to de Sitter (dS) spacetime (\textit{i.e.} inflation) into domain-wall solutions that asymptote to Anti de Sitter (AdS) spacetime. Because of this correspondence, it seems natural to apply the methods of the AdS/CFT correspondence to describe cosmology (and in particular to describe inflation). \\

\noindent
AdS/CFT is the conjectured correspondence (formulated by Maldacena in~\cite{Maldacena:1997re}) between theories of gravity in AdS${}_{d+1}$ and conformal field theories (CFTs) in $d$-dimensions. Actually, the correspondence can also be extended outside of the conformal region~\cite{deBoer:2000cz,deHaro:2000vlm,Skenderis:2002wp,Papadimitriou:2004rz} and in particular the deformation of the CFT (usually described by an RG flow induced by an operator $\mathcal{O}(x)$) is interpreted as a deformation of the AdS geometry. As a consequence, a natural interpretation of the $\beta$-function formalism of Chapter~\ref{chapter:beta} arises in this framework. Moreover, as AdS/CFT sets a weak/strong duality, in this context it is possible to define new models for early time cosmology where gravity is strongly coupled. \\

\noindent We begin this Chapter by discussing the ideas that led to the application of holography to cosmology and we present the general procedure to implement this description. In particular, in Sec.~\ref{sec_holography:AdS_dS_correspondence} we show the formal resemblance between the inflationary universe and the evolution of a scalar field in a nearly AdS spacetime. In Sec.~\ref{sec_holography:AdS_CFT} we then present the AdS/CFT correspondence of Maldacena and we discuss its application to the case of nearly AdS spacetime. In Sec.~\ref{sec_holography:Holographic_inflation} we discuss the holographic interpretation of inflation and of the $\beta$-function formalism for inflation presented in Chapter~\ref{chapter:beta}. Finally, in Sec.~\ref{sec_holography:Holographic_strong} we discuss the possibility of computing cosmological observables using the holographic QFT. In particular, we discuss the possibility of considering models where a weak gravity description is not viable. For technical details on dS and AdS spacetime see Appendix~\ref{appendix_GR:general} (in particular Sec.~\ref{appendix_GR:dS_AdS}) and for details on the construction of CFTs we refer to Appendix~\ref{appendix_CFT:Conformal_field_theories}.

\section{The holographic universe.}
\label{sec_holography:holographic_universe}
The holographic principle was firstly formulated by Susskind in~\cite{Susskind:1994vu} and inspired by the works of `t Hooft~\cite{'tHooft:1993gx}, Thorn~\cite{Thorn:1991fv} and by his own work~\cite{Susskind:1993aa}. The holographic principle states that the information on the dynamics of a system can be contained on its boundary. Clearly this principle was inspired by the well known problem of the entropy of a black hole studied by Bekenstein~\cite{Bekenstein:1972tm,Bekenstein:1973ur,Bekenstein:1974ax} and Hawking~\cite{Bardeen:1973gs,Hawking:1974sw}. In particular, in the referenced works the authors showed that the entropy of a black hole does not scale as its volume, but rather it scales as its surface. \\

\noindent Clearly this framework offers fascinating new interpretations of gravitational phenomena. In this context the formulation of the AdS/CFT correspondence of Maldacena~\cite{Maldacena:1997re} was a striking improvement. In particular, as the two sides of the correspondence are related by a weak/strong duality, the existence of a consistent formulation of the theory is always ensured.\footnote{More details on the work of Maldacena are given in Sec.~\ref{sec_holography:AdS_CFT}.} A further development came with the works of Mc Fadden and Skenderis~\cite{McFadden:2009fg,McFadden:2010na} who proposed the application of holography to cosmology. In particular they suggested the possibility of using an holographic description to address the case of the inflationary universe. \\

\begin{figure}[t]
\includegraphics[width=1.\textwidth]{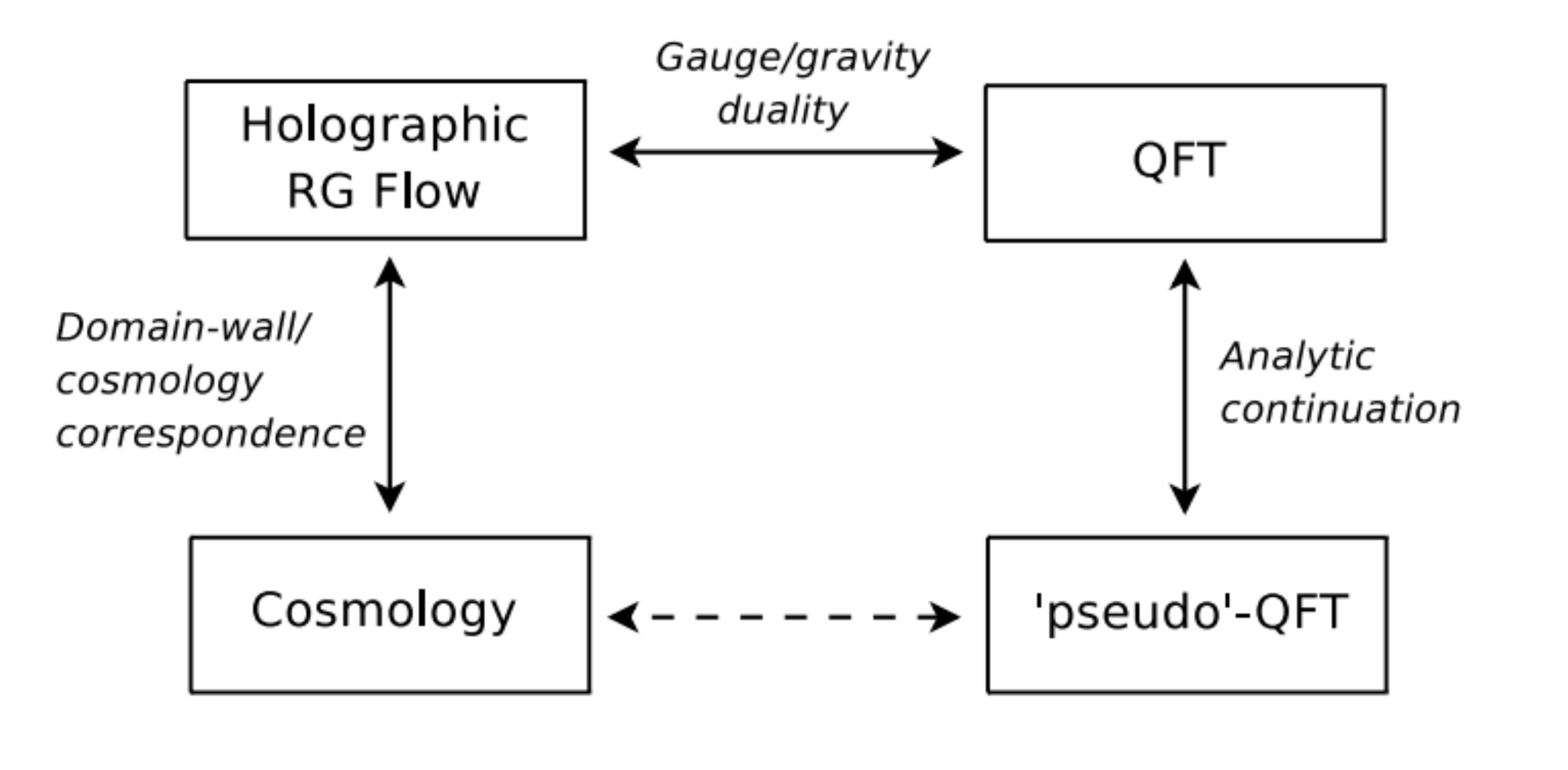}
\caption{Schematic representation of the realization of the Cosmology/'pseudo'-QFT correspondence.\label{fig_holography:scheme_holography}}
\end{figure} 

\noindent As pointed out in~\cite{McFadden:2009fg,McFadden:2010na}, in order to implement the holographic description of the inflationary universe it is necessary to specify the dual `pseudo'-QFT. A schematic idea of the steps necessary to realize this procedure is shown in Fig.~\ref{fig_holography:scheme_holography}. The first step is provided by the domain-wall/cosmology correspondence. This states that every cosmological solution for a single minimally coupled scalar field can be mapped into a corresponding domain-wall solution. The details of the procedure to set this relation are explained in Sec.~\ref{sec_holography:AdS_dS_correspondence}. The crucial point in this association is that a deformation of a nearly dS spacetime (that takes place during inflation) is mapped into the deformation of a nearly AdS spacetime (on the corresponding domain-wall solution). As explained in Sec.~\ref{sec_holography:AdS_CFT}, in the AdS/CFT framework the deformation of the AdS geometry is interpreted as the RG flow of dual three-dimensional QFT. As in the context of QFT the RG flow is usually described in terms of a $\beta$-function, this mapping offers a natural interpretation\footnote{Consistently with the discussion of Chapter~\ref{chapter:beta}, the (A)dS geometry that corresponds to inflation, is realized when the dual QFT becomes conformal \textit{i.e.} in correspondence of a zero of the dual $\beta$-function. More details on this step are given in Sec.~\ref{sec_holography:Holographic_inflation}.} for the formalism introduced in Chapter~\ref{chapter:beta}. Finally, performing a last analytical continuation, it is possible to obtain the QFT that is dual to the original cosmological solution. This is usually called a `pseudo'-QFT as we only have an operational definition of this theory.\\

\noindent The application of holography to early time cosmology has several interesting consequences. In particular we present two remarkable results:
\begin{itemize}
   \item In this new framework we can introduce some alternative interpretations of the physical phenomenon. A clear example of this possibility has already been introduced in Chapter~\ref{chapter:beta}. Indeed the introduction of a $\beta$-function to describe the inflating universe is not fortuitous. As argued in the previous paragraph, by applying holography to cosmology it is natural to describe the deformation of (A)dS geometry (\textit{i.e.} inflation) in terms of an RG flow in the neighborhood of a fixed point. As usual this description is developed in terms of RG equations and thus we are directly led to the introduction of a $\beta$-function. More details on this procedure are given in Sec.~\ref{sec_holography:Holographic_inflation}.

   \item Holography allows to study models where gravity is strongly coupled. In the standard framework for inflation we assume that gravity is weakly coupled at early times and this ensures the possibility of neglecting higher order terms for gravity. Clearly in the limit where these terms cannot be neglected the standard description is no longer valid. Using holography, these cases can be studied in terms of the dual QFT. As a consequence we can introduce a whole new class of models to describe the physics of the inflationary universe. More details on the realization of these theories is left to Sec.~\ref{sec_holography:Holographic_strong}.
 \end{itemize} 

\noindent In the rest of this Chapter we present a detailed analysis of the procedure to apply holography to cosmology and we illustrate an example to clarify the consequences of this process.

\section{Domain-wall/Cosmology correspondence.}
 \label{sec_holography:AdS_dS_correspondence}
\noindent In this Section we follow the treatment of Mc Fadden and Skenderis in~\cite{McFadden:2010na} and we present an explicit realization of the correspondence between domain-wall solutions and cosmological solutions. This equivalence holds both in the unperturbed case and in presence of perturbations around the background evolution. We begin our treatment in Sec.~\ref{sec_holography:AdS/dS_mapping} by considering the first of these two cases and then we move to the second one. After exhibiting the correspondence, in Sec.~\ref{sec_holography:AdS_dS_observables} we give a review of the observable quantities that are relevant for our treatment.

\subsection{The correspondence.}
\label{sec_holography:AdS/dS_mapping}
To show that each cosmological solution has a domain-wall analogous, we start by considering the action for a scalar field $\varphi$ with canonical kinetic term and which is minimally coupled with gravity that as usual is described by a standard Einstein-Hilbert term:
\begin{equation}
  \label{eq_holography:defining_action_dimensionful}
    \mathcal{S} = \int\mathrm{d}t\mathrm{d}^3x \sqrt{|g|}\left( \frac{R}{2\kappa^2} -  \tilde{X} - V(\varphi) \right) \ ,
  \end{equation}
where as usual we use the convention $\textrm{d}s^2 = -\textrm{d}t^2 + a^2(t)\textrm{d}\vec{x}^2$ and we have defined $\tilde{X} \equiv g^{\mu\nu}\partial_{\mu} \varphi \partial_{\nu} \varphi / 2$ and $\kappa^2 \equiv 8 \pi G_N$. As a first step we collect the $\kappa^{-2}$ factor and we introduce the dimensionless scalar field $\Phi$ defined as $\Phi \equiv \kappa \varphi$. After these manipulations the action reads:
\begin{equation}
  \mathcal{S} = \frac{1}{\kappa^2}\int\mathrm{d}t\mathrm{d}^3x \sqrt{|g|}\left( \frac{R}{2} - X - \kappa^2 V(\varphi) \right) \ ,
\end{equation}
where $X$ is now defined in terms of the dimensionless field $\Phi$ \textit{i.e.} $X \equiv g^{\mu\nu}\partial_{\mu} \Phi \partial_{\nu} \Phi / 2$. As multiplying the whole action by a constant factor does not affect the equations of motion for the system, we express the action as:
\begin{equation}
  \label{eq_holography:defining_action}
    \mathcal{S}= - \frac{\eta}{\kappa^2} \int\mathrm{d}r\mathrm{d}^3x \sqrt{|g|}\left( \frac{R}{2} - X - \kappa^2 V(\Phi) \right),
  \end{equation}
where we have introduced the constant $\eta$ and a new `time' coordinate $r$ whose definitions are given in the following. In particular, in the rest of this Section we will show that this action can be used to produce a consistent treatment for both the cosmology and the domain-wall case\footnote{In this Section we are only considering Euclidean domain-walls. Indeed it is possible to show that the case of Lorentzian domain-walls is equivalent. In particular these can be recovered by performing an analytical continuation of one of the spatial coordinates~\cite{Skenderis:2006jq, Skenderis:2006fb}.}. In a first instance we assume that our system is homogeneous and that the spatial slices of our $d+1$ dimensional spacetime are flat. It should be clear that both cosmological and domain-wall solutions can satisfy these requirements. Under these assumptions, a general ansatz that solves the equations of motion associated with the action of Eq.~\eqref{eq_holography:defining_action} can be expressed as:
 \begin{equation}
  \label{eq_holography:unperturbed_solutions}
    \textrm{d}s^2 = \eta \textrm{d}r^2 + a^2(r)\textrm{d}\vec{x}^2, \qquad \qquad \Phi = \phi(r),
  \end{equation}
where in the case of cosmology we have $\eta = -1$ and $r$ is identified with cosmic time (that as usual is denoted with $t$). On the contrary, in the case of Euclidean domain-wall solutions, we have $\eta = +1$ and $r$ is identified with the radial coordinate (that we denote with $u$). From these expressions, it should be clear that for $a^2(r) \simeq \exp\{ 2 C r \}$ where $C>0$ is a constant (dimensionful) factor (basically the inverse of the curvature radius), the background metric matches with the metric of dS${}_{4}$ and of AdS${}_{4}$ given respectively by Eq.~\eqref{appendix_GR:dS_metric} and by Eq.~\eqref{appendix_GR:metric}. The dS${}_{4}$ spacetime associated with inflation is therefore mapped into a AdS${}_{4}$ spacetime on the domain-wall side of the correspondence. \\

\noindent 
To give an explicit proof of the equivalence between the cosmological and the domain-wall solutions, let us express the equation of motion associated with the action~\eqref{eq_holography:defining_action}. Einstein equations read:
\begin{equation}
  3H^2 = \frac{\dot{\phi}^2}{2} - \eta \kappa^2 V \ , \qquad \qquad -2\dot{H} = \dot{\phi}^2 \ ,
\end{equation}
where dots are used to denote derivatives with respect to $r$. The equation of motion for the homogeneous scalar field $\phi$ reads:
\begin{equation}
  \ddot{\phi} + 3 H \dot{\phi} - \eta \kappa^2 V_{,\phi} = 0 \ .
\end{equation}
It is then clear that every cosmological solution for a model with potential $V(\phi)$ is equivalent to a domain-wall solution for a model with potential $-V(\phi)$. In analogy with the treatment of Chapter~\ref{chapter:beta} under the reasonable assumption of a \emph{piecewise monotonic} field we can use the Hamilton-Jacobi formulation of Salopek and Bond~\cite{Salopek:1990jq} to express the equations of motion. In this setting we invert $\phi(r)$ and we use the field $\phi$ as the variable to describe the evolution of the system. In this framework, we can directly follow the same treatment as Chapter~\ref{chapter:beta}, and, with analogous notation, we can express the Hubble parameter as $-2 H(\phi) \equiv W(\phi)$. In this formulation the equations of motion for the system simply read:
 \begin{equation}
  \label{eq_holography:eom}
  \frac{\dot{a}}{a} = - \frac{W(\phi)}{2} \ , \qquad \qquad \dot{\phi} = W_{,\phi} \ , \qquad \qquad 2 \eta \kappa^2 V(\phi) =  W^{ 2}_{,\phi} - \frac{3 }{2}W^2 \ , 
  \end{equation}
where $W_{,\phi} \equiv \textrm{d} W(\phi)/\textrm{d}\phi$. Again it is important to stress that, fixing $V(\phi) = const$, the two sides of the correspondence asymptote to a dS${}_{4}$ and an AdS${}_{4}$ manifold respectively. On the cosmological side, the flow of the system away from the dS${}_4$ configuration corresponds to the inflationary phase. On the corresponding domain-wall solution, we can apply the AdS/CFT correspondence and interpret this flow as an RG flow of the dual QFT. Before proceeding with our treatment, it is also interesting to notice that an inversion in the sign of the potential is completely equivalent to an analytical continuation that maps $\kappa^2 $ into $\bar{\kappa}^2 = - \kappa^2 $. In the rest of this Chapter we use the continuation of $\kappa$. In the following sections we provide an explanation for this choice. \\

\noindent As stated at the beginning of this Section, the correspondence between cosmological solutions and domain-wall solutions can also be extended to the perturbations. A general treatment of this problem is presented in Appendix~\ref{appendix_perturbations:Cosmological_perturbations} where we give a description that is valid for both the cases of cosmological and domain-wall perturbations. For the scope of this chapter, we only need to show that there exists a correspondence between the cosmological and domain-wall perturbations. As a consequence, we only need to check that the explicit expression for the equations of motion for the linearized cosmological perturbations are equivalent. As discussed in Appendix~\ref{appendix_perturbations:Cosmological_perturbations}, to describe the perturbations we have to expand the metric $g_{\mu\nu}(r,\vec{x})$ and the scalar field $\Phi(r,\vec{x})$ as:
\begin{equation}
  g_{\mu\nu}(r,\vec{x}) = {}^{(0)}g_{\mu\nu}(r) + \delta g_{\mu\nu}(r,\vec{x}) \ , \qquad \qquad \Phi(r,\vec{x}) =  {}^{(0)} \bar{\phi}(r) + \delta \bar{\phi} (r,\vec{x}) \ , 
\end{equation}
after fixing the gauge to remove the unphysical degrees of freedom, the perturbations can be described in terms of $\zeta$ comoving curvature perturbation and of $\gamma_{ij}$ transverse traceless part of the spatial metric. The equations of motion for the spatial Fourier transform of these two quantities are computed in Appendix~\ref{appendix_perturbations:Cosmological_perturbations}. In particular, these equations read:
\begin{eqnarray}
  \label{eq_holography:scalar_eom}
  \ddot{\tilde{\zeta}} + \left( 3 H + \frac{\dot{\epsilon_H}}{\epsilon_H} -2 \frac{\dot{c}_s}{c_s}\right) \dot{\tilde{\zeta}} - \frac{ \eta k^2 }{a^2(r)} \tilde{\zeta}&=& 0,\\
  \label{eq_holography:tensor_eom}
  \ddot{\tilde{\gamma}}_{ij} + 3 H \dot{\tilde{\gamma}}_{ij} - \frac{ \eta k^2 }{a^2(r)} \tilde{\gamma}_{ij} &=& 0,
  \end{eqnarray}
where $\epsilon_H \ \equiv -\dot{H}/H^2$ is the first slow roll parameter, $c_s$ is the speed of sound (defined in Eq.~\eqref{eq_inflation:speed_of_sound}) and $\vec{k}$ is the comoving wavevector of the perturbation. Notice that for the models discussed in this Chapter we have $c_s^2 = 1$. Similarly to the case of the background solution, the equivalence is manifestly realized by defining an analytical continuation that maps $k^2$ into $ \bar{k}^2 = - k^2$. \\

\noindent The correspondence between cosmologies and domain-walls for the background solutions and for linear perturbations is then realized by:
\begin{equation*}
  \begin{aligned}
  \textbf{Cosmology} \ \text{\textit{i.e.}} \  \eta = &-1:  \\
  \text{Solution in terms of:} \  &\kappa^2, k^2.
  \end{aligned} \qquad  \iff  \qquad  
\begin{aligned}
\textbf{Domain-wall} \ \text{\textit{i.e.}} \  \eta = & \ 1: \\
\text{Solution in terms of:} \ &\bar{\kappa}^2 ,\bar{k}^2.
  \end{aligned} 
\end{equation*}
Notice that $\bar{k}^2 = -k^2$ can both be satisfied by $\bar{k} = i k $ and $\bar{k} = -i k $ and therefore to completely specify the domain-wall/cosmology correspondence, we have to remove this degeneracy. As we discuss in the following, the correct choice is $\bar{k} = -i k $. In particular this choice corresponds to imposing the correct boundary conditions for the domain-wall solution.

\subsection{Observable quantities.}
\label{sec_holography:AdS_dS_observables}
Let us start this Section by giving a convenient expression for the cosmological scalar and tensor power spectra. These quantities have been computed in Appendix~\ref{appendix_perturbations:observables}, leading to the expressions of Eq.~\eqref{appendix_perturbations:scalar_power_spectrum} and Eq.~\eqref{appendix_perturbations:tensor_power_spectrum_final} respectively. To proceed with our treatment it is useful to define $ \tilde{\zeta}_{k}(r) $ (mode functions that solve the equations of motion for $\tilde{\zeta}$) and $\tilde{\gamma}_{k}(r)$ (mode functions for the tensor perturbations)\footnote{Given $e_{ij}$ polarization vector of the tensor perturbation $\tilde{\gamma}_{ij}(r)$, we define $ \tilde{\gamma}_{k,ij}(r) \equiv e_{ij} \tilde{\gamma}_{k}(r) $.}. As discussed in Appendix~\ref{appendix_perturbations:observables}, in the case of cosmology it is possible to show that the action (for $c_s^2 =1 $) is:
\begin{equation}
  \mathcal{S} = \frac{1}{\kappa^2} \int\textrm{d} t \textrm{d}^3 \vec{k} \left[ \left( a^3 \ \epsilon_H \right) \dot{\tilde{\zeta}}^2(t,\vec{k}) - \left(\frac{\eta k^2 }{a^2 }\right) \tilde{\zeta}^2(t,\vec{k}) \right] \ , 
\end{equation}
gives the correct equations of motion~\eqref{eq_holography:scalar_eom} for $\tilde{\zeta}$. Following a similar procedure we can define a similar action to describe the case of tensor perturbations. Using these actions, we can compute the canonical momenta: 
\begin{equation}
\label{eq_AdS_dS:canonical_momenta_cosmo}
   \tilde{\Pi}_{k}^{(\tilde{\zeta})} = \frac{2 a^3 \epsilon_H }{ \kappa^2} \dot{\tilde{\zeta}}_{k} \ , \qquad \qquad \qquad  \tilde{\Pi}_{k}^{(\tilde{\gamma})} = \frac{a^3 }{4 \kappa^2} \dot{\tilde{\gamma}}_{k}  \ .
 \end{equation} 
Imposing the canonical commutation relations, we get the Wronskian conditions:
\begin{equation}
\label{eq_holography:wronskian_condition}
  i = \tilde{\zeta}_{k} \tilde{\Pi}_{k}^{(\tilde{\zeta})*} - \tilde{\Pi}_{k}^{(\tilde{\zeta})} \tilde{\zeta}_{k}^* \ ,  \qquad \qquad \frac{i}{2} = \tilde{\gamma}_{k} \tilde{\Pi}_{k}^{(\tilde{\gamma})*} - \tilde{\Pi}_{k}^{(\tilde{\gamma})} \tilde{\gamma}_{k}^* \ .
\end{equation}
In order to have a better connection with the holographic analysis, it is useful to express the scalar and tensor power spectra in terms of the two linear response functions $E(k)$ and $\Omega(k)$ defined as:
\begin{equation}
  \label{eq_AdS_dS:linear_response_functions}
  \tilde{\Pi}_{k}^{(\tilde{\zeta})} \equiv \Omega(k) \  \tilde{\zeta}_{k} , \qquad \qquad \tilde{\Pi}_{k}^{(\tilde{\gamma})} \equiv E(k) \  \tilde{\gamma}_{k} \ .
\end{equation}
Using these definitions, we can express the Wronskian conditions as:
\begin{equation}
\label{eq_holography:response_functions_cosmology}
\begin{aligned}
  i & = |\tilde{\zeta}_{k} |^2 \left[ \Omega(k)^* - \Omega(k) \right] \ , \qquad \longrightarrow \qquad |\tilde{\zeta}_{k} |^2 = \frac{-1}{2 \text{Im}\left[\Omega(k)\right]}\\
   \frac{i}{2} & = |\tilde{\gamma}_{k} |^2 \left[ E(k)^* - E(k) \right] \ , \qquad \longrightarrow \qquad |\tilde{\gamma}_{k} |^2 = \frac{-1}{4 \text{Im}\left[E (k)\right]} \ ,
  \end{aligned}
\end{equation}
where, given complex number $z$, we use $\text{Im}[z] \equiv - i (z - z^*)/2 $ to denote the imaginary part. As discussed in Appendix~\ref{appendix_perturbations:observables}, to compute the scalar and tensor power spectra we first need the late time (\textit{i.e.} superhorizon) behavior of $|\tilde{\zeta}_{k} |^2 $ and $|\tilde{\gamma}_{k} |^2 $ and then we should evaluate these functions at the time when they re-enter the horizon \textit{i.e.} at $k \tau = 1$ in the case of scalar perturbations and at $k\tau = 1 $ for tensor perturbations. Let us define $\Omega_{(0)}(k)$ and $E_{(0)}(k)$, late time (\textit{i.e.} superhorizon) response functions. Substituting into Eq.~\eqref{appendix_perturbations:scalar_power_spectrum} and Eq.~\eqref{appendix_perturbations:tensor_power_spectrum_final} the scalar and tensor power spectra respectively read:
\begin{equation}
\label{eq_holography:power_spectra_cosmo}
\begin{aligned}
 & \left. \Delta^2_s (k)\right|_{k = a H} =  \frac{-k^3}{4 \pi^2 \text{Im}\left[\Omega_{(0)}(k)\right]}  \ , \\
  & \left.  \Delta^2_t (k)  \right|_{k = a H } =   \frac{ -k^3}{2\pi^2 \text{Im}\left[E_{(0)}(k)\right]}  \ .
  \end{aligned}
\end{equation}
Most of the relevant observable quantities for cosmology that are interesting for the scope of this work can be expressed in terms of $\Delta^2_s (k)$ and $\Delta^2_t (k)$. \\

\noindent
We should then discuss the correspondent quantities for the domain-wall solutions. As a first step we should discuss how the initial condition of a Bunch-Davies vacuum translates in the case of domain-wall solution. As discussed in Appendix~\ref{appendix_perturbations:Cosmological_perturbations}, we impose the early-time behavior of the cosmological solution to be $\sim \ \exp \{ -ik \tau \}$. Using the analytically continued variable $\bar{k} = -i k$, its analogous in the case of the domain-wall spacetime is $\sim \ \exp \{ \bar{k} \tau \}$. As the early-time behavior is fixed at $\tau \rightarrow -\infty$, the choice of a Bunch-Davies vacuum on the cosmological side translates into an exponentially decaying solution in the interior of the euclidean AdS spacetime. This condition clearly ensures the regularity of the solution in the interior of the euclidean AdS spacetime. This regularity is a prerequisite to perform the holographic analysis and this clearly justifies our choice of continuing $k$ into $\bar{k} = -i k$.\\

\noindent
At this point we can consider the case of the domain-wall solution. In analogy with the case of the cosmology, the canonical momenta are defined as:
\begin{equation}
\label{eq_AdS_dS:canonical_momenta_wall}
   \tilde{\bar{\Pi}}_{\bar{k}}^{(\tilde{\zeta})} = \frac{2 a^3 \epsilon_H }{\bar{\kappa}^2} \dot{\tilde{\zeta}}_{\bar{k}} \ , \qquad \qquad  \tilde{\bar{\Pi}}_{\bar{k}}^{(\tilde{\gamma})} = \frac{a^3 }{4 \bar{\kappa}^2} \dot{\tilde{\gamma}}_{\bar{k}}  \ .
 \end{equation} 
Notice that to respect the Wronskian condition (in particular to be consistent with the signs), the response functions are defined as:
\begin{equation}
  \label{eq_AdS_dS:linear_response_functions_wall}
  \tilde{\bar{\Pi}}_{\bar{k}}^{(\tilde{\zeta})} \equiv - \bar{\Omega}(\bar{k}) \  \tilde{\zeta}_{\bar{k}} , \qquad \qquad \tilde{\bar{\Pi}}_{\bar{k}}^{(\tilde{\gamma})} \equiv - \bar{E}(\bar{k}) \  \tilde{\gamma}_{\bar{k}} \ ,
\end{equation}
where the minus sign is thus arising from the continuation $\bar{\kappa}^2 = - \kappa^2$. Comparing the definition of the cosmological response function (Eq.~\eqref{eq_AdS_dS:linear_response_functions}) with the definition of the domain-wall response function (Eq.~\eqref{eq_AdS_dS:linear_response_functions_wall}), it should be clear that:
\begin{equation}
  \bar{\Omega}(\bar{k}) =  \bar{\Omega}(- i k ) = \Omega(k) \ , \qquad \qquad  \bar{E}(\bar{k}) =  \bar{E}(- i k ) = E(k) \  ,
\end{equation}
and, using the Wronskian conditions, we get:
\begin{equation}
\begin{aligned}
  i & = -|\tilde{\zeta}_{\bar{k}} |^2 \left( \bar{\Omega}^* -  \bar{\Omega} \right) \ , \qquad \longrightarrow \qquad |\tilde{\zeta}_{\bar{k}} |^2 = \frac{1}{2 \text{Im}\left[\Omega(\bar{k})\right]} \ , \\
  \frac{i}{2} & = - |\tilde{\gamma}_{\bar{k}} |^2 \left( \bar{E}^* - \bar{E} \right) \ ,  \qquad \longrightarrow \qquad |\tilde{\gamma}_{\bar{k}} |^2 = \frac{1}{4 \text{Im}\left[E(\bar{k})\right]} \ .
  \end{aligned}
\end{equation}
It is interesting to notice the presence of a minus sign with respect to the analogous expression for the cosmological solutions (Eq.~\eqref{eq_holography:response_functions_cosmology}). Finally the scalar and tensor power spectra can be expressed as:
\begin{equation}
\begin{aligned}
\label{eq_holography:power_spectra_wall}
  \left. \Delta^2_s (\bar{k})\right|_{\bar{k} = a H } & =  \left. \frac{ - i \bar{k}^3}{4 \pi^2 \text{Im}\left[\bar{\Omega}_{(0)}(\bar{k})\right]} \right|_{\bar{k} = a H} \ ,  \\
  \left.  \Delta^2_t (\bar{k})  \right|_{\bar{k} = a H } & =   \left. \frac{ -i \bar{k}^3}{2\pi^2 \text{Im}\left[\bar{E}_{(0)}(\bar{k})\right]} \right|_{\bar{k} = a H } \  .
  \end{aligned}
\end{equation}
In the next Sections of this Chapter (for the explicit expressions see Sec.~\ref{sec_holography:Holographic_strong}) we explain how we can compute the domain-wall power spectra in terms of the dual three-dimensional QFT using the AdS/CFT correspondence.

\section{AdS/CFT.}
 \label{sec_holography:AdS_CFT}
At the beginning of 20th century Quantum Mechanics (QM) has been introduced to give an explanation to microscopic phenomena. The unification of QM and Special Relativity (SR) has led to the formulation of QFT that provide a adequate description of Electroweak and Strong interactions. On the contrary, the geometrical theory of gravitation, proposed by Einstein in 1915 under the name of GR, provides an elegant framework to describe the physics of the large scales. To produce a coherent description of Quantum Gravity (QG), it seems natural to attempt to formulate GR with the language of QFTs. Unfortunately this procedure presents several considerable difficulties and thus different paths, such as loop quantum gravity and string theory, have been attempted. In this context the AdS/CFT correspondence of Maldacena~\cite{Maldacena:1997re} has provided an extremely useful framework to have a deeper understanding of QG and string theory.\\

\noindent
AdS/CFT is a marvelous realization of the holographic principle that conjectures the existence of a weak/strong duality between CFTs and theories of gravity in AdS spacetime. The idea of AdS/CFT originates from the observation that both CFTs in four dimensions and theories of gravity in $AdS_5$ are theories with $SO(2,4)$ symmetry\footnote{For details see Appendix~\ref{appendix_CFT:Conformal_field_theories} and Appendix~\ref{appendix_GR:dS_AdS}}. Another crucial observation that suggests the possibility of relating these two sets of theories comes out by analyzing the structure of $AdS_5$ spacetime. In particular, as discussed in Appendix~\ref{appendix_GR:dS_AdS}, the boundary of this manifold is conformally equivalent to $\mathbb{R}^{1,3}$, \textit{i.e.} Minkowski spacetime that is the spacetime that is normally used to construct QFTs and CFTs. Enforced by these observations the AdS/CFT correspondence conjectures a dual equivalence between a theory of gravity in the bulk of $AdS_5$ and a CFT on its boundary.\\

\noindent For the scopes of this discussion it is useful to introduce the notion of 't Hooft limit and the notion of string tension:
\begin{itemize}
   \item It is known that a $U(N)$ gauge theory is characterized by two dimensionless parameters: its coupling constant $g$, and $N$, rank of the gauge group. As discussed by `t Hooft in~\cite{'tHooft:1973jz}, for pure gauge theories it is useful to combine these two parameters into an effective coupling $\lambda \equiv g N$. The `t Hooft limit is defined as $g\rightarrow 0$, $N\rightarrow \infty $ with $\lambda $ fixed. In this limit it is possible to define a $1/N$ expansion of the theory. Feynmann diagrams can be divided into planar and non-planar diagrams and in particular we find that non-planar diagrams are suppressed by $1/N$ factors according to their topology~\cite{'tHooft:1973jz}.

   \item The basic object of string theory are strings with a certain string tension defined as $T = 1/(2\pi\alpha^\prime)$. The tension has the dimension of a mass squared meaning that $\alpha^\prime$ has the dimension of a length squared and we can therefore define $l_s$ string length as $\alpha^\prime = l_s^2$. Notice that the limit $\alpha^\prime \ll 1$ corresponds to $1 \ll T$ that is the limit of infinite string tension in which strings can be well approximated by point-like particles.
 \end{itemize}  

\noindent Inspired by these observations, Maldacena has shown in~\cite{Maldacena:1997re} that the large $N$ limit of a $\mathcal{N} = 4 $ $U(N)$ Super-Yang-Mills (SYM) theory is equivalent to type IIB strings in a $AdS_5 \times S_5$ spacetime. Moreover, given $R_A$, ``radius" of the $AdS_5$ spacetime\footnote{More on its definition is said in Appendix~\ref{appendix_GR:general}, in particular see Sec~\ref{appendix_GR:dS_AdS}.}, Maldacena has shown that $(R_A/l_s)^4 \propto \lambda$. Remarkably this relation implies that if the CFT happens to be strongly coupled \textit{i.e.} $ 1 \ll \lambda $, we also have $l_s \ll R_A$. In this regime the string theory reduces to the case of a QFT in fixed AdS background. This proves that computations can always be performed in one of the two sides of the correspondence either for $ 1 \ll \lambda $ and for $ \lambda \ll 1$. In this sense the AdS/CFT correspondence is a weak/strong duality.\\

\noindent Before concluding this Section it is also useful to stress one more property of the AdS/CFT correspondence. Let us consider AdS${}_{d+1}$ in terms of the Poincar\'e coordinates\footnote{In these coordinates the radial coordinate $u$ is expressed in terms of a new coordinate $z$ as:
\begin{equation}
  \frac{u}{R_A} = - \ln\left(\frac{z}{R_A} \right) \ .
\end{equation}
As a consequence, the boundary of AdS ($u \rightarrow \infty$) is approached for $z \rightarrow 0$.} defined in Eq.~\eqref{appendix_GR:poincaree_metric}. The CFT is defined on the conformal $d$-dimensional Minkowski boundary of AdS${}_{d+1}$ spacetime that is reached for $z \rightarrow 0$. Let us call $d_P$, the proper distance between two point on the boundary in terms of Poincar\'e coordinates, and $d_{M}$ the distance between the same points in terms of the Minkowski coordinates. These quantities are related as $d_P = d_{M} L/z$. A similar but inverse relation holds also for the energies $E_M = E_P L/z$. This relation implies that the radial coordinate should be interpreted as the energy scale of the dual field theory. In fact at fixed $E$ (energy for the gravity theory) the UV limit of the CFT is obtained when we consider the region close to the boundary $z \rightarrow 0 $ of AdS${}_{d+1}$. Conversely the IR limit is obtained when we approach the region close to the horizon corresponding to $z \rightarrow 0 $. As CFTs are (classically) scale invariant, once we define the theory at a certain value of $z$, the theory is specified at all scales. \\

\noindent
As widely discussed in the literature~\cite{deBoer:2000cz,deHaro:2000vlm,Skenderis:2002wp,Papadimitriou:2004rz}, the AdS/CFT correspondence can be extended outside of the conformal regime. As usual, the deformation of a CFT is described as a RG flow induced by some operator. In this context, the deformation of an asymptotically AdS spacetime is interpreted as the RG flow of the dual QFT in the neighborhood of a RG fixed point. In this Section we give the formulation of the AdS/CFT correspondence (in Sec.~\ref{sec_holography:AdS_CFT_formulation}), we show one pedagogical example (in Sec.~\ref{sec_holography:AdS_CFT_twopointfunction}) and finally (in Sec.~\ref{sec_holography:RG_flows}) we discuss the application of AdS/CFT to non conformal theories. The latter will then be relevant for the holographic interpretation of inflation presented in the next Section.\\

\subsection{Formulation of the correspondence.}
\label{sec_holography:AdS_CFT_formulation}
AdS/CFT formulates a map between the observables of the theories on the two sides of the correspondence. In a QFT observable quantities are expressed in terms of the $n$-point functions of the operators of the theory. To compute these quantities it is useful to introduce $Z[J]$ generating functional of the theory. For example let us consider the case of a scalar field theory. In this case $Z[J]$ can be defined as the functional integral:
\begin{equation}
  \label{eq_holography:generating_functional}
  Z[J] \equiv \int \mathcal{D}\phi \exp\left\{ i \left( \mathcal{S}[\phi] + \int \textrm{d}^d x J(x) \phi(x) \right) \right\},
\end{equation}
where $\mathcal{S}[\phi]$ is the action for the scalar field and $J(x)$ is a classical source coupled to the scalar field. It is well known that in this formalism the $n$-point functions for scalar field theory can be expressed as:
\begin{equation}
  \langle \phi(x_1) \phi(x_2) \dots \phi(x_n) \rangle = \frac{1}{Z_0} \left. \frac{\delta}{\delta J(x_1)}  \frac{\delta}{\delta J(x_2)} \dots  \frac{\delta}{\delta J(x_n)} Z[J] \right|_{J = 0},
\end{equation}
where we have defined $Z_0 \equiv \left.  Z[J] \right|_{J = 0}$. Given $ Z[J]$ it is also useful to define $W[J]$ generating functional of connected Green's functions as:
\begin{equation}
  \label{eq_holography:connected_functional}
  Z[J] \equiv \exp\{ W[J] \} \ .
\end{equation}
Clearly it is possible to generalize this formalism to the more general case of a QFT with some operator $\mathcal{O}(x)$ that is coupled to a classical source $J(x)$ through a term $\mathcal{O}(x)J(x)$. \\

\noindent AdS/CFT states that fields in AdS${}_{d+1}$ are associated with operators in the $d$-dimensional CFT through some boundary coupling. As both theories have $SO(2,d)$ symmetry, the AdS${}_{d+1}$ field and the corresponding CFT operator must have the same $SO(2,d)$ quantum numbers. For simplicity let us restrict to the case of a scalar field in AdS${}_{d+1}$. As discussed in the following paragraphs generalizations to different cases can be found by using the symmetries of the two theories. As $\phi(u,x)$ is an $SO(2,d)$ scalar, its restriction on the boundary of AdS${}_{d+1}$ can be coupled with some dual scalar operator $\mathcal{O}(x)$ living on the boundary of AdS${}_{d+1}$. In particular they can be coupled through a term $\phi_0(x)\mathcal{O}(x)$ where $\phi_0(x) = \left. \phi(u,x) \right|_{u \rightarrow \infty} $. The AdS/CFT correspondence states that the generating functional for the correlation functions for $\mathcal{O}(x)$ is given by~\cite{Aharony:1999ti,Zaffaroni:2000vh}:
\begin{equation}
\label{eq_holography_correspondence}
    \exp \left\{ W_{CFT} [\phi_0(x)] \right\}  =  Z_{CFT}[\phi_0(x)] =  Z_{AdS_{d+1}} \left[\phi_0(x) \right] \simeq \exp \left\{ -\mathcal{S}_{AdS_{d+1}} \left[\phi_0(x) \right] \right\},
\end{equation}
where $ Z_{AdS_{d+1}} \left[\phi_0(x) \right]$ and $\mathcal{S}_{AdS_{d+1}} \left[\phi_0(x) \right]$ are respectively used to denote the partition function and the action of the gravity theory evaluated on a solution of the classical equation of motion that satisfies :
\begin{equation}
\label{eq_holography_bd_condition}
  \left. \phi(u,x^\mu)\right|_{u \rightarrow \infty} = \phi_0 (x^\mu). 
\end{equation}
Notice that on the left hand side of Eq.~\eqref{eq_holography_correspondence} we have a functional depending on an arbitrary configuration for the $d$-dimensional field $\phi_0(x)$ while on the right hand side we have the partition function of gravity theory with the boundary condition of Eq.~\eqref{eq_holography_bd_condition}. It is also interesting to point out that the approximation:
\begin{equation}
 \ln \left\{  Z_{AdS_{d+1}} \left[\phi \rightarrow \phi_0(x) \right] \right\}  \simeq -\mathcal{S}_{AdS_{d+1}} \left[\phi \rightarrow \phi_0(x) \right] , 
\end{equation}
corresponds to ignoring corrections depending on $\alpha^\prime$. Using Eq.~\eqref{eq_holography_correspondence}, we can express a general correlation function of $\mathcal{O}(x)$ as:
\begin{equation}
\label{eq_holography_correlators}
  \langle \mathcal{O}(x_{1}) \mathcal{O}(x_{2}) \dots\mathcal{O}(x_{n}) \rangle = \frac{1}{Z_0} \left. \frac{\delta}{\delta \phi_0(x_{1})}  \frac{\delta}{\delta \phi_0(x_{2})} \dots  \frac{\delta}{\delta \phi_0(x_{n}) } Z[\phi_0] \right|_{\phi_0 = 0}.
\end{equation}
Before considering generalizations of this case it is important to make some remarks on the meaning of Eq.~\eqref{eq_holography_correspondence}:
\begin{enumerate}
  \item \label{comments_AdS/CFT_1} It is important to stress that on the gravity side of the correspondence we are considering a solution of the classical equations of motion. This implies that we are considering an on-shell configuration for the scalar field. On the other hand on the CFT side of the correspondence we have not imposed this condition and thus the theory is off-shell. 

  \item \label{comments_AdS/CFT_renormalization}
  In general it is not possible to directly evaluate the gravity action on an on-shell configuration because it typically diverges. To be able to define properly the CFT generating functional we thus need to define a renormalized gravity action $\mathcal{S}^{reg}_{AdS_{d+1}}$ by following some procedure of holographic renormalization.

  \item \label{comments_AdS/CFT_2}  
  The equations of motion\footnote{For an explicit computation see Appendix~\ref{appendix_GR:dS_AdS} (in particular Sec.~\ref{appendix_GR:AdS_dynamics}).} for a massive scalar field $\phi(u,x)$ with mass $m$ in AdS${}_{d+1}$ admits two solutions of the form:
\begin{equation}
  \left. \phi(u,x) \right|_{u \rightarrow \infty} \simeq \left. e^{ - u \Delta_\pm /R_A } \phi_{\pm}(x) \right|_{u \rightarrow \infty} \ ,
\end{equation}
where as usual $g_{ab} = \text{diag}(1 , e^{2u/R_A} , \dots ,e^{2u/R_A} )$ so that $u \rightarrow \infty$ corresponds to the boundary of AdS${}_{d+1}$ spacetime and where we have defined:
\begin{equation}
\label{eq_holography:Delta_pm}
    \Delta_{\pm} = \frac{d}{2} \left( 1 \pm \sqrt{1  + \frac{ 4 R_{A}^{ \ 2} m^2}{d^2}} \right) \ .
\end{equation}
Notice that the $\Delta_+$ solution diverges in the interior ($u \rightarrow - \infty$) of AdS${}_{d+1}$ and thus it must be discarded. It should be clear that $\Delta_- < \Delta_+$ and thus the leading contribution to $\phi(u,x) $ in a neighborhood of the boundary is always carried by the $\Delta_-$ solution. Notice that if\footnote{Notice that as discussed in appendix~\ref{appendix_GR:AdS_spacetime}, AdS spacetime may support a negative mass squared until the BF bound~\cite{Breitenlohner:1982jf} is satisfied. In this case $\Delta_- > 0 $ and thus the corresponding solution goes to zero on the boundary.} $m^2>0$ we have $\Delta_- < 0$ and thus to get a consistent expression for generating functional of the CFT we need a procedure to regularize this quantity. The most natural choice is to extract the divergent multiplicative factor by considering:
\begin{equation}
 \phi_{Reg}(x) \propto \left. e^{ u \Delta_- /R_A} \phi(u,x) \right|_{u \rightarrow \infty}  \ . 
\end{equation}
The correct definition of $\phi_{Reg}(x)$ is given in Appendix~\ref{appendix_GR:general} (see Eq.~\eqref{appendix_GR:regularized_field}). As a consequence $\phi_{Reg}(x)$ should be identified with the inverse Fourier transform of the function $\tilde{\mathcal{A}}(k)$ (see Eq.~\eqref{appendix_GR:tilde_phi_expansion}). Notice that this quantity depends on $\Delta_-$. 

\item \label{comments_AdS/CFT_renormalized}
As discussed in the comment~\ref{comments_AdS/CFT_renormalization} it may be necessary to define a regularization procedure in order to obtain the renormalized action $\mathcal{S}^{reg}_{AdS_{d+1}}$ that should be used to define the correct CFT generating functional. Similarly, in comment~\ref{comments_AdS/CFT_2} we have argued that this action should be expressed in terms of the regularized field $\phi_{Reg}(x)$. In terms of these quantities we can express Eq.~\eqref{eq_holography_correlators} as:
\begin{equation}
\label{eq_holography_renorm_correlators}
  \langle \mathcal{O}(x_{1}) \mathcal{O}(x_{2}) \dots\mathcal{O}(x_{n}) \rangle =  \left. \frac{\delta^{(n)}  \mathcal{S}^{reg}_{AdS_{d+1}}[\phi]}{\delta \phi_{Reg}(x_{1}) \ \delta \phi_{Reg}(x_{2}) \ \dots  \ \delta \phi_{Reg}(x_{n}) } \right|_{\phi_{Reg} = 0}.
\end{equation}

\item \label{comments_AdS/CFT_3} Let us consider a scalar operator $\mathcal{O}$, that is dual to a scalar field $\phi$. The boundary action $\mathcal{S}_{bd}$ that describes their coupling has the form:
\begin{equation}
  \mathcal{S}_{bd} \sim \left. \int \textrm{d}^{d}x \sqrt{|\gamma|} \phi(u,x) \mathcal{O}(u,x)\right|_{u\rightarrow \infty},
\end{equation}
where $\gamma$ is the determinant of the metric induced on the boundary of AdS${}_{d+1}$. As $\sqrt{|\gamma|} =  e^{u d /R_A}$ and $ \left. \phi(u,x) \right|_{u \rightarrow \infty} \propto \left. e^{ - u \Delta_-/R_A} \phi_{Reg}(x) \right|_{u \rightarrow \infty}$ , the action reads:   
\begin{equation}
  \mathcal{S}_{bd} \sim  \left. \int \textrm{d}^{d}x \    e^{ u \Delta_+/R_A } \phi_{Reg}(x)  \mathcal{O}(u,x) \right|_{u\rightarrow \infty} \ ,
\end{equation}
where we have used Eq.~\eqref{appendix_GR:delta_relations}. In order to make this action finite and independent on $u $ when we send $u \rightarrow \infty$, we should thus require:
\begin{equation}
\label{eq_holography_O_Scaling}
\left. \mathcal{O}(u,x)  \right|_{u \rightarrow \infty} \propto e^{ - u \Delta_+ /R_A} \mathcal{O}(x).
\end{equation}
Notice that the multiplying factor $ e^{ - u \Delta_+/R_A }$ in front of the operator $\mathcal{O}(x) $ can be seen as the effect of the transformation properties of the QFT operator as we change $u$. As discussed at the beginning of this Section, a transformation of the coordinate $u$ is interpreted as a scale transformation for the QFT. In this context it is thus natural to interpreted $\Delta_+$ as the mass scaling dimension of the operator $\mathcal{O}(x)$ dual to the scalar field $\phi$. 
\end{enumerate}

\noindent As we have already anticipated during this Section, AdS/CFT does not only work for scalar fields and scalar operators. Actually it is possible to generalize the above treatment to introduce a relationship between general representations of $SO(2,d)$. Indeed the coupling should be consistent with the assumption of $SO(2,d)$ symmetry and thus a bulk field can only be coupled with a CFT operator carrying the same $SO(2,d)$ quantum numbers. For example, given $\mathcal{L}_{CFT}$, lagrangian density for a boundary CFT with scalar $\mathcal{O}(x)$ vector $A_{\mu}(x)$ and tensor $T_{\mu\nu}(x)$ operators, the natural linearized couplings to some sources $ \phi_0(x) , J_0^{\mu}(x), g_0^{\mu\nu}(x), \dots $ have the form:
\begin{equation}
  \mathcal{L}_{CFT} + \phi_0 \mathcal{O} + J_0^{\mu} A_{\mu} + g_0^{\mu\nu}T_{\mu\nu} +\dots \ .
\end{equation}
Notice that this expression implies that a gauge field of the $d$-dimensional CFT may be naturally coupled to an AdS${}_{d+1}$ current. Similarly the stress-energy tensor of the CFT may be naturally coupled with the AdS${}_{d+1}$ metric. A generalization of Eq.~\eqref{eq_holography:Delta_pm} to set relationships between the masses of the bulk fields and the dimensions of CFT operators can be found in~\cite{Aharony:1999ti,Freedman:1999gp,Ferrara:1998ej}. Before concluding this section it is also interesting to point out that given $F_{\mu\nu}$ (field strength of the CFT) it is possible to construct the scalar operator $\mathcal{O} \equiv Tr[F_{\mu\nu} F^{\mu\nu}]$ that may be coupled with some scalar field $ \phi_0 $. In this framework, it is thus natural to interpret the scalar field $ \phi_0 $ as the coupling constant of the boundary CFT. 

\subsection{Two point correlation function in AdS/CFT.}
\label{sec_holography:AdS_CFT_twopointfunction}
\noindent The core statement of AdS/CFT thus is that an off-shell CFT in $d$ dimensions corresponds to an on-shell theory of gravity in $d+1$ dimensions. This is a general feature of all the AdS/CFT inspired correspondences. In this Section we consider the pedagogic example of the holographic computation of the two point function for an operator $\mathcal{O}$, dual to a scalar field $\phi(u,x)$.\\

\noindent Our starting point is the action for a scalar field in AdS${}_{d+1}$. Up to an overall constant factor the action can be expressed as:
 \begin{equation}
  \label{eq_holography:scalar_action}
    \mathcal{S}= \int\mathrm{d}u \ \mathrm{d}^d x \ \sqrt{g}\left( \frac{  g^{ab} }{2 }  \partial_a \phi \partial_b \phi + V(\phi) \right) \ .
  \end{equation}
Notice that this action can be obtained from the action of Eq.~\eqref{eq_holography:defining_action} by restoring the dimension for the scalar field. Using the definition of $\tilde{\phi}(u,k^\mu)$ spatial Fourier transform of $\phi(u,x^\mu)$:
\begin{equation}
  \phi(u, x^\mu) = \int \frac{\mathrm{d}^d k}{(2 \pi)^{d/2}} e^{i x^\mu k_\mu} \tilde{\phi}(u, k^\mu) \ ,
\end{equation}
and, assuming $V(\phi) = m^2 \phi^2 /2 $, we can express the action of Eq.~\eqref{eq_holography:scalar_action} as:
 \begin{equation}
\label{eq_holography:scalar_action_fourier}
 \begin{aligned}
    \mathcal{S}= \frac{1}{2} \int\mathrm{d}u \ \mathrm{d}^d k_1 \mathrm{d}^d k_2 \ & \delta^{(d)}(k_1 + k_2) \ e^{du/R_A} \ \left[  \dot{\tilde{\phi}}_{k_1} \dot{\tilde{\phi}}_{k_2} +  \right. \\
    & \left. - (k_1 \cdot k_2) \ e^{-2u/R_A} \ \tilde{\phi}_{k_1} \tilde{\phi}_{k_2} + m^2 \tilde{\phi}_{k_1} \tilde{\phi}_{k_2} \right] \ ,
    \end{aligned}
  \end{equation}
  where dots are used to denote derivatives with respect to $u$ and where, to lighten the notation, we have defined $\phi_{k_1} \equiv \tilde{\phi}(u,k_1^{ \ \mu}) $, $\phi_{k_2} \equiv \tilde{\phi}(u,k_2^{ \ \mu}) $ and $k_1 \cdot k_2 \equiv \delta_{\mu \nu} \ k_1^\mu k_2^\nu $ . Integrating by parts the first term we get:
 \begin{equation}
\label{eq_holography:scalar_action_fourier_2}
 \begin{aligned}
    \mathcal{S}&= \frac{1}{2} \int \left[ \ \mathrm{d}^d k_1 \mathrm{d}^d k_2 \  \delta^{(d)}(k_1^{ \ \mu} + k_2^{ \ \mu}) \ e^{du/R_A} \ \tilde{\phi}_{k_1} \dot{\tilde{\phi}}_{k_2} \right]_{u\rightarrow - \infty}^{u \rightarrow \infty} + \\
    & +\frac{1}{2} \int\mathrm{d}u \ \mathrm{d}^d k_1 \ e^{du/R_A} \ \tilde{\phi}_{k_1} \left[ - \ddot{\tilde{\phi}}_{-k_1} - \frac{ d \ \dot{\tilde{\phi}}_{-k_1} }{R_A}+ k_1^{\ 2}  e^{-2u/R_A} \tilde{\phi}_{-k_1} + m^2  \tilde{\phi}_{-k_1} \right] \ .
    \end{aligned}
  \end{equation}
As we are considering a configuration of $\tilde{\phi}$ that satisfies the classical equation of motion, the second line is equal to zero\footnote{Notice that this is a peculiar property of quadratic potentials. In a more general case this result is not holding and it is thus necessary to follow a different procedure.}. As discussed in the Appendix~\ref{appendix_GR:AdS_dynamics}, an explicit solution for a scalar field in AdS${}_{d+1}$ is proportional to a modified Bessel function of second kind and therefore the expression for $\tilde{\phi}_{k_2}$ is given in Eq.~\eqref{appendix_GR:tilde_phi_expansion_1}. As we are interested in solutions that are regular in the interior of AdS${}_{d+1}$, we can set $\tilde{\mathcal{B}}(k_2) = 0$ and keep only the second term of Eq.~\eqref{appendix_GR:tilde_phi_expansion}. Approaching the boundary this solution grows exponentially and thus the leading contribution to Eq.~\eqref{eq_holography:scalar_action_fourier_2} is given by:
 \begin{equation}
\label{eq_holography:scalar_action_fourier_3}
    \mathcal{S} \simeq \frac{1}{2} \int \left. \ \mathrm{d}^d k_1 \mathrm{d}^d k_2 \  \delta^{(d)}(k_1^{ \ \mu} + k_2^{ \ \mu}) \ e^{du/R_A} \ \tilde{\phi}_{k_1} \dot{\tilde{\phi}}_{k_2} \right|_{u \rightarrow \infty} \  .
  \end{equation}
We can thus follow the procedure described in Appendix~\ref{appendix_GR:AdS_dynamics} and define the coordinate $ z =\equiv R_A e^{-u/R_A}$ so that the action reads:
\begin{equation}
  \label{eq_holography:boundary_action_new}
  \mathcal{S} \propto  \int \left. \  \mathrm{d}^d k_1 \mathrm{d}^d k_2 \  \delta^{(d)}(k_1^{ \ \mu} + k_2^{ \ \mu})  \ k_2^{ \ d - 1} \ (z k_2)^{-d+1} \  \tilde{\phi}_{k_1} \partial_z \tilde{\phi}_{k_2} \right|_{z \rightarrow 0} \ .
\end{equation}
We can proceed by defining $\theta = k_2 z$, so that the action reads:
\begin{equation}
  \label{eq_holography:boundary_action_theta_new}
  \mathcal{S} \propto  \int \left. \  \mathrm{d}^d k_1 \mathrm{d}^d k_2 \  \delta^{(d)}(k_1^{ \ \mu} + k_2^{ \ \mu})  \ k_2^{ \ d } \ \theta^{-d+1} \  \tilde{\phi}_{k_1} \tilde{\phi}_{k_2} \partial_\theta \left(\ln \tilde{\phi}_{k_2} \right) \right|_{\theta \rightarrow 0} \ .
\end{equation}
At this point we can use the expansion of $\tilde{\phi}_k$ in terms of the Bessel functions (see Eq.~\eqref{appendix_GR:regular_bessel_approx}) to express the action as:
\begin{equation}
  \label{eq_holography:boundary_action_theta}
  \begin{aligned}
  \mathcal{S} \propto \int \ & \mathrm{d}^d k_1 \mathrm{d}^d k_2 \  \delta^{(d)}(k_1^{ \ \mu} + k_2^{ \ \mu})  \ k_2^{\ d} \ \theta^{-d+1} \ \tilde{\phi}_{k_1} \tilde{\phi}_{k_2} \left[ \frac{\Delta_-}{\theta}  + \right.  \\
  &  \left. \left. + 2 D_1 \theta + \dots + 2\alpha D_\alpha \theta^{2\alpha - 1} \ln(\theta)+ D_\alpha \theta^{2\alpha - 1} + o(\theta^{2\alpha - 1}) \right] \right|_{\theta \rightarrow 0} \ ,
   \end{aligned}
\end{equation}
where $D_1, \dots, D_\alpha$ are constant factors (depending on the expansion of the Bessel function) and $\alpha = d/2 - \Delta_- $. It is now crucial to stress that Eq.~\eqref{eq_holography_renorm_correlators} implies that the two point function can be expressed as:
\begin{equation}
\label{eq_holography:two_point_func_defintion}
   \langle \tilde{\mathcal{O}}(k_1) \tilde{\mathcal{O}}(k_2) \rangle =  \left. \frac{\delta^2 \mathcal{S}^{reg}}{\delta \tilde{\phi}_{Reg,k_1} \delta \tilde{\phi}_{Reg,k_2} } \right|_{\tilde{\phi}_{Reg}= 0, \theta \rightarrow 0} \ ,
\end{equation}
and thus to compute this quantity we need to regularize the action~\eqref{eq_holography:boundary_action_theta} by subtracting all the divergent terms\footnote{Actually these terms are not relevant for our analysis because in a QFT computation they correspond to local divergent terms that can be reabsorbed by local counterterms.}. Notice that Eq.~\eqref{eq_holography:two_point_func_defintion} implies that to perform this calculation we only need terms proportional to $ \tilde{\mathcal{A}}(k_1) \tilde{\mathcal{A}}(k_2)$. As a first step, we have to express $\tilde{\phi}_k$ in terms of $\tilde{\phi}_{Reg,k}$. In particular, the leading contribution to $ \tilde{\phi}_k $ in the limit of $\theta \rightarrow 0$ can be expressed (see Eq.~\eqref{appendix_GR:tilde_phi_expansion_theta}) as:
\begin{equation}
\label{eq_holography:asymptotic_phi_exp}
    \left. \tilde{\phi}_k \right|_{\theta \rightarrow 0} \simeq k^{- \Delta_-} \theta^{\Delta_-} \ \tilde{\mathcal{A}}(k) \ .
\end{equation}
We can thus substitute into Eq.~\eqref{eq_holography:boundary_action_theta} to get:
\begin{equation}
  \label{eq_holography:boundary_action_theta_2}
  \begin{aligned}
  \mathcal{S} \propto \int \ & \mathrm{d}^d k_1 \mathrm{d}^d k_2 \  \delta^{(d)}(k_1^{ \ \mu} + k_2^{ \ \mu})  \ k_2^{ \ d -2 \Delta_- }  \ \tilde{\mathcal{A}}(k_1) \tilde{\mathcal{A}}(k_2) \left[ \Delta_- \theta^{-2 \alpha} + \right .  \\
  & + \left. \left. 2 D_1 \theta^{2- 2 \alpha} + \dots + 2\alpha D_\alpha  \ln(\theta)+ D_\alpha + \mathcal{O}(\theta) \right] \right|_{\theta \rightarrow 0} \ .
   \end{aligned}
\end{equation}
To define the regularized action, we should subtract all terms that diverge for $\theta \rightarrow 0$. Once this operation is performed, the part of the regularized action that we need for this computation simply reads:
\begin{equation}
  \label{eq_holography:boundary_reg_action}
  \mathcal{S}^{reg} \propto \int \  \mathrm{d}^d k_1 \mathrm{d}^d k_2 \  \delta^{(d)}(k_1^{ \ \mu} + k_2^{ \ \mu})  \ k_2^{ \ 2 \alpha }  \ \tilde{\mathcal{A}}(k_1) \tilde{\mathcal{A}}(k_2) \left[ D_\alpha + \mathcal{O}(\theta) \right]_{\theta \rightarrow 0} \ ,
\end{equation}
where we have also used $d - 2 \Delta_-= 2 \alpha$. Notice that this quantity is well defined for $\theta \rightarrow 0$. Once we have neglected the subleading terms, we can take two functional derivatives of Eq.~\eqref{eq_holography:boundary_reg_action} in order to get:
\begin{equation}
  \label{eq_holography:functional_derivative_action}
 \frac{\delta^2 \mathcal{S}^{reg}}{\delta \tilde{\phi}_{Reg,k_1} \delta \tilde{\phi}_{Reg,k_2} } = \frac{\delta^2 \mathcal{S}^{reg}}{\delta \tilde{\mathcal{A}}(k_1) \delta \tilde{\mathcal{A}}(k_1) }  =  \frac{R_A^{\ d-1}}{2} \delta^{(d)}(k_1^{ \ \mu} + k_2^{ \ \mu}) \ k_2^{ \ 2 \alpha }  D_\alpha \ .
\end{equation}
Substituting into Eq.~\eqref{eq_holography:two_point_func_defintion}, the two point function reads:
\begin{equation}
   \langle \tilde{\mathcal{O}}(k_1) \tilde{\mathcal{O}}(k_2) \rangle \propto \delta^{(d)}(k_1^{ \ \mu} + k_2^{ \ \mu}) k^{2\alpha } \ ,
\end{equation}
where we have dropped all the constant multiplying factors. To get an explicit expression for $\langle \mathcal{O}(x_1) \mathcal{O}(x_2) \rangle$, we can then express $\mathcal{O}(x_1)$ and $ \mathcal{O}(x_2)$ in terms of their Fourier transforms and use $2 \alpha = 2 \Delta_+ - d$: 
\begin{equation}
\label{eq_holography:holographic_two_point_function}
\begin{aligned}
  \langle \mathcal{O}(x_1) \mathcal{O}(x_2) \rangle \propto & \int \frac{\textrm{d}^d k}{(2\pi)^d} e^{- i k ( x_1 - x_2)} \ k^{2\Delta_+ - d }  \propto \frac{1}{\left| x_1 - x_2 \right|^{2 \Delta_+}} \ .
  \end{aligned}
\end{equation}
As discussed in Appendix~\ref{appendix_CFT:Conformal_field_theories}, the two point function for CFT operators can directly be obtained by using the symmetries of the theory. It should be clear that the result obtained with the holographic analysis matches with one obtained with the CFT calculation. In particular the matching between Eq.~\eqref{eq_holography:holographic_two_point_function}, and Eq.~\eqref{appendix_CFT:CFT_two_point_function} is made manifest by imposing $\Delta_1 = \Delta_2 = \Delta_+$. Notice that this result is consistent with the discussion of comment~\ref{comments_AdS/CFT_3} of the previous Section (in particular see Eq.~\eqref{eq_holography_O_Scaling}) \textit{i.e.} the scaling dimension of the $\mathcal{O}(x_1) $ dual to $\phi$ (with scaling dimension $\Delta_-$) is equal to $\Delta_+$.\\

\noindent
It is interesting to notice that using Eq.~\eqref{eq_holography_renorm_correlators} the one-point function for the dual operator $\tilde{\mathcal{O}}$ can be expressed as:
\begin{equation}
\label{eq_holography:one_point_func_def}
  \langle \tilde{\mathcal{O}}(k) \rangle = \left. \frac{\delta \mathcal{S}^{reg}_{AdS_{d+1}}}{\delta \tilde{\phi}_{Reg,k}} \right|_{\tilde{\phi}_{Reg}=0,\theta\rightarrow 0} .
\end{equation}
Let us compute the explicit expression for this quantity. We start by using the action of Eq.~\eqref{eq_holography:scalar_action_fourier_3} to compute $\tilde{\Pi}_{k}$, canonical momentum conjugated to $\tilde{\phi}_{k}$:
\begin{equation}
  \tilde{\Pi}_{k} \equiv \frac{\partial \mathcal{L}}{\partial ( \partial_u \tilde{\phi}_{k} )}  = \frac{1}{2} \left. e^{du/R_A} \ \tilde{\phi}_{-k} \right|_{u \rightarrow \infty} .
\end{equation}
We can thus use Eq.~\eqref{appendix_GR:tilde_phi_expansion} to get:
\begin{equation}
\label{eq_holography:tilde_pi_expansion}
  \tilde{\Pi}_k(u)  \simeq ( k R_A )^{\Delta_+} \tilde{\mathcal{A}}(-k) e^{u \Delta_+/R_A} + ( k R_A )^{\Delta_-}\tilde{\mathcal{B}}(-k)e^{ u\Delta_-/R_A}
\end{equation}
We can thus proceed by expressing the action~\eqref{eq_holography:scalar_action_fourier_3} in terms of $\tilde{\Pi}_k(u)$:
\begin{equation}
\label{eq_holography:scalar_action_momentum}
    \mathcal{S} \propto  \int \left. \  \mathrm{d}^d k_1 \mathrm{d}^d k_2 \ \delta^{(d)}(k_1^{ \ \mu} + k_2^{ \ \mu}) \ \tilde{\Pi}_{-k_1} \dot{\tilde{\phi}}_{k_2} \right|_{u \rightarrow \infty} \ .
  \end{equation}
Following the procedure carried out to compute the two point function we define the action in terms of $\theta = R_A k_2 e^{-u/R_A}$. We can then substitute the expression for $\dot{\tilde{\phi}}_{k_2}$ obtained by using the expansion of $\tilde{\phi}_{k_2}$ in terms of the Bessel functions given in Eq.~\eqref{appendix_GR:regular_bessel_approx}:
\begin{equation}
  \label{eq_holography:boundary_action_pi}
  \begin{aligned}
  \mathcal{S}  \propto  \int \  \mathrm{d}^d k_1 \mathrm{d}^d k_2 & \delta^{(d)}(k_1^{ \ \mu} + k_2^{ \ \mu}) \  \tilde{\Pi}_{-k_1} \tilde{\phi}_{k_2} \left[ \Delta_- + \right. \\
  &  \left. \left. + 2 D_1 \theta^{2} + \dots + 2\alpha D_\alpha \theta^{2\alpha} \ln(\theta)+ D_\alpha \theta^{2\alpha } + \mathcal{O}(\theta^{2\alpha}) \right] \right|_{\theta \rightarrow 0} \ .
   \end{aligned}
\end{equation}
As $\tilde{\phi}_{Reg} = \mathcal{A}(k)$, it should be clear from Eq.~\eqref{eq_holography:one_point_func_def} that the only terms that are relevant for this computation are terms linear in $\mathcal{A}(k)$. By using Eq.~\eqref{appendix_GR:tilde_phi_expansion} and Eq.~\eqref{eq_holography:tilde_pi_expansion}, we can then express the relevant part of the action as:
\begin{equation}
\label{eq_holography:scalar_action_momentum_2}
\begin{aligned}
    \mathcal{S}^{reg} \propto  \int \ \mathrm{d}^d k_1 \mathrm{d}^d k_2 & \ \delta^{(d)}\left(k_1^{ \ \mu} + k_2^{ \ \mu}\right) \times \\
    & \times \left[( k R_A )^{2\Delta_+} \tilde{\mathcal{A}}(-k_1)\tilde{\mathcal{B}}(k_2) + ( k R_A )^{2\Delta_-}\tilde{\mathcal{A}}(k_2)\tilde{\mathcal{B}}(-k_1) \right] \ ,
\end{aligned}
  \end{equation}
where we have already dropped all the divergent and subleading terms. We can finally take the functional derivative of this action with respect to $\tilde{\mathcal{A}}(k_2)$ and use Eq.~\eqref{eq_holography:one_point_func_def} to express the one-point function as:
\begin{equation}
\label{eq_holography:one_point_func}
  \langle \tilde{\mathcal{O}}(k) \rangle \propto \tilde{\mathcal{B}}(k_2) .
\end{equation}
This implies that the one-point function of the operator $\tilde{\mathcal{O}}$ is directly given by the regular part of the subdominant term in $\tilde{\phi}(u)$. As already explained in this Section, the term proportional to $\tilde{\mathcal{B}}(k_2)$ diverges like $e^{-u\Delta_+/R_{A}}$ in the interior ($u \rightarrow - \infty$) of AdS${}_{d+1}$. In order to get a regular solution, we thus have to set this term to zero. It should be clear that this is consistent with the CFT computation that actually gives $\langle \tilde{\mathcal{O}}(k) \rangle = \langle \mathcal{O}(x) \rangle = 0$. \\

\noindent
Before concluding this Section it is worth mentioning that different methods could have been used to define the regularized field and the regularized action. A method commonly used in the literature~\cite{Aharony:1999ti,Zaffaroni:2000vh} is based on the definition of a cut-off $\epsilon >0 $ so that $\epsilon \leq w$, the computations are then carried out by imposing the boundary condition:
\begin{equation}
 \lim_{w \rightarrow \epsilon } \tilde{\phi}(w, k^\mu) = \tilde{\phi}_0(k^\mu) = \frac{(k w)^{d/2} K_\alpha(wk)}{(k \epsilon )^{d/2} K_\alpha(k \epsilon )} \ ,
\end{equation}
and finally the cut-off is sent to zero.

\subsection{Non conformal theories and RG flows.}
\label{sec_holography:RG_flows}
Let us conclude this Section by discussing the possibility of extending AdS/CFT to cases where we do not have exact conformal symmetry on the QFT side. Of course this corresponds to a non exact AdS geometry on the gravity side. Let us start our treatment by considering the action for a dimensionless scalar field $\Phi$ in $d+1$ dimensions:
 \begin{equation}
  \label{eq_holography:action_RG_flow}
    \mathcal{S} = \mathcal{S}_g + \mathcal{S}_m =\bar{\kappa}^{-d+1}\int\mathrm{d}u\mathrm{d}^d x \sqrt{|g|}\left( \frac{R}{2} - \frac{g^{ab}}{2} \partial_a \Phi \partial_a \Phi \ + \bar{\kappa}^{d-1} V(\Phi) \right).
  \end{equation} 
Notice that for $d = 3$ this action is directly obtained by the action of Eq.~\eqref{eq_holography:defining_action} imposing $\eta = 1$ and $\kappa^2 = - \bar{\kappa}^2 $, \textit{i.e.} considering the domain-wall side of the domain-wall/cosmology correspondence. The metric can thus be expressed as:
\begin{equation}
  \textrm{d} s^2 = du^2 + e^{2 A(u)} \delta_{\mu \nu}\textrm{d}x^\mu \textrm{d} x^\nu \ ,
\end{equation}
where $A(u)$ is a generic function of $r$. Notice that fixing $A(u) = u/R_{A}$ we can once again recover the case of AdS${}_{d+1}$. As usual, the dynamics of the system is described by Einstein equations: 
\begin{equation}
\label{eq_holography:Einstein_eq_RG}
    G_{a b} \equiv R_{a b} - \frac{1}{2} g_{a b} R = \partial_a \Phi \partial_a \Phi - g_{ab} \left[  \frac{g^{ab}}{2} \partial_a \Phi \partial_a \Phi \ - \bar{\kappa}^{d-1} V(\Phi) \right] =  \frac{2}{\sqrt{|g|}} \frac{\delta \mathcal{S}_m}{\delta g^{ab}} \equiv T_{a b}. 
 \end{equation} 
Assuming the field $\Phi$ to be homogeneous (\textit{i.e.} it only depends on $u$), we get:
\begin{equation}
\label{eq_holography:Einstein_RG_flow}
  \frac{d(d-1)}{2}  \dot{A}^{2} =  \frac{\dot{\Phi}^{2}}{2}  + \bar{\kappa}^{d-1} V(\Phi) , \qquad \qquad   - \frac{d-1}{2}\left[2 \ddot{A} +d  \dot{A}^{2} \right] =  \frac{\dot{\Phi}^{2}}{2}  - \bar{\kappa}^{d-1} V(\Phi) ,
\end{equation}
where dots are used to denote differentiation with respect to $u$. We can then take the sum and the difference of these two equations to get:
\begin{equation}
\label{eq_holography:Einstein_RG_flow_2}
  \dot{\Phi}^2 = -(d-1) \ddot{A} \ , \qquad \qquad  2 \bar{\kappa}^{d-1} V = (d-1) \left[ \ddot{A} +d  \dot{A}^{2} \right] .
\end{equation}
It is also useful to compute the equation of motion for the scalar field:
\begin{equation}
  \label{eq_holography:eom_scalar}
  \ddot{\Phi} + d \dot{A} \dot{\Phi} + \bar{\kappa}^{d-1}  \frac{\partial V}{\partial \Phi} = 0 \
\end{equation}
As already anticipated, the AdS${}_{d+1}$ solution is recovered by fixing $A(u) = u/R_{A}$ that directly gives:
\begin{equation}
\label{eq_holography:recover_AdS}
 \bar{\kappa}^{d-1} V(\Phi) = \frac{d(d-1)}{2 R_{A}^{ \ 2}}  , \qquad  \dot{\Phi} = 0  .
\end{equation}
It should be clear that this is the only configuration that solves the system~\eqref{eq_holography:Einstein_RG_flow_2} assuming $A(u) = u/R_{A}$. Moreover, it should also be clear that this actually corresponds to the case of a static empty universe with a negative cosmological constant described in Appendix~\ref{appendix_GR:AdS_spacetime}. In the general case we should consider the variations of the AdS${}_{d+1}$ geometry due to the field's back-reaction. \\

\noindent
As we are interested in solutions that are close to the AdS${}_{d+1}$ case, we start by assuming that the potential $V(\Phi)$ has an extremum at $\Phi = \Phi_*$ and we assume that its expansion in the neighborhood of this extremum reads:
\begin{equation}
\label{eq_holography:potential_expansion}
 \bar{\kappa}^{d-1} V(\Phi_*) =  \frac{d(d-1)}{2 R_{A}^{ \ 2}} + \frac{m^2}{2} (\Phi-\Phi_*)^2 + \dots \ .
\end{equation}
Without loss of generality, we proceed with our treatment assuming $\Phi_* = 0$. Let us consider the ansatz:
\begin{equation}
\label{eq_holography:field_scale_expansion}
  \Phi = 0 + \delta\Phi + \dots \ , \qquad A(u) = \frac{u}{ R_{A}} + \delta A \ .
\end{equation}
Substituting the expression for the potential of Eq.~\eqref{eq_holography:potential_expansion} and the ansatz of Eq.~\eqref{eq_holography:field_scale_expansion} into Eq.~\eqref{eq_holography:Einstein_RG_flow_2} and Eq.~\eqref{eq_holography:eom_scalar}, we obtain the system: 
\begin{eqnarray}
\label{eq_holography:Einstein_perturbative_1}
 && \ddot{\delta \Phi} + d\left (\frac{1}{R_{A}} +\dot{\delta A}\right) \dot{\delta \Phi} - m^2 \delta \Phi = 0 , \\
\label{eq_holography:Einstein_perturbative_2}
 && \ddot{\delta A} = \dot{\delta \Phi}^2 \ , \\
  \label{eq_holography:eom_perturbative}
&& (d-1)\left[\ddot{\delta A} + d \left(\frac{2}{R_{A}}\dot{\delta A} + \dot{\delta A}^2 \right) \right] =  m^2 \delta \Phi^2  \ .
\end{eqnarray}
This system can then be solved perturbatively for $\delta \Phi$ and $\delta A$. Neglecting higher order corrections in $\delta \Phi$ and $\delta A$, the solution for this system is:
\begin{equation}
\begin{aligned}
  \delta \Phi &= C_1 e^{-u \Delta_{-} /R_{A}} + C_2 e^{-u \Delta_{+} /R_{A}} \ ,\\
   \delta A &= - \frac{1}{d-1} \left[ \frac{C_1^2}{4} e^{-2u\Delta_{-}/R_{A}} +\frac{C_2^2}{4} e^{-2u\Delta_{+}/R_{A}} +\frac{2 C_1C_2 \Delta_{+} \Delta_{-} }{d^2} e^{-u d/R_{A}} \right]\ , 
\end{aligned}
\end{equation}
where $C_1$ and $C_2$ are constant factors and $\Delta_{\pm}$ are the usual $\Delta_\pm$ defined in Eq.~\eqref{eq_holography:Delta_pm}. It should be clear that $C_1$ and $C_2$ correspond to the regular parts of the scalar field in AdS${}_{d+1}$. \\

\noindent
As discussed in Appendix~\ref{appendix_GR:AdS_dynamics}, if we restore the spatial dependence in $\Phi$,  $C_1$ and $C_2$ must be replaced by some functions of $x^\mu$, say $C_1 f(x)$ and $C_2 g(x)$. In particular these two functions must be related to the modified Bessel functions. Comparing with Eq.~\eqref{appendix_GR:regular_bessel_approx}, it should be clear that the Fourier transform of these two quantities must satisfy $\tilde{f}(k) = \tilde{A}(k)$, $\tilde{g}(k) = \tilde{B}(k)$. As explained in the previous sections, $\tilde{A}(k)$ is associated with the source for the dual operator $\mathcal{O}$ and $\tilde{B}(k)$ is associated with the 1-point function of $\mathcal{O}$. Solutions proportional to $\exp\{-u\Delta_{+}/R_{A} \}$ are thus introducing a non-zero vacuum expectation value for the operator $\mathcal{O}$ (proportional to $\tilde{B}(k)$). On the other hand, solutions proportional to $\exp\{-u\Delta_{-}/R_{A} \}$ correspond to deformations of the original CFT due to the introduction of a term proportional to $A(k)\mathcal{O}$ in the action of the theory.\\

\noindent Assuming the solution to be regular in the interior of AdS${}_{d+1}$ we should set $C_2 = 0$ (as usual $2\Delta_{-} \leq d \leq 2\Delta_{+}$). Under this assumption the expression for $ \Phi$ and $ A$ read:
\begin{equation}
\label{eq_holography:asymptotic_flow}
   \Phi \simeq C_1 e^{-u \Delta_{-} /R_{A}} \ , \qquad  
    A \simeq \frac{u}{R_A} -   \frac{C_1^2}{4(d-1)} e^{-2u\Delta_{-}/R_{A}} \ . 
\end{equation}
Notice that assuming $\Delta_- < 0$, the fixed point of the potential is reached for $u \rightarrow -\infty$ (\textit{i.e.} for $ \Phi \rightarrow 0$ ). Conversely, for $u \rightarrow \infty$ the ratio $ R_A \delta A /u$ grows exponentially and thus in this regime the perturbative solution is no longer valid. It should be clear that for $\Delta_->0$ the behavior of the solution is reversed. The fixed point of the potential is thus approached for $u \rightarrow \infty$ and the validity of the perturbative expansion is broken for $u \rightarrow -\infty$. \\

\noindent
The deformation of the dual CFT and the corresponding breaking of conformal invariance is associated with a breaking of the scale invariance that induces a running for the parameters of the theory. In QFT this process is usually described in terms of RG equations. An RG flow on the QFT side thus corresponds to the presence of a field that modifies the AdS${}_{d+1}$ geometry on the gravity side. It is interesting to notice that in this context it is natural to identify the energy scale on the CFT side with the coordinate $u$. In this framework the region close to boundary (\textit{i.e.} $u \rightarrow \infty$) of AdS${}_{d+1}$ corresponds to the UV region on the QFT side. On the contrary the interior of the AdS${}_{d+1}$ spacetime (\textit{i.e.} $u \rightarrow -\infty$), corresponds to the IR region. As already discussed in the previous sections, given the mass $m^2$ of the scalar field, the scaling dimension of the dual operator $\mathcal{O}$ is equal to $\Delta_+$. Remembering that AdS${}_{d+1}$ may support a negative mass squared until the BF condition $- d^2/(2R_A)^2 \leq m^2$ is satisfied and using $d - \Delta_+ = \Delta_-$, we identify three different kind of deformations for the original CFT:
\begin{itemize}
  \item Negative mass: $d/2 \leq \Delta_+ < d$, $ 0  < \Delta_- \leq d/2 $. The deformation is called \textbf{relevant}. The fixed point is reached in the UV and the perturbations grow exponentially towards the IR.
  \item Positive mass: $d < \Delta_+ $, $ \Delta_- < 0 $. The deformation is called \textbf{irrelevant}. The fixed point is reached in the IR and the perturbations grow exponentially towards the UV.
  \item Zero mass: $ \Delta_+ = d$, $ \Delta_- = 0$. The deformation is called \textbf{marginal} and higher order terms are required in order  to understand the asymptotic behavior.
\end{itemize}
These definitions actually come from QFT and statistical mechanics, where these techniques are normally used to study the RG flows from the UV towards the IR. Notice that in the case of a negative mass the deformation is not affecting the theory in UV but it is affecting it in the IR. In this sense we should regard the deformation as relevant. On the contrary, in the case of irrelevant deformations, we have a positive mass squared term and the situation is reversed \textit{i.e.} the UV is strongly affected and the IR is not affected. In the case of marginal deformations higher order correction are necessary in order to understand the asymptotic behavior of the theory.

 \section{Holographic inflation.}
 \label{sec_holography:Holographic_inflation}
 In this Section we discuss the possibility of describing the inflating universe by means of a RG equation in the framework of AdS/CFT correspondence. As already argued in the previous sections of this chapter, the Domain-wall/Cosmology correspondence sets an equivalence between the inflating universe and the dynamics of a scalar field in a nearly (Euclidean) AdS spacetime. Once this mapping is realized, it seems reasonable to use techniques coming from AdS/CFT to get an alternative description of the system. In particular, the modification of the AdS${}_{d+1}$ geometry due to the presence of the scalar field is interpreted as an RG flow on the QFT side of the correspondence. In this framework it is thus natural to describe the inflating universe using an RG equation, giving strong theoretical supports to the introduction of the $\beta$-function formalism for inflation discussed in Chapter~\ref{chapter:beta}. \\

\noindent
An holographic description of inflation offers several advantages. For example it is worth mentioning that:
\begin{itemize}
  \item The choice of the initial conditions conditions for inflation can be interpreted from a different point of view. 
  \item The inflationary potential can be discussed in terms of the dual QFT operators, leading to a classification in terms of relevant, irrelevant and marginal operators.
  \item The gauge side of the duality may give hints for the definition of a UV complete theory for inflation.
  \item As we explicitly show in Sec.~\ref{sec_holography:Holographic_strong}, in this framework it is possible to define a new class of models where gravity is strongly coupled. While these models cannot be described using standard techniques, it is possible to define a consistent analysis in terms of the dual theories.
\end{itemize}

\noindent
We start this Section by expressing the holographic RG flow in terms of the superpotential and of the $\beta$-function introduced in Sec.\ref{chapter:beta}. Once the dynamics is expressed in terms of this formalism, we show the correspondence between power spectra in cosmology and correlators in QFT. Finally we discuss the properties of the dual operators associated with the universality classes introduced in~\cite{Binetruy:2014zya}.

\subsection{Superpotential formalism and holographic RG flow.}
As discussed in Chapter~\ref{chapter:inflation} and in Chapter~\ref{chapter:beta}, inflation may be realized by considering homogeneous dimensionful scalar field $\phi$ in a FLRW background. Assuming that $\phi$ has a canonical kinetic term, and that is minimally coupled with gravity, which is described by a standard Einstein-Hilbert term its action reads:
\begin{equation}
  \label{eq_holography:action}
    \mathcal{S}= \int\mathrm{d}t\mathrm{d}^3x \sqrt{|g|}\left( \frac{R}{2 \kappa^2} -  X - V(\phi) \right).
  \end{equation}
As explained at the beginning of this Chapter, after some simple algebraic manipulations, this action can be mapped into the action Eq.~\eqref{eq_holography:defining_action}. In particular, this can be realized by defining the a new `time' coordinate $r$, a dimensionless field $\Phi$, and by expressing the metric as:
\begin{equation}
\label{eq_holography:metric}
  \textrm{d} s^2 = \eta dr^2 + e^{2 A(r)} \delta_{\mu \nu}\textrm{d}x^\mu \textrm{d} x^\nu \ ,
  \end{equation}
where for $\eta = -1$ we have $r=t$ and for $\eta = 1$ we have $r = u$. It should be clear that after these manipulations, we can follow the discussion of Sec.~\ref{sec_holography:AdS/dS_mapping} and define an equivalent description of the system in terms of a domain-wall solution. With this procedure, it is then possible to apply the discussion of Sec.~\ref{sec_holography:RG_flows} (with $d=3$). However, in a first instance, we proceed by describing the system in terms of the dimensionful field $\phi$. This prescription is chosen in order to reproduce the equations obtained in Chapter~\ref{chapter:beta}. Once the $\beta$-function formalism is recovered, we finally express the system in terms of $\Phi$, and we interpret our results in light of the discussion of Sec.~\ref{sec_holography:RG_flows}.\\

\noindent Given the action of Eq.~\eqref{eq_holography:action} expressed in terms of $r$, and using the parametrization of Eq.~\eqref{eq_holography:metric} for the metric, the equations of motion for our system are:
\begin{eqnarray}
\label{eq_holography:RG_flow_beta_1}
 (d-1)A_{,rr} & = &  - \kappa^2 \phi_{,r}^2 \  , \\ 
  \label{eq_holography:RG_flow_beta_2}
   (d-1) \left[ A_{,rr} +d  A_{,r}^{\ 2} \right]  & = & - 2 \eta \kappa^2 V  \ , \\
  \label{eq_holography:eom_scalar_beta}
    \phi_{,rr} + d A_{,r} \phi_{,r} - \eta \frac{\partial V}{\partial \phi} &=& 0 \ ,
\end{eqnarray}
where as usual ${}_{,r}$ denotes differentiation with respect to $r$. We proceed by introducing the superpotential $W(\phi)$, as a solution of the non-linear equation:
\begin{equation}
\label{eq_holography:superpotential_beta}
  - 2 \eta \kappa^2 V(\phi) =  \left[ \frac{d}{d-1} W^{\ 2}(\phi)  - \frac{W^2_{,\phi}(\phi)}{\kappa^2} \right] \ ,
\end{equation}
so that dynamics is completely specified by:
\begin{equation}
A_{,r} = -\frac{W(\phi)}{(d-1) } \ , \qquad  \qquad \qquad  \phi_{,r} =  \frac{W_{,\phi}(\phi)}{\kappa^2} \ .
\end{equation}
In analogy with the treatment of Chapter~\ref{chapter:beta}, we can thus introduce the holographic $\beta$-function as:
\begin{equation}
  \beta(\phi) \equiv \kappa \frac{\textrm{d} \phi}{\textrm{d}\ln a} = \kappa \frac{\textrm{d} \phi}{\textrm{d}A} = \kappa \frac{\phi_{,r}} {A_{,r} } =  - \frac{d-1}{\kappa} \frac{W_{,\phi}}{W}  .
\end{equation}
Finally we describe the system in terms of the dimensionless scalar field $\Phi$. After the field redefinition the holographic $\beta$-function reads:
\begin{equation}
\label{eq_holography:beta_dimensionless}
  \bar{\beta}(\Phi) \equiv  \beta(\phi(\Phi)) = \kappa \frac{\textrm{d} \phi}{\textrm{d}\Phi} \frac{\textrm{d} \Phi}{\textrm{d}\ln a} = \frac{\textrm{d} \Phi}{\textrm{d}\ln a} = -( d-1) \frac{W_{,\Phi}}{W} \ .
\end{equation}
As explained in Chapter~\ref{chapter:beta}, an exact dS geometry is realized at $\beta(\Phi) = 0$, and the departure from this configuration actually corresponds to a phase of inflation. Let us describe this system in terms of his domain-wall realization. Let us assume that the potential can be expanded as in Eq.~\eqref{eq_holography:potential_expansion} and let us assume that the field $\Phi$ is regular in the interior of AdS${}_{d+1}$. Under this assumptions $\Phi$ and $A$ can thus be expressed as in Eq.~\eqref{eq_holography:asymptotic_flow}. Substituting into Eq.~\eqref{eq_holography:beta_dimensionless} we get:
\begin{equation}
\label{eq_holography:holographic_beta}
\bar{\beta}(\Phi) = \frac{\Phi_{,r}} {A_{,r} } = - \frac{\Delta_- C_1 \exp\left(-u \Delta_- /R_A\right) }{1 + \frac{C_1^2 \Delta_-}{2(d-1)}\exp\left(-2 u \Delta_- /R_A\right)}  = - \frac{\Delta_- \Phi}{1 +\frac{\Delta_-}{2(d-1)} \Phi^2} .
\end{equation} 
As explained in Sec.~\eqref{sec_holography:RG_flows}, the regime of validity for the perturbative solution depends on the sign of $\Delta_-$. If the deformation is irrelevant, \textit{i.e.} $\Delta_-<0$, the perturbative solution is valid in the IR and in this regime the $\beta$-function reads $\bar{\beta}(\Phi) \simeq - \Delta_- \Phi $. Notice that this corresponds to linear class introduced in Chapter~\ref{chapter:beta}. Consistently, approaching the fixed point expansion of the potential~\eqref{eq_holography:potential_expansion} matches with Eq.~\eqref{eq_beta:potential_linear}. Conversely, for $\Delta_->0$ the deformation is relevant and the perturbative solution only holds in the UV. Again the asymptotic expression is given by $\bar{\beta}(\Phi) \simeq - \Delta_- \Phi $. \\

\noindent Consistently with our expectations, depending on the sign of $\Delta_-$, a fixed point of the $\beta$-function is obtained either approaching the horizon ($u \rightarrow -\infty$) or the boundary ($u \rightarrow \infty$). As extensively explained in Chapter~\ref{chapter:beta}, approaching a zero of the $\beta$-function, corresponds to a spacetime geometry that approaches an (A)dS configuration. Correspondingly, the dual QFT approaches conformal invariance. The departure from a nearly $(A)dS_{3+1}$ geometry (that is typical of inflation) is thus translated into a departure from a scale-invariant regime for the dual QFT. As usual this process is described by means of a RG equation and thus the appearance of a $\beta$ function to describe the inflating universe is not fortuitous. \\

\noindent Notice that in this picture inflation starts at $u \rightarrow -\infty$ that corresponds to the interior of $(A)dS_{3+1}$,  and thus to the low energy (or IR for infrared) limit for the dual QFT.  During inflation the scale factor grows and tends to become infinitely large in the limit $u \rightarrow \infty$. This is actually reached when we approach the boundary of $(A)dS_{3+1}$, that corresponds to the high energy (or UV for ultraviolet) limit for the dual QFT. In this picture an inflating universe is thus corresponding to an (inverse) RG flow from the IR towards the UV. Notice that a conventional RG flow from the UV towards the IR corresponds to a shrinking universe. \\

\noindent As already discussed in Sec.\eqref{sec_holography:RG_flows}, in the language of the dual QFT, this RG flow is induced by an operator $\mathcal{O}(x)$, dual to the scalar field $\Phi$, that is characterized by its scaling dimension $\Delta_+$. A classification of the operators $\mathcal{O}(x)$ in terms of their scaling dimensions, is thus equivalent to a classification of the dual inflationary models. In particular, starting from the action of Eq.~\eqref{eq_holography:action}, we can define the dimensionless field $\Phi$ that is described by the action of Eq.~\eqref{eq_holography:defining_action}. Given an inflationary potential $V(\Phi)$, with a fixed point at $\Phi = \Phi_*$, we can define the mass $m^2$ of the inflaton as:
 \begin{equation}
m^2 \equiv \left. \kappa^2 \frac{\textrm{d}^2 V(\Phi)}{\textrm{d}\Phi^2}\right|_{\Phi =  \Phi_*} \ .
\end{equation}
Notice that a positive mass for the inflaton in cosmology corresponds to a negative mass squared for the corresponding domain-wall solution. Finally we can use Eq.~\eqref{eq_holography:Delta_pm}, to compute the scaling dimension of $\mathcal{O}(x)$.

\subsection{Identifying the dual theories.}
As explained in the previous section, the scaling dimension of the dual operator is associated with the mass of the inflaton. To compute this quantity we should thus specify the inflationary potential, whose parametrization in terms of the superpotential is given by Eq.~\eqref{eq_holography:superpotential_beta}. Using the definition of the holographic $\beta$-function, given in Eq.~\eqref{eq_holography:beta_dimensionless}, it is thus trivial to get:
\begin{equation}
  \kappa^2 V(\Phi) = -\eta \, \frac{ d}{2 (d-1)} W^2(\Phi) \left[ 1 - \frac{\bar{\beta}^2(\Phi)}{d(d-1)} \right] \ ,
\end{equation}
so that the lowest order expression for the mass of the field $\Phi$ is given by:
\begin{equation}
  m^2 \simeq \left. \eta \, \frac{2 d }{(d-1)^2} W^2(\Phi)\left( - \frac{2 \bar{\beta}^2(\Phi)}{d-1} + \bar{\beta}_{, \Phi} + \frac{\bar{\beta}\bar{\beta}_{, \Phi \Phi}}{d} \right)\right|_{\Phi =  \Phi_*} \ .
\end{equation}
Finally we can thus use this expression to identify the dual theories corresponding to the universality classes defined in Sec.~\ref{sec_beta:universality_classes}. We can start by considering the Linear class \textbf{Ia(1)}, that actually is special. As in this case the $\beta$-function is linear in $\Phi$, its first derivative is a constant. For the models of this class, the mass of the dual operator is thus given by:
\begin{equation}
\label{sec_holography:dual_mass_linear}
  m^2 \simeq \frac{2 d }{(d-1)^2} W_{*}^{\, 2} \bar{\beta}_{1} \ > 0,
\end{equation}
where we have defined $W_{*} \equiv W(\Phi_*)$. As in the Linear class $\Phi_* = 0$, we can use Eq.~\eqref{eq_beta:superpotential_linear} to get:
\begin{equation}
  W_{*} = W_{\textrm{f}} \exp \left[ \frac{\beta_1}{ 4} (\kappa\phi_{\textrm{f}})^{2} \right] \ . 
\end{equation}
Eq.~\eqref{sec_holography:dual_mass_linear} implies that for the models of the linear class, the mass of the dual operator is positive. As a consequence, using the classification of Sec.~\ref{sec_holography:RG_flows}, we can conclude that the corresponding dual operator is irrelevant. On the contrary, it is easy to show that for all the other classes defined in Sec.~\ref{sec_beta:universality_classes}, we always get $m^2 = 0$. We can thus conclude that the corresponding dual operators are marginal. \\

\noindent 
Before concluding this Section it is worth mentioning that a detailed analysis of the holographic interpretation of the Exponential class \textbf{II} of Sec.~\ref{sec_beta:universality_classes} has been produced by Kiritsis in~\cite{Kiritsis:2013gia}. In particular, these models are referred as Asymptotically-Flat (Free) Inflationary Models (AFIM), and one of their properties is:
\begin{equation}
  \label{eq_holography:zero_couplings}
  \left. \frac{\textrm{d}^n V(\Phi)}{\textrm{d}\Phi^n}\right|_{\Phi =  \Phi_*} = 0 \ , \qquad \qquad \forall n \in \mathbb{N} \ , 
\end{equation}
that actually is the condition that defines the Asymptotic-Flatness. Actually this condition is not only valid for the Exponential class \textbf{II} of Sec.~\ref{sec_beta:universality_classes} but also for all the other large field models discussed in Sec.~\ref{sec_beta:universality_classes}. However, we should point out that only in the case of the Exponential class \textbf{II} the derivatives of the potential are \emph{exponentially suppressed}. For more details on the holographic interpretations of the universality classes of Sec.~\ref{sec_beta:universality_classes} see~\cite{Bourdier:2013axa} and~\cite{Binetruy:2014zya}.

 \section{Holographic spectra and strongly coupled theories.}
 \label{sec_holography:Holographic_strong}
In this Section, we focus our analysis on the discussion of the QFTs that are dual to the cosmological/domain-wall solutions discussed so far. The first step is to recognize the operators that are dual to the comoving curvature perturbation $\zeta$ and to the metric perturbation $\gamma_{ij}$. Once these operator are identified, we compute the corresponding two-point function. Finally, we discuss the possibility of considering theories where gravity is strongly coupled. \\

\noindent
From now on, we restrict to the case of $d = 3$, \textit{i.e.} to the case of standard cosmology in four dimensions which corresponds to a three-dimensional QFT. As a first step we should start by identifying the dual operators for $\zeta$ and $\gamma_{ij}$. For this purpose, it is useful to remember that the domain-wall metric is expressed as:
\begin{equation}
  \label{eq_holography:AdS_metric}
    \textrm{d}s^2 = \textrm{d}u^2 + g_{ij}(u,\vec{x})\textrm{d}x^i \textrm{d}x^j \ .
\end{equation}
By definition, the stress-energy tensor is given by the variation of the action with respect to the metric (see Eq.~\eqref{appendix_GR:Stress_energy_tensor}). It is thus natural to interpret $ g_{ij}(u,\vec{x}) $ as the source for the stress-energy tensor $ T_{ij}(u,\vec{x})$ of the three-dimensional dual QFT. Before proceeding with our discussion it is also useful to mention that in general $g_{ij}(u,\vec{x})$ can be expressed as~\cite{deHaro:2000vlm,Papadimitriou:2004ap,Papadimitriou:2004rz,McFadden:2010na,McFadden:2010vh}:
\begin{equation}
  \label{eq_holography:metric_radial_formalism}
    g_{ij}(u,\vec{x})= e^{2u /R_A} \left[ g_{(0)ij}(\vec{x}) + e^{- 2u /R_A} g_{(2)ij}(\vec{x}) + e^{- 3 u/R_A} g_{(3)ij}(\vec{x}) + \dots \right] \ .
\end{equation}
In the limit $u\rightarrow \infty$ (\textit{i.e.} on the boundary of AdS) the leading term is thus given by the first coefficient in the expansion $ \textit{i.e.} g_{(0)ij}(\vec{x})$. Similarly, the coefficient $g_{(3)ij}(\vec{x})$ will give the leading (finite) contribution, to the three dimensional metric in the interior of AdS. \\

\noindent
At this point, it is important to notice that $\gamma_{ij}(u,\vec{x})$ only depends on the transverse traceless part of $g_{ij}(u,\vec{x})$ and conversely $\zeta(u,\vec{x})$ only depends on the trace of $g_{ij}(u,\vec{x})$ (for the definitions of $\zeta$ and $\gamma_{ij}$ see Appendix~\ref{appendix_perturbations:Cosmological_perturbations}). The projections of $\zeta(u,\vec{x})$ and $g_{ij}(u,\vec{x})$ on the three-dimensional slices at constant $u$, thus depend on the trace and on the transverse traceless part of $g_{ij}(u,\vec{x})$ respectively. As a consequence, we can conclude that $\zeta$ plays the role of the source for trace of $T_{ij}$ and $\gamma_{ij}$ plays the role of the source for transverse traceless part of $T_{ij}$.\\
 
\noindent
In order to express the holographic scalar and tensor power spectra, we should thus compute the two-point function for the three dimensional stress-energy tensor $ T_{ij}$. In particular we have to compute the two-point function of $\tilde{T}_{ij}$, Fourier transform of $T_{ij}$. On general grounds, this can be expressed as:
\begin{equation}
  \label{eq_holography:Stress_energy_tensor_general}
  \langle \tilde{T}_{ij}(\vec{\bar{k}}_1) \tilde{T}_{lk}(\vec{\bar{k}}_2)\rangle = \delta^{(3)}(\vec{\bar{k}}_1 + \vec{\bar{k}}_2) \left[ A(\bar{k}_1) \Pi_{ijlk} + B(\bar{k}_1) \pi_{ij} \pi_{lk} \right] \ ,
\end{equation}
where $\Pi_{ijlk}$ is the three-dimensional transverse traceless operator that is expressed in terms of the transverse projector $\pi_{ij}$ as:
\begin{equation}
  \Pi_{ijlk} \equiv \frac{1}{2} \left( \pi_{ik} \pi_{lj} + \pi_{il} \pi_{kj}+ \pi_{ij}\pi_{lk} \right) \ , \qquad \qquad \pi_{ij} \equiv \delta_{ij} - \frac{\bar{k}_i \bar{k}_j}{\bar{k}^2} \ .
\end{equation}
With an accurate analysis\footnote{This analysis is usually performed in terms of the so-called ``radial Hamiltonian formulation''. For more details on this formalism see~\cite{Papadimitriou:2004rz}.} it is possible to identify the different terms that contribute to $T_{ij}$. As a consequence, it is possible to set a direct relation between the response function (defined in Eq.~\eqref{eq_AdS_dS:linear_response_functions}) and the functions $A$ and $B$ appearing in Eq.~\eqref{eq_holography:Stress_energy_tensor_general}. While this analysis goes beyond the scope of this work, we report the final result~\cite{McFadden:2009fg,McFadden:2010na,McFadden:2010vh}:
\begin{equation}
  \label{eq_holography:Holographic_spectra}
    A(\bar{k}) = 4 \bar{E}_{(0)}(\bar{q}) \ , \qquad \qquad  B(\bar{k}) = \frac{1}{4} \bar{\Omega}_{(0)}(\bar{q}) \ ,
\end{equation}
where the subscripts $(0)$ denote that, according to the expansion of $g_{ij}$ shown in Eq.~\eqref{eq_holography:metric_radial_formalism}, these are the contribution to $\bar{E}(\bar{q})$ and to $\bar{\Omega}(\bar{q})$ that are independent of $u$ in the limit $u \rightarrow \infty $. Notice, that in general the complete expressions for $\bar{E}(\bar{q})$ and $\bar{\Omega}(\bar{q})$ are divergent on the boundary and thus they should be regularized before taking the limit $u \rightarrow \infty $.\\

\noindent
Finally, we conclude this Chapter by discussing the possibility of considering inflationary models where gravity is strongly coupled. So far we have considered the case of weakly coupled gravity, where inflation can be described in terms of a scalar field coupled to gravity. Moreover, the theory of cosmological perturbations and its application to compute the inflationary spectra (presented in detail in Appendix~\ref{appendix_perturbations:Cosmological_perturbations}) are based on the assumption that the problem can be treated perturbatively. In this case, the holographic analysis is mostly interesting from a theoretical point of view, as it offers new interpretations of the problem. However, as the standard computations are well-defined, in practice it only reproduces the results obtained with the standard methods, and thus is not adding new details to characterize the system. However, this dramatically changes when we consider models where gravity is strongly coupled. In this case, not only the assumption of perturbativity but also the description of the system in terms of a scalar field coupled to the metric may no longer be valid. As in this case, the standard techniques of cosmological perturbation theory cannot be applied, a different description is required. Remarkably, because of the strong/weak duality, this case can be consistently treated in term of the holographic approach described in this Chapter. In particular, assuming a smooth transition from the phase with strongly coupled gravity to the usual cosmology, the primordial power spectra can be specified by the holographic analysis.\\

\noindent
As explained in Sec.~\ref{sec_holography:AdS_dS_observables}, the primordial power spectra can be expressed (see Eq.~\eqref{eq_holography:power_spectra_cosmo}) in terms of the response function (defined in Eq.~\eqref{eq_AdS_dS:linear_response_functions}). Once these quantities are computed for the dual theory (through Eq.~\eqref{eq_holography:Stress_energy_tensor_general}), we can thus use Eq.~\eqref{eq_holography:Holographic_spectra} to specify the cosmological observables. In principle, in order to carry out this procedure, we should start by specifying a theory of gravity (string theory), we should then define its dual QFT (the `pseudo'-QFT) and finally we can compute the response functions. However, following the approach of~\cite{McFadden:2009fg,McFadden:2010na} it is also possible to reverse the problem. In particular, it is possible to discuss whether a dual QFT may predict observable quantities that are compatible with current observations. \\

\noindent 
An example of dual model is proposed in~\cite{McFadden:2009fg,McFadden:2010na}: a three-dimensional $SU(N)$ Yang-Mills theory with a certain number of gauge field, scalars, conformal scalars and fermions. Once the stress-energy tensor of the theory is determined, the quantities $ A(\bar{k})$ and $ B(\bar{k})$ can be computed. While the details on the predictions depend on the explicit choice for the parameters of the model, an interesting model-independent result is obtained \textit{i.e.}:
\begin{equation}
  \label{eq_holography:running_holographic}
  \alpha_s = \frac{\textrm{d} n_s}{\textrm{d} \ln k} \simeq - (n_s - 1) \ ,
\end{equation}
where $n_s$ and $\alpha_s$ are respectively the scalar-spectral index and the running. As a consequence, these models are extremely different from slow-roll models where, as explained in Chapter~\ref{chapter:inflation}, $\alpha_s$ is expected to be higher order in the slow-roll parameters. Notice that the predictions of this class of models (given in Eq.~\eqref{eq_holography:running_holographic}) are thus in tension with the Planck constraints on the running~\cite{Ade:2015xua,Ade:2015lrj}.

 {\large \par}}
{\large 
\chapter{Some generalizations.}
\label{chapter:generalized_models}

\horrule{0.1pt} \\[0.5cm] 

\begin{abstract} 

\noindent 
In this chapter, we discuss the application of the $\beta$-function formalism for inflation introduced in Chapter~\ref{chapter:beta}, to some generalized models of inflation. In particular, we discuss the application of the formalism to models with non-standard kinetic terms, and to models where the inflaton has a non-minimal coupling with gravity. As discussed in~\cite{Pieroni:2015cma}, some hints on the procedure to generalize the formalism, actually come from models where, via a conformal transformation and a field redefinition, it is possible to reduce to the case discussed in Chapter~\ref{chapter:beta}. 
\end{abstract}

\horrule{0.1pt} \\[0.5cm] 

\noindent
As explained in Chapter~\ref{chapter:beta}, the $\beta$-function formalism for inflation proposes a reasonable method to classify inflationary models. In particular, this framework naturally provides a scheme to define universality classes for inflationary models. It is important to stress that, in analogy with statistical mechanics, these universality classes should be considered as sets of theories that share a common scale invariant limit. As with this formalism, we are not working in terms of single models but rather in terms of universality classes, the results are more general than the ones obtained in the standard framework. Therefore, the formulation of the problem in terms of the $\beta$-function formalism is not a simple rewriting but rather a generalization. In Chapter~\ref{chapter:beta}, we have defined the $\beta$-function formalism for models that implement the simplest realization of inflation. However, there are several reasons to think that in order to define a well-motivated model for inflation, it is necessary to go beyond this minimal setup. \\

\noindent
As explained in Chapter~\ref{chapter:inflation} (in particular in Sec.~\ref{sec_inflation:large_eta_prob}), after the first proposal of a concrete inflationary model \textit{i.e.} the chaotic model of Linde~\cite{Linde:1983gd}, it was realized that radiative corrections may spoil the conditions to achieve inflation. This problem (the $\eta$-problem), actually defines a severe threat against inflationary model building. For this reason, several strategies (see Sec.~\ref{sec_inflation:eta_solutions}) was attempted in order to solve the $\eta$-problem. In particular, in this Chapter we focus on models that arise when we explore two of these possibilities: the embedding of the inflation in a UV-complete theory and the possibility of considering models that go beyond the standard slow-roll approximation. As discussed in Sec.~\ref{sec_inflation:generalized_models}, the simplest realization of inflation is actually based on several assumptions. In particular, in these models we usually assume that inflation is driven by a single-field (the inflaton) with canonical kinetic terms that is minimally coupled with gravity, which as usual is described by a standard Einstein-Hilbert term. In the definition of a realistic model for inflation, which also considers the UV-completion of the theory, we may relax one (or more) of these assumptions. In this Chapter we focus on two of these possibilities: models with non-standard kinetic terms (introduced in Sec.~\ref{sec_inflation:Non_standard_kinetic}) and non-minimal coupling between the inflaton and gravity (introduced in Sec.~\ref{sec_inflation:Non_minimal_coupling}). \\

\noindent
As the $\beta$-function formalism offers a natural method to generalize results (from single model to universality classes) obtained in the standard framework, it seems interesting to discuss its extension to the models considered in this Chapter. Moreover, as this formalism offers a simple analysis of the evolution of the Universe during inflation, its application to generalized models can be useful to have a deeper comprehension of the inflationary regime.\\

\noindent
In this Chapter we start with a general introduction on non-standard kinetic terms (Sec.~\ref{sec_inflation:Non_standard_kinetic}) and on non-minimal couplings (Sec.~\ref{sec_inflation:Non_minimal_coupling}). After this introduction, in Sec.~\ref{sec_generalized:beta_k_inflation} and In Sec.~\ref{sec_inflation:beta_non_minimal_coupling} we discuss the application of the $\beta$-function formalism to these generalized inflationary models. Finally, to give more details on the topic, we report \textit{in extenso} the complete paper~\cite{Pieroni:2015cma} at the end of the Chapter (from Sec.~\ref{sec_generalized:paper_introduction} to Sec.~\ref{sec_generalized:Appendix}).

\section{General considerations on non-standard kinetic terms.}
\label{sec_inflation:Non_standard_kinetic}
Generalized kinetic terms may both appear when we discuss the embedding of the inflation in a UV-complete theory and when we consider models that go beyond the standard slow-roll approximation. As explained in Sec.~\ref{sec_inflation:generalized_models}, models with non-standard kinetic terms may naturally arise both in supergravity (kinetic term defined through the K\"ahler potential) and in string theory (low energy description of the D-branes). Moreover, as noticed by Armendariz-Picon, Damour and Mukhanov in~\cite{ArmendarizPicon:1999rj}, if a non-standard kinetic term is allowed, inflation can also be realized beyond the slow-roll approximation. In this section we present a review of the main characteristics of models with non-standard kinetic terms. After a general discussion of the modified inflationary evolution, we focus our attention on an explicit example. \\

\noindent
In general, the action for a homogeneous classical scalar field $\phi(t)$ with a non-standard kinetic term can be expressed as:
\begin{equation}
\label{eq_generalized:action}
    S=\int\mathrm{d}^4x\sqrt{-g}\left(\frac{R}{2\kappa^2}+ p(\phi,X)\right) \ ,
  \end{equation}
where as usual $X \equiv g^{\mu \nu} \partial_{\mu}\phi \partial_{\nu}\phi / 2$ and $p(\phi,X)$ is a general function of the fields. As usual we assume a FLRW metric ($\textrm{d}s^2 = -\textrm{d}t^2 + a(t)^2 \textrm{d}\vec{x}^2 $) so that $X$ reduces to $X = - \dot{\phi}^2 / 2$. Notice that the standard case, discussed in Chapter~\ref{chapter:beta} is recovered by imposing $p = -X - V$.\\

\noindent
We start by deriving the expressions for the energy density $\rho$ and pressure $p$ associated with the scalar field $\phi$ that is described by the action of Eq.~\eqref{eq_generalized:action}:
  \begin{equation}
  \label{eq_generalized:general_prho} 
  p = p(\phi,X) \ , \qquad \qquad \rho(\phi,X) = -\dot{\phi}^2 p_{,X}-p = 2 X p_{,X} - p  \ , 
  \end{equation}
where $p_{,X} \equiv \partial p / \partial X$. As usual the dynamics of the system is described by:
\begin{equation}
\label{eq_generalized:Friedmann}
  H^2=\frac{\kappa^2}{3} \rho \ , \qquad\qquad -2 \dot{H} =\kappa^2 ( \rho+p ) = 2 \kappa^2 X p_{,X} \ . 
\end{equation}
Once again, we notice that for $p = -X - V$ we recover the equations of Chapter~\ref{chapter:inflation} and of Chapter~\ref{chapter:beta}. As usual, the condition to achieve inflation is a nearly constant $H$, and a nearly negligible $\dot{H}$:
\begin{equation}
 - \frac{\dot{H}}{H^2} \ll 1 \ .
\end{equation} 
Using Eq.~\eqref{eq_generalized:Friedmann}, we can thus express this quantity in terms of $p$ and $\rho$ (for completeness we also give the expression in terms of $\dot{\phi}$) as:
\begin{equation}
\label{eq_generalized:inflationcond}
 - \frac {\dot H}{H^2}= \frac{3}{2}\frac{\rho + p}{\rho} = -  p_{,X} \dot{\phi}^2 \frac{\kappa^2}{ 2H^2} \ . 
\end{equation}
As usual a phase of inflation is thus taking place when $p\simeq-\rho$. In the simplest realization for inflation (where we have $p_{,X} = -1$), this regime is actually reached when the kinetic energy \textit{i.e.} $\dot{\phi}^2 $ becomes much smaller than $H^2$. However, if we consider theories with $p_{,X}\neq -1 $ and $p_{,X} \sim 0$, inflation can also be realized at a finite value for $\dot{\phi}$~\cite{ArmendarizPicon:1999rj}. While in the standard models inflation is basically driven by the potential, in this case inflation is actually driven by the kinetic term. Models where inflation is realized at a finite value for $X$ are usually referred to as ``k-inflation''~\cite{ArmendarizPicon:1999rj}.\\

\noindent
An interesting feature of models with non-standard kinetic terms, is that the speed of sound of the model can be $c_s^2 \neq 1$. As usual, the speed of sound is defined (according with the definition of Eq.~\eqref{eq_inflation:speed_of_sound}) as:
\begin{equation}
\label{eq_generalized:speedofsound}
  c_s^2 \equiv \left . \frac{\delta p }{\delta \rho}\right|_{\delta \phi = 0} = \frac{p_{,X}}{\rho_{,X}} = \frac{p_{,X}}{p_{,X}+2Xp_{,XX}} = \frac{1}{1+2 \partial \ln p,_X/ \partial \ln X} \ . 
\end{equation}
where we have used Eq.~\eqref{eq_generalized:general_prho} to express $\rho$ in terms of $p$, $X$ and $p_{,X}$. While for standard kinetic terms ($p,_X = -1$) we simply have $c_s^2= 1$, in the generalized models discussed in this section, this condition is clearly relaxed. It is also useful to introduce the parameter $\Sigma$ defined~\cite{Chen:2006nt,Chen:2010xka} as:
\begin{equation}
  \label{eq_generalized:sigma_def}
  \Sigma \equiv X p_{,X} + 2 X^2 p_{,XX} = \frac{\epsilon_{H} H^2}{c_s^2} \ 
\end{equation}
where $\epsilon_H$ is the usual first slow-roll parameter $\epsilon_H \equiv - \dot{H}/H^2$ and where we have used $ X p_{,X} = \epsilon_H H^2 $. Notice that with this definition, the (dimensionless) scalar power spectrum at horizon crossing, derived in Appendix~\ref{appendix_perturbations:Cosmological_perturbations} (see Eq.~\eqref{appendix_perturbations:scalar_power_spectrum_final}), can be expressed as:
\begin{equation}
\label{eq_generalized:power_spectra_final}
\left.   \Delta^2_s (k,\tau) \right|_{\tau = (k c_s)^{-1}}  =   \frac{ 1 }{8 \pi^2 } \frac{H^2 \kappa^2}{  c_s \ \epsilon_H } = \frac{ \kappa^2 }{8 \pi^2 } \frac{\Sigma}{ \epsilon_H^2} \ .
\end{equation}

\noindent
As already introduced in Chapter~\ref{chapter:inflation} (see Sec.~\ref{sec_inflation:generalized_models}), models with non-standard kinetic terms may predict sizable non-Gaussianities. In order to discuss non-Gaussianities it is customary to express the bispectrum (\emph{i.e.} the three point correlation function) as:
\begin{equation}
  \label{eq_generalized:fnl_definition}
  B(k_1,k_2,k_3) \equiv f_{NL} F(k_1,k_2,k_3) \ ,
\end{equation}
where $f_{NL}$ is the so-called ``nonlinearity parameter'' (introduced in~\cite{Babich:2004gb}) that corresponds to the amplitude of the bispectrum (Plack constraints~\cite{Ade:2015lrj,Ade:2015ava} on this quantity are reported in Sec.~\ref{sec_pseudoscalar:non_gaussianities}) and $F(k_1,k_2,k_3)$ tells the shape of the bispectrum in momentum space. It is possible to show~\cite{Chen:2006nt,Chen:2010xka} that for single field models, the leading contributions to $f_{NL}$ (which are called $f_{NL}^{\lambda}$ and $f_{NL}^c$) can be expressed as:
\begin{equation}
  \label{eq_generalized:leading_fnl}
  f_{NL}^{\lambda} = \frac{5}{81} \left( \frac{1}{c_s^2} - 1 -\frac{2 \lambda}{\Sigma}\right) \ , \qquad f_{NL}^{c} = \frac{35}{108} \left( \frac{1}{c_s^2} - 1 \right) \ ,
\end{equation}
where $c_s^2$ is the speed of sound $\Sigma$ is the parameter introduced in Eq.~\eqref{eq_generalized:sigma_def}, and $\lambda$ is defined as:
\begin{equation}
  \lambda \equiv  X^2 P_{,XX} + \frac{2}{3}  X^3 P_{,XXX} \ .
\end{equation}
Notice that in the case of standard kinetic terms (\textit{i.e.} for $P_{,X} =-1$) we have $\lambda = 0$ and $c_s^2 =1$ so that both these contributions are equal to zero. Higher order contributions are at least linear in the slow-roll parameters~\cite{Chen:2006nt,Chen:2010xka}. \\

\noindent
A powerful theorem on the generation of non-Gaussianities was proved by Creminelli and Zaldarriaga in~\cite{Creminelli:2004yq} in the so-called ``squeezed limit'' \textit{i.e.} when one of the three momenta goes to zero. In this work they have proved that:
\begin{equation}
  \lim_{k_1 \rightarrow 0} \langle \zeta(\tau,\vec{k}_1) \zeta(\tau,\vec{k}_2)\zeta (\tau,\vec{k}_3) \rangle = (2 \pi)^3 \delta^{(3)}(\vec{k}_1 + \vec{k}_2 + \vec{k}_3) (1 -n_s) \mathcal{P}_s(k_1) \mathcal{P}_s(k_3) \ ,
\end{equation}
where $\mathcal{P}_s(k)$ is the usual (dimensionful) power spectrum (defined in Eq.~\eqref{appendix_perturbations:scalar_power_spectrum_definition}). As $(1 -n_s)$ is linear in the slow-roll parameters, this quantity is expected to be extremely small. As a consequence, a direct detection of non-Gaussianities in the squeezed limit would rule out the simplest (single-field) realization of inflation.\\

\noindent
We conclude this section by presenting an explicit example of inflationary model with non-standard kinetic term: the case of Dirac-Born-Infeld (DBI) inflation proposed by Eva Silverstein and David Tong in~\cite{Silverstein:2003hf}. This model belongs to the category of brane inflation motivated by string theory. In particular, this model is the string description of a $U(N)$ $\mathcal{N} = 4$ SuperYang-Mills (SYM) theory in the limit of strong coupling\footnote{The effective description of this theory in the limit of strong coupling was derived in~\cite{Aharony:1999ti} using the AdS/CFT correspondence of Maldacena~\cite{Maldacena:1997re}. More details on AdS/CFT are given in Chapter~\ref{chapter:holographic_universe}, in particular see Sec.~\ref{sec_holography:AdS_CFT}.}. In the case of DBI inflation, inflation is driven by a single scalar field $\phi$ which is described by an effective lagrangian~\cite{Silverstein:2003hf,Alishahiha:2004eh}: 
\begin{equation}
\label{eq_generalized:pressure_DBI}
p(X,\phi) = \frac{1}{f(\phi)} \left[ 1 - \sqrt{1 + 2f(\phi) X}\right] - V(\phi) \ ,
\end{equation}
where $f(\phi)=\lambda/\phi^4$ and $V(\phi)$ is the inflaton field potential. In this model, the velocity $\dot{\phi}$ of the field $\phi$ is bounded by the condition $ \lambda \dot{\phi}^2/ \phi^4 < 1 $ \textit{i.e.} the argument of the square root should be positive. As a consequence, the dynamics of the model is strongly affected by this condition in the regime $\phi \simeq 0 $. Let us show that this models is suitable to describe inflation. Using Eq.~\eqref{eq_generalized:pressure_DBI}, we directly get:
\begin{equation}
\label{eq_generalized:pX_DBI}
  p_{,X} = -\frac{1}{\sqrt{1 + 2f(\phi) X}} \ .
\end{equation}
The energy density for this model can then be computed using Eq.~\eqref{eq_generalized:general_prho}:
\begin{equation}
  \rho = \frac{1}{f} \left( -1 + \frac{1}{\sqrt{1 + 2f(\phi) X}} \right) + V \ .
\end{equation}
Finally we can express $(p+\rho)/\rho$ as:
\begin{equation}
  \frac{p+\rho}{\rho} = -\frac{2 X f}{1 + f V\sqrt{1 + 2f(\phi) X} - \sqrt{1 + 2f(\phi) X} }  \ .
\end{equation}
where again we have used Eq.~\eqref{eq_generalized:general_prho} to express $p+\rho = 2 X p_{,X}$. It is also interesting to compute the speed of sound associated with this model. Substituting Eq.~\eqref{eq_generalized:pX_DBI} into Eq.~\eqref{eq_generalized:speedofsound} we directly get:
\begin{equation}
  c_s^2 = 1 + 2f(\phi) X \ .
\end{equation}
Notice that inflation can be realized with $f V(\phi)$ and $f \lesssim 1/(2X)$ (that corresponds to $c_s^2 < 1$). As a consequence, the speed of sound for this model is varying with time. Notice that as $c_s^2 \neq 1 $ (and also as $p_{,XX}$ and $p_{,XXX}$ are non zero) DBI models may generate sizable non-Gaussianities\footnote{As it is possible to see from Eq.~\eqref{eq_generalized:leading_fnl}.} (For a discussion of this topic see for example~\cite{Silverstein:2003hf,Alishahiha:2004eh,Creminelli:2003iq,Babich:2004gb,Chen:2006nt,Chen:2010xka}).

\section{General considerations on non-minimal coupling.}
\label{sec_inflation:Non_minimal_coupling}
As explained in the discussion of Sec.~\ref{sec_inflation:generalized_models}, non-minimal couplings may naturally arise in the context of supergravity and string theory and also as radiative corrections in the context of QFT in curved spacetime~\cite{birrell1984quantum}. While these models are already interesting on their own (as natural extensions of the simplest realization of inflation of Sec.~\ref{sec_inflation:simplest_Inflation}) particular interest came during the last years when it was realized that they both offer a natural mechanism to flatten the inflationary potential (allowing to consider the standard model Higgs field as the inflaton~\cite{Bezrukov:2007ep,Bezrukov:2009db}) and that in some case they lead to the existence of universal attractors (such as the ``$\alpha$-attractors''~\cite{Kallosh:2013yoa,Kallosh:2014rga} or the attractor at strong coupling~\cite{Kallosh:2013tua} which we discuss in the following) for inflationary predictions. For these reasons, in order to define a concrete model for inflation, it is proper to keep this possibility into account. \\

\noindent
Let us start by considering the general action that describes a homogeneous scalar field $\phi(t)$ that is non-minimally coupled with gravity:
    \begin{equation}
  \label{eq_generalized:non_minimal_coupling_action}
    S =\int\mathrm{d}^4x\sqrt{-g_J}\left( \frac{\Omega(\phi)}{2\kappa^2}R_J - X_J - V_J(\phi) \right) \ ,
  \end{equation}
where as usual $X_J \equiv g_J^{\mu \nu}\partial_{\mu} \phi \partial_{\nu} \phi/2$, $V_J(\phi)$ is the inflationary potential (on which we are not imposing any constraint) and $\Omega(\phi)$ is defined as: 
  \begin{equation}
\label{eq_generalized:omega_def}
  \Omega(\phi) = 1 + \xi f(\phi) \ ,
  \end{equation}
where $\xi$ is a (dimensionless) coupling constant and $f(\phi)$ is a generic function of $\phi$. Notice that this parametrization is quite general as at this point we are not imposing any constraint on the explicit expressions for $f(\phi)$ and $V_J(\phi)$. It should also be stressed that $\xi = 0$ corresponds to the standard case where the scalar field is minimally coupled with gravity.\\

\noindent
Notice that in this frame, gravity is not described by a standard Einstein-Hilbert term. As a consequence, this can be considered as the Jordan frame formulation of the model (explaining the choice of the $J$ subscripts). In order to recover the standard Einstein-Hilbert term for gravity, we can proceed by performing a conformal transformation \textit{i.e.}:
  \begin{equation}
\label{eq_generalized:metric_conformal}
  g_{J \, \mu\nu} \rightarrow g_{\mu\nu} = \Omega(\phi) g_{J \, \mu\nu} \ ,
  \end{equation}
so that the action can be expressed in terms of the new metric $g_{\mu\nu}$ as:
    \begin{equation}
  \label{eq_generalized:non_minimal_coupling_einstein_action}
    S=\int\mathrm{d}^4x\sqrt{-g}\left( \frac{1}{2\kappa^2}R -F(\phi)X -\bar{V}(\phi) \right) \ ,
  \end{equation}
defining the Einstein frame formulation of the theory. The terms $F(\phi)$ and $\bar{V}(\phi)$ appearing in Eq.~\eqref{eq_generalized:non_minimal_coupling_einstein_action} are defined as:
  \begin{equation}
\label{eq_generalized:generalF_V_def}
    F(\phi) \equiv \Omega^{-1} + \frac{3}{2} \left( \frac{\mathrm{d} \ln \Omega}{\mathrm{d} \phi}\right)^2, \qquad \qquad \qquad \bar{V}(\phi) \equiv \frac{V_J (\phi)}{\Omega(\phi)^2}. 
  \end{equation}
In the Einstein frame, the non-minimal coupling disappears but conversely we have a non-standard kinetic term for the inflaton. As a consequence, these models can be seen as a particular case of the parametrization discussed in Sec.~\ref{sec_inflation:Non_standard_kinetic}. Notice that via a field redefinition~\cite{Pieroni:2015cma}, it is possible to describe the problem in terms a scalar field $\varphi$:
\begin{equation}
\label{eq_generalized:def_varphi_minimal}
\left( \frac{\mathrm{d} \varphi}{ \mathrm{d} \phi} \right)^2 \equiv F(\phi) \ ,
\end{equation} 
that by definition has a standard kinetic term. It is worth stressing that for these models the leading contribution (Eq.~\eqref{eq_generalized:leading_fnl}) to $f_{NL}$ (defined in Eq.~\eqref{eq_generalized:fnl_definition}) are thus expected to be zero ($c_s^2 = 1$ and $\lambda = 0$)! These models are thus generating a negligible amount of non-Gaussianities. \\ 

\noindent
As already argued in this section, an interesting feature of these generalized models, is that under particular conditions, they may lead to the appearance of a set of universal attractors. In this context, it is worth mentioning the class of models defined by Kallosh, Linde in~\cite{Kallosh:2013xya, Kallosh:2013hoa, Kallosh:2013daa} and subsequently generalized by Kallosh, Linde and Roest in~\cite{Kallosh:2013yoa,Kallosh:2014rga}, which lead to the appearance of the well known ``$\alpha$-attractors''. Another interesting class of models has been recently proposed by Linde, Kallosh and Roest~\cite{Kallosh:2013tua}, which lead to existence of a universal attractor at strong coupling. In order to present an explicit example, we consider the particular expression for $V_J(\phi)$ discussed in~\cite{Kallosh:2013tua}:
\begin{equation}
\label{eq_generalized:def_potential_attractor}
V_J(\phi) = \lambda^2 f^2(\phi) \ ,
\end{equation}
where $f(\phi)$ is a generic function of $\phi$ and $\lambda$ is a constant. This parametrization, is actually motivated by the possibility of defining a natural supergravity embedding~\cite{Kallosh:2010ug} for these models. In the limit of small coupling ($\xi \ll 1 $), both $\Omega(\phi)$ and $F(\phi)$ approach one and the simplest realization of inflation is recovered. As a consequence, fixing an explicit parametrization for $f(\phi)$ in this limit, corresponds to fixing the potential for the theory. Clearly, different choices for $f(\phi)$, correspond to different predictions for $n_s$, scalar spectral index, and $r$, tensor-to-scalar ratio. However, as pointed out in~\cite{Kallosh:2013tua}, this picture changes if consider the limit of a strong coupling $1 \ll \xi$. In particular, it is possible to show~\cite{Kallosh:2013tua,Pieroni:2015cma} that in this regime the expression for $N$, number of e-foldings, simply reads:
  \begin{equation}
\label{eq_generalized:e_foldings_non_minimal}
    N(\phi) \simeq \frac{3}{4} \xi f(\phi). 
  \end{equation}
Moreover, it is also possible to show~\cite{Kallosh:2013tua,Pieroni:2015cma} that in this limit, the expressions for $n_s$ and $r$ are:
\begin{equation}
\label{eq_generalized:ns_r_attractor_non_minimal}
n_s = 1 - \frac{2}{N}\ , \qquad \qquad \qquad r = \frac{12}{N^2} \ .
\end{equation}
It is important to stress that these results are independent on the explicit choice for $f(\phi)$. As a consequence, we have a set of theories that share a single asymptotic behavior in the limit of $1 \ll \xi$. This is usually referred to as the universal attractor at strong coupling.

\section{$\beta$-function formalism for non-standard kinetic terms.}
\label{sec_generalized:beta_k_inflation}
In this Section, we discuss the extension of the $\beta$-function formalism for inflation to models with non-standard kinetic terms. As explained in Chapter~\ref{chapter:beta}, in this framework the cosmological evolution of a scalar field in its potential is described in terms of a Renormalization Group (RG) equation. In particular, inflation is interpreted as the RG flow away from a repulsive (IR) fixed point. The nearly scale invariant regime in the neighborhood of this fix point, naturally provides a scheme to define universality classes for inflationary models. In order to define the $\beta$-function formalism, we should start by following the usual Hamilton-Jacobi approach. In this framework, the field $\phi$ is used as a clock for the system, and we introduce the superpotential 
\begin{equation}
\label{eq_generalized:superpotential_k_inflation}
  W(\phi) \equiv -2H(\phi) \ .
\end{equation}
At this point, we can thus use Eq.~\eqref{eq_generalized:Friedmann} to get:
      \begin{equation}
      - \kappa^2 p_{,X} \, \dot{\phi} = W_{,\phi} \ . \label{eq_generalized:kequphidot}
    \end{equation}
Notice that this equation is similar in form to Eq.~\eqref{eq_beta:phisuperpot}, but in practice these two equations are extremely different. In the case of canonical kinetic terms ($p_{,X}=-1$), this equation directly sets a one-to-one correspondence between $\dot{\phi}$ and $W_{,\phi}$. On the contrary, if $p_{,X}$ is a general function of $X$ and $\phi$, in order to express all the dynamics in terms of $\phi$, we first need to solve this equation and specify an explicit expression for $\dot{\phi}$. While in general it is possible to consider cases where a global solution does not exist, a local solution should always exist. It is interesting to point out that all the cases described in this Chapter are actually described by this equation. Moreover, we should stress that for models where $p_{,X}$ only depends on $\phi$ (such as the ones of Sec.~\ref{sec_inflation:Non_minimal_coupling} and of~\cite{Pieroni:2015cma}), we can always redefine the field as in Eq.~\eqref{eq_generalized:def_varphi_minimal} in order to recover the standard case described by in Chapter~\ref{chapter:beta}.  \\ 

\noindent
We can proceed with our treatment by using Eq.~\eqref{eq_generalized:general_prho} and Eq.~\eqref{eq_generalized:Friedmann} to express the Hamilton-Jacobi equation:
\begin{equation}
  \frac{3}{4}W^2(\phi) - \frac{ W^2_{,\phi}(\phi)}{(-p_{,X}) \kappa^2}= - 2 p \ .\label{eq_generalized:kequpot}
\end{equation}
As explained above, models with a non-minimal coupling between the inflaton and gravity presented in Sec.~\ref{sec_inflation:Non_minimal_coupling}, always admit a description in terms of a new field $\varphi$ that can naturally be described as in Chapter~\ref{chapter:beta}. As a consequence, in order to define a generalizations of the $\beta$-function formalism to models with non-standard kinetic terms, we can follow an analogous of the procedure carried out in this case~\cite{Pieroni:2015cma} and define the $\beta$-function as:
\begin{equation}
  \beta(\phi)\equiv\kappa\left(-p_{,X}\right)^{1/2}\frac{\textrm{d} \phi}{\textrm{d} \ln a}= - 2 \kappa \left(-p_{,X}\right)^{1/2} \frac{\dot{\phi}}{W} = -\frac{2}{\kappa}\left( - p_{,X}\right)^{-1/2}\frac{W_{,\phi}}{W} \ .
\end{equation}
To check the consistency of this definition we may start by computing the explicit expressions for both the energy density $\rho$ and the pressure $p$:
\begin{equation}
  \label{eq_generalized:prho_beta}          
  \rho =  \frac{3}{4\kappa^2} W^2 \ , \qquad \qquad p = -  \frac{3}{4\kappa^2} W^2 (1 - \beta^2/3) \ .
\end{equation}
Notice that these equations are equivalent to the ones derived in Chapter~\ref{chapter:beta}. Moreover, it is crucial to stress that with this definition for $\beta$, the equation of state can be expressed as: 
\begin{equation} 
  \frac{\rho+p}{\rho}= \frac{\beta^2(\phi)}{3} \ .  \label{eq_generalized:master}
\end{equation}
This equation shows that once again it is the $\beta$-function that controls inflation. In particular, once again inflation is realized in the vicinity of a zero of this function. As Eq.~\eqref{eq_generalized:master} and Eq.~\eqref{eq_generalized:prho_beta} are identical to the ones derived in Chapter~\ref{chapter:beta}, the discussion in terms of the universality classes may also be properly defined in this generalized case. More details on the application of $\beta$-function formalism will be presented in an upcoming work~\cite{new_paper} where an accurate treatment of the topic is carried out.

\section{$\beta$-function formalism for non-minimal couplings.}
\label{sec_inflation:beta_non_minimal_coupling}
The application of the $\beta$-function formalism for inflation to models where the inflaton is non-minimally coupled with gravity has been discussed in~\cite{Pieroni:2015cma} (which is reproduced \emph{in extenso} at the end of this Chapter). In this paper, the problem is discussed in terms of the new field $\varphi$ (see Eq.~\eqref{eq_generalized:def_varphi_minimal}), that has both a minimal coupling with gravity and a standard kinetic term. As a $\beta$-function for $\varphi$ can be defined following the procedure described in Chapter~\ref{chapter:beta}, the formalism can be naturally extended to the case when the problem is described in terms of the field $\phi$. Once the formalism is set, its application to a certain set of models is discussed. For example, we re discuss the appearance of the universal attractor at strong coupling of~\cite{Kallosh:2013tua} in terms of the $\beta$-function formalism. In particular, we find that the appearance of the attractor can be explained in terms of the mechanism of interpolation discussed in Sec.~\ref{sec_beta:comoposite_classes}. \\

\noindent
Let us explain the main ideas that lead to this conclusion (for the complete treatment see the paper \textit{in extenso}). In order to explain the appearance of the attractor, we start by focusing on the model of Kallosh Linde and Roest~\cite{Kallosh:2013tua} \textit{i.e.} by parameterizing $V_J$ as in Eq.~\eqref{eq_generalized:def_potential_attractor}. At this point, we proceed by formulating the problem in terms of the field $\varphi$ which is defined according to Eq.~\eqref{eq_generalized:def_varphi_minimal}. Setting $\kappa =1 $ to lighten the notation, the $\beta$-function can be defined as:
\begin{equation}
    \label{eq_generalized:beta_approx_varphi_minim}
    \beta(\varphi ) \equiv \frac{\textrm{d} \varphi}{\textrm{d} \ln a} \simeq - \frac{\mathrm{d} \ln V(\varphi)}{\mathrm{d} \varphi} =  - 2 \frac{\tilde{f}_{,\varphi}(\varphi)}{\tilde{f}(\varphi) \left[1 + \xi \tilde{f}(\varphi) \right]} ,
\end{equation}  
where $\tilde{f}(\varphi) \equiv f(\phi(\varphi))$ and the potential $V(\varphi)$ is defined as:
\begin{equation}
   V(\varphi) \equiv \bar{V}(\phi(\varphi)) = \frac{V_J (\phi(\varphi))}{\Omega^2 (\phi(\varphi)) } \ , 
\end{equation}
where as usual $\Omega = 1 + \xi f(\phi)$. From the definition of $F(\phi)$ given in Eq.~\eqref{eq_generalized:generalF_V_def}, it is easy to show that in the limit of strong coupling we get:
\begin{equation}
  \label{eq_generalized:F_strong_minim}
  F(\phi) \simeq \frac{3}{2}\left( \frac{f_{,\phi} (\phi)}{ f(\phi)} \right)^2 \ . 
\end{equation}
As a consequence, by integrating Eq.~\eqref{eq_generalized:def_varphi_minimal} we can express $f(\phi)$ as:
\begin{equation}
  \label{eq_generalized:def_varphi_strong_minim}
  f(\phi(\varphi)) = \tilde{f}(\varphi) = f_{\mathrm{f}}\exp \left[  \sqrt{\frac{2}{3}} (\varphi - \varphi_{\mathrm{f}} ) \right] \ .
\end{equation}
Substituting this expression into Eq.~\eqref{eq_generalized:beta_approx_varphi_minim} we find:
\begin{equation}
\label{eq_generalized:explicit_beta_strong_minim}
\beta(\varphi) = -\sqrt{\frac{8}{3}} \frac{1}{\xi f_{\mathrm{f}}} \exp \left[ - \sqrt{\frac{2}{3}} (\varphi - \varphi_{\mathrm{f}} ) \right] \ .
\end{equation}
At this point, some explicit parameterizations for $f(\phi)$ can be fixed in order to show the interpolation from weak to strong coupling (where the attractor is approached). An explicit realization of the mechanism of interpolation is shown\footnote{These are the same figures presented in the paper \emph{in extenso} updated with the data of Planck 2015~\cite{Ade:2015lrj}.} in Fig.~\ref{fig_generalized:linear} and in Fig.~\ref{fig_generalized:log} where we consider $f(\phi) = \phi^\alpha$ for some values of $\alpha$.\\

\begin{figure}[htb!]
\centering

{\includegraphics[width=0.85 \columnwidth]{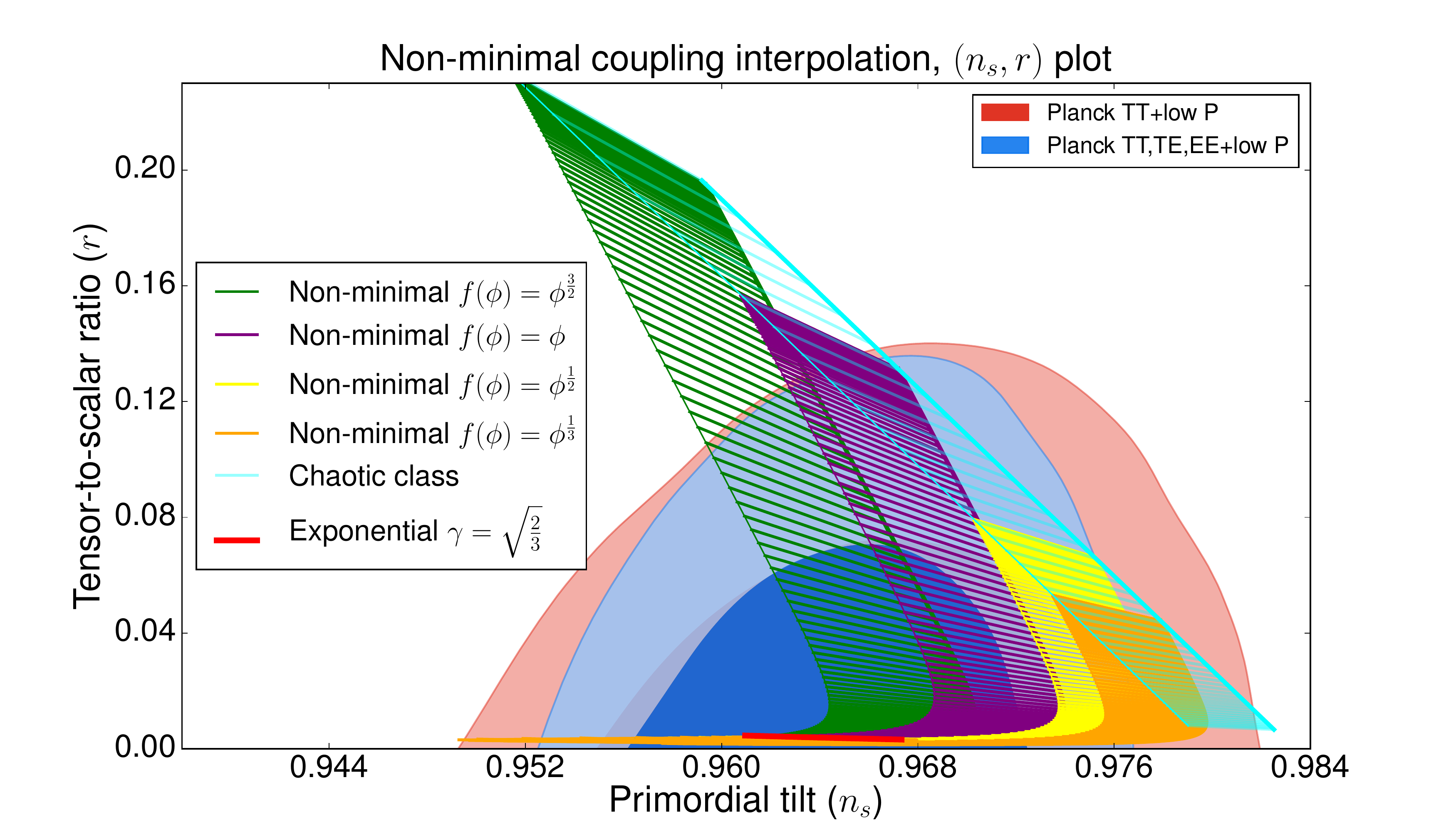}}\\
\caption{   \label{fig_generalized:linear} \footnotesize Numerical predictions of $n_s$ and $r$ for the non-minimally coupled models are compared with Chaotic class (with some values for $\alpha$ in the range $[0.1,3]$) and with the Exponential class with $\gamma = \sqrt{2/3}$. The results are presented with the constraints on $n_s$ and $r$ set by Planck 2015 base\_r TT+low P (red) and Planck base\_r TT,TE,EE+low P~\cite{Ade:2015lrj}. }
\end{figure}

\begin{figure}[htb!]
\centering

{\includegraphics[width=0.85 \columnwidth]{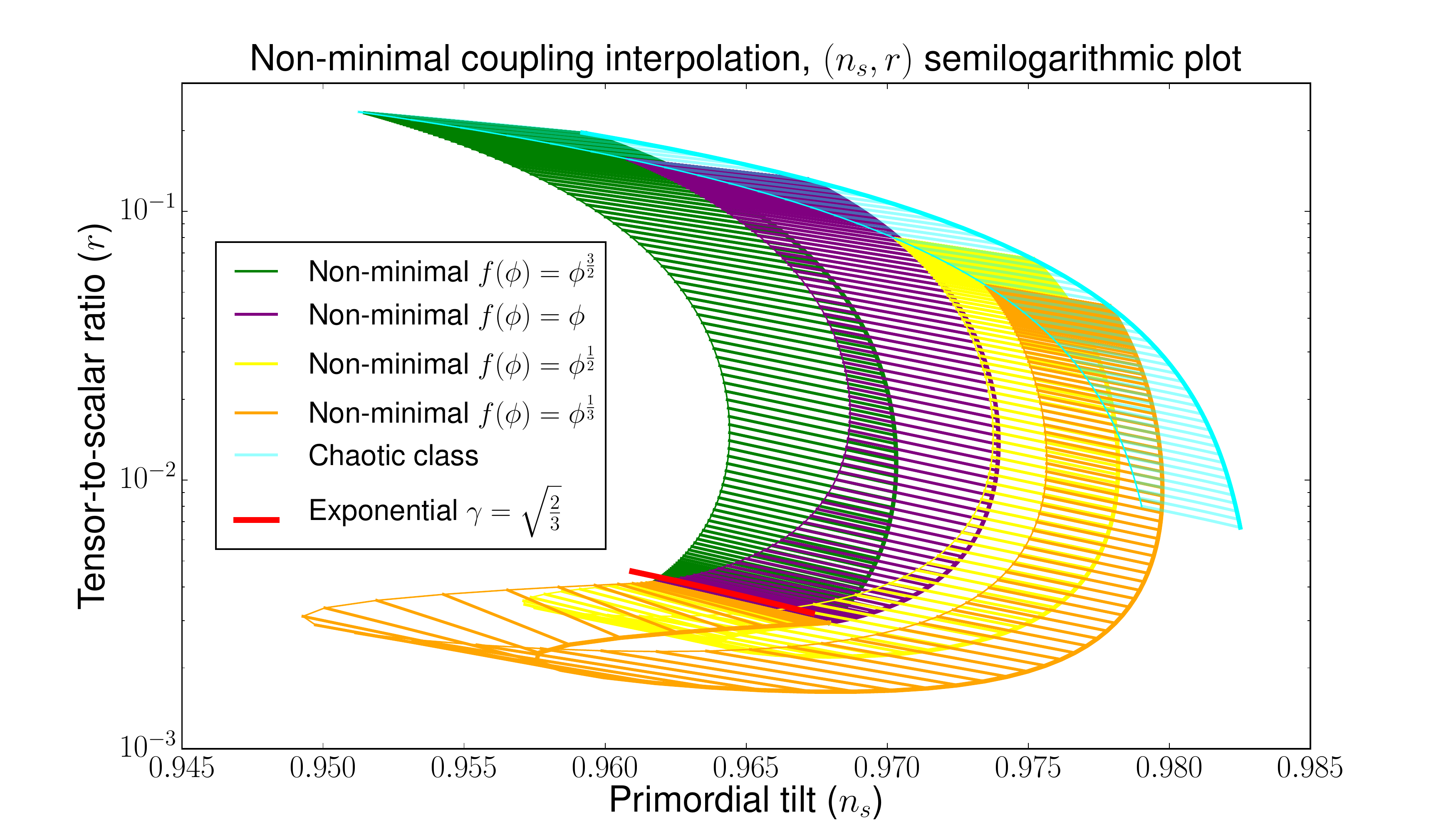}}\\
\caption{   \label{fig_generalized:log}\footnotesize  Numerical predictions of Fig.~\ref{fig_generalized:linear} presented in a semilogarithmic plot.}
\end{figure}

\noindent
The analysis of~\cite{Pieroni:2015cma} is not limited to the model described in~\cite{Kallosh:2013tua} but it is extended to a wider set of theories. For example the application to the $\alpha$-attractors of Kallosh Linde and Roest~\cite{Kallosh:2013yoa,Kallosh:2014rga} is discussed. Moreover, the introduction of a further functional freedom in the model is considered. After this generalization, we re discuss the model and we focus on the possibility of preserving the attractor. We find that in general, the attractor is not stable under these generalizations and the universality is thus broken.\\

\noindent
The $\beta$-function formalism proves to be suitable for the analysis carried out in~\cite{Pieroni:2015cma}. In particular, it offers a powerful method to perform a critical analysis of the conditions required to preserve the attractor at strong coupling. Remarkably, we are able to prove that, when an additional functional freedom is introduced in the model, weak assumptions are required in order to preserve the attractor. Moreover, we show that in the case where the universal attractor of Kallosh Linde and Roest is evaded, we can define a more general set attractors.

\clearpage

\begin{center}
{\Large {\textbf{
$\beta$-function formalism for inflationary models with a non minimal coupling with gravity.}}}
 
 ~\\
\vskip 3cm

{  \textbf{M. Pieroni$^{ab}$ } }\\
~\\
~\\
{\em ${}^a$ Laboratoire AstroParticule et Cosmologie,
Universit\'e Paris Diderot } \\
{\em ${}^b$ Paris Centre for Cosmological Physics, F75205 Paris Cedex 13}\\

\end{center}

\vskip 1cm
\centerline{ {\textbf{ Abstract}}}{ \noindent We discuss the introduction of a non minimal coupling between the inflaton and gravity in terms of our recently proposed $\beta$-function formalism for inflation. Via a field redefinition we reduce to the case of minimally coupled theories. The universal attractor at strong coupling has a simple explanation in terms of the new field. Generalizations are discussed and the possibility of evading the universal attractor is shown. }

\indent

\vfill

\date{\normalsize\today} 
\newpage

\section{Introduction.}
\label{sec_generalized:paper_introduction}
Inflation is the most suitable extension of standard cosmology to solve the horizon, monopoles and flatness problems. The Planck mission~\cite{Ade:2015lrj} and other cosmological observations help to fix several constraints on the general mechanism driving this phenomenon. The chaotic model~\cite{Linde:1983gd} with potential $V(\phi) = \lambda \phi^4$ with a non-minimal coupling of the scalar field with gravity $\frac{\xi \phi^2}{2} R$ has been recently proposed by Bezrukov and Shaposhnikov~\cite{Bezrukov:2007ep} as a natural extension of the Standard Model in order to include inflation. For a large number $N$ of e-folding this model gives predictions for the scalar spectral index and the tensor to scalar ratio:
\begin{equation}
\label{eq_generalized:ns-r-attractor}
n_s = 1 - \frac{2}{N}, \qquad \qquad \qquad r = \frac{12}{N^2}.
\end{equation}
Assuming that $N\sim50-60$ we find numerical values in good agreement with Planck data. As Starobinsky model~\cite{Starobinsky:1980te} and many other inflationary models are also predicting similar values for $n_s$ and $r$ it is important to define a systematic classification in order to avoid this degeneracy. Some proposals to explain this degeneracy have been formulated by Mukhanov~\cite{Mukhanov:2013tua} and Roest~\cite{Roest:2013fha}. In this spirit we recently proposed a $\beta-$function formalism for inflation~\cite{Binetruy:2014zya}. This new approach is based on the idea of providing universality classes of models of inflation by relying on the approximate scale invariance during the inflationary epoch. This suggestion has a deep connection with the idea proposed by McFadden and Skenderis~\cite{McFadden:2010na} of applying the holographic principle to describe the inflationary Universe. In the language of the well known (A)dS/CFT correspondence of Maldacena~\cite{Maldacena:1997re}, the asymptotic de Sitter spacetime is dual to a (pseudo) Conformal Field Theory. In this framework the equations describing the cosmological evolution are thus interpreted as holographic Renormalization Group (RG) equations for the corresponding QFT~\cite{Kiritsis:2013gia}. This correspondence suggests that universality classes for inflationary models should be defined in terms of the Wilsonian picture of fixed points (exact deSitter solutions), scaling regions (inflationary epochs), and critical exponents (scaling exponents of the power spectra related with the slow-roll parameters). It is important to stress that, in analogy with statistical mechanics, these universality classes should be considered as sets of theories that share a common scale invariant limit. As results obtained in this framework are not only valid for particular models but for whole sets of theories, it should be clear that they should be conceived as more general than the ones obtained using the standard methods. \\

\noindent
In this paper we discuss inflationary models where a scalar field is non-minimally coupled with gravity. A discussion of this topic has been recently proposed by Linde, Kallosh and Roest~\cite{Kallosh:2013tua} in terms of the standard picture of defining inflationary models by identifying the inflationary potentials and they proved the existence of a universal attractor at strong coupling. Both to have a deeper comprehension of the inflationary regime and to produce a further generalization of the results presented in~\cite{Kallosh:2013tua}, it is interesting to treat theories for scalar field with a non-minimal coupling with gravity in terms of the $\beta-$function formalism. In Sec.~\ref{sec_generalized:paper_model_definition}, we present a model of a scalar field with a non-minimal coupling with gravity. In Sec.~\ref{sec_generalized:paper_beta_function} we formulate the problem in terms of the $\beta-$function formalism and we present the weak and strong limits. In Sec.~\ref{sec_generalized:paper_general_case}, we consider a more general class of models by relaxing an assumption on the expression for the potential. In this context we prove that it is possible to evade the universal attractor and that other attractors can be reached. In Sec.~\ref{sec_generalized:paper_conclusions}, we finally present our conclusions.

\section{Setting up the model.}
\label{sec_generalized:paper_model_definition}
The simplest action to describe the inflating universe consists of a the standard Einstein-Hilbert term to describe gravity plus the action for a homogeneous scalar field in curved spacetime\footnote{We use the convention $ds^2 = dt^2 - a(t)^2 (dr^2 + r^2 d\Omega^2)$}:
\begin{equation}
  \label{eq_generalized:minimal-action}
    S =\int\mathrm{d}^4x\sqrt{-g}\left( -  \frac{1}{2\kappa^2}R +  X - V_J(\phi) \right),
  \end{equation}
where $X\equiv g^{\mu \nu}  \partial_\mu \phi \partial_\nu \phi /2 = \dot{\phi}^2 /2 $ is the standard kinetic term for a homogeneous scalar field. Let us consider a generalization of this action to include a non-minimal coupling between the scalar field and gravity. In this paper, we follow the proposal of~\cite{Kallosh:2013tua}, and we consider the action:
\begin{equation}
  \label{eq_generalized:non-minimal-action2}
    S =\int\mathrm{d}^4x\sqrt{-g}\left( -  \frac{\Omega(\phi)}{2\kappa^2}R +  X - V_J(\phi) \right).
  \end{equation}
As gravity is not described by a standard Einstein-Hilbert term, this should be considered as the Jordan frame formulation of the model. Notice that we have not imposed any constraint on the explicit expression of $V_J(\phi)$. Let us consider: 
\begin{equation}
\label{eq_generalized:omega}
  \Omega(\phi) = 1 + \xi f(\phi),
  \end{equation}
where $\xi$ is the coupling constant and $f(\phi)$ is a function of $\phi$. It is again interesting to stress that this parametrization is quite general as we are not imposing any constraint on the explicit expression for $f(\phi)$. It should also be stressed that $\xi = 0$ corresponds to the standard case of a scalar field minimally coupled with gravity.\\

\noindent
It is well known that by means of a conformal transformation i.e.
  \begin{equation}
\label{eq_generalized:metric}
  g_{\mu\nu} \rightarrow \Omega(\phi)^{ -1} g_{\mu\nu},
  \end{equation}
we can recover the standard Einstein-Hilbert term for gravity. The action in terms of the new metric can be expressed as:
    \begin{equation}
  \label{eq_generalized:non-minimal-action}
    S=\int\mathrm{d}^4x\sqrt{-g}\left( - \frac{1}{2\kappa^2}R +  F(\phi)X - \bar{V}(\phi) \right),
  \end{equation}
where $F(\phi)$ and $\bar{V}(\phi)$ are defined by:
  \begin{equation}
\label{eq_generalized:generalF}
    F(\phi) \equiv \Omega^{-1} + \frac{3}{2} \left( \frac{\mathrm{d} \ln \Omega}{\mathrm{d} \phi}\right)^2, \qquad \qquad \qquad \bar{V}(\phi) \equiv \frac{V_J (\phi)}{\Omega(\phi)^2}. 
  \end{equation}
This is usually known as the Einstein frame formulation for the theory. From now on, we impose $\kappa^2 = 1 $ to simplify the notation. As discussed in~\cite{Kallosh:2013tua}, it is interesting to consider the particular expression for $V_J(\phi)$:
\begin{equation}
V_J(\phi) = \lambda^2 f^2(\phi).
\end{equation}
This parametrization is motivated by the possibility of defining a natural supergravity embedding~\cite{Kallosh:2010ug} for this class of models. It is important to stress that in the limit of small coupling $\xi \ll 1 $, both $\Omega(\phi)$ and $F(\phi)$ are close to one. In this limit, fixing an explicit parametrization for $f(\phi)$ we are directly fixing the potential for the theory. At this point it should be clear that in this regime different choices for $f(\phi)$ correspond to different predictions for $n_s$, scalar spectral index, and $r$, tensor to scalar ratio. As discussed in~\cite{Kallosh:2013tua}, it is interesting to consider the limit of a strong coupling $1 \ll \xi$. It possible to show that in this regime the expression for $N$, number of e-foldings, simply reads:
  \begin{equation}
\label{eq_generalized:e-foldings}
    N(\phi) \simeq \frac{3}{4} \xi f(\phi). 
  \end{equation}
It is also possible to show that in this limit, the expressions for $n_s$ and $r$ are simply given by Eq.~\eqref{eq_generalized:ns-r-attractor}. It is important to stress that this result is independent on the explicit choice for $f(\phi)$. As different theories share the same asymptotic behavior in the limit of $1 \ll \xi$, this proves the existence of a universal attractor at strong coupling. In the rest of this work we will focus both on the interpretation of this attractor in terms of the $\beta$ function formalism of~\cite{Binetruy:2014zya}, and on the possibility of extending these results for more general classes of models. In particular, in Sec.~\ref{sec_generalized:paper_general_case}, we will discuss the consequences of choosing a different parameterization for $V_J(\phi)$ i.e. 
\begin{equation}
\Omega(\phi) = 1 + \xi f(\phi) ,\qquad \qquad \qquad V_J(\phi) = \lambda^2 g^2(\phi) ,
\end{equation}
with $f(\phi) \neq g(\phi)$.

\section{$\beta$-function formalism.}
\label{sec_generalized:paper_beta_function}
Let us consider the model described in Sec.~\ref{sec_generalized:paper_model_definition}. By means of a field redefinition it is possible to reduce to the problem of a scalar field with a canonically normalized kinetic terms. In particular let us define\footnote{ Notice that this definition may drive to an ambiguity as it implies $d\varphi / d \phi = \pm \sqrt{F(\phi)}$. To solve this problem we simply have to select a solution and be consistent with our choice. In this paper we will consider the $+$ solution. Clearly an equivalent treatment can be achieved in terms of the $-$ solution. } the new field $\varphi$ as:
\begin{equation}
\label{eq_generalized:def_varphi}
\left( \frac{\mathrm{d} \varphi}{ \mathrm{d} \phi} \right)^2 = F(\phi).
\end{equation} 
By definition the kinetic term of $\varphi$ is canonically normalized and thus we can directly follow the procedure discussed in~\cite{Binetruy:2014zya}. Assuming that the time evolution of the scalar field $\varphi(t)$ is \emph{piecewise monotonic} we can invert to get $t(\varphi)$ and use the field as a clock. Under this assumption we can thus describe the dynamics of the system in terms of the Hamilton-Jacobi approach of Salopek and Bond~\cite{Salopek:1990jq}. In this framework we define $W(\varphi) \equiv -2H(\varphi)$, that satisfies $\dot{\varphi} = W_{,\varphi} (\varphi)$ and also 
\begin{equation}
  \label{eq_generalized:superpotential} 
  2 V(\varphi) = \frac{3}{2} W^2(\varphi) - \left[W_{,\varphi} (\varphi)\right]^2.
 \end{equation}
The latter expression leads to call the function $W(\varphi)$ \emph{superpotential} because of a similar parameterization in the context of supersymmetry. In analogy with QFT we define:
\begin{equation}
  \label{eq_generalized:beta_varphi} 
  \beta(\varphi) \equiv \frac{\mathrm{d} \varphi}{\mathrm{d} \ln a} = - 2\frac{\mathrm{d} \ln W(\varphi)}{\mathrm{d} \varphi}.
 \end{equation}
It is important to notice that the equation of state for the scalar field in terms of $\beta$ reads:
\begin{equation}
  \label{eq_generalized:eq_of_state} 
 \frac{p+ \rho}{\rho} = \frac{\beta^2 (\varphi)}{3}.
 \end{equation}
This expression for the equation of state implies that an inflationary epoch is associated with the neighborhood of a zero of $\beta(\varphi)$. In fact, by specifying a parametrization for $\beta(\varphi)$, we are fixing the evolution of the system (or equivalently the RG flow) close to a fixed point. As a single asymptotic behavior can be reached by several models, the parametrization of $\beta(\varphi)$ is not simply specifying a single inflationary model but rather a set of theories sharing a scale invariant limit. In particular, using the language of statistical mechanics, we are specifying a \emph{universality class} for inflationary models. It is important to stress that in this framework all the informations on the inflationary phase are thus enclosed in the parametrization of $\beta(\varphi)$ in terms of the critical exponents. Substituting Eq.~\eqref{eq_generalized:beta_varphi} into Eq.~\eqref{eq_generalized:superpotential} we express the potential $V(\varphi)$ as:
  \begin{equation} 
\label{eq_generalized:potential_1}
            V(\varphi)=\frac{3 W^{2}(\varphi)}{4}\left[1-\frac{\beta^{2}(\varphi)}{6}\right].
    \end{equation}
During an inflationary epoch $\beta(\varphi)$ must be close to zero and thus at the lowest order, we can approximate\footnote{As Eq.~\eqref{eq_generalized:beta_varphi} implies: \begin{equation*}
W(\varphi) = W_\textrm{f} \exp \left[ - \int_{\varphi_f}^{\varphi} \frac{\beta(\hat{\varphi})}{2} d\hat{\varphi} \right], \end{equation*} to be consistent with this approximation we need: \begin{equation*} \left| \beta(\varphi)\right|^2 \ll \left| \int_{\varphi_f}^{\varphi} \beta(\hat{\varphi}) d\hat{\varphi} \right|. \end{equation*}
In the rest of this paper we will consider explicit expressions for $\beta(\varphi)$ that satisfy this requirement.} Eq.~\eqref{eq_generalized:potential_1} with: $V(\varphi) \sim \frac{3}{4} W(\varphi)^2$. From Eq.~\eqref{eq_generalized:eq_of_state}, we can notice that $\beta^2 /2 $ is equal to the first slow-roll parameter $\epsilon_H = - \dot{H}/H^2$. In the slow-rolling regime the $\beta$-function formalism is thus equivalent to the horizon-flow approach of Hoffman and Turner~\cite{Hoffman:2000ue, Kinney:2002qn, Liddle:2003py, Vennin:2014xta}. In this limit we can thus express $\beta(\varphi)$ as:
\begin{equation}
    \label{eq_generalized:beta_approx_varphi}
    \beta(\varphi ) \sim - \frac{\mathrm{d} \ln V(\varphi)}{\mathrm{d} \varphi} =  - 2 \frac{\tilde{f}_{,\varphi}(\varphi)}{\tilde{f}(\varphi) \left[1 + \xi \tilde{f}(\varphi) \right]} ,
\end{equation}  
where $\tilde{f}(\varphi) \equiv f(\phi(\varphi))$. Characterizing the system in terms of $\varphi$ helps to have a deeper comprehension of this model and leads to an interpretation of the attractor at strong coupling. In the rest of this section we discuss the limits of large and small $\xi$ and we present an explicit example to understand the interpolation between these two regimes.

\subsection{Strong and weak coupling limits.}
\label{sec_generalized:paper_strong}
In the strong coupling limit we have $1 \ll \xi $ and thus the lowest order approximation for~\eqref{eq_generalized:beta_approx_varphi} simply reads:
\begin{equation}
  \label{eq_generalized:beta_strong}
    \beta(\varphi ) \simeq  - \frac{2}{\xi} \frac{\tilde{f}_{,\varphi} (\varphi )}{ \tilde{f}^2 (\varphi)}  .  
\end{equation} 
Using Eq.~\eqref{eq_generalized:generalF} we can get the lowest order expression for $F(\phi)$ i.e.
\begin{equation}
  \label{eq_generalized:F_strong}
  F(\phi) \simeq \frac{3}{2}\left( \frac{f_{,\phi} (\phi)}{ f(\phi)} \right)^2. 
\end{equation}
We can substitute Eq.~\eqref{eq_generalized:F_strong} into Eq.~\eqref{eq_generalized:def_varphi} and integrate to get:
\begin{equation}
\label{eq_generalized:def_varphi_strong}
f(\phi(\varphi)) = \tilde{f}(\varphi) = f_{\mathrm{f}}\exp \left[  \sqrt{\frac{2}{3}} (\varphi - \varphi_{\mathrm{f}} ) \right],
\end{equation}
where we defined $f_{\mathrm{f}} \equiv \tilde{f}(\varphi_{\mathrm{f}})$. It is crucial to notice that in the limit $1 \ll \xi$ the expression for $f(\varphi)$ does not depend on the explicit choice for $f(\phi)$! As shown in Eq.~\eqref{eq_generalized:beta_strong}, the expression of $\beta(\varphi)$ in the limit of a strong coupling is only depending on $\tilde{f}(\varphi)$. We can thus substitute Eq.~\eqref{eq_generalized:def_varphi_strong} into Eq.~\eqref{eq_generalized:beta_strong} to get:
\begin{equation}
\label{eq_generalized:explicit_beta_strong}
\beta(\varphi) = -\sqrt{\frac{8}{3}} \frac{1}{\xi f_{\mathrm{f}}} \exp \left[ - \sqrt{\frac{2}{3}} (\varphi - \varphi_{\mathrm{f}} ) \right].
\end{equation}
It is important to stress that at the end of inflation $1 + p/\rho $ is close to one and thus Eq.~\eqref{eq_generalized:eq_of_state} implies that $\left| \beta(\varphi_{\mathrm{f}}) \right| \sim 1 $. Eq.~\eqref{eq_generalized:explicit_beta_strong} then leads to $f_{\mathrm{f}} \sim \sqrt{8/3} / \xi$ that can be substituted into Eq.~\eqref{eq_generalized:explicit_beta_strong} to conclude that: 
\begin{equation}
\label{eq_generalized:beta_strong_ff}
\beta(\varphi) = -\exp \left[ - \sqrt{\frac{2}{3}} (\varphi - \varphi_{\mathrm{f}} ) \right].
\end{equation}
It is then clear that the expression for $\beta(\varphi)$, in the limit of big $\xi$, is independent on the explicit choice for $f(\phi)$. As the dynamics of the system during the inflationary phase is completely specified by $\beta(\varphi)$, this directly leads to the universality. In particular, we notice that $\beta(\varphi)$ approaches the exponential class of~\cite{Binetruy:2014zya}. This universality class is entirely determined by a single critical exponent, denoted with $\gamma$, that in this case is equal to the $\sqrt{2 /3 }$ factor in the exponential of~\eqref{eq_generalized:beta_strong_ff}. As discussed in~\cite{Binetruy:2014zya}, the scalar spectral index and the tensor to scalar ratio are given by: 
        \begin{eqnarray} 
          \label{eq_generalized:ns_strong}
            n_{s} - 1 &\simeq& - \frac{2}{N}, \\
            \label{eq_generalized:r_strong}
            r  &\simeq& \frac{8}{\gamma^2 N^2} = \frac{12}{N^2} .
    \end{eqnarray}
As expected, these results are in perfect agreement with the ones discussed in~\cite{Kallosh:2013tua}. In this framework, the independence of $\beta(\varphi)$ on $f(\phi)$ directly leads to the universality for the values of $n_s$ and $r$. In particular, in terms of the $\beta$-function formalism, the appearance of a universal attractor at strong coupling corresponds to the flow of the system into a particular universality class. \\

\noindent
For completeness we can also express $N(\varphi)$ as:
\begin{equation}
\label{eq_generalized:efold_ff}
N(\varphi) =  - \int_{\varphi_f}^{\varphi} \frac{1}{\beta(\hat{\varphi})} d\hat{\varphi} =  \sqrt{ \frac{3}{2}}  \left\{ \exp \left[  \sqrt{\frac{2}{3}} \left( \varphi - \varphi_{\mathrm{f}} \right)     \right]  - 1 \right\} .
\end{equation}
Substituting Eq.~\eqref{eq_generalized:efold_ff} into Eq.~\eqref{eq_generalized:beta_strong_ff} we find that choosing values for $N$ in the range $ [50,60]$ we get $\beta \in [-0.024,-0.02]$. Notice that in the limit of strong coupling, the dynamics in terms of $\varphi$ does not depend on $\xi$. It is also interesting to point out that in this case the asymptotic fixed point is reached for $N \rightarrow \infty$ that corresponds to $\varphi \rightarrow \infty$. \\

\noindent
It is interesting to point out that $\xi = 0$ corresponds to a minimal coupling between the inflaton and gravity. In this case we can again use the equations derived in Sec.~\ref{sec_generalized:paper_beta_function}, but the expression for $\tilde{f}(\varphi)$ given by Eq.~\eqref{eq_generalized:def_varphi_strong} does not hold. As a consequence we are not expecting to obtain a model independent expression for $\beta(\varphi)$ and thus results will be model dependent. In particular, by choosing particular parameterizations for $\beta(\varphi)$, we can reproduce the universality classes introduced in~\cite{Binetruy:2014zya}. As in the limit of a weak coupling we are just introducing a small variation with respect to the case of $\xi = 0$, we will only obtain a little departure from the standard results. In particular it is possible to prove that in the limit of a weak coupling the lowest order expressions for $n_s$ and $r$ correspond to the ones presented in~\cite{Kallosh:2013tua}.

\subsection{An explicit example.}
\label{sec_generalized:paper_example}
In this section we present an example to be have a better understanding of the transition from the weak to the strong coupling limits. In particular, we consider some particular models by specifying an explicit expression for $f(\phi)$. From Eq.~\eqref{eq_generalized:beta_varphi} and Eq.~\eqref{eq_generalized:potential_1}, it should be clear that $\beta(\varphi) \sim - 2\alpha/\varphi$ simply gives:
\begin{equation}
W(\varphi) = W_\textit{f} \left (\frac{\varphi}{\varphi_\textit{f}} \right)^{\alpha}, \qquad \qquad V(\varphi)  = \frac{3 W_\textit{f}^2}{4} \left (\frac{\varphi}{\varphi_\textit{f}} \right)^{2\alpha} = V_\textit{f} \left (\frac{\varphi}{\varphi_\textit{f}} \right)^{2\alpha}.
\end{equation}
This clearly corresponds to the well known case of chaotic inflation~\cite{Linde:1983gd}. As in the limit of small $\xi$ we have $\varphi \sim \phi$ and $\beta(\varphi) \sim - 2 \tilde{f}_{,\varphi}(\varphi)/\tilde{f}(\varphi)$, to obtain this expression for $\beta(\varphi)$ we simply choose $f(\phi) = \phi^\alpha$. It is well known that in this case the lowest order predictions for the $n_s$, scalar spectral index, and $r$, tensor to scalar ratio, are given by:
  \begin{equation} 
  \label{chaotic_predictions}
    n_{s} \simeq 1 - \frac{1+\alpha}{N}, \qquad \qquad              r  \simeq \frac{8\alpha}{N}.
\end{equation}
On the contrary, the strong limit predictions have been discussed in Sec~\ref{sec_generalized:paper_strong}, and these are given by Eq.~\eqref{eq_generalized:ns_strong} and Eq.~\eqref{eq_generalized:r_strong}. Variating the value of $\xi$, we expect to shift from the model dependent regime to the universal attractor at strong coupling. Numerical results for our choice for $f(\phi)$ are shown in Fig.~\ref{Fig_generalized:figure1} and Fig.~\ref{Fig_generalized:figure2}. In this particular case the parametrization of $\beta(\varphi)$ is completely specified by the value of the critical exponent $\alpha$. Once this constant is fixed, we can compute numerical predictions as a function of $N$, number of e-foldings. in Fig.~\ref{Fig_generalized:figure1} and Fig.~\ref{Fig_generalized:figure2} we use different colors to plot models associated with a different values for $\alpha$. The solid black lines in the plot of Fig.~\ref{Fig_generalized:figure2} are used to follow the variation of $\xi$ while the values of $\alpha$ and $N$ are fixed. The thick line corresponds to $ N = 60 $ and the thin one corresponds to $N = 50$. Numerical results are compared with the ones obtained for the chaotic class i.e. $\beta(\varphi) = - \alpha / \varphi$ with some values for $\alpha$ in the range $[0.1,3]$ and for the exponential class i.e. $\beta(\varphi) = - \exp \left[ - \gamma \varphi \right]$ with $\gamma = \sqrt{2/3}$. In limit of a weak coupling numerical predictions match with the chaotic class while in the strong coupling limit we approach the predicted exponential class attractor. In the intermediate region we have a whole set of valid inflationary models that actually interpolate between the two fundamental classes.\\

\begin{figure}[htb!]
\centering

{\includegraphics[width=1 \columnwidth]{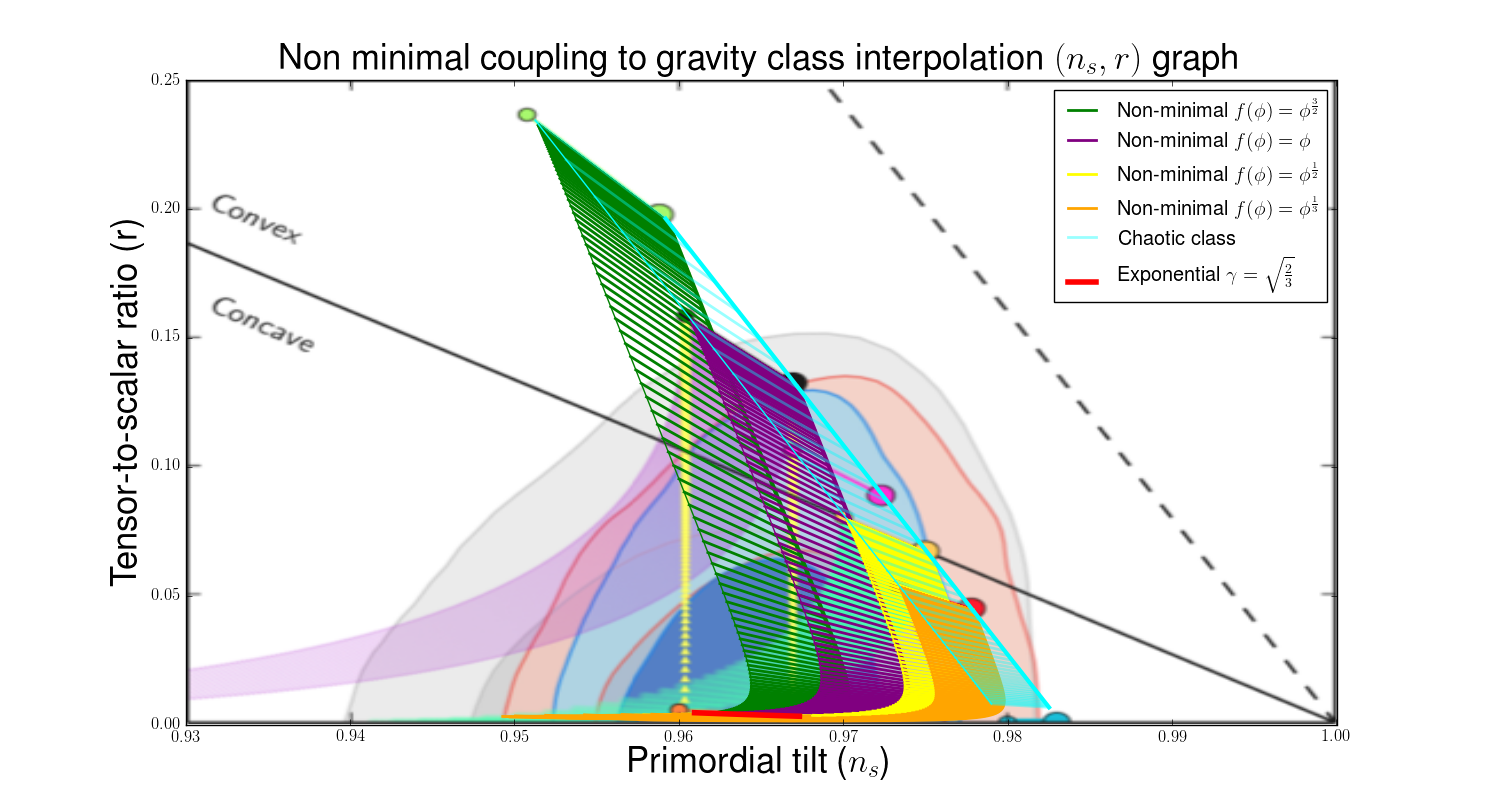}}\\
\caption{   \label{Fig_generalized:figure1} \footnotesize Numerical predictions for $n_s$ and $r$ for the non-minimally coupled models are compared with Chaotic class with some values for $\alpha$ in the range $[0.1,3]$ and exponential class with $\gamma = \sqrt{2/3}$. The results are presented with the famous Planck $(n_s,r)$ graph as a background~\cite{Ade:2015lrj}. In particular we have Planck 2013 (gray contours), Planck TT+lowP (red contours), and Planck TT,TE,EE+lowP (blue contours). 
}
\end{figure}

\begin{figure}[htb!]
\centering

{\includegraphics[width=0.95 \columnwidth]{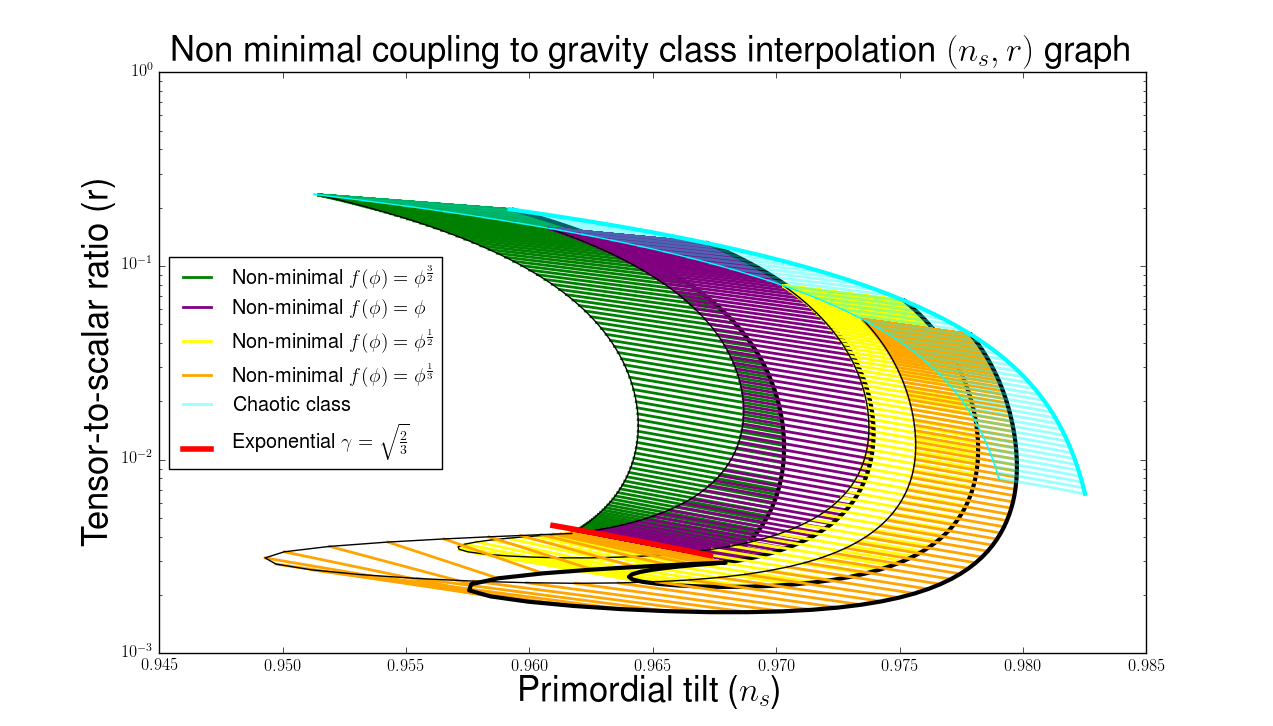}}\\
\caption{   \label{Fig_generalized:figure2}\footnotesize  Numerical predictions of Fig.~\ref{Fig_generalized:figure1} are presented in a semilogarithmic plot.}
\end{figure}

\noindent
This behavior is quite similar to the mechanism of interpolation discussed in~\cite{Binetruy:2014zya}. In these paper we have shown that, by introducing a new scale $f$, it is possible to construct a $\beta$-function that approaches different universality classes as we consider different values for $f$. In particular we have shown that a small and a large field regime can be reached. For example let us consider a model with a scalar field $\chi$ and let us assume that $\beta(\chi) = g(\chi)$ is the $\beta$-function for this model. As we are interested in studying an inflationary stage, the system is close to zero of the $\beta$-function. Without loss of generality we assume that $\beta(\chi = 0) = 0$. Finally we consider a model with $\beta$-function defined by $\beta(\chi) = \epsilon g(\chi)$ where $\epsilon \ll 1$ is a constant. It should be clear that this system inflates for all the values for $\chi$ such that $\beta(\chi) = \epsilon g(\chi) \ll 1$ and thus we can have inflation even for $g(\chi) \sim 1$. As a matter of fact the small field regime is stretched and it can be reached even for bigger values for $\chi$. In the case of a scalar field with a non-minimally coupling with gravity the role of the scale $f$ appears to be played by the coupling $\xi$. In particular this can be shown expressing the $\beta$-function in terms of $\phi$:
\begin{equation}
\label{eq_generalized:beta_phi}
 \bar{\beta}(\phi) = \beta(\varphi(\phi)) = - \left( \frac{\mathrm{d} \phi}{\mathrm{d} \varphi} \right) \frac{\mathrm{d} \ln \bar{V}(\phi)}{\mathrm{d} \phi}.
\end{equation} 
Using Eq.~\eqref{eq_generalized:generalF} and Eq.~\eqref{eq_generalized:F_strong} we express the $\beta$-function in the limit of strong coupling as:
\begin{equation}
\label{eq_generalized:final_beta_phi}
 \bar{\beta}(\phi) = - \sqrt{\frac{8}{3}} \frac{\phi^{-\alpha}}{\xi}.
\end{equation} 
It should be clear that a zero of this function is reached for $1 \ll \phi^\alpha$. However, by choosing a large value for $\xi$, it is still possible to have inflation in the limit of $\phi^\alpha \ll 1$.
In particular, consistently with~\cite{Kallosh:2013tua}, this mechanism allows the production of cosmological perturbations in the regime $ \phi^\alpha \ll 1$. By choosing a large value of $\xi$ we have thus stretched the asymptotic large field regime so that it can be obtained even for small values of $\phi$.

\section{A more general discussion on non-minimal coupling.}
\label{sec_generalized:paper_general_case}
In this section we are interested in discussing a more general parametrization for the model for a scalar field with a non minimal coupling with gravity. In particular we follow the proposal of~\cite{Kallosh:2013tua} and we start by considering the same lagrangian of Eq.~\eqref{eq_generalized:minimal-action} i.e.
 \begin{equation}
  \label{eq_generalized:non_minimal_action_2}
    S =\int\mathrm{d}^4x\sqrt{-g}\left( -  \frac{\Omega(\phi)}{2\kappa^2}R +  X - V_J(\phi) \right),
  \end{equation}
but we introduce an additional functional freedom in the model i.e. 
\begin{equation}
\Omega(\phi) = 1 + \xi f(\phi) ,\qquad \qquad \qquad V_J(\phi) = \lambda^2 g^2(\phi) ,
\end{equation}
where both $f(\phi)$ and $g(\phi)$ are generic functions of $\phi$. As argued in~\cite{Kallosh:2013tua}, by studying the system in terms of $\phi$, it seems reasonable to assume that small variations of the potential should not affect the occurrence of the attractor. In the following we will prove that in general this conclusion does not appear to hold.\\

\noindent
By means of a conformal transformation we recover the Einstein frame action of Eq.~\eqref{eq_generalized:non-minimal-action}. In this case the expressions for $\Omega(\phi)$ and $V(\phi)$ are given by:
\begin{equation}
\label{eq_generalized:omega_V}
\Omega(\phi) = 1 + \xi f(\phi) , \qquad \qquad \bar{V}(\phi) = \frac{\lambda^2 g^2(\phi) }{\Omega^2}.
\end{equation}
Following the procedure defined in Sec.~\ref{sec_generalized:paper_beta_function}, we describe the system in terms of a new field $\varphi$ with a canonically normalized kinetic term. Substituting Eq.~\eqref{eq_generalized:generalF} into Eq.~\eqref{eq_generalized:def_varphi} we get: 
\begin{equation}
\label{eq_generalized:general_vaphi}
\left( \frac{\mathrm{d} \varphi}{\mathrm{d} \phi} \right)^2  = F(\phi) = \frac{1 + \xi f + \frac{3}{2} \xi^2 f^2_{,\phi} }{(1 + \xi f(\phi))^2}.
\end{equation}
The expression for the $\beta$-function associated with the system reads: 
\begin{equation}
\label{eq_generalized:beta_general}
\beta(\varphi) \sim - \frac{\mathrm{d} \ln V(\varphi)}{\mathrm{d} \varphi} = - 2 \left( \frac{\tilde{g}_{,\varphi}(\varphi)}{\tilde{g}(\varphi)} - \frac{\xi \tilde{f}_{,\varphi}(\varphi)}{1 + \xi \tilde{f}(\varphi)} \right), 
\end{equation}
where in analogy with $\tilde{f}(\varphi)$, we defined $\tilde{g}(\varphi) = g(\phi(\varphi))$. Without loss of generality we can parameterizations $\tilde{g}(\phi)$ as: 
\begin{equation}
\label{eq_generalized:eq_parametization}
\tilde{g}(\varphi) = \tilde{f}(\varphi) \tilde{h}(\varphi),
\end{equation}
where $\tilde{h}(\varphi)$ is a generic function of $\varphi$. It is important to stress that we are not specifying an explicit expression for $\tilde{h}(\varphi)$ and thus we can produce a quite general description of the problem. We can substitute Eq.~\eqref{eq_generalized:eq_parametization} into Eq.~\eqref{eq_generalized:beta_general} to get:
\begin{equation}
\label{eq_generalized:beta_general_2}
\beta(\varphi) = - 2 \left\{ \frac{\tilde{h}_{,\varphi}(\varphi)}{\tilde{h}(\varphi)} +  \frac{ \tilde{f}_{,\varphi}(\varphi)}{\tilde{f}(\varphi) \left[ 1 + \xi \tilde{f}(\varphi) \right]} \right\}.
\end{equation}
It is easy to notice that in the case of $\tilde{h}_{,\varphi}(\varphi)/\tilde{h}(\varphi) = 0$, this equation is exactly equal to Eq.~\eqref{eq_generalized:beta_approx_varphi}. In particular, in the limit of strong coupling $\xi$ Eq.~\eqref{eq_generalized:beta_general_2} simply reads:
\begin{equation}
\label{eq_generalized:beta_general_3}
\beta(\varphi) = - 2 \left[ \frac{\tilde{h}_{,\varphi}(\varphi)}{\tilde{h}(\varphi)} +  \frac{ \tilde{f}_{,\varphi}(\varphi)}{ \xi \tilde{f}^2(\varphi) } \right].
\end{equation}
It is interesting to point out that choosing $\tilde{f}(\varphi) = \tilde{g}(\varphi)$ or equivalently $\tilde{h}(\varphi) = 1$, the zero order term in $1/\xi$ is set equal to zero. Under this assumption the expression for $\beta(\varphi)$ is thus dominated by the first order term $1/\xi$. In particular the $\beta$-function is simply given by Eq.~\eqref{eq_generalized:beta_strong} and thus we reduce to the case discussed in Sec.~\ref{sec_generalized:paper_strong}. Relaxing the assumption of $\tilde{h}(\varphi) = 1$, we can consider the case of a zero order term different from zero. As an inflationary stage corresponds to $\beta(\varphi) \rightarrow 0$ any choice of $\tilde{h}(\varphi)$ that satisfies: 
\begin{equation}
\label{eq_generalized:eq_evasion}
 \frac{ \tilde{f}_{,\varphi}(\varphi)}{ \xi \tilde{f}^2(\varphi) } \ll \frac{\tilde{h}_{,\varphi}(\varphi)}{\tilde{h}(\varphi)} \rightarrow 0,
\end{equation}
describing a viable inflationary model. As no other restriction has been imposed on the choice for $\tilde{h}(\varphi)$, we can immediately conclude that in general the attractor at strong coupling can be evaded. In the Sec.~\ref{sec_generalized:paper_general_example} we present an explicit example to discuss the conditions to preserve the attractor at strong coupling. In Sec.~\ref{sec_generalized:paper_further_generalizations} we show that models defined via further generalizations of the action Eq.~\eqref{eq_generalized:non_minimal_action_2} are still included in this class and we investigate the characterization of the $\alpha$-attractors of Kallosh and Linde~\cite{Kallosh:2013hoa,Kallosh:2013daa,Kallosh:2013yoa,Kallosh:2015lwa} in terms of our formalism. Some other examples of the parametrization for $\tilde{h}(\varphi)$ are discussed in Appendix~\ref{sec_generalized:Appendix}.

\subsection{Polynomial expansion.}
\label{sec_generalized:paper_general_example}
Let us assume that $f(\phi)$ and $g(\phi)$ admit a Taylor expansion in terms of $\phi$:
\begin{equation}
\label{eq_generalized:Taylor}
f(\phi) = \sum_{i = 0}^\infty f_i \phi^i, \qquad \qquad g(\phi) = \sum_{i = 0}^\infty g_i \phi^i .
\end{equation}
Let us restrict to the case of both $f(\phi)$ and $g(\phi)$ vanishing for a certain value of $\phi$. By means of a field redefinition we can fix $f_0 = g_0 = 0$. Without loss of generality we can also rescale $\lambda $ and $\xi$ to impose $f_1 = g_1 = 1$. The case $f(\phi) = g(\phi)$ has been discussed in Sec.~\ref{sec_generalized:paper_strong}, and in particular we have shown that under this assumption it is possible to choose $\xi$ such that $\phi \ll 1$. As the first order terms of Eq.~\eqref{eq_generalized:Taylor} are imposed to be equal and high orders in terms of $\phi$ are expected to be negligible, it would be reasonable to conclude that the attractor at strong coupling should be preserved. Surprisingly, expressing the dynamics in terms of $\varphi$, it is possible to show that the attractor at strong coupling may be evaded! Let us fix a particular expression for $f(\phi)$ and $g(\phi)$, in particular we choose:
 \begin{equation}
\label{eq_generalized:explicit_expansion}
f(\phi) = \phi, \qquad \qquad g(\phi) =\phi + g_{n+1} \phi^{n+1} = \phi (1 +g_{n+1} \phi^{n} ) .
\end{equation}
In the strong coupling limit, the lowest order approximation for Eq.~\eqref{eq_generalized:general_vaphi} simply reads:
\begin{equation}
  \label{eq_generalized:example_beta_strong}
    \left( \frac{ \textit{d} \varphi}{ \textit{d} \phi} \right)^2 =  F(\phi) \simeq \frac{3}{2}\left( \frac{f_{,\phi} (\phi)}{ f(\phi)} \right)^2.  
\end{equation} 
We can integrate Eq.~\eqref{eq_generalized:example_beta_strong} to get an explicit expression for $\tilde{f}(\varphi)$:
\begin{equation}
\label{eq_generalized:example_varphi}
\tilde{f}(\varphi) = \tilde{f}_{\mathrm{f}}\exp \left[ \sqrt{\frac{2}{3}} (\varphi - \varphi_{\mathrm{f}} ) \right].
\end{equation} 
Finally we substitute into Eq.~\eqref{eq_generalized:explicit_expansion} to get:
\begin{eqnarray}
\label{eq_generalized:example_phi_varphi}
\phi &=& f(\phi) = \tilde{f}(\varphi) = \tilde{f}_{\mathrm{f}}\exp \left[ \sqrt{\frac{2}{3}} (\varphi - \varphi_{\mathrm{f}} ) \right], \\
\label{eq_generalized:explicit_g}
\tilde{g} (\varphi) &=& \tilde{f}_{\mathrm{f}}\exp \left[  \sqrt{\frac{2}{3}} (\varphi - \varphi_{\mathrm{f}} ) \right] \left\{ 1 + g_{n+1} \tilde{f}_{\mathrm{f}}^{\ n} \exp \left[  \sqrt{\frac{2}{3}} n (\varphi - \varphi_{\mathrm{f}} ) \right] \right\},
\end{eqnarray}
where $\tilde{g} (\varphi)  = g(\phi(\varphi))$. It should be clear that this corresponds to: 
\begin{equation}
\label{eq_generalized:example_h}
 \tilde{h}(\varphi) =  1 + g_{n+1} \tilde{f}_{\mathrm{f}}^{\ n} \exp \left[  \sqrt{\frac{2}{3}} n (\varphi - \varphi_{\mathrm{f}} ) \right].
\end{equation}
Using Eq.~\eqref{eq_generalized:beta_general_2} we can then compute the explicit expression for $\beta(\varphi)$: 
\begin{equation}
\label{eq_generalized:explicit_beta}
 \beta(\varphi) = -\sqrt{\frac{8}{3}}  \left\{ \frac{n g_{n+1} \tilde{f}^{\ n}_{\mathrm{f}} \exp \left[  \sqrt{\frac{2}{3}} n (\varphi - \varphi_{\mathrm{f}} ) \right]  }  {1 + g_{n+1} \tilde{f}^{\ n}_{\mathrm{f}}\exp \left[  \sqrt{\frac{2}{3}} n (\varphi - \varphi_{\mathrm{f}} ) \right] }  +  \frac{1  }{1 + \xi  \tilde{f}_{\mathrm{f}}\exp \left[  \sqrt{\frac{2}{3}} (\varphi - \varphi_{\mathrm{f}} ) \right] } \right\}.
\end{equation}
Notice that the first term on the right hand side of Eq.~\eqref{eq_generalized:explicit_beta} gives the zero order contribution in $1/\xi$ while the second term on the right hand side of Eq.~\eqref{eq_generalized:explicit_beta} is a first order term in $1/\xi$. It is important to stress that imposing $ g_{n+1} = 0$, is equivalent to fix $f(\phi) = g(\phi)$. As discussed in Sec.~\ref{sec_generalized:paper_strong}, in this case the inflationary regime is reached for large positive values for $\varphi$ and $\beta(\varphi)$ is approximated by Eq.~\eqref{eq_generalized:beta_strong_ff}. On the contrary when $ g_{n+1} \neq 0$, the second term on the right hand side of Eq.~\eqref{eq_generalized:explicit_beta} is negligible with respect to the first one\footnote{The consistency of this assumption is discussed in the following.}. Under this assumption $\beta(\varphi)$ can be approximated as:
\begin{equation}
\label{eq_generalized:explicit_beta_2}
 \beta(\varphi) \sim -\sqrt{\frac{8}{3}}  \left\{ \frac{ n g_{n+1} \tilde{f}^{\ n}_{\mathrm{f}} \exp \left[  \sqrt{\frac{2}{3}} n (\varphi - \varphi_{\mathrm{f}} ) \right]  }  {1 + g_{n+1} \tilde{f}^{\ n}_{\mathrm{f}}\exp \left[  \sqrt{\frac{2}{3}} n (\varphi - \varphi_{\mathrm{f}} ) \right] } \right\}.
\end{equation}
In this case the zero of $\beta(\varphi)$ that corresponds to the inflationary phase is thus reached for large negative values for $\varphi$. In this regime the expressions for $\beta(\varphi)$ and $N(\varphi)$ are:
\begin{eqnarray}
\label{eq_generalized:explicit_beta_3}
 \beta(\varphi) &\sim& -\sqrt{\frac{8}{3}} n g_{n+1} \tilde{f}^{\ n}_{\mathrm{f}} \exp \left[  \sqrt{\frac{2}{3}} n (\varphi - \varphi_{\mathrm{f}} ) \right] ,  \\
\label{eq_generalized:efold_explicit}
N(\varphi) &=& - \frac{3}{4} \frac{1}{n^2 g_{n+1} \tilde{f}^{\ n}_{\mathrm{f}}}  \left\{ \exp \left[ - \sqrt{\frac{2}{3}} n \left( \varphi - \varphi_{\mathrm{f}} \right)     \right]  - 1 \right\}.
\end{eqnarray}
To ensure $0 \leq N(\varphi)$, we impose $0 < - f^n_{\mathrm{f}} g_{n+1}$. Following the same procedure of Sec.~\ref{sec_generalized:paper_strong}, we also impose the condition $\left| \beta(\varphi_{\mathrm{f}}) \right| \sim 1 $ to fix the value of $\beta(\varphi)$ at the end of inflation: 
\begin{equation}
\label{eq_generalized:condition}
 \left| \beta(\varphi_{\mathrm{f}}) \right| = \left| \sqrt{\frac{8}{3}} n g_{n+1} \tilde{f}^{\ n}_{\mathrm{f}}  \right| \sim 1.
\end{equation}
As $n$ and $g_{n+1}$ are expected to be of order one, we can conclude that $ f_{\mathrm{f}}$ is expected to be of order one too. Finally, using Eq.~\eqref{eq_generalized:condition} we express $\beta(\varphi)$ and $N(\varphi)$ as:
\begin{eqnarray}
\label{eq_generalized:explicit_beta_final}
 \beta(\varphi) &\sim& \exp \left[  \sqrt{\frac{2}{3}} n (\varphi - \varphi_{\mathrm{f}} ) \right], \\
 \label{eq_generalized:explicit_efold_final}
 N(\varphi) &=& \sqrt{\frac{3}{2 n^4}} \left\{ \exp \left[ - \sqrt{\frac{2}{3}} n \left( \varphi - \varphi_{\mathrm{f}} \right)     \right]  - 1 \right\}.
\end{eqnarray}
The expression for $\beta(\varphi)$, in the limit of big $\xi$, thus depends on $n$ and this leads to the evasion from the universality. In particular, $\beta(\varphi)$ approaches the exponential class of~\cite{Binetruy:2014zya} with $\gamma = n \sqrt{2 /3 }$. The corresponding expression for the scalar spectral index and for the tensor to scalar ratio are given by: 
\begin{equation}
\label{eq_generalized:ns_r_evaded_attractor}
n_s = 1 - \frac{2}{N}, \qquad \qquad \qquad r = \frac{12}{n^2 N^2}.
\end{equation}
It is interesting to notice that the attractor at strong coupling of~\cite{Kallosh:2013tua} can be reproduced by imposing $n = 1$. Actually it is possible to go further and prove that the attractor can be recovered under some more general condition. As argued during this section, the inflationary phase is reached for $\varphi \ll -1$ and this corresponds to $\phi \ll 1$. In this regime higher order corrections to the expression of $g(\phi)$:
 \begin{equation}
\label{eq_generalized:generalized_expansion}
f(\phi) = \phi, \qquad \qquad g(\phi) =\phi + g_{n+1} \phi^{n+1} + \sum_{i = n+2 }^\infty g_i \phi^i  ,
\end{equation}
are not producing significant changes in the lowest order expressions for $\beta(\varphi)$ and $N(\varphi)$ given in Eqs.~\eqref{eq_generalized:explicit_beta_final},\eqref{eq_generalized:explicit_efold_final}. In particular this implies that assuming $g_2 \neq 0$, the dominant contribution to the expression of $\beta(\varphi)$ is fixed by the term with $n=1$. This condition is thus sufficient to preserve the attractor\footnote{Notice that this is a specific feature of the parametrization of Eq.~\eqref{eq_generalized:generalized_expansion}. As discussed in Sec.~\ref{sec_generalized:paper_general_case}, the attractor can be evaded under the quite general condition of Eq.~\eqref{eq_generalized:eq_evasion}. Some explicit examples of the evasion are presented in the appendix~\ref{sec_generalized:Appendix}.}. Conversely, other attractors are find for different values of $1<n$. \\

\noindent
To conclude this section we discuss the consistency of the assumption that the second term on the right hand side of Eq.~\eqref{eq_generalized:explicit_beta} is negligible with respect to the first one. To be sure that this term is subdominant from the end of inflation up to the production of cosmological perturbations we need: 
\begin{equation}
\label{eq_generalized:attractor_condition}
  \frac{1  }{1 + \xi  \tilde{f}_{\mathrm{f}}\exp \left[  \sqrt{\frac{2}{3}} (\varphi_{\mathrm{H}} - \varphi_{\mathrm{f}} ) \right] }  \ll  \exp \left[  \sqrt{\frac{2}{3}} n (\varphi_{\mathrm{H}} - \varphi_{\mathrm{f}} ) \right]   \ll 1,
\end{equation} 
where $\varphi_{\mathrm{H}}$ is the value of $\varphi$ at the production of cosmological perturbation. Using the expression for $N(\varphi)$ given by Eq.~\eqref{eq_generalized:explicit_efold_final} it is clear that Eq.~\eqref{eq_generalized:attractor_condition} satisfied if $N_{\mathrm{H}}^2/\xi \ll 1$.

\subsection{$\alpha$-attractors.}
\label{sec_generalized:paper_further_generalizations}
It is interesting to notice that the class of models described in this section also includes further generalizations of the lagrangian of Eq.~\eqref{eq_generalized:non_minimal_action_2}. In particular some of these generalizations have been presented in~\cite{Galante:2014ifa} and~\cite{Kallosh:2014laa}. Following the proposal of~\cite{Galante:2014ifa} we consider the general Jordan frame action\footnote{$\kappa^2$ is set equal to 1.} to describe a homogeneous scalar field with a non-minimal coupling with gravity:
  \begin{equation}
  \label{eq_generalized:general_action_jordan}
    S=\int\mathrm{d}^4x\sqrt{-g}\left( -  \Omega(\phi)\frac{R}{2} + K_J(\phi) X - V_J(\phi) \right).
  \end{equation}
As usual we perform a conformal transformation:
  \begin{equation}
\label{eq_generalized:conformal}
  g_{\mu\nu} \rightarrow \Omega(\phi)^{ -1} g_{\mu\nu},
  \end{equation}
to get the Einstein frame formulation of the theory: 
    \begin{equation}
  \label{eq_generalized:general_action_einstein}
    \mathcal{L}_E = - \frac{R}{2} +  F(\phi)  X - V(\phi) ,
  \end{equation}
  where we defined $F(\phi)$ and $V(\phi)$ as:
  \begin{equation}
  F(\phi) \equiv \left[ \frac{K_J (\phi)}{\Omega(\phi)} + \frac{3}{2} \left(  \frac{\textrm{d} \ln \Omega(\phi)}{\textrm{d}\phi}\right)^2 \right] \qquad \qquad V(\phi) = \frac{V_J(\phi)}{\Omega^2(\phi)}.
  \end{equation}
It is clear that the cases discussed in the previous sections can be recovered simply imposing $K_J (\phi) = 1$. Again we can define a new field $\varphi$:
    \begin{equation}
  \label{eq_generalized:general_varphi_def}
    \left(  \frac{\textrm{d} \varphi}{\textrm{d}\phi}\right)^2  \equiv F(\phi) = \left[ \frac{K_J (\phi)}{\Omega(\phi)} + \frac{3}{2} \left(  \frac{\textrm{d} \ln \Omega(\phi)}{\textrm{d}\phi}\right)^2 \right] ,
  \end{equation}
that has a canonically normalized standard kinetic term. In particular the lagrangian for this field simply reads:
    \begin{equation}
  \label{eq_generalized:general_varphi_einstein}
    \mathcal{L}_E = - \frac{R}{2} + \frac{(\partial \varphi)^2}{2} - \tilde{V}(\varphi) ,
  \end{equation}
  where $\tilde{V}(\varphi) $ is defined as $\tilde{V}(\varphi) = V(\phi(\varphi))$. As in terms of the canonically normalized field $\varphi$ the three functional dependence are merged into $\tilde{V}(\varphi)$, the model construction reduces to fixing a particular parametrization for this function. Using Eq.~\eqref{eq_generalized:beta_varphi}, and the lowest order approximation $V(\varphi) \sim \frac{3}{4} W^2(\varphi)$ we can finally express the $\beta$-function as:
    \begin{equation}
  \label{eq_generalized:beta_varphi_def}
   \beta(\varphi) \sim -  \frac{\textrm{d} \ln \tilde{V}(\varphi)}{\textrm{d}\varphi} .
  \end{equation}
Again the whole dynamics of the model is thus fixed by the parametrization of the $\beta$-function. As different choices for $\Omega(\phi), K_J(\phi)$ and $V_J(\phi)$ lead to the same expression for $\beta$, this explains the possibility for degeneracies to arise.\\

\noindent
Several models described by the action of Eq.~\eqref{eq_generalized:general_action_jordan} has been presented in~\cite{Galante:2014ifa} and~\cite{Kallosh:2014laa}. In this paper we consider the $\alpha$-attractors of~\cite{Kallosh:2015lwa} as an interesting example for this class of models. In particular let us consider the case of \emph{T-models}~\cite{Kallosh:2013hoa}. \emph{T-models} can be described in terms of the action of Eq.~\eqref{eq_generalized:general_action_jordan} by fixing:  
\begin{equation}
\label{eq_generalized:alpha_attrators}
  \left(  \frac{\textrm{d} \varphi}{\textrm{d}\phi}\right)^2 = F(\phi) = \left( 1 - \frac{\phi^2}{6 \alpha}\right)^{-2} \qquad \qquad \qquad V(\phi) = \frac{m^2}{2} \phi^2.
  \end{equation}
Using Eq.~\eqref{eq_generalized:alpha_attrators} we define the canonically normalized field and using Eq.~\eqref{eq_generalized:omega_V} and Eq.~\eqref{eq_generalized:eq_parametization} we can compute the explicit expression for $h(\phi)$. In particular we get:
 \begin{equation}
\phi = \sqrt{6\alpha} \tanh\left(\frac{\varphi}{\sqrt{6\alpha}} \right), \qquad \qquad \qquad h(\phi) \sim \phi.
  \end{equation}
Finally we can use Eq.~\eqref{eq_generalized:beta_general_3} to compute the explicit expression for the $\beta$-function:
 \begin{equation}
 \label{eq_generalized:beta_alpha_attractors}
 \beta(\varphi) = - \sqrt{\frac{2}{3 \alpha}} \left[ \frac{1 - \tanh^2 \left( \frac{\varphi}{\sqrt{6 \alpha}} \right)}{\tanh \left( \frac{\varphi}{\sqrt{6 \alpha}} \right)} \right] \sim -  \exp \left[ - \sqrt{\frac{2}{3 \alpha}} ( \varphi-\varphi_f) \right].
  \end{equation}
Eq.~\eqref{eq_generalized:beta_alpha_attractors} implies that the $\beta$ function for \emph{T-models} falls in the exponential class of~\cite{Binetruy:2014zya}. As already discussed in this paper, the predictions for $n_s$ and $r$ are thus given by:
\begin{equation}
\label{eq_generalized:ns_r_alpha_attractor}
n_s = 1 - \frac{2}{N}, \qquad \qquad \qquad r = \frac{8}{\gamma^2 N^2} = \frac{12 \alpha}{N^2}.
\end{equation}
Similar conclusions can be draw for the other models for $\alpha$-attractors presented in~\cite{Kallosh:2013hoa}.

\section{Conclusions.}
\label{sec_generalized:paper_conclusions}

In the analysis of this paper, by means of a conformal transformation and of a field redefinition, we have discussed the problem of inflationary models with a non-minimal coupling with gravity in terms of a single field with a canonically normalized kinetic term. In particular we have shown, the application of the $\beta-$function formalism, helps to understand the asymptotic behavior of the system during the inflationary phase. In this framework, the fall of the system into the attractor is interpreted as the approach of a universality class. In this sense, the formulation of the problem in this framework, should not be seen as a simple rewriting of the results obtained with standard methods, but on the contrary it should be considered as a further generalization.\\

\noindent
The $\beta-$function formalism appears to be extremely useful when we investigate the stability of the attractor at strong coupling under generalizations of the theory. In particular, once we have defined the $\beta-$function associated with our system, it has been easy to identify the dominant contribution to characterize inflation. Specifically, in Sec.~\ref{sec_generalized:paper_general_case}, we have discussed the possibility of introducing an additional functional freedom in the model. In this case the behavior of the system is dominated by the zeroth order term that conversely was set equal to zero in the treatment followed in Sec.~\ref{sec_generalized:paper_beta_function}. As in general this term can be chosen arbitrarily, it leads to the possibility of evading the attractor at strong coupling. A critical review of the conditions required to preserve the attractor at strong coupling has been presented and the existence of different attractors has been shown. \\

\noindent
The further generalization discussed in~\cite{Galante:2014ifa} and~\cite{Kallosh:2014laa} have been presented. In these works it was shown that a slight modification of the theory may lead to the existence of other attractors. Indeed for these models an analogous of the treatment presented in this paper can be carried out and it leads to similar conclusions. In particular we have presented the application of our formalism to the case of the $\alpha$-attractors of~\cite{Kallosh:2015lwa}. A further generalization of the formalism proposed in~\cite{Binetruy:2014zya} can be also useful to have a deeper understanding of more general models with a non-standard kinetic term or with more scalar fields\cite{Kaiser:2013sna}. It seems reasonable to suppose that in analogy with the case of the non-minimal coupling, the $\beta$-function formalism can be coherently applied to these models as well.

\section*{ Acknowledgements}
I would like to thank Nathalie Deruelle, David Kaiser, Andrei Linde and Diderik Roest for their suggestions. In particular I would like to thank Pierre Bin\'etruy for all the useful discussions that lead to the production of this work. I acknowledge the financial support of the UnivEarthS
Labex program at Sorbonne Paris Cit\'e (ANR-10-LABX-0023 and ANR-11-IDEX-0005-02) and the Paris Centre for Cosmological Physics.

\newpage
\section*{Appendix.}
\section{Some explicit examples.}
\label{sec_generalized:Appendix}
In Sec.~\ref{sec_generalized:paper_general_case}, we discussed the consequences of introducing a further functional freedom in the Jordan frame forumalation of the model. In particular we considered:
\begin{equation}
  \label{eq_generalized:non_minimal_action_3}
    S =\int\mathrm{d}^4x\sqrt{-g}\left( -  \frac{\Omega(\phi)}{2\kappa^2}R +  X - V_J(\phi) \right),
  \end{equation}
where $\Omega(\phi)$ and $V_J (\phi)$ have been defined as:
\begin{equation}
\Omega(\phi) = 1 + \xi f(\phi) ,\qquad \qquad \qquad V_J(\phi) = \lambda^2 g^2(\phi).
\end{equation}
After the usual conformal transformation we recover the Einstein frame formulation of the theory. By means of a field redefinition we are finally able to describe the system in terms of a field with a canonically normalized kinetic term. The strong coupling expression for $\tilde{f}(\varphi)$ is fixed by Eq.~\eqref{eq_generalized:general_vaphi} and thus the model definition reduces to fixing an explicit expression for $g(\phi)$. In this appendix we consider some parameterizations for $\tilde{g}(\varphi)$ to study the possibility of preserving and evading the attractor. In particular we show that different universality classes can be reached.
\begin{itemize}
\item \textbf{Exponential.}
Let us consider:
\begin{equation}
 \tilde{g}(\varphi) =  \exp\left[ \sqrt{\frac{2}{3}} \left( \varphi - \varphi_{\mathrm{f}} \right) - \frac{e^{-\alpha (\varphi - \varphi_{\mathrm{f}})}}{\alpha}  \right].
\end{equation}
It is straightforward to derive the lowest order expression for $\beta(\varphi)$:
\begin{equation}
 \beta(\varphi) \sim - \frac{2 e^{-\alpha (\varphi - \varphi_{\mathrm{f}})}}{\alpha}. 
\end{equation}
This expression for $\beta(\varphi)$ corresponds to the exponential class presented in~\cite{Binetruy:2014zya}. In this case the predictions for $n_s$ and $r$ are:
  \begin{eqnarray} 
            n_{s} &\simeq& 1 - \frac{2}{N}, \\
            r & \simeq& \frac{8}{\alpha^2 N^2}.
    \end{eqnarray}
It may be interesting to notice that the attractor at strong coupling of~\cite{Kallosh:2013tua} can only be reproduced for $\alpha = \sqrt{2/3}$. For any other value for $\alpha$ the attractor is evaded.

\item \textbf{Chaotic.}
Let us consider:
\begin{equation}
 \tilde{g}(\varphi) = \left( \varphi - \varphi_{\mathrm{f}} \right)^\alpha \exp\left[ \sqrt{\frac{2}{3}} \left( \varphi - \varphi_{\mathrm{f}} \right) \right]
\end{equation}
clearly $ \tilde{g}_{,\varphi}(\varphi) / \tilde{g}(\varphi) = \sqrt{2/3} + \alpha /( \varphi - \varphi_{\mathrm{f}})$ that gives the lowest order expression:
\begin{equation}
\label{eq_generalized:beta_general_ex}
\beta(\varphi) = \frac{- 2\alpha}{ \varphi - \varphi_{\mathrm{f}}}.
\end{equation}
This case corresponds to the Chaotic class discussed in~\cite{Binetruy:2014zya} and gives:
  \begin{eqnarray} 
            n_{s} &\simeq &1 - \frac{2+a}{2N}, \\
            r & \simeq & \frac{4a}{N^2}.
    \end{eqnarray}
In this case the attractor at strong coupling is clearly evaded.

\item \textbf{Polynomial.}
Let us consider:
\begin{equation}
 \tilde{g}(\varphi) = \left\{  1 + \frac{P_1(\varphi)}{\exp \left[ \sqrt{\frac{2}{3}} \left( \varphi - \varphi_{\mathrm{f}} \right) \right]} \right\} \exp\left[ \sqrt{\frac{2}{3}} \left( \varphi - \varphi_{\mathrm{f}} \right) \right]
\end{equation}

where $P_1(\varphi)$ is a polynomial in $\varphi$. It is possible to prove that in this case the lowest order expression for $\beta(\varphi)$ reads:
\begin{equation}
\beta(\varphi) \sim \frac{P_2(\varphi)}{  \tilde{f}(\varphi)},
\end{equation}
where $P_2(\varphi)$ is a polynomial in $\varphi$. It is possible to show that at the lowest order the expressions for $n_s$ and $r$ are:
        \begin{eqnarray} 
            n_{s} - 1 &\simeq& - \frac{2}{N}, \\
            r & \simeq& \frac{12}{N^2}  .
    \end{eqnarray}
In this case the attractor is always preserved independently on the explicit expression for $P_1(\varphi)$. 
\end{itemize}

 {\large \par}}
{\large 
\chapter{An inflationary landscape.}
\label{chapter:pseudoscalar}

\horrule{0.1pt} \\[0.5cm]

\begin{abstract} 

\noindent 
In this Chapter we discuss the framing of inflation in a realistic landscape for early time cosmology. In this context we discuss the possibility of coupling the inflaton to some other particles. As we explain through this Chapter, this may change dramatically several features of inflation giving rise to several observational consequences. Our discussion is focused on the case of a pseudo-scalar inflaton. In this case, the generic coupling to any Abelian gauge field may strongly affect the background dynamics and give rise to a strong enhancement of the scalar and tensor power spectra. The main observational consequences are then discussed.
\end{abstract}

\horrule{0.1pt} \\[0.5cm] 

\noindent
Inflation is nowadays accepted as a cornerstone of modern cosmology. As explained through this work, its simplest realization in terms of single field models appears to fit with the cosmological observations at CMB scales. However, while the main mechanisms that drive inflation are basically understood, we are still far from the definition of a realistic model of inflation that is consistent with our knowledge of the fundamental interactions. In particular, a convincing description of the interactions between the inflaton and the particles that are framed in the Standard Model (SM) of particle physics still lacks. In this Chapter we discuss a realization of inflation where other particles (gauge fields) are present. In this context, we present some of the main consequences of this generalized landscape on the inflationary predictions. We show that in this modified framework there are significant consequences both on the background dynamics and on the perturbations. Remarkably, we show that it is possible to induce an exponential enhancement of the scalar and tensor power spectra at small scales. This mechanism may lead to several observational consequences such as the production of observable Primordial Gravitational Waves (GW)~\cite{Cook:2011hg,Barnaby:2011qe, Barnaby:2011vw,Domcke:2016bkh}, the presence of a non-Gaussian component in the scalar power spectrum~\cite{Barnaby:2011qe,Barnaby:2011vw,Anber:2012du,Barnaby:2010vf,Linde:2012bt,Domcke:2016bkh}, the generation of Primordial Black Holes (PBH)~\cite{Linde:2012bt,Domcke:2016bkh} and the generation of $\mu$-distortions~\cite{Domcke:2016bkh}. \\

\noindent In this Chapter we proceed as follows. We start by discussing (in Sec.~\ref{sec_pseudoscalar:mechanism}) the consequences of the introduction of a generalized coupling between a \emph{pseudo-scalar} inflaton and some \emph{Abelian} gauge fields. In Sec.~\ref{sec_pseudoscalar:signatures}, we discuss some of the possible observational consequences and the possible constraints that they can put on these models. In Sec.~\ref{sec_pseudoscalar:analytical}, we produce a model-independent discussion of the shape of the scalar and tensor spectra and in Sec.~\ref{sec_pseudoscalar:models} we discuss the different models that are grouped in classes. In this Chapter we pay particular attention on a class of models that according to the choice of~\cite{Domcke:2016bkh} is called Starobinsky-like potentials\footnote{Pseudo-scalar fields effectively describing the Starobinsky-like model of inflation can be considered in the context of supergravity by employing a shift-symmetry in the K\"ahler potential~\cite{Kawasaki:2000yn} (see for example~\cite{Dall'Agata:2014oka,Kallosh:2010ug,Kallosh:2010xz}). On the other hand, these models may be difficult to obtain from string theory. In this context, the coupling $\phi F \tilde{F}$ is associated with the presence of a pseudo-anomalous $U(1)$ and thus the pseudo-scalar field is an axion. Few more details on the embedding in supergravity are given in Sec.~\ref{sec_pseudoscalar:discussion}. However, a detailed discussion of the UV completion of the models considered in this Chapter goes beyond the scope of this work.}. These model actually correspond to the exponential class of Chapter~\ref{chapter:beta}, but in this case the field is a pseudo-scalar. These models actually appear to be the most promising for what concerns the production of observable GW. In order to have a better understanding of the observational prospects, we also present two (parameter) scan plots for these models.

\section{Pseudo-scalar inflaton in the presence of gauge fields.}
\label{sec_pseudoscalar:mechanism}
In this Section we discuss the case of a \emph{pseudo-scalar} inflaton $\phi$ in presence of some Abelian gauge fields. In particular, we are interested in discussing this problem when we introduce a generic higher-dimensional coupling between the inflaton and the gauge fields\footnote{Notice that in the context of an Effective Field Theory (EFT) the introduction of this term is perfectly consistent and rather unavoidable.}. As we show in the following, the presence of this term in the theory produces an instability, which leads to an exponential enhancement of the gauge fields~\cite{Turner:1987bw, Garretson:1992vt, Anber:2006xt}. As a consequence, the presence of the gauge fields induces a back-reaction both on the background dynamics~\cite{Anber:2009ua,Barnaby:2011qe,Barnaby:2011vw} and on the perturbations~\cite{Anber:2012du,Linde:2012bt}. The main effect on the background dynamics is the introduction of a new friction term. Such a term is sourced by the gauge fields and dominates the last part of the evolution. At the same time, the gauge fields act as a source both for scalar and tensor perturbations, leading to an amplification of the spectra at small scales. Remarkably, under particular conditions that depend on the parameters of the model, it is possible to generate a signal in the observable range of direct GW detectors.

\subsection{Background field equations.}
\label{sec_pseudoscalar:background_solution}
We start our treatment by considering the action~\cite{Turner:1987bw, Garretson:1992vt, Anber:2006xt,Anber:2012du,Anber:2009ua,Barnaby:2011qe,Barnaby:2011vw,Linde:2012bt} for a pseudo-scalar inflaton $\phi$, that is non-minimally coupled to a certain number $\mathcal{N}$, of Abelian gauge fields $A_\mu^a$ associated to $U(1)$ gauge symmetries:
\begin{equation}
\label{eq_pseudoscalar:action_pseudoscalar}
\mathcal{S}= \int \textrm{d}^4 x \sqrt{|g|} \left[\frac{R}{2 \kappa^2} -\frac{1}{2} \partial_\mu \phi \partial^\mu \phi - V(\phi) - \frac{1}{4} F^a_{\mu \nu} F_a^{\mu \nu} - \frac{\alpha^a}{4 \Lambda} \phi F^a_{\mu \nu} \tilde{F}_a^{\mu \nu} \right ]\ ,
\end{equation}
where consistently with the rest of this work, the background metric is expressed as $g_{\mu\nu} = \textrm{diag}(-1 ,a^2(t),a^2(t),a^2(t))$. Let us explain in detail all the terms appearing in this action. The first two terms of this action are respectively the kinetic term and the potential for the inflaton. The $F^a_{\mu \nu}$ terms are the usual field strength tensors for the gauge fields\footnote{As usual in gauge theories, the field strength is defined as the commutator of two covariant derivatives. Given a gauge field $A_{\mu}$, and the coupling constant $g$, the covariant derivative and the field strength tensor are defined as:
\begin{equation}
	D_{\mu} \equiv \partial_\mu - g A_{\mu}  \ , \qquad \qquad  F_{\mu \nu} \equiv - \frac{1}{g} \left[ D_{\mu} , D_{\nu} \right] = \partial_\mu A_\nu - \partial_\nu A_\mu - g \left[ A_{\mu} , A_{\nu} \right] .
\end{equation}
In the case of Abelian gauge field we have $\left[ A_{\mu} , A_{\nu} \right] = 0$.}. On the other hand, the term $\tilde{F}_a^{\mu \nu}$ corresponds to the dual field strength tensor defined as:
\begin{equation}
	\label{eq_pseudoscalar:dual_tensor}
	\tilde{F}_a^{\mu \nu} \equiv \epsilon^{\mu\nu\rho\sigma} F_{a \rho \sigma} \equiv \frac{1}{2} \frac{\varepsilon^{\mu\nu\rho\sigma}}{\sqrt{|g|}} F_{a \rho \sigma} \ ,
\end{equation}
where $\varepsilon^{\mu\nu\rho\sigma}$ is the Levi-Civita symbol. The constant $\Lambda$ is a mass scale, that as usual in the framework of EFT, is introduced in order to suppress higher-dimensional operators of the theory. The dimensionless constants $\alpha^a$, are the coupling constants that parametrize the strength of the interactions between the inflaton and the gauge fields. Notice that a term proportional to $F^a_{\mu \nu} \tilde{F}_a^{\mu \nu}$, may be expressed as a total derivative and thus it does not affect the dynamics. On the contrary this result does not hold for the higher-dimensional term $\alpha^a \phi F^a_{\mu \nu} \tilde{F}_a^{\mu \nu}/ 4 \Lambda$. For simplicity, in the following we consider $\alpha^a = \alpha$ for all $a = \{1,2,..\mathcal{N}\}$. \\

\noindent
Let us start by computing the background equations of motion for the inflaton $\phi(t)$ and for the gauge fields $A_\mu^a(t,x)$. Without loss of generality, we assume $ \phi > 0, \ V_{,\phi}(\phi)>0, \ \dot{\phi} < 0$ and we choose to describe the problem in the Coulomb gauge ($A_0^a = 0$, $\partial^\mu A_\mu^a(t,x) = 0$). Under these assumptions the equations of motion can be expressed as:
\begin{align}
\label{eq_pseudoscalar:eq_motion}
\ddot \phi + 3 H \dot{\phi} + \frac{\partial V}{\partial \phi} & = \frac{\alpha}{2\Lambda} \frac{\varepsilon^{\mu\nu\rho\sigma}}{\sqrt{|g|}} \langle \partial_\mu A_\nu \partial_\rho A_\sigma \rangle \equiv \frac{\alpha}{\Lambda} \langle \vec{E}^a \cdot \vec{B}^a \rangle \ ,\\
\frac{d^2}{d \tau^2}\vec{A}^a - \nabla^2 \vec{A}^a - \frac{\alpha}{\Lambda} \frac{\textrm{d} \phi}{\textrm{d} \tau} \nabla \times \vec{A}^a & = 0 \ , \label{eq_pseudoscalar:eq_motionA}
\end{align}
where dots are used to denote derivatives with respect to cosmic time $t$, conversely we denote with $\tau$ the conformal time defined as $\textrm{d}t \equiv a\textrm{d}\tau  $ and with $\vec{\nabla}$ the ordinary $3$-dimensional gradient operator. The brackets $\langle \cdot \rangle$ are used to denote the spatial mean of the scalar product between the vectors $\vec{E}^a$ and $\vec{B}^a$, ``electric'' and ``magnetic'' fields, appearing in Eq.~\eqref{eq_pseudoscalar:eq_motion}, that are the defined as:
\begin{equation}
	\label{eq_pseudoscalar:electric_magnetic}
	\vec{E}^a \equiv -\frac{1 }{a^2} \frac{\textrm{d} \vec{A}^a}{\textrm{d} \tau} = -\frac{1 }{a} \frac{\textrm{d} \vec{A}^a}{\textrm{d} t} \ , \qquad \qquad  \vec{B}^a \equiv \frac{1}{a^2} \vec{\nabla} \times \vec{A}^a  \ .
\end{equation}
Finally we write Friedmann equation:
\begin{equation}
\label{eq_pseudoscalar:friedmann}
3 H^2 \kappa^{-2} = \frac{1}{2} \dot{\phi}^2 + V(\phi) + \frac{1}{2} \langle \vec{E}^{a \, 2}  +  \vec{B}^{a \, 2}\rangle  \ .
\end{equation}
This set of equations is completely specifying the background dynamics. By solving it, we can thus study the evolution of a model where the inflaton interacts with the gauge fields. \\

\noindent
In general the solution of Eq.~\eqref{eq_pseudoscalar:eq_motion}, Eq.~\eqref{eq_pseudoscalar:eq_motionA}, and Eq.~\eqref{eq_pseudoscalar:friedmann} does not exist analytically. However, as we show in the following, an analytical solution for Eq.~\eqref{eq_pseudoscalar:eq_motionA} exists if we assume $\dot{ \phi} $ to be slowly varying. Indeed this is a reasonable assumption during inflation. Once this solution is found, we can substitute it into Eq.~\eqref{eq_pseudoscalar:eq_motion} and into Eq.~\eqref{eq_pseudoscalar:friedmann} and study the back-reaction. In Fourier transform, the equations of motion for the gauge fields can be expressed as:
\begin{equation}
\label{eq_pseudoscalar:eq_motionA_fourier}
   \frac{\textrm{d}^2 \ \vec{A}^{a}(\tau,\vec{k})}{\textrm{d} \tau^2} \ + \ \vec{k}^2 \vec{A}^{a}   \ + \  i \frac{\alpha}{\Lambda} \frac{\textrm{d}\phi}{\textrm{d} \tau } \vec{k} \times \vec{A}^{a}   =  \ 0 \ .
\end{equation}
Assuming $\vec{k}$ to be parallel to $\hat{x}$, we can then proceed by defining the two helicity vectors $\vec{e}_{\pm} = (\hat{y} \pm i \hat{z})/\sqrt{2}$ and expressing the gauge field as $\vec{A} = A_{+} \vec{e}_{+} + A_{-} \vec{e}_{-}$. Using this parametrization, the gauge fields $\vec{A}^{a}$ and the cross product $\vec{k} \times \vec{A}^{a}$ can be expressed as:
\begin{equation}
\vec{A}^{a} =  \vec{e}_{\pm} A^{a}_{\pm} \ ,  \qquad \qquad  \qquad \vec{k} \times \vec{A}^a =  A^{a}_{\pm} \vec{k} \times \vec{e}_{\pm} = \mp i A^{a}_{\pm} |\vec{k}| \vec{e}_{\pm} \ .
\end{equation}  
The equation of motion for the Fourier transform of the gauge fields then reads:
\begin{equation}
\label{eq_pseudoscalar:eq_motionA2}
  \frac{\textrm{d}^2 \ A^{a}_{\pm}(\tau,\vec{k})}{\textrm{d} \tau^2}  + \left[ k^2 \pm 2k  \frac{\xi}{\tau} \right]A^{a}_{\pm}(\tau,\vec{k}) =  \ 0 \ , 
  \end{equation}
where, we have introduced the parameter $\xi$ defined as:
\begin{equation}
\xi \equiv \frac{\alpha |\dot{\phi}|}{2 \Lambda H} \ .
\label{eq_pseudoscalar:xi}
\end{equation}
Notice that the cross-product in Eq.~\eqref{eq_pseudoscalar:eq_motionA_fourier} (that arises from the antisymmetric $\varepsilon$-tensor in $\tilde F_{\mu \nu}$) is turned into the $\pm$ in Eq.~\eqref{eq_pseudoscalar:eq_motionA2}. This leads to a tachyonic instability in the $A_+$ mode (for $\dot{\phi} < 0$) that induces an exponential growth for the vector field. It is possible to show~\cite{Anber:2009ua,Barnaby:2010vf,Barnaby:2011vw} that for $(8 \xi)^{-1} \lesssim k/(aH) \lesssim 2 \xi$, the growing mode may be well approximated by:
\begin{equation}
A_+^a \simeq \frac{1}{\sqrt{2k}} \left( \frac{k}{2 \xi a H}\right)^{1/4} e^{ \pi \xi - 2 \sqrt{2 \xi k/(a H)}} \ .
\end{equation}
To introduce the back-reaction into the equation of motion for the scalar field and into Friedmann equation, we should compute the integrals~\cite{Barnaby:2011vw}:
\begin{equation}
\begin{aligned}
	\langle \vec{E}^a \cdot \vec{B}^a \rangle & = - \frac{1}{4 \pi a^2} \int_{0}^{\infty} \textrm{d} k \ k^3 \frac{\textrm{d}}{\textrm{d}\tau} \left| A_+^a \right|^2 \ , \\
	  \frac{1}{2}\langle \vec{E}^{a \, 2} + \vec{B}^{a \, 2} \rangle &  = \frac{1}{4 \pi a^2} \int_{0}^{\infty} \textrm{d} k^2 \left[  \left| A_+^{\prime \, a} \right|^2 + k^2 \left| A_+^a \right|^2\right] \ .
\end{aligned} 
\end{equation}
An analytic expression for these quantities was derived in~\cite{Anber:2009ua}. In particular, these integrals can be expressed as:
\begin{equation}
\langle \vec{E}^a \cdot \vec{B}^a \rangle \simeq \mathcal{N} \cdot \   2.4 \cdot 10^{-4} \frac{H^4}{\xi^4} e^{2 \pi \xi} \ , \quad \frac{1}{2} \langle \vec{E}^{a \, 2} + \vec{B}^{a \, 2} \rangle  \simeq \mathcal{N} \cdot \   1.4 \cdot 10^{-4} \frac{H^4}{\xi^3} e^{2 \pi \xi} \ .
\end{equation}
It is important to stress that these expressions do not hold for too small values of $\xi$, but only for $\xi \gtrsim 1$. On the other hand, for small values of $\xi$ we should use~\cite{Anber:2009ua}:
\begin{equation}
\langle \vec{E}^a \cdot \vec{B}^a \rangle \simeq \mathcal{N} \frac{H^4}{\xi^4} e^{2 \pi \xi} \frac{1}{2^{21} \pi^2} \int_0^{8 \xi}
  x^7 e^{-x} dx \ ,
\end{equation}
but actually in this regime back-reaction is almost negligible. It is possible to show (see for example~\cite{Barnaby:2011vw}) that while back-reaction on the Friedmann equation are fairly negligible through the whole evolution, the back-reaction on the equation of motion for the scalar field cannot be neglected in the last part of the evolution. This back-reaction is introducing an additional friction term that has an exponential dependence on $\xi$. As this parameter is proportional to $\dot{\phi}$, it increases towards the end of inflation. Correspondingly the new friction term is significantly slowing down the last part of the evolution. As we explain in detail in Sec.~\ref{sec_pseudoscalar:analytical}, this effect induces a shift of the region in the potential that can be probed by CMB observations. It is crucial to stress that the gauge fields are not changing the total number $N$ of e-foldings, and CMB is still generated at $N_\text{CMB} \simeq 60$. However, they are introducing a part at the end of the evolution that is dominated by the friction.

\subsection{Scalar and tensor perturbations.}
\label{sec_pseudoscalar:scalar_tensor_perturbations}
As explained in Chapter~\ref{chapter:inflation} and as extensively discussed in Appendix~\ref{appendix_perturbations:Cosmological_perturbations}, to describe the perturbations around the homogeneous background, we should decompose the inflaton field and the metric as:
\begin{equation}
	\Phi(t,\vec{x}) = \phi(t) + \delta \phi(\vec{x},t) \ , \qquad \qquad G_{\mu \nu}(\vec{x},t) = g_{\mu\nu}(t) + \delta g_{\mu \nu} (\vec{x},t) \ .
\end{equation}
Using this parametrization, we can derive the equations of motion for the scalar and tensor fluctuations and then the scalar and tensor power spectra. 

\subsubsection{Scalar perturbations.}
Our starting point is the linearized equation of motion for the scalar field perturbations~\cite{Anber:2006xt,Anber:2009ua,Barnaby:2010vf,Barnaby:2011vw,Barnaby:2011qe} that reads\footnote{This equation is obtained with a procedure similar to the one described in Appendix~\ref{appendix_perturbations:Cosmological_perturbations} (see Sec.~\ref{appendix_perturbations:eom_scalar}). While this Eq.~\eqref{eq_pseudoscalar:scalfluc} is expressed in terms of $\delta \phi $, the treatment of Appendix~\ref{appendix_perturbations:Cosmological_perturbations} is carried out in terms of $\zeta$. The only difference between these two cases is the term that is due to the presence of the gauge fields.}:  
\begin{equation}
 \delta \phi^{\prime \prime} + 2 a(\tau) H  \delta \phi^{\prime}+ \left[ - \nabla^2 + a^2(\tau) V_{,\phi \phi}(\phi) \right] \delta\phi = - \frac{\alpha}{\Lambda} a^2(\tau) \delta[\vec{E}^a \cdot \vec{B}^a]  \ ,
\label{eq_pseudoscalar:scalfluc}
\end{equation}
where primes denote derivatives with respect to conformal time $\tau$ and where we have defined~\cite{Anber:2009ua}:
\begin{equation}
\label{eq_pseudoscalar:gauge_field_contrib}
\delta[\vec{E}^a \cdot \vec{B}^a] = [\vec{E}^a \cdot \vec{B}^a - \langle \vec{E}^a \cdot \vec{B}^a \rangle]_{\delta \phi = 0} + \frac{\partial \langle \vec{E}^a \cdot \vec{B}^a \rangle}{\partial \dot{\phi}}\frac{ \delta \phi^{\prime} }{a} \ .
\end{equation}
Notice that the gauge fields are acting as a source for the scalar perturbations. As $\langle \vec{E}^a \cdot \vec{B}^a \rangle$ only depends on $\phi$ through $\xi$, we can show that:
\begin{equation}
 	\frac{\partial \langle \vec{E}^a \cdot \vec{B}^a \rangle}{\partial \dot{\phi}} = \frac{\partial \langle \vec{E}^a \cdot \vec{B}^a \rangle}{\partial \xi} \frac{\partial \xi }{\partial \dot{\phi}} = 2 \pi \ \langle \vec{E}^a \cdot \vec{B}^a \rangle \cdot \left(  - \frac{\alpha}{2 \Lambda H } \right) \ ,
\end{equation} 
where the minus sign comes from the choice of having $\dot{\phi} < 0$. Finally, we can express the equation of motion for the scalar field perturbations as:
\begin{equation}
 \ddot{ \delta \phi} + 3  H b \dot{ \delta \phi} +  \left[ - \frac{\nabla^2}{a^2(t)} + V_{,\phi \phi}(\phi) \right] \delta\phi = - \frac{\alpha}{\Lambda} \delta_{\vec{E}^a \cdot \vec{B}^a} \ ,
\label{eq_pseudoscalar:scalfluc_cosm}
\end{equation}
where we have defined $b$ and $\delta_{\vec{E}^a \cdot \vec{B}^a}$ as:
\begin{equation}
b \equiv 1 - 2 \pi \xi \frac{\alpha \langle \vec{E}^a \cdot \vec{B}^a \rangle}{3 \Lambda H \dot{\phi}} \ , \qquad \qquad \delta_{\vec{E}^a \cdot \vec{B}^a} \equiv \left[\vec{E}^a \cdot \vec{B}^a - \langle \vec{E}^a \cdot \vec{B}^a \rangle \right]_{\delta \phi = 0} .
\end{equation} 
In general, this differential equation for $\delta \phi$ is not easy to be solved, but approximated solutions can be found both in the weak and strong gauge fields regimes. In the weak field regime\footnote{It is possible to show~\cite{Barnaby:2011qe,Linde:2012bt} that this approximation holds for: 
\begin{equation}
2 \pi \xi \alpha \frac{\langle \vec{E}^a \cdot \vec{B}^a \rangle}{3 H \dot{\phi}} \ll 1 \ .	
\end{equation}} we approximate the system at the linear order~\cite{Barnaby:2010vf,Barnaby:2011vw} in $\delta \phi$ and in the gauge fields \textit{i.e.} we set $b = 1$. However, this approximated solution is not valid in the last part of the evolution~\cite{Barnaby:2011qe,Linde:2012bt} where, because of the exponential growth of the gauge fields, this term dominates the dynamics. As suggested by Linde~\cite{Linde:2012bt}, that actually follows a proposal of Barnaby, Pajer and Peloso~\cite{Barnaby:2011qe}, in the strong gauge fields regime we can approximate the left-hand side of Eq.~\eqref{eq_pseudoscalar:scalfluc_cosm} by only considering its second term. With this approximation, the equation of motion reads: 
\begin{equation}
3 b H \dot{ \delta \phi} =  - \frac{\alpha}{\Lambda}  \delta_{\vec{E}^a \cdot \vec{B}^a}\ .
\label{eq_pseudoscalar:scalfluc_2}
\end{equation}
Moreover, as we can approximate $\dot{ \delta \phi}$ with $H  \delta \phi$, this expression can be used to get an estimate of the amplitude of the scalar power spectrum at small scales. \\

\noindent
In order to compute the scalar power spectrum, we first need to quantize the scalar perturbations. The detailed procedure to perform this calculation can be found in~\cite{Anber:2009ua,Barnaby:2010vf,Barnaby:2011qe}. In particular, we should first compute the Green's function for Eq.~\eqref{eq_pseudoscalar:scalfluc} without sources, and we should then integrate it with the source. Once this procedure is performed, the scalar power spectrum is expressed as:
\begin{equation}
	\frac{H^2}{\dot{\phi}^2}\langle \delta \phi(\tau,\vec{x}) \, \delta \phi(\tau,\vec{y})  \rangle  \simeq  \langle\zeta(\tau,\vec{x}) \, \zeta(\tau,\vec{y})  \rangle   \equiv  \int\frac{\textrm{d}^3 \vec{k}}{4 \pi} \frac{\Delta^2_s(k)}{k^3} e^{- i \vec{k} (\vec{x} - \vec{y})} \ ,
\end{equation}
where the brackets $\langle \cdot \rangle$ denote the mean value over all the statistical realizations of the system. An estimate of the result in the strong gauge field regime can be obtained~\cite{Anber:2009ua,Linde:2012bt} using the approximation of Eq.~\eqref{eq_pseudoscalar:scalfluc_2}:
\begin{equation}
	\langle \zeta^2(x)  \rangle  \simeq \left( \frac{\alpha}{3 \Lambda b H \dot{\phi} }\right)^2 \langle  (\delta_{\vec{E}^a \cdot \vec{B}^a} )^2\rangle \simeq \left( \frac{\alpha  }{3 \Lambda b H \dot{\phi}}\right)^2 \mathcal{N} \langle \vec{E} \cdot \vec{B} \rangle^2 = \left( \frac{\alpha \langle \vec{E}^a \cdot \vec{B}^a \rangle/ \sqrt\mathcal{N}}{3 \Lambda b H \dot{\phi} }\right)^2  ,
\end{equation}
where $\langle \vec{E} \cdot \vec{B} \rangle $ is the value of $\langle \vec{E}^a \cdot \vec{B}^a \rangle$ for $\mathcal{N} =1$. Notice that the term $\langle  (\delta_{\vec{E}^a \cdot \vec{B}^a} )^2\rangle$ carries a $\mathcal{N}$ factor\footnote{As explained in~\cite{Anber:2009ua}, the contributions of the different gauge fields is expected to add incoherently. This is easily explained if we think at the Feynmann diagrams associated with the two-point function $\langle 0| \hat{\delta \phi} \hat{\delta \phi} | 0 \rangle $. In particular, each gauge field carries a one-loop contribution to this two-point function.} into scalar power spectrum. As the power spectrum is expected to be nearly constant in the last part of the evolution~\cite{Linde:2012bt} (that moreover is expected to give the dominant contribution to this integral) an estimate of the power spectrum\footnote{As we discuss in Sec.~\ref{sec_pseudoscalar:signatures} and in Sec.~\ref{sec_pseudoscalar:discussion}, the consequences of this assumption are actually relevant for the discussion of the PBH bounds.} is obtained by approximating:
\begin{equation}
	\langle \zeta^2(x) \rangle  = \int \frac{\textrm{d}^3 \vec{k} }{4\pi}  \frac{\Delta^2_s (k)}{ k^3} \simeq \mathcal{O}(1) \Delta^2_s (k) \simeq \Delta^2_s (k) \ .
\end{equation}
As in the regime where the gauge fields contribution is negligible the power spectrum is dominated by the standard vacuum amplitude, the complete expression for the scalar power spectrum reads~\cite{Linde:2012bt,Domcke:2016bkh}:
\begin{equation}
\Delta^2_s(k) = \Delta^2_s(k)_\text{vac} + \Delta^2_s(k)_\text{gauge} = \left(\frac{H^2}{2 \pi |\dot{\phi}|}\right)^2 + \left( \frac{\alpha \langle \vec{E}^a\cdot \vec{B}^a \rangle/ \sqrt\mathcal{N}}{3 \Lambda b H \dot{\phi}} \right)^2 .
\label{eq_pseudoscalar:scalar}
\end{equation}
With a proper choice of the parameters of the model we can thus ensure that at large scales, \textit{i.e.} at the scales probed through CMB observations, the gauge field contribution is negligible and Eq.~\eqref{eq_pseudoscalar:scalar} reduces the standard scale-invariant power spectrum of inflation. On the contrary, at small scales, \textit{i.e.} in the last part of inflation, the gauge fields dominate and the spectrum is well approximated by~\cite{Linde:2012bt,Domcke:2016bkh}:
\begin{equation}
\Delta^2_s(k ) \simeq \frac{1}{\mathcal{N} (2 \pi \xi)^2} \ .
\label{eq_pseudoscalar:scalar_strong}
\end{equation}
Notice that in presence of several $U(1)$ the power spectrum is suppressed at small scales. As we discuss in Sec.~\ref{sec_pseudoscalar:signatures}, this result is actually relevant for the discussion of some experimental bounds. Before concluding this Section we also point out that the interactions between the inflaton and the gauge fields induce non-zero non-Gaussianities. However, as discussed in the following, these non-Gaussianities are strongly constrained at CMB scales~\cite{Ade:2015lrj}.

\subsubsection{Tensor perturbations.}
The procedure to derive the tensor power spectrum is similar to the one carried out for the scalar power spectrum. As usual tensor fluctuations are described by the transverse traceless part of the spatial metric perturbations. The starting point is now given by the linearized Einstein equation~\cite{Maggiore:1900zz} in the presence of the gauge fields\footnote{Except for the term that is due to the presence of the gauge fields, this equation corresponds to Eq.~\eqref{appendix_perturbations:tensor_eom_gamma} expressed in terms of the conformal time $\tau$.}:
 \begin{equation}
\frac{d^2 h_{ij}}{d \tau^2} + 2 \frac{d \ln a}{d \tau} \frac{d h_{ij}}{d \tau} - \Delta h_{ij} = 2\kappa^2 \Pi_{ij}^{\mu \nu} T_{\mu \nu} \ ,
\label{eq_pseudoscalar:greens}
 \end{equation}
 where $\Pi^{ij}_{\mu \nu}$ is the transverse, traceless projector and as usual $T_{\mu \nu}$ denotes the matter energy-momentum tensor that sources the GW. Similarly to the case of scalar fluctuations, in order to find a solution for Eq.~\eqref{eq_pseudoscalar:greens} we should start by computing the Green's function $G_k(\tau, \tau_1)$ for the corresponding homogeneous differential equation and we should then use:
 \begin{equation}
\tilde{h}_{ij}(\vec{k},\tau) = 2\kappa^2 \int d\tau_1 G_k(\tau, \tau_1) \Pi_{ij}^{ab}(\vec{k}) T_{ab}(\vec{k}, \tau_1)\ ,
 \end{equation}
where as usual $\tilde{h}_{ij}(\vec{k},\tau)$ is the spatial Fourier transform of $h_{ij}(\vec{x},\tau)$. Once this solution is found, we can proceed by computing the spectra $\Delta^2_{t,+}$ and $\Delta^2_{t,\times}$ for the two polarizations\footnote{Similarly to Sec.~\ref{sec_pseudoscalar:background_solution}, we assume $\vec{k} $ to be parallel to $\hat{x}$. Notice that left polarization of~\cite{Barnaby:2011qe, Barnaby:2011vw} corresponds to the $\times$ polarization of our convention.} ($+,\times$) of the GW. To decompose $h_{ij}(\vec{k},\tau)$ in terms of the two polarizations we can use the projector:
\begin{equation}
	\Pi_{ij,+/\times} \equiv e_{i}^{\pm} \, e_{j}^{\pm} \ ,
\end{equation}
The tensor spectrum $\Delta^2_{t}$ is then given by $\Delta^2_{t} = \Delta^2_{t,\times} + \Delta^2_{t,+}$. As the gauge fields contribution to $\Delta^2_{t,+}$ is suppressed (for more details see for example~\cite{Barnaby:2011qe, Barnaby:2011vw}) by a $10^{-3}$ factor with respect to the gauge fields contribution to $\Delta^2_{t,\times}$, when we compute $\Delta^2_{t} = \Delta^2_{t,\times} + \Delta^2_{t,+}$ we approximate $\Delta^2_{t,+} $ with its vacuum contribution. The normalized density of GW at present time can thus be expressed as:
\begin{equation}
\Omega_{GW} \equiv \frac{\Omega_{R,0}}{24} \Delta^2_{t} \simeq \frac{1}{12} \Omega_{R,0} \left(  \frac{\kappa H}{ \pi} \right)^2 \left(1 + 4.3 \cdot 10^{-7} \mathcal{N} \frac{\kappa^2 H^2}{ \xi^6} e^{4 \pi \xi}\right)\ ,
\label{eq_pseudoscalar:OmegaGW}
\end{equation}
where $\Omega_{R,0} = 8.6 \cdot 10^{-5}$ denotes the radiation energy density today and, consistently with the notation of this work, $\kappa^{-1} \simeq 2.4 \cdot 10^{18}\,$~GeV denotes the reduced Planck mass. To lighten the notation, in the following we set $\kappa^2 = 1$. Similarly to the case of the scalar power spectrum, the first term in the second parenthesis of Eq.~\eqref{eq_pseudoscalar:OmegaGW} \textit{i.e.} the $1$, is the usual vacuum contribution from inflation. On the other hand the second term is the one that is due to the presence of the gauge fields. \\

\noindent 
At this point, it is important to stress that as one of the two polarizations ($\Delta^2_{t,+}$) is suppressed with respect to the other ($\Delta^2_{t,\times}$), the generated GW signal is expected to be chiral. This is a rather unusual characteristics for GW signals and in particular this is a rare feature for a GW background. Because of this peculiarity, this signal can be distinguished from the one produced by other sources.\\

\noindent
As in the context of direct GW observations it is customary to express quantities in terms of the frequency $f = k / (2 \pi)$, it is useful to introduce the relation\footnote{To derive this relation we use $k = aH$ and we assume $H(t) \simeq H(t_{CMB})$ so that we have:
\begin{equation}
	\ln\left(\frac{k}{0.002\,\text{Mpc}^-1}\right) - \ln\left(\frac{k_{CMB}}{0.002\,\text{Mpc}^{-1}}\right) = \ln\left(\frac{a_E}{a(t_CMB)}\right) + \ln\left(\frac{a(t)}{a_E}\right) \ ,
\end{equation}
where $a_E$ is the value of the scale factor at the end of inflation. We can then use $e^{-N} = a/a_E$ (Eq.~\eqref{eq_inflation:e_folds_problems}) and $\ln(2\pi \times 100\,\text{Hz}/(0.002\,\text{Mpc}^{-1}))\simeq 44.9$.} between $f$ and the number of e-foldings $N$~\cite{Barnaby:2011qe,Domcke:2016bkh}:
\begin{equation}
N = N_\text{CMB} + \ln \frac{k_\text{CMB}}{0.002 \text{ Mpc}^{-1}} - 44.9 - \ln\frac{f}{10^2 \text{ Hz}} \ ,
\label{eq_pseudoscalar:Nf}
\end{equation}
where $k_\text{CMB} = 0.002 \text{ Mpc}^{-1}$ and $N_\text{CMB} \simeq 50 - 60$. Following the convention used throughout this work, the number of e-foldings $N$ decreases during inflation, reaching $N=0$ at the end of inflation.\\

\noindent Finally, it is interesting to point out that large scalar perturbations on small scales may happen to source sizable second order tensor perturbations~\cite{Baumann:2007zm}. However, for all the models considered in this Chapter, these effects are subdominant compared to the leading order GW contribution.

\section{Experimental bounds and observable signatures.}
\label{sec_pseudoscalar:signatures}
The models considered in this Chapter are a rather natural extension of the simplest realization of inflation. Moreover, as the non-minimal coupling that we are considering can only be present if the inflaton is a pseudo-scalar, a detection of an effect generated by these particular models would be extremely relevant to probe the microphysics of inflation. As discussed in~\cite{Domcke:2016bkh}, the modified models considered in this Chapter give rise to several signatures that can be detected in experimental observations. While the existing bounds can be used to set constraints on the parameters of the model, theoretical predictions are important to guide future experiments towards the detection of new physics. Both for these reasons and in order to frame the mechanism discussed in Sec.~\ref{sec_pseudoscalar:mechanism} in the context of modern cosmology, it is thus interesting to give a detailed review of the existing experimental constraints. \\

\noindent
We start our discussion by presenting the constraints that are set by CMB observations. In particular, we start by discussing the COBE normalization, the constraints on $n_s$, $r$, $\alpha_s$, the non-Gaussianity bound and the constraints on the so-called $\mu$-type distortions in the CMB. We then discuss the constraint on the number of additional massless degrees of freedom in our Universe, the possible generation of PBH, and the possible generation of primordial magnetic fields at the end of inflation.

\subsection{COBE normalization and Planck constraints.}
\label{sec_pseudoscalar:COBE_Planck}
As discussed in Sec.~\ref{sec_pseudoscalar:mechanism}, the presence of the gauge fields modifies the scalar and tensor power spectra. However, these modifications should not affect the scales that are actually probed by CMB observations, where tight constraints are set on the spectra~\cite{Ade:2015xua,Ade:2015ava,Ade:2015lrj}. In particular, we report the COBE normalization and the constraints on $n_s$, $r$ and $\alpha_s$:

\begin{itemize}
	\item \textbf{COBE Normalization}: This constraint sets the value of the scalar power spectrum at the CMB scales. In particular we have~\cite{Ade:2015lrj}: 
	\begin{equation}
	\label{eq_pseudoscalar:COBE}
		 \left. \Delta^2_s\right|_{N_{CMB}} = (2.21 \pm 0.07) \cdot 10^{-9} \ .
	\end{equation}
	As explained in Chapter~\ref{chapter:inflation}, this corresponds to fixing a constraint on the inflationary potential \textit{i.e.} on the scale of inflation.

	\item \textbf{Planck measurements}: As already explained through this work, CMB measurements can be used to set constraints on the scalar-spectral index $n_s$, the running of the spectral index $\alpha_s$ and on the tensor-to-scalar ratio $r$ (defined in Chapter~\ref{chapter:inflation}) at CMB scales. The constraints on these parameters set from the Planck mission~\cite{Ade:2015lrj} read (at $68\%$ CL for $n_s$ and $\alpha_s$, $95\%$ CL for $r$):
	\begin{equation}
		\label{eq_pseudoscalar:PLANCK}
		n_s = 0.9645 \pm 0.0049\ , \qquad \alpha_s = -0.0057 \pm 0.0071\ , \qquad r < 0.10 \ .
	\end{equation}
	As we show in the following sections, given the shape of the potential, these constraints actually set a limit on the length of the strong gauge field regime.
\end{itemize}

\subsection{Non-Gaussianities.}
\label{sec_pseudoscalar:non_gaussianities}
As already stated in~\ref{sec_pseudoscalar:mechanism} and as widely discussed in~\cite{Anber:2012du,Barnaby:2010vf,Barnaby:2011vw,Barnaby:2011qe,Linde:2012bt}, the presence of the gauge fields induces a non-zero non-Gaussian component in the power spectrum. Non-Gaussianities are strongly constrained at the CMB scales for example by Planck measurements~\cite{Ade:2015lrj,Ade:2015ava}. In particular, Planck constrains the dominant non-Gaussian contribution\footnote{The dominant non-Gaussian contribution is the equilateral contribution. This corresponds to the case where the three-momenta appearing in the bispectrum (defined in Eq.~\eqref{eq_inflation:def_bispectrum}) satisfy $|k_1| \simeq |k_2| \simeq |k_3|$.} $f_{NL}^\text{equil}$, to be $|f_{NL}^\text{equil}| < |-4 \pm 43|$ at 68$\%$ CL. As discussed in~\cite{Barnaby:2010vf,Barnaby:2011vw,Barnaby:2011qe,Linde:2012bt}, the gauge field contribution to the three-point function can be expressed as:
\begin{equation}
f_{NL}^\text{equil} \simeq 6.16 \cdot 10^{-16} e^{6 \pi \xi}/\xi^9 \ .
\label{eq_pseudoscalar:fNL}
\end{equation}
To derive this equation, we have reasonably assumed that at CMB scales $\Delta^2_s$ is governed by the vacuum contribution. Eq.~\eqref{eq_pseudoscalar:fNL} directly implies:
\begin{equation}
	\label{eq_pseudoscalar:NG}
		\xi_{CMB} = \frac{\alpha}{2 \Lambda} \left|\frac{\dot{\phi}}{H}\right|_{N=N_{CMB}} \lesssim 2.5 \ , 
	\end{equation}
at CMB scales ($95 \%$ CL, assuming Gaussian errors). It is crucial to stress that this bound only applies at CMB scales (\textit{i.e.} in the weak gauge field regime), while in the strong gauge field regime we have $f_{NL}^\text{equil} = - 1.3 \  \xi$~\cite{Anber:2012du}.

\subsection{$\mu$-type distortions in the CMB.}
\label{sec_pseudoscalar:mu_distortions}
As already mentioned in Chapter~\ref{chapter:introduction} (in particular in Sec.~\ref{sec_introduction:CMB}), the so-called ``$\mu$-type distortions'' are thermal distortions of the CMB spectrum from a pure black-body distribution. These distortion are basically due to some energy injection (for example due to particles decaying into photons) in the spectrum of photons between recombination and decoupling. In this epoch, the interactions between photons and matter are dominated by Thomson scattering that is not providing an efficient mechanism to smooth the deviation from a pure black-body spectrum. As a consequence, these distortions cannot be removed and they are effectively introducing a non-zero (frequency dependent) chemical potential in the distribution of CMB photons~\cite{Hu:1994bz,Pajer:2013oca,Meerburg:2012id}.\\

\noindent
It is interesting to notice that while CMB is mainly sensitive to scales around $k \simeq 10^{-2} \text{ Mpc}^{-1}$, $\mu$-type distortions are sensitive to the integrated scalar power spectrum in the range\footnote{This corresponds to a frequency range of $10^{-15}~\text{ Hz} \lesssim f \lesssim 10^{-9}~\text{ Hz}$~\cite{Meerburg:2012id}. For completeness we report the values of $T$ that corresponds to $z_i$ and $z_f$ \textit{i.e.} $T_i\simeq 4 \times 10^2$\,eV and $T_f\simeq 10\,$eV.} $50 \text{ Mpc}^{-1} \lesssim k \lesssim 10^4 \text{ Mpc}^{-1}$, : 
\begin{equation}
\mu \simeq  \int_{k_{D}(z_i)}^{k_{D}(z_f)} d\ln k  \; \Delta^2_s(k) \left[ e^{ - k/k_{D}(z)}\right]^{z_f}_{z_i} \ ,
\end{equation}
with $k_D = 4 \times 10^{-6} z^{3/2} \text{ Mpc}^{-1}$ and $z_i = 2 \times 10^6$ ($z_f = 5 \times 10^4$) denoting the redshift when the dominant inelastic (elastic) scattering processes for CMB photons freeze out. The current bound on $\mu$-distortions is set by the COBE / FIRAS constraints \textit{i.e.} $\mu < 6 \times 10^{-5}$~\cite{Fixsen:1996nj} at 95$\%$ CL, imposing a bound on the coupling parameter $\alpha$ comparable to the one from the limits on non-Gaussianities in the CMB ($\xi_* < 2.5$). The PIXIE experiment, with a forecasted sensitivity of $\mu \lesssim 2 \times 10^{-8}$~\cite{Kogut:2011xw}, is expected to improve this bound and actually it is expected to reach the level of the vacuum contribution in this frequency range.

\subsection{CMB and BBN bounds on primordial GWs.}
\label{sec_pseudoscalar:CMB_BBN}
CMB and Big Bang Nucleosynthesis (BBN) set constraints on the additional massless degrees of freedom \textit{i.e.} on the radiation density of the Universe. As in the models of our interest we are producing a large amount of massless degrees of freedom \textit{i.e.} GW, these constraints actually happen to be relevant for the scope of our work. In particular these constraints can be phrased in terms of the effective number of massless neutrino species $N_\text{eff}$ (SM value: $N_\text{eff} = 3.046$) as:
\begin{equation}
\int \textrm{d} (\ln f) \  \Omega_\text{GW} = \Omega_{R,0} \frac{7}{8} \left( \frac{4}{11}\right)^{4/3} (N_\text{eff} - 3.046) \ ,
\end{equation}
where for the CMB (BBN) bound\footnote{For more details on the derivation of these bounds see for example~\cite{Allen:1996vm,Meerburg:2015zua,Cabass:2015jwe}.} the integral is performed over all frequencies $ f \gtrsim 10^{-15}$~Hz ($f \gtrsim 10^{-10}$~Hz). The current CMB bound~\cite{Ade:2015xua} is $N_\text{eff} = 3.04 \pm 0.17$ and the current~\cite{Cyburt:2015mya} BBN bound is $N_\text{eff} = 3.28 \pm 0.28$. \\

\noindent
However, it is also interesting to point out that a recent paper from Riess et al.~\cite{Riess:2016jrr} argues for a higher effective number of massless neutrino species \textit{i.e.} $\Delta N_\text{eff} \simeq 0.4 - 1$.

\subsection{Primordial black holes.}
Black Holes (BHs) are extremely compact objects that are predicted by GR. A defining characteristics of a BH is that its gravity is so strong that particles within a certain surface around the BH (the horizon of the BH) cannot escape from the BH but they are forced to fall into it. In particular, this means that the escape velocity becomes larger than the speed of light, and thus no classical causal signal\footnote{This result is no longer valid in the context of Quantum Mechanics. In particular, this result is violated by the so-called Hawking radiation~\cite{Hawking:1974sw}, which leads to BH evaporation through thermal radiation. This effect has recently been observed by Steinhauer~\cite{Steinhauer:2014dra} in an analogue of a BH realized with cold-atoms.} is allowed to escape from the BH. It is customary to distinguish between BH that are formed through astrophysical processes, and PBH that on the other hand are formed in very early times through cosmological processes. \\

\noindent
As discussed in Sec.~\ref{sec_pseudoscalar:scalar_tensor_perturbations}, the introduction of other particles during inflation may lead to an amplification of the scalar spectrum at small scales~\cite{Anber:2009ua,Barnaby:2010vf,Barnaby:2011qe}. This strong enhancement of the fluctuations at small scales (far beyond the CMB observable scales) may induce a local increase of the density that can cause the matter to collapse leading to the formation of a distribution of PBH~\cite{Linde:2012bt}. The models discussed in Sec.~\ref{sec_pseudoscalar:mechanism} thus provides a natural mechanism to generate a distribution of PBH. The non-observation of PBHs can be used to set some constrains on the fraction of energy going into PBHs at their formation, as a function of the PBH mass. As a first step we can thus divide the PBHs into three categories according to their mass:
\begin{itemize}
	\item PBHs with masses smaller than $10^{15}$~g. These PBHs have already evaporated and they can be detected observing their entropy production in the early universe.
	\item PBHs with masses around $10^{15}$~g. These PBHs would be evaporating today and thus they would leave signals in $\gamma$-rays.
	\item PBHs with masses bigger than $10^{15}$~g. As these PBHs would still be stable, they can be searched for in lensing and GW experiments. 
\end{itemize}
Constraints over a wide mass range have been collected in~\cite{Carr:2009jm}. While it is worth mentioning constraints on the Hawking evaporation and on the present day gravitational effects, the strongest bounds actually are obtained from CMB anisotropies~\cite{Carr:2009jm}.\\

\noindent
An estimate of the fraction of PBHs was given by Linde in~\cite{Linde:2012bt}. As usual we define $\zeta \equiv - H\delta\phi / \dot{\phi}$ and thus $\zeta_c \simeq 1$ corresponds to the critical value leading to black hole formation. Given $P(\zeta)$, probability distribution for $\zeta$, we can express $b$, fraction of space that can collapse and form a PBH, as:
\begin{equation}
\label{sec_pseudoscalar:integral_PBH}
b = \int_{\zeta_c}^\infty P(\zeta) d\zeta .
\end{equation}
The probability distribution of $\zeta$ was computed in~\cite{Linde:2012bt} using the estimate of $\zeta$ in the strong gauge field regime (given in Sec.~\ref{sec_pseudoscalar:mechanism}) \textit{i.e.} $\zeta \propto \delta_{\vec{E}^a \cdot \vec{B}^a}$. As the perturbations of the gauge fields are nearly Gaussian~\cite{Linde:2012bt}, we can conclude that it is possible to express $\zeta$ in terms a Gaussian distributed field $g$ as $\zeta = g^2 - \langle g^2 \rangle$ and we are thus able to compute the integral of Eq.~\eqref{sec_pseudoscalar:integral_PBH}. \\

\noindent
The typical mass of the PBHs, depends on the scale of the fluctuations. In particular, given the scale of the scalar perturbations (and hence a corresponding value of $N$), we can estimate~\cite{Linde:2012bt} the mass using:
\begin{equation}
M_{PBH} = \frac{4 \pi}{H} e^{a N},
\end{equation}
where $a = \{2,3\}$ is a coefficient that depends on the efficiency of reheating. Finally, using this formula, the constraints on $b$ collected in~\cite{Carr:2009jm} \textit{i.e.} $b \lesssim 10^{-28} - 10^{-5} $, can be turned into constraints on the scalar power spectrum $\Delta^2_s \lesssim 1.3 \cdot 10^{-4} - 5.8 \cdot 10^{-3}$~\cite{Linde:2012bt}\footnote{Note that this constraint is considerably stronger compared to the one obtained by assuming Gaussian fluctuations, $\Delta^2_s \lesssim 10^{-2}$.}. Since the PBH bound is strong for relatively light PBHs, this puts a strong constraint on the amplitude of the scalar perturbations at the end of inflation. Using the approximation of $\Delta^2_s$ in the strong gauge field regime given in Eq.~\eqref{eq_pseudoscalar:scalar_strong}, this constraint can be used to directly put a lower bound on $\xi_{\text{max}}$ \textit{i.e.} $ \xi_{\text{max}} \gtrsim 14/\sqrt\mathcal{N}$.\\

\noindent
It should be stressed that the calculations performed in this Section are based on the strong gauge field regime. As a consequence, the approximations performed in calculating this bound are estimated to account for up to an order one factor~\cite{Linde:2012bt}. Moreover (as we discuss in the following Sections), the large amplitude of the scalar perturbations reached in this regime indicates that higher orders in the perturbative expansion may not be completely negligible. Ignoring these terms induces a further theoretical error in this regime. For these reasons, while the value of the bound derived in~\cite{Linde:2012bt} is shown in the plots of this Chapter, models that violate this bound by an order one factor are still considered as viable.\\

\subsection{Primordial magnetic fields.}
As a pseudo-scalar inflaton should couple to all the $U(1)$ gauge fields in the theory, it can also couple to the SM electromagnetic one. The gauge field production discussed in this work could hence generate primordial magnetic fields. The generation of primordial magnetic fields in the model of our interest has been widely discussed in literature, see e.g.~\cite{Durrer:2010mq, Anber:2006xt,Caprini:2014mja, Fujita:2015iga,Green:2015fss}. The result of these analysis is that, the magnetic fields generated with this mechanisms turn out to be too weak to provide the seeds for the observed fields in galaxies and clusters.

\section{Analytical estimates.}
\label{sec_pseudoscalar:analytical}
In this Section we discuss the predictions of these models for different inflationary potential. In particular, as in the existing literature~\cite{Turner:1987bw, Garretson:1992vt, Anber:2006xt,Anber:2012du,Anber:2009ua,Barnaby:2011qe,Barnaby:2011vw,Linde:2012bt} the focus has always been put on chaotic potentials, we are interested in extending the analysis to a broader class of models. Following the discussion of Chapter~\ref{chapter:beta}, instead of specifying a single inflationary model, we can produce a more general description of the problem by specifying classes of models. In particular, (according with the discussion of Chapter~\ref{chapter:beta}) a large set of single-field slow-roll models can be recovered~\cite{Mukhanov:2013tua, Binetruy:2014zya} by parameterizing  the first slow-roll parameter $\epsilon$ as:
\begin{equation}
\label{eq_pseudoscalar:Nparameterization}
\epsilon_H \simeq \epsilon_V \simeq  \frac{\beta_p}{N^p}  + \mathcal{O}(1/N^{p+1})\ ,
\end{equation}
where $\beta_p$ is a positive constant, $p$ is a constant bigger than $1$ and as usual the slow-roll parameters are defined as:
\begin{equation}
\label{eq_pseudoscalar:epsilon_HV}
\epsilon_H = \frac{\dot{\phi}^2}{2 H^2}\ , \qquad \qquad \epsilon_V = \frac{1}{2} \left( \frac{V_{,\phi}}{V} \right)^2 \ .
\end{equation}
Notice that with this parametrization we are only specifying the asymptotic behavior, without fixing the explicit expression for the potential $V(\phi)$. In particular we are not fixing conditions on the part of the potential that is relevant for reheating. As a consequence this procedure makes the analysis more general. \\

\noindent
It is interesting to notice that this parametrization also happens to be particularly convenient for our analysis. In particular, since the parameter $\xi$ (defined Eq.~\eqref{eq_pseudoscalar:xi}) that governs the exponential enhancement of the gauge fields is expressed as:
\begin{equation}
\xi \propto \sqrt{\epsilon_H} \ ,
\label{eq_pseudoscalar:xigrowth}
\end{equation}
with the parametrization of Eq.~\eqref{eq_pseudoscalar:epsilon_HV}, we have a direct control over $\xi$. As we see in the following this is extremely useful both to produce a description of the background evolution and of the scalar and tensor power spectra. At this point it is also crucial to stress that while $\epsilon_H$ and $\epsilon_V$ basically coincide during the first part of the evolution, \textit{i.e.} at CMB scales, they are different in the last part of the evolution, \textit{i.e.} when the gauge fields dominate. In particular, when the gauge fields dominate, $\epsilon_V$ may also be bigger than $1$! On the contrary, while $\epsilon_H \ll 1 $ is a defining condition to have inflation, this condition cannot be avoided and as usual inflation ends at $\epsilon_H \simeq 1 $. \\

\noindent
 In this Section, we use the parametrization of Eq.~\eqref{eq_pseudoscalar:Nparameterization}, to study the evolution of $\xi$ and thus the scalar and tensor power spectra. As a starting point for our analysis, we use $dN = - H \  dt$ to express the equation of motion for the scalar field given in Eq.~\eqref{eq_pseudoscalar:eq_motion} as:
\begin{equation}
	\label{eq_pseudoscalar:approx_eq_motion_2}
	- \phi_{, N} + \frac{V_{,\phi}}{V}  = \mathcal{N} \  \frac{2.4}{9} \cdot 10^{-4} \left( \frac{\alpha}{\Lambda} \right) \frac{V}{\xi^4} e^{2 \pi \xi} \ .
\end{equation}
In the evolution of the system we can distinguish three different regimes:
\begin{itemize}
	\item \textbf{Weak gauge fields (A)}: The gauge fields are subdominant and the evolution is similar to the case of standard single-field slow-roll inflation models. 
	\item \textbf{Intermediate region (B)}: The friction term due to the gauge fields overcomes the standard Hubble friction.
	\item \textbf{Strong gauge fields (C)}: The back-reaction (the friction term that grows exponentially with $\xi$) dominate the evolution.
\end{itemize}
In order to lower the notation, the analysis of this Section is performed with $\mathcal {N} = 1$. However, we reintroduce this number in the equations that are relevant for the treatment of the following sections.\\

\noindent
A schematic idea of the evolution of $\phi$ and $\xi$ is shown in Fig.~\ref{fig_pseudoscalar:phi_xi_schematic}. The dashed gray line in the left plot of Fig.~\ref{fig_pseudoscalar:phi_xi_schematic} shows the background evolution of $\phi$ when the interactions between the inflation and the gauge fields are set to zero \textit{i.e.} $\alpha = 0$. On the contrary solid blue line, shows the evolution in presence of the gauge fields. The meaning of all the relevant points shown in this figure is explained in the following.

\begin{figure}
	\includegraphics[width=1. \textwidth]{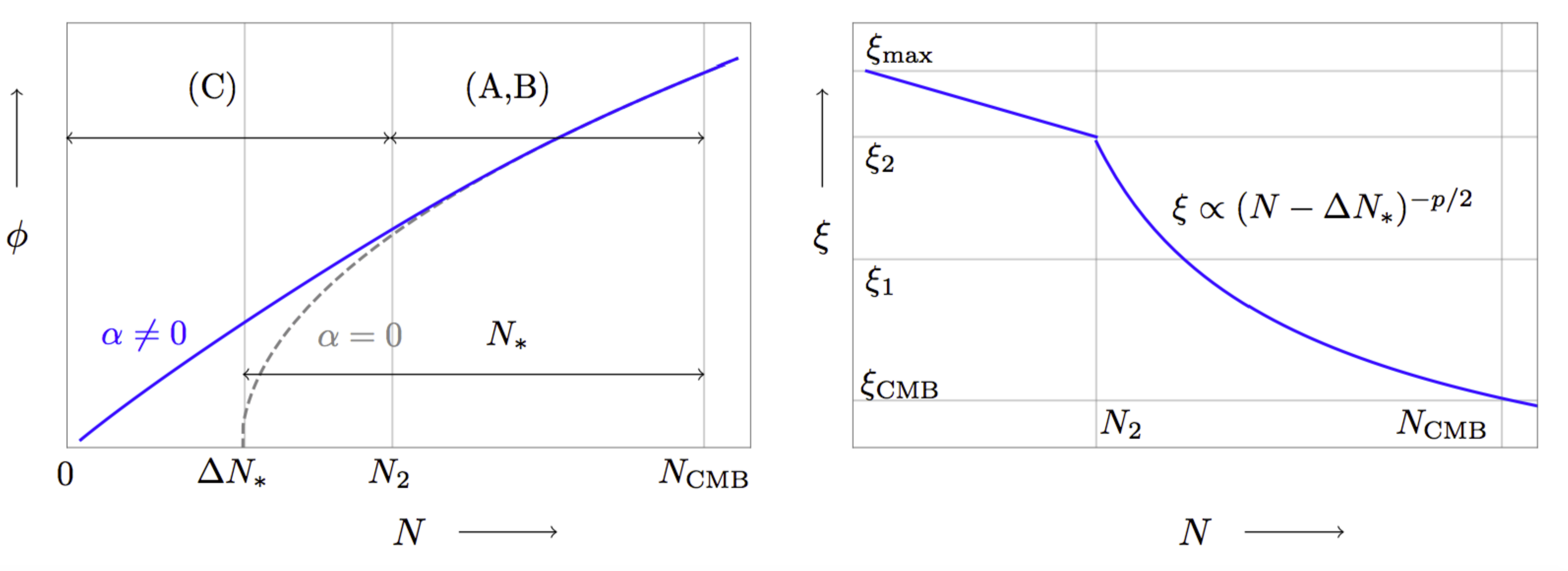}
	\caption{Schematic view of the evolution of the inflaton field $\phi$ (left panel) and of the parameter $\xi$ controlling the growth of the gauge fields (right panel) as a function of the number of e-foldings of inflation. \label{fig_pseudoscalar:phi_xi_schematic}}
\end{figure}

\subsection{The three regimes. \label{sec_pseudoscalar:analytic1}}
Before discussing in detail the three regimes, we should explain a crucial issue of the modified dynamics. Let us start by defining $N_\text{CMB} \simeq 50 - 60$, number of e-foldings between the moment when CMB scales exit the horizon and the end of inflation. As it is possible to see from Fig.~\ref{fig_pseudoscalar:phi_xi_schematic}, the back-reaction (present for $\alpha \neq 0$) are actually slowing down the last part of the evolution. In practice, this results in a shift of the point of the potential probed by the CMB. As a consequence, while in the complete evolution this point still corresponds to $N_\text{CMB}\simeq 50 - 60$, in the field space with $\alpha = 0$, this point should be associated to a different (lower) value $N_* < N_\text{CMB}$ of e-foldings. As a consequence, it is natural to define $\Delta N_*$ as the difference between these two numbers. In the following, we use $\Delta N_*$ in order to quantify the shift of the CMB point due to the presence of the gauge fields.

\subsubsection{Weak gauge fields (A).}
\label{sec_pseudoscalar:weak_gauge}

\noindent In this regime the equation of motion for the inflaton can be approximated as:
\begin{equation}
- \phi_{, N} + \frac{V_{,\phi}}{V} \simeq 0 \ .
\label{eq_pseudoscalar:motion0}
\end{equation}
In order to specify the end of the first regime, it is useful to find the expressions for the scalar and tensor power spectra when the gauge fields are weak. In this regime the spectra are well approximated by their vacuum contributions and thus we have:
\begin{equation}
\Delta^2_s\bigg|_{N_*} = \frac{H^2}{8 \pi^2 \epsilon_V} \bigg|_{N_*} \ , \qquad \qquad \Omega_{GW} = \frac{\Omega_{R,0} H^2}{12 \pi^2}  = \frac{4}{3} \Delta^2_s \epsilon_V \Omega_{R,0} \bigg|_{N_*} \ ,
\label{eq_pseudoscalar:vacuumamplitudes}
\end{equation}
where, as explained at the beginning of this Section, $N_* =N_{CMB} - \Delta N_*$. Using the expression for the GW spectrum of Eq.~\eqref{eq_pseudoscalar:OmegaGW}, we can thus set the end of the weak regime, at a value $\xi_1$ at which the gauge field contribution to the GW spectrum is of the same order of the vacuum contribution. This value can actually be defined using:
\begin{equation}
\xi < \xi_1 \quad \text{with} \quad \frac{e^{4 \pi \xi_1}}{\xi_1^6} = \left(4.3 \cdot 10^{-7} H_1^2 \right)^{-1} \ .
\label{eq_pseudoscalar:analytical_xi1}
\end{equation}
Notice that this value nothing particular happens in the evolution of $\xi$ (see the schematic plot of $\xi$ shown in Fig.~\ref{fig_pseudoscalar:phi_xi_schematic}).To get an estimate of the value of $\xi_1$ we first need an estimate of $H_1$. As $H$ is nearly constant during inflation, this can be obtained by using the parametrization of Eq.~\eqref{eq_pseudoscalar:Nparameterization} and neglecting the shift $\Delta N_*$:
\begin{equation}
H_1^2 \simeq \pi^2  \Delta^2_s \bigg|_{N_*} \cdot  \frac{8 \beta_p}{(N_\text{CMB})^p} \ .
\end{equation}
Substituting into Eq.~\eqref{eq_pseudoscalar:analytical_xi1} we can thus get $\xi_1$. Notice that given the value of $\xi_\text{CMB}$ (\textit{i.e.} the value of $\xi$ at CMB scales), we can translate a value of $\xi$ into a value for $N$ using:
\begin{equation}
\frac{\xi}{\xi_\text{CMB}} = \left( \frac{N_\text{CMB} - \Delta N_*}{N - \Delta N_*} \right)^{p/2} \ .
\label{eq_pseudoscalar:xigrowth2}
\end{equation}
In particular this equation can be used to compute the value of $N_1$. Finally, using Eq.~\eqref{eq_pseudoscalar:Nf}, we can also get the corresponding frequency.

\subsubsection{Intermediate region (B).}
\label{sec_pseudoscalar:intermediate_gauge}

\noindent In the regime the gauge field contribution dominates in the GW spectrum but not on the background dynamics. In particular, the additional friction in Eq.~\eqref{eq_pseudoscalar:eq_motion} is small with respect to the standard Hubble friction. Until this condition is satisfied, both the parametrization of Eq.~\eqref{eq_pseudoscalar:Nparameterization}, and Eq.~\eqref{eq_pseudoscalar:xigrowth2} are holding. Notice that the latter implies that in this regime both the scalar and tensor power spectra are strongly blue. This regime comes to an end when the gauge field friction term in Eq.~\eqref{eq_pseudoscalar:approx_eq_motion_2} overcomes the Hubble friction. This actually happens for a value $\xi_2$ that satisfies:
\begin{equation}
\xi_1 < \xi_2, \quad 
\frac{ e^{2 \pi \xi_2}}{\xi_2^{5} }\simeq \left( \frac{\alpha}{\Lambda} \right)^{-2} \left[ 0.4 \cdot 10^{-4} H^2 \right]^{-1}\ .
\label{eq_pseudoscalar:knee_equation}
\end{equation}
To solve this equation we need to specify the (model dependent) value of $H$ at this point. As already explained, the values of $\xi \in [\xi_1,\xi_2]$ can be translated into values for $N$ (and consequently of $f$) using Eqs.~\eqref{eq_pseudoscalar:xigrowth2} (and~\eqref{eq_pseudoscalar:Nf}).\\

\subsubsection{Strong gauge fields (C).}
\label{sec_pseudoscalar:strong_gauge}

\noindent In the last part of the evolution, \textit{i.e.} for $\xi_2 < \xi$, the non-linear gauge field friction term becomes dominant. In this regime we have $|\phi_{,N}| \ll |V_{, \phi}/V|$ and thus Eq.~\eqref{eq_pseudoscalar:approx_eq_motion_2} can be approximated as:
\begin{equation}
 \frac{V_{,\phi}}{V}  \simeq \frac{0.8}{3} \cdot 10^{-4} \left( \frac{\alpha}{\Lambda} \right) \frac{V}{\xi^4} e^{2 \pi \xi} \ .
 \label{eq_pseudoscalar:eomC}
\end{equation}
As the left-hand side of this equation is expected to be slowly variating, $\xi$ can grow at most logarithmically\footnote{A useful hint on the evolution of the system is given by considering the analogy with the classical problem of an object falling in some medium. As the friction increases with velocity, we expect the velocity $\phi_{,N} \simeq \xi$ to approach an asymptotic value. Moreover, it is interesting to notice that in this regime the friction term is stronger than required by the usual assumptions made in slow-roll inflation. We can thus safely neglect the acceleration term $\ddot \phi$, with respect to the other terms in the equation of motion for $\phi$.}. This regime actually lasts until the end inflation, \textit{i.e.}\ as long as $\epsilon_H = |\dot H|/H^2 < 1$. It is interesting to notice that this bound can be saturated, yielding to an upper bound for $\xi$. In particular, using $3 H^2 = V$ it is easy to get $\dot H \simeq V_{,\phi} \dot{\phi}/(6 H)$. This corresponds to $V_{,\phi} \lesssim 3 H^2 \alpha/(\xi \Lambda)$ that can be substituted into Eq.~\eqref{eq_pseudoscalar:eomC} to get:
\begin{equation}
\xi < \xi_\text{max} \ , \quad \frac{e^{2 \pi \xi_\text{max}}}{\xi_\text{max}^3} \lesssim \frac{3}{\mathcal{N} \cdot 2.4 \cdot 10^{-4} H^2} \ ,
\label{eq_pseudoscalar:xi_end}
\end{equation}
where we have also reintroduced $\mathcal{N}$. As in general we are interested in cases where $\xi_\text{max} > 1$, this equation implies that low-scale models of inflation, that typically corresponds to models with $p>2$ in the parametrization of Eq.~\eqref{eq_pseudoscalar:Nparameterization}, allow for larger values of $\xi$ and hence for stronger effects due to the presence of gauge fields. Notice that Eq.~\eqref{eq_pseudoscalar:xi_end} does not depend of $\alpha$. This implies that once the number of gauge fields $\mathcal{N}$ and the parametrization of Eq.~\eqref{eq_pseudoscalar:Nparameterization} are fixed, models with different values of $\alpha \neq 0$ give the same value $\xi_{max}$. As we discuss in Sec.~\ref{sec_pseudoscalar:scalar_tensor}, this bound is extremely useful to get hints on the shape of the scalar and tensor power spectra at small $N$.\\

\noindent
To conclude this Section we can finally discuss the dynamics in the strong gauge field regime. As already explained in this Section, in this regime $\xi$ is approximately constant. Moreover, as $\dot{\phi} = - \phi_{,N} H $ and $\xi \propto |\dot{\phi}|/H $ in this regime we also have $\phi_{,N}$ approximately constant. As a consequence we can approximate $\phi(N)$ as:
\begin{equation}
\label{eq_pseudoscalar:phi_last}
\phi \simeq \bar{\phi}_{,N} N + \phi_0 \ ,
\end{equation}
where we have defined $|\bar{\phi}_{,N}| \equiv 2 \bar{\xi} \Lambda/\alpha $ with $\bar{\xi} \equiv (\xi_\text{max} + \xi_2)/2 $  and $\phi_0$ denotes value of the $\phi$ at the end of inflation\footnote{The evolution is approximated by a uniform motion. The quantity $|\bar{\phi}_{,N}|$ can be interpreted as a mean velocity for the scalar field and $\phi_0$ can be interpreted as the initial position.}. A good approximation of $\phi_0$ can be determined by using $\epsilon_V = 1$. This formula we be used to get and estimate of the amount of e-foldings that the system spends in the strong gauge field regime:
\begin{equation}
\label{eq_pseudoscalar:N_estimate}
N_2 = (\phi_2 - \phi_0) \frac{\alpha}{2 \Lambda \bar{\xi}} \ ,
\end{equation}
Again the corresponding value of the frequency can be computed using Eq.~\eqref{eq_pseudoscalar:Nf}. It is crucial to stress that using this equation we can also determine $\Delta N_*$. For this purpose we start by computing $N_2$ and $\phi_2$, and proceed by computing $N_2^0$, number of e-foldings elapsed between $\phi_2$ and $\phi_0$ for $\alpha = 0$, so that $\Delta N_*$ simply reads:
\begin{equation}
\Delta N_* = N_2 - N_2^0\ ,
\label{eq_pseudoscalar:N2}
\end{equation}
This value can finally be substituted into the analytical expressions of Sec.~\ref{sec_pseudoscalar:intermediate_gauge} and Sec.~\ref{sec_pseudoscalar:weak_gauge} to get analytical estimates of the relevant points in scalar and tensor power spectra.

\subsection{The scalar and tensor spectra.}
\label{sec_pseudoscalar:scalar_tensor}
In this Section we discuss some of the main features of the scalar and tensor spectra and in particular we focus on the GW spectrum. As a first step, we notice that at CMB scales the gauge field contribution to the scalar power spectrum can be fixed to be subdominant, so that the spectrum is nearly scale-invariant around an amplitude of $\Delta^2_s \simeq 2.2 \cdot 10^{-9}$. Moreover, for all the models described by the parametrization of Eq.~\eqref{eq_pseudoscalar:Nparameterization}, the scalar-spectral index and the tensor-to-scalar ratio can be expressed as:
\begin{equation}
n_s \simeq 1 - \frac{\mathcal{O}(1)}{N_*}\ , \qquad \qquad r \simeq \frac{\mathcal{O}(1)}{N_*^p}
\label{eq_pseudoscalar:ns_r_N}
\end{equation}
where the $\mathcal{O}(1)$ factors depends on the choice of $\beta_p$ and $p$. As explained in the previous section, the gauge fields induce a shift in the point of the part of the potential that is probed by CMB observations in particular we have $N_* = N_\text{CMB} - \Delta N_* < N_\text{CMB}$. This implies that for a given model, we expect a smaller value for $n_s$ and a larger value of $r$ with respect to the case with $\alpha = 0$. As anticipated in Sec.~\ref{sec_pseudoscalar:signatures}, the observed values of $n_s$ and $r$ can thus be used to impose an upper bound on $\Delta N_*$. Notice that using Eq.~\eqref{eq_pseudoscalar:N2}, this can be turned into a constrain on the maximum value of $N_2$. Consequently using Eq.~\eqref{eq_pseudoscalar:N_estimate}, this can be used to set an upper bound on $\alpha/\Lambda$. \\

\noindent
We can proceed by discussing the consequences of the bound on $\xi_{max}$ given by Eq.~\eqref{eq_pseudoscalar:xi_end} on the shape of the spectra at small $N$. As discussed in Sec.~\ref{sec_pseudoscalar:mechanism}, in the strong gauge field regime the scalar power spectrum is given by Eq.~\eqref{eq_pseudoscalar:scalar_strong}, \textit{i.e.} it is proportional to $1/(\mathcal{N}\xi^2)$. As Eq.~\eqref{eq_pseudoscalar:xi_end} implies that small scale models give larger values for $\xi_{max}$, we find that the scalar power spectrum at small scales is suppressed in low-scale models of inflation. It is also crucial to notice that substituting Eq.~\eqref{eq_pseudoscalar:xi_end} into Eq.~\eqref{eq_pseudoscalar:OmegaGW}, we find that for fixed $\beta_p$ and $p$, we also have an absolute upper bound on the GW spectrum $\Omega_{GW}$:
\begin{equation}
\Omega_{GW} h^2 \lesssim 2.4 \cdot 10^{-5} \mathcal{N}^{-1} \ ,
\label{eq_pseudoscalar:OmegaMax}
\end{equation}
Notice that both the this bound and the one on the scalar power spectrum are derived on rather feeble assumptions, and in practice they can be seen as model independent. It is also interesting to stress that since this bound is saturated at the end of inflation, \textit{i.e.} at $N = 0$, this value of $\xi$ is approached at a universal value of the frequency:
\begin{equation}
f_\text{max} \simeq 3.6 \cdot 10^8~\text{Hz} \ ,
\end{equation}
that is directly obtained by substituting $N= 0$ into Eq.~\eqref{eq_pseudoscalar:Nf}.\\

\begin{figure}[h]
\centering
	\includegraphics[width=0.65 \textwidth]{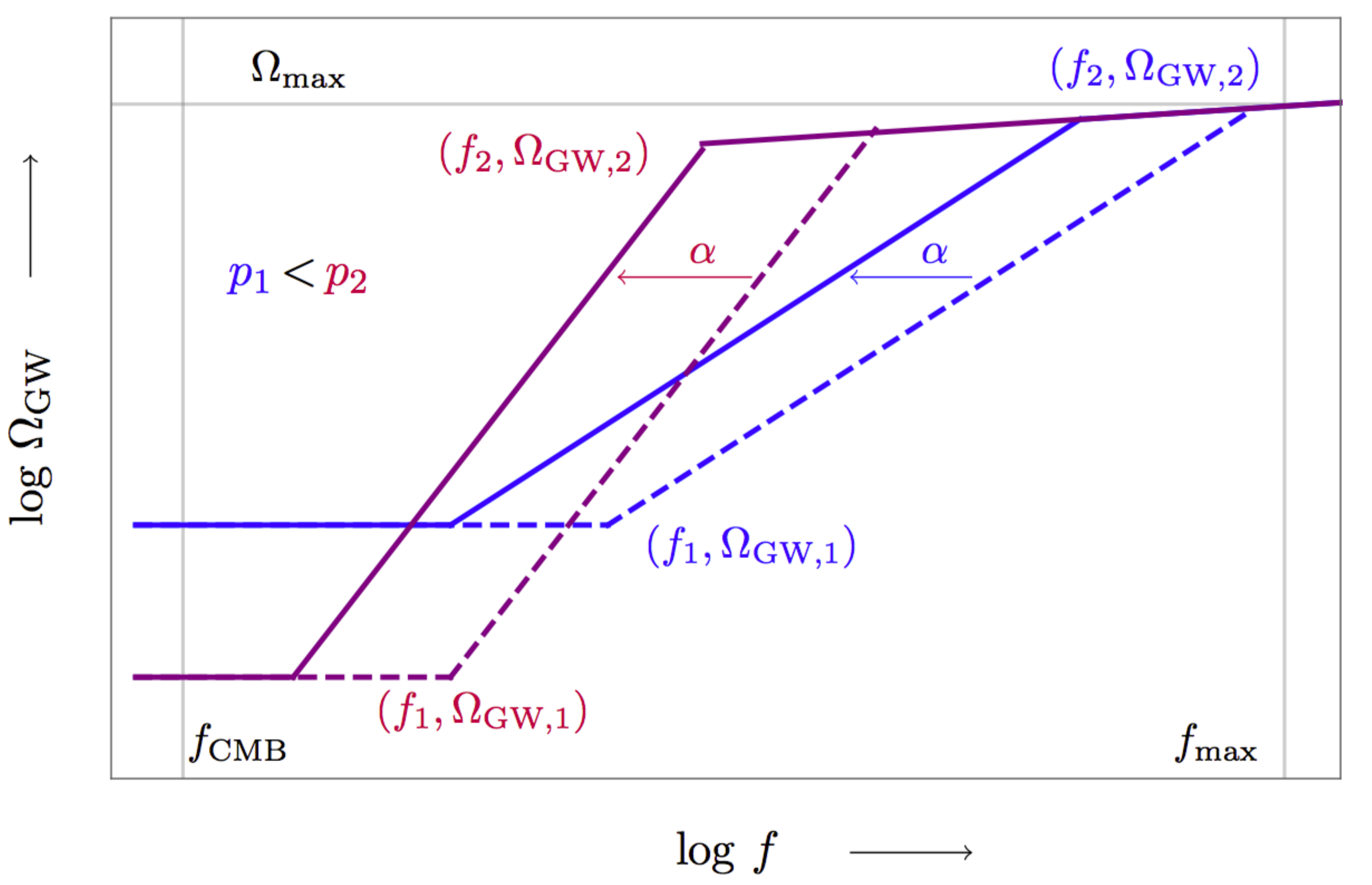}
	\caption{Schematic view of the gravitational wave spectrum for two different values of $p$ in Eq.~\eqref{eq_pseudoscalar:Nparameterization} and for two different values of the coupling $\alpha/\Lambda$ between the inflaton and the gauge field. \label{fig_pseudoscalar:Omega_schematic} }
\end{figure}

\noindent
As explained in Sec.~\ref{sec_pseudoscalar:mechanism}, the contribution of the gauge fields to GW spectrum is basically controlled by the parameter $\xi$. In the weak field regime, that actually corresponds to CMB scales or alternatively to very small frequencies, the GW spectrum is basically governed by the first slow-roll parameter that is proportional to $\beta_p/N_\text{CMB}^p$. On the contrary, in strong gauge fields regime, that corresponds to the last part of the evolution or alternatively to very large frequencies, the universal value $\Omega_\text{max}$ is slowly approached. In between, \textit{i.e.} in the intermediate regime, the spectrum has steep increase, that is actually governed by the $1/N^{p/2}$ growth of $\xi$ from $\xi = \xi_1$ to $\xi = \xi_2$. Models with higher value of $p$ in Eq.~\eqref{eq_pseudoscalar:Nparameterization}, correspond to low-scale models for inflation giving a smaller values of $H_1$. As a consequence, while these models have a smaller vacuum amplitude, they produce a steeper increase between $f_1$ and $f_2$, due to the faster growth of $\xi$. As a result, for these models, the plateau in the GW spectrum, corresponding to an approximately constant value of $\xi$, starts at smaller values of the frequency. We can thus conclude that models with a lower (vacuum) value for the tensor-to-scalar ratio $r = 16 \epsilon$ are expected to produce a larger GW signal in this setup. \\

\noindent
A schematic representation of the expected GW spectrum for two models with different values of $p$ is shown in Fig.~\ref{fig_pseudoscalar:Omega_schematic}. In particular, $p_1$ (blue) is fixed to be smaller than $p_2$ (purple). In this figure, the dashed curves have the same values of $\beta_p$ and $p$ and smaller value of $\alpha/\Lambda$ with respect to the corresponding solid curves. As it is possible to see from Fig.~\ref{fig_pseudoscalar:Omega_schematic}, the second parameter that considerably affects the spectrum is $\alpha/\Lambda$, coupling between the gauge field and the inflaton, that can actually be related to $\xi_{CMB}$. It should be clear from Eq.~\eqref{eq_pseudoscalar:xigrowth2}, that reducing this values corresponds to reducing $N_1$ (and correspondingly $N_2$) to smaller values, \textit{i.e.}\ shifting $f_1$ and $f_2$ to higher frequencies. \\

\noindent
Finally, we may notice that the slow increase of $\Omega_\text{GW}$ between $f_2$ and $f_\text{max}$, is directly related to the slow increase of $\xi$ in the strong gauge field regime. We can thus compare Eq.~\eqref{eq_pseudoscalar:knee_equation} and Eq.~\eqref{eq_pseudoscalar:xi_end} to get:
\begin{equation}
 \frac{e^{2 \pi \xi_2}}{\xi_2^3} = \frac{\phi_{,N}^2(N_2)}{2} \frac{e^{2 \pi \xi_\text{max}}}{\xi_\text{max}^3} \ ,
\end{equation}
where we have used the definition of $\xi$ in terms of $\alpha/\Lambda$ and $\phi_{,N}$. An estimate of the value of $\phi_{,N} (N_2)$ can be obtained using the parametrization of Eq.~\eqref{eq_pseudoscalar:Nparameterization}, and thus we may notice that low-scale models (while allowing for an earlier and steeper growth of the spectrum) are also predicting a smaller value for $\xi_2$. \\

\noindent
We can finally conclude this Section by summarizing the effects of the different parameters on the spectra. The parameter $p$ both affects the vacuum amplitude and the slope of the increase in the scalar and tensor spectra. The coupling $\alpha/\Lambda$, shifts the spectrum horizontally and finally $\beta_p$ affects the vacuum amplitude and thus vertically shifts the spectra. Using these parameters we can discuss the detectability of the GW signal.

\section{Some explicit models.}
\label{sec_pseudoscalar:models}
As introduced in Sec.~\ref{sec_pseudoscalar:analytical}, the mechanism discussed in Sec.~\ref{sec_pseudoscalar:mechanism} has widely been discussed~\cite{Turner:1987bw, Garretson:1992vt, Anber:2006xt,Anber:2012du,Anber:2009ua,Barnaby:2011qe,Barnaby:2011vw,Linde:2012bt} in the case of Chaotic potentials (that corresponds to $p = 1$ in the parametrization of Eq.~\eqref{eq_pseudoscalar:Nparameterization}). In order to extend the analysis to a broader class of models~\cite{Domcke:2016bkh}, in this Section we consider models that correspond to different values of $p$. As already explained in this Chapter, fixing a value for $p$ does not correspond to fix a model but rather it corresponds to specifying a class of models. \\

\noindent
Notice that for all the models considered in this Section we set $\mathcal{N} \neq 1$. The consequences of the relaxation of this condition are discussed in Sec.~\ref{sec_pseudoscalar:discussion}\\

\noindent 
To starting with this analysis, it is useful to compute the potentials that correspond to different values of $p$ and $\beta_p$. For this purpose we start by recalling the approximate relationship between the potential $V(\phi)$ and the number of e-foldings $N$ derived in Sec.~\ref{chapter:beta}:
\begin{equation}
	\label{eq_pseudoscalar:models_N}
	\frac{\textrm{d} N}{\textrm{d} \phi} \simeq \left( \frac{\textrm{d} \ln V(\phi)}{\textrm{d} \phi}  \right)^{-1} \ .
\end{equation}
By differentiating Eq.~\eqref{eq_pseudoscalar:Nparameterization} and substituting into Eq.~\eqref{eq_pseudoscalar:models_N} we can thus get the differential equation: 
\begin{equation}
	\label{eq_pseudoscalar:models_differential}
	\epsilon_{V,\phi} = - \frac{p}{\sqrt{2} \beta_p^{ \ \frac{1}{p}}} \epsilon_V^{\ \frac{p+2}{2p}} \ .
\end{equation}
Notice that the case $p = 2$ is special and it should be distinguished from the other cases. Once the solution of this equation is found, the corresponding potential is determined by:
\begin{equation}
	\label{eq_pseudoscalar:epsilon_V_rel}
	\epsilon_V = \frac{1}{2} \left( \frac{\textrm{d} \ln V(\phi)}{\textrm{d}\phi } \right )^2 \ .
\end{equation}
As a consequence we find:
\begin{itemize}
	\item $p = 2$.
	As in this case Eq.~\eqref{eq_pseudoscalar:models_differential} reduces to:
	\begin{equation}
		\epsilon_{V,\phi} = - \frac{2}{\sqrt{2 \beta_p } } \epsilon_V \ ,
	\end{equation}
	the solution is given by:
	\begin{equation}
		\label{eq_pseudoscalar:models_epsilon_starobinsky}
		\epsilon_V \simeq \exp\left( -\sqrt{\frac{2}{\beta_p}} \phi \right) \ .
	\end{equation}
	Finally we can use Eq.~\eqref{eq_pseudoscalar:epsilon_V_rel} to compute the corresponding potential. As already explained in Chapter~\ref{chapter:beta}, this result can be obtained if we consider models in the Exponential class with potential:
	\begin{equation}
		\label{eq_pseudoscalar:models_starobinsky_potentials}
		V(\phi) \simeq V_0 \left(1 - e^{- \gamma \phi}\right)^2 .
	\end{equation}
	Notice that for $\gamma = \sqrt{2/3}$ this asymptotic behavior matches with the standard Starobinsky\footnote{It is however crucial to stress that the standard Starobinsky model describes \emph{scalar} particles. For this reason we refer to these models as Starobinsky-like class.} model~\cite{Starobinsky:1980te}. Following the definitions of~\cite{Domcke:2016bkh}, in this whole Chapter, we refer to this class \textit{i.e.} to plateau-like potentials of Chapter~\ref{chapter:inflation} as Starobinsky-like class. 
	
	\item $p \neq 2$.
	In this case the solution is given by:
	\begin{equation}
	\label{eq_pseudoscalar:models_epsilon_general}
	\epsilon_V \simeq \left( -\frac{(p-2)}{\sqrt{8}\beta_p^{\ \frac{1}{p}}} \phi \right)^{\frac{2p}{p-2}}.
	\end{equation}
	We can thus get the explicit expressions for the potentials corresponding to some given values of $\beta_p$ and $p$. In particular, as already explained in Chapter~\ref{chapter:beta}, it is easy to show that $p=1$ corresponds to chaotic models~\cite{Linde:1983gd} with potential:
	\begin{equation}
		\label{eq_pseudoscalar:models_chaotic_potentials}
		V (\phi)= V_0 \ \phi^q  \ .
	\end{equation}
	Similarly, we can show that Hilltop models~\cite{Boubekeur:2005zm} introduced in Chapter~\ref{chapter:inflation}, correspond to bigger values of $p$. The potentials for these models are:
	\begin{equation}
	\label{eq_pseudoscalar:models_hilltop_potentials}
	V(\phi) =  V_0 \left[1 - \left (\frac{\phi}{v} \right)^q \right]^2 \ ,
	\end{equation}
	where, consistently with the discussion of Chapter~\ref{chapter:beta}, $p = 2 (q-1)/(q-2)$. 
\end{itemize}
Using this classification we can finally discuss the background dynamics and the perturbations associated with the different models. Notice that once $p$ is set, the Chaotic and Starobinsky-like models are basically specified by three parameters (four if we also consider models with $\mathcal{N} \neq 1$) \textit{i.e.} $\alpha/\Lambda , V_0 $ and $\beta_p$, where the dependence on $\beta_p$ is expressed in terms of the parameters $\gamma$ and $q$. On the other hand, in the Hilltop models the value of $\beta_p$ is basically set by the lowest order approximation of Eq.~\eqref{eq_pseudoscalar:models_hilltop_potentials} and we have thus introduced the energy scale $v$. The parameter of the models are chosen by using the analytical estimates of Sec.~\ref{sec_pseudoscalar:analytical}. In particular, we choose the parameters that maximize the GW signal without violating the constraints of Sec.~\ref{sec_pseudoscalar:signatures}. More details on the methods used to perform these estimates are given in the Appendix of~\cite{Domcke:2016bkh}.

\subsection{Numerical results.}
\label{sec_pseudoscalar:numerical_results}

\begin{figure}[ht!]
\centering
\subfloat[][\emph{Quadratic model with $\alpha/\Lambda = 35$ and $V_0~=~1.418~\cdot 10^{-11}$}.]
{\includegraphics[width=.45\columnwidth]{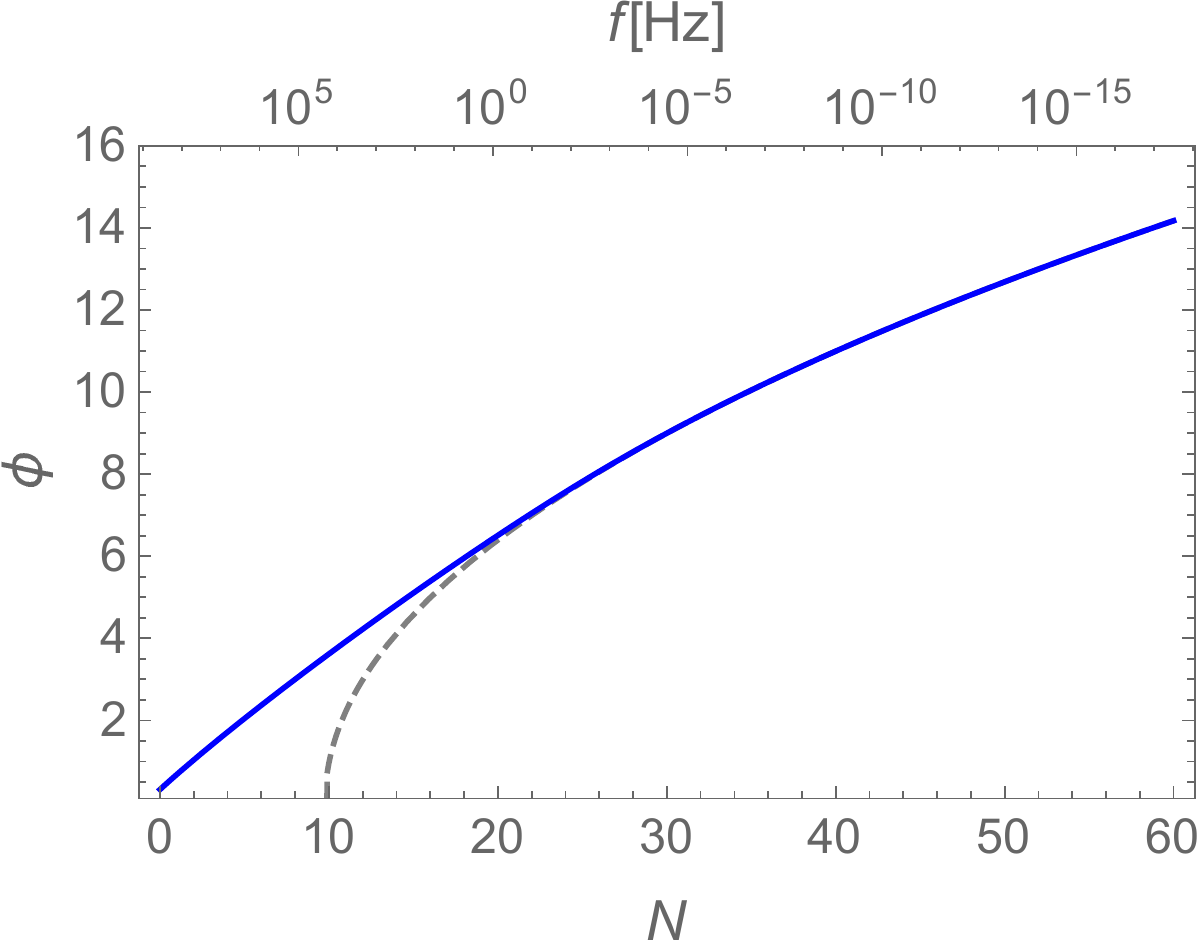}} \quad
\subfloat[][\emph{Starobinsky-like model with $\alpha /\Lambda = 75 $, $\gamma~=~0.3$, $V_0 = 1.17 \cdot 10^{-9}$}.]
{\includegraphics[width=.45\columnwidth]{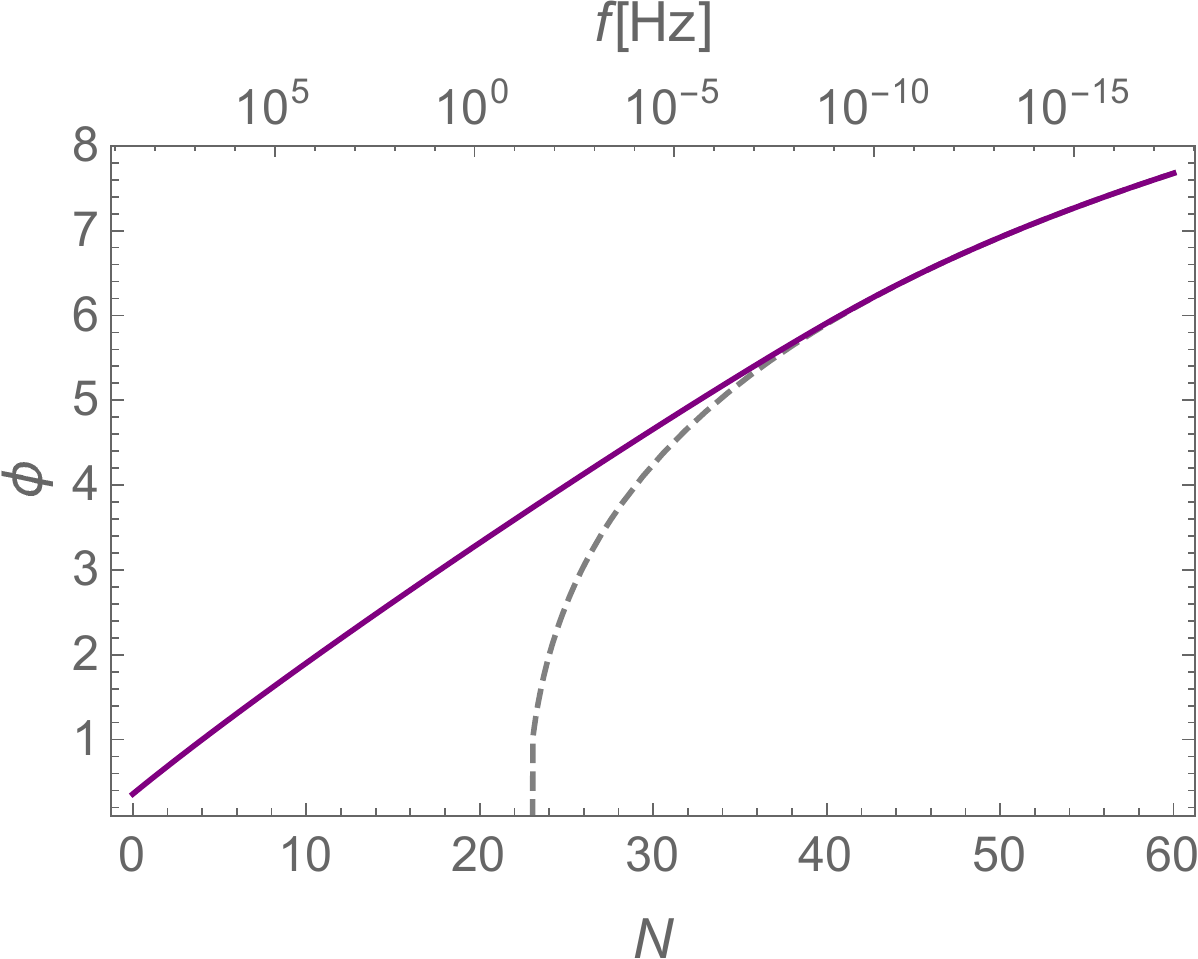}} \\
\subfloat[][\emph{Hilltop model with $q =4$, $\alpha/\Lambda =2000$, $v =0.1$ and $V_0 = 1.0 \cdot 10^{-21}$ }.]
{ \hspace{-0.1cm}\includegraphics[width=.485\columnwidth]{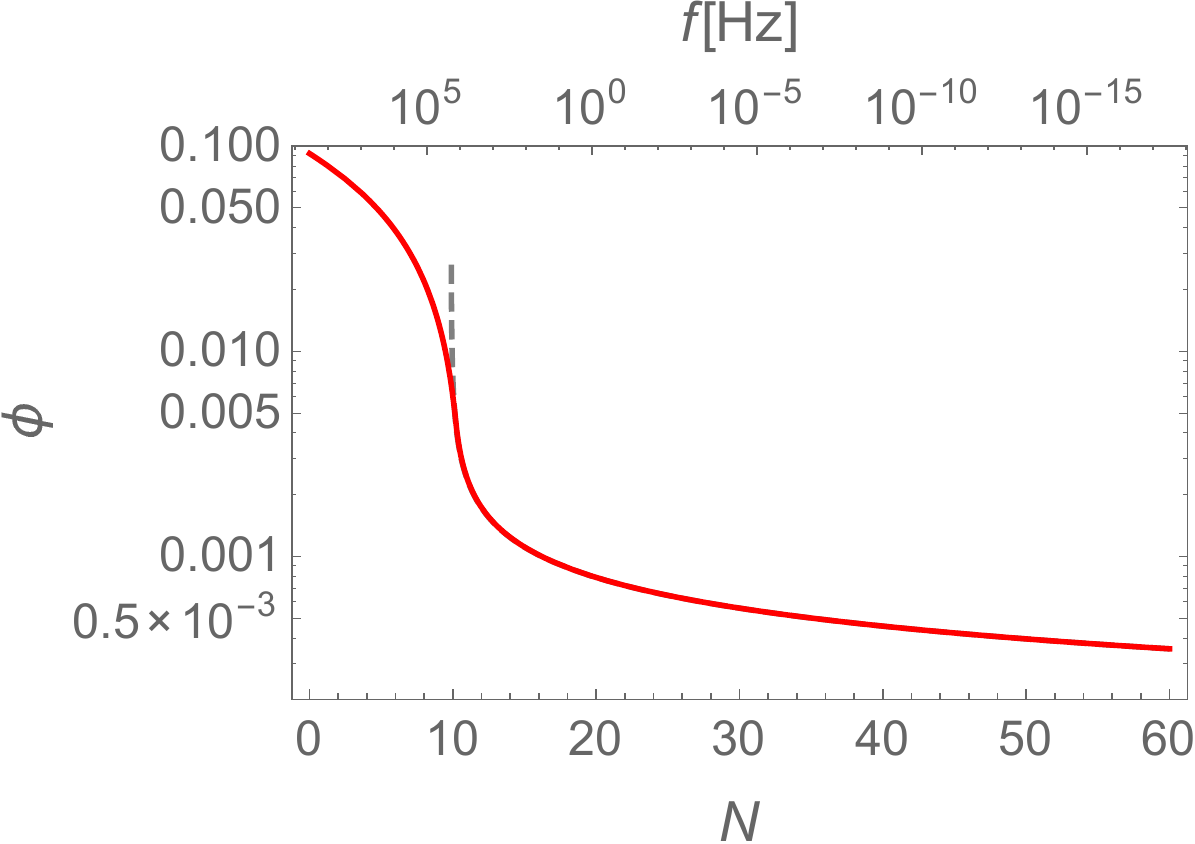}} \quad
\subfloat[][\emph{Hilltop model with $q =3$, $\alpha/\Lambda =2000$, $v =0.1$ and $V_0 = 3.6 \cdot 10^{-18}$ }.]
{ \includegraphics[width=.468\columnwidth]{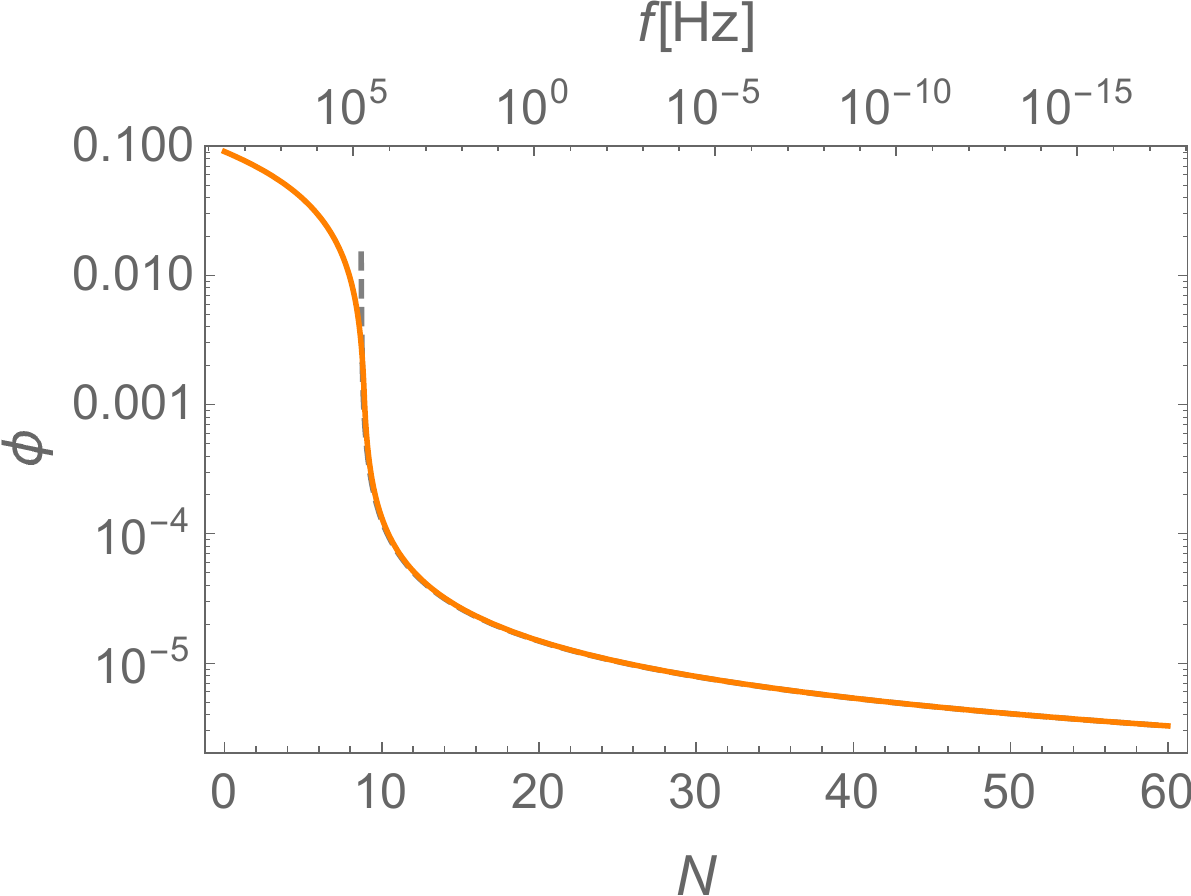}}
\caption{Evolution of inflaton field $\phi$ as a function of $N$ with (solid line) and without (dashed line) the non-minimal interaction with the gauge fields. \label{fig_pseudoscalar:phi_comparison} }
\end{figure}

\noindent
The evolution of the scalar field $\phi$, for models with $p = 1,2,3,4$ at a fixed parameter point\footnote{Here we have set $N_\text{CMB} = 60$. As discussed in Sec.~\ref{sec_pseudoscalar:discussion}, there is a degeneracy between the choice of $N_\text{CMB}$ and $\alpha/\Lambda$.} is shown in Fig.~\ref{fig_pseudoscalar:phi_comparison}. Notice that these plots (as well as all the other plots shown in this Chapter) have been obtained by numerically solving the complete equation of motion give in Eq.~\eqref{eq_pseudoscalar:approx_eq_motion_2}. As expected, the additional friction terms due to the presence of the gauge field, only affects the last part of inflation and actually slows down the evolution.\\

\begin{figure}[h]
\centering
{\includegraphics[width=.6\columnwidth]{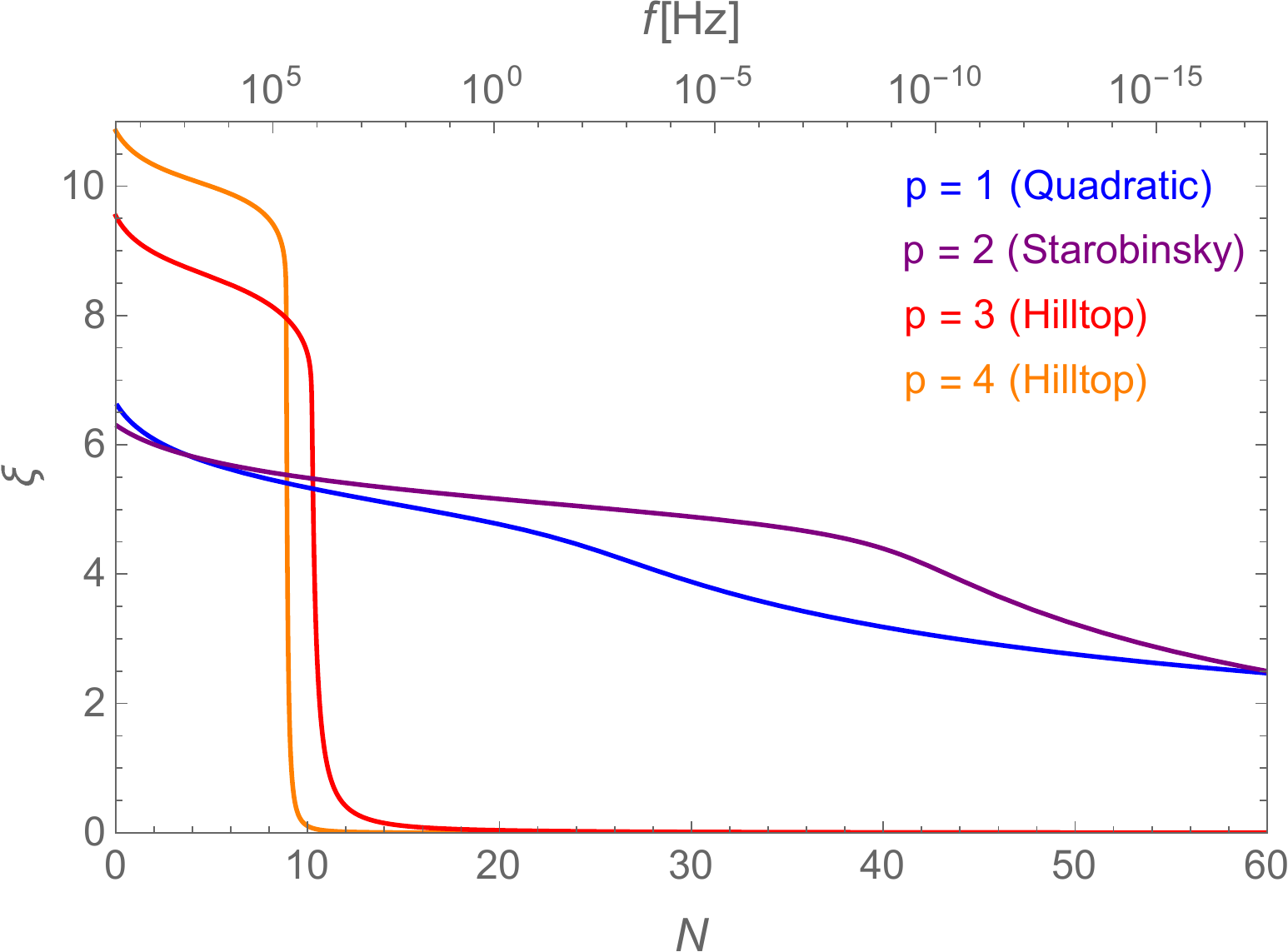}}
\caption{Evolution of the parameter $\xi$ governing the strength of the gauge interactions for models with different values of $p$ as defined in Eq.~\eqref{eq_pseudoscalar:Nparameterization}. The parameter choices for all of these models are the same of Fig.~\ref{fig_pseudoscalar:phi_comparison}. \label{fig_pseudoscalar:xi_all} }
\end{figure}

\noindent
In Fig.~\ref{fig_pseudoscalar:xi_all} we show the evolution of the parameter $\xi$ for all the models with $p = 1,2,3,4$. Notice that as expected the plots of $\xi$ for all these models are approximately resembling the plot of Fig.~\ref{fig_pseudoscalar:phi_xi_schematic}. Models with higher values of $p$ (that correspond to low-scale inflationary models) give a bigger value for $\xi_{\text{max}}$. We can also notice that the values of $\xi_\text{CMB}$ for the different models are respecting the condition of Eq.~\eqref{eq_pseudoscalar:NG}. \\

\noindent
In Fig.~\ref{fig_pseudoscalar:DeltaS_all} we show the scalar power spectra for the models of Fig.~\ref{fig_pseudoscalar:phi_comparison} and of Fig.~\ref{fig_pseudoscalar:xi_all}. We can immediately notice that the parameters of the models are fixed in order to fit the COBE normalization at $N \simeq 60$ and moreover, in order to respect the Planck constraints of Eq.~\eqref{eq_pseudoscalar:PLANCK}, all of the spectra are nearly flat at CMB scales. In agreement with the estimate of Eq.~\eqref{eq_pseudoscalar:scalar_strong}, the value of $\Delta^2_s$ on small scales is proportional to $\xi^{-2}$. In particular, the Hilltop models are predicting a smaller value of $\Delta^2_s$ at small scales. It is fair to point out that, when we restrict to the case $\mathcal{N} = 1$, all these models are in tension with the estimate of the PBH bound given by Linde in~\cite{Linde:2012bt}. However, the discrepancy is only by a $\mathcal{ O}(1)$ factor, that can actually be addressed by taking into account the theoretical uncertainties in the PBH bound. Moreover, as we show in the following (see Sec.~\ref{sec_pseudoscalar:several_gauge_fields} and in particular Fig.~\ref{fig_pseudoscalar:several_gauge_plots}), considering models with $\mathcal{N} \neq 1$, it is actually possible to produce observable GW, while respecting this bound. Notice that as predicted by Eq.~\eqref{eq_pseudoscalar:vacuumamplitudes}, models with $p =3,4$ presents a much steeper decrease in the first part of the evolution with respect the other models.\\

\begin{figure}
\centering
{\includegraphics[width=.7\columnwidth]{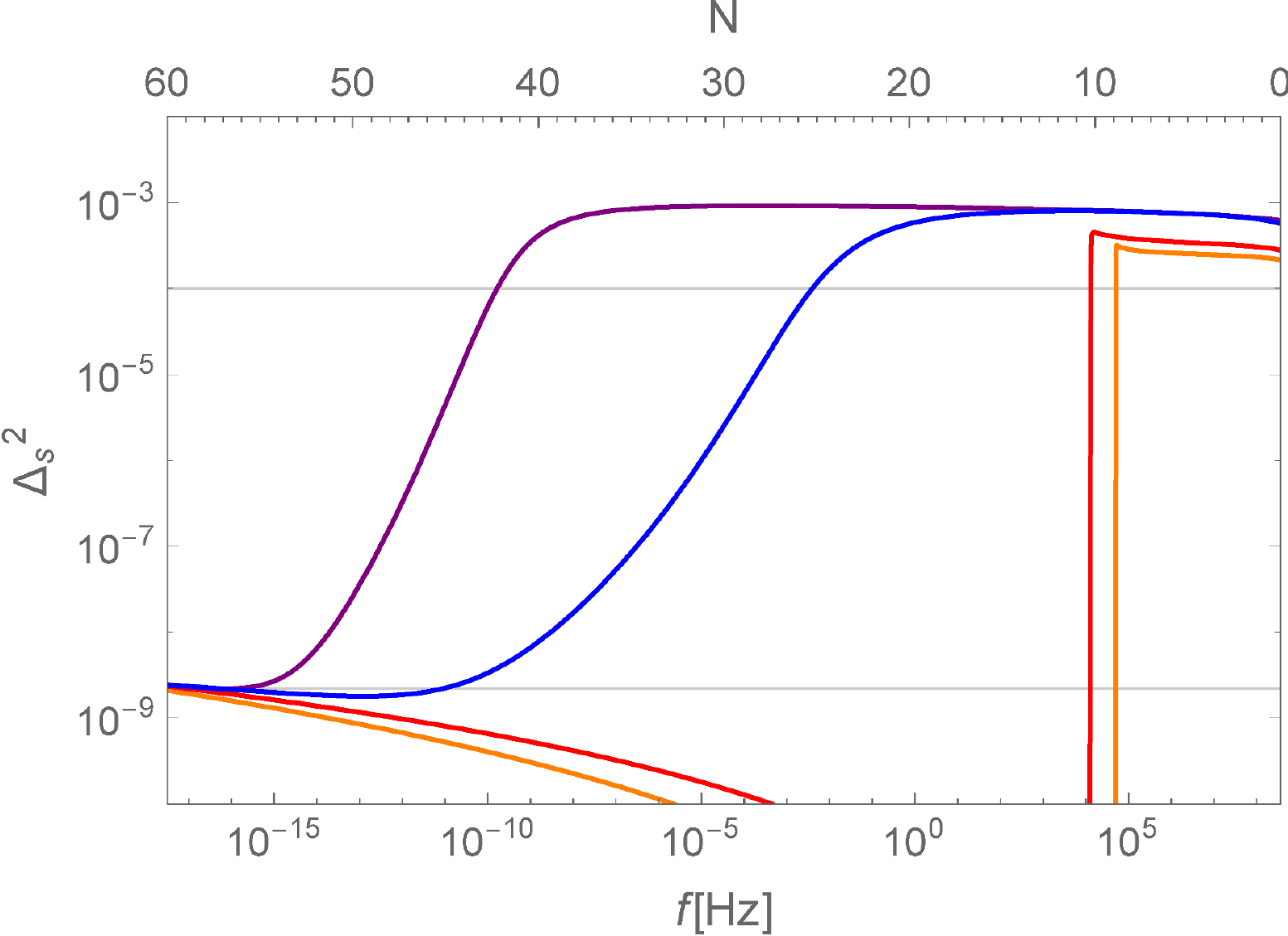}}
\caption{Power spectrum of scalar perturbations for all the models with the same parameters and color code of Fig.~\ref{fig_pseudoscalar:xi_all}. The upper horizontal line estimates the PBH bound, the lower one indicates the COBE normalization. \label{fig_pseudoscalar:DeltaS_all} }
\end{figure}

\begin{figure}
\centering
{\includegraphics[width=.7\columnwidth]{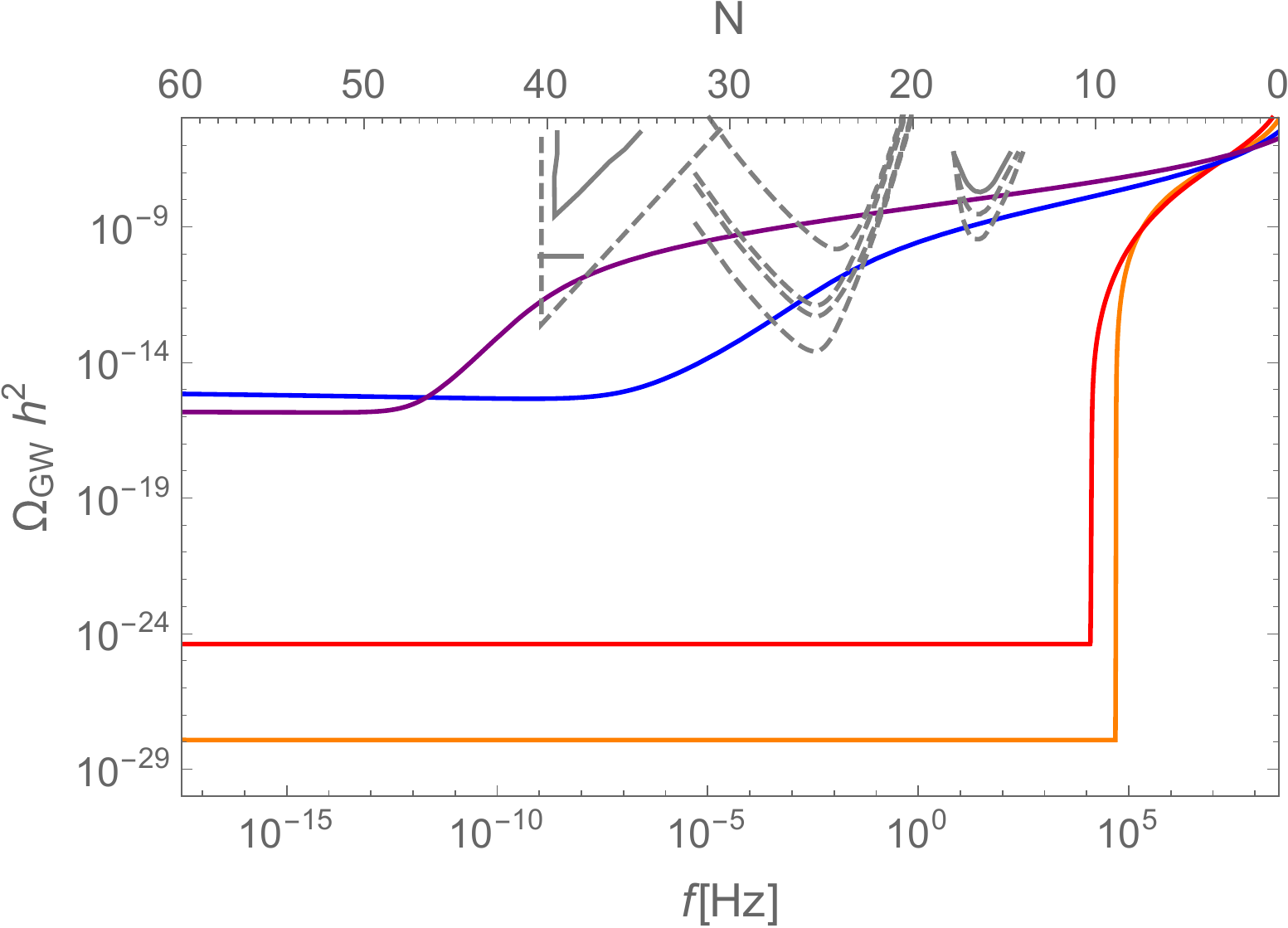}}
\caption{Gravitational wave spectrum for all the models with the same parameters and color code of Fig.~\ref{fig_pseudoscalar:xi_all}. We are also showing the sensitivity curves for (from left to right): milli-second pulsar timing, eLISA, advanced LIGO. Current bounds are denoted by solid lines, expected sensitivities of upcoming experiments by dashed lines. See main text for details. \label{fig_pseudoscalar:GW_all} }
\end{figure}

\noindent
The GW spectra for the models considered in this work are shown in Fig.~\ref{fig_pseudoscalar:GW_all}. Again we notice that the shape of the spectra are approximatively reproducing the schematic behavior shown in Fig.~\ref{fig_pseudoscalar:Omega_schematic}. In particular, all the curves present two abrupt changes in the slope. In Fig.~\ref{fig_pseudoscalar:Omega_schematic} we compare the GW spectra with the sensitivity curves of present (solid lines) and future (dashed lines) direct GW detectors. The first set of curves on the left represents the millisecond pulsar timing arrays covering frequencies around $10^{-10}$~Hz. In particular we show the constraint depicted in Ref.~\cite{Smith:2005mm}, the update from EPTA~\cite{vanHaasteren:2011ni} and the expected sensitivity of SKA~\cite{Kramer:2004rwa}. The two other sets are respectively space-based GW interferometers in the milli-Hz range (eLISA~\cite{Caprini:2015zlo}) and ground-based detectors sensitive at a few 10 Hz (LIGO/VIRGO~\cite{TheLIGOScientific:2016wyq}). The sensitivity curves for eLISA, correspond to the four configurations listed in Tab.~\ref{tab:lisa}. For LIGO, we depict the current bound O1:2015-16, as well as the expected sensitivities for the runs O2:2016-17 an O5:2020-22. \\

\begin{table}[h]
\centering
\begin{tabular}{llccc}
name & full name & number of arms & armlength [Gm] & lifetime [yr] \\ \hline
C1 & L6A5M5N2 & 3 & 5 & 5 \\
C2 & L6A1M5N2& 3 & 1 & 5 \\
C3 & L4A2M5N2 & 2 & 2 & 5 \\
C4 & L4A1M2N1& 2 & 1 & 2 
\end{tabular}
\caption{Configurations of the planned space-based GW mission eLISA considered in this paper.}
\label{tab:lisa}
\end{table}

\noindent
As it is possible to see from Fig.~\ref{fig_pseudoscalar:GW_all}, the parameter choice of~\cite{Domcke:2016bkh} leads the Quadratic and the Starobinsky-like model to generate a GW signal that can both be observed at eLISA and at advanced LIGO. It is interesting to notice that for the parameter choice of~\cite{Domcke:2016bkh}, the Starobinsky-like model also happens to produce a GW spectrum that can be observed by the milli-second pulsar timing. On the other hand, the two Hilltop models produce a GW signal that is well outside the observable windows for all of these experiments. The main reason for this result is, that the value of $n_s$ predicted by these models, even with $\alpha / \Lambda = 0$, is smaller than the value measured by Planck and reported in Eq.~\eqref{eq_pseudoscalar:PLANCK}. As the introduction of the gauge field is effectively reducing the value of $n_s$, to limit the decrease of $n_s$ we can only allow for gauge field production in the very last part of inflation.\\

\noindent
As already stated during this Section, the parameter choices for the models were guided by the estimates of Sec.~\ref{sec_pseudoscalar:analytical}\footnote{Details on the procedure to use the estimates to choose the parameters of the models are given in~\cite{Domcke:2016bkh}.}. In particular, we have used these estimates to maximize the GW signal without violating the observational constraints of Eq.~\eqref{eq_pseudoscalar:COBE}, Eq.~\eqref{eq_pseudoscalar:PLANCK} and Eq.~\eqref{eq_pseudoscalar:NG}. As a consequence, it is actually possible to reduce the GW signal by variating the parameters of the models. For example, by reducing the value of $\alpha/\Lambda$, we can shift the rise of the spectrum at larger frequencies. This actually results in reducing the predicted signal at a given frequency. A numerical scan of the parameter space for the case of Starobinsky-like model is presented in Sec.~\ref{sec_pseudoscalar:discussion}. We choose to represent this particular class of models, whose potential is shown in Eq.~\eqref{eq_pseudoscalar:models_starobinsky_potentials}, as it appears to be the most promising for what concerns the generation of observable GW signatures. Similar plots can actually be produced for all the classes discussed in this work.

\section{Discussion.}
\label{sec_pseudoscalar:discussion}
The analysis performed in this Chapter has revealed some universal features in the models described by the action of Eq.~\eqref{eq_pseudoscalar:action_pseudoscalar} and it shed light on the parameter dependencies and degeneracies. As discussed through this Chapter, the condition $\epsilon_H \simeq 1$ at the end of inflation, induces a universal feature in the GW spectrum at large frequencies. In particular, we find that the amplitude of the GW at the very end of inflation does not depend on the underlying model of inflation and moreover it happens to be insensitive to variations of the coupling parameter $\alpha/\Lambda$. However, as explained through this Chapter, the ratio $\alpha/\Lambda$ can be used to shift the increase of the GW spectrum. In particular, by reducing the value of $\alpha/\Lambda$ we can shift the increase towards larger values of the frequency, pushing the signal out of the expected range for future detectors. \\

\noindent
A useful classification of the different inflationary models is obtained by using the parametrization of Eq.~\eqref{eq_pseudoscalar:Nparameterization}. Remarkably, we find that low-scale inflation models ($p=3,4$), which give a small tensor-to-scalar ratio, induce a steeper increase of the GW spectrum. While this feature would suggest that these models are more likely to produce a detectable GW signal, in practice the CMB constraints only allow the gauge fields to affect the very last part of the evolution, giving rise to an undetectable signal. This is basically due to the modifications of $n_s$ and $r$ induced by the gauge fields (explained in Sec.~\ref{sec_pseudoscalar:scalar_tensor}). In particular, as low-scale inflation models predict a fairly small value of $n_s$ (and the presence of the gauge field tends to reduce this value) their production can only be allowed at the very end of inflation. \\

\noindent
As we have shown in Sec.~\ref{sec_pseudoscalar:models}, if $\alpha/\Lambda$ is sizable, the signals produced by models with $p=1$ and $p=2$ are expected to be detected (or conversely ruled out) by the upgraded versions of LIGO/VIRGO. In the case of a positive detection, the upcoming eLISA mission would potentially help in differentiating between these two cases, as well as constraining the value of $\alpha/\Lambda$. Moreover, the Starobinsky-like model ($p=2$) is also producing a signal that can be observed by millisecond pulsar-timing arrays. As among the models discussed in this Chapter, the pseudo-scalar Starobinsky-like model appears to the most promising from the point of view of possible direct GW detections, in the following we focus our discussion on this particular case.

\subsection{Scan Plots.}
\noindent
The CMB constraints and all the other experimental bounds presented in Sec.~\ref{sec_pseudoscalar:signatures}, provide a powerful method to constrain the models presented in Sec.~\ref{sec_pseudoscalar:models}. In particular, these constraints can be gathered into scan plots for the parameters of the models. In the following we show two of these plots for the Starobinsky-like model of Sec.~\ref{sec_pseudoscalar:models}. To produce these plots, we explore the parameter space spanned by $\alpha/\Lambda$ and $\gamma$, by solving numerically the equation of motion~\eqref{eq_pseudoscalar:approx_eq_motion_2} for the inflaton field, iteratively fixing the value of $V_0$ in order to respect the COBE normalization of Eq.~\eqref{eq_pseudoscalar:COBE}.\\

\begin{figure}
\centering
\subfloat[][\emph{LIGO plot}.]
{\includegraphics[width=.45\columnwidth]{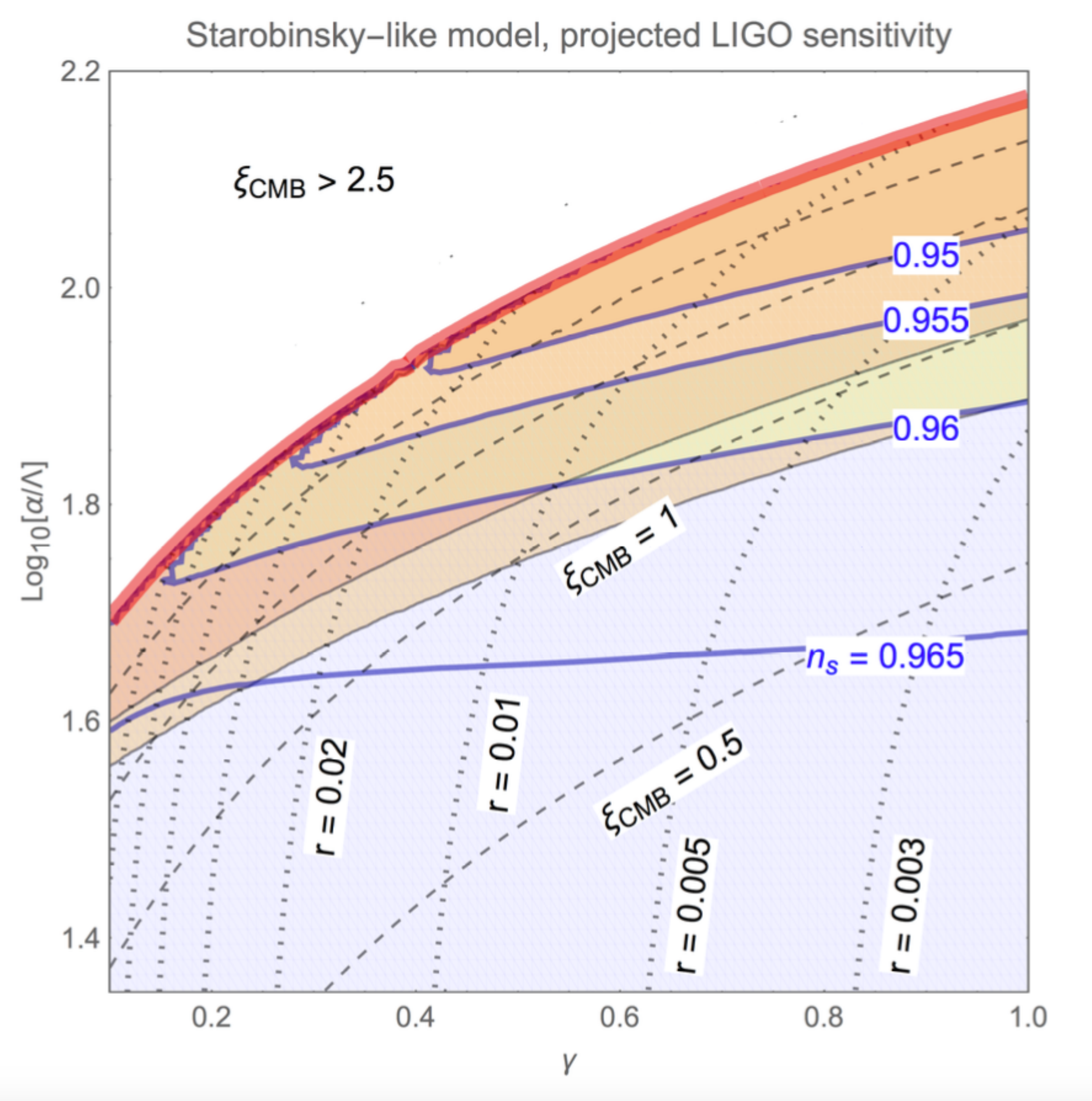}} \quad
\subfloat[][\emph{eLISA plot}.]
{\includegraphics[width=.45\columnwidth]{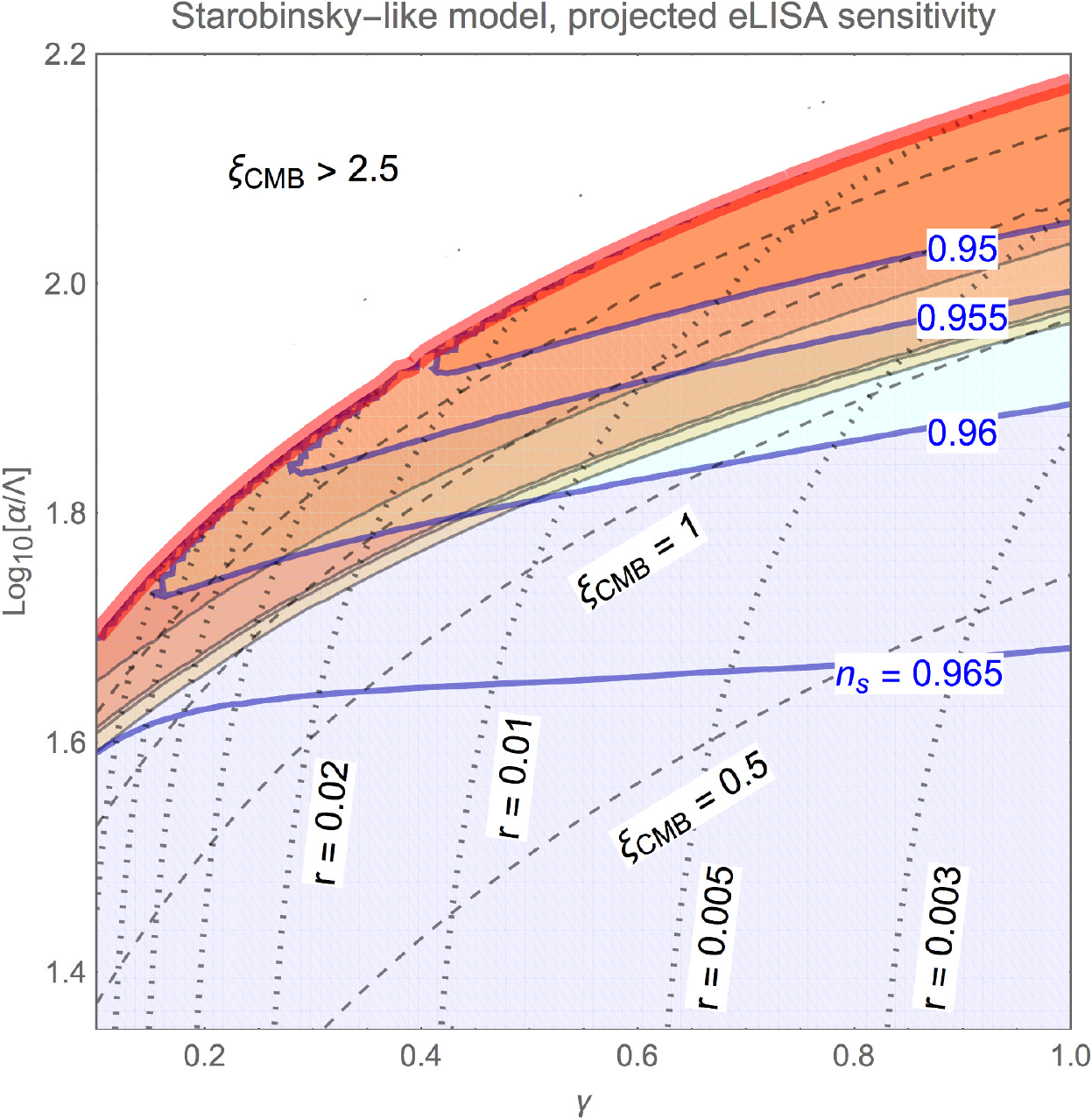}} 
\caption{Plot of the $(\alpha/\Lambda, \gamma)$ parameter space for the Starobinsky-like model with contour lines for $n_s$ (solid blue), $r = \{0.003, 0.005, 0.1, 0.2, 0.3, \dots \}$ (dotted) and $\xi_\text{CMB} = \{ 0.5, 1, 1.5, \dots \}$ (dashed). The orange shaded regions denote the projected sensitivity for advanced LIGO in the O2 and O5 run (left panel) and for eLISA in the C1 - C4 configurations (right panel). \label{fig_pseudoscalar:scan_plots} }
\end{figure}

\noindent
Fig.~\ref{fig_pseudoscalar:scan_plots} shows constraints from CMB measurements ($\xi_\text{CMB}$, $n_s$, $r$) as well as constraints and the projected sensitivity of direct gravitational wave detectors (eLISA and LIGO/VIRGO). The solid blue lines correspond to fixed values for $n_s$, dotted lines correspond to fixed values for $r$ and dashed lines correspond to constant values for $\xi_{CMB}$. The upper bound $\xi_{CMB} \simeq 2.5$ is marked by the red line. The orange shaded regions correspond to the observable regions for LIGO (left panel, evaluated at $50\,$Hz, runs O1, O2 and O5 as detailed in Sec.~\ref{sec_pseudoscalar:models}) and LISA (right panel, evaluated at $0.01\,$Hz, configurations C1 - C4 as detailed in Sec.~\ref{sec_pseudoscalar:models}). Remarkably, the current constraint on $\xi_\text{CMB}$ approximately coincides with the recently published data on LIGO run O1~\cite{TheLIGOScientific:2016wyq}. Moreover, for $\gamma \gtrsim 0.2$ this also corresponds to the line in parameter space above which a too large value for $n_s$ is achieved.\\

\noindent
It is interesting to notice that measurements are starting to probe the viable parameter space of this model, and they are beginning to corner the parameter space from different directions: searches for non-Gaussianities in the CMB and direct gravitational wave detection probe the region of large $\alpha/\Lambda$, searches for GWs in the CMB constrain the small $\gamma$ region. Clearly a more precise measurement of $n_s$, could further narrow down the viable range for $\alpha/\Lambda$. As already stated at the end of Sec.~\ref{sec_pseudoscalar:models}, similar plots can actually be produced for all the other classes discussed in this work. In particular, in the case of the Hilltop models ($p\geq 2$), the corresponding parameter space is spanned by $\alpha/\Lambda$ and $v$. However, for $n_s > 0.9$, these models feature an unobservable GW signal both for eLISA and LIGO. Moreover, the tensor to scalar ratio is typically unobservable and the spectral index is smaller than the observed value.\\

\begin{figure}
\centering
\includegraphics[width=.45\columnwidth]{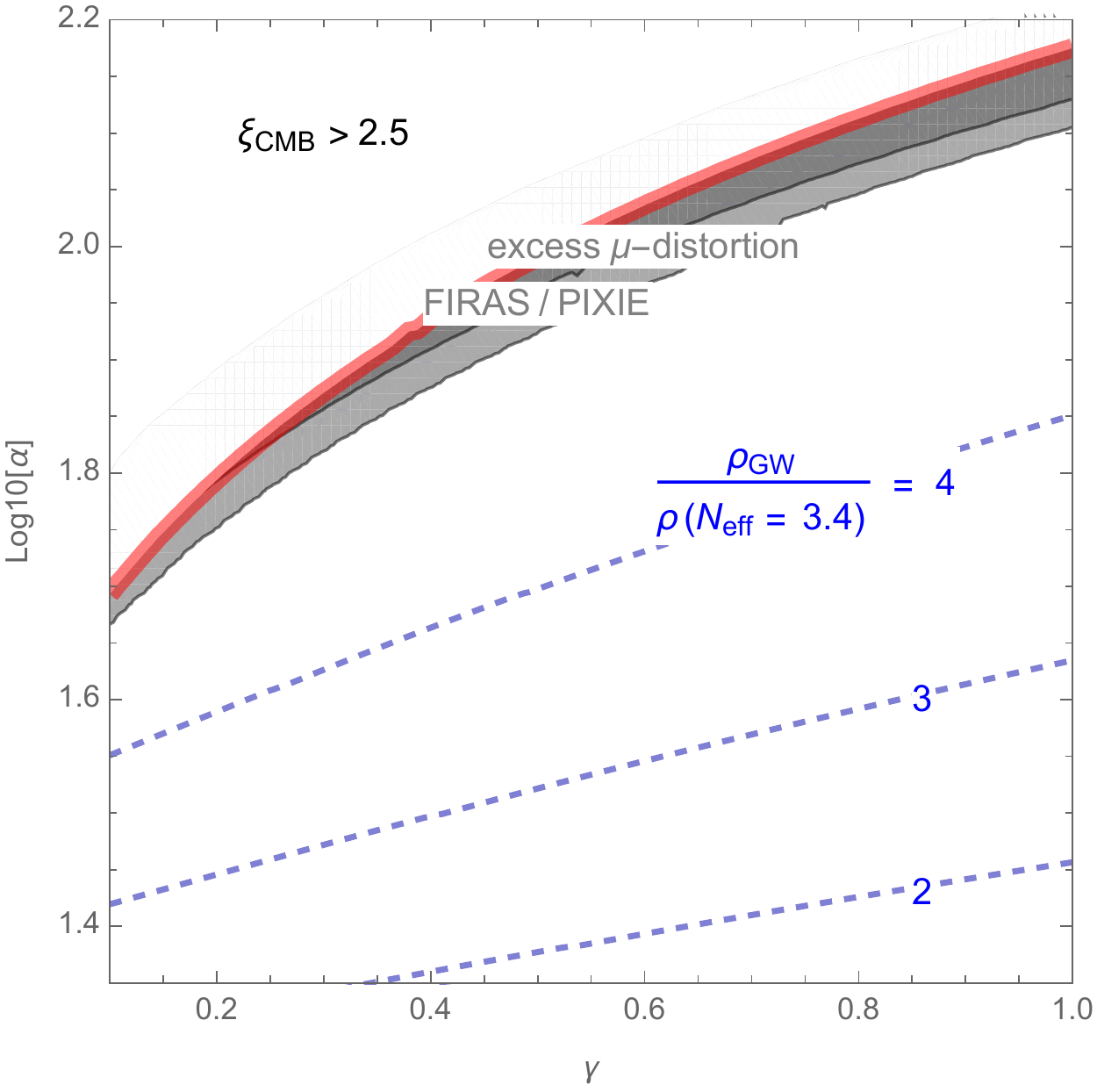}
\caption{
Plot of the $(\alpha/\Lambda, \gamma)$ parameter space for the Starobinsky-like model. The dashed blue lines denote the factor by which $N_\text{eff}$ exceeds the current 95$\%$ CL bound of~\cite{Ade:2015xua}. Dark gray colored regions denote the constraints the amount of $\mu$-distortions that are excluded by FIRAS (COBE) at 95$\%$ CL. The light gray region shows the expected sensitivity of PIXIE.\label{fig_pseudoscalar:mudis} }
\end{figure}

\noindent
The plot of Fig.~\ref{fig_pseudoscalar:mudis}, shows the constraints on $N_{eff}$, and on the generation of $\mu$-distortions. The dashed blue lines denote the factor by which the value of $N_{eff}$ (that is related to the fraction of energy carried by GW) exceeds the current $95\%$ CL bound set by Planck~\cite{Ade:2015xua}. The dark gray colored region shows the constraints set by FIRAS (COBE) and the light gray region shows the expected sensitivity of PIXIE for an excess in the amount of $\mu$-distortions above the vacuum contribution (based on the current $95\%$ CL region for $\Delta^2_s$ and $n_s$). In particular, the gray shaded regions indicates the region of the parameter space where the predictions for the $\mu$-distortions generated by the Starobinsky-like model, exceed the expected vacuum contribution for $f \leq 10^{-9}\,$Hz. Actually it is possible to show that among the models considered in this Chapter, only the Starobinsky-like model features an increase over the vacuum contribution in this frequency range. Finally, the red line is again corresponding to the upper bound on non-Gaussianities at CMB scales \textit{i.e.} $\xi_\text{CMB} \simeq 2.5$ and thus the region on the top left is excluded by the non-Gaussianity bound. \\

\noindent
As it is possible to see from the plot of Fig.~\ref{fig_pseudoscalar:mudis}, most of the models in the considered parameter space exceed the $95 \%$ CL region for $N_{eff}$ by an $\mathcal{O}(1)$ factor. As we discuss in the following this tension may be resolved by a better understanding of the theoretical uncertainties in the strong gauge field regime or by increasing the number $\mathcal{N}$ of $U(1)$ gauge groups in the theory (see Fig.~\ref{fig_pseudoscalar:several_gauge_plots}). It is also fair to point out that if the result of Riess et al.~\cite{Riess:2016jrr} ($\Delta N_\text{eff} \simeq 0.4 - 1$) Is confirmed, it could help to resolve the present tension.

\subsection{Several gauge fields.}
\label{sec_pseudoscalar:several_gauge_fields}
As stated in Sec.~\ref{sec_pseudoscalar:models}, all the numerical plots shown in this Chapter have been produced assuming $\mathcal{N} = 1$. However, at energy scales where inflation takes place, several $U(1)$ gauge fields (and also larger non-Abelian gauge groups) may be present. As we explain in the following, the case of non-Abelian gauge fields is more complicated and it requires an accurate analysis. On the other hand, the case of several $U(1)$ can be treated with the formulas derived in Sec~\ref{sec_pseudoscalar:mechanism}. As explained in Sec~\ref{sec_pseudoscalar:mechanism}, the predictions for the scalar and tensor power spectra for models with $\mathcal{N} \neq 1$ are given by Eq.~\eqref{eq_pseudoscalar:scalar} and Eq.~\eqref{eq_pseudoscalar:OmegaGW}. The result of a direct numerical evaluation for the cases with $\mathcal{N} = 3,5,10$ is shown in Fig.~\ref{fig_pseudoscalar:several_gauge_plots}.\\

\begin{figure}[ht]
\centering
\subfloat[][\emph{Scalar power spectra}.]
{\includegraphics[width=.44\columnwidth]{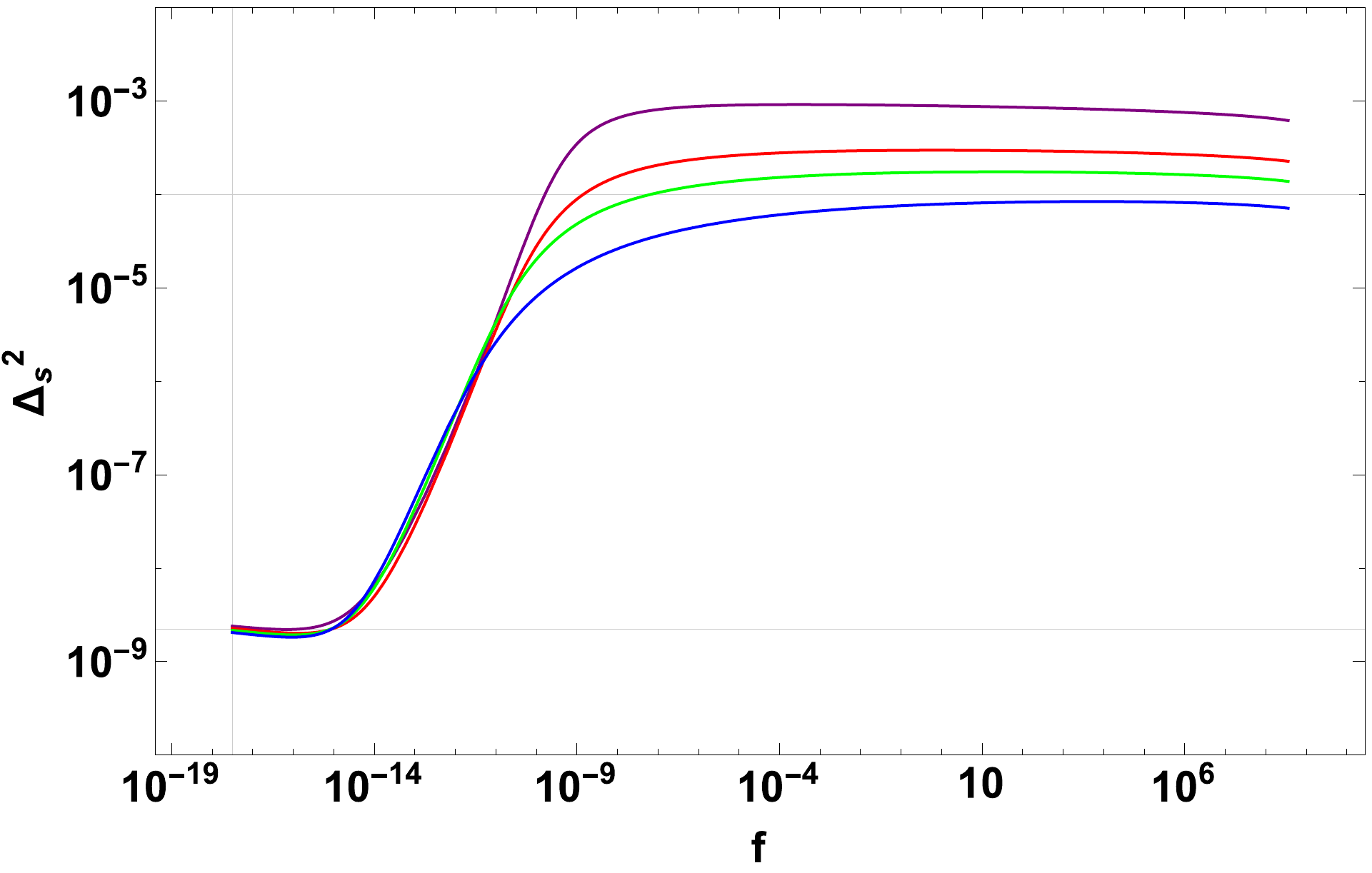}} \quad
\subfloat[][\emph{Tensor power spectra}.]
{\includegraphics[width=.5\columnwidth]{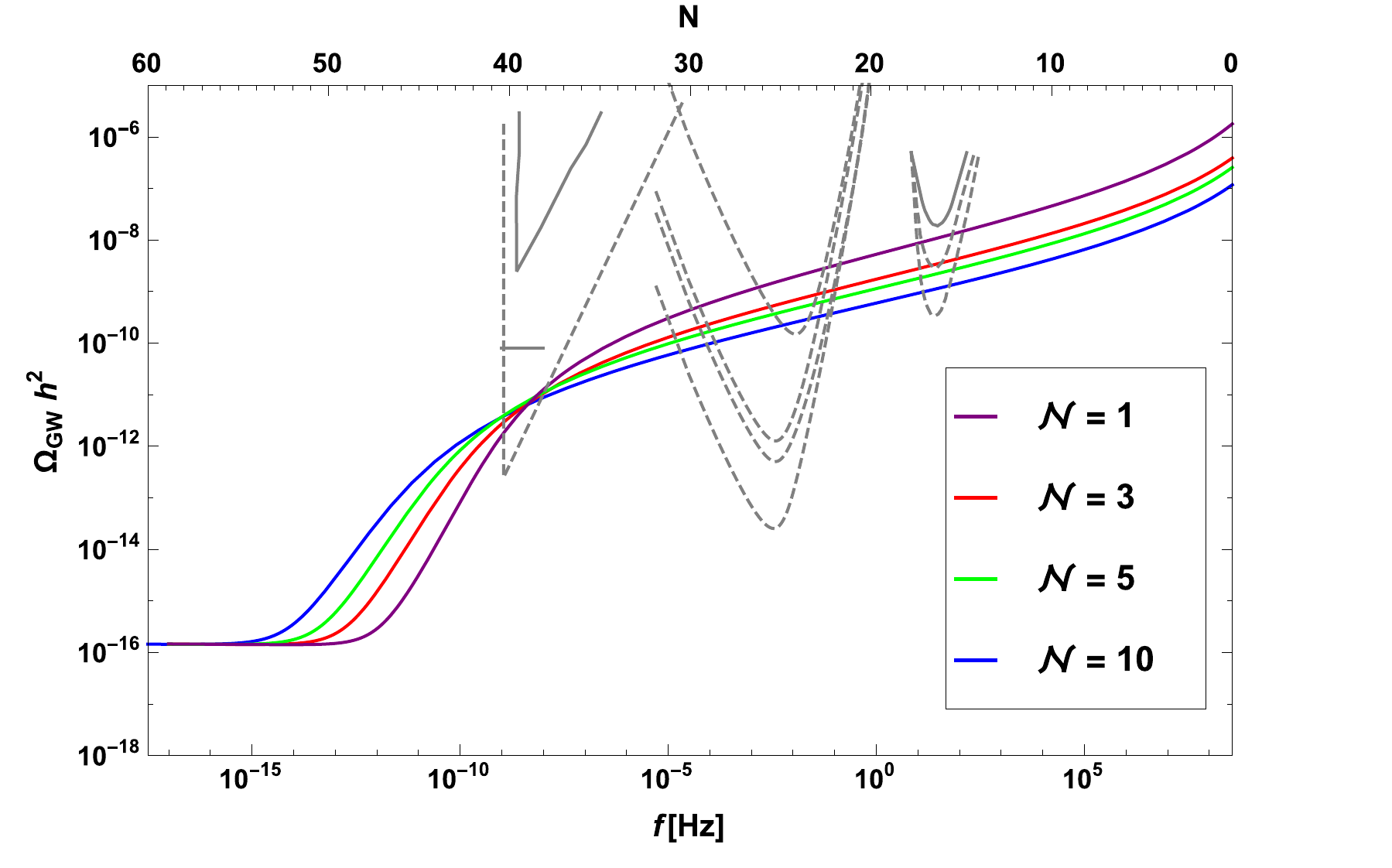}} 
\caption{Plot of scalar and tensor power spectra for Starobinsky-like models with parameters: $\alpha/\Lambda = 75, \  \gamma = 0.3, \ V_0 = 1.17 \cdot 10^{-9}$ for $\mathcal{N} = 1$ (purple), $\mathcal{N} = 3$ (purple), $\mathcal{N} = 5$ (red), $\mathcal{N} = 10$ (blue). The upper horizontal line in the plot on the left corresponds to the PBH bound and the lower one is the COBE normalization. In the plot for the tensor spectra we show the sensitivity curves for (from left to right): milli-second pulsar timing, eLISA, advanced LIGO. More details on these curves are given in Sec.~\ref{sec_pseudoscalar:models}. \label{fig_pseudoscalar:several_gauge_plots} }
\end{figure}

\noindent
As expected, by increasing the number of the gauge fields we are only affecting the last part of the evolution without spoiling the dynamics at early times. For all the models considered in the two plots, the predictions for scalar and tensor power spectra at CMB scales match with the results obtained in the case $\mathcal{N} = 1$. Moreover, the plot on the left shows that the estimate on the late time behavior for the scalar power spectra given in Eq.~\eqref{eq_pseudoscalar:scalar_strong} appears to be extremely accurate. As expected, the order one tension between the spectrum and the PBH bound is solved for the model with $\mathcal{N} = 10$. \\

\noindent
On the contrary, while looking at Eq.~\eqref{eq_pseudoscalar:OmegaGW} we would have naively expected an amplification of the GW spectrum as we increase the value of $\mathcal{N}$, Fig.~\ref{fig_pseudoscalar:several_gauge_plots} clearly shows a different behavior. While in a first phase, larger values of $\mathcal{N}$, induce a faster increase in the spectra, this growth actually lasts for a shorter period. In particular, the GW amplitude at later times is found to be suppressed. An explanation to this phenomenon may be be provided by reasoning on the way the gauge fields affect the dynamics. Several $U(1)$ will offer several channels for the decay of the inflaton, and thus in a first phase this will lead to a higher gauge field density and correspondingly to an enhancement of the GW spectra. However, this process will also accelerate the occurrence of the gauge field dominated regime, where the exponential growth of the GW spectrum is shut off. In fact this can be explained by considering Eq.~\eqref{eq_pseudoscalar:OmegaGW} and Eq.~\eqref{eq_pseudoscalar:eomC}. In the strong gauge field regime the dynamics is set by Eq.~\eqref{eq_pseudoscalar:eomC} and thus we have roughly $e^{2 \pi \xi}/\xi^4 \propto 1/\mathcal{N}$. As a consequence the parameter $\xi$ that enters exponentially in Eq.~\eqref{eq_pseudoscalar:OmegaGW} is suppressed for larger values of $\mathcal{N}$. It is worth mentioning that this peculiar feature, naturally provides a method to ease the tension with the $N_\text{eff}$ bound shown in the plot of Fig.~\ref{fig_pseudoscalar:mudis}. In particular for models with $\mathcal{N} \gtrsim 5$ the present tension is completely removed. \\

\noindent
 As already mentioned in this Section, an obvious extension of the framework discussed in this Chapter is to consider the coupling of the pseudo-scalar inflaton to non-Abelian gauge groups. A major difference between this case and the case discussed in this Chapter is that Eq.~\eqref{eq_pseudoscalar:eq_motionA} would not be valid anymore. In particular this equation should be modified in order to take into account the self-interactions between the gauge fields. As a consequence, as long as the amplitudes of the gauge fields are small, the system is expected to behave similarly to its Abelian analogous. However, as the exponential growth sets in, the non-Abelian nature of the gauge fields becomes important. In particular, the self-interactions may generate an effective mass term that can spoil the instability.\\

\noindent
 Similar situations have been studied in lattice simulations for explosive gauge field production through preheating, both for the case of a parametric resonance~\cite{Enqvist:2015sua} and of a tachyonic instability~\cite{Skullerud:2003ki}, finding that the non-Abelian interaction terms lead to a redistribution of the mode population towards higher values of $k$. In addition, effective mass terms may shut off the tachyonic instability prematurely. These arguments indicate that the GW production should be less efficient in the non-Abelian case. Similar questions have been addressed in the setup of so called chromo-natural inflation~\cite{Adshead:2012kp}. In this case, a coupling of a pseudo-scalar inflaton to non-Abelian gauge fields with a non-vanishing homogeneous vacuum expectation value can lead to a similar production of a chiral GW background, see e.g.~\cite{Dimastrogiovanni:2012ew,Obata:2016tmo}. However, since the simple estimates of Sec.~\ref{sec_pseudoscalar:mechanism} no longer apply, to draw conclusions on this case we should carry out a quantitative analysis that is beyond the scope of the current work.

\subsection{Inflationary model building and reheating.}
As discussed in this Chapter, the introduction of other particles and in particular of a non-minimal coupling between the inflaton and the gauge fields has several consequences on inflation. In particular, as the gauge fields slow down the last part of the evolution, inducing a shift in the point of the potential that is probed by CMB, a major effect is the reduction of the spectral index $n_s$ and the increase of the tensor-to-scalar ratio $r$. This is a universal feature of this setup and in particular the quantitative effect is shown in Eq.~\eqref{eq_pseudoscalar:ns_r_N}. This may move inflationary models with a too flat (red) spectrum, such as e.g.\ supersymmetric hybrid inflation~\cite{Linde:1993cn,Binetruy:1996xj,Halyo:1996pp} with $n_s \simeq 0.98$, right into the sweet spot of the Planck.\\

\noindent
 Notice that the observed reduction of $n_s$, parameterized by a reduction of the effective number of e-foldings $N_* = N_\text{CMB} - \Delta N_*$ for a given inflation model, is degenerate with the uncertainties in the reheating phase which determine $N_\text{CMB}$, number of e-foldings at which the scales that are presently observable in the CMB left the horizon. For a given inflation model, the parameter $\alpha$ hence allows to shift the predictions for a given model in the $n_s$ - $r$ plane along the lines with different values of $N$.\\

\noindent
The analysis proposed in this Chapter was carried out with a phenomenological approach. In particular, we have classified inflationary models using the parametrization of Eq.~\eqref{eq_pseudoscalar:Nparameterization}, and then we have studied in detail some well-known representative examples. A powerful tool on the way to embed these models in a top-down approach has been put forward in Ref.~\cite{Linde:2012bt}. A common strategy to define inflationary models in supergravity, is to invoke a shift-symmetry~\cite{Kawasaki:2000yn} to protect the flatness of the inflationary trajectory. By imposing this symmetry for the imaginary part (instead of the real part that is used to get scalar fields) of the scalar component of the inflaton superfield, we realize the desired pseudo-scalar inflaton. A further concrete realization in supergravity, based on a superconformal symmetry, can be obtained from Ref.~\cite{Buchmuller:2013zfa} (see also~\cite{Kallosh:2013xya} for related work).\\

\noindent
In this Chapter, and more in general in this work, we have mainly focused our interest on the study of Inflation, and on the consequences that it has on our Universe. However, as explained in Chapter~\ref{chapter:inflation}, another crucial phase takes place at the end of inflation \textit{i.e.} reheating, in which the inflaton decays filling the Universe with matter and radiation. In particular, the discussion of reheating is important in order to understand the fate of the produced gauge fields after the end of inflation. In the simplest case, the $U(1)$ gauge group is identified with SM hypercharge. In this case, the large abundance of gauge fields produced by the $\phi F \tilde{F}$ interaction during inflation as well as in the inflaton decay after the end of inflation, will quickly populate the thermal bath~\cite{Barnaby:2011qe}. This suggests a very efficient reheating mechanism with an equation of state of $\omega \simeq 1/3$. Further implications of such a coupling to SM gauge groups are the presence of primordial magnetic fields and even a possible contribution to baryogenesis. For more details on this topic see for example the recent works~\cite{Kusenko:2014uta, Anber:2015yca, Adshead:2015jza}. However, it is also possible to imagine more complicated scenarios. In particular, these gauge fields may be associated with additional $U(1)$ symmetries (actually these are expected from the point of view of string theory at the energy scales of inflation), that are spontaneously broken after the end of inflation\footnote{In this case cosmic strings will be produced. Their non-observation in the CMB constrains the symmetry breaking scale to be around or below the GUT scale.}. Depending on their couplings to the SM, the corresponding gauge bosons will decay into SM particles\footnote{An interesting example for such an additional $U(1)$ with couplings to the SM is the $U(1)_{B-L}$, with $B$-$L$ denoting the difference of baryon and lepton number, see~\cite{Buchmuller:2012wn,Buchmuller:2013dja} for possible further implications for early universe cosmology. } or into some hidden sector, contributing either to reheating or to dark matter.

\subsection{Theoretical uncertainties.}
As already anticipated in this Chapter, several theoretical uncertainties may affect the predictions in the high-frequency regime. As bounds on the experimental side are improving rapidly (in particular direct GW detection at interferometers and improved $N_\text{eff}$ measurements through the next generation of CMB experiments), quantifying and improving on the theory uncertainties becomes crucial. In this Section we give a review of some of the main sources of these uncertainties. In particular we focus on the uncertainties due to: (1) the quantum treatment of the perturbations, (2) the incorporation of the transfer function which modifies the spectra. Before discussing these topics in detail, we also mention another source of uncertainties: the possibility of having a decay of the gauge fields into particles (say $X$) that are charged under the corresponding gauge group. Until this process occurs (\textit{i.e.} until $\langle \vec{E}^2 + \vec{B}^2 \rangle > m_X^2$) we actually have a loss of the energy of the gauge sector that may affect the shape of the spectra. For details see also~\cite{Kobayashi:2014zza}. \\
 
\begin{enumerate}
	\item \textbf{Quantum corrections.}\\
	The possible breakdown of a perturbative analysis for large values of $\xi$ has recently been discussed in~\cite{Ferreira:2015omg,Peloso:2016gqs}. To clarify this point, we stress that while we assume perturbativity in the inflaton and on tensor fluctuations, the gauge field production is clearly a non-perturbative process. We do not attempt a perturbative analysis of the gauge field, but we actually work with the classical, non-perturbative background solution. The requirement of perturbativity for the inflaton fluctuations imposes: 
	\begin{equation}
	\delta \phi \lesssim \Lambda/\alpha \ .
	\end{equation} 
	During most of the evolution, this condition is fulfilled\footnote{Notice that to obtain Eq.~\eqref{eq_pseudoscalar:perturbativity_bound} we use~\cite{Linde:2012bt}:
	\begin{equation}
	\langle \zeta(x)^2 \rangle \simeq \mathcal{O}(1) \Delta^2_s(k) \ .	
	\end{equation} A similar analysis can be carried out for tensor fluctuations. With $\langle h (x)^2 \rangle  \simeq C \Delta^2_t(k)$, where $C$ is a constant factor, and using Eq.~\eqref{eq_pseudoscalar:OmegaMax}, we can show that perturbativity is ensured for $C \lesssim 10^{5}$ .}. However, we should also notice that towards the end of the evolution, we can use the asymptotic expressions for $\Delta^2_s$ and $\phi_{,N}$ to get:
	\begin{equation}
	\label{eq_pseudoscalar:perturbativity_bound}
	\langle \delta \phi^2 \rangle  \simeq \frac{\dot{\phi}^2}{H^2} \Delta^2_s \simeq \phi_{,N}^2 \frac{1}{(2 \pi \xi)^2} =  \left(\frac{\Lambda}{\alpha \pi}\right)^2 \ . 
	\end{equation}
	Hence perturbativity is merely ensured by a factor of $1/\pi$, implying a potentially significant theoretical uncertainty in the asymptotic value of the scalar power spectrum~\cite{Ferreira:2015omg}. The analysis of~\cite{Peloso:2016gqs} finds that perturbativity is ensured as long as $\xi \lesssim 4.7$. Notice that for this value of $\xi$, we are well inside the regime where the tensor spectrum is dominated by GWs sourced by the gauge fields. While this effect should not affect dramatically the conclusions on the generation of observable GW, it can be relevant in the discussion of the mechanism that leads to the generation of a distribution of PBHs. In particular, a detailed analysis in the strong back-reaction regime can be important in order to ease the tension between the predicted scalar spectra and the PBH bound. It is fair to point out that a full analysis in the strong back-reaction regime probably requires a lattice study of the non-perturbative system. \\

	\item \textbf{Transfer function.}\\
	\noindent
	To calculate the scalar and tensor power spectra that are observable at present time, we actually need two ingredients: the scalar and tensor power spectra at the time of creation, \textit{i.e.} when they exited the horizon during inflation, and the transfer function, which encodes the red-shift of the spectra from the moment they re-enter horizon until today. The transfer function for modes $k$ that re-enter the horizon during the radiation dominated regime is roughly given by~\cite{Buchmuller:2013lra,Domcke:2016bkh}:
	\begin{equation}
	T_k^2 \simeq \Omega_{R,0} \frac{g_*^k}{g_*^0} \left(\frac{g_{*,s}^0}{g_{*,s} ^k} \right)^{4/3} \ ,
	\label{eq_pseudoscalar:transferfunc}
	\end{equation}
	were $g_*$ ($g_{*,s}$) counts the effective degrees of freedom entering the energy density (entropy) of the thermal bath and again $\Omega_{R,0} = 8.6 \cdot 10^{-5}$ denotes the radiation energy density today. The superscript indicates evaluation at $t_k$ when the mode $k$ re-enters the horizon or today $(t_0)$, respectively. For modes entering earlier, during the reheating phase, the transfer function depends on the respective equation of state. In particular, for a matter dominated reheating phase, there is a suppression factor of $(k_{RH}/k)^2$. Eq.~\eqref{eq_pseudoscalar:scalar} and Eq.~\eqref{eq_pseudoscalar:OmegaGW} actually assume instantaneous reheating, or a reheating phase with $\omega = 1/3$\footnote{Possible changes in the degrees of freedom of the thermal bath, due to e.g.\ supersymmetry breaking, are also omitted in Eq.~\eqref{eq_pseudoscalar:OmegaGW}.}. If this condition is violated we have a suppression of the spectrum for frequencies larger than $f_\text{rh} \simeq 0.4~\text{Hz } (T_\text{RH}/10^7~\text{GeV})$, with $T_\text{RH}$ denoting the reheating temperature~\cite{Turner:1993vb,Seto:2003kc,Nakayama:2008ip,Buchmuller:2013lra}. This effect can both be relevant for the discussion of the observable scalar and tensor spectra. In particular, for GUT-scale models of inflation (such as Starobinsky-like models) this may hide a potential signal from the LIGO band, but typically not from the eLISA band which is located at lower frequencies.\\
\end{enumerate}

 {\large \par}}
{\large 
\chapter{Conclusions and future perspectives.}
After a general introduction on standard cosmology (in Chapter~\ref{chapter:introduction}), we have focused on the study of inflation(in Chapter~\ref{chapter:inflation}). In particular we have produced a comprehensive discussion of the topic that has covered both the simplest realization of inflation (in terms of a single slow-rolling field) and several generalizations of this minimal picture. \\

\noindent
A main part of this work is the definition (in Chapter~\ref{chapter:beta}) of a new formalism to describe inflation. This approach is based on the application of the Hamilton-Jacobi formalism to inflation. Under the reasonable assumption of a piecewise monotonic field, the field is used as the clock to describe the evolution of the system. In this context, inflation is described by a system of first order differential equations which have a formal resemblance with RG equations in quantum field theory (QFT). In the case of inflation, the role of the renormalized coupling is played by the inflaton field, and the role of renormalization scale is played by the scale factor. Exact scale invariance is realized in correspondence with a zero of the cosmological $\beta$-function. In this point, the geometry of the Universe approaches a de Sitter (dS) spacetime.  \\

\noindent
In this framework inflation is associated with the slow motion of the field away from a repulsive fixed point. In analogy with QFT, the RG flow is parametrized in terms of the newly defined $\beta$-function. With this method, it is quite natural to define a set of universality classes for inflationary models. In analogy with statistical mechanics, these universality classes must be considered as sets of theories that share a common scale invariant limit under the process of RG flow. In this sense the question of degeneracy between inflationary models has partially been explained. \\

\noindent
As explained in Chapter~\ref{chapter:holographic_universe}, the possibility of describing inflation as an RG flow is not fortuitous. By performing an analytical continuation, cosmological solutions that asymptote to dS can be mapped into domain-wall solutions that asymptote to Anti de Sitter (AdS). Because of this correspondence, it is possible to carry out an holographic description (in particular using AdS/CFT) of inflation. In this context, inflation corresponds to an RG flow for the dual (3-dimensional) QFT. The definition of a $\beta$-function to describe this RG flow thus naturally emerges. \\

\noindent
While in Chapter~\ref{chapter:beta} we have defined the $\beta$-function formalism for the simplest realization of inflation, in Chapter~\ref{chapter:generalized_models} we have discussed the application of the formalism to broader classes of models. In particular, we have discussed two generalizations of the simplest realization of inflation: models with non-standard kinetic terms and models where the inflaton is non-minimally coupled with gravity. We have shown that it is possible to generalize the formalism in order to describe these generalized models. Remarkably, the formalism proves to be particularly useful in these cases. While the standard description (in terms of the potential) may be misleading when applied to generalized models, the $\beta$-function formalism perfectly fits and provides a powerful tool to identify the leading contributions that affect the dynamics during inflation. \\

\noindent
As the $\beta$-function formalism presents several advantages with respect to the standard description, it is worth considering the possibility of applying this methods to discuss other cases that have not been treated in this work. For example, it would be interesting to discuss its application to multi-field models. As the formalism provides a method to identify the leading contribution that drives inflation, it would provide a tool to discuss the couplings between the different fields that may be present during inflation. While in this work we have mainly discussed early time cosmology, it would also be interesting to explore the case of late time cosmology. By analogy with the case of inflation, the accelerated expansion due to Dark Energy (DE) can be interpreted as an RG flow towards an attractive fixed point in the future. As a consequence, it would be interesting to discuss the application of the $\beta$-function formalism to study models for quintessence. In particular, after the definition of a set of universality classes for models of quintessence, it would be interesting to study the take over of quintessence over the ordinary matter in terms of the $\beta$-function formalism. \\

\noindent
In Chapter~\ref{chapter:pseudoscalar} we have discussed a realistic landscape for early time cosmology. As other particles are expected to be present during inflation, we have discussed the consequences of allowing for particle production in the last part of the inflation. In particular we have considered the case of a pseudo-scalar inflaton with a non-minimal coupling with some gauge fields. We have shown that the strong production of particles back-reacts of the system resulting in an additional friction term that affects the last part of inflation. Moreover, we have shown that particle production is not only affecting the background dynamics but it also act as a source term for perturbations leading to an exponential enhancement of the scalar and tensor spectra. \\

\noindent
In this context we have discussed a wide set of possible observational consequences such as: the possibility of generating a (chiral) gravitational wave (GW) background in the observable range for Advanced LIGO/VIRGO and eLISA, the possibility of generating a distribution of primordial black holes (PBH), the reduction of the spectral index $n_s$, the enhancement of the tensor-to-scalar ratio $r$, the generation of non-Gaussianities and the generation of $\mu$-distortions. To carry out this analysis we have used a classification of inflationary models in terms of universality classes.\\

\noindent 
While the analysis proposed in Chapter~\ref{chapter:pseudoscalar} has clarified the main features that affect the dynamics of a  pseudo-scalar inflaton with a non-minimal coupling with some gauge Abelian fields, several intriguing topics are still open. For example, an accurate discussion of the embedding of these models in some high energy theory would be an extremely interesting topic for future work. Another interesting question concerns the extension of this mechanism to models where the inflaton is coupled to non-Abelian gauge fields (such as the standard model $SU(2)_L$ and $SU(3)_C$). In particular, it would be interesting to understand whether the interactions between the gauge fields may generate an effective mass term capable of spoiling the tachyonic instability. As the $\phi F \tilde{F}$ term introduces a decay mechanism for the inflaton, an accurate study of reheating (where other spectator fields may be present) would be also an interesting topic for future works. Finally, as binary systems of BH are important GW sources, a study of the generated distribution of PBH may be extremely interesting for the study of sources for GW astronomy.\\

 {\large \par}}

\appendix
\setcounter{secnumdepth}{3}

{\large\chapter{A compendium on general relativity.}
 \label{appendix_GR:general}
General relativity (GR) is the theory of gravitational interactions introduced by Einstein in 1915. In this theory, the presence of matter and energy induces a modification in the structure of the spacetime. In this Appendix we introduce the typical quantities required to perform computations in GR, we express Einstein Equations in terms of these quantities and we derive a solution of these equations in the weak gravitational field approximation.

\section{The basic equations of GR.}
 \label{appendix_GR:General_equations}
As already stated in the above paragraph, in GR matter and energy affect the structure of the spacetime inducing curvature. As a consequence, the natural framework to describe GR is given by differential geometry. A formal introduction of the differential geometry formalism is beyond the scope of this work and it can actually be found in all standard GR textbooks, see for example~\cite{wald:1984}.\\

\noindent A first problem to face in curved spacetime concerns the definition of derivation. To solve this problem we introduce the Christoffel symbols\footnote{Because of the similarities with the case of gauge theories, the Christoffel symbols are also known as affine connections.} that $\Gamma^\mu_{ \ \nu \rho}$ can be expressed in terms of the spacetime metric $g_{\mu\nu}$ as:
\begin{equation}
\Gamma_{\mu\nu}^{\rho}\equiv\frac{1}{2}g^{\rho\eta}(\partial_{\mu}g_{\nu\eta}+\partial_{\nu}g_{\mu\eta}-\partial_{\eta}g_{\mu\nu}) \ .\label{appendix_GR:Christoffels}
\end{equation}
In terms of these quantities we define the covariant derivative operator $\nabla_{\mu}$, whose action over vector fields $v^{\nu}$ and dual vector fields $w_\nu$ is respectively given by:
\begin{align}
\label{appendix_GR:Covariant_vector}
\nabla_{\mu}v^{\nu} & = \partial_{\mu}v^{\nu}+\Gamma_{\mu\rho}^{\nu}v^{\rho} \ , \\
\label{appendix_GR:Covariant_dual}
\nabla_{\mu}w_{\nu} & = \partial_{\mu}w_{\nu} - \Gamma_{\mu\nu}^{\rho}v^{\rho} \ .
\end{align}
It is also useful to give an explicit expression of the action of the covariant derivative $\nabla_\sigma $ over a tensor field $T^{  \mu_1, \dots, \mu_n }_{ \nu_1, \dots, \nu_n }$ :
\begin{equation}
\begin{aligned}
\label{appendix_GR:Covariant_tensor}
\nabla_\sigma  T^{\mu_1, \dots, \mu_m }_{\nu_1, \dots, \nu_n }  = \frac{\partial T^{\mu_1, \dots, \mu_m }_{\nu_1, \dots, \nu_n } }{\partial x^\sigma} &+ \Gamma^{\mu_1 }_{\rho \sigma} \  T^{\sigma, \dots, \mu_m }_{\nu_1, \dots, \nu_n } + \dots + \Gamma^{\mu_m }_{\rho \sigma} \ T^{\mu_1, \dots, \sigma }_{\nu_1, \dots, \nu_n } + \\
&  - \Gamma^{\rho }_{\nu_1 \sigma} \  T^{\mu_1, \dots, \mu_m }_{\rho, \dots, \nu_n } - \dots - \Gamma^{\rho}_{\nu_n \sigma} \ T^{\mu_1, \dots, \mu_m }_{\nu_1, \dots, \rho }  \ .
\end{aligned}
\end{equation}
The Riemann Tensor is then defined by the action as the commutator of two covariant derivatives over a dual vector as:
\begin{equation}
(\nabla_{\mu}\nabla_{\nu}-\nabla_{\nu}\nabla_{\mu})\ w_{\eta}=R_{\ \mu\nu\eta}^{\sigma}w_{\sigma}\label{appendix_GR:Riemann_tensor_def}
\end{equation}
The Riemann tensor can be expressed in terms of Christoffel symbols as: 
\begin{equation}
 \label{appendix_GR:Riemann_tensor}
   R^{\sigma}_{\ \mu \rho \nu} = \partial_\rho \Gamma^\sigma_{\ \mu  \nu} -  \partial_\mu  \Gamma^\sigma_{\ \rho \nu} +   \Gamma^\sigma_{\ \alpha \rho} \Gamma^\alpha _{\ \mu  \nu} -  \Gamma^\sigma_{\ \alpha  \mu} \Gamma^\alpha _{\ c \nu} \ .
\end{equation}
By contracting indexes we define the Ricci Tensor:
\begin{equation}
\label{appendix_GR:Ricci_tensor}
  R_{\mu\nu}\equiv R_{\ \mu \sigma\nu}^{\sigma} = \partial_\sigma \Gamma^\sigma_{\ \mu  \nu} -  \partial_\mu  \Gamma^\sigma_{\ \sigma \nu} +   \Gamma^\sigma_{\ \sigma \alpha} \Gamma^\alpha_{\ \mu  \nu} -  \Gamma^\sigma_{\ \alpha \mu} \Gamma^\alpha_{\ \sigma \nu} \ , 
\end{equation}
and the Ricci Scalar:
\label{appendix_GR:Ricci_scalar}
\begin{equation}
  R \equiv g^{\mu\nu} R_{\mu\nu} \ .
\end{equation}
The Einstein tensor $G_{\mu\nu}$ is then defined as 
\begin{equation}
\label{appendix_GR:Einstein_tensor}
G_{\mu\nu}=R_{\mu\nu}-\frac{1}{2} R \ g_{\mu\nu}
\end{equation}
With an explicit computation it is possible to show that:
\begin{equation}
\nabla^{\mu}G_{\mu\nu}=0\label{appendix_GR:derivative_eintstein}
\end{equation}
The matter stress-energy tensor $T_{\mu \nu}$ can be defined as:
\begin{equation}
\label{appendix_GR:Stress_energy_tensor}
T_{\mu \nu } \equiv -\frac{2}{\sqrt{|g|}} \frac{\delta \mathcal{S}_m}{\delta g^{\mu \nu}} \ ,  
\end{equation}
where $\mathcal{S}_m$ is the part of the action that describes matter. Finally we can write Einstein Equations, that is a system of ten non-linear partial differential equations: 
\begin{equation}
G_{\mu\nu}+\Lambda g_{\mu\nu}=8\pi G_N T_{\mu\nu}\label{appendix_GR:eintstein_equations} \ ,
\end{equation}
where we have also introduced  a cosmological constant term $\Lambda$.

\section{Weak gravitational field equations.}
\label{appendix_GR:GW}
It is interesting to derive the lowest order approximation of Einstein Equations~\ref{appendix_GR:eintstein_equations} in the limit of weak gravitational field. In this limit we approximate the metric as:
\begin{equation}
g_{\mu\nu}=\eta_{\mu\nu}+h_{\mu\nu} \qquad \qquad |h_{\mu\nu}|\ll1\label{appendix_GR:linear_approx} \ ,
\end{equation}
where $\eta_{\mu\nu}$ is the usual Minkowski metric. Notice that  at the linear order in $h_{\mu\nu}$, indexes are raised and lowered using $\eta^{\mu\nu}$ and $\eta_{\mu\nu}$. At this order the inverse metric $g^{\mu\nu}$ is simply given by 
\begin{equation}
g^{\mu\nu}=\eta^{\mu\nu}-h^{\mu\nu}+O\left(h^{2}\right) \ ,
\end{equation}
and Christoffel symbols read:
\begin{equation}
\Gamma_{\ \mu\nu}^{\sigma}=\frac{1}{2}\eta^{\sigma\phi}\{\partial_{\mu}h_{\nu\phi}+\partial_{\nu}h_{\mu\phi}-\partial_{\phi}h_{\mu\nu}\}\label{appendix_GR:Christoffel_linear} \ .
\end{equation}
Substituting into Eq.~\eqref{appendix_GR:Riemann_tensor} we get the linearized Riemann tensor:
\begin{equation}
R_{\ \mu\nu\theta}^{\sigma}  = \frac{1}{2}\left[ \partial_{\nu}\eta^{\sigma\phi} \left(\partial_{\mu}h_{\theta\phi}+\partial_{\theta}h_{\mu\phi}-\partial_{\phi}h_{\mu\theta}\right)-\partial_{\mu}\eta^{\sigma\phi}\left(\partial_{\theta}h_{\nu\phi}+\partial_{\nu}h_{\theta\phi}-\partial_{\phi}h_{\theta\nu}\right) \right] \ ,
\end{equation}
and the linearized Ricci tensor $R_{\mu\theta}$:
\begin{equation}
R_{\mu\theta}  =  \frac{1}{2}\left( \partial^{\phi}\partial_{\theta}h_{\mu\phi}-\Box h_{\mu\theta}-\partial_{\mu}\partial_{\theta}h+\partial^{\nu}\partial_{\mu}h_{\theta\nu} \right) \ ,
\end{equation}
where we have defined $h\equiv\eta^{\nu\phi}h_{\nu\phi}$ and $\Box \equiv \eta^{\mu \nu} \partial_\mu \partial_\nu $ is the usual flat Minkowski spacetime d'Alembertian operator. Contracting with the inverse metric $\eta^{\mu \nu}$ we get the Ricci scalar:
\begin{equation}
R=\partial^{\phi}\partial^{\mu}h_{\mu\phi}-\Box h , 
\end{equation}
and substituting into Eq.~\eqref{appendix_GR:Einstein_tensor} we can finally get the linearized Einstein tensor: 
\begin{eqnarray}
G_{\mu\nu} & = & \frac{1}{2}\left[\partial^{\phi}(\partial_{\nu}h_{\mu\phi}+\partial_{\mu}h_{\nu\phi})-\Box h_{\mu\nu}-\partial_{\mu}\partial_{\nu}h-\eta_{\mu\nu}(\partial^{\phi}\partial^{\sigma}h_{\sigma\phi}-\Box h)\right] \ .
\end{eqnarray}
Setting the cosmological constant to zero the linearized Einstein equations read:
\begin{equation}
16\pi T_{\mu\nu}= \partial^{\phi}(\partial_{\nu}h_{\mu\phi}+\partial_{\mu}h_{\nu\phi})-\Box h_{\mu\nu}-\partial_{\mu}\partial_{\nu}h-\eta_{\mu\nu}(\partial^{\phi}\partial^{\sigma}h_{\sigma\phi}-\Box h)\label{appendix_GR:Linearized_EE}
\end{equation}
In order to solve Eq.~\eqref{appendix_GR:Linearized_EE}, we should remove the degeneracy due to the gauge freedom. For this purpose we properly choose a gauge fixing procedure. For the scope of our inquiry it is useful to describe the problem in the harmonic gauge:
\begin{equation}
\partial^{\nu}(h_{\mu \nu }-\frac{1}{2}\delta_{\mu \nu } h)=0\label{appendix_GR:Harmonic_gauge} \ .
\end{equation}
It is also useful to express Eq.~\eqref{appendix_GR:Linearized_EE} in terms of the traceless field:
\begin{equation}
\bar{h}_{\mu\nu}=h_{\mu\nu}-\frac{1}{2}\eta_{\mu\nu}h \ .
\end{equation}
In terms of this quantity the linearized Einstein equations and the harmonic gauge condition respectively read:
\begin{eqnarray}
\Box  \bar{h}^{\mu\nu} & = & -16\pi T_{\mu\nu}\label{appendix_GR:Linearized_EE_Harmonic_T} \ , \\
\partial^{\mu} \bar{h}_{\mu\nu} & = & 0\label{appendix_GR:Linearized_Harmonic_T} \ .
\end{eqnarray}
Notice that this is standard Minkowski space wave equation, describing waves that are propagating at the speed of light, in presence of a source. These waves are called \emph{gravitational waves} (GW).\\

\noindent
It is crucial to notice that Eq.~\eqref{appendix_GR:Linearized_Harmonic_T} is not uniquely fixing the gauge. Indeed Eq.~\eqref{appendix_GR:Linearized_Harmonic_T} is invariant under a gauge transformation:
\begin{equation}
\label{appendix_GR:Harmonic_gauge_transf}
\bar{h}_{\mu\nu} \rightarrow \bar{h}^\prime_{\mu\nu} =  \bar{h}_{\mu\nu}  + \partial_\mu \xi_\nu
\end{equation}
where $\xi_{\mu}(x)$ are harmonic functions \textit{i.e.} they satisfies $\Box \xi_{\nu}(x) = 0 $. By choosing these four functions the gauge is uniquely fixed. We can thus proceed with a counting of the total number of physical degrees of freedom (d.o.f.) associated with the system. As we started with a $4 \times 4$ symmetric matrix the initial number of d.o.f. was equal to $10$. Choosing the coordinates and we can remove $4$ of these d.o.f. and by fixing the $\xi_{\mu}(x)$ functions we can remove $4$ more. The system can thus be described in terms of independent $2$ d.o.f. that are identified with the two polarizations of the GW. \\

\noindent We conclude this Appendix by showing the explicit expression for a GW propagating in a given direction. Let us start by considering the ansatz:
\begin{equation}
\label{appendix_GR:GW_anstaz}
\bar{h}_{\mu\nu} (x)= A_{\mu\nu} \exp \left( - i \eta_{\rho \sigma } k^{\rho} x^{\sigma} \right) .
\end{equation}
We start by using the harmonic gauge condition:
\begin{equation}
\label{ }
 A_{\mu\nu} k^{\mu} = 0 \ ,
\end{equation}
and choosing $\vec{k}$ to be along the $x$ axis \textit{i.e.} $k^{\mu} = (k,k,0,0)$ that gives:
\begin{equation}
A_{0 \nu} + A_{1 \nu}  = 0 \ .
\end{equation}
We proceed by fixing the four functions $\xi_{\mu}(x)$ in order to set
\begin{equation}
A_{0 \nu} = A_{1 \nu}  = 0 \ .
\end{equation}
By construction $A_{\mu\nu}$ is traceless \textit{i.e.} $A_{yy} + A_{zz} = 0$ and symmetric \textit{i.e.} $A_{yz} =  A_{zy}$. Defining $h_{+} \equiv A_{yy}  \exp \left( - i \eta_{\rho \sigma } k^{\rho} x^{\sigma} \right) $ and $h_{\times} \equiv A_{yz}  \exp \left( - i \eta_{\rho \sigma } k^{\rho} x^{\sigma} \right) $, the ansatz of Eq.~\eqref{appendix_GR:GW_anstaz} can be expressed as:
\begin{equation}
\label{appendix_GR:GW_anstaz_x}
\bar{h}_{\mu\nu} (x) = 
\begin{pmatrix}
   0   &  0 & 0 & 0  \\
    0   &  0 & 0 & 0  \\
     0   &  0 & h_{+} & h_{\times}  \\
      0   &  0 & h_{\times} & -h_{+}  \\
\end{pmatrix}
\end{equation}
The two physical quantities $h_{+}$ and $h_{\times}$ are usually referred as the plus and cross polarizations of the GW.

\section{Spaces of constant curvature:dS and AdS spacetimes.}
 \label{appendix_GR:dS_AdS}
In this section we introduce and describe some general features of de Sitter (dS) and Anti de Sitter (AdS) spacetimes. These are spaces of constant scalar curvature whose definition is extremely relevant for cosmology and more generally for modern theoretical physics. The definition of these two spaces naturally arises in general relativity as they correspond to the solutions of a maximally symmetric empty space in presence of positive and negative cosmological constants respectively. Let us start by considering the generic action for a system containing some matter described by an action $\mathcal{S}_{m}$, gravity (described by a standard Einstein-Hilbert term) and a cosmological constant $\Lambda$ in $(d+1)$-dimensions:
  \begin{equation}
  \label{appendix_GR:defining_action}
    \mathcal{S}_{TOT} = \mathcal{S}_{m} +\frac{1}{2 \kappa^{d-1}} \int\mathrm{d}^{d+1}x \sqrt{|g|}\left(R - 2 \Lambda   \right),
  \end{equation}
where $\kappa^2 \equiv 8 \pi G_N$. As discussed in Appendix~\ref{appendix_GR:General_equations}, this system can be described in terms of the Einstein Equations~\ref{appendix_GR:eintstein_equations} :
\begin{equation}
  \label{appendix_GR:general_EE}
  R_{\mu \nu} - \frac{R}{2} g_{\mu \nu} + \Lambda g_{\mu \nu} = \kappa^{d-1} T_{\mu \nu} \ ,
\end{equation}
where $  R_{\mu \nu} $ is the Ricci tensor, $R$ is the Ricci scalar and $T_{\mu \nu} $ is the stress-energy tensor. Let us contract this equation with the metric $g_{ab}$ to get: 
\begin{equation}
  \label{appendix_GR:general_EE_trace}
\frac{1 - d}{2} R  + \Lambda (d+1) = \kappa^{d-1} T \ ,
\end{equation}
where we have defined $T \equiv g^{\mu \nu } T_{\mu \nu }$. Let us restrict to the case of an empty universe \textit{i.e.} $T_{\mu \nu} = 0$. In this case Eq.~\eqref{appendix_GR:general_EE} and Eq.~\eqref{appendix_GR:general_EE_trace} simply read:
\begin{equation}
\label{appendix_GR:dS_AdS_EE}
  R = 2 \frac{d+1}{d-1} \Lambda \ , \qquad  R_{\mu \nu } =  \frac{2}{d-1} \Lambda  g_{\mu \nu}  \ .
\end{equation}
Let us assume that the metric is mostly positive\footnote{In this case $\eta_{ab}$, metric of $(d+1)$-dimensional Minkowski spacetime is:
\begin{equation}
   \eta_{\mu \nu} = \text{diag} (-1 , \underbrace{+1 , \dots , +1}_{d } ) \ .
 \end{equation} } with signature $(1,d)$, and let us also assume that the space is maximally symmetric \textit{i.e.} that the metric is diagonal. In this case we can use Eq.~\eqref{appendix_GR:dS_AdS_EE} to restrict to three cases:
 \begin{itemize}
   \item $\Lambda = 0 \ \iff \ R = 0$, in this case we also have $R_{\mu \nu} = 0 $ that implies a constant metric. This corresponds to \textbf{Minkowski} spacetime.  
   \item $0 < \Lambda \ \iff \ 0 < R$, this space is usually called \textbf{de Sitter} (dS) spacetime. 
   \item $\Lambda < 0 \ \iff \ R < 0$, that is \textbf{Anti de Sitter} (AdS) spacetime. 
 \end{itemize}
Let us introduce a length scale $l>0$, that actually corresponds to the curvature radius, and a parameter $\eta$ such that $\eta = -1$ if $0 < \Lambda$ and $\eta = 1$ if $\Lambda > 0$. It is convenient to parametrize $\Lambda $ as:
\begin{equation}
\label{appendix_GR:cosmological_constant_sign}
\Lambda = - \eta \frac{d(d-1)}{2 l^2} \ .
\end{equation} 
It is possible to show that for spaces of constant curvature, the Riemann tensor~\ref{appendix_GR:Riemann_tensor}, can be expressed as:
\begin{equation}
  \label{appendix_GR:Riemann_tensor_cc}
  g_{\mu \sigma } R^{\sigma}_{ \ \alpha \nu \beta} = R_{\mu \alpha \nu \beta} = \frac{1}{l^2} \left( g_{\mu \nu } g_{\alpha \beta } - g_{\alpha \nu}g_{\beta \mu}  \right) .
\end{equation}
In the rest of this Appendix we will present some details on dS and $AdS$ spacetime that we will use throughout this work.

\subsection{De Sitter spacetime.} 
\label{appendix_GR:dS_spacetime}
As discussed in the introduction of this Appendix, dS spacetime is the maximally symmetric space with constant positive curvature $0 < R$. Let us consider a $(d+2)$-dimensional Minkowski spacetime with interval:
\begin{equation}
  \textrm{d} s^2 = - \textrm{d} x_0^2 + \sum_{i = 1}^{d + 1} \textrm{d} x_i^2 \ , 
\end{equation}
the dS${}_{d+1}$ spacetime can be thought as a $(d+1)$-dimensional embedded hyperboloid:
\begin{equation}
\label{appendix_GR:dS_hyperboloid}
  l^2 = -  x_0^2 + \sum_{i = 1}^{d + 1}  x_i^2 \ .
\end{equation}
To parametrize this surface it is possible to introduce a set of coordinates $(t,\theta_1,\dots,\theta_{d-1},\phi)$ with $t \in (-\infty,\infty)$, $\theta_i \in [0,\pi]$, $\phi \in [0,2\pi] $ such that:
\begin{equation}
  x_0 = l \sinh(t/l), \qquad \qquad x_i =  l \cosh(t/l) \chi_i \ ,
\end{equation}
where $\chi_i$ are the normal vectors $\hat{x}_i$ defined in terms of the angles $\theta_i$ and $\phi$ as:
\begin{equation}
\label{appendix_GR:dS_solid_angle}
  \chi_1 = \cos \theta_1 , \qquad  \chi_2 = \sin \theta_1 \cos  \theta_2 , \qquad \dots \qquad \chi_{d+1} = \sin \theta_1 \dots \sin \theta_{d-1} \sin \phi \ .
\end{equation}
In terms of these quantities the induced spacetime interval on dS${}_{d+1}$ reads:
\begin{equation}
\textrm{d} s^2 = - \textrm{d} t^2 + l^2 \cosh^2 (t/l) \textrm{d} \Omega_{d}^{\ 2} \ , 
\end{equation}
where $\Omega_{d}^{\ 2}$ is used to denote the $d$-dimensional solid angle. From this expression and from the definition of $(t,\theta_1,\dots,\theta_{d-1},\phi)$, it should be clear that dS${}_{d+1}$ has the same topology as $\mathbb{R} \times S^{d}$. It is also interesting to stress that defining a new time coordinate $\eta$ as:
\begin{equation}
  \tan \left(\frac{\eta}{2 l} \right) = \tanh \left(\frac{t}{2 l} \right) \ ,
\end{equation}
and using the identities:
\begin{equation}
  \tan(x/2) = \frac{\sin(x)}{1 + \cos(x)} \ , \qquad \qquad  \cosh(x)=\cosh^2(x/2) + \sinh^2(x/2) , 
\end{equation}
it is possible to show that the spacetime interval reads:
 \begin{equation}
   \textrm{d} s^2 = \frac{1}{\cos^2 (\eta/l) } \left( - \textrm{d} \eta^2 + l^2 \textrm{d} \Omega_{d}^{\ 2} \right) \ . 
 \end{equation}
Notice that $\eta \in [-\frac{\pi}{2},\frac{\pi}{2}]$ and thus this new time coordinate is compact. As the angles $\theta_i$ and $\phi$ are compact too, we have defined a compact set of coordinates to describe dS${}_{d+1}$.These are the coordinates used to define the Penrose diagram\footnote{A Penrose diagrams is a finite size two-dimensional diagram that is used to represent higher dimensional manifolds. In particular Penrose diagrams are used to study the causal structure and the infinities of a spacetime. The introduction of Penrose diagrams is not required for the studies presented in this thesis and thus lies beyond the scope of this work. For some references see for example~\cite{book:15251,Mukhanov:991646}.} of dS${}_{d+1}$ and to study its structure at the infinity. \\

\noindent For the purpose of this work it is also useful to consider one more set of coordinates to parametrize dS${}_{d+1}$. Let us define the coordinates $(t,\tilde{x}_1, \dots \tilde{x}_d)$ with $t \in (-\infty,\infty)$, $\tilde{x}_i \in (-\infty,\infty)$ such that:
\begin{equation}
  x_0 = l \sinh(t/l) + r^2 \frac{e^{t/l}}{2 l}, \qquad  x_1 =  l \cosh(t/l) - r^2 \frac{e^{t/l}}{2 l} , \qquad  x_i = \tilde{x}_i e^{t/l} \ ,
\end{equation}
where we have defined $r^2 \equiv \sum_{i = 1}^{i = d} \tilde{x}_i^{ \ 2}$ . It is possible to show that in terms of these coordinates the spacetime interval reads:
 \begin{equation}
 \label{appendix_GR:dS_metric}
   \textrm{d} s^2 =  - \textrm{d} t^2 + e^{2 t/l} \left( \textrm{d} \tilde{x}_1^{ \ 2} + \dots + \textrm{d} \tilde{x}_d^{ \ 2} \right) \ . 
 \end{equation}
Notice that in the limit of $\dot{H}/H^2 \ll 1$ \textit{i.e.} for nearly constant $H$, we can identify the scale $l$ with the Hubble radius $R_H = H^{-1}$, so that the FLRW metric of Eq.~\eqref{eq_intro:FLRW} has exactly this shape. As extensively discussed in Chapter~\ref{chapter:inflation}, the inflating Universe is thus a nearly dS${}_{4}$ spacetime. \\

 \noindent 
 Before concluding this section we can give the relationships between the Ricci scalar, the cosmological constant $\Lambda $ and the constant $l$ introduced in Eq.~\eqref{appendix_GR:dS_hyperboloid}. Using the dS${}_{d+1}$ metric defined in Eq.~\eqref{appendix_GR:dS_metric}, the definitions of the previous section and Eq.~\eqref{appendix_GR:dS_AdS_EE} we have:
 \begin{equation}
  R = \frac{d(d+1)}{l^2} \ , \qquad \qquad  \Lambda = \frac{d(d-1)}{2 l^2} \ .
 \end{equation}

\subsection{Anti de Sitter spacetime.} 
\label{appendix_GR:AdS_spacetime}
Analogously to dS spacetime, $AdS$ is the maximally symmetric space with constant negative curvature $R < 0$. Again it is useful to introduce the AdS${}_{d+1}$ as a manifold embedded in a higher dimensional space. Let us consider the $(d+2)$-dimensional flat space with interval:
\begin{equation}
  \textrm{d} s^2 = - \textrm{d} x_{-1}^{\ 2} - \textrm{d} x_0^2 + \sum_{i = 1}^{d} \textrm{d} x_i^2 \ , 
\end{equation}
The AdS${}_{d+1}$ spacetime is defined as the the $(d+1)$-dimensional hyperboloid:
\begin{equation}
\label{appendix_GR:AdS_hyperboloid}
 - R_{A}^2 = - x_{-1}^{ \ 2}  - x_0^2 + \sum_{i = 1}^{d }  x_i^2 \ ,
\end{equation}
where the constant $R_{A} $ is sometimes called the `radius' of AdS${}_{d+1}$ spacetime. Notice that AdS${}_{d+1}$ has isometry group $O(2,d)$. Similarly to the case of dS${}_{d+1}$, to parametrize this surface we introduce a set of coordinates $(\tau,\rho,\theta_1,\dots,\theta_{d-2},\phi)$ with $\rho \in (-\infty,\infty)$, $\tau \in [0,2\pi] , \theta_i \in [0,\pi], \phi \in [0,2 \pi]$, such that:
\begin{equation}
x_0 = R_{A} \cosh(\rho) \cos(\tau), \qquad x_{-1} = R_{A} \cosh(\rho) \sin(\tau) \ , \qquad x_i = R_{A} \chi_i \sinh (\rho) \ ,
\end{equation}
where the $\chi_i$ are defined in Eq.~\eqref{appendix_GR:dS_solid_angle} and thus they clearly satisfy $ \sum_{i = 1}^{i = d} \chi_i^{ \ 2} = 1$. In terms of these coordinates the spacetime interval reads:
\begin{equation}
  \textrm{d} s^2 = R_{A}^2 \left[ \textrm{d}\rho^2 -\cosh^2 (\rho) \textrm{d} \tau^2 + \sinh^2 (\rho) \textrm{d} \Omega_{d-1}^{\ 2} \right] \ ,
\end{equation}
where $\Omega_{d-1}^{\ 2}$ is used to denote the $(d-1)$-dimensional solid angle. It is crucial to stress that this parametrization shows the presence in AdS${}_{d+1}$ of closed timelike curves creating problems with the causal structure. The origin of this problem is that AdS${}_{d+1}$ has the topology of $\mathbb{R}^{d} \times S^1$ that is not simply connected. To solve this problem, one should instead consider the \emph{universal covering} of AdS${}_{d+1}$ obtained by unwrapping the $S^1$ into $\mathbb{R}$ \textit{i.e.} extending $\tau$ to $\tau \in (-\infty,\infty)$. This new space has the topology of $\mathbb{R}^{d+1}$ that is simply connected and does not contain closed timelike curves. In this whole work when we refer to AdS${}_{d+1}$ space we are actually referring to its universal covering.\\

\noindent A useful coordinate set to parametrize the AdS${}_{d+1}$ space are the \emph{Poincar\'e} coordinates $(\sigma,t,y_1,\dots,y_{d-1})$, with $\sigma \in [0,\infty)$, $t \in (-\infty,\infty) , y_i \in(-\infty,\infty) $ defined as:
\begin{equation}
\label{appendix_GR:AdS_poincare}
\begin{aligned}
x_{-1} = \frac{\sigma }{R_{A}} t  ,\qquad & \qquad x_0 = \frac{R_{A}^2}{2\sigma }\left[ 1 + \frac{\sigma ^2}{R_{A}^4} \left(R_{A}^2 + \vec{y}^2 - t^2 \right) \right] \ , \\
x_i = \frac{\sigma }{R_{A}} y_i  ,\qquad & \qquad x_{d} =  \frac{R_{A}^2}{2\sigma }\left[ 1 - \frac{\sigma ^2}{R_{A}^4} \left(R_{A}^2 - \vec{y}^2 + t^2 \right) \right] \ .
\end{aligned}
\end{equation}
It is possible to show that in terms of these coordinates the spacetime interval reads:
\begin{equation}
  \textrm{d} s^2 = - \frac{\sigma^2}{R_{A}^2} \textrm{d} t^2 + \frac{R_{A}^2}{\sigma^2} \textrm{d} \sigma^2 + \frac{\sigma^2}{R_{A}^2} \textrm{d} \vec{y}^2 \ .
\end{equation}
Notice that for $\sigma \rightarrow \infty$ that corresponds to the boundary of AdS${}_{d+1}$, the metric blows up. As the metric on this hypersurface is well defined after a conformal transformation, this is usually referred as the conformal boundary of AdS${}_{d+1}$. On the contrary, the timelike killing vector $\frac{\partial}{\partial t}$ has zero norm on the hypersurface at $\sigma  = 0$. This hypersurface is thus an horizon. \\

\noindent
It is useful to introduce some other forms of the Poincar\'e coordinates. We start by introducing $z \equiv R_{A}^2 / \sigma $ so that the spacetime interval reads:
\begin{equation}
\label{appendix_GR:poincaree_metric}
  \textrm{d} s^2 = \frac{R_{A}^2}{z^2} \left( \textrm{d} z^2 - \textrm{d} t^2  +  \textrm{d} \vec{y}^2 \ \right) .
\end{equation}
Notice that with this parametrization the boundary is approached for $z \rightarrow 0$ and the horizon for $z \rightarrow \infty$. Moreover Eq.~\eqref{appendix_GR:poincaree_metric} clearly shows that the constant $z$ slices of AdS${}_{d+1}$ are conformal to $\mathbb{R}^{d}$ equipped with a metric with signature $(1,d-1)$ \textit{i.e.} to a $d$-dimensional Minkowski spacetime. \\

\noindent
An alternative description can be given in terms of the coordinate $u/R_{A} = - \ln(z/R_{A})$ such that: 
\begin{equation}
  \textrm{d} s^2 = \textrm{d} u^2 - e^{2 u/R_{A}} \left( \textrm{d} t^2  - \textrm{d} \vec{y}^2 \ \right)  .
\end{equation}
 Notice that with this parametrization the boundary of AdS spacetime is approached for $u \rightarrow \infty$ and the horizon is approached for $u \rightarrow -\infty$. Notice that for $1 \ll u$, the spacetime interval asymptotes to the spacetime interval of a $d$-dimensional Minkowski spacetime.\\

\noindent
It is also useful to introduce the Euclidean version of AdS${}_{d+1}$ spacetime. This is obtained by performing the analytical continuation defined by $x^0 \equiv i t$. In terms of this new coordinates the spacetime interval reads:
 \begin{equation}
  \label{appendix_GR:metric}
    \textrm{d}s^2 =  \textrm{d}u^2 + e^{2u/R_{A}} \delta_{\mu \nu } \textrm{d} x^\mu \textrm{d} x^\nu \ ,
  \end{equation}
where $\delta_{\mu \nu}$ is the standard Kronecker delta and $\mu,\nu = 0, 1, \dots, d$.\\

\noindent
Before concluding this section we compute the expressions for the Ricci scalar $R $ and for the cosmological constant $\Lambda$ in terms of $R_{A}$. Similarly to the case of dS${}_{d+1}$, we can get:
 \begin{equation}
  R = - \frac{d(d+1)}{R_{A}^2} \ , \qquad \qquad  \Lambda = - \frac{d(d-1)}{2 R_{A}^2} \ .
 \end{equation}

\subsection{Scalar fields in AdS spacetime.}
 \label{appendix_GR:AdS_dynamics}
\noindent Let us consider the case of a scalar field in a fixed AdS${}_{d+1}$ Euclidean spacetime. The metric of this space is thus given by Eq.~\eqref{appendix_GR:metric}. The action for a scalar field in AdS${}_{d+1}$ is simply given by:
 \begin{equation}
  \label{appendix_GR:action}
    \mathcal{S}= \int\mathrm{d}u \ \mathrm{d}^d x \ \sqrt{g}\left( \frac{  g^{ab} }{2 }  \partial_a \phi \partial_b \phi + V(\phi) \right).
  \end{equation}
The equation of motion for $\phi$ is computed by setting $\delta_\phi \mathcal{S} = 0$, where $\delta_\phi \mathcal{S}$ denotes the variation of $\mathcal{S}$ with respect to $\phi$ \textit{i.e.}
\begin{equation}
  \label{appendix_GR:eom_curved}
  \frac{1}{ \sqrt{g}} \partial_a \left[ \sqrt{g} \  g^{ab} \ \partial_b \phi \right] - V_{, \phi} (\phi) = 0.
\end{equation}
Using the parametrization for $g_{ab}$ given by Eq.~\eqref{appendix_GR:metric} we obtain:
\begin{equation}
\label{appendix_GR:eom_AdS}
    \ddot{\phi} + \frac{d \  \dot{\phi}}{R_{A}} + e^{ - 2 u /R_{A}} \square \phi  - V_{, \phi} (\phi) = 0, 
\end{equation}
where we use dots to denote derivatives with respect to the radial coordinate $u$ and the differential operator $\square$ is defined as $\square \equiv \partial^\mu \partial_\mu$. Let us first restrict to the case of a homogeneous scalar field with potential $V(\phi) = m^2/2$. In this case Eq.~\eqref{appendix_GR:eom_AdS} simply reduces to:
\begin{equation}
\label{appendix_GR:eom_AdS_homo}
    \ddot{\phi} + \frac{d \  \dot{\phi}}{R_{A}}  - m^2 \phi = 0.
\end{equation}
It is clear that all the solution of this equation can be expressed as $\phi(u) \simeq C e^{ - \Delta u/R_{A}} $, where $C$ is a constant dimensionful factor and $\Delta$ has to satisfy:
\begin{equation}
\label{appendix_GR:AdS_Delta_1}
    \Delta ( \Delta - d )  - m^2  R_{A}^{ \ 2} = 0.
\end{equation}
This equation obviously has two solutions:
\begin{equation}
\label{appendix_GR:AdS_Delta_2}
     \Delta_{\pm} \equiv  \frac{d}{2} \left( 1 \pm \sqrt{1  + \frac{ 4 R_{A}^{ \ 2} m^2}{d^2}} \right) .
\end{equation}
Notice that Eq.~\eqref{appendix_GR:AdS_Delta_2} implies that AdS spacetime can support scalar fields with a negative mass squared until the condition $- d^2/ (2R_A )^2 \leq m^2 $, is satisfied. This relationship is known as Breitenlohner-Freedman (BF) bound for the mass~\cite{Breitenlohner:1982jf}. It is also interesting to point out that Eq.~\eqref{appendix_GR:AdS_Delta_2} implies the useful relation:
 \begin{equation}
 \label{appendix_GR:delta_relations}
  \Delta_-  + \Delta_+ = d  \ .
 \end{equation}
 It is important to stress that Eq.~\eqref{appendix_GR:AdS_Delta_2} also implies that:
\begin{itemize}
  \item $m^2 = 0$ directly corresponds to $\Delta_- = 0$, $\Delta_+ = d$.
  \item A negative mass squared term corresponds to $ 0 < \Delta_- < \frac{d}{2}$, $ \frac{d}{2} < \Delta_+ < d $.
  \item A positive mass squared term corresponds to $\Delta_- < 0 $, $ d < \Delta_+ $. 
\end{itemize}
\noindent It is worth to point out that $\Delta_{+}$ is always positive and thus the solution $\phi = \phi_+ e^{ -u \Delta_+/R_{A}} $ is always exponentially suppressed in the neighborhood of the boundary and exponentially growing in the interior of AdS${}_{d+1}$ spacetime. Notice that if we want the the scalar field to be regular in the interior of AdS${}_{d+1}$, we should discard this solution. It is also useful to point out that independently on the value of $m^2$ we have $ \Delta_- < \Delta_+$. This clearly implies that in the neighborhood of the boundary (\textit{i.e.} $u \rightarrow \infty$) the leading contribution to $\phi (u)$ is always carried by the $\Delta_-$ solution, and thus we have:
\begin{equation}
     \left. \phi(u) \right|_{u \rightarrow \infty} \simeq \left. \phi_- e^{ - u\Delta_-/R_{A}} \right|_{u \rightarrow \infty} \ .
\end{equation} 
Notice that a scalar field in AdS${}_{d+1}$ with a negative mass squared term is exponentially growing in the neighborhood of the boundary.\\

\noindent
We can now consider the general case by reintroducing the spatial dependence into the scalar field $\phi$. As usual $\tilde{\phi}(u, k^\mu)$, Fourier transform of $\phi(u,x^\mu)$ over the spatial coordinates, is defined by:
\begin{equation}
  \phi(u, x^\mu) = \int \frac{\mathrm{d}^d k}{(2 \pi)^{d/2}} e^{i k^\mu x_\mu} \tilde{\phi}(u, k^\mu).
\end{equation}
We can then substitute into Eq.~\eqref{appendix_GR:eom_AdS} to get the equations of motion for a mode at fixed $k^\mu$:
\begin{equation}
\label{appendix_GR:complete_eom_fourier}
    \ddot{\tilde{\phi}} + \frac{d \  \dot{\tilde{\phi}}}{R_{A}}   - (  m^2 + e^{-2 u/R_{A} } k^2 )  \tilde{\phi} = 0. 
\end{equation}
As in the limit of $u \rightarrow \infty$, \textit{i.e.} in neighborhood of the boundary, the term depending on $k^2$ is exponentially suppressed, it should again be possible to express the asymptotic solution as:
\begin{equation}
    \tilde{\phi}_0(k^\mu) \equiv \left. \tilde{\phi}(u,k^\mu) \right|_{u \rightarrow \infty} \propto \left. e^{ - u \Delta_-/R_{A}} \ \tilde{\phi}_{Reg} (k^\mu) \right|_{u \rightarrow \infty} ,
\end{equation} 
where $\tilde{\phi}_{Reg} (k^\mu) $ is a finite function of $k^\mu$ that does not depends on $u$. The precise definition of $\tilde{\phi}_{Reg} (k^\mu)$ is given in the following (see Eq.~\eqref{appendix_GR:regularized_field}). For the precise expression for $\tilde{\phi}_0(k^\mu)$ in terms of $\tilde{\phi}_{Reg} (k^\mu)$ see Eq.~\eqref{appendix_GR:tilde_phi_expansion}.\\

\noindent To get an explicit expression for $\tilde{\phi}_{Reg} (k^\mu)$ it is useful to express the equation of motion for the scalar field $\phi$ in the Poincar\'e coordinates using $z$. As $z \equiv R_{A}e^{- u/R_{A}}$, we can thus express Eq.~\eqref{appendix_GR:complete_eom_fourier} as:
\begin{equation}
\label{appendix_GR:complete_eom_fourier_2}
    z^{d+1} \partial_z \left[ z^{-d + 1} \partial_z \tilde{\phi} (z,k^\mu) \right] - (  m^2 R_{A}^{ \ 2} + z^2 k^2 ) \tilde{\phi}(z,k^\mu) = 0.
\end{equation}
Notice that with this parametrization the boundary is reached for $z \rightarrow 0$ and in this limit the dependence on $k^2$ disappears. We already know that in this limit the equation for $\tilde{\phi}$ admits two solutions proportional to $ z^{\Delta_\pm}$ respectively. We can stress once again that the dominating contribution in the neighborhood of the boundary is carried by the $\Delta_-$ solution. We can proceed with our treatment by defining the dimensionless parameter $\theta \equiv z k$, and the field $y = \theta^{-d/2} \tilde{\phi}$, so that the equation of motion for $y$ reads:
 \begin{equation}
\label{appendix_GR:complete_eom_fourier_3}
    \theta^2 \frac{\textrm{d}^2}{\textrm{d} \theta^2} y + \theta \frac{\textrm{d}}{\textrm{d} \theta} y - (
    m^2 R_{A}^{ \ 2} + d^2/4 + \theta^2) y = 0 \ .
\end{equation}
Eq.~\eqref{appendix_GR:complete_eom_fourier_3} is a modified Bessel's equation with $\alpha^2 = m^2 R_{A}^{ \ 2} + d^2/4 = ( \Delta_+ -d/2)^2 $. The solutions of Eq.~\eqref{appendix_GR:complete_eom_fourier_3} are a combination of $I_\alpha(\theta)$ and $K_\alpha (\theta)$, modified Bessel functions of the first and second kind respectively. A general solution for $\tilde{\phi} (\theta , k)$, can thus be expressed as:
\begin{equation}
\label{appendix_GR:tilde_phi_expansion_1}
   \tilde{\phi}(\theta , k) \simeq \tilde{\mathcal{B}}(k) \theta^{d/2} I_\alpha(\theta) + \tilde{\mathcal{A}}(k) \theta^{d/2} K_\alpha (\theta) \ ,
\end{equation}
where $\tilde{\mathcal{B}}(k)$ and $\tilde{\mathcal{A}}(k)$ are two functions of $k$ with the same dimension of $ \tilde{\phi}(\theta , k)$. As the expansions for $I_\alpha (\theta)$ and $K_\alpha (\theta)$ at $\theta = k z \rightarrow 0$ are:
\begin{equation}
\begin{aligned}
\label{appendix_GR:regular_bessel_approx}
 I_\alpha (\theta) & \simeq \theta^{\alpha} \left(1+ C_1 \theta^2 + \dots \right) \ , \\
  K_\alpha (\theta) & \simeq \theta^{-\alpha} \left(1+ D_1 \theta^2 + \dots + D_\alpha \theta^{2\alpha}  + D_\alpha \theta^{2\alpha} \ln(\theta) + \dots \right) \ , 
 \end{aligned}
\end{equation}
where $C_1,D_1,D_\alpha$ are constants. Notice that except for the term proportional to $D_\alpha$, these expansions only contain terms like $D_n \ \theta^{2n}$ with $n$ integer. Finally we can substitute these into Eq.~\eqref{appendix_GR:tilde_phi_expansion_1} to get:
\begin{equation}
\label{appendix_GR:tilde_phi_expansion_theta}
   \tilde{\phi}(\theta,k) \simeq \tilde{\mathcal{B}}(k) \theta^{\Delta_+} + \tilde{\mathcal{A}}(k) \theta^{\Delta_-} \ .
\end{equation}
It should thus be clear that the leading contribution to the regular part of $\tilde{\phi}(\theta,k)$ for $\theta \rightarrow 0 $ is given by $\tilde{\mathcal{A}}(k)$. As a consequence we can set:
\begin{equation}
\label{appendix_GR:regularized_field}
  \tilde{\phi}_{Reg} (k^\mu) = \tilde{\mathcal{A}}(k) \ .
\end{equation}
It is also useful to express this equations in terms of $u$:
\begin{equation}
\label{appendix_GR:tilde_phi_expansion}
   \tilde{\phi}(u , k) \simeq \ \tilde{\mathcal{B}}(k)(kR_A)^{\Delta_+} e^{-u\Delta_+/R_A} + \tilde{\mathcal{A}}(k)(kR_A)^{\Delta_-} e^{-u\Delta_-R_A} \ .
\end{equation}
Notice that these two functions have the same dimension of $ \tilde{\phi}(\theta , k)$. Once again $\Delta_- < \Delta_+ $ implies that approaching the boundary the leading contribution is the given by the term proportional to $\tilde{\mathcal{A}}(k) $. Moreover if $m^2 < 0 $, we get $\Delta_- < 0$, and thus the field is diverging in the neighborhood of the boundary.

{\large \par}}
{\large \chapter{Cosmological perturbations.}
 \label{appendix_perturbations:Cosmological_perturbations}
In this Appendix we present a slight generalization of the standard cosmological perturbation theory. In particular we want to perform a description that can also be applied to the case of perturbations on domain-wall solutions. The case of standard cosmological perturbation theory has been widely treated in the literature. In particular, it is worth mentioning the works of Bardeen~\cite{Bardeen:1980kt,Bardeen:1983qw}, Mukhanov, Feldman and Brandenberger~\cite{Mukhanov:1990me}, Kodama and Sasaki~\cite{Kodama01011984}. On the other hand, the interest in the perturbations of domain-wall solution is motivated by its applications in the context of holography~\cite{McFadden:2009fg, McFadden:2010na}. For this reason, in this Appendix we will use the same parameterization used in Sec.~\ref{sec_holography:AdS_dS_correspondence}. In particular, the metric is given by:
\begin{equation}
  \textrm{d}s^2 = \eta \textrm{d}r^2 + a^2(t) \textrm{d}\vec{x}^2 \ , 
\end{equation}
where for $\eta = -1$ we identify $r$ with the cosmic time $t$ and $\eta = 1$ we identify $r$ with the radial coordinate $u$ of the domain-wall. Using this parameterization, we can express the action as:
 \begin{equation}
  \label{appendix_perturbations:action}
    \mathcal{S}=-\frac{\eta}{\kappa^2} \int\mathrm{d}r\mathrm{d}^3x \sqrt{|g|}\left( \frac{R}{2} + p(X,\Phi) \right),
  \end{equation}
where $p(X,\Phi)$ is a generic function of $X$ and $\Phi$, and as usual we have defined $X \equiv g^{\mu \nu} \partial_\mu \Phi \partial_\nu \Phi /2$. In the following, we choose to work with dimensionless fields. To recover the standard cosmological perturbation theory, we just pick the $\eta = -1 $ case and we express the problem in terms of dimensionful fields. \\

\noindent
In order to carry out the analysis of this Appendix (and as usual in the context of cosmological perturbation theory), we expand the scalar field and the metric as:
\begin{equation}
\label{appendix_perturbations:general_expansion}
  g_{\mu\nu}(r,\vec{x}) = {}^{(0)}g_{\mu\nu}(r) + \delta g_{\mu\nu}(r,\vec{x}) \ , \qquad \qquad \Phi(r,\vec{x}) =  {}^{(0)}\phi(r) + \delta \phi (r,\vec{x}) \ .
\end{equation}
With this parameterization we are thus separating the homogeneous background from the space-dependent perturbations. \\

\noindent
In this appendix we proceed as follows: we express the metric and matter perturbations in terms of gauge invariant quantities in Sec.~\ref{appendix_perturbations:metric_perturbations} and Sec.~\ref{appendix_perturbations:scalar_perturbations}. In Sec.~\ref{appendix_perturbations:eom_scalar} and Sec.~\ref{appendix_perturbations:eom_tensor} we present the explicit derivation of the equations of motion for the scalar and tensor perturbations respectively. In Sec.~\ref{appendix_perturbations:quantization} we show the procedure to quantize the cosmological perturbations and finally in Sec.~\ref{appendix_perturbations:observables} we derive the expression for the observable quantities.

\section{Metric perturbations.}
\label{appendix_perturbations:metric_perturbations}
To start with our treatment we need to specify the metric in terms of the background and of its small perturbations as:
\begin{equation}
  \label{appendix_perturbations:perturbations}
  \textrm{d}s^2 =\eta\left[1 + 2\varphi(r,\vec{x}) \right] \textrm{d}r^2 + 2 a^2(r)  B_i(r,\vec{x}) \textrm{d}r \textrm{d}x^i + a^2(r) \left[\delta_{ij} + h_{ij}(r,\vec{x}) \right] \textrm{d}x^i \textrm{d}x^j \ .
  \end{equation}
Notice that this expression is consistently introducing ten degrees of freedom for the metric perturbations. We can proceed by decomposing the vector $B_i$ and the spatial metric $h_{ij}(r,\vec{x})$ as:
 \begin{equation}
 \begin{aligned}
  \label{appendix_perturbations:perturbed_metric}
  B_i(r,\vec{x}) & = \partial_i \nu(r,\vec{x}) + \nu_i(r,\vec{x}) \ , \\
 h_{ij}(r,\vec{x}) & = - 2 \psi(r,\vec{x}) \delta_{ij} + 2\partial_i \partial_j \chi(r,\vec{x}) + (\partial_i \omega_j(r,\vec{x}) + \partial_j \omega_i(r,\vec{x})) + \gamma_{ij}(r,\vec{x}) \ ,
 \end{aligned} 
  \end{equation}
where the vectors $\nu_i(r,\vec{x})$ and $\omega_i(r,\vec{x})$ are transverse and the tensor $\gamma_{ij}(r,\vec{x})$ is transverse traceless\footnote{As a consequence, $\gamma_{ij}(r,\vec{x})$ only contains two degrees of freedom. The perturbations described by $\gamma_{ij}(r,\vec{x})$ correspond to propagating GWs (discussed in Appendix~\ref{appendix_GR:GW}).}. These conditions on  $\nu_i(r,\vec{x}) , \omega_i(r,\vec{x})$ (transverse) and on $\gamma_{ij}(r,\vec{x})$ (transverse, traceless) must be imposed to ensure the correct number of degrees of freedom for the perturbations. \\

\noindent
At this point it is useful to apply the decomposition theorem that states that metric perturbations can be divided into a \emph{scalar}, a \emph{vector} and a \emph{tensor} contribution. Moreover, the theorem states that these three types of perturbations are evolving independently. We can then express:
\begin{equation}
\begin{aligned}
\label{appendix_perturbations:perturbations_metric}
   \textrm{d}s^2_{S} &=\eta\left[1 + 2\varphi(r,\vec{x}) \right] \textrm{d}r^2 + a^2(r)  \partial_i \nu(r,\vec{x})  \textrm{d}r \textrm{d}x^i + \\
   & + a^2(r) \left\{ \delta_{ij} \left[1 - 2 \psi(r,\vec{x}) \right] + 2\partial_i \partial_j \chi(r,\vec{x})  \right\} \textrm{d}x^i \textrm{d}x^j , \\
    \textrm{d}s^2_{V} &=\eta \textrm{d}r^2 + 2 a^2(r) \nu_i(r,\vec{x})  \textrm{d}z \textrm{d}x^i + a^2(r) \left[\delta_{ij} + (\partial_i \omega_j(r,\vec{x}) + \partial_j \omega_i(r,\vec{x})) \right] \textrm{d}x^i \textrm{d}x^j , \\
     \textrm{d}s^2_{T} &=\eta \textrm{d}r^2 + a^2(r) \left[\delta_{ij} + \gamma_{ij}(r,\vec{x}) \right] \textrm{d}x^i \textrm{d}x^j .
\end{aligned}
\end{equation}
We can now show that it is possible to use a gauge fixing procedure to reduce the number of physically relevant functions. For this purpose let us consider the change of coordinates defined by:
\begin{equation}
\label{appendix_perturbations:gauge}
  x^{\mu} \rightarrow \tilde{x}^{\mu} = x^{\mu} + \xi^\mu, \qquad \qquad \text{with} \qquad \qquad \xi^i = \partial^i \xi + V^i,
\end{equation}
where the vector $V^i$ is transverse. As usual, the transformation of the metric is simply defined by:
\begin{equation}
\label{appendix_perturbations:metric_transformation}
 g_{\mu \nu} \rightarrow g_{\mu \nu} = \tilde{g}_{\rho \sigma} \frac{\partial \tilde{x}^{\rho}}{\partial x^{\mu}} \frac{ \partial \tilde{x}^{\sigma}}{ \partial x^{\nu}} = \tilde{g}_{\mu \nu} + \tilde{g}_{\mu \sigma} \partial_\nu \xi^\sigma +\tilde{g}_{\sigma \nu} \partial_\mu \xi^\sigma . 
\end{equation}
Let us assume that $\xi^\mu$ is a small quantity of the order of the perturbations. Under this assumption we can express Eq.~\eqref{appendix_perturbations:metric_transformation} at first order as:
\begin{equation}
\begin{aligned}
\label{appendix_perturbations:metric_first}
   g_{00} & = \tilde{g}_{00} +2 \tilde{g}_{00} \partial_0 \xi^0 \ , \\
   g_{0i} & = \tilde{g}_{0i} + \tilde{g}_{00} \partial_i \xi^0 +\tilde{g}_{ij} \partial_0 \left( \partial^i \xi + V^i\right),\\
   g_{ij} & = \tilde{g}_{ij} + 2 \tilde{g}_{i k } \partial_j \left(\partial^k \xi + V^k\right).
\end{aligned}
\end{equation}
Notice that $\tilde{g}_{0i}$ is first order in the perturbations (it is zero at the zero order). As a consequence, terms like $\tilde{g}_{0i} (\partial^i \xi + V^i )$ are second order in the perturbations and thus they can be neglected. We can proceed by using Eq.~\eqref{appendix_perturbations:perturbations_metric} to get the transformations for the metric perturbations\footnote{Notice that $\tilde{g}_{ij} = a^2(r + \xi^0) [\delta_{ij } + h_{ij}] \simeq a^2(r) [\delta_{ij } + h_{ij}] + 2 a(r) \dot{a}(r) \xi^0 \delta_{ij}$, where we have neglected second order terms and we have used a dot to denote a derivative with respect to $r$. }:
\begin{equation}
\begin{gathered}
\label{appendix_perturbations:perturbations_transformations}
   \varphi \rightarrow \varphi - \partial_0 \xi^0 , \qquad \psi \rightarrow \psi + H \xi^0, \qquad \chi \rightarrow \chi - \xi, \qquad \nu \rightarrow \nu - \frac{\partial_0 \xi}{2} - \frac{\eta \xi^0}{2 a^2},  \\
   \nu_i \rightarrow \nu_i - \frac{\partial_0 V_i}{2} \qquad \qquad \omega_i \rightarrow \omega_i - V_i\\
   \gamma_{ij} \rightarrow \gamma_{ij}.
\end{gathered}
\end{equation}
It should then be clear that it is possible to define some gauge invariant combinations of the perturbations:
\begin{equation}
\begin{aligned}
\label{appendix_perturbations:invariant_variables}
   \Psi & \equiv \psi + 2 \eta a^2 H  \left( \nu - \frac{\partial_0 \chi}{2}\right) , \\ 
   \Upsilon & \equiv \varphi - 2 \eta a^2 \left[ 2 H \left( \nu - \frac{\partial_0 \chi}{2}\right) + \frac{\textrm{d}}{\textrm{d}t}\left( \nu - \frac{\partial_0 \chi}{2}\right) \right] , \\
    \Xi_i & \equiv \nu_i - \frac{\partial_0 \omega_i}{2}  \\
     \gamma_{ij} & \equiv  \gamma_{ij}.
\end{aligned}
\end{equation}
As these quantities are gauge invariant they represent the relevant degrees of freedom of the problem. In particular we are left with 2 scalar, 2 vector and 2 tensor degrees of freedom. The same result could have directly been derived from Eq.~\eqref{appendix_perturbations:perturbations_transformations}, by fixing a particular gauge. In particular an appropriate gauge to describe scalar perturbations is the so called \emph{Newton gauge} with $\chi = \nu = 0$. From Eq.~\eqref{appendix_perturbations:perturbations_transformations} it should be clear that the choice for $\xi^0$ and $\xi$ that realizes this gauge fixing is equivalent to the parameterization of Eq.~\eqref{appendix_perturbations:invariant_variables}. Finally we can express the first order gauge invariant perturbations as:
\begin{equation}
\begin{aligned}
\label{appendix_perturbations:perturbations_metric_invariant}
   \textrm{d}s^2_{S} &=\eta\left[1 + 2\Upsilon(r,\vec{x}) \right] \textrm{d}z^2 + a^2(r)  \delta_{ij} \left[1 - 2 \Psi (r,\vec{x}) \right]  \textrm{d}x^i \textrm{d}x^j , \\
    \textrm{d}s^2_{V} &=\eta \textrm{d}z^2 + 2 a^2(r) \Xi_i(r,\vec{x})  \textrm{d}z \textrm{d}x^i + a^2(r) \delta_{ij} \textrm{d}x^i \textrm{d}x^j , \\
     \textrm{d}s^2_{T} &=\eta \textrm{d}z^2 + a^2(r) \left[\delta_{ij} + \gamma_{ij}(r,\vec{x}) \right] \textrm{d}x^i \textrm{d}x^j .
\end{aligned}
\end{equation}
Vector perturbations are not generated during inflation and thus for the purposes of this work we can set them to zero. In the rest of this appendix we discuss the production of scalar and tensor perturbations and in particular we compute the observable quantities.

\section{Matter perturbations.}
\label{appendix_perturbations:scalar_perturbations}
Let us consider the action of Eq.~\eqref{appendix_perturbations:action}, (for later convenience we rename $\bar{\Phi}$ the scalar field). The background solution plus its perturbations are expressed as:
\begin{equation}
  \bar{\Phi}(r,\vec{x}) =  {}^{(0)} \bar{\phi}(r) + \delta \bar{\phi}(r,\vec{x}).
\end{equation}
As usual the stress-energy tensor is defined as:
\begin{equation}
\label{appendix_perturbations:stress_energy_definition}
 T_{\mu \nu} \equiv -\frac{2}{\sqrt{|g|} } \frac{\delta \mathcal{S}_m}{\delta g^{\mu \nu}} = \frac{\eta}{\kappa^2} \left[ \frac{p + \rho}{2 X} \partial_\mu \bar{\Phi} \partial_\nu \bar{\Phi} - g_{\mu \nu} p \right] \ ,
\end{equation}
where $ \mathcal{S}_m$ is the action for matter (\textit{i.e.} for the scalar field) and where we have used $\rho \equiv 2 X p_{, X} - p$. It is useful to define: 
\begin{equation}
\label{appendix_perturbations:velocity}
  U_{\mu} \equiv \frac{\partial_\mu \bar{\Phi}}{\sqrt{2 \eta X}},
\end{equation}
so that $T^\mu_{\ \nu}(r,\vec{x})$ can be expressed as:
\begin{equation}
\label{appendix_perturbations:stress_energy_expression}
 T^\mu_{ \ \nu}(r,\vec{x}) = \frac{1}{\kappa^2} \left[ (p + \rho) g^{\mu \alpha} U_\alpha U_\nu - \eta \delta^{\mu}_{\ \nu} p \right] \ . 
\end{equation}
Notice that for $\eta = -1$ this actually matches with the standard expression for the stress-energy tensor of a perfect fluid at rest and in thermodynamic equilibrium (shown in Eq.~\eqref{eq_intro:energy_content_tensor}). In analogy with the treatment of Sec.~\ref{appendix_perturbations:metric_perturbations}, we proceed by expressing the stress-energy tensor as:
\begin{equation}
\label{appendix_perturbations:stress_energy_perturbations}
 T^\mu_{ \ \nu}(r,\vec{x}) = {}^{(0)}T^{\mu}_{ \ \nu} (r)+ \delta  T^\mu_{ \ \nu}(r,\vec{x}),
\end{equation}
where $ \delta  T^\mu_{ \ \nu} (r,\vec{x})$ is the linear perturbation around the background solution ${}^{(0)}T^{\mu}_{ \ \nu} (r)$. Using Eq.~\eqref{appendix_perturbations:velocity} and the hypothesis of a homogeneous ${}^{(0)} \bar{\phi}$, we can easily obtain ${}^{(0)}U_{\mu} = (1,0,0,0)$. The background stress-energy tensor thus reads:
\begin{equation}
  \label{appendix_perturbations:stress_energy_0}
  {}^{(0)}T^{\mu}_{ \ \nu} =  \frac{\eta }{\kappa^2} \textrm{diag}\left[{}^{(0)}  \rho, - {}^{(0)}p,- {}^{(0)}p,- {}^{(0)}p \right] \ .
\end{equation}
On the other hand, the most general expression for $\delta  T^\mu_{ \ \nu}(r,\vec{x})$ is given by:
\begin{equation}
\label{appendix_perturbations:general_stress_pert}
  \delta  T^\mu_{ \ \nu} = \frac{1}{\kappa^2} \left[  {}^{(0)}\left( U^{\mu} U_\nu \right)  \left(\delta \rho + \delta p\right)+ {}^{(0)}\left( \rho + p\right) \left({}^{(0)}U_\nu \delta U^{\mu}+ {}^{(0)} U^{\mu}\delta U_\nu \right) - \eta \delta^\mu_{\ \nu} \delta p - \eta \Pi^\mu_{\ \nu} \right] \ ,
\end{equation}
where we have introduced the tensor $\Pi^\mu_{\ \nu} (r,\vec{x})$ to represent the contribution from \emph{anisotropic stresses}. It is important to point out that the spatial part of $\Pi^\mu_{\ \nu}$ is traceless, as any general contribution $\Pi^i_{\ i} \neq 0$ can be reabsorbed in the isotropic part of $T^i_{ \ j}$, namely ${}^{(0)}p$. Without loss of generality, we can also impose $U^{\mu}\Pi_{\mu \nu} = 0$. To get the explicit expression for $\delta U_\mu$ we start by using:
\begin{equation}
  U_{\mu}U_{\nu} g^{\mu \nu} = \eta, \qquad \rightarrow  \qquad 2 \eta \delta U_{0} + \delta g^{00} = 0, \qquad \rightarrow  \qquad \delta U_{0} = \varphi \ ,
\end{equation}
where $\varphi$ is part of the metric perturbations defined in Eq.~\eqref{appendix_perturbations:perturbations}. In order to lighten the notation, in the rest of this Appendix the homogeneous part of the scalar field $\bar{\Phi}$ is simply denoted with $ \bar{\phi}$ \emph{i.e.}  we drop the ${}^{(0)}$ superscript. We proceed by using the definition of $U_\mu$ to get $\delta U_i \equiv \partial_i [ \delta \bar{\phi}/\partial_0 \bar{\phi}] $. This directly gives:
\begin{equation}
\begin{aligned}
  U_{\mu} & = \left[ 1 + \varphi, \partial_i \left( \frac{\delta \bar{\phi}}{\partial_0 \bar{\phi}}\right) \right] \ , \\
  \delta U^{\mu}  & = \left[  - \eta \varphi, \ - 2 \eta B^i + \frac{\delta^{ij}}{a^2 (r)} \partial_j \left( \frac{ \delta \bar{\phi}}{\partial_0 \bar{\phi} }  \right)\right] \ .
 \end{aligned}
\end{equation}
Finally the can substitute into Eq.~\eqref{appendix_perturbations:general_stress_pert} to get:
\begin{equation}
  \begin{aligned}
 \delta T^{0}_{\ 0 } & = \frac{\eta }{\kappa^2} \delta \rho \ ,\\
   \delta T^{i}_{\ 0 }  & = \ \frac{ {}^{(0)}(p+\rho)}{\kappa^2} \left[ - 2 \delta^{ij} B_j + \frac{\delta^{ij} }{a^2 (r)} \partial_j \left( \frac{ \delta \bar{\phi}}{\partial_0 \bar{\phi} }  \right) \right] \ , \\
  \delta T^{0}_{\ i } & = \frac{\eta }{\kappa^2} \ {}^{(0)}(p+\rho) \partial_i \left( \frac{\delta \bar{\phi}}{\partial_0 \bar{\phi}}\right) \ , \\
   \delta T^{i}_{\ j } & = - \frac{\eta}{\kappa^2} \left[ \delta^i_{\ j} \ \delta p + \Pi^{i}_{\ j} \right] \ .
\end{aligned}
\end{equation}
As discussed in Sec.~\ref{appendix_perturbations:metric_perturbations}, we should then proceed by fixing the gauge. For this purpose, we should then study the transformation properties of $T^\mu_{ \ \nu} (r,\vec{x})$. Considering the change of coordinates defined by Eq.~\eqref{appendix_perturbations:gauge}, at the linear order $T^\mu_{ \ \nu}$ transforms as:
\begin{equation}
  T^\mu_{ \ \nu} (r,\vec{x}) \longrightarrow T^\mu_{ \ \nu} (r,\vec{x}) = \tilde{T}^\alpha_{ \ \beta} \frac{\partial x^\mu}{\partial \tilde{x}^\alpha} \frac{\partial \tilde{x}^\beta}{\partial x^\nu} = \tilde{T}^\mu_{ \ \nu} - \tilde{T}^\alpha_{ \ \nu}(\partial_\alpha \xi^\mu) + \tilde{T}^\mu_{ \ \beta} (\partial_\nu \xi^\beta) \ .
\end{equation}
In particular this implies:
\begin{equation}
  {}^{(0)}T^\mu_{ \ \nu} = {}^{(0)}\tilde{T}^\mu_{ \ \nu} \ , \qquad \delta T^\mu_{ \ \nu} = \delta \tilde{T}^\mu_{ \ \nu} - \delta \tilde{T}^\alpha_{ \ \nu}(\partial_\alpha \xi^\mu) + \delta \tilde{T}^\mu_{ \ \beta} (\partial_\nu \xi^\beta) \ .
\end{equation}
Using these relations we can thus prove that:
\begin{equation}
\begin{aligned}
\label{appendix_perturbations:transformation_stress_energy}
  \delta T^{0}_{\ 0 } & = \delta \tilde{T}^{0}_{\ 0 } \ , \\
   \delta T^{0}_{\ i } & =  \delta \tilde{T}^{0}_{\ i }+\frac{\eta}{\kappa^2} \ {}^{(0)}( \rho + p) \ \partial_i \xi^0\ ,\\
 \delta T^{i}_{\ 0 } & =  \delta  \tilde{T}^{i}_{\ 0 } - \frac{\eta}{\kappa^2} \ {}^{(0)}( \rho + p) \ \partial_0 \xi^i \ ,\\
 \delta T^{i}_{\ j }  & =   \delta \tilde{T}^{i}_{\ j } \ .
\end{aligned}
\end{equation}
In analogy with Sec.~\ref{appendix_perturbations:metric_perturbations}, we decomposed the vector $\xi^i $ into a scalar $\xi$ and a transverse vector $V^i$. Using the transformation properties of $B_i$, derived in the previous section we can finally show that:
\begin{equation}
\begin{gathered}
\label{appendix_perturbations:matter_transformation}
   \delta \rho \rightarrow \delta \rho , \qquad \Pi^i_{ \ j } \rightarrow \Pi^i_{ \ j }, \qquad \delta p \rightarrow \delta p, \qquad \delta \bar{\phi} \rightarrow \delta \bar{\phi} - \xi^ 0 \partial_0 \bar{\phi} 
\end{gathered}
\end{equation}
As the gauge has been fixed in Sec.~\ref{appendix_perturbations:metric_perturbations}, we redefine the scalar field in \emph{Newton gauge} as:
\begin{equation}
  \Phi = \phi(r) +\delta \phi(r,\vec{x}) \equiv \bar{\Phi} - \xi^ 0 \partial_0 \bar{\phi} = \bar{\phi} +\delta \bar{\phi} - \xi^ 0 \partial_0 \bar{\phi}.
\end{equation}
Notice that in terms the new field $\Phi$, in Newton gauge we have:
\begin{equation}
\begin{aligned}
\label{appendix_perturbations:Stress_0i}
 \delta T^{0}_{\ 0 }  & = \frac{\eta }{\kappa^2} \delta \rho  \ , \\
 \delta T^{0}_{\ i } & = \frac{\eta}{\kappa^2} {}^{(0)}( \rho + p) \partial_i \left( \frac{ \delta \phi}{\partial_0 \phi }  \right) \ .
 \end{aligned}
  \end{equation}
Notice that Eq.~\eqref{appendix_perturbations:Stress_0i} only depends on scalar quantities. During inflation the contribution from anisotropic stress $\Pi^{i}_{\ j}$ is negligible and thus we can set $\Pi^{i}_{\ j} = 0$. It is also possible to prove that $\Pi^{i}_{\ j}$ acts as a source for terms proportional to $\Psi - \Upsilon$. For the scope of this work we can thus fix $\Psi = \Upsilon$ in Eq.~\eqref{appendix_perturbations:perturbations_metric_invariant}. 

\section{Equations of motion for scalar perturbations.}
\label{appendix_perturbations:eom_scalar}
As usual the evolution of the system is described by Einstein equations which are derived by taking the variation action of Eq.~\eqref{appendix_perturbations:action} with respect to $g^{\mu \nu}$:
\begin{equation}
\label{appendix_perturbations:einstein_eq}
 G_{\mu \nu} \equiv R_{\mu \nu} - \frac{1}{2} g_{\mu \nu} R = -\eta \kappa^2 T_{\mu \nu}, \qquad  \longrightarrow    \qquad G^{\mu}_{\ \nu} \equiv R^{\mu}_{\ \nu} - \frac{1}{2} \delta^{\mu}_{\ \nu} R = -\eta \kappa^2 T^{\mu}_{\ \nu} \ .
\end{equation}
We start by considering the case of scalar perturbations. To specialize these equations to the case of our interest we start by computing $\delta G^{\mu}_{\ \nu} $. As discussed in Sec.~\ref{appendix_perturbations:metric_perturbations} and in Sec.~\ref{appendix_perturbations:scalar_perturbations}, we can express the background metric and its scalar perturbations as:
\begin{equation}
  \textrm{d}s^2_{S} =\eta\left[1 + 2\Upsilon (r,\vec{x}) \right] \textrm{d}r^2 + a^2(r)  \delta_{ij} \left[1 -2 \Upsilon (r,\vec{x}) \right]  \textrm{d}x^i \textrm{d}x^j.
\end{equation}
To express Eq.~\eqref{appendix_perturbations:einstein_eq} at the linear order, we first have to computing the Christoffel symbols (defined in Eq.~\eqref{appendix_GR:Christoffels}). It is possible to show that at the lowest order the only non-zero components are\footnote{Accordingly to the notation used in the rest of this work, we denote with dots derivatives with respect to the time or radial coordinate $r$ and with ${}_{,i}$ derivatives with respect to $x^i$.}:
\begin{equation}
  \label{appendix_perturbations:Christoffel_0}
  {}^{(0)} \Gamma^0_{\ ij} = -\eta\delta_{ij}  a\dot{a}, \qquad \qquad {}^{(0)} \Gamma^i_{\ 0j} = \Gamma^i_{\ j0} = \delta^i_j \frac{\dot{a}}{a}.
\end{equation}
On the other hand, at the linear order we get:
\begin{equation}
  \begin{aligned}
  \label{appendix_perturbations:Christoffel_1}
  &\delta \Gamma^0_{\ 00} = \dot{\Upsilon} \ ,  \qquad \qquad \delta \Gamma^0_{\ ij} =  \eta \delta_{ij} a^2  \dot{\Upsilon}  \ ,\\
 & \delta  \Gamma^0_{\ 0i} =\Upsilon_{,i} \ ,\qquad  \qquad \delta \Gamma^i_{\ 00} = -\eta \frac{\delta^{ij}\Upsilon_{,j}}{a^2} \ ,  \\
&\delta  \Gamma^i_{\ 0j}= - \delta^i_j\dot{\Upsilon} \ , \qquad \delta \Gamma^i_{\ jk} = \delta^{il} \left(\delta_{jk}\Upsilon_{,l} - \delta_{jl}\Upsilon_{,k} - \delta_{kl}\Upsilon_{,j} \right) \ .
  \end{aligned}
\end{equation}
We can thus proceed by computing the Ricci tensor (defined in Eq.~\eqref{appendix_GR:Ricci_tensor}). At the lowest order the only non-zero components are:
\begin{equation}
  \label{appendix_perturbations:Ricci_tensor_0}
  {}^{(0)} R_{00} = -3 \frac{\ddot{a}}{a} \ , \qquad \qquad {}^{(0)} R_{ij} = -\eta \delta_{ij} \left( a\ddot{a} + 2 \dot{a}^2 \right) \ .
\end{equation}
At the linear order we get:
\begin{equation}
  \begin{aligned}
  \label{appendix_perturbations:Ricci_tensor_1}
  &\delta R_{00} =3 \ddot{\Upsilon} +9 H \dot{\Upsilon} - \eta \frac{\Delta \Upsilon}{a^2} \ , \\
  &\delta R_{0i} = 2 H \Upsilon_{,i} + 2 \dot{\Upsilon_{,i}} \, \\
&   \delta R_{ij} = - \delta_{ij} \left[   \Delta \Upsilon + \eta a^2 (7 H \dot{\Upsilon} + \ddot{\Upsilon}) +4\eta \Upsilon (2 \dot{a}^2 + a \ddot{a}) \right] \ ,
  \end{aligned}
\end{equation}
where $\Delta = \partial_i\partial^i $ is the standard laplacian operator. We can then proceed by computing the Ricci scalar (defined in Eq.~\eqref{appendix_GR:Ricci_scalar}). The lowest and linear order are respectively given by:
\begin{equation}
  \label{appendix_perturbations:Ricci_scalar_expanded}
  \begin{aligned}
  {}^{(0)} R &= -6\eta\left(H^2 + \frac{\ddot{a}}{a}\right) \ , \\
\delta R &= 2 \eta\left[- 3 \ddot{\Upsilon}  +15 H\dot{\Upsilon} +6 \Upsilon\left(H^2 + \frac{\ddot{a}}{a}\right)  + \frac{\eta}{a^2} \Delta \Upsilon \right] \ .
  \end{aligned}
\end{equation}
Finally we can get the lowest order expression for the Einstein tensor:
\begin{equation}
\label{appendix_perturbations:Einstein_tensor_0}
  {}^{(0)} G^{0}_{\ 0} = 3 \eta H^2 \ , \qquad \qquad  {}^{(0)}G^{i}_{\ j} = \eta \left( H^2 + 2 \frac{\ddot{a}}{a}\right).
\end{equation}
Substituting these expressions and the background stress-energy tensor of Eq.\eqref{appendix_perturbations:stress_energy_0} into Eq.~\eqref{appendix_perturbations:einstein_eq}, we recover the standard Friedmann equations:
\begin{equation}
  \label{appendix_perturbations:Friedmann}
  3 \eta H^2 = - {}^{(0)}\rho \ , \qquad \qquad  2 \eta \dot{H} = {}^{(0)}(p + \rho) \ .
\end{equation}
On the other hand, it is possible to show that the evolution of the perturbations is completely specified by:
\begin{eqnarray}
\label{appendix_perturbations:einstein_perturbed_00}
   \delta G^{0}_{\ 0} &=& \delta T^{0}_{\ 0} \ ,\\
\label{appendix_perturbations:einstein_perturbed_0i}
   \delta G^{0}_{\ i} &=& \delta T^{0}_{\ i} \ .
\end{eqnarray}
The explicit expressions for $ \delta G^{0}_{\ 0} $ and $ \delta G^{0}_{\ i}$ are:
\begin{equation}
   \begin{aligned}
   \label{appendix_perturbations:Einstein_1}
   \delta G^{0}_{\ 0} &= - 2 \eta \left( 3 H^2 \Upsilon + \eta \frac{\Delta \Upsilon }{a^2} + 3 H \dot{\Upsilon} \right) \ ,\\
   \delta G^{0}_{\ i} &= 2 \eta \left( H \Upsilon_{,i} + \dot{\Upsilon}_{,i} \right) \ .
   \end{aligned}
 \end{equation} 
Before substituting Eq.~\eqref{appendix_perturbations:Einstein_1} into Eq.~\eqref{appendix_perturbations:einstein_perturbed_00} and into Eq.~\eqref{appendix_perturbations:einstein_perturbed_0i}, we also need to express $\delta T^{0}_{\ 0} = \delta \rho$ as a function of known quantities. For this purpose we should start by using:
\begin{equation}
  \delta \rho = {}^{(0)} \left( \frac{\partial  p}{\partial  X}\right) \delta X +  {}^{(0)} \left( \frac{\partial  p}{\partial  \phi}\right) \delta \phi 
\end{equation}
From now on we denote $p_{,X} \equiv \partial  p/\partial  X $ and $p_{,\phi} \equiv \partial  p/\partial  \phi$. We can then use the lowest order expressions:
\begin{equation}
  {}^{(0)}\dot{\rho} = - 3 H \ {}^{(0)}(p +\rho) \ , \qquad  {}^{(0)} p_{,\phi} = {}^{(0)} \left( \frac{\dot{\rho} - p_{,X} \dot{X}}{\dot{\phi}} \right) \ ,
\end{equation}
and we can express $\delta X$ in terms of $ \Upsilon$ and $\delta\phi$ as:
\begin{equation}
  \delta X = - 2 \  {}^{(0)} \dot{X} \left( - \Upsilon + \frac{\delta\dot{\phi}}{\dot{\phi}}\right) \ .
\end{equation}
As a consequence, $\delta T^{0}_{\ 0} $ can be expressed as:
\begin{equation}
\label{appendix_perturbations:deltarho}
\delta T^{0}_{\ 0} = - 3 H \ {}^{(0)}(p +\rho) \left( \frac{\delta \phi}{\dot{\phi}}\right)+ {}^{(0)}\left(\frac{p +\rho}{c_s^2}\right) \left[ -\Upsilon + \frac{\textrm{d}}{\textrm{d} r} \left( \frac{\delta \phi}{\dot{\phi}}\right)\right],
\end{equation}
where we have defined the speed of sound:
\begin{equation}
  \label{appendix_perturbations:speedofsound}
  c_s^2 \equiv {}^{(0)}\left(\left. \frac{\delta p }{\delta \rho} \right|_{\delta\phi = 0} \right)=  {}^{(0)}\left( \frac{p_{,X}}{\rho_{,X}} \right) = {}^{(0)}\left(  \frac{p + \rho}{2 X \rho_{,X}} \right).
\end{equation}
Finally we can substitute Eq.~\eqref{appendix_perturbations:deltarho}, Eq.~\eqref{appendix_perturbations:Stress_0i} and Eq.~\eqref{appendix_perturbations:Einstein_1} into Eq.~\eqref{appendix_perturbations:einstein_perturbed_00} and Eq.~\eqref{appendix_perturbations:einstein_perturbed_0i} to get:
\begin{eqnarray}
\label{appendix_perturbations:perturbations_1}
  \frac{\delta \phi}{\dot{\phi}} & =&  \frac{2 \eta }{{}^{(0)}(p +\rho)} \left( H \Upsilon + \dot{\Upsilon} \right) = \frac{1}{\epsilon_H} \left[ \frac{\Upsilon}{H} + \frac{\dot{\Upsilon}}{H^2} \right] \ , \\
  \label{appendix_perturbations:perturbations_2}
  \frac{\textrm{d}}{\textrm{d} r} \left( \frac{\delta \phi}{\dot{\phi}}\right) &=& \left[ 1 - \frac{2  }{a^2} \ {}^{(0)}\left(\frac{c_s^2}{p+\rho} \right) \Delta \right] \Upsilon = \left[ 1 - \left(\frac{\eta c_s^2}{ a^2 H^2 \epsilon_H} \right) \Delta \right] \Upsilon\ . 
\end{eqnarray}
Where we have used Eq~\eqref{appendix_perturbations:Friedmann} and the standard definition of $\epsilon_H \equiv {}^{(0)}\left( -\dot{H}/H^2 \right)$. We can then define the new fields $\xi$ and $\zeta$ as:
\begin{equation}
\label{appendix_perturbations:xi_zeta_definition}
  \xi \equiv \frac{a \Upsilon}{H}, \qquad \qquad \zeta \equiv H \left( \frac{\delta \phi}{\dot{\phi}} \right) + \Upsilon = \frac{\xi}{a} ,
\end{equation}
so that Eq.~\eqref{appendix_perturbations:perturbations_1} and Eq.~\eqref{appendix_perturbations:perturbations_2} read:
\begin{equation}
\label{appendix_perturbations:perturbation_eom}
  \dot{\xi} = a \epsilon_H \ \zeta \ , \qquad \qquad \dot{\zeta } = - \frac{\eta c_s^2}{a^3 \epsilon_H} \Delta \xi. 
\end{equation}
Let us take the spatial Fourier transform of this equation and let us consider the evolution of a single mode at fixed wave vector $\vec{k}$. Substituting the derivative of the second equation with respect to $u$ into the first one we finally obtain:
\begin{equation}
\label{appendix_perturbations:scalar_eom_zeta}
\ddot{\tilde{\zeta}} + \left(3H + \frac{\dot{\epsilon}_H}{\epsilon_H} -2\frac{\dot{c_s}}{c_s} \right) \dot{\tilde{\zeta}} - \frac{\eta k^2 c_s^2}{a^2} \tilde{\zeta} =  0 \ ,
\end{equation}
where $\tilde{\zeta}(r,\vec{k})$ is the spatial Fourier transform of $\zeta(r,\vec{x})$. Notice that the equation of motion for $\tilde{\zeta}(r,\vec{k})$ could have been derived by taking the variation with respect to $\tilde{\zeta}(r,\vec{k})$ of the action:
\begin{equation}
\label{appendix_perturbations:definition_action_zeta}
  \mathcal{S} = \frac{\eta}{\kappa^2} \int\textrm{d} r \textrm{d}^3 \vec{k} \left[ \left(\frac{a^3 \ \epsilon_H }{c_s^2 }\right) \dot{\tilde{\zeta}}^2(r,\vec{k}) + \left(\eta a \epsilon_H k^2 \right) \tilde{\zeta}^2(r,\vec{k}) \right] \ , 
\end{equation}
so that the canonical momentum associated with $\tilde{\zeta}$ is $\tilde{\Pi}^{(\tilde{\zeta})} = 2 \epsilon_H \ a^3 \dot{\tilde{\zeta}}/(c_s^2 \kappa^2)$. As in the following sections we will proceed with the quantization of cosmological perturbations,\footnote{At this point it is useful to stress that both $\zeta$ and $v$ are dimensionless and thus $\tilde{\zeta}$ and $\tilde{v}$ have the dimensions of a length cube. Correspondingly, $\tilde{\Pi}^{(\tilde{\zeta})}$ and $\tilde{\Pi}^{(\tilde{v})}$ are dimensionless.} it is useful to define the problem in terms of a new coordinate $\tau$ and of a new field $\tilde{v}$, so that the corresponding canonical momentum is $\tilde{\Pi}^{(\tilde{v})} = \tilde{v}^\prime/(\eta \kappa^2)$. This can be realized by defining a new coordinate $\tau$ such that $\textrm{d}/\textrm{d}\tau = a \textrm{d}/\textrm{d}r$ (notice that for cosmology this corresponds to the standard definition of conformal time) and the canonically-normalized Mukhanov variable $v$: 
\begin{equation}
\label{appendix_perturbations:definition_of_y}
  y \equiv \frac{a \sqrt{\eta \ {}^{(0)}(p+\rho)}}{c_s H } =  \frac{a \sqrt{ 2 \epsilon_H}}{c_s}, \qquad \qquad v \equiv y \zeta,
\end{equation}
so that the equation of motion for $\tilde{v}_{\vec{k}}(\tau) \equiv \tilde{v}(\tau ,\vec{k})$, spatial Fourier transform of $v(\tau,\vec{x})$, reads:
\begin{equation}
  \label{appendix_perturbations:scalar_eom_v_1}
\tilde{v}_{\vec{k}}^{\prime \prime} + \left( - \eta c_s^2 k^2  - \frac{y^{\prime \prime}}{y} \right) \tilde{v}_{\vec{k}} = 0 \ ,
\end{equation}
where we have used primes to denote derivatives with respect to $\tau$.\\

\noindent Before proceeding with our treatment, it is interesting to notice that defining the analytically continued variables: 
\begin{equation}
\label{appendix_perturbations:analytical_continuation}
  \bar{k} = - ik \ , \qquad \qquad \bar{\kappa}^2 = -\kappa^2 \ , 
\end{equation} 
a cosmological solution described in terms of $k$ and $\kappa$ can be mapped into a domain-wall solution described in terms of $\bar{k}$ and $\bar{\kappa}$. In the following Sections we present the quantization of the perturbations in the case $\eta = -1$ \textit{i.e.} in the case of cosmology. The quantization for the case of the corresponding domain-wall can be obtained by applying the analytical continuation of Eq.~\eqref{appendix_perturbations:analytical_continuation}. Some details on the interpretation of this analytical continuation are given in Sec.~\ref{sec_holography:AdS_dS_observables}. \\

\noindent
To proceed with our treatment it is useful to notice that Eq.~\eqref{appendix_perturbations:scalar_eom_v_1} is a differential equation that describes the evolution of an harmonic oscillator with time-dependent frequency. This equation has two different regimes:
\begin{itemize}
  \item \textbf{Short wavelength} ($ y^{\prime \prime}/y \ll c_s^2 k^2 $) , the solution of Eq.~\eqref{appendix_perturbations:scalar_eom_v_1} is approximatively given by $\tilde{v}_{\vec{k}} \propto \exp (i k c_s \tau) $.
  \item \textbf{Long wavelength} ($ c_s^2 k^2 \ll y^{\prime \prime}/y $), the solution of Eq.~\eqref{appendix_perturbations:scalar_eom_v_1} is approximatively given by $\tilde{v}_{\vec{k}} \propto y $.
\end{itemize}
At this point we should stress that at the lowest order $y^{\prime \prime}/y \simeq a^{\prime \prime}/a \simeq a^2H^2$. Moreover, while $H$ is nearly constant during inflation, $a$ grows exponentially. This implies that during inflation perturbations are generated at short wavelength, \textit{i.e.} much smaller then the horizon\footnote{As in our treatment we have kept $c_s^2$, the relevant length scale for scalar perturbations is given by the sound horizon. If we fix $c_s^2 = 1$, as in the case of tensor perturbations, the perturbations propagates at the speed of light and this corresponds to the standard horizon.}, they grow until they cross the horizon, entering the long wavelength regime and freezing out. On the contrary, during radiation and matter dominated epochs we have $\ddot{a}<0$. As a consequence, during these phases the modes at long wavelength may re enter the horizon and thus they can be observed.

\section{Equations of motion for tensor perturbations.}
\label{appendix_perturbations:eom_tensor}
So far we have only discussed the case of scalar perturbations, we should then discuss vector and tensor perturbations. Vector perturbations  are not expected to be produced during inflation\footnote{Moreover, it is possible to show that even if they are produced, they would quickly decay because of the expansion of the universe.}, and thus they are not relevant for the scope of this work. On the contrary, tensor perturbations correspond to primordial gravitational waves and thus the treatment of this case is extremely important for our discussion. Let us follow the same procedure carried out for scalar perturbations. At the linear order the Christoffel symbols are:
\begin{equation}
\begin{aligned}
  \label{appendix_perturbations:Christoffel_tensor}
  &\delta \Gamma^0_{\ 00} = \delta \Gamma^0_{\ 0i} = 0 \ , \qquad \qquad \delta \Gamma^0_{\ ij} =  \eta \left(a^2  H \gamma_{ij} +\frac{a^2 \dot{\gamma}_{ij}}{2}\right)\ ,\\
& \delta  \Gamma^i_{\ 0j} = \frac{\delta^{il} \dot{\gamma}_{ lj}}{2} \ , \qquad  \delta \Gamma^i_{\ jk} = \frac{\delta^{il}}{2} \left( \gamma_{l j , k} + \gamma_{l k , j} -  \gamma_{j k , l} \right) \ .
  \end{aligned}
\end{equation}
The $00$ and $ij$ components of the Ricci tensor $R_{\mu\nu} = {}^{(0)} R_{\mu\nu} + \delta R_{\mu\nu}$ are:
\begin{equation}
  \label{appendix_perturbations:Riemann_tensor}
  R_{00} = -3\frac{\ddot{a}}{a} , \qquad R_{ij} = -\eta g_{ij} \left( \frac{\ddot{a}}{a} + 2 H^2 \right)  -\eta a^2 \left[  \frac{3}{2} H \dot{\gamma}_{ ij} + \frac{\ddot{\gamma}_{ij}}{2} \right] - \frac{\Delta \gamma_{ ij}}{2},
\end{equation}
where we have defined $g_{ij} = a^2 (\delta_{ij} + \gamma_{ij})$. As $\gamma_{ij}$ is traceless, it is easy to show that $\delta R$, linear order of the Ricci scalar for tensor perturbations, is vanishing. This directly implies $R = {}^{(0)}R$. As $\delta R$ is equal to zero, we can use Eq.~\eqref{appendix_perturbations:einstein_eq} to get $\delta G^{i}_{\ j} = \delta R^{i}_{\ j}$.
Using Eq.~\eqref{appendix_perturbations:Riemann_tensor} we can then derive the expression for $ R^{i}_{\ j} = g^{ik} R_{kj} = {}^{(0)} R^{i}_{\ j} +  \delta R^{i}_{\ j}$ :
\begin{equation}
 R^{i}_{\ j} = -\eta \left[ \left( \frac{\ddot{a}}{a} + 2 H^2 \right) g^{ik} g_{kj} + \frac{3}{2} a^2 H g^{ik} \dot{\gamma}_{kj} + g^{ik} \frac{\ddot{\gamma}_{kj}}{2} a^2  \right] - g^{ik} \frac{\Delta \gamma_{ ij}}{2}.
\end{equation}
As the term proportional to $g^{ik} g_{kj} = \delta^{i}_{ \ j} $ is a zero order term, we can directly get:
\begin{equation}
  \delta G^{i}_{\ j} =  - \eta \delta^{ik} \frac{\ddot{\gamma}_{ij}}{2} - \eta \frac{3 H}{2} \delta^{ik} \dot{\gamma}_{ij} - \delta^{ik} \frac{ \Delta \gamma_{ ij}}{2 a^2} 
\end{equation}
As the right hand side of Eq.~\eqref{appendix_perturbations:einstein_eq} has to be set equal to zero, the evolution of tensor perturbations is thus described by:
\begin{equation}
\label{appendix_perturbations:tensor_eom_gamma}
  \ddot{\tilde{\gamma}}_{ij} + 3 H \dot{\tilde{\gamma}}_{ij}  - \frac{ \eta k^2}{ a^2}  \tilde{\gamma}_{ ij}= 0
\end{equation}
where $\tilde{\gamma}(z,\vec{k})$ is the spatial Fourier transform of $\gamma(r,\vec{x})$. As $\gamma_{ij} $ is transverse traceless it only contains two independent physical degrees of freedom $\tilde{h}_+(z,\vec{k})$ and $\tilde{h}_{\times}(z,\vec{k})$ (that correspond to the two polarizations of the GW). Eq.~\eqref{appendix_perturbations:tensor_eom_gamma} can thus be expressed as:
\begin{equation}
\label{appendix_perturbations:tensor_eom_h}
  \ddot{\tilde{h}}_{\alpha} + 3 H \dot{\tilde{h}}_{\alpha}  - \frac{\eta k^2}{a^2}  \tilde{h}_{\alpha}= 0,
\end{equation}
where $\alpha = + , \times$. Following the procedure discussed in Sec.~\ref{appendix_perturbations:eom_scalar}, we can then define the coordinate $\tau$ and the canonically-normalized fields $v_\alpha$ as:
\begin{equation}
\frac{ \textrm{d}}{\textrm{d}\tau} \equiv a \frac{\textrm{d}}{\textrm{d}z},  \qquad \qquad v_\alpha \equiv \frac{a h_\alpha}{2},
\end{equation}
and again we proceed by restricting to the case of cosmological perturbations \textit{i.e.} we fix $\eta = -1$, to get: 
\begin{equation}
  \label{appendix_perturbations:scalar_eom_v_a}
\tilde{v}_{\alpha,\vec{k}}^{ \ \prime \prime} + \left( k^2 - \frac{a^{\prime \prime}}{a}  \right) \tilde{v}_{\alpha,\vec{k}} = 0 \ .
\end{equation}
The interpretation of this equation is analogous to the one of Eq.~\eqref{appendix_perturbations:scalar_eom_v_1}, given at the end of Sec.~\ref{appendix_perturbations:eom_scalar}. However we should stress that in this case we have two modes at fixed $k^2$, corresponding to the two polarizations of the GW, and we have no $c_s^2$ as the GW propagate at the speed of light.

\section{Quantization of the perturbations.}
\label{appendix_perturbations:quantization} 
In the final part of Sec.~\ref{appendix_perturbations:eom_scalar} and of Sec.~\ref{appendix_perturbations:eom_tensor} we have reparameterized the scalar and tensor perturbations in order to get differential equations for a time-dependent harmonic oscillators. The reason for this choice is that in this particular case we can define a consistent procedure to quantize inspired by the flat spacetime case. Let us start with a brief review of the standard procedure in order to extend it to the case of our interest.

\subsection{Scalar field theory in flat spacetime.}
\label{appendix_perturbations:quantization_flat} 
Let us consider a scalar field $\Phi(t,\vec{x})$ in a flat $4$-dimensional spacetime with potential $V(\Phi) = \frac{m^2}{2}$. The hamiltonian for this scalar field simply reads:
\begin{equation}
\label{appendix_perturbations:scalar_field_hamiltonian}
  H = \int \textrm{d}^3 \vec{x} \left[ \frac{\Pi^2}{2} - \frac{(\vec{\nabla} \Phi)^2}{2} + \frac{m^2 \Phi^2}{2} \right] \ ,
\end{equation}
where $\vec{\nabla}$ is the gradient and where $\Pi(t,\vec{x}) \equiv \delta \mathcal{L}/\delta (\partial_t \Phi)$ is the canonical momentum associated with the field $\Phi(t,\vec{x})$. The promotion of this field to an operator is realized by imposing the usual canonical commutation relations\footnote{In this part we work in the Schr\"{o}dinger picture and thus all the operators are time independent. The time dependence will be restored in the next paragraph when we will furnish the expression for the field in terms of the annihilation and creation operators.}:
\begin{equation}
\begin{aligned}
   \left[ \hat{\Phi}(t,\vec{x}), \hat{\Phi}(t,\vec{x}^\prime) \right] &=  0 \ , \\
  \left[ \hat{\Pi}(t,\vec{x}),\hat{\Pi}(t,\vec{x}^\prime) \right] = 0 \ , \\
     \left[ \hat{\Phi}(t,\vec{x}),\hat{\Pi}(t,\vec{x}^\prime) \right] & = i \delta^{3} (\vec{x} - \vec{x}^\prime) \ .
   \end{aligned}
 \end{equation} 
To proceed with our treatment it is useful to describe the system in terms of its spatial Fourier transform. The hamiltonian can be expressed in terms of $\tilde{\Pi}(t,\vec{k})$ and $\tilde{\Phi}(t,\vec{k})$ spatial Fourier transform of $\Pi(t,\vec{x})$ ,$\Phi(t,\vec{x})$ as:
\begin{equation}
 \tilde{H} = \int \textrm{d}^3 \vec{k} \left[ \frac{\tilde{\Pi}(t,\vec{k})\tilde{\Pi}(t,-\vec{k})}{2} + (k^2 + m^2) \frac{\tilde{\Phi}(t,\vec{k})\tilde{\Phi}(t,-\vec{k})}{2} \right] \  .
\end{equation}
Promoting the fields to operators the hamiltonian reads:
\begin{equation}
  \hat{\tilde{H} }= \frac{1}{2} \int \textrm{d}^3 \vec{k} \left[ \frac{\hat{\tilde{\Pi}}_{\vec{k}} \hat{\tilde{\Pi}}_{\vec{k}}^\dagger + \hat{\tilde{\Pi}}_{\vec{k}}^\dagger \hat{\tilde{\Pi}}_{\vec{k}}}{2} + (k^2 + m^2) \frac{ \hat{\tilde{\Phi}}_{\vec{k}} \hat{\tilde{\Phi}}_{\vec{k}}^\dagger + \hat{\tilde{\Phi}}_{\vec{k}}^\dagger \hat{\tilde{\Phi}}_{\vec{k}} }{2} \right] \ ,
\end{equation}
where $\hat{\tilde{\Pi}}_{\vec{k}} \equiv \hat{\tilde{\Pi}} (t,\vec{k})$ and $\tilde{\Phi}_{\vec{k}} \equiv \hat{\tilde{\Phi}} (t,\vec{k})$ and where we have imposed the reality of $\tilde{\Phi}_{\vec{k}}$ and $\tilde{\Pi}_{\vec{k}}$ to get $\tilde{\Pi}_{-\vec{k}} = \tilde{\Pi}_{\vec{k}}^\dagger $ and $\tilde{\Phi}_{-\vec{k}} = \tilde{\Phi}_{\vec{k}}^\dagger $. Notice that the canonical commutation relations in terms of $\tilde{\Pi}(t,\vec{k})$ and $\tilde{\Phi}(t,\vec{k})$ read: 
\begin{equation}
   [\hat{\tilde{\Phi}}_{\vec{k}} , \hat{\tilde{\Phi}}_{\vec{k}^\prime} ] =  0 \ , \qquad [\hat{\tilde{\Pi}}_{\vec{k}},\hat{\tilde{\Pi}}_{\vec{k}^\prime}] = 0 , \qquad [\hat{\tilde{\Phi}}_{\vec{k}} ,\hat{\tilde{\Pi}}_{\vec{k}^\prime}] = i \delta^{3} (\vec{k} + \vec{k}^\prime) \ .
 \end{equation} 
As usual the field and the canonical momentum can be expressed in terms of a set of annihilation and creation operators:
\begin{equation}
  \hat{\tilde{\Phi}}_{\vec{k}} = \frac{1}{\sqrt{2 \omega_{\vec{k}} }}  \left ( \hat{a}_{\vec{k}} + \hat{a}_{-\vec{k}}^\dagger \right) \ , \qquad  \hat{\tilde{\Pi}}_{\vec{k}} = - i \sqrt{ \frac{\omega_{\vec{k}}}{2 }} \left ( \hat{a}_{\vec{k}} - \hat{a}_{-\vec{k}}^\dagger \right) \,
\end{equation}
where $\omega_{\vec{k}} \equiv \sqrt{\vec{k}^{ \ 2} + m^2}$ is the frequency of the mode with wavevector $\vec{k}$. Notice that the operators $\hat{a}_{\vec{k}}$, $\hat{a}_{\vec{k}}^\dagger$ satisfy the commutation relations\footnote{Notice that the commutation relations imply that both $\hat{a}_{\vec{k}}$ and $\hat{a}_{\vec{k}}^\dagger$ have the dimension of a length to the $3/2$.}:
\begin{equation}
\label{appendix_perturbations:commutation_relations_annhi_crea}
   [ \ \hat{a}_{\vec{k}_1}\ ,\hat{a}_{\vec{k}_2} \ ] =  [ \ \hat{a}_{\vec{k}_1}^\dagger \ ,\hat{a}_{\vec{k}_2}^\dagger \ ] = 0 , \qquad [ \ \hat{a}_{\vec{k}_1},\hat{a}_{\vec{k}_2}^\dagger \ ] = \delta^{3} (\vec{k}_1 - \vec{k}_2) \ ,
 \end{equation}
In terms of these operators the hamiltonian reads:
\begin{equation}
  \hat{\tilde{H}} = \frac{1}{2}  \int \textrm{d}^3 \vec{k}  \ \omega_{\vec{k}} \left(   \ \hat{a}_{\vec{k}}^\dagger \ \hat{a}_{\vec{k}} + \hat{a}_{\vec{k}} \ \hat{a}_{\vec{k}}^\dagger  \right) =   \frac{\delta^{3}(0)}{2} \int \textrm{d}^3 \vec{k}  \ \omega_{\vec{k}} + \ \int \textrm{d}^3 \vec{k}  \ \omega_{\vec{k}} \  \hat{a}_{\vec{k}}^\dagger \ \hat{a}_{\vec{k}} .
\end{equation}
The first term corresponds to the infinite vacuum energy that should thus be subtracted in order define the energy levels of the system. Once the vacuum state $|0 \rangle $ that satisfies $\hat{a}_{\vec{k}} | 0 \rangle = 0$ for all $\vec{k}$ (and also $\langle 0 | \hat{a}_{\vec{k}}^\dagger = 0 $ for all $\vec{k}$) is defined, we are directly lead to the construction of the Fock space of states.  \\

\noindent
Finally we can switch to the Heisenberg picture with time-dependent operators. It is crucial to stress that even in this picture the annihilation and creation operators $\hat{a}_{\vec{k}}$, $\hat{a}_{\vec{k}}^\dagger$ are \emph{time independent}. Once we restore the time dependence, the expression for the field $\hat{\Phi}(x,t)$ reads:
\begin{equation}
\label{appendix_perturbations:scalar_field_operator}
  \hat{\Phi}(t,x) = \int  \frac{\textrm{d}^3 \vec{k} }{(2 \pi)^{3/2}} \left( e^{i \vec{k} \vec{x}} \ f(\omega_{\vec{k}}) \ \hat{a}_{\vec{k}}   +  e^{-i \vec{k} \vec{x}} f^*(\omega_{\vec{k}}) \ \hat{a}_{\vec{k}}^\dagger  \right ) \ ,
\end{equation}
where we have defined\footnote{Notice that for this choice the Wronskian condition:
\begin{equation}
\label{appendix_perturbations:wronskian_condition}
  f(\omega_{\vec{k}}) \frac{\textrm{d} f^*(\omega_{\vec{k}})}{\textrm{d} t} -  \frac{\textrm{d} f(\omega_{\vec{k}})}{\textrm{d} t} f^*(\omega_{\vec{k}}) = i \ ,
\end{equation} 
imposed by the canonical commutation relations, is automatically satisfied.}
\begin{equation}
\label{appendix_perturbations:scalar_field_modes}
   f(\omega_{\vec{k}}) = \frac{1 }{\sqrt{2\omega_{\vec{k}}} } \exp\{ -i \omega_{\vec{k}} \ t \}  \ .
 \end{equation}
It is also crucial to stress that $f(\omega_{\vec{k}})$ is the configuration that minimizes the energy of a mode at fixed $\vec{k}$. As the vacuum state is defined as the state with minimal energy, it is reasonable to choose the modes that keep the vacuum energy minimized. For this reason, the modes of Eq.~\eqref{appendix_perturbations:scalar_field_modes} can be used to define a proper description for fluctuations over Minkowski spacetime. Notice also that Minkowski spacetime is invariant under time translations and the hamiltonian for the system is time independent. This implies that the vacuum state is well defined at all times and the construction of the Fock space is always consistent. Of course this condition is no longer satisfied when we consider a time dependent background.

\subsection{Scalar field theory in curved spacetime.}
\label{appendix_perturbations:curved_quantization}
In a curved spacetime the notion of a vacuum state that is invariant under time translation is lost and thus a different procedure should be implemented. The evolution of the background implies that the frequencies are time-dependent and in order to define a vacuum state we need to fix a given time $t_0$. Once $t_0$ is fixed, the whole set of states at that time is defined and consequently it does exist a unique vacuum state that satisfies $\hat{a}_{\vec{k}} | 0 \rangle_{t_0} = 0$ for all $\vec{k}$. Let us define $ | 0 \rangle_{t_1} = 0$, vacuum state at different time $t_1$, let $\hat{b}_{\vec{k}}$ and $\hat{b}_{\vec{k}}$ be the annihilation and creation operators and let us assume for simplicity that $t_0 < t_1$. The crucial point is that in a curved spacetime, $| 0 \rangle_{t_1}$ in general does not correspond to the state obtained by performing the time evolution of the state $| 0 \rangle_{t_0}$. As a consequence, even if we prepare the system in its vacuum state at a given time $t_0$, particles will be generated by the evolution of the system towards a later time $t_1$. Let us show this mechanism in detail.\\

\noindent 
As discussed in the previous sections, Eq.~\eqref{appendix_perturbations:scalar_eom_v_1} and Eq.~\eqref{appendix_perturbations:scalar_eom_v_a} describe time-dependent harmonic oscillators. From the definition of $\tau $ it is possible to show that during inflation we have $a(\tau) \simeq - 1/(\tau H )$. With this approximation Eq.~\eqref{appendix_perturbations:scalar_eom_v_1} reads:
\begin{equation}
  \label{appendix_perturbations:scalar_eom_v_approx}
\tilde{v}_{\vec{k}}^{\prime \prime} + \left( c_s^2 k^2  - \frac{2}{\tau^2} \right) \tilde{v}_{\vec{k}} = 0 \ .
\end{equation}
Notice that the solutions $\tilde{v}_{\vec{k}} \ $ of Eq.~\eqref{appendix_perturbations:scalar_eom_v_approx} can be expressed as a linear superposition of the two functions:
\begin{equation}
\label{appendix_perturbations:solution_eom}
  \tilde{v}^+_{\vec{k}} = \frac{\kappa}{\sqrt{2 k c_s}} \left( 1 + \frac{i}{k c_s \tau} \right) \exp \left\{ i k c_s \tau \right\} \ , \qquad \tilde{v}^-_{\vec{k}} = \frac{\kappa}{\sqrt{2 k c_s}} \left( 1 - \frac{i}{k c_s \tau} \right) \exp \left\{ -i k c_s \tau \right\} \ .
\end{equation}
In analogy with the case of Minkowski spacetime, at a fixed time $\tau_0$ we can define a vacuum state $|0 \rangle_{\tau_0}$ and the corresponding annihilation and creation operators $\hat{a}_{\vec{k}} $ and $\hat{a}_{\vec{k}} ^\dagger \ $. Generalizing Eq.~\eqref{appendix_perturbations:scalar_field_operator}, we can thus express the field $\hat{v}(\tau,\vec{x}) $ as:
\begin{equation}
\label{appendix_perturbations:curved_field_operator}
  \hat{v}(\tau,\vec{x}) = \int  \frac{\textrm{d}^3 \vec{k} }{(2 \pi)^{3/2}} \left( e^{i \vec{k} \vec{x}} \ \tilde{v}_{\vec{k},\tau_0}(\tau) \ \hat{a}_{\vec{k}}   +  e^{-i \vec{k} \vec{x}} \ \tilde{v}^*_{\vec{k},\tau_0}(\tau) \ \hat{a}_{\vec{k}}^\dagger \  \right ) \ ,
\end{equation}
Notice that the operators $\hat{a}_{\vec{k}} $ and $\hat{a}_{\vec{k}} ^\dagger \ $ \footnote{Notice that to satisfy the analogous of the commutation relations of Eq.~\eqref{appendix_perturbations:commutation_relations_annhi_crea}, this operators must have the dimension of a length to the $3/2$. To be consistent with dimensional analysis the function $ \tilde{v}_{\vec{k},\tau_0}(\tau)$ must have the dimension of a length to the $3/2$ too.} respectively annihilate and create modes at a frequency $\omega_{\vec{k},\tau_0}$ that has been fixed at time $\tau_0$. At a different time $\tau_1 > \tau_0$ we have a different vacuum state $|0 \rangle_{\tau_1}$ and consequently we have a different set of annihilation and creation operators $\hat{b}_{\vec{k}} $ and $\hat{b}_{\vec{k}} ^\dagger $. If we express the field $\hat{v}(\tau,\vec{x}) $ in terms of these quantities we get:
\begin{equation}
\label{appendix_perturbations:curved_field_operator_2}
  \hat{v}(\tau,\vec{x}) = \int  \frac{\textrm{d}^3 \vec{k} }{(2 \pi)^{3/2}} \left( e^{i \vec{k} \vec{x}} \ \tilde{v}_{\vec{k},\tau_1}(\tau) \ \hat{b}_{\vec{k}}   +  e^{-i \vec{k} \vec{x}} \ \tilde{v}^*_{\vec{k},\tau_1}(\tau) \ \hat{b}_{\vec{k}}^\dagger  \right ) \ .
\end{equation}
As Eq.~\eqref{appendix_perturbations:scalar_eom_v_approx} is linear, it is possible to express $\tilde{v}_{\vec{k},\tau_1}(\tau)$ and $\tilde{v}^*_{\vec{k},\tau_1}(\tau)$ as a linear combination of $\tilde{v}_{\vec{k},\tau_0}(\tau)$ and $\tilde{v}^*_{\vec{k},\tau_0}(\tau)$ in terms of some time-dependent coefficients $\alpha_{\vec{k}} $ and $\beta_{\vec{k}} $ :
\begin{equation}
\label{appendix_perturbations:redefine_coefficients}
  \tilde{v}_{\vec{k},\tau_1}(\tau) = \alpha_{\vec{k}} \  \tilde{v}_{\vec{k},\tau_0}(\tau) + \beta_{\vec{k}}  \  \tilde{v}^*_{\vec{k},\tau_0}(\tau) \ ,  \qquad \tilde{v}^*_{\vec{k},\tau_1}(\tau) = \beta_{\vec{k}}^*   \  \tilde{v}_{\vec{k},\tau_0}(\tau) + \alpha_{\vec{k}}^*  \  \tilde{v}_{\vec{k},\tau_0}^* (\tau) \ .
\end{equation}
As the expansion of $\hat{v}(\tau,\vec{x})$ given in Eq.~\eqref{appendix_perturbations:curved_field_operator} should match with the expansion of Eq.~\eqref{appendix_perturbations:curved_field_operator_2}, we can express $\hat{b}_{\vec{k}} $ and $\hat{b}_{\vec{k}} ^\dagger $ as:
\begin{equation}
\hat{b}_{\vec{k}} = \alpha_{\vec{k}}^*  \  \hat{a}_{\vec{k}} - \beta_{\vec{k}}^*  \  \hat{a}_{\vec{k}}^\dagger \ , \qquad \qquad \hat{b}_{\vec{k}}^\dagger = - \beta_{\vec{k}} \  \hat{a}_{\vec{k}} + \alpha_{\vec{k}}  \ \hat{a}_{\vec{k}}^\dagger \ .
\end{equation}
If we start at time $\tau_0$ with the system in the vacuum state $|0 \rangle_{\tau_0}$ and we compute the number of particles $N \equiv \int  \textrm{d}^3 \vec{k} \  \hat{b}_{\vec{k}}^\dagger \ \hat{b}_{\vec{k}} $ at a time $\tau_1 > \tau_0$ we thus get:
\begin{equation}
\label{appendix_perturbations:number_of_particles}
  {}_{\tau_0}\langle0 | \int  \textrm{d}^3 \vec{k} \  \hat{b}_{\vec{k}}^\dagger \ \hat{b}_{\vec{k}} \ |0 \rangle_{\tau_0} = |\beta_{\vec{k}}|^2 \ ,
\end{equation}
implying that the number of particles at a time $\tau_1$ is non vanishing. In the context of cosmology this implies that given the coupling of the inflaton with gravity, QM produces fluctuations from the vacuum. \\

\noindent
Notice that in the short wavelength regime ($a^2 H^2 \simeq 1/\tau^2 \ll k^{ 2} c_s^2$) the solutions of Eq.~\eqref{appendix_perturbations:scalar_eom_v_approx} are plane waves. As in this regime the system matches the case of a scalar field in a flat spacetime, initial condition for our system can naturally imposed. Notice that the short wavelength regime is regime is reached for $1 \ll k^{ 2} c_s^2 \tau^2$ \textit{i.e.} $ \tau \rightarrow - \infty$. As in the case of flat spacetime the modes are defined by Eq.~\eqref{appendix_perturbations:scalar_field_modes}, in the case of curved spacetime we require $\tilde{v}_{\vec{k}} \ (\tau) $ to satisfy the initial condition:
\begin{equation}
\label{appendix_perturbations:initial_condition_ds}
 \lim_{\tau \rightarrow -\infty } \tilde{v}_{\vec{k}} \ (\tau) = \kappa f(\omega_{\vec{k}}) = \frac{\kappa }{\sqrt{2\omega_{\vec{k}}} } \exp\{ -i \omega_{\vec{k}} \tau \} \ .
\end{equation}
Notice that we have introduced the factor $\kappa$ in order to be consistent with dimensional analysis. Imposing the condition of Eq.~\eqref{appendix_perturbations:initial_condition_ds}, corresponds to fixing the vacuum state for our theory in the infinite past. This choice for the vacuum state of a quantum field theory in curved spacetime is known as the Bunch-Davies vacuum. As $ \tilde{v}_{\vec{k}} \ $ can be expressed as a linear superposition of the functions $ \tilde{v}_{\vec{k}}^+$ and $ \tilde{v}_{\vec{k}}^-$ defined in Eq.~\eqref{appendix_perturbations:solution_eom}, it is trivial to prove that:
\begin{equation}
\label{appendix_perturbations:BD_modes}
  \tilde{v}_{\vec{k}} \ (\tau) = \frac{\kappa}{\sqrt{2 k c_s}} \left( 1 - \frac{i}{k c_s \tau} \right) \exp \left\{ -i k c_s \tau \right\} .
\end{equation}
In terms of this quantity we will characterize the correlations function that define the observables of our theory. 

\section{Observable quantities.}
\label{appendix_perturbations:observables} 
As usual for QFT, a set of physically observable quantities is given by the correlators. As widely discussed in literature, the gauge invariant quantity that should be used to define observable quantities is the scalar field $\zeta(\tau,x)$ defined in Eq.~\eqref{appendix_perturbations:xi_zeta_definition}. In terms of this quantity we can thus compute the correlators for cosmological perturbations. It is useful to remind that $\zeta(\tau,x)$ can be expressed in terms of $v(\tau,x)$ by using the definition of Eq.~\eqref{appendix_perturbations:scalar_eom_v_1}. Following the discussion of Sec.~\ref{appendix_perturbations:curved_quantization}, we can define the quantization of our theory and fix the vacuum for our theory to be the Bunch-Davies vacuum $| BD \rangle \equiv |0 \rangle$. 
This choice is required to fix well defined initial conditions for our system. At this point we can settle the expansion of $ \hat{v}(\tau,\vec{x})$ in terms of annihilation and creation operators by using Eq.~\eqref{appendix_perturbations:curved_field_operator} and compute the correlators of our theory. \\

\noindent
From the expansion of $ \hat{v}(\tau,\vec{x})$ defined in Eq.~\eqref{appendix_perturbations:curved_field_operator}, it should be clear that the one-point function $\langle 0| \hat{\zeta}(\tau,x) |0\rangle$ is equal to zero. On the contrary, the two-point function is defined as:
\begin{equation}
\begin{aligned}
\label{appendix_perturbations:two_point_scalar}
   \langle 0| \hat{\zeta}(\tau,\vec{x}_1) \hat{\zeta}(\tau,\vec{x}_2) |0 \rangle & = \frac{1}{y^2}  \langle 0| \  \int \frac{\textrm{d}^3 \vec{k}_1 \textrm{d}^3 \vec{k}_2 }{(2 \pi)^3} \ \tilde{v}_{\vec{k}_1} \tilde{v}_{\vec{k}_2}^*  \hat{a}_{\vec{k}_1}  \hat{a}_{\vec{k}_2}^\dagger e^{-i (\vec{k}_1\vec{x}_1 - \vec{k}_2\vec{x}_2)}  \ |0 \rangle \\
    & = \frac{1}{y^2} \  \int \frac{\textrm{d}^3 \vec{k} }{(2 \pi)^3} \ \tilde{v}_{\vec{k}} \  \tilde{v}_{\vec{k}}^* e^{-i \vec{k} (\vec{x}_1 - \vec{x}_2)}  \  .
   \end{aligned} 
 \end{equation} 
As this the quantities in the integral depend only on $k = |\vec{k}|$, we can express $\textrm{d}^3 \vec{k} = k^2\textrm{d}k \textrm{d}\Omega$ where $\textrm{d}\Omega$ denotes the solid angle ($\int \textrm{d}\Omega = 4 \pi$). We can thus express Eq.~\eqref{appendix_perturbations:two_point_scalar} as:
\begin{equation}
\label{appendix_perturbations:two_point_scalar_final}
 \langle 0| \hat{\zeta}(\tau,\vec{x}_1) \hat{\zeta}(\tau,\vec{x}_2) |0 \rangle =  \int \frac{\textrm{d}k }{2 \pi^2} \ k^2 \  \tilde{\zeta}_{\vec{k}} \  \tilde{\zeta}_{\vec{k}}^* e^{-i \vec{k} (\vec{x}_1 - \vec{x}_2)}  \ .
 \end{equation}
As we have non-zero fluctuations only for $\vec{k}_1 = \vec{k}_2$ it is useful to define the scalar power spectrum:
\begin{equation}
  \label{appendix_perturbations:scalar_power_spectrum_definition}
   \langle 0 | \hat{\tilde{\zeta}}(\tau,\vec{k}_1) \hat{\tilde{\zeta}}(\tau,\vec{k}_2) | 0 \rangle \equiv (2 \pi)^3 \delta^{(3)}(\vec{k}_1 + \vec{k}_2) \mathcal{P}_s(k_1,\tau) = (2 \pi)^3 \delta^{(3)}(\vec{k}_1 + \vec{k}_2) \frac{ \Delta_s^2(k_1,\tau)}{4 \pi k_1^3} \ ,
\end{equation}
where the normalization of the dimensionless power spectrum $\Delta_s^2 $, was chosen in order to have:
\begin{equation}
  \label{appendix_perturbations:scalar_power_spectrum_normalization}
  \langle \zeta(\tau,\vec{x}_1) \zeta(\tau,\vec{x}_2) \rangle = \int \frac{\textrm{d} k }{k} \Delta^2_s (k,\tau) \  e^{-i \vec{k}_1\cdot(\vec{x}_1 - \vec{x}_2)}.
\end{equation}
Comparing Eq.~\eqref{appendix_perturbations:scalar_power_spectrum_normalization} with Eq.~\eqref{appendix_perturbations:two_point_scalar} we can finally express $\Delta^2_s (k,\tau)$ as:
\begin{equation}
\label{appendix_perturbations:scalar_power_spectrum}
  \Delta^2_s (k,\tau) = \frac{k^3}{2 \pi^2} |\tilde{\zeta}_{\vec{k}}|^2 = \frac{k^3}{2 \pi^2} \frac{ |\tilde{v}_{\vec{k}}|^2 }{y^2} =  \frac{\kappa^2}{8 \pi^2 a^2 \epsilon_H c_s \tau^2 } \left( 1 + k^2 c_s^2 \tau^2 \right)  \ ,
\end{equation}
where we have substituted Eq.~\eqref{appendix_perturbations:BD_modes} and we have used the definition of $y$ given in Eq.~\eqref{appendix_perturbations:definition_of_y}. As already discussed in the previous sections, during inflation quantum fluctuations are generated at small scales where $ 1 \ll k c_s \tau $, they grow until they cross the horizon and they freeze on superhorizon scales where $k c_s \tau \ll 1$. These fluctuations are then observed at later times when they re-enter the horizon. As a consequence, to compute the power spectrum for modes that are presently observable at CMB scales, we should first evaluate Eq.~\eqref{appendix_perturbations:scalar_power_spectrum} at $ k c_s \tau \ll 1 $, and then we should set $\tau = (k c_s)^{-1}$ so that we get:
\begin{equation}
\label{appendix_perturbations:scalar_power_spectrum_final}
 \left.   \Delta^2_s (k,\tau)  \right|_{\tau = (k c_s)^{-1}} =  \left.  \frac{\kappa^2}{8 \pi^2 a^2 \epsilon_H c_s \tau^2 }  \right|_{\tau = (k c_s)^{-1}} = \frac{ \kappa^2 H^2 }{8 \pi^2 c_s \epsilon_H} \ ,
\end{equation}
where we have used $(k c_s)^{-1} = \tau \simeq - 1/(a H )$. \\

\noindent
The quantization of tensor perturbations can be realized by following an analogous of the procedure carried out in Sec.~\ref{appendix_perturbations:curved_quantization} for the case of scalar perturbations. In this case we get:
\begin{equation}
  \label{appendix_perturbations:tensor_vacuum}
  \langle 0| \hat{h}_\alpha(\tau,x_1) \hat{h}_\alpha(\tau,x_2) |0 \rangle = \  \int \frac{\textrm{d}^3 \vec{k} }{(2 \pi)^3} \ |\tilde{h}_{\alpha, \vec{k}}|^2 e^{-i \vec{k} (\vec{x}_1 - \vec{x}_2)} \ ,
\end{equation}
where, in order to compute the vacuum expectation value, we express the operators  $\hat{h}_\alpha(\tau,x_1)$ associated with the two polarizations of the GW, in terms of the canonically normalized operators $\hat{\tilde{v}}_{\alpha, \vec{k}} \equiv a \hat{\tilde{h}}_{\alpha, \vec{k}} / 2$. Notice that in this case, we should sum over the two polarizations of the GW. In analogy with the definition of Eq.~\eqref{appendix_perturbations:scalar_power_spectrum_definition}, we can introduce the tensor power spectrum as:
\begin{equation}
  \label{appendix_perturbations:tensor_power_spectrum_definition}
    \langle \hat{h}_\alpha(\tau,\vec{k}_1) \hat{h}_\alpha(\tau,\vec{k}_2) \rangle  \equiv (2 \pi)^3 \delta^{(3)}(\vec{k}_1 + \vec{k}_2) \mathcal{P}_t(k_1,\tau) = (2 \pi)^3 \delta^{(3)}(\vec{k}_1 + \vec{k}_2) \frac{ \Delta_t^2(k_1,\tau)}{4 \pi k_1^3} \ .
\end{equation}
Notice that $\Delta_t^2$ is defined as the sum over the two polarizations. The normalization of the (dimensionless) tensor power spectrum $\Delta_t^2(k)$, is chosen to respects the analogous of Eq.~\eqref{appendix_perturbations:scalar_power_spectrum_normalization}. With this normalization, the tensor power spectrum reads:
\begin{equation}
\label{appendix_perturbations:tensor_power_spectrum}
\Delta^2_t (k,\tau) \equiv  \frac{2 k^3}{ \pi^2} \frac{ |\tilde{v}_{\alpha, \vec{k}}|^2 }{a^2}  =   \frac{ 2 k^3}{ \pi^2} \frac{1}{a^2} \frac{\kappa^2}{ k } \left( 1 + \frac{1}{k^2 \tau^2} \right) = \frac{2 \kappa^2 H^2}{\pi^2} \left( 1 + k^2 \tau^2 \right) \ ,
\end{equation}
where we have substituted Eq.~\eqref{appendix_perturbations:BD_modes} with $c_s^2 = 1$, we have summed over the two polarizations of the GW and in the last step we have also used $\tau \simeq - 1/(a H )$. Finally, for scales that can be probed by CMB experiments ($ k \tau \ll 1 $) we have:
\begin{equation}
\label{appendix_perturbations:tensor_power_spectrum_final}
\left.   \Delta^2_t (k,\tau) \right|_{\tau = k^{-1}} = \frac{2 \kappa^2 H^2 }{\pi^2  }  \ ,
\end{equation}
where as usual we have evaluated the spectrum at horizon crossing ($\tau k  =1$).\\

\noindent
Before concluding this Appendix, it is worth spending some words on the possibility of generating non-Gaussianities\footnote{For a comprehensive review of the topic see for example~\cite{Chen:2010xka}.}. Clearly, our starting point is again given by action of Eq.~\eqref{appendix_perturbations:action}. The main difference with respect to the case discussed in this Appendix, is that the perturbative expansion should not be performed up to the linear (\emph{second} order for the \emph{bispectum} defined in Sec.~\ref{sec_inflation:generalized_models}) in the perturbations. After this expansion is performed, the system should again be described in terms of the gauge invariant quantities $\zeta$ and $\gamma_{ij}$. At this point, we can then compute for example the three-point function:
\begin{equation}
  \langle \zeta(\tau,\vec{x}_1) \zeta(\tau,\vec{x}_2)\zeta (\tau,\vec{x}_3) \rangle \ ,
\end{equation}
that in general is expected to be non-zero.

 {\large \par}}
{\large \chapter{Conformal Field Theories.} \label{appendix_CFT:CFT}
 \noindent In this appendix we present a general introduction to Conformal Field Theories (CFTs). We start our treatment by discussing conformal transformations and deriving the generators of the conformal group. Studying the algebra of the conformal group we can finally define a CTF. 

 \section{Conformal transformations and conformal group.}
 \label{appendix_CFT:Conf_transf_and_group}
\noindent A conformal transformation is defined as a change of coordinates $x^\mu \rightarrow x^{\prime \ \mu(x)}$ such that the metric changes accordingly with:
\begin{equation}
\label{appendix_CFT:Conformal_transformation}
   g^\prime_{\mu \nu} (x^\prime) = \Omega^{2}(x) g_{\mu\nu}(x),
\end{equation}
where $\Omega(x)$ is a generic function of the spacetime coordinates. It should be clear that this particular set of transformations preserves angles between vectors but it does not preserve distances. In order to construct a QFT which is invariant under conformal transformation, we should study the structure of the conformal group and find Casimir operators to label the states of the theory. For this purpose let us restrict to the case of a flat $d$-dimensional space-time with metric $g_{\mu\nu}(x) = \eta_{\mu \nu}$ with a given signature $(p,q)$.\footnote{Notice that this choice for the signature implies $\mu,\nu = -p+1, -p+2, \dots, q-1 , q$.} We can then consider the infinitesimal transformation:
\begin{equation}
  \label{appendix_CFT:infinitesimal_conformal}
   \begin{cases} & x^{\prime \ \mu}(x) = x^\mu + \epsilon^\mu(x), \\
   & \Omega(x) = 1 + \omega(x)/2.
   \end{cases}
\end{equation}
We can then write the usual metric transformation and impose Eq.~\eqref{appendix_CFT:Conformal_transformation} to get:
\begin{equation}
\label{appendix_CFT:infinitesimal_metric}
\Omega^{2}(x) g_{\mu\nu}(x) = g^\prime_{\mu \nu} (x^\prime) = g_{\rho \sigma}(x) \frac{\partial x^{\prime \rho}}{\partial x^\mu} \frac{\partial x^{\prime \sigma}}{\partial x^\nu}. 
\end{equation}
Substituting Eq.~\eqref{appendix_CFT:infinitesimal_conformal} into Eq.~\eqref{appendix_CFT:infinitesimal_metric} we can then obtain the first order equation:
\begin{eqnarray}
  \label{appendix_CFT:infinitesimal_generators}
   \partial_\mu \epsilon_\nu + \partial_\nu \epsilon_\mu &=& \eta_{\mu \nu} \ \omega(x), \\
   2 \partial^\mu \epsilon_\mu &=& d \ \omega(x) ,
   \label{appendix_CFT:infinitesimal_generators_trace}
   \end{eqnarray}
  where Eq.~\eqref{appendix_CFT:infinitesimal_generators_trace} is obtained by taking the trace of Eq.~\eqref{appendix_CFT:infinitesimal_generators}.
Substituting Eq.~\eqref{appendix_CFT:infinitesimal_generators_trace} into Eq.~\eqref{appendix_CFT:infinitesimal_generators} we get: 
 \begin{equation}  
   \label{appendix_CFT:infinitesimal_generators_2}
  \partial_\mu \epsilon_\nu + \partial_\nu \epsilon_\mu = \frac{2 }{d} \eta_{\mu \nu} \ \partial^\rho \epsilon_\rho(x).
\end{equation}
As widely discussed in literature, the 2-dimensional case is special as Eq.~\eqref{appendix_CFT:infinitesimal_generators_2} has an infinite number of solutions\footnote{In this case, by performing a Wick rotation over the time-like coordinate, Eq.~\eqref{appendix_CFT:infinitesimal_generators_2} reduces to:
\begin{eqnarray}
  \partial_0 \epsilon_0 &=& \partial_1 \epsilon_1 , \\
   \qquad \qquad  \partial_0 \epsilon_1 &=& - \partial_1 \epsilon_0 .
  \end{eqnarray}
  These are the well known Cauchy-Riemann equations and thus the solutions of this equations are the holomorphic functions on the plane.}. To find an explicit solution in the case with $d \neq 2$ we start by differentiating Eq.~\eqref{appendix_CFT:infinitesimal_generators_2} with respect to $x^\sigma$:
  \begin{equation}  
   \label{appendix_CFT:infinitesimal_generators_3}
  \partial_\sigma \partial_\mu \epsilon_\nu + \partial_\sigma \partial_\nu \epsilon_\mu = \frac{2 }{d} \eta_{\mu \nu} \ \partial_\sigma \partial^\rho \epsilon_\rho(x).
\end{equation} 
We can then subtract two permutations over the indexes to get:
\begin{equation}
  \label{appendix_CFT:infinitesimal_generators_final_1}
     \left( - \eta_{\mu\nu} \partial_\sigma  +  \eta_{\mu\sigma } \partial_\nu +  \eta_{\sigma \nu} \partial_\mu \right) \frac{\partial^\rho \epsilon_\rho }{d} = \partial_\mu \partial_\nu \epsilon_\sigma .
\end{equation}
Finally, taking a second derivative with respect to $x^\sigma$ and multiplying by $\eta^{\mu\nu} $ we finally get:
\begin{equation}
   \label{appendix_CFT:infinitesimal_generators_final_2}
  \frac{(d-1)}{d} \square \partial^\sigma \epsilon_\sigma = 0.
\end{equation}
Eq.~\eqref{appendix_CFT:infinitesimal_generators_final_2} clearly implies that $\epsilon_\mu(x)$ is at most quadratic in the coordinates. As a consequence, a general solution for $\epsilon_\mu (x)$ can be expressed as:
\begin{equation}
\label{appendix_CFT:generators}
   \epsilon_\mu(x) = a_\mu + b_{\mu \nu} x^\nu + c_{\mu \nu \rho} x^\nu x^\rho ,
\end{equation}
where $a_\mu, b_{\mu \nu} , c_{\mu \nu \rho}$ are constants and $c_{\mu \nu \rho} = c_{\mu \rho \nu}$. This expression for $\epsilon_\mu(x) $, parameter of the infinitesimal transformation, is obtained by combining all the different transformations of the conformal group. To produce a systematic classification of these transformations we start by splitting $b_{\mu \nu}$ into its symmetric and antisymmetric parts:
\begin{equation}
\label{appendix_CFT:splitting_b}
  b_{\mu \nu} = \lambda \eta_{\mu \nu} + m_{\mu \nu} .
\end{equation}
Finally we can consider separately all the different contributions:
\begin{itemize}
\item The term $a_\mu$ simply induces a \textbf{translation}.\\ It is well known that the corresponding generator is the momentum $P_\mu=-i\partial_\mu $.
\item The term $ m_{\mu \nu } x^\nu$ induces a \textbf{rotation}.\\ As usual its generator is the angular momentum $M_{\mu\nu} = i(x_\mu\partial_\nu-x_\nu\partial_\mu)$
\item The term $\lambda \eta_{\mu \nu } x^\nu$ induces a \textbf{scale transformation}. \\ This transformation is generated by the dilatation operator $ D=-ix_\mu\partial_\mu$ 
\item The term $c_{\mu \nu \rho} x^\nu x^\rho$ induces a \textbf{Special Conformal Transformation (SCT)}.\\ We can substitute Eq.~\eqref{appendix_CFT:generators} into Eq.~\eqref{appendix_CFT:infinitesimal_generators_final_1} and define $c_\mu \equiv c^\rho_{ \ \rho \mu} / d$ to have a better expression for these transformations:
\begin{equation}
\label{appendix_CFT:SCT}
x^\prime_\mu = x_\mu+2(x\cdot c)x_\mu-(x\cdot x)c_\mu .  
\end{equation} 
The generator for these transformation can be expressed as:
\begin{equation}
\label{appendix_CFT:SCT_generator}
K_\mu=-i(2x_\mu x^\nu\partial_\nu-(x\cdot x)\partial_\mu) .  
\end{equation} 
\end{itemize}
To have a deeper understanding of the structure of the conformal group we can study the algebra of its generators. We can first compute the commutation relations:
\begin{equation}
\label{appendix_CFT:algebra}
\begin{aligned}
&[ D,K_{\mu}] = -iK_{\mu }, \qquad \qquad [ D,P_{\mu} ] = iP_{\mu }, \qquad \qquad [ K_{\mu },P_{\nu} ] =  2i\eta _{\mu \nu }D-2iM_{\mu \nu }, \\
&[ P_{\rho },M_{\mu \nu} ] =  i(\eta _{\rho \mu }P_{\nu }-\eta _{\rho \nu }P_{\mu }), \qquad \qquad [ K_{\mu },M_{\nu \rho} ] =  i(\eta _{\mu \nu }K_{\rho }-\eta _{\mu \rho }K_{\nu }) ,\\
& [ D,M_{\mu \nu} ] =  0,\qquad [ M_{\mu \nu },M_{\rho \sigma} ]= i(\eta _{\nu \rho }M_{\mu \sigma }+\eta _{\mu \sigma }M_{\nu \rho }-\eta _{\mu \rho }M_{\nu \sigma }-\eta _{\nu \sigma }M_{\mu \rho }).
\end{aligned}
\end{equation}
It is crucial to notice that the standard mass operator $P^\mu P_\mu$ does not commute with other generators and thus it is not a Casimir operator. In particular as consequence of the invariance of the theory under rescaling, $P^\mu P_\mu$ does not commute with $D$. It is then clear that in this framework energy does not provide a good way to label the states. Before discussing the procedure to construct CFTs, it is interesting to proceed with the study of the conformal group. In particular we can then compute the number $N$ of generators. In $d$-dimension we have $d$ translations, one dilatation, $d(d-1)/2$ rotations and $d$ special conformal transformations. Adding up all these contributions we get $N = (d+2)(d+1)/2$. Noticing that $N$ is equal to the number of possible rotations in a $(d+2)$-dimensional space, we define:
\begin{align}
    J_{\mu \nu}=M_{\mu\nu}, \qquad&\qquad J_{-p,\mu}=\frac{1}{2}(P_\mu-K_\mu),\\
    J_{-p,q+1}=D, \qquad&\qquad J_{q+1,\mu}=\frac{1}{2}(P_\mu+K_\mu).
  \end{align}
It is then clear that the generators fit in a $(d+2)\times(d+2)$ anti-symmetric matrix:
\begin{equation}
J_{ab} = \left( \begin{array}{ccc}
0 & \frac{1}{2}(P_\nu-K_\nu) & D \\
-\frac{1}{2}(P_\mu-K_\mu) & M_{\mu \nu } & \frac{1}{2}(P_\mu+K_\mu) \\
-D & -\frac{1}{2}(P_\nu+K_\nu) & 0 \end{array} \right) , 
\end{equation}
where $a,b = -p,-p+1, \dots,q,q+1$. Using the commutation relations of Eq.~\eqref{appendix_CFT:algebra} we can compute the commutator:
\begin{equation}
[J_{ab},J_{lk}]=i ( \eta_{al} J_{bk} +  \eta_{bk} J_{al}  -  \eta_{ak} J_{bl}  -  \eta_{bl} J_{ak} ),
\end{equation}
where we defined the metric $\eta_{ab} \equiv diag(-1,\dots,-1,1,\dots,1)$ with signature $(p+1,q+1)$. The generators of the conformal group thus satisfy the same algebra of the generators of $SO(p+1,q+1)$, isometry group of a pseudo-Euclidean space with signature $(p+1,q+1)$.

\section{Conformal field theories.}
\label{appendix_CFT:Conformal_field_theories}
\noindent As usual the symmetry group is used to find the proper labeling for the states of the theory. As the symmetry fixes a specific structure for the states, it also imposes constraints on the energy-momentum tensor and on the predictions of the theory. Let us consider in detail the procedure to construct the spectrum. As argued in Sec.~\ref{appendix_CFT:Conf_transf_and_group}, the mass generator does not provide a proper method to label the states of a CFT. Conversely, the dilatation operator commutes with the generator of the angular momentum and thus these operators can be simultaneously diagonalized. In the standard approach, known as \emph{radial quantization}, we label the states using $\Delta$, \emph{conformal dimension}, defined by:
\begin{equation}
\label{appendix_CFT:conformal_dimension_label}
D \left| \Delta, l \right \rangle= - i\Delta \left| \Delta, l \right \rangle.
\end{equation}
In this picture the dilatation operator is also the generator of the unitary evolution of the theory. To proceed with this construction we use $P_\mu$, generator of the momentum, and $K_\mu$, generator of SCT, as the raising and lowering operators respectively. To specify the spectrum of the theory we define a \emph{vacuum state} $\left| 0 \right >$, as the state that is annihilated by all the conformal generators. Finally it is useful to introduce the concept of \emph{primary states}. We define a state $\left| \Delta, l \right >$, eigenstate of the dilatation operator with eigenvalue $- i \Delta$, \emph{primary state} if $\left| \Delta, l \right >$ is annihilated by the lowering operator $K_\mu$. Notice that a primary state can be naturally associated with a \emph{primary operator} $\mathcal{O}_{\Delta,l}(0)$ whose action on the vacuum state is defined by:
\begin{equation}
\label{appendix_CFT:primary_operator}
\mathcal{O}_{\Delta,l}(0) \left| 0 \right \rangle= \left| \Delta, l \right \rangle.
\end{equation}
This association makes manifest the equivalence between the description in terms of states and the description in terms of operators. Notice that the primary operator has been defined at $x = 0 $. The transformation properties of $\mathcal{O}_{\Delta,l}(0)$ can be derived using the commutation relations of Eq.~\eqref{appendix_CFT:algebra}. In particular it is interesting to point out that:
\begin{equation} 
\label{appendix_CFT:translation_primary}
[P_\mu, \mathcal{O}_{\Delta,l}(x)] = −i\partial_\mu \mathcal{O}_{\Delta,l}(x).
\end{equation}
Using Eq.~\eqref{appendix_CFT:translation_primary} and Eq.~\eqref{appendix_CFT:algebra} we can derive the action of all the generators of conformal algebra on the local operator $\mathcal{O}_{\Delta,l}(x)$. Given the correspondence between states and operators this permits to completely determine the spectrum of the theory.\\ 

\subsection{n-point functions.}
\noindent As argued at the beginning of this section, conformal invariance fixes several conditions on the predictions of the theory. Let us compute some correlation functions to show these constraints explicitly. To be general let us consider a CFT in a $d$-dimensional space with signature $(p,q)$. For simplicity we consider spinless objects and thus our primary operators are simply denoted with $\mathcal{O}_{\Delta}(x)$. It is useful to point out that by definition $\mathcal{O}_{\Delta}(x)$ transforms as:
\begin{equation} 
\label{appendix_CFT:transformation_property_primary}
 \mathcal{O}_{\Delta}(x) \rightarrow   \mathcal{O}_{\Delta}(x) = \left| \frac{\partial x^\prime}{\partial x}\right|^{\Delta/d} \  \mathcal{O}^\prime_{\Delta}(x^\prime)  \ ,
\end{equation}
where $\left| \frac{\partial x^\prime}{\partial x}\right|$ denotes the Jacobian of the coordinate transformation. Let us use this remark to give an explicit derivation of the one-point and of the two-point functions.\\

\noindent
\textbf{1-point function.} We can directly use Eq.~\eqref{appendix_CFT:transformation_property_primary} to get: 
\begin{equation}
\label{appendix_CFT:one_point_function}
\langle \mathcal{O}^\prime_{\Delta}(x_0^\prime) \rangle=  \left| \frac{\partial x^\prime}{\partial x}\right|^{-\Delta/d}_{x = x_0}  \langle \mathcal{O}_{\Delta}(x_0)\rangle.
\end{equation}
The expectation value is invariant under translations. As in this case $\left| \frac{\partial x^\prime}{\partial x}\right| = 1$, this result can not depend on $x_0$. Furthermore, the 1-point function should also be invariant under scale transformations and this directly implies :
\begin{equation}
\label{appendix_CFT:one_point_function_2}
\langle \mathcal{O}_{\Delta}(x_0)\rangle = \delta_{\Delta,0}.
\end{equation}
These conditions clearly imply that the only operator with a non-vanishing 1-point function is the identity. \\

\noindent
\textbf{2-point function.}
 Using the transformation properties of Eq.~\eqref{appendix_CFT:transformation_property_primary} we have: 
\begin{equation}
\label{appendix_CFT:two_point_function}
\langle \mathcal{O}^\prime_{\Delta_1}(x_1^\prime) \mathcal{O}^\prime_{\Delta_2}(x_2^\prime)\rangle =  \left| \frac{\partial x^\prime}{\partial x}\right|^{-\Delta_1/d}_{x = x_1}  \left| \frac{\partial x^\prime}{\partial x}\right|^{-\Delta_2/d}_{x = x_2}  \langle \mathcal{O}_{\Delta_1}(x_1) \mathcal{O}_{\Delta_2}(x_2)\rangle.
\end{equation}
The invariance under translations implies that the two-point function can only depend on the difference $x_1 - x_2$. As the theory is also invariant under rotations the result can depend on $x_{12} = |x_1 - x_2|$. To lower the notation we define $F(x_{12}) \equiv \langle \mathcal{O}_{\Delta_1}(x_1) \mathcal{O}_{\Delta_2}(x_2)\rangle $. We can then proceed by imposing the invariance under dilatations:
\begin{equation}
\label{appendix_CFT:two_point_function_dilatation}
F(x_{12}) = \lambda^{\Delta_1 + \Delta_2} F(\lambda x_{12}) \qquad \longrightarrow \qquad F( x_{12})=\frac{C}{|x_1 - x_2|^{ \Delta_1 + \Delta_2}},
\end{equation}
where $C$ is a constant factor depending of $d$ and $\Delta$. The next step consists in imposing the invariance under SCT. Using Eq.~\eqref{appendix_CFT:SCT} we can show that for these transformations the Jacobian read:
\begin{equation}
\label{appendix_CFT:SCT_jacobian} 
\left| \frac{\partial x^\prime}{\partial x}\right| = \left(1 - 2c \cdot x+c^2 x^2 \right)^{-d}.
\end{equation}
For $x_1$ and $x_2$ we define $\gamma_{i} = \left(1 - 2c \cdot x_{i}+c^2 x_{}^2 \right)^{-d} $ where $i = 1,2$ respectively. Moreover it is possible to show that under a SCT we have:
\begin{equation}
\label{appendix_CFT:SCT_modulus}
|x_1 - x_2|^{ \Delta_1 + \Delta_2}  \qquad\longrightarrow  \qquad|x_1^\prime - x_2^\prime|^{ \Delta_1 + \Delta_2} = \frac{|x_1 - x_2|^{ \Delta_1 + \Delta_2}}{(\gamma_1\gamma_2)^{ (\Delta_1 + \Delta_2)/2 } }
\end{equation}
Finally we can use Eq.~\eqref{appendix_CFT:SCT_jacobian} and Eq.~\eqref{appendix_CFT:SCT_modulus} to appreciate the action of a SCT on the expression for the two-point function given by Eq.~\eqref{appendix_CFT:two_point_function_dilatation} :
\begin{equation}
\label{appendix_CFT:two_point_function_SCT}
\frac{C}{|x_1 - x_2|^{ \Delta_1 + \Delta_2}} = \frac{ (\gamma_1\gamma_2)^{ (\Delta_1 + \Delta_2)/2 } }{\gamma_1^{\Delta_1} \gamma_2^{\Delta_2}}   \frac{C}{|x_1 - x_2|^{ \Delta_1 + \Delta_2}} .
\end{equation}
As this equation can only be satisfied for $\Delta_1 = \Delta_2$ we can conclude that:
\begin{equation}
\label{appendix_CFT:CFT_two_point_function}
\langle \mathcal{O}_{\Delta_1}(x_1) \mathcal{O}_{\Delta_2}(x_2)\rangle \  = \frac{C}{|x_1 - x_2|^{\Delta_1 + \Delta_2}} \delta_{\Delta_1,  \Delta_2}
\end{equation}

\noindent
\textbf{3-point function.}
The case of the three-point function can be treated similarly to the case of the two-point function. Again we can use the symmetries of the theory to get:
\begin{equation}
\label{appendix_CFT:CFT_three_point_function}
\langle \mathcal{O}_{\Delta_1}(x_1) \mathcal{O}_{\Delta_2}(x_2) \mathcal{O}_{\Delta_3}(x_3)\rangle \  = \frac{D}{x_{12}^{\ \ \Delta_1 + \Delta_2- \Delta_3} x_{23}^{ \ \ \Delta_3 + \Delta_2- \Delta_1} x_{31}^{\ \ \Delta_1 + \Delta_3- \Delta_2}},
\end{equation}
where we defined $x_{23}$ and $x_{31}$ we in analogy to $x_{12}$.\\

\noindent
\textbf{$n$-point function.} In the general case with $3 < n$, the $n$-point function is not completely fixed by conformal symmetry. Every case then require an independent treatment.

\subsection{Stress-energy tensor.}
Conformal invariance also has implications on the shape of the energy-momentum tensor of the theory. As usual we impose invariance under a conformal transformation $x_\mu \rightarrow x_\mu(x) + \epsilon_\mu(x)$. Noether's theorem implies the existence of a conserved current:
\begin{equation}
\label{appendix_CFT:Noether_CFT}
j_\mu = T_{\mu\nu}\epsilon^\nu, 
\end{equation}
where the symmetric tensor $T_{\mu\nu}$ is called energy-momentum tensor. As a first step let us consider the case of a constant $\epsilon_\mu$. As $j_\mu $ is conserved, it is trivial to show that $ \partial^\mu T_{\mu\nu} = 0$. In the case where the parameter of the transformation $\epsilon_\mu (x)$, depends on the spacetime coordinate we get: 
\begin{equation}
\label{appendix_CFT:Noether_conservation}
0 = \partial^\mu j_\mu = T_{\mu\nu} \partial^\mu \epsilon^\nu = \frac{ T_{\mu\nu}}{2} \left( \partial^\mu \epsilon^\nu + \partial^\nu \epsilon^\mu \right) = \frac{ T_\mu^{\ \mu}}{d} \left( \partial^\sigma \epsilon_\sigma \right)
\end{equation}
where we used the symmetry of the energy-momentum tensor and Eq.~\eqref{appendix_CFT:infinitesimal_generators}. As this equation holds for every $\epsilon_\mu (x)$ we conclude that the energy-momentum tensor of a CFT is traceless. It is important to stress that this conclusion holds for classical theories. In particular QM introduces the so called "conformal anomaly" that breaks conformal symmetry and the stress-energy tensor acquires a non-vanishing trace.
{\large \par}}

\bibliographystyle{hunsrt}
\bibliography{Thesis}


\end{document}